# SEARCH FOR EXTRA-TERRESTRIAL PLANETS: THE DARWIN MISSION - TARGET STARS AND ARRAY ARCHITECTURES

---

Dissertation

zur Erlangung des akademischen Grades

DOCTOR RERUM NATURALIUM

an der

NATURWISSENSCHAFTLICHEN UNIVERSITAET

der KARL-FRANZENS UNIVERSITAET GRAZ

INSTITUT FUER GEOPHYSIK, ASTROPHYSIK UND METEOROLOGIE

vorgelegt von

Mag. Dipl. Ing. Lisa Kaltenegger

begutachtet von

Univ-Prof. Dr. Arnold Hanslmeier

und

Univ.-Prof. Dr. Theo Neger

*To my parents*
*who taught me curiosity about this world and to follow my dreams*

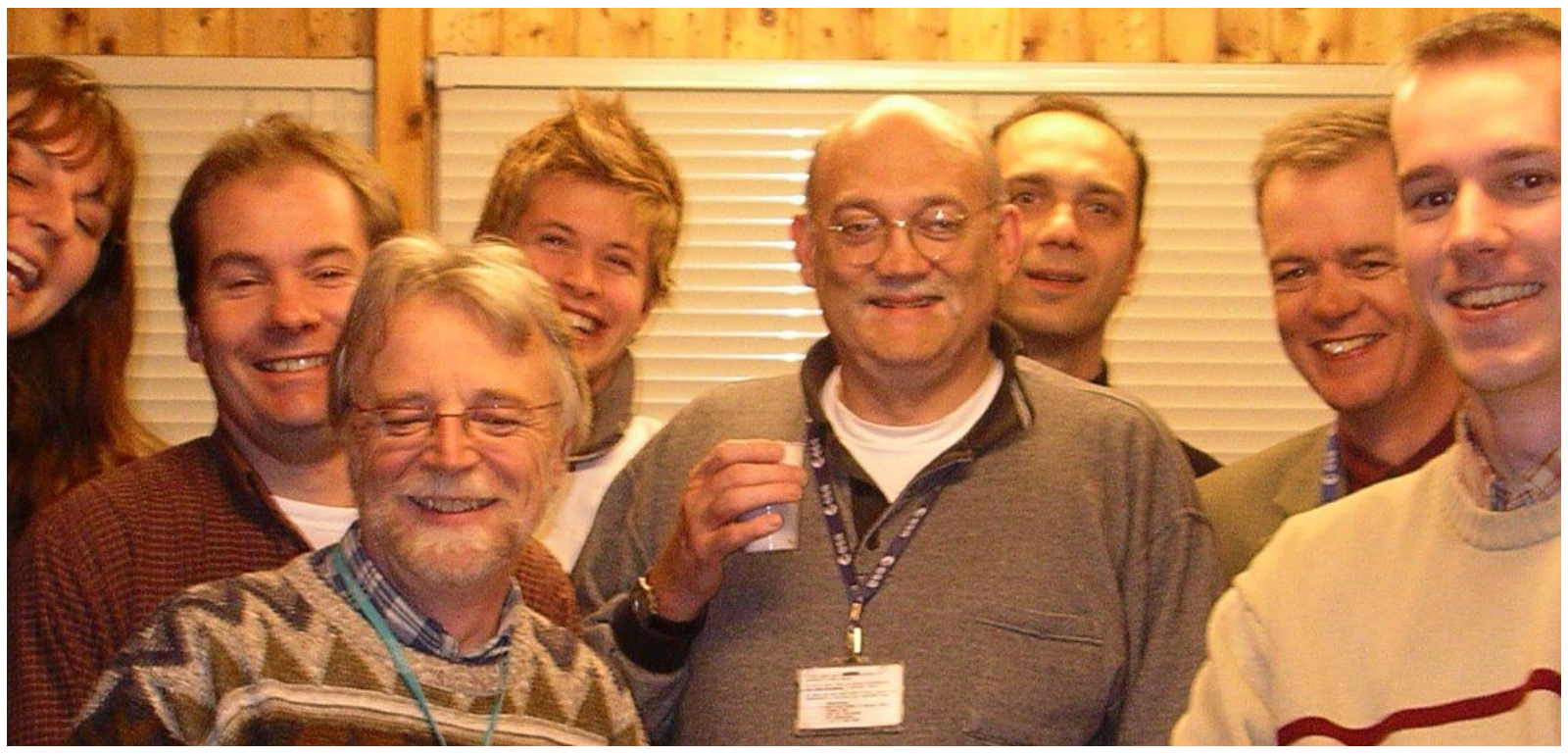

*most of the DARWIN team: Lisa, Roland, Lothar, Mikael, Malcolm, Luigi, Anders, Olivier*

***Special Thanks to***

*Anders and Malcolm for their enthusiasm, constructive comments, motivations, time to discuss and iterate ideas and their support*

*The DARWIN team for the enthusiasm and spirit even if times got too busy to sleep much*

*Prof Hanslmeier for his interest in new fields of research, his support and for giving me the possibility to undertake my research on this fascinating topic*

*Frank Selsis for providing comments and teaching me much about atmospheres*

*Oliver Absil, Wes Traub, Adam Burrows and Jim Kasting for interesting discussions*

*Mikael and the Dc0 corridor for filling work in the small office with fun & laughter*

*Christian and Luigi for always having 5min to discuss*

*my friends for locating me anywhere in the world to pass by for a cup of coffee & a talk wherever I happen to be*

*Maria for getting this document printed 1000km away & for sharing her breakfast*

*Kotska and Rachel for always having a cup of tea for me, whatever the hour*

*Iris and Claudia for making me smile even when I did not get much sleep, a smile that always came with cyber-caffeine*

*the Dekanat at the KF-University Graz for accepting unconventional forms of transmission of documents and for their friendly information service*

List of acronyms

| | |
|---|---|
| AO | Adaptive optics |
| AU | Astronomical unit |
| BB | Black body |
| BC | Beam combining |
| BCS | Beam combining scheme |
| BD | Brightness distribution |
| CHZ | Contiuously habitable zone |
| DNA | Desoxy-ribonucleine acid |
| EGP | Extrasolar giant planet |
| ESA | European space agency |
| FOV | Field of View |
| FT | Fourier transform |
| FWHM | Full Width Half Minimum |
| GENIE | Ground-based European Nulling Interferometer Experiment |
| Gpa | Giga-pascal |
| Gyr | Giga-year |
| Hd | Habitable distance |
| HST | Hubble space telescope |
| HZ | Habitable zone |
| IAC | Institute of Astrophysics of the Canary Islands |
| IO | Integrated optics |
| IR | Infrared |
| LZ | Local zodiacal |
| L2 | Second Lagrangian point |
| M | Mass |
| MBI | Multi beam injection |
| MM | Modulation Map |
| MMZ | Modified Mach Zehnder |
| Mpa | Mega-Pascal |
| Myr | Mega-year |
| NIR | Near infrared |
| NM | Nulled output map |
| OPD | Optical path delay |
| PAL | Present atmospheric level |
| Pc | Parsec |
| PSF | Point Spread Function |
| RNA | Ribonucleine acid |
| SNR | Signal to noise ratio |
| SWG | Science working group |
| T | Transmission factor |
| TM | Transmission maps |
| TOA | Time of arrival |
| TPF | Terrestrial planet finder |
| VLTI | Very large telescope Interferometer |
| VRE | Vegetation red edge |
| WFE | Wave front error |

## ABSTRACT

The DARWIN mission is an Infrared free flying interferometer mission based on the new technique of nulling interferometry. Its main objective is to detect and characterize other Earth-like planets, analyze the composition of their atmospheres and their capability to sustain life, as we know it. DARWIN is currently in definition phase.

This PhD work that has been undertaken within the DARWIN team at the European Space Agency (ESA) addresses two crucial aspects of the mission. Firstly, a DARWIN target star list has been established that includes characteristics of the target star sample that will be critical for final mission design, such as, luminosity, distance, spectral classification, stellar variability, multiplicity, location and radius of the star. Constrains were applied as set by planet evolution theory and mission architecture. The catalogue contains nearby stars that might harbor planets that are potentially habitable to complex life. On the basis of theoretical studies, the angular separations of potential habitable planets from their parent stars for the target systems have been established. The resulting target list allows to model realistic observation scenarios for nulling interferometry and translates into architectural constraints for the mission.

Secondly, a number of alternative mission architectures have been evaluated on the basis of interferometer response as a function of wavelength, achievable modulation efficiency, number of telescopes and starlight rejection capabilities. The study has shown that the core mission goals should be achievable with a lower level of complexity as compared to the current baseline configuration.

# 1 Extrasolar Planet Search, formation of planets and characteristics

*"There cannot be more worlds than one (Artmowicz 1999). "*

*Really?*

Our solar system exists - this is an irrefutable fact. Our Galaxy is about $10^{10}$ years old, roughly 100000 light years across and on average has a star every light-year (Annis, 1999). Billions of galaxies exist in the universe, each containing billions of stars. Given the size of our universe, the odds are excellent that our solar system is not unique. But as long as we do not know what the conditions for the formation of planetary systems like ours are the laws of probability are difficult to use to claim the existence of millions of such systems.

Other solar systems may be very different from ours. Until now the results of the search for extra-solar planets are biased by observation techniques. Given the sensitivity of the current techniques, our observations show a selection effect for high mass companions, close to their parent star. When the sensitivity of search methods increase we will find out if extrasolar planets really are different in structure, mass and distribution form the planets of our solar system.

*"The first discoveries have been full of surprises and we should be prepared for the unexpected (Schneider 1998)".*

We do not have a clear understanding yet, but we have the techniques to investigate.

## 1.1 Introduction

*"Our Sun is a very common and ordinary star. There is really nothing to distinguish it from the myriad of other similar stars in this region of the Galaxy. Yet, the Sun possesses a marvelous system of nine diverse planets."(Marcy, 1999)*

Are there any other planetary systems like ours out there? Until 1991 we just knew one planetary system, our own. Now we have 116 detected planetary candidates and a magnificent scope of different planetary systems to imagine.

The first planetary system found was a pulsar planetary system with 3 Earth-size planetary bodies, discovered in 1991 by Wolszcan (Wolszcan, 1997) using precise pulsar timing. Until now it still remains unclear how planets can form around a neutron star. Unlike this, the detection of a giant planet in 1995 by Mayor and Queloz (Mayor, 1995) around a sun-like stars fits into a now modified planetary evolution theory. The detection proved that we have the technology to start answering the question whether other planetary systems like our own exist. Until 1999 the detection was not unambiguous, as the planetary candidates had only been detected with one method, radial velocity search. In 1999 a planetary candidate was confirmed with a different detection method, transit detection (Charbonneau, 2001). The detection clarified that these objects are real, not mimicked by unknown variations in the host stars as well as that their composition is that of a giant planet.

The thesis provides an overview of and a connection between the different aspects included in extrasolar planet search. The main focus is DARWIN, a proposed ESA mission to detect extrasolar planets. DARWIN is a space based IR interferometer, consisting of a number of free flyers to be launched after 2014. We investigate different mission architectures, establish the characteristics from its target star catalogue as well as simulations of observation scenarios. The search for life on extrasolar planets with the DARWIN mission concept is based on the assumption that one can screen extrasolar planets for habitability spectroscopicaly. In that context DARWIN also has the unique capability to investigate physical properties and composition of a broader diversity of planets to understand the formation of planets and interpret potential biosignature.

## 1.2 Stellar and sub-stellar objects

Evidence for the existence of planetary systems around other stars become accessible via direct observation of extended (R=100-1000 Astronomical Units (AU)) circumstellar dust disks around stars, like the disk around β-Pictoris where a tilt between inner and outer disk can be explained by the formation of a planet. The planetary companion was inferred from warpage found in the disk thought to be caused by the gravitation of a planetary object (Burrows, 1999). The disks are detected via their dust radiation at far infrared wavelengths (12µm - 1.3mm). The detection of visible stellar companions and the effects of invisible companions on the primary star also provide evidence.

The difference between stars and sub-stellar objects is the ability of stellar bodies to ignite stable thermo-nuclear fusion in the interior at the end of the contracting phase. This happens at a central temperature of about $10^6$ K. Here the thermal pressure is

$$P_{th} = nkT \tag{1.2.1}$$

which is maintained by the nuclear reaction. It balances the gravitational pressure due to the weight given by equation (1.2.2) (Sicardy, 1998):

$$P_{grav} = \frac{GM^2}{R^4} \tag{1.2.2}$$

*G* and *k* are the Gravitational and Boltzmann constants, *n* the number density, *T* the central temperature and *R* the radius of the body. At high densities the degeneracy pressure due to the electrons becomes important. Electrons have half-integral spins and must obey the Pauli exclusion principle that requires that they fill up the lowest available energy state and are forbidden to occupy identical quantum energy states. Electrons in the higher energy levels contribute to degeneracy pressure because they cannot be forced into the filled lower states (Oppenheimer, 1998). The degeneracy pressure does not depend on the temperature. It is given by

$$P_{\deg} \propto n^{5/3} \propto \frac{M^{5/3}}{R^5} \tag{1.2.3}$$

It scales with $\rho^{5/3}$ and becomes important when it approximately equals the ideal gas pressure $\rho T$ (Oppenheimer, 1998). Our Sun has a density of about 1.5g/cm$^3$, the central density is about 150g/cm$^3$. The degeneracy pressure catches up with the gravitational pull at some *R* as the dependences are $R^{-5}$ and $R^{-4}$. From the Virial theorem $T \propto M/R$ we can express the temperature at that point:

$$P_{deg} = P_{grav} \tag{1.2.4}$$

$$R \sim M^{-1/3} \tag{1.2.5}$$

$$T \propto \frac{M}{R} \propto M^{4/3} \tag{1.2.6}$$

If $T \geq 10^7$ K the body ignites the nuclear reaction before the degeneracy pressure catches up with the thermal pressure. The thermal pressure $P_{th}$ then maintains the star. If not, the body stops its contraction eventually before the nuclear reaction starts (Baraffe, 1997). This takes longer than $10^{10}$ years for a Jupiter mass object (Burrows, 1999). As $T_{crit} \propto M^{4/3}$, the limit is only mass dependent and classically estimated to: $M_{limit} \sim 0.08 M_{solar} \sim 80 M_J$ (Sicardy, 1998) where *solar* refers to the Sun and *J* refers to Jupiter. Absorption lines of chemical elements like methane and water that exist at temperatures lower than about 1500K are indicators for sub-stellar objects. The mass luminosity relation for bodies with masses of about 10% of the solar mass is given by

$$L \propto M^{2.7} \tag{1.2.7}$$

The reason for the drastic change of this relation is due to the different opacity sources at work (Massey, 1999).

### 1.2.1 Brown Dwarfs and extrasolar planets

Sub-stellar objects are divided into Brown Dwarfs (BD) between 80 $M_J$ and 13 $M_J$, and planets with masses below 13 $M_J$ (Rebolo, 1999). The definition is somehow arbitrary and chosen on base of the ability of stable fusion of different elements. Fusion influences the evolution of sub-stellar objects drastically. An alternative definition of planetary objects is to define them as bodies that cannot fuse chemical elements. Additional planets are thought to be objects that only exist in orbit around their parent stars. A very young Jupiter sized object has recently been found free floating in space (Rebolo, 2003) but it is not clear yet whether it is a remnant of a disrupted BD formation or a body formed in similar ways like our giant planets. The age determination is critical for these objects to fit models to the detected brightness and thus constrain the mass. Most planetary evolution models are not valid for very young ages of planets, thus results have to be viewed critically.

Planets are sometimes also defined through their formation mechanism that is thought to be different from BD. Objects that are born in the interstellar medium in a manner similar to the processes by which stars form are named BD while objects that form in protostellar disks by processes that may differ from those of star formation are called planets. BD are defined as objects with masses smaller than the minimum mass required for sustained thermonuclear fusion of hydrogen, but larger than the minimum mass required for the fusion of deuterium and Li. Deuterium is a more fragile nuclear species than Li, it is burnt in sub-stellar-mass objects with masses above 12-15 $M_J$. The light-hydrogen burning is eventually extinguished as it cools. Surface energy losses are never completely compensated by thermonuclear burning in the core; surface and central temperature are never stabilized. Independent of the character of the pressure support, a BD contracts along a Hayashi track and is convective. BDs never stop decreasing their luminosity.

## 1.3 Evolution of substellar mass objects

Evolution models for substellar objects are very important to be able to predict the spectra of extrasolar planetary objects and thus to optimize the detection probability as well as finally interpret the detected signal. Burrows et al. (Burrows, 2001) and Baraf et al. (Baraf, 2003) summarize evolutionary models of substellar mass objects.

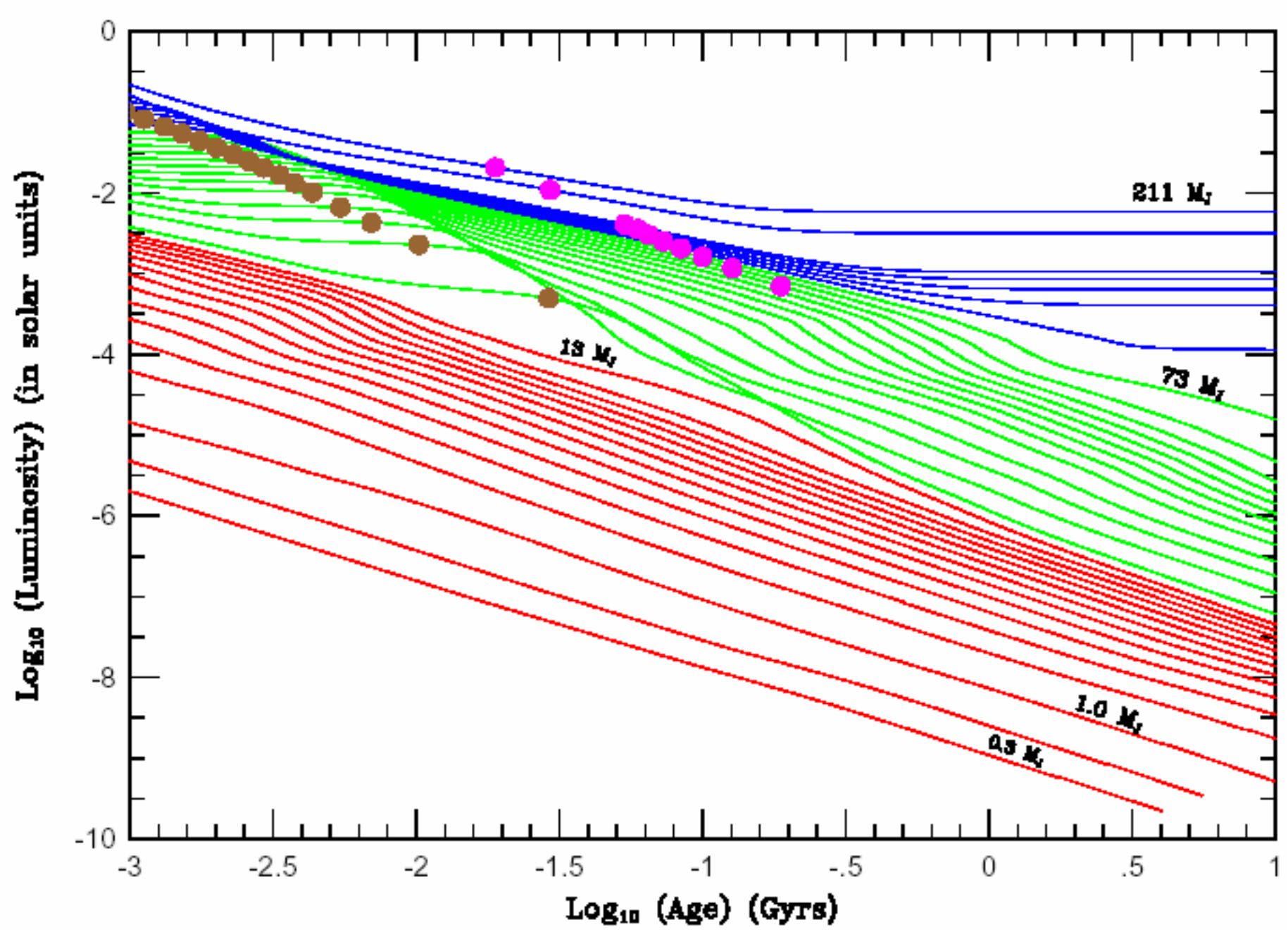

**Fig. 1: Evolution and luminosity of isolated solar metallicity red dwarf stars and sub-stellar objects versus time in years after formation (Burrows, 2001)**

Stars, BDs and planets are shown as blue, green and red curves in fig.1 to fig.3 respectively. In this model BD are defined as bodies that burn deuterium but do not light hydrogen and planets as bodies that do not burn more than 50% of deuterium (Burrows, 1998C). The lowest three curves correspond to the mass of Saturn, $0.5M_J$ and the mass of Jupiter respectively. The early luminosity plateaus in fig.1 are due to deuterium burning given an initial fraction of D of $2\ 10^{-5}$. A deuterium abundance of up to $3\ 10^{-5}$ does not lead to a Deuterium-burning main sequence (Burrows, 2001).

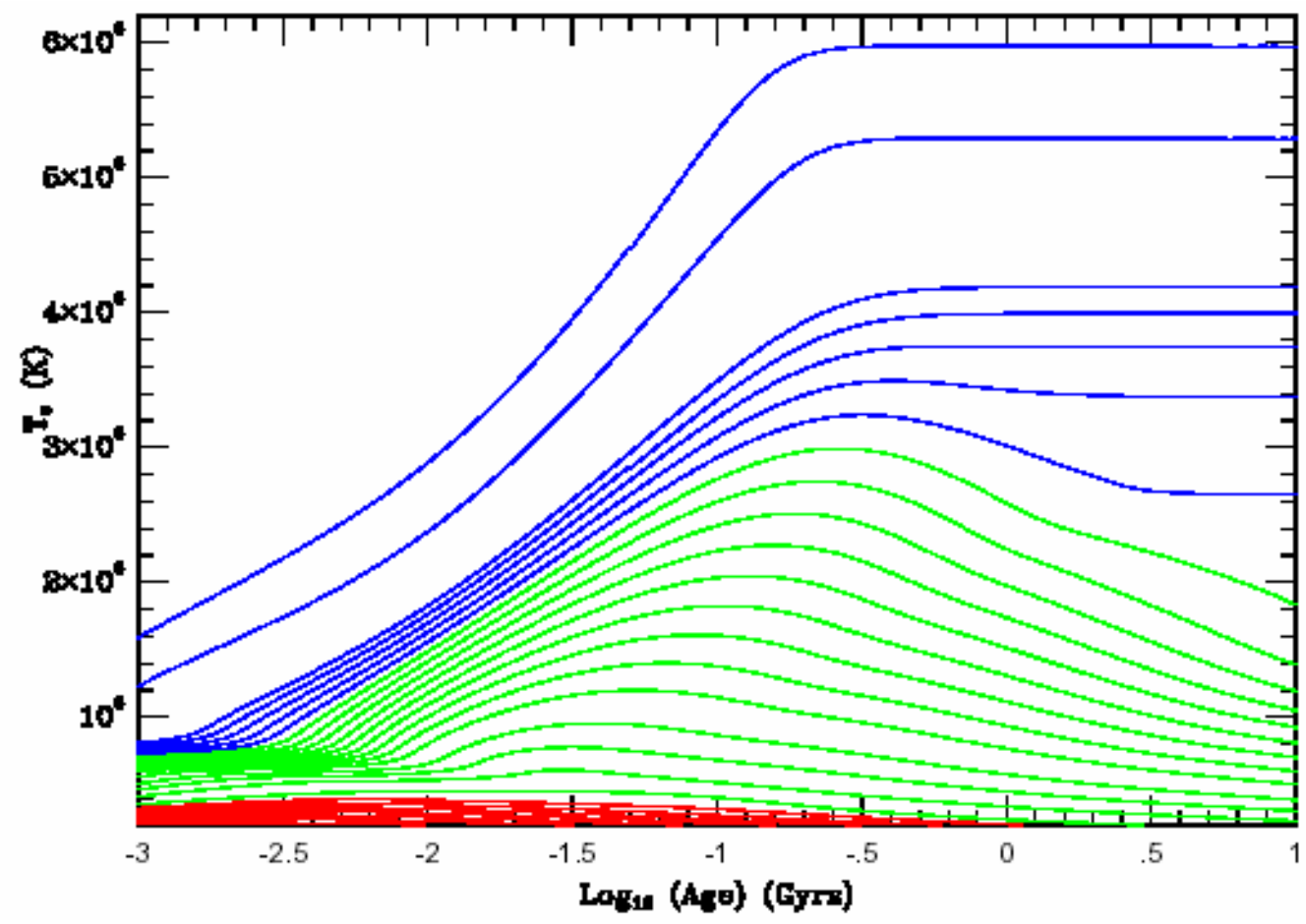

**Fig. 2: Temperature of substellar mass objects vs. the log of age in Gyr (Burrows, 2001)**

Fig.2 shows the evolution of temperature with age for objects with masses between 211 and $0.3M_J$. The late-time cooling phases for substellar objects follow approximate power laws. Fig.3 shows the evolution of radius with age and its non-monotonic dependence on mass. Early in the evolution, the radius is always a decreasing function of age. However at later times the dependence of mass upon radius inverts, with the less massive substellar-mass objects having the larger radii.

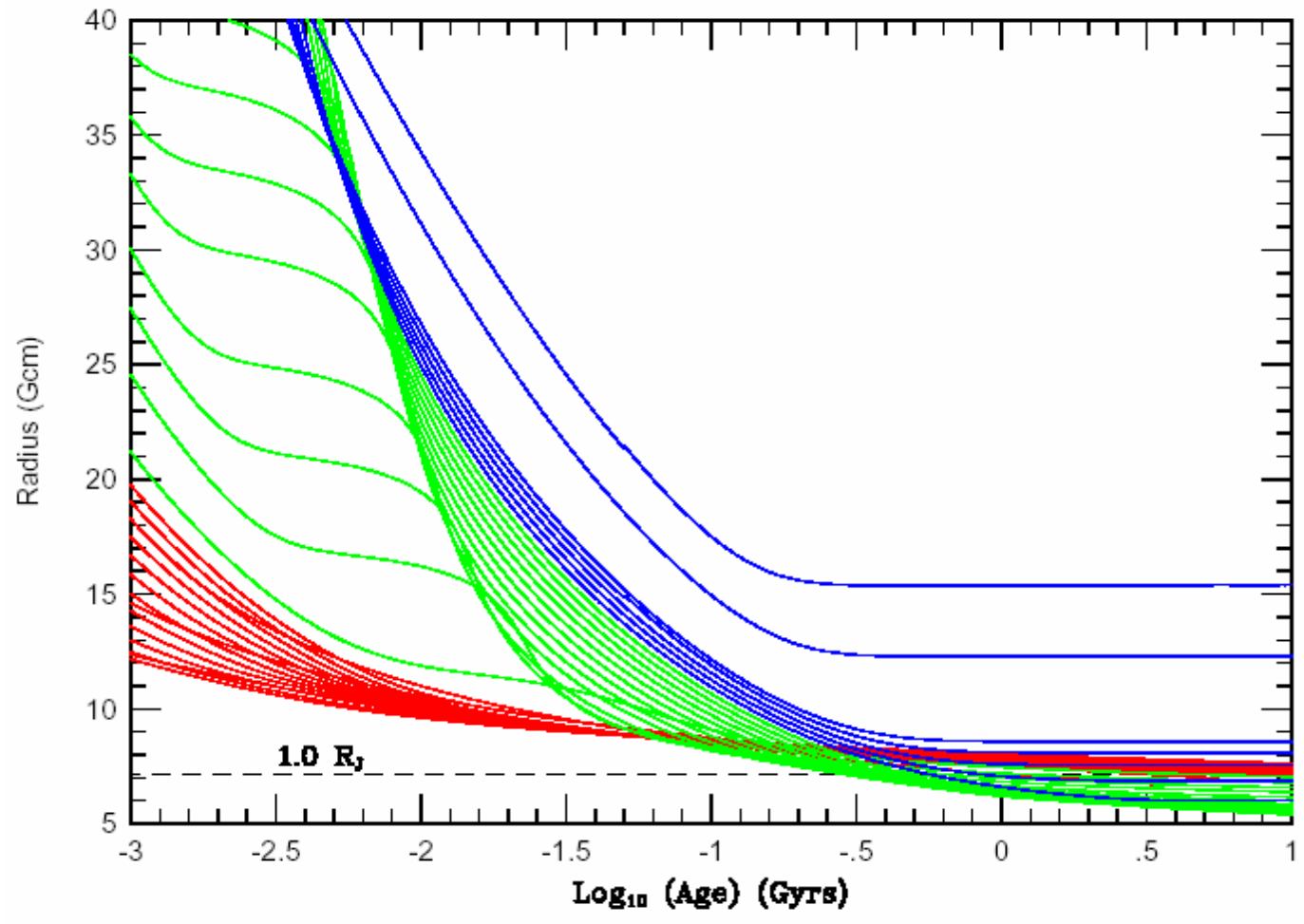

**Fig. 3: Radius in units of $10^9$cm of substellar mass objects vs. the log of age in Gyr (Burrows, 2001)**

This effect is very important to determine the radius of a detected planet from its luminosity without knowing the exact age of the object. For a cold substellar object, the peak radius is at a mass of about $4M_J$ but for a broad range of masses from $0.3M_J$ to $80M_J$, the older radii are independent of mass to within about 30%. It is the consequence of two competing effects in the equation of state: The Coulomb effect that would set a fixed density and interparticle distance scale at about 1Å leading to the relation: $R \propto M^{1/3}$, and the electron degeneracy effect leading to the classic relationship for low mass BDs: $R \propto M^{-1/3}$. The two effects render the radius about constant over roughly two orders of magnitude in mass near the radius of Jupiter.

As the mass ranges from $1M_J$ to $80M_J$ along the full cold BD and Extrasolar Giant Planet (EGP) branch, the radius changes by no more than 50% and is always near $R_J$=8.26 $10^7$m ~ $0.12R_{solar}$ allowing a rough estimate of the EGP radius for calculations over a wide mass range. Therefore the temperature is a good indicator of their luminosity.

### 1.3.1 Atmosphere models and spectroscopic characteristics

Bumps, seen in the cooling curves in fig.1 of objects from 0.03 $M_{solar}$ to 0.08 $M_{solar}$, between $10^{-4}$ $L_{solar}$ and $10^{-3}$ $L_{solar}$ and between $10^8$ to $10^9$ years, are due to silicate and iron grain formation for temperatures between 2500K and 1300K. Grain and cloud models are problematic and not yet fully understood. Important grains are perovskite ($CaTiO_3$) for high mass BD and various Ti oxides for lower mass objects. They do not regulate the opacity itself in the optical to near-IR region, but rob the photosphere of its TiO (Kirkpatrick 1998). Therefore the strong TiO bands seen in higher temperature objects vanish.

Methane and lithium lines in its spectrum can determine the sub-stellar nature of an object. The presence of Li at the surface of a throughout convective object can be used to set a lower limit of the internal temperature of a stellar object. It can be observed as Li I at 670.8 nm. Li burns very quickly at temperatures of about 2.5 million degrees (Liebert, 1998). Its survival means that hydrogen burning is not taking place or that the object is too young to have burnt its Li. An object with mass below 65 $M_J$ should preserve its initial Li content for its entire lifetime (Rebolo, 1998B). Clouds or dust on the surface of low temperature objects add uncertainty to the expected $T_{crit}$ and $L_{bol}$ between stars and sub-stellar objects. Including dust in atmospheric models will influence the minimum mass for H and Li burning, due to the condensation of grains and their opacity. For surface temperatures less than 1600K the Li may also condense into clouds (Burrows, 1998B). The larger the gravity is, the larger the pressure that favors the formation of grains at cool temperatures. For older BD this could limit the validity of the Li-test. In young star formation regions, even stars can show Li. For clusters, the Li test can determine the luminosity and temperature of the sub-stellar boundary (Basri, 1998).

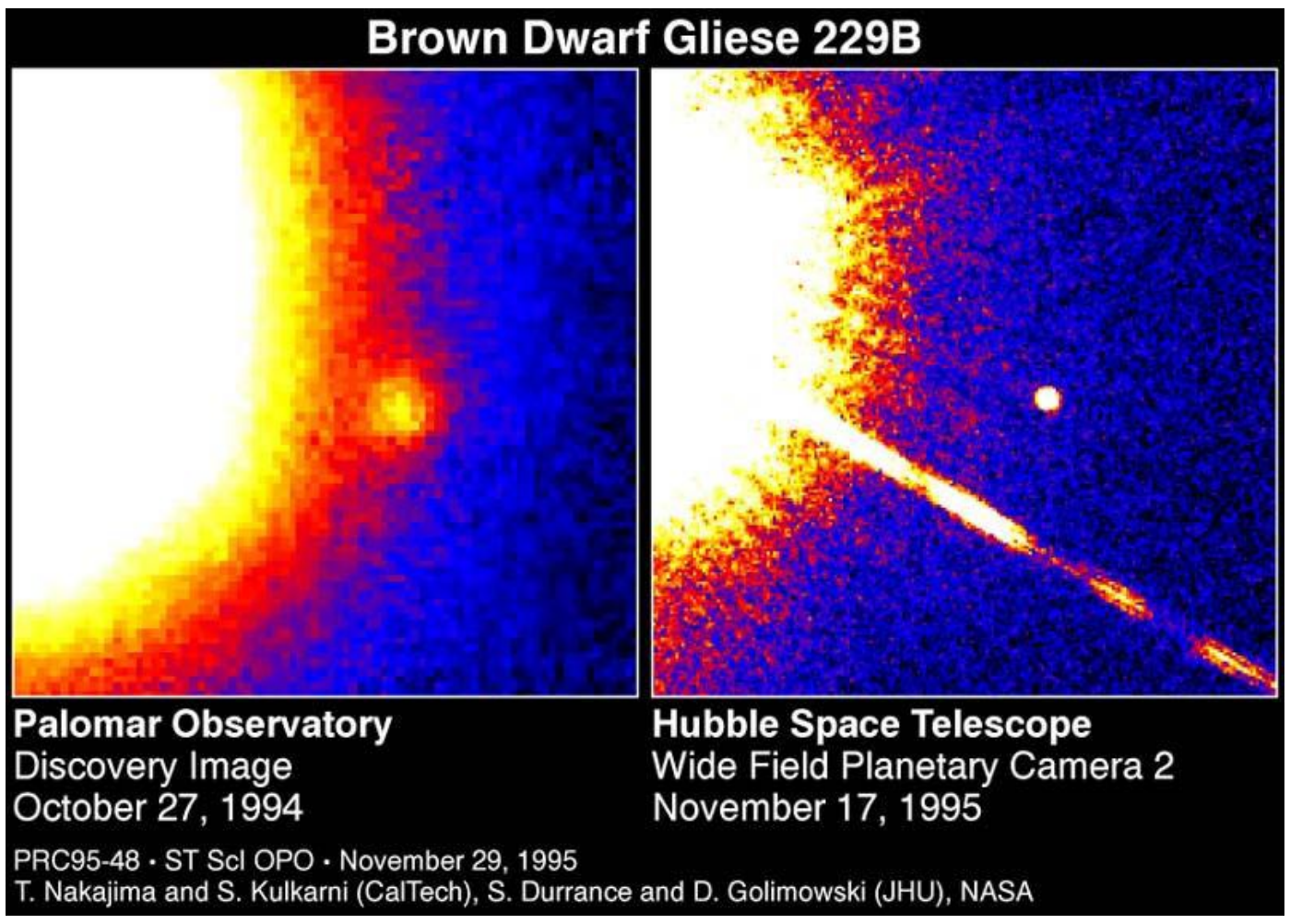

**Fig. 4: Gl229B, the first detected BD (http://www.stsci.edu/hst, 1995)**

Nakajima et al. (Nakajima, 1995) showed that the star Gliese 229 has a companion with a luminosity less than $10^{-5}$ $L_{solar}$ (Oppenheimer, 1998), a BD. It orbits its parent star at a projected distance of about 44AU and has approximately 20 to 50 times the mass of Jupiter (http://www.jywinc.com, 1998). Its spectrum looks remarkably like Jupiter's with major features of methane and water.

Methane absorption bands also dominate the 1.0-2.5 µm spectral region of Titan. At 0.89 µm, 1.6-1.8 µm, 2.1 µm and 2.2 µm Gl229 B shows these bands as well, confirming its BD nature (Oppenheimer, 1995). Fig.4 shows a Hubble Space Telescope image of Gliese 229 (left) and its substellar companion (point source right). The diagonal line is a diffraction spike produced by scattered starlight in the telescope's optical system. An Earth-like planet would be 50 times closer to the primary star center and 1.000.000 times fainter, far from detectable in this set-up.

The search for BD in nearby young star formation regions led to the detection of luminous BDs, as they are young objects. These objects could be benchmark objects for the proposed GENIE mission that will demonstrate nulling interferometry on the ground at the VLTI and investigate DARWIN targets. To demonstrate its capability test cases have to be established to validate the instrument. Detecting a young BD can be used as such a test case. The young embedded cluster candidates are both warmer and more luminous. The disadvantage of detecting BD in nearby young star formation regions is that one has to cope with dust extinction at infrared wavelengths (Liebert, 1998). Searching for companions has one main drawback, their proximity to a bright parent star. That makes spectroscopy and photometry hard. On the other hand if a sub-stellar object is gravitationally bound to another object, a direct dynamic measurement of its mass is possible (Hawkins, 1998). The measurements may become extremely difficult for large separations, e.g. for Gl229 as the projected separation is about 44AU. In the absence of a direct way to measure the masses of these companions the spectroscopic tests to determine the mass become crucial.

### 1.3.2 Jupiter and Gl229 B

The benchmark object Gl229 B, a companion to the M1V star Gl229 was discovered as a result of an optical chronograph and infrared direct imaging survey of nearby stars. The primary star is sufficient dim and the BD lies at a large distance so the spectrum of the companion is seen with minimum reflected light from the primary. In the optical the presence of a deep water-band at 0.925 µm and the drastic increase of flux toward the near infrared down to 1 µm show. The sharp absorption at 825 nm and 894 nm can be identified as the low excitation ground state transition of CsI. From 1 µm to 2.5 µm water vapor bands and methane bands dominate the spectrum. Carbon monoxide exists in non-equilibrium abundance. Using grainless model atmospheres reproduces the spectroscopic properties in the near infrared well and confirms the presence of methane bands. These models lead to an effective temperature of 900-1000K and a mass of 30-50$M_J$. Chemical equilibrium implies that grain formation must occur in such cool photospheres but none of the predicted effects is shown in Gl299 B (Oppenheimer, 1998). This suggests that dust clouds may have formed below the photosphere.

For objects with $T_{eff} \leq 1300$ K, the major opacity sources are $NH_3$, $CH_4$, $H_2O$ and $H_2$. Below 400K water clouds form at or above the photosphere, below 200K ammonia clouds form as seen in Jupiter. The photosphere is defined here as the region where $T=T_{eff}$. The appearance of water clouds brightens the planet in reflected red and IR light. At about 1500 - 2000K the standard refractories as iron, spinel and silicate condense out into grain clouds, which lower the $T_{eff}$ and luminosity by their large opacity (Burrows, 1998B). Cloud decks of many different compositions at many different temperature levels are expected. The four giant planets in our solar system have surface reflectivity influenced to varying extend by a combination of cloud and gas opacity sources. Scattered light from the primary star and thermal emission of both absorbed stellar light and the planet's own internal energy contribute to a planet's spectrum. Planets reflect best near 0.5µm where Rayleigh scattering dominates the reflected flux. The J, H and M bands are the primary bands to search for cold sub-stellar objects. As consequence of the increase of atmospheric pressure with decreasing $T_{eff}$, the anomal blue J-K and H-K colors get bluer not redder. The suppression of K by $H_2$ and $CH_4$ features is largely responsible for this anomal blueward trend with decreasing mass and $T_{eff}$. Another idea to determine the atmospheric signature of extrasolar planets is to look for the dissociation and ionization products of the escaping molecules of the planet that partially occult the star, if its orbital inclination is close to 90 degree. Planets close to their parent stars might show evaporation due to thermal escape, non-thermal evaporation by the stellar wind and UV flux and gravitational transfer to the massive companion (Coustenis, 1997).

### 1.3.3 Extrasolar Giant Planets

We review atmospheric models for EGP here in anticipation of their direct detection from the ground and from space. Spectral measurements are the key to unlocking the structural and atmospheric characteristics of EGPs, as well as to determining the true differences between giant planets and BDs.

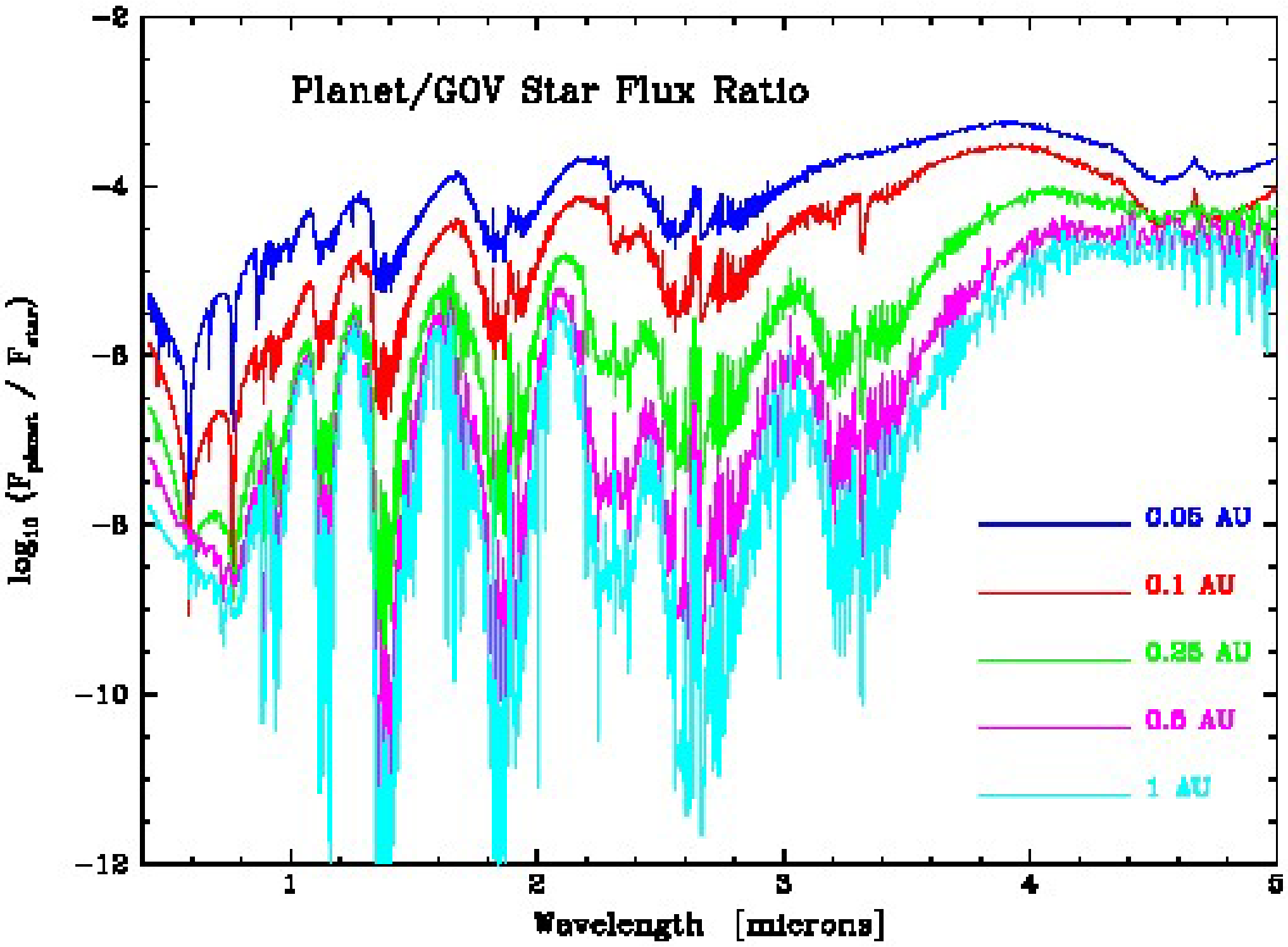

**Fig. 5: Theoretical planet/star flux ratio of self consistent EGP reflection spectra including heating by stellar irradiation from 0.4µm to 5.0µm with orbital distance from 0.05 to 1.0AU around a G0 V star. This set of models does not include clouds (Burrows, 2003)**

Fig.5 shows the theoretical planet/star flux ratio of self-consistent EGP reflection spectra including heating by stellar irradiation from 0.4 µm to 5.0 µm with orbital distance from 0.05 to 1.0AU around a G0 V star. This set of models does not include clouds (Sudarsky 2002) that are suspected to have a big influence on the reflected flux of a planet. Rayleigh scattering at short wavelengths, methane features for the more distant (hence, cooler) objects, Na-D and K I features at 0.589 µm and 0.77 µm, respectively, and water (steam) features can be seen. Without Rayleigh scattering, the flux shortward of 0.6$\mu$m would be 3-6 orders of magnitude lower. The results of atmospheric simulations are very important for calculations of the integration time needed for planetary spectroscopy. Depending on the planetary flux the mission lifetime of a space mission that searches for Earth-like extrasolar planets like DARWIN can be optimized to investigate a significant sample of possible host stars.

For precursor missions from the space (a proposed SMART3) or from the ground (GENIE) the high reflectance of EGP is very interesting as it should allow investigation of EGP atmosphere before the space-interferometers are ready to be launched, what gives us an excellent opportunity to investigate the atmospheric models and adjust any differing parameters as well as understand EGP formation and evolution. Burrows et al. calculated the expected flux of the closest EGP systems. Fig.8 shows the results for ε eridani b (3.3AU). The star-planet angular separation for this system is about 1 arcsec. The fluxes at longer wavelength in the mid-IR are expected to be even higher (Burrows, 2003).

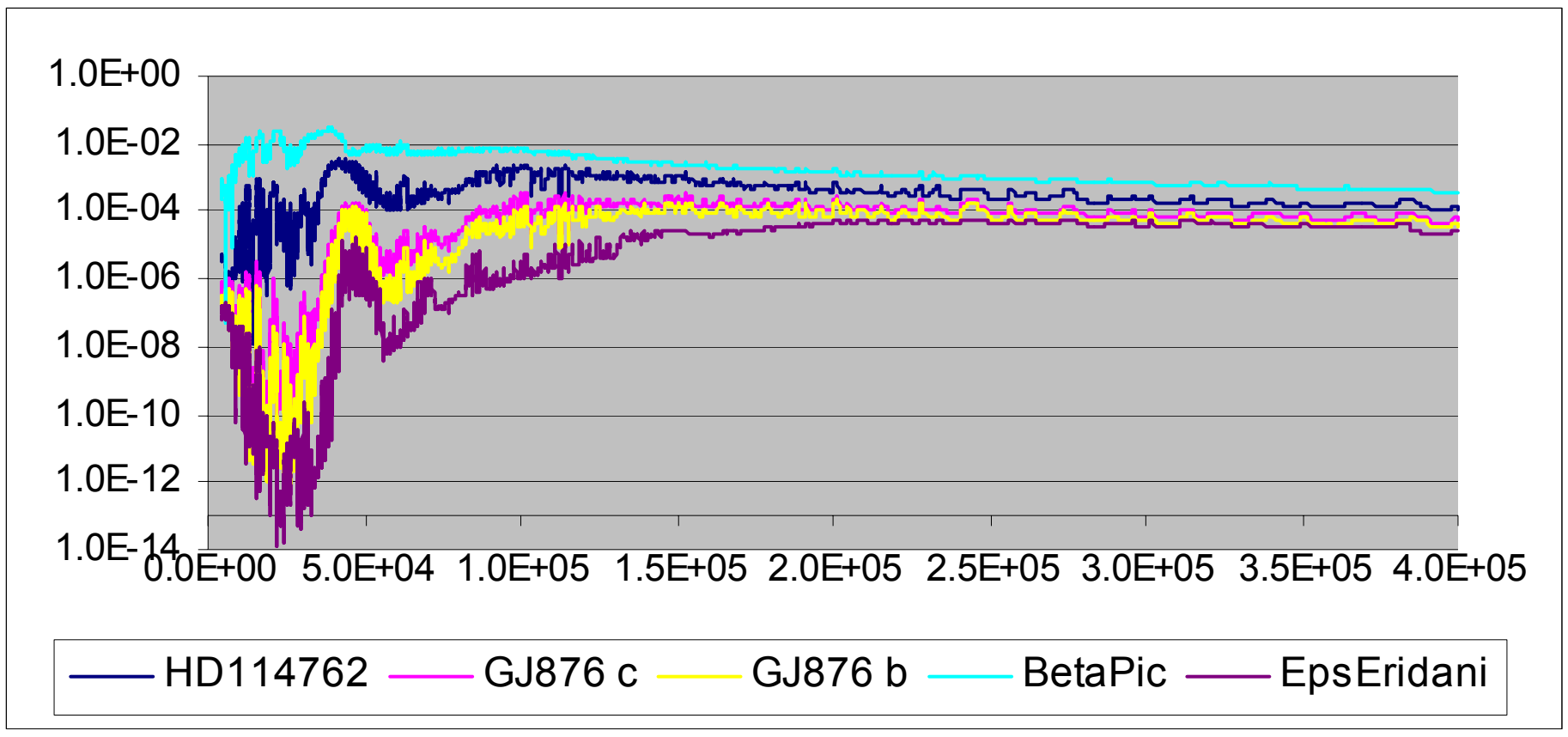

**Fig. 6: Calculated flux from some detected EGP (Burrows, 2004)**

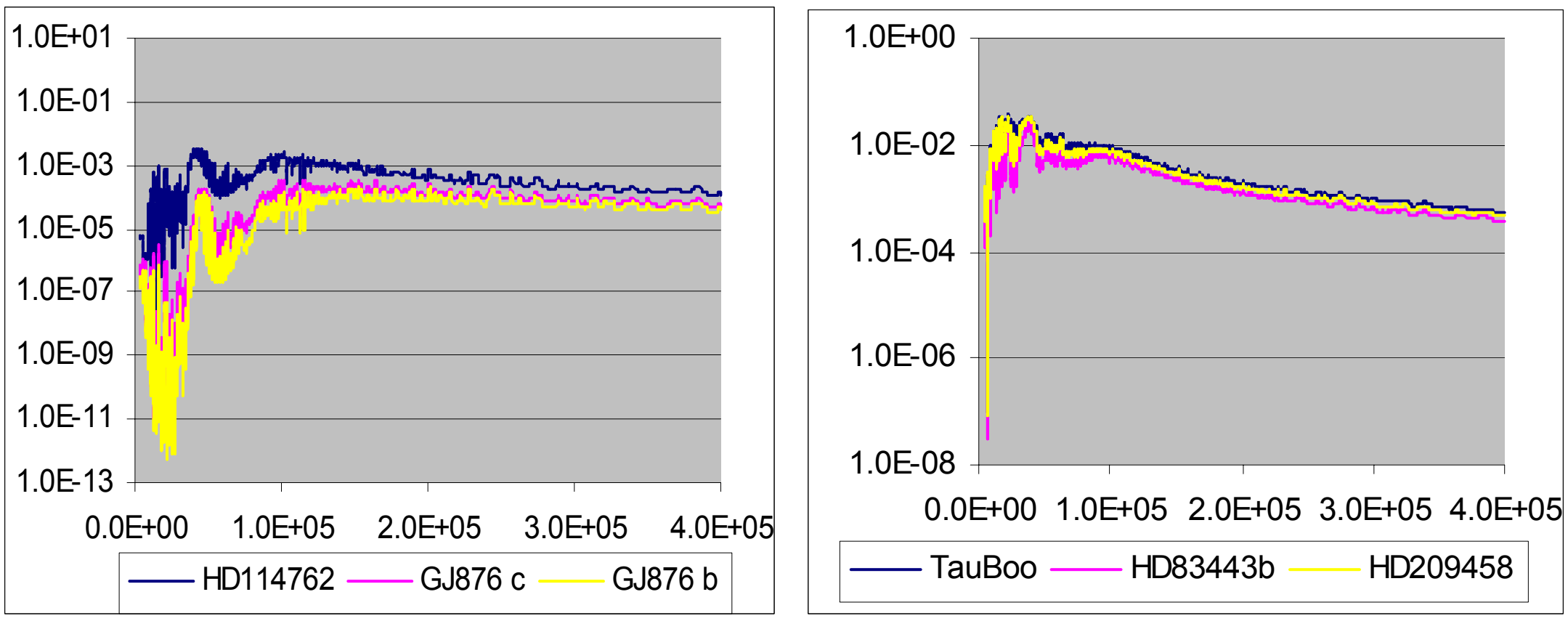

**Fig. 7: Calculated flux from some detected EGP (Burrows, 2004)**

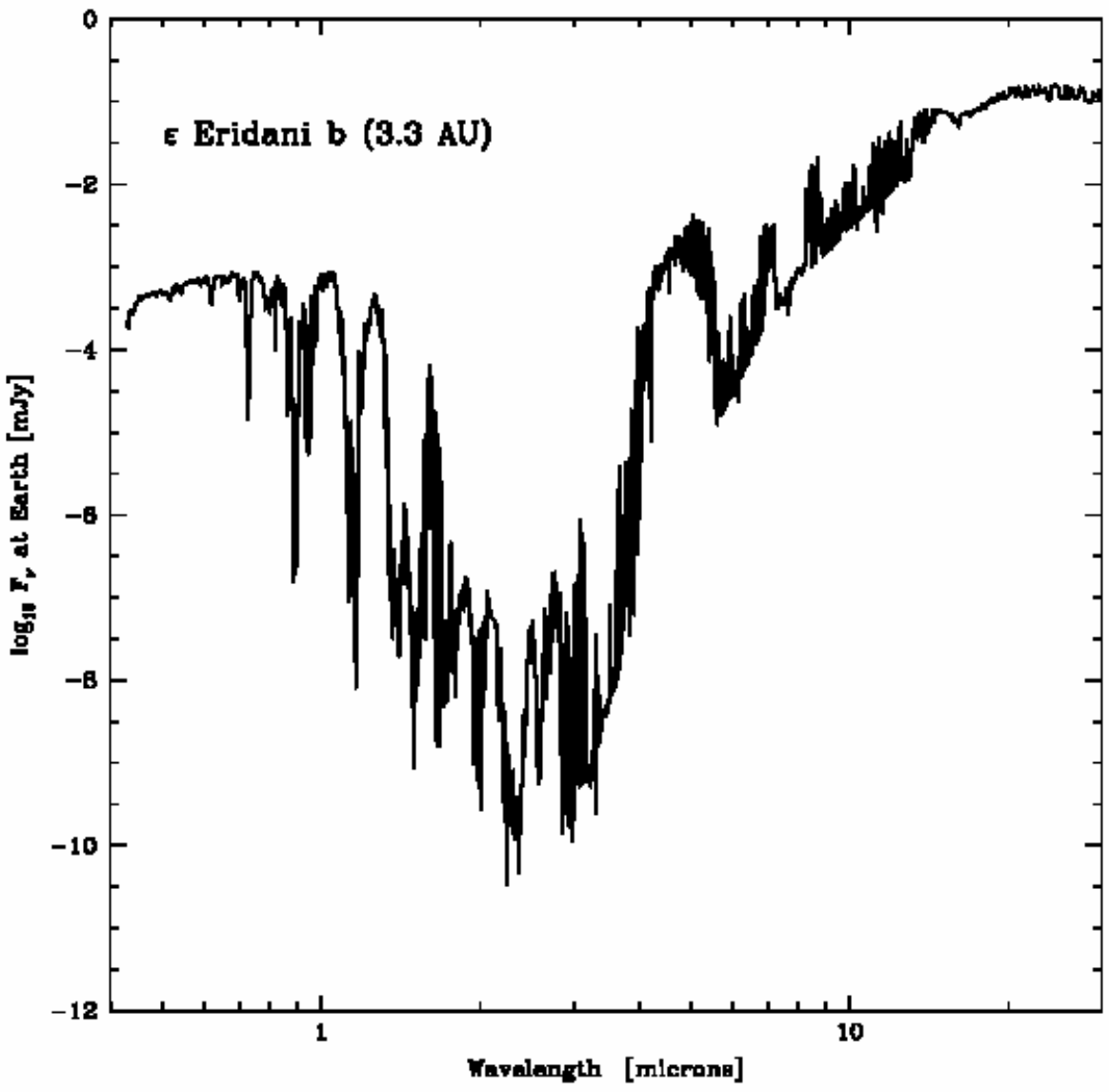

**Fig. 8: Simulated spectra of the planet orbiting epsilon Eridani at 3.3AU (Burrows, 2003).**

## 1.4 Planetary system formation

*"Extrapolating from the one planetary system known to orbit a main sequence star to a model of the variety of planetary systems which may be present throughout the galaxy is a daunting challenge surely fraught with pitfalls (Lissauer, 1995)."*

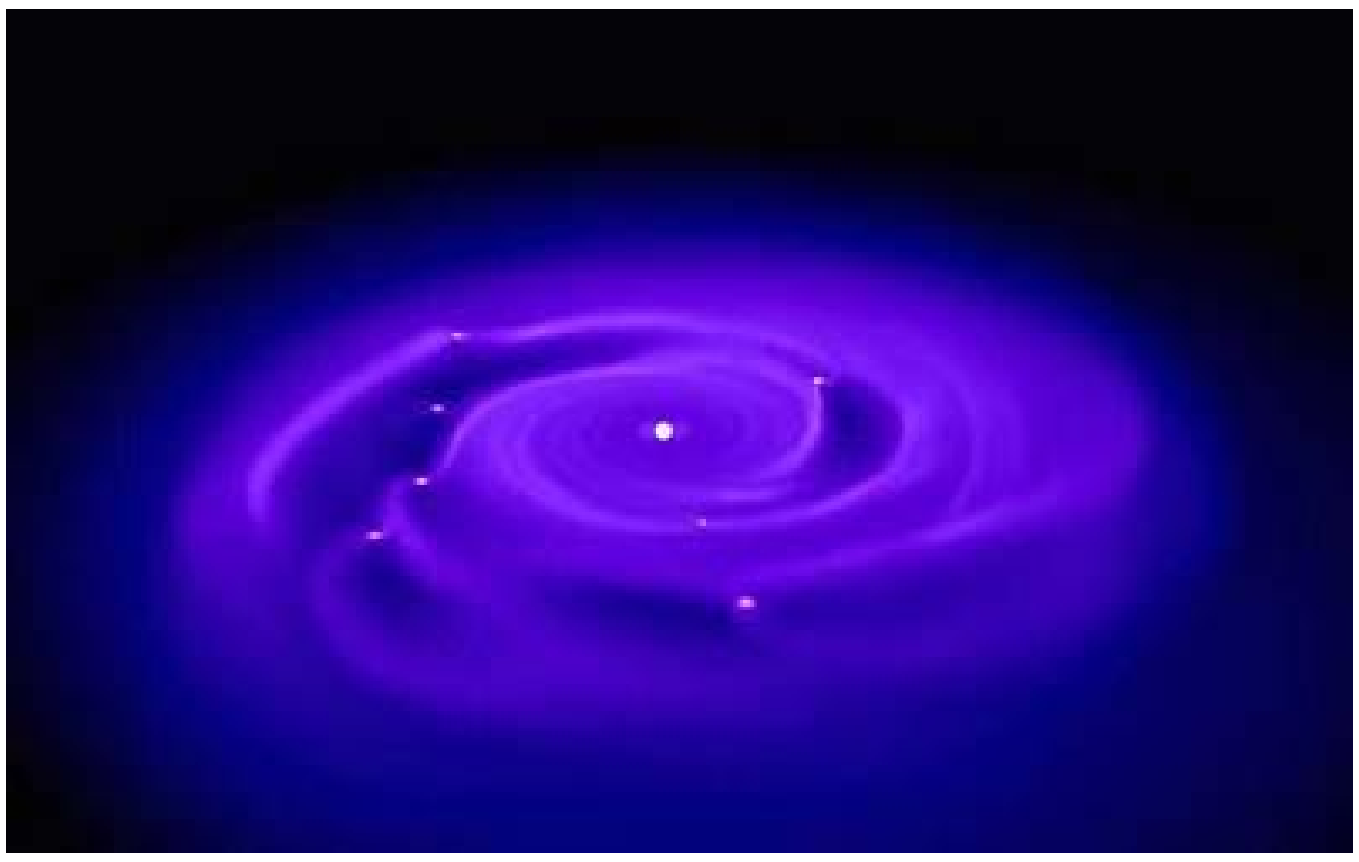

**Fig. 9: Simulations of planetary formation (Mayer, 2003)**

The pre-1996 theory of planetary formation did not, as a rule, consider the formation of super massive planets of multiple Jupiter masses, let alone ones in eccentric orbits (Artymowicz 1998). The common knowledge was that planetary orbits start circular and remain nearly circular forever. Theory also predicted a maximum mass a planet can achieve, based on the gap opening criteria in the standard solar nebula, thereby offering a simple explanation of Jupiter's mass in our system as the limiting mass case. Extension or modification was required for interpretation of giant planets of several Jupiter masses as planets.

The central star's evolution also influences the fate of planetary systems. Stars with masses below $\sim 0.8M_{solar}$ attain their largest size during their formation stage. They stay compact over the Hubble time and allow this same timescale for the lifetime of the surrounding planets. Stars with masses between $\sim 0.8M_{solar}$ and $10M_{solar}$ expand to $\sim 1000R_{solar}$ in the carbon oxygen core growth stage, absorbing the inner planets, like in the case of the solar system. At young star ages spectroscopic evidence of active disk accretion diminishes. Why? One possibility is that such disks form planetary systems. Fig.8 shows optical images of β-Pictoris where a tilt between inner and outer disk can be explained by the formation of a planet, even so the result is still debated. The planetary companion was inferred from warpage found in the disk thought to be caused by the gravitation of a planetary object (Burrows, 1999). The lower image of our solar systems puts the dimensions into context with the extent of our solar system.

Planetary systems and sub-stellar objects orbiting stars could form by different theories (Bodenheimer, 1997) (Boss, 2003) (Mayer, 2003) (Klahr, 2003) (Kaltenegger, 99).

1. Direct fragmentation during the collapse of a rotating interstellar cloud under gravity: This forms a system of small mass ratios e.g. 1:10.
2. Fragmentation of a collapsing cloud into a multiple system of various masses, followed by the capture of a low mass fragment by a high mass fragment through gravitational interaction.
3. Gravitational instability in an equilibrium disk, which has condensed out of the collapse, gives a structure, consisting of a central star, an accretion disk in equilibrium, surrounded by infalling gas. The disk accretes matter until it is gravitationally unstable, then it forms fragments, producing a sub-stellar companion in orbit around the star.
4. The disk forms as in the previous scenario. Small dust particles accrete by collision into planetesimals then into planetary cores of several Earth-masses. The gas accretion onto the core becomes a runaway process and builds up a giant planet.

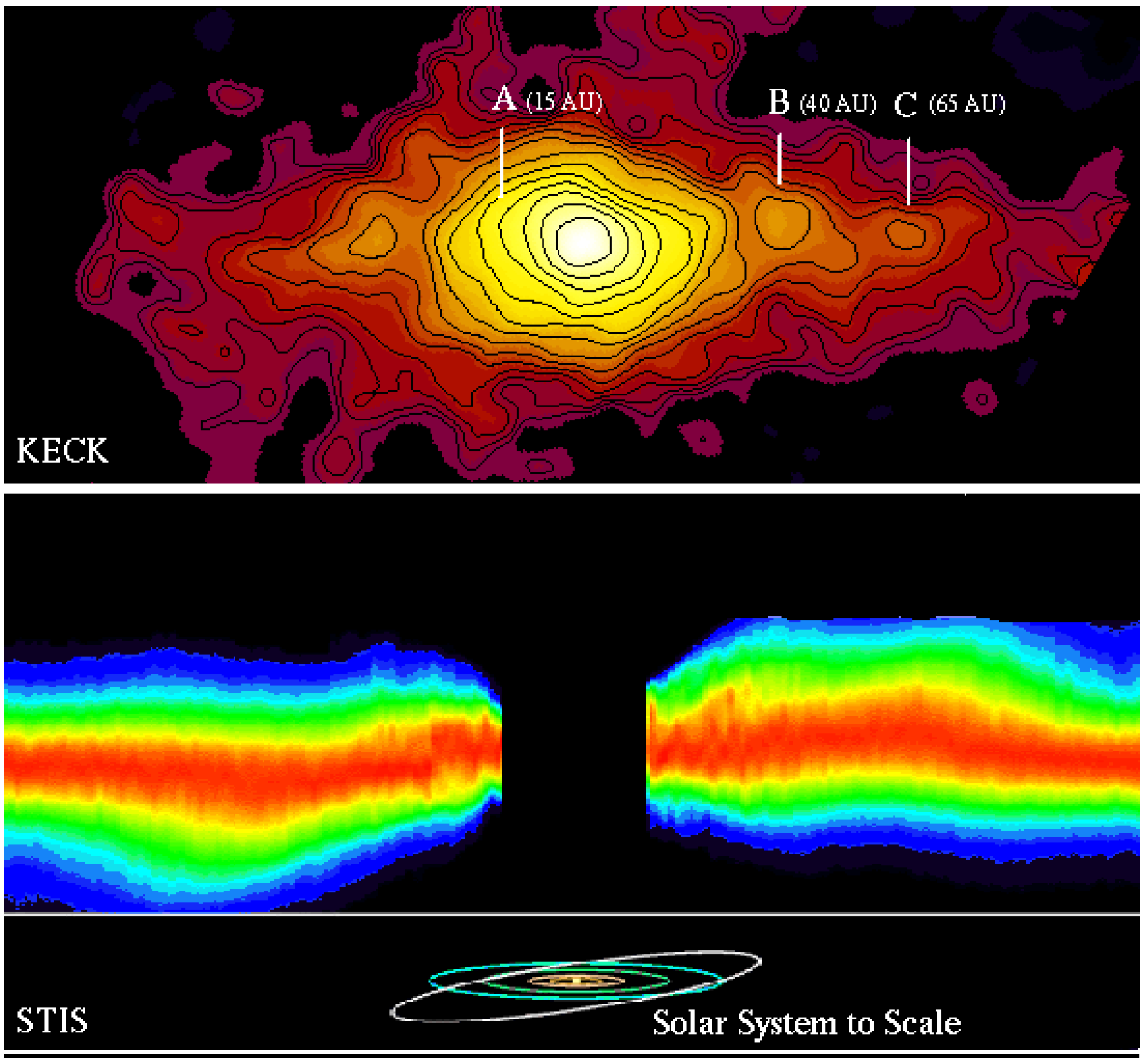

**Fig. 10: β-Pictoris, a disk that is thought to have formed a planetary body because of observations of a tilt between the inner and outer disk (upper panel 18µm observations Keck) (http://www.stsci.edu/hst, 2000).**

For the first two processes the minimum mass that is obtained is about $0.007M_{solar}$ while the disk is assumed to remain isothermal during the collapse. The gravitational instability scenario has the advantage that the formation time for the giant planet is short, consistent with observational constrains like the lifetime of disks. It does not explain the structure of Neptune and Uranus. The accretion time increases rapidly with the size of the planet's orbit. It shows the relation between the outer limit of the solar system and the age of the planetary system. N-body calculations of cluster growth in proto-planetary disks are shown by Kempf et al. 1999 (Kempf, 1999). The inner limit of planet formation is the distance where evaporation of silicate grains takes place, $a_{si}$. For a solar type star this distance is given by equation (1.4.1). The result is given in AU.

$$a_{Si} \propto 0.03\left(\frac{M_*}{M_{solar}}\right)^2 \tag{1.4.1}$$

The planetary candidates around 51Peg, τBoo and υAnd appear to be close to this boundary, Mercury is not (Tutukov, 1998). Using the given equations we can determine the time needed to form planets at the silicate evaporation limit. The distance at which the ice shells of dust grains are evaporated may determine the inner boundary of giant planets. The maximum mass for a planet formed by accretion in a disk due to the tidal truncation conditions is given by equation (1.4.2.) and (1.4.3.) (Bodenheimer, 1997):

$$M_p < \frac{40b}{\omega a^2} M_* \tag{1.4.2}$$

$$M_p < \frac{3c_s^3}{(\omega a)^3} M_* \qquad (1.4.3)$$

where $b$ is the viscosity, ω the orbital frequency, $c_s$ the sound speed and $a$ the distance from the star. The basic problem with this theory is that in minimum mass solar nebulae the accretion times for Jupiter and Saturn are too long, over $10^7$ years (Bodenheimer, 1997). The minimum solar nebular is calculated by augmenting the condensed element portions in the planets in the solar system, which leads to a value of about 0.01 to 0.02 $M_{solar}$. The distribution of the semi-major orbital axes for extrasolar planets is limited on one hand by the distance inside which dust particles are vaporized by the radiation of the central star and on the other hand by the distance inside which large planets can accrete due to asteroid collisions over the lifetime of the central star (Tutukov, 1998). Planetary configurations that are unstable evolve in simulations into more stable states via ejection and mergers.

Planets around pulsars could form from first generation objects that survived the explosion or second-generation objects formed around the neutron star. If they are first generation objects that survived the explosion, the definition of planets should be reconsidered, as their characteristics are likely to be very different from a "first generation" planet. Surviving first generation objects would have been much more massive objects that had part of their material stripped off by a supernova explosion.

### 1.4.1 Formation theories of our solar system

The nearly planar and almost circular orbits of the planets in our solar system argue for planetary formation within flattened circumstellar disks (Lissauer, 1998). In astrophysical models such disks are a byproduct of star formation from a collapsing molecular cloud perhaps triggered by external events such as nearby supernovae. The spectra of many young stars are much broader than a single temperature black body (Lin, 1998). This suggests the presence of a circumstellar disk, as each source radiates over a wide range of temperatures.

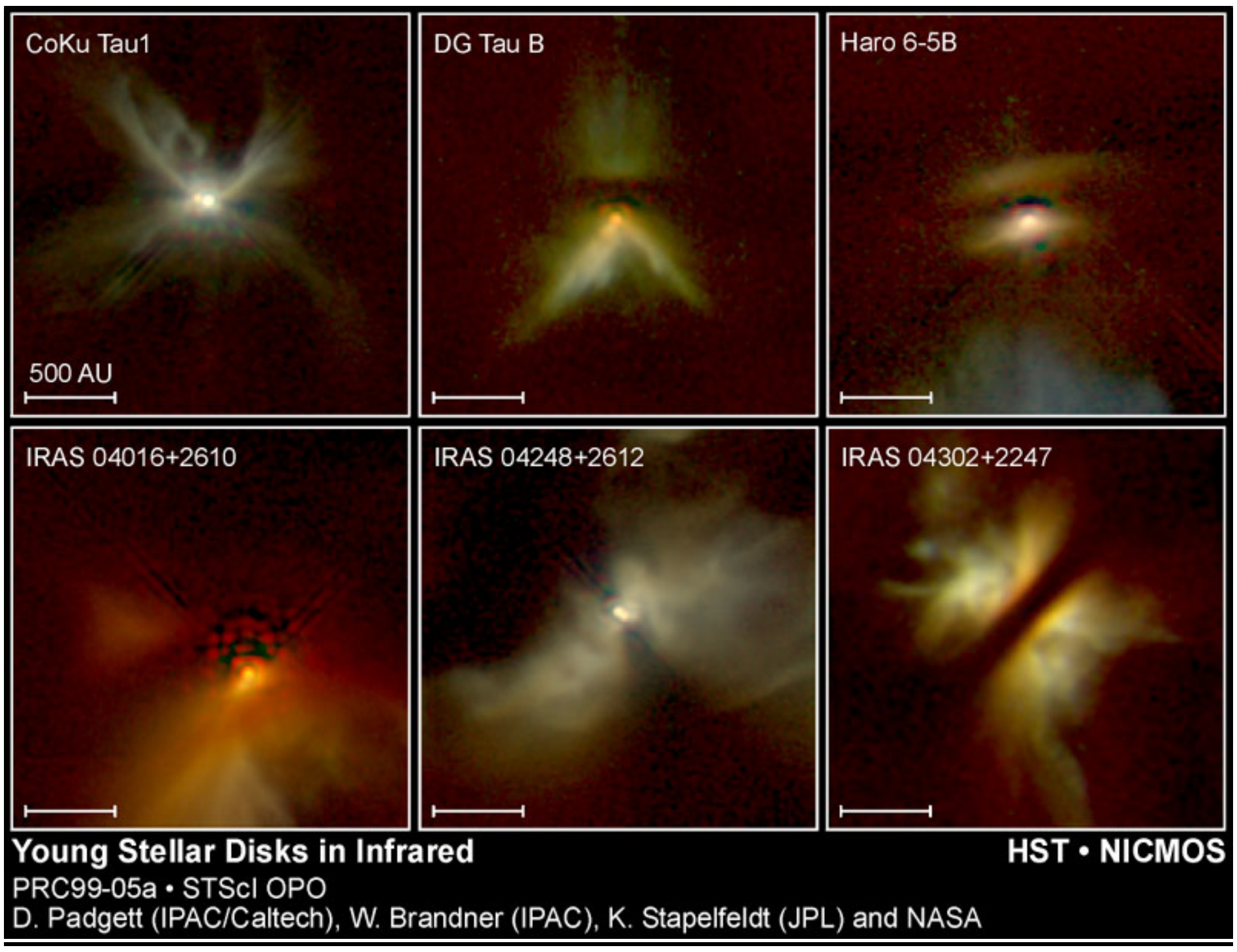

**Fig. 11:Young Stellar disks observed in the IR by the HST (http://www.stsci.edu/hst, 1999)**

T-Tauri stars are thought to be in the process of planet formation because of their youth and the frequent presence of circumstellar disks sufficiently massive for the formation of our solar system's planets. The fraction of stars showing such IR-excess decreases with stellar age and suggests a lifetime of about $10^6$ to 3 $10^7$ years. Even a very slow rotating cloud core has far too much angular momentum, more than 6 $10^{16}m^2/s$, to collapse down to a stellar object (Lin, 1998). A significant fraction of that material forms a rotationally supported disk orbiting the pressure-supported protostar. While matter located on the rotation axis can move directly to the center, matter in the equatorial plane at the surface of the sphere will be at the outermost radius of a disk after collapse (Papaloizous 1998). The formation of a protostellar disk through the collapse of a molecular cloud core takes $10^5$ to $10^6$ years. During the early stage the disk has a significant mass compared to the central star. Strong disk winds may exist. A very good overview of disk and planetary formation is given by Papaloizous et al. 1998 (Papaloizous 1998).

Condensates or remnant interstellar grains exist in a distance from the parent star where it is cool enough to remain in solid form. During the infall stage the disk is active and probably highly turbulent due to the mismatch of the specific angular momentum of the gas hitting the disk with that, required to maintain Keplerian rotation. The dynamics of small solid bodies within a protoplanetary disk are not well understood as the interactions of the gaseous components of the disk affect its dynamics. Gas grains rotate with the gas, large solid bodies orbit at Keplerian motion and medium sized particles move at an intermediate speed of those. Thus large and medium sized bodies are subject to a headwind from the gas that removes angular momentum from the particles, causing them to spiral inward toward the central star (Lin, 1998).

Smaller particles should drift less rapidly because the headwind they face is not as strong. Large particles drift less because they have a greater mass-to-surface area ratio. For kilometer sized and large solid bodies, the primary perturbations on the Keplerian orbits are mutual gravitational interactions and physical collisions that lead to accretion of planetesimals. The most massive planetesimals in the swarm have the largest gravitationally enhanced collision cross section and accrete almost everything they collide with. If the random velocity of most planetesimals in the swarm remains smaller than the escape speed, the largest bodies grow extremely rapidly, accreting most of the small bodies in their gravitational reach. At this point that runaway growth phase stops. As the eccentricities of planetary embryos are pumped up by long-range mutual gravitational perturbations, slow growth can continue. For solid material another loss mechanism exists, impact vaporization, when its velocity exceed 10 to 20 km/s (Lissauer, 1995) if the disk is not sufficiently dense. At low gas density, the majority of the high velocity vapor may become ionized and removed by the star's wind or condense into particles so small that they are blown out of the system by the radiation pressure from the star (Bodenheimer 1997).

Massive and dense planets far enough from the parent star not to be too deep within its gravitational potential well, can eject material into interstellar space as planetary masses and random velocity grows. Oort Cloud comets are believed to be icy planetesimals that were ejected with a speed close to the solar system escape velocity. The ultimate size and spacing of gas giant planets provide a major source of uncertainty in modeling the diversity of planetary systems. A potential hazard to planetary system formation is radial decay of planetary orbits resulting from interaction with disk material (Lissauer, 1998). If Jupiter were in an eccentric orbit, Earth and Mars would likely be gravitationally scattered out of the solar system. In most planetary systems, all the planets should revolve around the central star in the same direction as a result of their origin within a rotating disk.

## 1.5 EGP formation

There appear to be only two distinct possibilities for forming gas giant planets, namely from the "bottom up" (core accretion), or from the "top down" (disk instability). Most of the work on giant planet formation has been performed in the context of the core accretion mechanism, so its strengths and weaknesses are better known than those of the disk instability mechanism, which has only recently been subjected to serious investigation. The conventional explanation for the formation of gas giant planets, core accretion, presumes that a gaseous envelope collapses upon a roughly 10 Earth-mass, solid core that was formed by the collision accumulation of planetary embryos orbiting in a gaseous disk. The more radical explanation, disk instability, hypothesizes that the gaseous portion of protoplanetary disks undergoes a gravitational instability, leading to the formation of self-gravitating clumps, within which dust grains coagulate and settle to form cores (Boss, 2003). This scenario should show eccentric orbits as the rapid collapsing cloud breaks up into fragments moving on ecliptical orbits (Boss, 1998B). They may also form from rocky cores of about 10 Earth-masses followed by hydrodynamical accretion of gas as predicted by core accretion. In this second scenario the core of a jovian planet is thought to grow by collisional growth and accretion of solid planetesimals in the disk. The speed of this process should scale roughly as the square of the metallicity because collision rate proceed at density squared (Marcy, 1999).

It is uncertain whether gas giant planet formation is common; most protoplanetary disks may dissipate before solid planet cores can grow large enough to gravitationally trap substantial quantities of gas. The growth times predicted by current models are similar to the estimated lifetime of the gaseous protoplanetary disk. Assuming that the protoplanet can accrete gas efficiently, it will first take in gas in the neighborhood of its orbit until it fills its Roche radius $R_R$:

$$R_R = \left( \frac{M_p}{3M_*} \right)^{1/3} a_p \tag{1.5.1}$$

where $a_p$ gives the radius of the assumed circular orbit of the planet. The mass of the protoplanet is then given by (Papaloizous 1998):

$$M_p = \left( \frac{\pi r^2 \Sigma}{M_*} \right)^{3/2} M_* \tag{1.5.2}$$

$\pi r^2 \Sigma$ gives the characteristic disk mass within radius $r$ and with $\Sigma$ being the surface density. With $\Sigma$ about 200g/cm$^2$ and a distance of 5.2AU, this gives a protoplanet mass about 0.4M$_J$ (Trilling, 1997). With $\Sigma$ about 1g/cm$^2$ and a distance of 5AU, this gives a protoplanet mass about $10^{27}$g (Papaloizous 1998). In the core accretion scenario several planets appear from a dusky disk and remain on about circular orbits. The other scenario is a proto-stellar cloud collapsing in two, dividing its energy and angular momentum between the star and a unique more massive companion in a random way, thus allowing high eccentric orbits. The disk instability mechanism requires a relatively massive, cool disk and may occur early in the evolution of the protoplanetary disk. To avoid being swallowed by the protosun, giant protoplanets must form toward the end of the main protostellar accretion phase. Then most of the disk material has been added to the protostar but the disk is still massive enough to be unstable at star ages around 0.1 Myr to 1 Myr (Trilling, 1998).

Massive planets that are observed near stars indicate that either the giant planets are formed at a minimum distance of about 4AU and then migrate toward the primary, or a massive telluric planet formed in situ, or that the present understanding of giant planet formation is not complete (Podolak, 2003). Why is it difficult to form giant planets at 0.05AU? In many nebular models the temperature is too high at this distance from the star to allow for condensation of solid particles and there is insufficient mass in the inner region to form a Jupiter size planet. Did the planet migrate to that site? How could migration stop before the star engulfed the planet? The formation of a telluric planet in situ would require a massive protoplanetary disk. Also, the processes that halt the accretion of gas and determine the final masses of giant planets existing are not well understood. Nonetheless giant planets are more likely to form in high surface mass density, long lived protoplanetary disks. Close in giant planets are found to be rare objects, as current detection technique is sensitive enough to detect them. Giant planets are predicted to form from accretion of dust particles resulting in a solid core of a few Earth-masses, accompanied by a very low mass gaseous envelope (Bodenheimer, 1997). Due to further accretion of gas and solids, the mass of the envelope increases faster than that of the core until a crossover mass is reached. Runaway gas accretion involves little accretion of solids. The dissipation of the nebula stops the accretion. The planet contracts and cools at constant mass to the present state.

### 1.5.1 Terrestrial planet formation

Terrestrial planets are believed to form from interstellar dust by steady accumulation of smaller particles. At first accumulation is achieved through sticking, later aided by gravity between larger planetesimals and subsequently by gas accretion onto rocky cores (Bechwith, 1999) (Boss, 2003). In the earliest stage, particles are built up by coagulation and sticking as they bump into one another in the disk. Brownian motion, setting, turbulence and orbital migration can evolve small micro sized bodies to meter size bodies through collision in about $10^4$ years. Brownian motion collisions are only relevant for small, micro sized grains because of the inverse mass dependence.

$$\Delta v_{th} \propto \sqrt{\frac{kT}{\mu}} \tag{1.5.3}$$

with

$$\mu = \frac{m_1 m_2}{m_1 + m_2} \tag{1.5.4}$$

where $k$ is the Boltzmann constant, $T$ the temperature, $v$ the velocity and $m$ the mass of the particles. After about another $10^4$ years the bodies can be kilometer size. Beyond 1 km, gravitational attraction causes the planetesimals to grow by pair wise collisions. Later runaway growth sets in and planets will be present after about 1Myr to 10Myr though collision accumulation of successively larger planetesimals and planetary embryos (Papaloizous 1998).

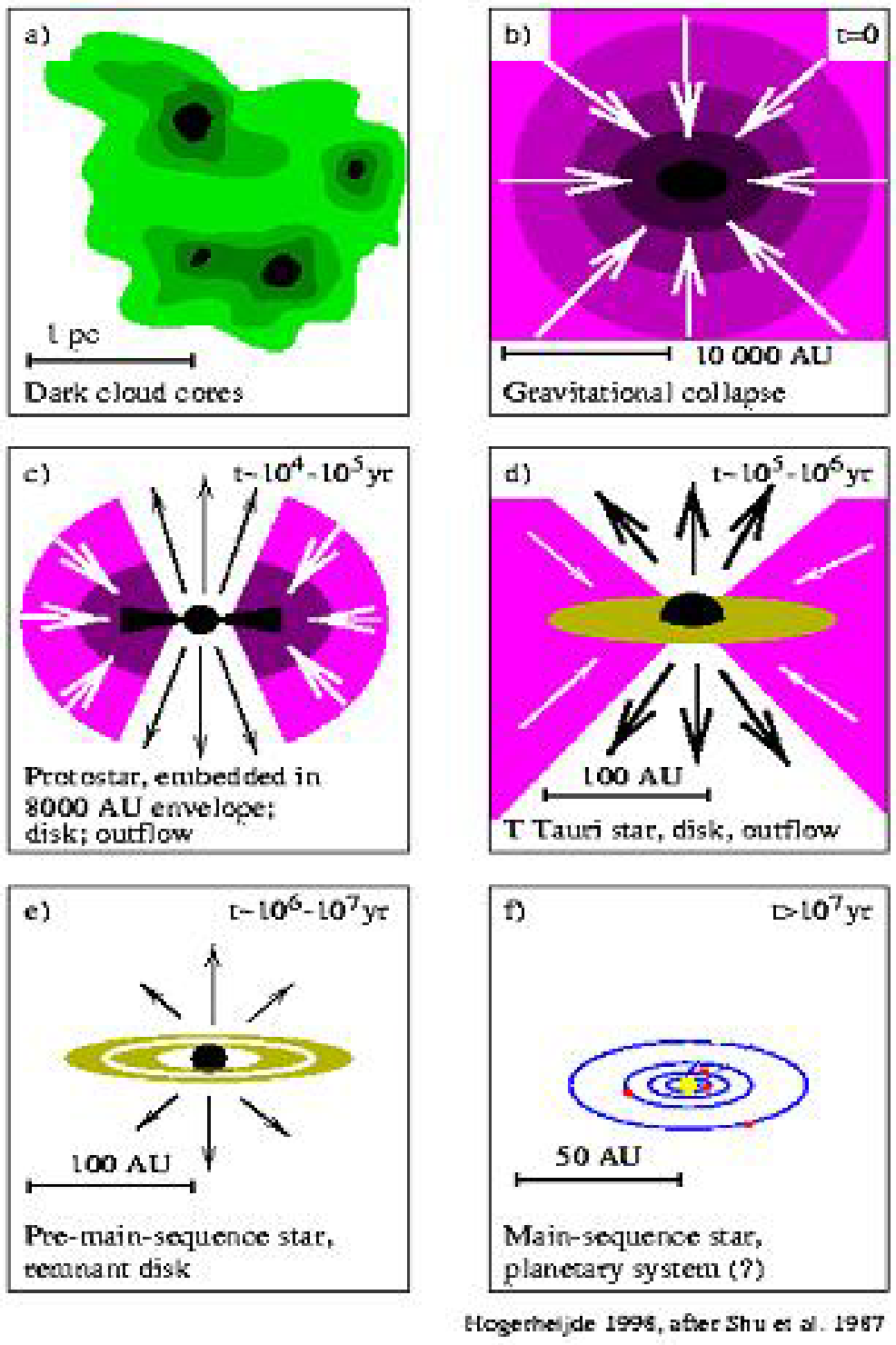

**Fig. 12: Planet formation scenario: grain growth through particle-particle interaction, growth of planetesimals by two body interaction aided by gravity and the accretion of gas via gravitational attraction (Fridlund 2000)**

The growth process is governed by the sticking probability and the strength of adhesion force holding the aggregation together (Bechwith, 1999). The sticking probability depends on the collision velocity, masses, shape and material properties of the colliding particles. The sticking probability increases significantly for irregularly shaped dust grains. Van der Waals forces are thought responsible for the sticking of the small particles. Growth from 1 mm to 1 km size is poorly understood.

*"It is not clear why rocks of centimeter or meter size would stick together when they collide at speeds of meters per second. Terrestrial rocks certainly do not." (Marcy, 1999)*

Numerical simulations by Godon et al. 1999 (Godon, 1999) suggest that heavy dust particles rapidly concentrate in the cores of anticyclonic vortices in a disk that formed our solar system, increasing the density of centimeter size grains and favoring the formation of larger scale objects. These are capable of triggering a gravitational instability. The change in the Keplerian velocity of the flow in the disk, due to the anticyclonic motion in the vortex, induces a net force towards the center of the vortex. As a consequence the concentration of dust grains in the anticyclonic vortex becomes much larger than outside.

A challenge to the understanding of the giant planets formation mechanism comes from the direct experimental measurements of noble gases abundances performed by the Galileo probe in the atmosphere of Jupiter (Owen, 1999). Those measurements show that the abundance of Argon, Kripton and Xenon display the same enhancement respect to the solar abundance as other high Z elements. But Ar, Kr and Xe can only be trapped as volatiles in amorphous ice at temperatures below 30K (Bar-Nun, 1988). This is not consistent with the accretion of Jupiter by planetesimals formed in the region extending outward from Jupiter, where the temperature must have been cold enough for water ice to condense at around 160K, up to Neptune, at about 55 K. The planetesimals could have been formed during the early phases of collapse of the molecular cloud, Jupiter's core could have migrated from regions external to Pluto's orbit or the nebula could have been much colder than current models predict (Fridlund 2000). Also, while the formation of a large primary silicate planet cannot be ruled out, it seems probable that the majority of the detected EGPs are large gaseous, in analogy with Jupiter. The Doppler technique that has detected the known EGP so far provides only the minimum mass $M_{minp} = M_p sin(i)$, due to the uncertainty of the inclination $i$ of the planetary system to the line of sight. It is impossible to deduce the mass of planetary candidates from the stellar wobbles due to the unknown orientation of the companions' orbits with respect to the line of sight. A relatively low-mass planet in an edge on orbit or a more massive object in an orbit at a shallower angle to the line of sight can cause the same radial velocity variation. The probability to see a system with small inclination is high. If you hold a ball on top and bottom the probability for random distributed observer to see the top or the bottom is much smaller than the probability to see it inclined with a small angle.

Knowledge of the stellar mass combined with modeling the precise photometric transit measurements provides estimates of the basic system parameters such as the orbital inclination and planetary radius. The detected transit of the known short period EGP HD209458b (Charbonneau, 1999) clarified that we are observing giant planets by constraining the radius and thus the density of that object. Determining the orbital inclination removes the *sin(i)* dependency in the planetary mass estimate. The resulting mean density of about $0.27 gcm^{-3}$ confirms that the planet is a gas giant. The combined data (Charbonneau, 2000) (Henry, 2000) were fit with a transit model, yielding a determination of the stellar radius $R = 1.17 \pm 0.03 R_{solar}$, the planetary radius $R_p = 1.43 \pm 0.04 R_J$, and the orbital inclination $i = 86.1^o \pm 0.1^o$ (Schultz, 2003).

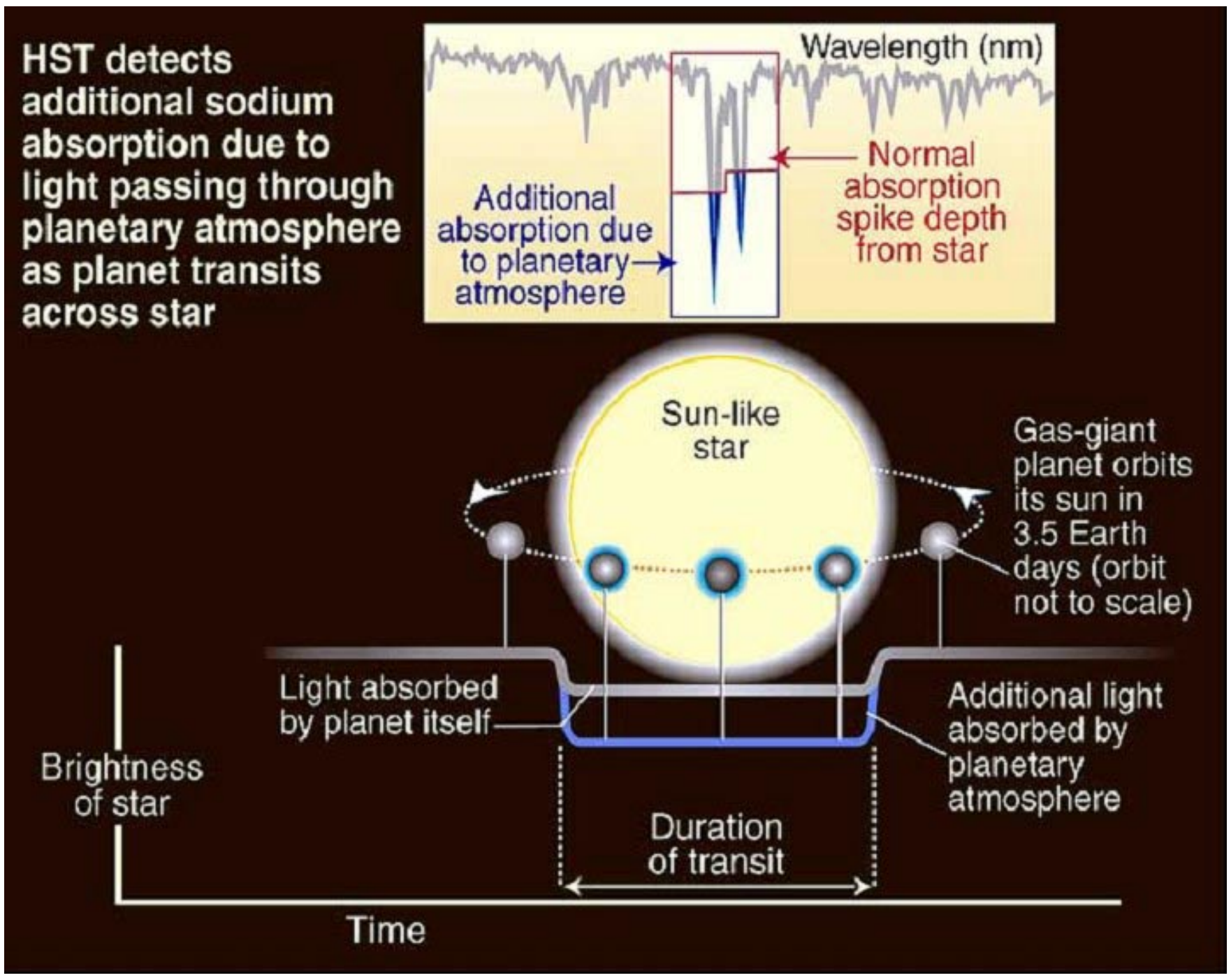

**Fig. 13: First detected EGP transit around HD209458 (Charbonneau, 2001)**

The demonstration that the depth of HD209458b's transit is wavelength-dependent (Hubbard, 2001) and that neutral sodium resides in its atmosphere (Charbonneau, 2002) is a stepping stone in the search for extrasolar planets and clarifies that the EGP close to their parent stars are indeed gas giants. In November 2001 the transmission spectrum of the planet was observed with the Hubble Space telescope. The planetary atmosphere blocks starlight at wavelengths where there are strong absorbers, thus planetary features are superimposed on the star's spectrum. Even though the transit only gave limited information, absorption of the trace element neutral sodium was detected. It can be used to confirm and constrain parameters in planetary evolution models. The temperature of the EGP was constrained to about 1100K in its orbits about 100 times closer to its sun than Jupiter with its 120K. The planet will receive a very different wavelength-distribution of starlight, peaking in the near infrared (Seager, 2002).

Transits exhibit a flat bottom minimum, with limb darkened ingress and egress, and repeat like clockwork. The fractional change in brightness is proportional to the fraction that is occulted by the planet and gives its size. A terrestrial planet should simply occult a fraction of the stellar light reducing the brightness in optical and infrared wavelengths, as it is cooler than the star. The atmosphere of larger gaseous planets may also cause absorption features during transits that could be measured with high-resolution spectroscopy. Multiple planets or moons could be detected through the periodic change they induce on the timing of the successive transit of the primary occulting body (Schneider, 1997). Transit search can give statistical numbers of planets around a large sample of stars and may also determine if rings and satellites are present. It is limited by the fact that the planet must be close to the star, in an orbit that is seen edge-on and by very small brightness variation. The observed star must also have a very stable luminosity. A solar type star will dim by 1% with duration of hours depending on the orbital radius of an EGP (Hale, 1994). Earth-size planets would dim the star by 0.01%. The photometric precision to measure an Earth transit requires a space platform (Marcy, 1999).

### 1.5.2 Disk gaps cleared by planets

Planet formation is believed to involve accretion from the surrounding disk of material. Gas accretion onto a planet leads to non-Keplerian flow pattern near the planet. The planet is subject to gravitational torques due to its interaction with the gas disk, which results in planetary migration (Lubow, 2003). A planet's motion in a disk excites density waves, both exterior and interior to the planet (Papaloizous 1998) (Delpopolo, 2003) (D'Angelo, 2003). These waves create a gap in the disk as the planet clears material from its orbit. Thus the inner disk loses angular momentum while the outer disk gains it. The tendency is therefore to form a gap. Even after gap formation there may still be some residual accretion between the disk and the protoplanet.

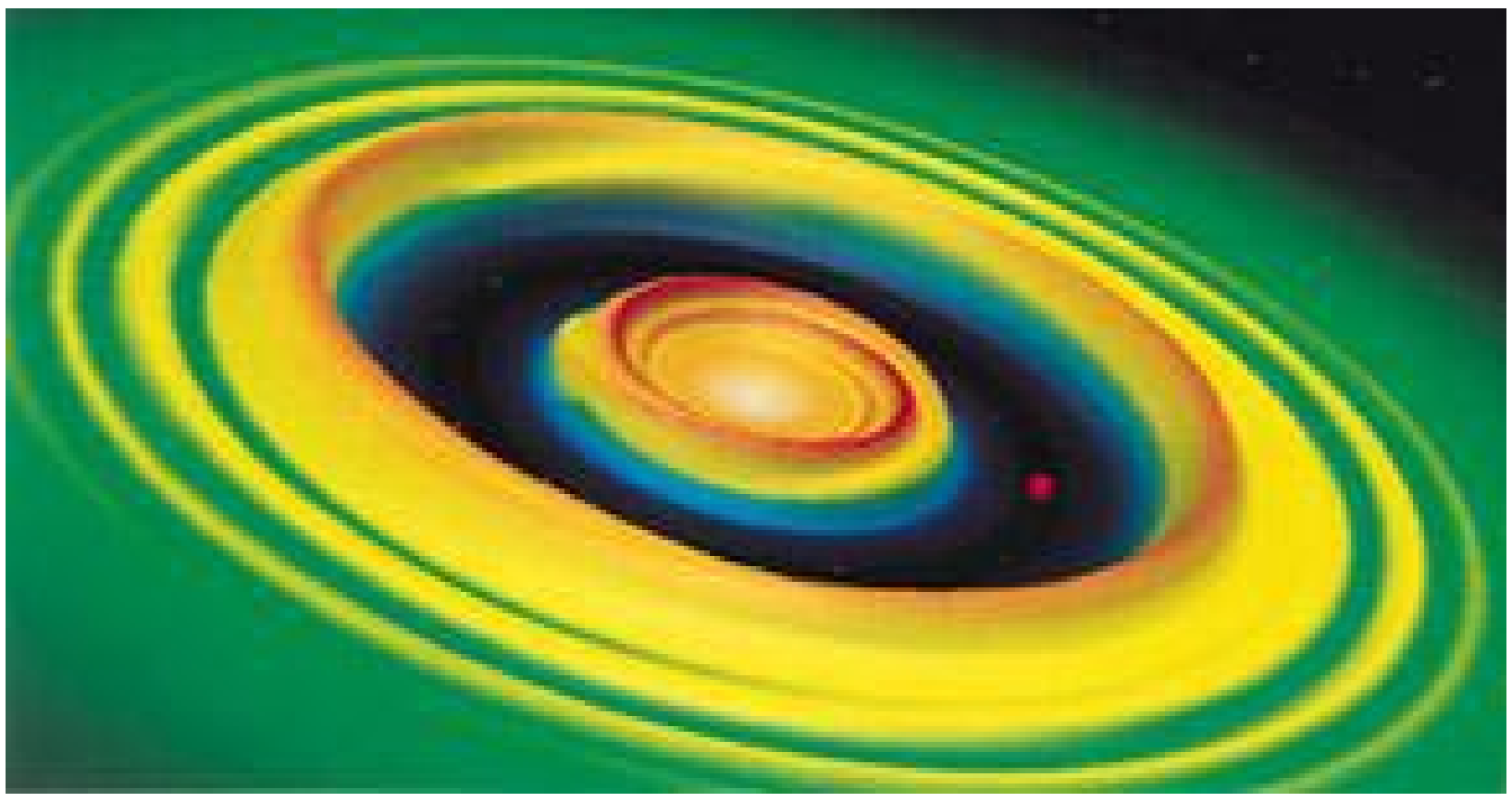

**Fig. 14: Simulations of a gap formed in planet accretion (Marcy, 1999)**

The ability of a planet to terminate its accretion of gas by clearing a gap around its orbit depends on the estimated viscosity of the protoplanetary disk.

### 1.5.3 Orbital migration

Gas giant planets are expected to form outside of a 4AU radius where condensation of ice grains can occur and where there is more disk mass. Thus the discovery of Jovian mass planets in eccentric orbits closer than 1AU prompted a cascade of theories on orbital migration, planet–planet dynamical interactions and planet–disk tidal interactions (Goldreich, 1980) (Lin, 1996) (Levison, 1998) (Murray, 2001) (Chiang, 2002) (Ford, 2003). Implicit in the concept of orbital migration is the expectation that not all migrating planets and planetesimals will park in stable orbits, some gas-depleted material will fall into the star and mix in the convective zone, possibly enriching the abundance of refractory elements. Lin et al. (Lin, 1998) suggested that the planet could be stopped on tenth of an AU from the star by tidal torques from the star counteracting the disk torques. A substantial reduction in disk torque could also occur when the planet is within a nearly empty zone near the star, as the disk might not extend down to the stellar radius. The existence of giant planets with orbital periods ranging from 15 days to three years casts doubt upon this model, as they are unlikely to have encountered a clear zone of the disk and feel negligible tidal torque from their star. Another model assumes gravitational scattering of multiple planetary bodies. One observable test of models that rely on gravitational scattering of one giant planet by another would be the presence of a second companion in a very high eccentricity, long period orbit. If giant planets migrate, terrestrial planets could generally be lost in this process. An alternative interpretation proposed by Black (Black, 1997) is that these objects are BD formed near their current location. It is likely that part of the EGP candidates are BD due to the uncertainty of the orbital inclination, but at least one planet has been confirmed by transit detection to be a real Jupiter mass object.

### 1.5.4 Theory of migration due to tidal disk forces

If giant planets form by core accretion, the initial formation of a solid core of at least 5 to 10 Earth-masses may be very difficult to achieve very close to the star due to the high temperature. The theory of tidal disk forces allows giant planets that form in circumstellar disks to migrate from their initial position. Radial migration could be caused by inward torques between the disk and the planet, outward torques between the planet and the spinning star and Roche lobe overflow and consequent mass loss from the planet (Papaloizous, 1998). A very general model that sums up the torques on a planet, finds its radial motion in the circumstellar disk, and distinguishes three classes of planets, is given by Trilling et al. (Trilling, 1998):

1. (I) planets that migrate inward too rapidly and disappear due to mass loss from the Roche lobe overflow,

2. (II) planets which migrate inward, loose some but not all of their mass and survive in very small orbits and
3. (III) planets that do not loose any mass during migration.

For a protoplanetary disk of 0.011 $M_{solar}$, a viscosity of 5 $10^{-3}$ and a distance of the planet of 5.2AU this results in an initial mass smaller than $3M_J$ for class (I) objects and larger than $4M_J$ for class (III) objects. The masses defined as limits between these classes of planets vary with the viscosity and mass of the disk. Additionally, for large planet to disk mass ratios, the planet clears such a large gap that the resonance between planet and disk is small, as for class (III) objects. This addition is needed to include Jupiter in this model. The planets in this model are not calculated as point masses. They have radii and internal structures that are calculated for every point through the planets evolution but mass transfer does not include the planetary core.

The first stage in the planet's migration is the inward migration stage for about $10^6$ years due to the disk interaction. The radial motion of the planet is inversely proportional to the mass of the planet, so that more massive planets move less rapidly. The gap formed by the planet is essential in determining the evolution of the system in this model. As the gas moves inward, the planet moves inward (Lin, 1998). Larger viscosities as well as larger disk mass cause smaller gaps and therefore faster inward migration. A planet whose orbital period is slower than the rotation of its parent star slows this rotation, as it is the case for the Moon-Earth system. Energy is dissipated within the star and as the star slows down, the planet must move outward to conserve angular momentum. When a migrating planet gets sufficiently close to its parent star, the Roche radius can be smaller than the planet's radius. The planet moves outside during transfer to conserve the angular momentum of the system. In case of a stable mass transfer it moves to a distance at which its planetary radius is equal to the Roche radius. Any subsequent inward motion induced by the circumstellar disk decreases the Roche radius further, which results in more mass loss for the planet. The Roche-lobe overflow may halt the migration of a planet, though the mechanism is difficult to implement. As long as the disk exists, mass loss continues, when it dissipates the mass loss stops. One explanation for the absence of super giant planets in our solar system includes an exterior event that disturbed the disk around our protosun. Since there was not sufficient mass left to accrete, Jupiter did not develop into a super giant and did not migrate (Queloz, 1998C).

### 1.5.5 Stability criteria for orbital configurations

Stability from an astronomical viewpoint implies that the system remains bound and that no ejection and merges of planets will occur and this is robust against small perturbations (Artymowicz, 1998). The ejection or merge of planets on unstable orbits could have been the final phase of planetary growth in our solar system as the planets orbit close enough to each other. The aspects that are best understood in stability of planetary orbits are those that can be studied by using the planar circular restricted 3-body problem. One planet is taken to be massless, the other has a mass much less than the star. A system with two planets on initially circular orbits will never experience close approaches and collisions if the ratio of the separation in orbital radii to the planets' semi-major axes $|\Delta a|$ fulfill equation 15.5.1 (Lissauer, 1998).

$$\frac{|\Delta a|}{a} > 2\sqrt{3}\,\frac{R_H}{a} = 23^{1/3}\left(\frac{M_p}{M_*}\right)^{1/3} \approx 2.4\left(\frac{M_1+M_2}{M_*}\right)^{1/3} \qquad (1.5.5.1)$$

where $a$ is the average semi-major axes and the Hill radius $R_H$ is given by:

$$R_H = \left(\frac{m_1+m_2}{3M_{solar}}\right)^{1/3} \frac{a_1+a_2}{2} \qquad (1.5.5.2)$$

Deterministic chaos is also important in the stability of planetary orbits, as trajectories tend to be chaotic whenever resonances overlap. To avoid the resonance overlap region of a planet the distance of a test particle must be bigger than

$$\frac{|\Delta a|}{a} > 1.5\left(\frac{M_P}{M_*}\right)^{2/7} \tag{1.5.5.3}$$

Equation 1.5.5.1 gives the generalization of the Hill-Jacobi exclusion zone and equation 1.5.5.3 the resonance overlap criterion of the restricted 3-body problem. These equations appear to be useful to determine the stability of our planetary system and the satellite systems of the giant planets. The minimum separation between two planets increases with planetary mass, $M_p$, but the dependency is far less than linear. Larger eccentricity, e, and inclination, i, require excess energy, give more freedom of motion and thus allow greater instability as determined by Hasegawa et al. (Hasegawa, 1990).

$$\frac{|\Delta a|}{a} > 12 + \frac{4}{3}\left(\frac{m_1 + m_2}{3M_{solar}}\right)^{2/3}\left(e^2 + i^2\right) \tag{1.5.5.4}$$

That protection mechanism works to stabilize the system. The mutual perturbation between the planets increase their eccentricity and therefore decrease their minimum approach distance at subsequent conjunction, this in turn produce an increase in $\Delta a$, increasing the distance of the closest approach, counteracting the increase in eccentricity. Unless bodies are locked in orbital resonance, eccentricity generally destabilizes systems because bodies approach each other more closely. Inclination usually increases stability because the bodies tend to be further apart. Perturbations from several planets add in a nonlinear, complicated manner that is not fully understood.

### 1.5.6 Environment of protoplanetary evolution

The effect clustering and vigorous massive stars may have on protoplanetary evolution is not well understood. McCaughrean (McCaughrean, 1997) investigated the Orion Nebula and M16, where photo evaporation by O and B stars may disrupt the star formation process and may influence the development of disks and hence planets. The outflow phase regulates the angular momentum removal and dissipation of the outer envelope and may define the final stellar mass. The outflow will stir up the surrounding molecular cloud on scales up to parsecs and therefore regulates further collapse of star formation in the cloud. Surveys in Giant Molecular clouds for young stars have shown that they are found in dense clusters with 10 to 1000 members. Many dense clusters contain OB stars. Their ionizing radiation and strong winds add another level of disruption to the low mass stars typically 0.1 to 1 pc away. This may influence the disk by truncating them and therefore setting their mass or as well destroying them trough ionization, wind and dynamical interaction. In the Orion Nebula disks can be directly observed, using a bright background nebular emission as a screen, against which disks may be detected as silhouettes. A small number of disks, relatively far from the OB star, have been found; a couple of them have faint ionized rims around the silhouettes. The interplanetary environment of external planetary systems is a function of the interaction of each star with the cloud surrounding it. It affects the planetary atmospheres and climate. Stellar wind flow, cloud density, temperature and velocity relative to the star regulate this interaction. The solar motion, about 17 pc per million years, combined with interstellar cloud motion driven by stellar evolution provides a constantly changing galactic environment for the Sun and solar system. This environment affects both outer and inner planets of the solar system.

### 1.5.7 Different observable features of planet formation scenarios

Giant planets that form by gravitational instability should form rapidly so that within a few hundred years of the onset of the instability the effect of the planet should be seen e.g. in the periodic movement of young stellar object that host it from 0.1 Myr on. If they form from core accretion, an observational movement should not be visible for 10-20 Myr (Boss, 1998). About 1 Myr is calculated for our solar system to form planetary embryos of 10 Earth-masses, and about 10 Myr to accrete the up to 300 Earth-masses of nebular gas from the massive envelopes of the giant planets in a core accretion scenario. These times are relative to the epoch after the protostar has formed and the nebular has become quiescent enough to allow dust grain growth to planetesimal size and take about 0.1 Myr to 10 Myr (Papaloizous 1998). Core accretion has severe difficulty in explaining the formation of the ice giant planets, unless two extra protoplanets are formed in the gas giant planet region and thereafter migrate outward. Recently, an alternative mechanism for ice giant planet formation has been proposed, based on observations of protoplanetary disks in the Orion nebula cluster: disk instability leading to the formation of four gas giant protoplanets with cores, followed by photoevaporation of the disk and gaseous envelopes of the protoplanets outside about 10AU by a nearby OB star, producing ice giants (Boss, 2003). In this scenario, Jupiter survives unscathed, while Saturn is a transitional planet. These two basic mechanisms have very different predictions for gas and ice giant extrasolar planets, both in terms of their frequency and epoch of formation, suggesting a number of astronomical tests which could determine the dominant mechanism for giant planet formation.

### 1.5.8 Stellar metallicity planet connection

It is not yet clear if supermetallicity is a factor that favors the formation of planets or simply enhances both the number and the quality of spectral lines, making it easier to detect small amplitude variations in the radial velocity and thus detecting orbiting planets. The formation in the core accretion scenario for jovian-like and terrestrial planets should be strongly dependent on the metallicity of the parent molecular cloud. Supporting the view that metallicity is an initial condition, high metallicity in a protoplanetary disk provides more raw materials: the grains that begin the process of planetesimal formation. Higher metallicity may also enhance cooling in the disk and hence condensation of gas which may facilitate grain formation and growth. The first generations of stars in our galaxy could not have developed planetary systems like our own because they lacked these planet building blocks. Some threshold metallicity is required for planet formation, but it is not clear what that threshold might be (Fisher, 2003). The EGP found so far share orbital characteristics but also anomal metallicities (Marcy, 2003). Is high metallicity a pre-request to form a giant planet with a small orbit or does the formation of such planets lead to metal enrichment of the parent stars?

At an age of about 30Myrs the convective region of the Sun was $0.03M_{solar}$ (Sackmann, 1993). The accretion of Jupiter at that time would have increased the surface [Fe/H] abundance about 0.03dex, an accretion of 20 Earth-masses of chondrit about 0.11dex. Accretion could also result in a high surface Li abundance. The systems could also simply have formed in a metal rich cloud clump, perhaps enriched by a local supernova. High metallicity in a disk results in a rapid buildup of rocky planetesimals while much of the H and He gas is still present, available to form gas giants. The two systems with the largest minimum masses, 70Vir and HD 114762, have the smallest metallicities, what may be indicating that their companions could be BDs (Gonzalez, 1997). Siess et al. (Siess, 1998) propose that strong correlation between high Li abundance and IR excess could be explained by the accretion of massive planets or BD by a giant star.

Results from the Doppler detections suggest that stars with super-solar metallicity form planets that migrate or scatter inward. Is planet formation in these systems so efficient that it destroys solar systems like our own? Does low metallicity retard planet formation? How does metallicity of the protoplanetary disk affect planet formation? The discovery of Jupiter-like planets due to the higher sensitivity of the Doppler search should help to answer some of these questions but just asking them shows that the answer could be different if you ask about solar-system equivalents or the existence of planets in general.

## *1.6 Detected EGPs*

As of Nov 2003, there are 116 planet candidates orbiting FGKM–type main sequence stars (Marcy, 2003) (Udry, 2003). All were discovered by detecting the wobble of the host star as inferred from precise Doppler measurements. Three primary observables can be deduced from the Doppler measurements: planet minimum mass ($M$ sin $i$), orbital period (equivalently semi-major axis, $a$, from Kepler's 3rd Law), and the orbital eccentricity, $e$. Approximately 2000 nearby FGKM main sequence and subgiant stars have been surveyed, including most such stars within 50pc that are brighter than V = 8mag. The Doppler planet surveys are less complete for M dwarfs because only Keck and VLT have sufficient aperture to achieve a precision of $3ms^{-1}$ (Marcy, 2003). The following statistics always uses the minimum planetary mass (*Mj sin i*) for calculations.

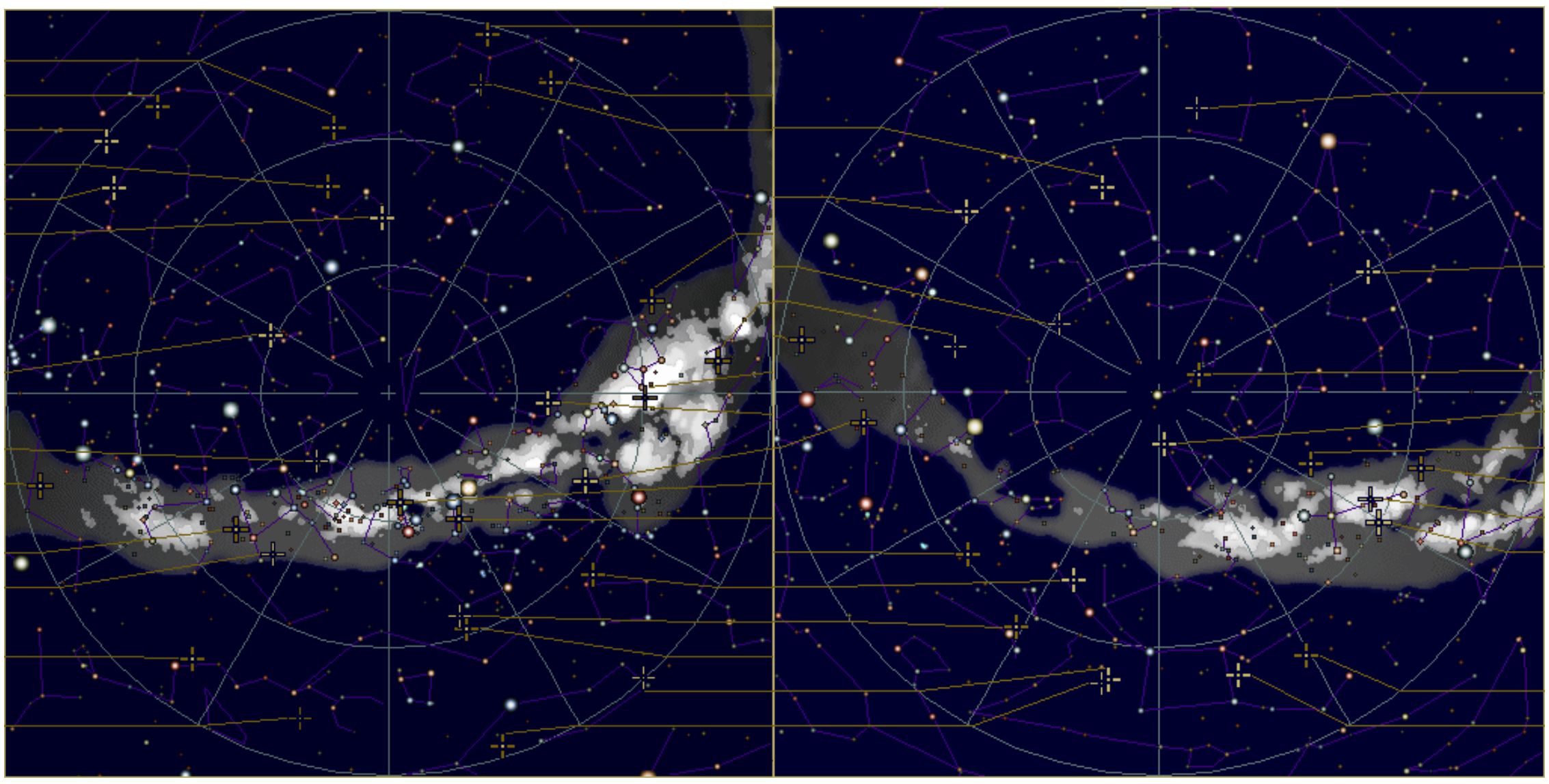

**Fig. 15: Position of some of the host stars on the northern and southern hemisphere (webplanetencyclopaedia, 2003)**

The positions of the stars harboring extrasolar planets in the sky have been compiled by several groups and can be checked for the northern and southern hemisphere (webplanetencyclopaedia, 2003) (webplanetquest, 2003). The position of some of the host star in the neighborhood of our sun can be seen in Fig.16.

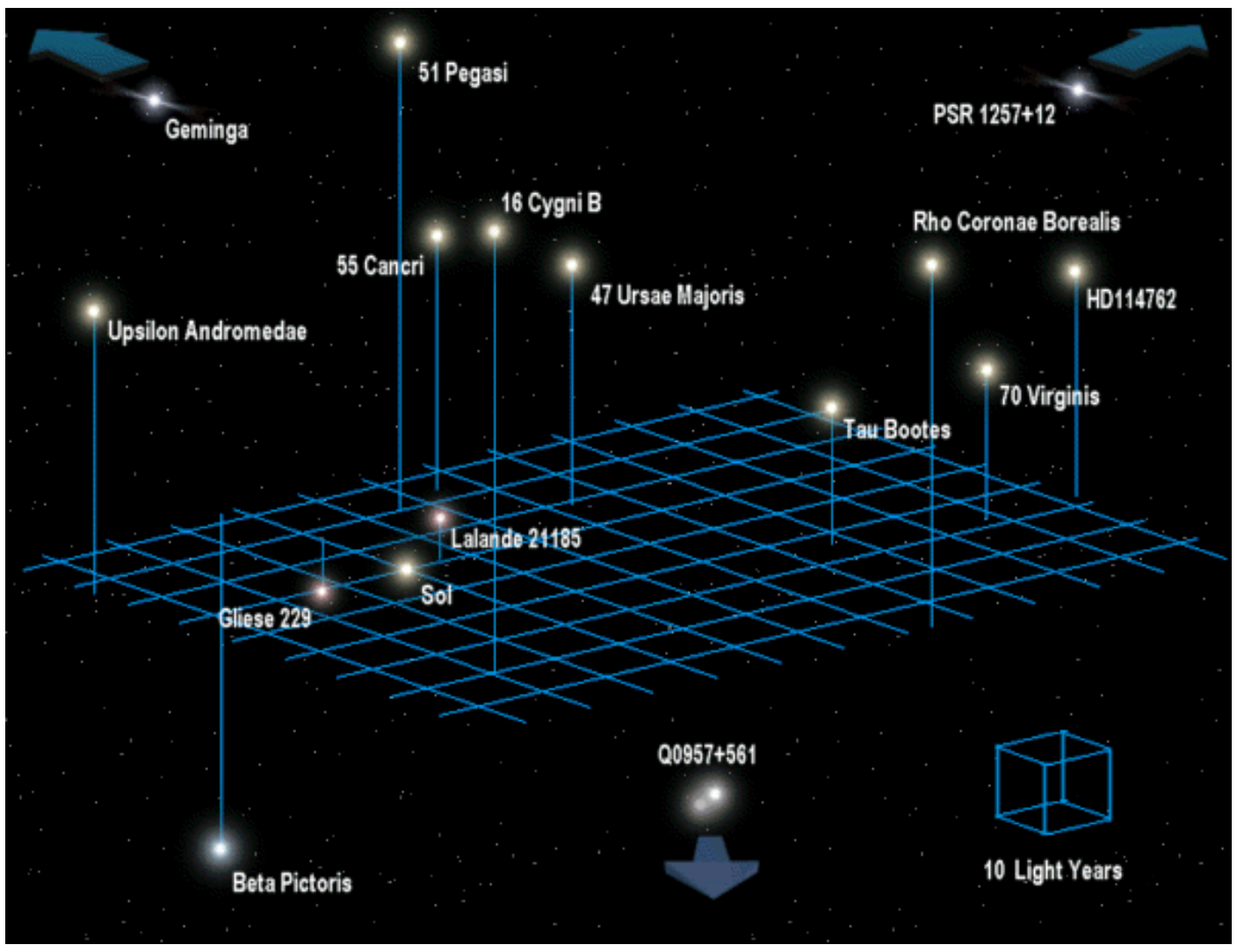

**Fig. 16: Position of some of the host star in comparison in the neighborhood of our sun (Kaltenegger, 1999).**

Despite the fact that massive planets are easier to detect, the mass distribution of detected planets is strongly peaked toward the lowest detectable masses. The period distribution is strongly peaked toward the longest detectable periods, even so short period planets are easier to detect.

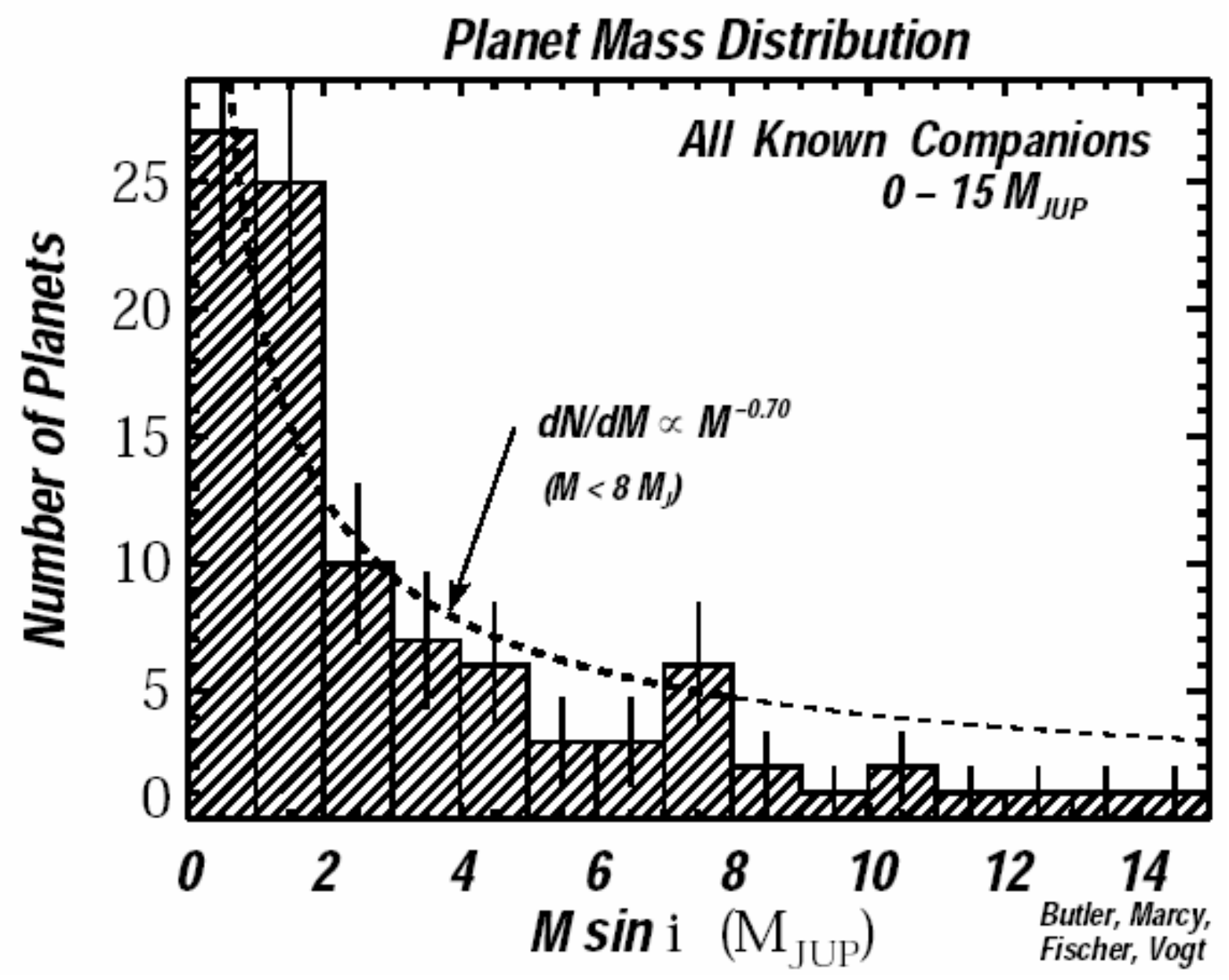

**Fig. 17: Mass histogram of 97 known extrasolar planets (Marcy, 2003)**

The majority of extrasolar planets have $M$ sin$i$ < 2$M_J$ and reside in distinctly non–circular orbits (Marcy, 2000). Updates of all known planets and their orbital parameters are provided on the web (webplanetencyclopaedia, 2003) (webplanetquest, 2003). The distribution of masses rise rapidly toward the lower masses, $dN \propto M^{-0.7}$, with the lowest $M$ sin $i$ being 0.12$M_J$ for HD49674. The distribution of orbit size reveals a minimum near $a$ = 0.3AU and an increasing number of planets in larger orbits. More extrasolar planets are known orbiting beyond 1AU than within, with signs of a large population beyond 3AU. There is an obvious lack of planets bigger than 4$M_J$ orbiting within 0.3AU, indicating that the migration mechanism is either inefficient or too efficient for them (Marcy, 2003). The orbital eccentricities of all extrasolar planets are spread nearly uniformly from $e$ = 0 to 0.7.

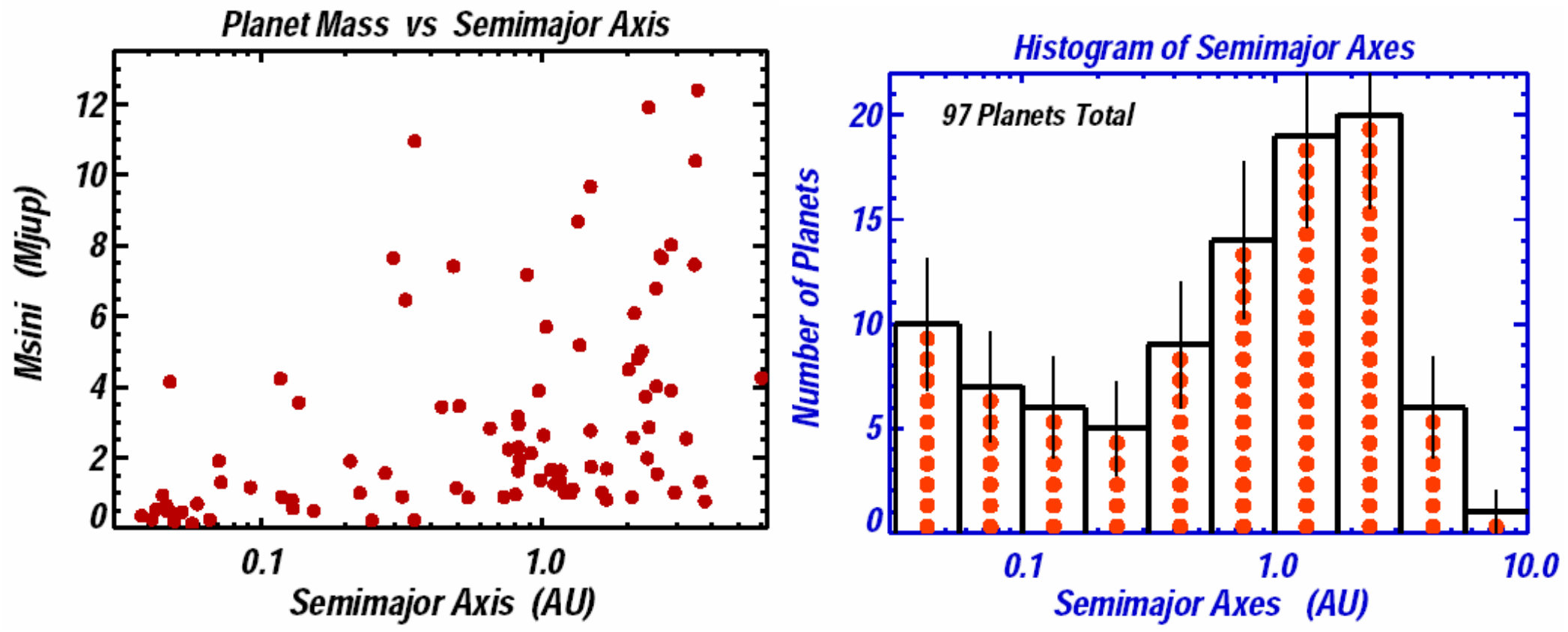

**Fig. 18: Mass versus semi major axis for known extrasolar planets (left) Distribution of semi-major axes among extrasolar planets showing a minimum of 0.3AU and a rise toward 3AU (right) (Marcy, 2003).**

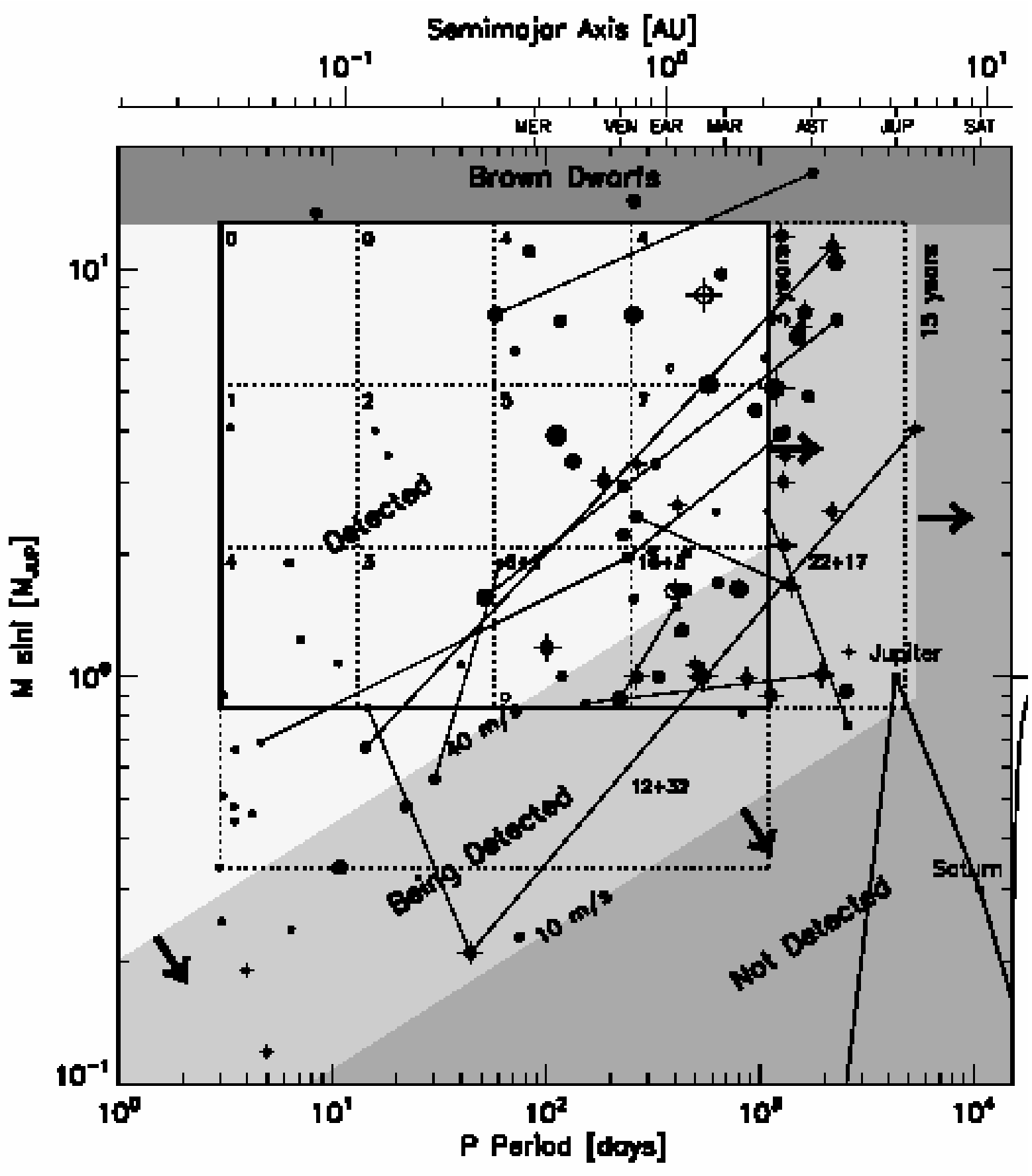

**Fig. 19: Mass as a function of period of the 101 extrasolar planets detected by August 2002 (Lineweaver, 2003)**

Radial velocity search is incomplete for orbits bigger than 3AU as the observation time for such systems is long. Until now we have been sampling only part of the extrasolar planet distribution – the only part that we are capable of sampling. If the sun were a target star in one of the Doppler surveys, no planet would have been detected around it.

### 1.6.1 Multiple planetary systems

10 multiple planetary systems are known so far, leading to the conclusion that multiple planetary systems are very common. Here only two multiple systems are selected to show some of the information their characteristics can give us concerning the planetary formation scenario.

### 1.6.2 Upsilon Andromedae

Chiang et al. (Chiang, 2003) call this system a Rosetta stone in planetary dynamics. It was the first multi–planet system discovered. Upsilon Andromedae is a sun-like star harboring at least three planetary companions. It is unique as it combines many of the surprising architectural features of extrasolar planetary systems found: a hot Jupiter, two planets on highly eccentric orbits, and a stellar companion. Planet-disk interactions seem critical to the emerging story. Fig. 20 shows the orbit fit of the three planets orbiting Upsilon Andromeda where the measured velocity is shown as dots and the best-fit model as a line (Marcy, 2004).

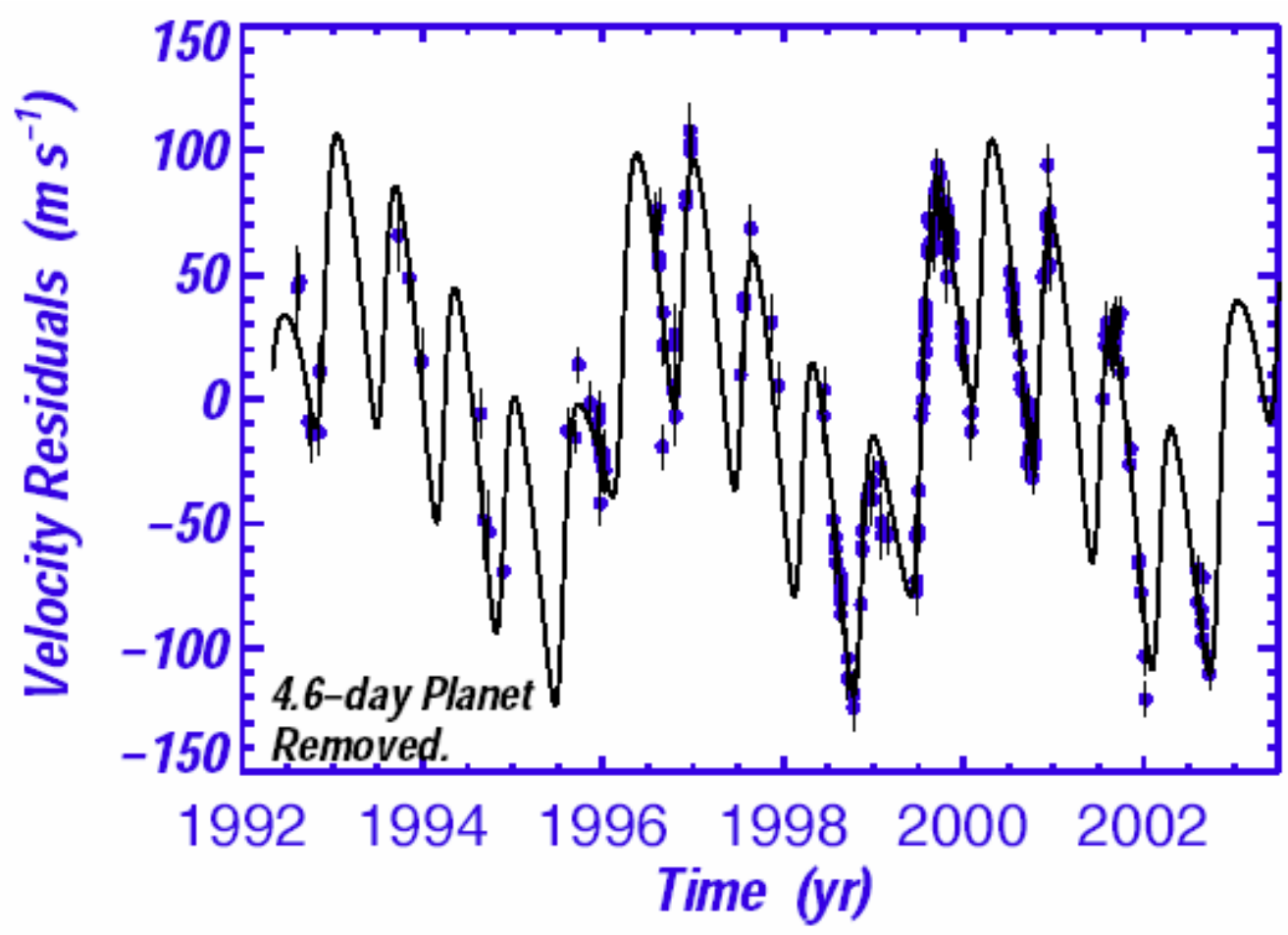

**Fig. 20: Orbit fit of the three planets orbiting Upsilon Andromeda (measured velocity: dots, best fit model: line) (Marcy, 2003)**

Fig. 21 shows the flux calculated in respect to its parent star for Upsilon Andromedae (Burrows, 2004). It shows that the detection of EGPs should be feasible even for low rejection of the stellar light.

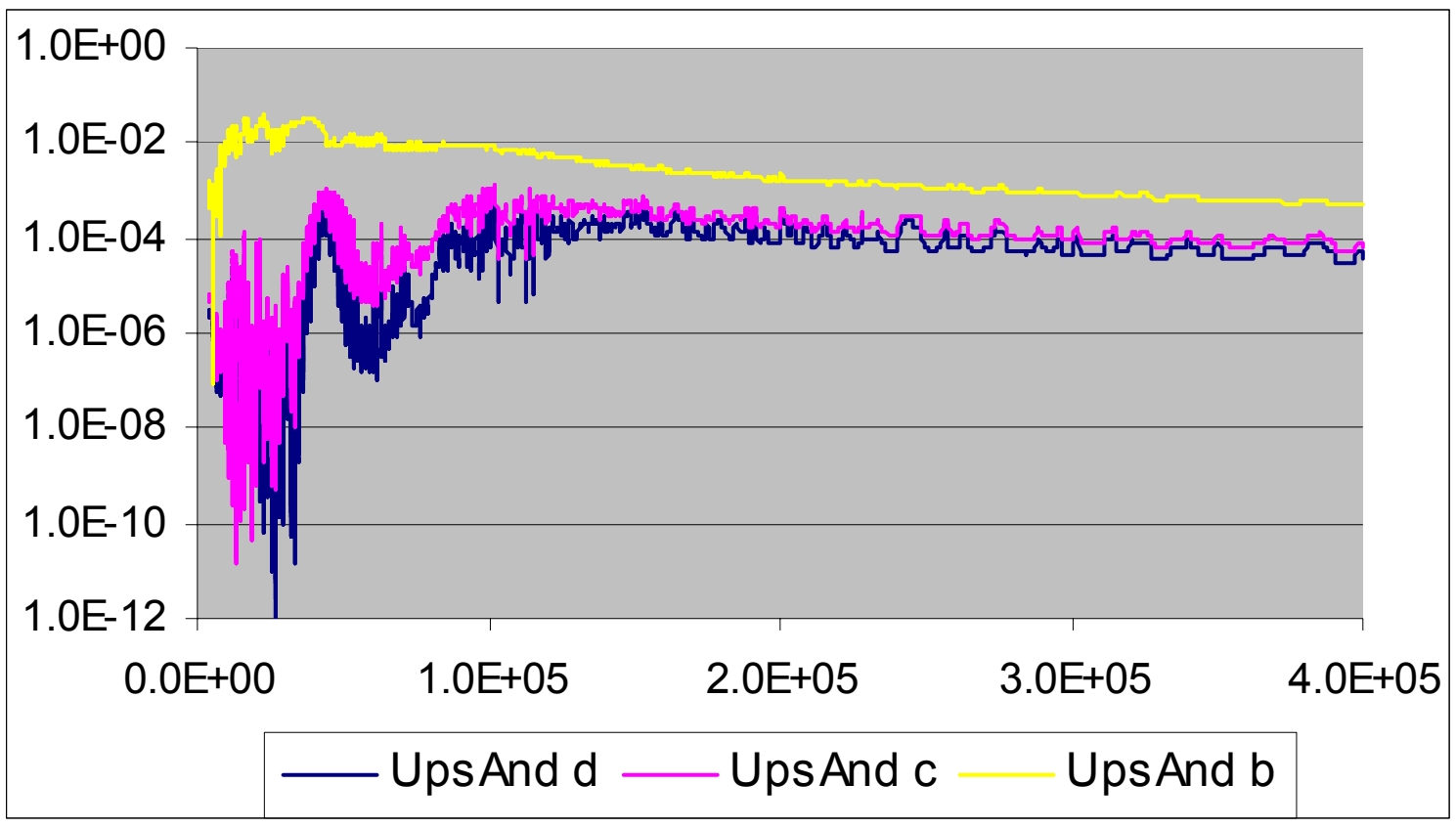

**Fig. 21: Calculated flux of the EGP in the UpsAnd system in respect to their host star**

**(data by Burrows, 2004)**

### 1.6.3 55 Cancri system

Recent studies of the 55 Cancri system suggest the existence of three planets with periods of about 15 days, 45 days, and 5500 days (Marcy, 2004). The inner two planets are near the 3:1 mean motion resonance. Theories exist that the two planets became trapped in the resonance while further from the star and migrated together (Novak, 2003). As the innermost planet began to dissipate energy through tides the planets broke out of the resonance. The outer planet in the 55 Cancri system is the first extrasolar planet discovered to orbit beyond 5AU from its star with a nearly circular orbit ($e < 0.2$). It has at least 4 time Jupiter's mass and the system with its two inner Jupiter–mass planets exhibits a very different architecture to our solar system (Marcy, 2003). A two–planet fit to 55 Cancri leaves residuals that exhibit a periodicity of 44 days, possibly caused by a third planetary companion. However, the rotation period of the star is 35–42 days (Henry, 2000). Thus a danger exists that the 44–day period in the velocities may be caused by stellar surface inhomogeneities rotating into and out of the view across the hemisphere.

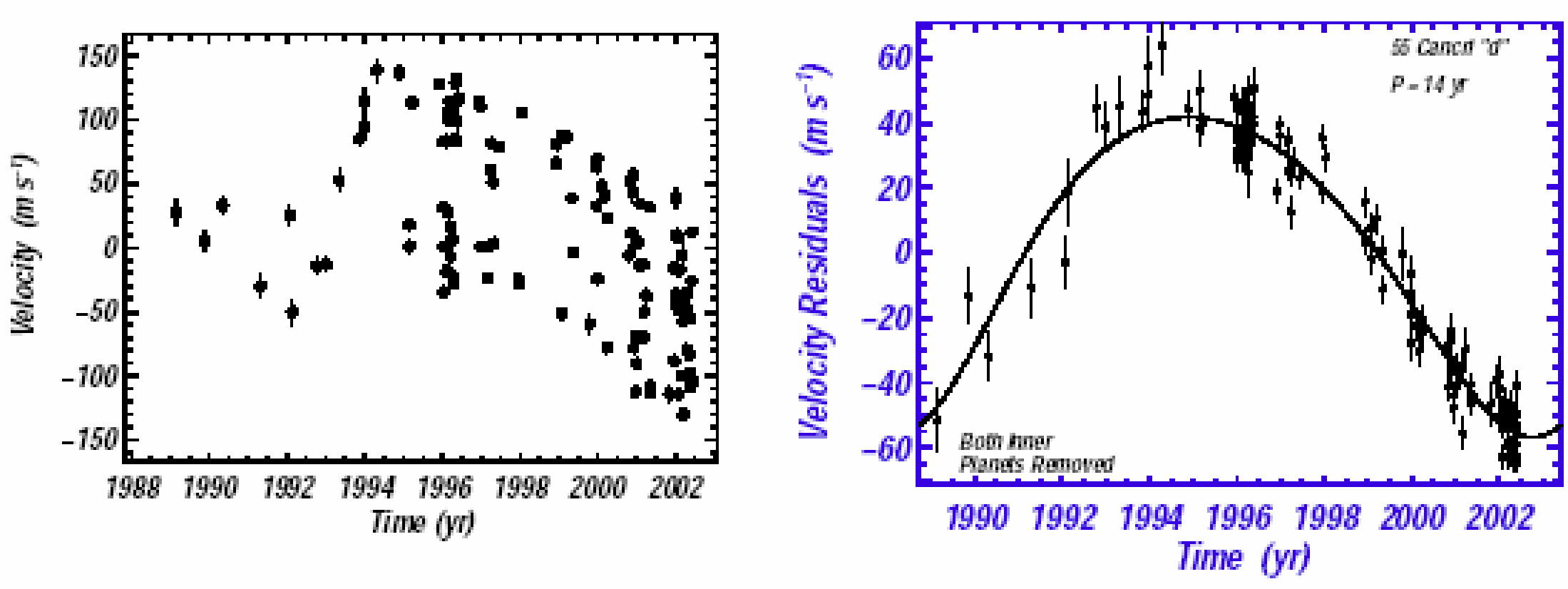

**Fig. 22: Raw velocity versus time data (left) of the three planets orbiting 55 Cancri (Marcy, 2004)**

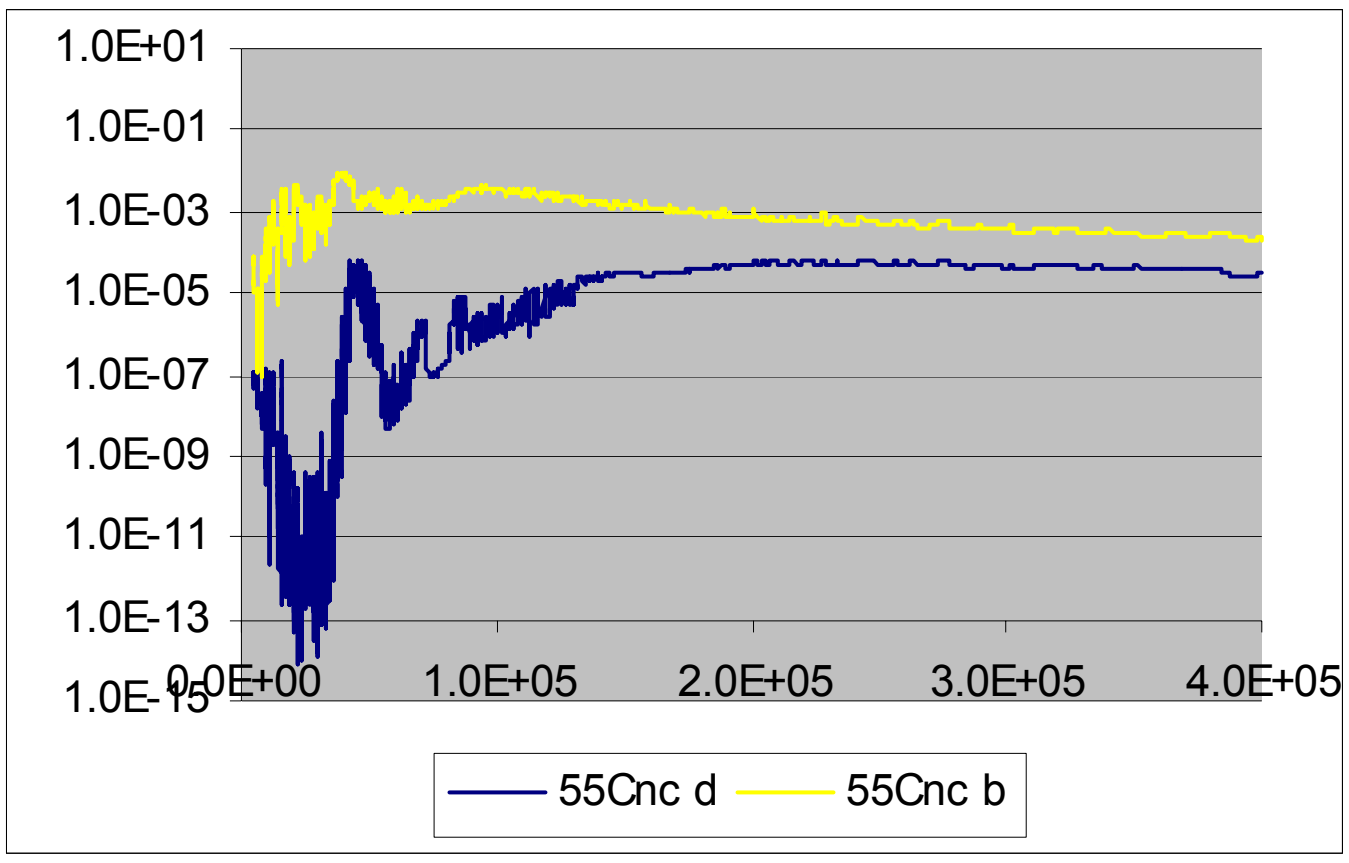

**Fig. 23: Calculated flux of the EGP in the 55Cnc system in respect to their host star**

**(data by Burrows, 2004)**

The disk around 55Cancri is relatively dark at 2.3µm consistent with absorption of light by methane ice on the dust particles. The disk inclination on the plane of the sky is measured as about 27 degrees (Trilling, 1998B). Assuming that the disk is coplanar with the planet, the planets' masses were determined.

### 1.6.4 Planets in multiple star systems

To date, several planets have been discovered in known multiple star systems. A fraction of stars known to host planets exhibit a drift in the systemic velocity indicating the presence of an additional distant companion (Fischer, 2001). These observations show that planets form and survive in certain types of multiple systems (Eggenberg, 2003). The different models of planet formation in multiple star systems are not yet conclusive.

# 2 Target selection for DARWIN

In this chapter we discuss the target star catalogue for DARWIN that was established from the HIPPARCOS catalogue in the scope of this PhD (Kaltenegger 2003). The characteristics of the stars in the target list influences the configuration of the interferometer architecture, as it will be optimized to detect planets in the habitable zone around the selected target stars. It also crucial to calculate the time needed to detect a planet as well as do spectroscopy to determine the atmosphere composition and thus the amount of targets observable during the mission lifetime. The target star list presented is used for the trade off studies of different mission architectures for the proposed DARWIN mission. During the DARWIN feasibility study (Fridlund, 2000), Leger and Ollivier had identified a preliminary list of target stars for the DARWIN mission from the Gliese star catalogue for first calculations that contained a smaller number of stars.

In preparation of the DARWIN mission we established a catalogue of stellar systems of nearby stars that are potentially habitable to complex life. The list was created from the Hipparcos catalogue by examining the information on distance, stellar variability, multiplicity, location and spectral classification. The DARWIN target star list includes 826 stars including and 625 stars excluding multiple systems tagged by HIPPARCOS and 488 stars excluding multiple system entries in the CCDM catalogue. The target stars will be observed by a ground based nulling experiment (GENIE) that will be implemented at the VLTI in 2006. A target list for DARWIN targets observable with GENIE has also been created in the scope of this PhD. The results of these observations will influence the final sample of stars selected for the DARWIN mission. Combined with theoretical studies on habitable zones, we calculated the angular separation of potential habitable planets from their parent stars, determining the equivalent distance of a 1AU, $a_{Hd}$, Earthlike planet for the target systems. To determine the equivalent habitable zone we use: 0.8 $a_{Hd}$ < d < 1.7 $a_{Hd}$, see section 6.4. Further modeling is needed to include the extent of the whole possible habitable zone, depending on atmospheric conditions, in the target catalogue. This catalogue will have to be modified when we learn about individual objects, the current analysis results in a list of 826 and 625 suitable target stars including and excluding multiple systems respectively, consistent with the selection criteria within 25pc of our sun. The data and important parameter for the DARWIN mission design are summarized here.

To model realistic observation scenarios for nulling interferometry, the luminosity, distance and radius of the star and the corresponding habitable distance (Hd) for an Earth-like planet has to be taken into account. The starlight suppression and thus planet detection as well as the optimization of the mission baseline depend critically on those parameters. The angular extent of the HZ is important, as the design of the interferometer has to be optimized to provide the highest throughput at that distance. The minimal and maximal value the instrument has to be able to cover is given in the sections on the different target star groups. Furthermore a realistic target catalogue compiled for DARWIN allows investigating the selected stars further and determining the best targets for the search for Earth-like planets. Using the SIMBAD interface, suitable targets for DARWIN were compiled from the Hipparcos main catalogue. Details of the DARWIN target catalogues that consists of F, G, K, M stars are shown below, further details on the closest stars can be found (Kaltenegger, 2003B). The target stars have B-V colors indicating main sequence stars and are additionally classified by the SIMBAD data archive as luminosity class V (main sequence) or undetermined classes.

Until now the starlight suppression was calculated using the Sun as standard at a distance of 10pc. It is a good estimate for first calculations. More detailed simulations will critically depend on correct data on the target stars. Optimizing the mission for a sun at 10pc, a star with a radius smaller than that of the close target stars, would thus overestimate the null depth achieved by DARWIN. Also the extent of the Hd depends critically on the stellar type. As DARWIN will be optimized to look for planets in the Hd around a star, the correct values should be included in meaningful simulations. Using G-type stars as standard would overestimate the angular extend of the Hd for K and M stars while underestimating it for F stars.

## 2.1 Selection Criteria

The selection criteria for the stars from the HIPPARCOS catalogue lead to a total of 826 stars. The criteria are listed below:

- distance smaller than 25 pc for F,G,K, M stars
- coordinates between –45 and +45 ecliptic declination (refers to the 45 degree cone in anti sun direction)
- B-V index for main sequence stars
  F: 0.25 < B-V < 0.58, G: 0.52 < B-V < 0.81, K: 0.74 < B-V < 1.40, M: 1.33 < B-V < 1.93
- Stellar luminosity class (main sequence: incl. also sample with undetermined luminosity class)
- magnitude smaller than 12 in V (HIPPARCOS is complete to up to $8^{th}$ magnitude) as DARWIN concentrates on brighter/closer star the completeness of the sample is not an issue as it only concerns dim stars that are ranked to lower priority in the star sample

The Hipparcos mission's typical parallax standard error of about 1mas allows very precise distance measurements. The standard error is about 20% or less for about 50000 stars. The luminosity calculations are based on the star's spectral type, and hence used to determine the habitable zone around the target star systems. The Hipparcos catalogue also includes accurate photometry (B-V uncertainty typically less than 0.02mag), proper motion data as well as information on variability and multiplicity. Using all this data we established a catalogue of well characterized stars that will be used as the DARWIN target star list. For many of the nearby stars data on multiplicity, precise stellar type, rotation, metallicity etc are still missing, thus modifications of the DARWIN catalogue will be implemented as such data becomes available.

The 45-degree cone in the selection criteria is given by constrains of the sunshields on each free flyer. The shields will be designed to permit a +/-45 degree cone of observation in anti-sun direction. That also influences the maximum time during which a target can be observed. If it is located in the ecliptic, it can be seen for ¼ of a year (as from entering to exiting the observation cone, it passes 90 degrees of a 360 degree orbit). The further away from the ecliptic it is located, the less time it can be observed. Fig.22 shows the dependence of the possible observation time to the coordinates of the source. From the graph it can be seen that the targets up to 42 degrees ecliptic latitude can at least be observed for 30 days, what should be sufficient to allow spectroscopy on the detected planets. The time needed to investigate the detected planets will depend on the composition of its atmosphere, the distance to its parent stars, the brightness ratio between the host star and the planet as well as the brightness of the planet. Different groups model these parameters. The conclusion of these models will influence the required observation time for each planet but should rarely exceed 1 month.

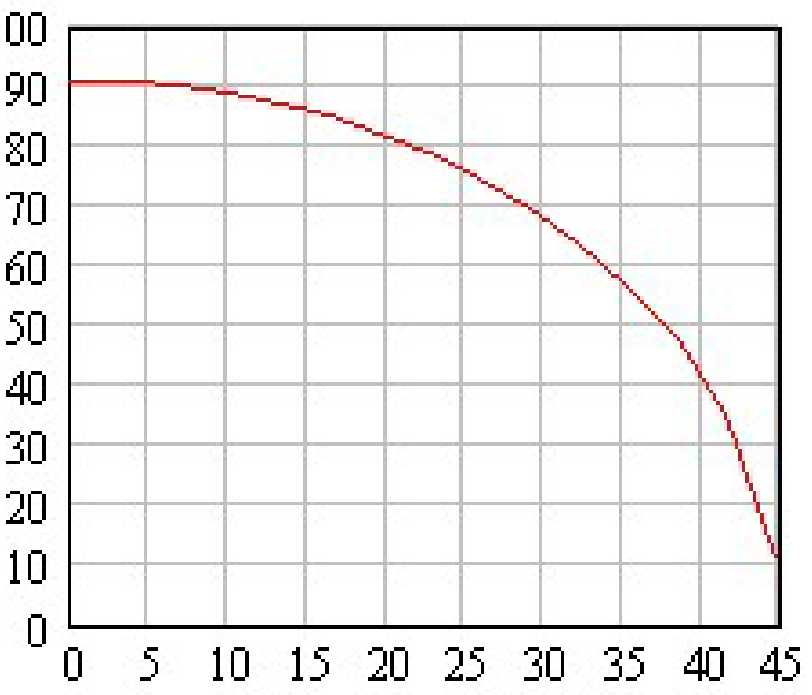

**Fig. 24: Maximum observation time in days depending on the ecliptic latitude of the star. Small difference from the formula used before for the industrial DARWIN study were found but the resulting function leads to similar results.**

The maximum time, $t_{max}$, a star can be observed depending on its ecliptic latitude $\upsilon$ is given by equation (2.1.1):

$$t_{max} = t_0 \cdot \cos\alpha \qquad (2.1.1)$$

$$\alpha = \arcsin\frac{\delta}{\theta} \qquad (2.1.2)$$

The equations relate the ecliptic latitude of the star $\delta$ to $\cos\alpha$, and thus the time the star will be seen in the circular FOV. $\theta$ is the opening angle of the +/- 45-degree cone in anti-sun direction. $t_0$ is the maximum time a star can be observed in the cone, given at ecliptic latitude 0. $\delta$=0 leads to an observation time of 365/4 = 91.25 days as such a target star can be seen for 90 degrees of the 360 degree orbit, thus ¼ of its orbit.

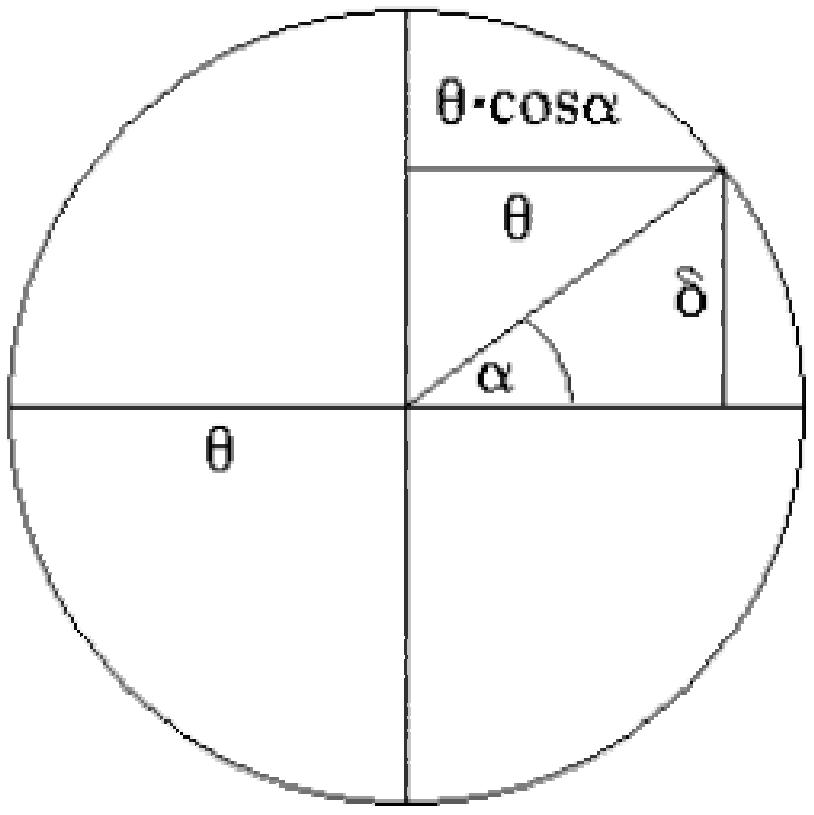

**Fig. 25: Sphere used to calculate the maximum observation time of a target star**

The total sky coverage is given by $\sin\upsilon$, a result from integration of the sphere segment over +/- 45 degrees

$$SkyCoverage = \frac{2\pi \int_{-90}^{+90} R_K \cos\theta \cdot \partial\theta \cdot R_K}{2\pi R_K^2} \qquad (2.1.3)$$

This results in 70.7% sky coverage for a +/-45 degrees cone. In Fig..24 the parts V1 and V2 show the volumes that cannot be observed per half sky sphere. Rotating the half sphere shows that the unobservable part of the sky is a cone in each of the two hemispheres covering together 29.3% of the sky. The DARWIN target stars are all chosen from the observable fraction of the sky.

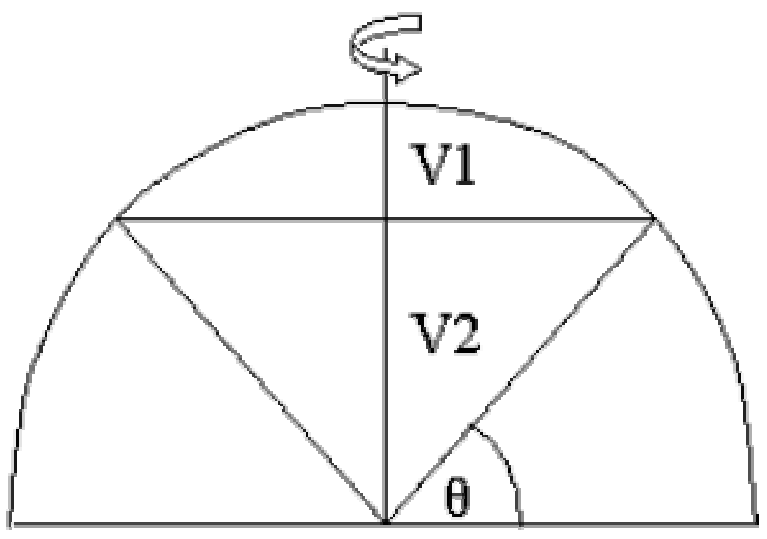

**Fig. 26: Sphere used to calculate the Sky Coverage of the 45-degree cone**

The calculations show that the limitation to a +/-45 degree cone in anti-sun direction places no severe constraint on the mission or limits its performance. The sky-coverage of 70.7% leaves us with a sample of 826 suitable target stars that can be observed for at least 30 days for up to +/-42 degrees.

## 2.2 Different issues discussed for star target selection

❑ A thought on priority of stellar types

Naturally G stars will turn out to be prime targets as they are similar to our sun and the likelihood that under similar conditions, similar planets with similar life-form showing similar atmospheric features evolve seems most likely. That will not exclude other star from the target catalogue but will make the G stars a natural prime target sample. Including other stellar types provides the unique opportunity to characterize planets around different host stars. EGP have been found around different stellar types thus the target sample should reflect those stellar classes. It is essential that we have a certain number of targets stars to derive conclusions for detections or non-detections. Even so the frequency of extra-solar terrestrial planets can only be guessed, we believe that a sample has to consist of about 100 to 150 stars to be able to derive conclusions.

Investigations by e.g. Lammer et al. have questioned whether M stars are good targets for habitable planets (Lammer, 2003). Also other groups have investigated the activity issue of the different stellar types as well as the possibility to keep atmospheres from freezing out, what might be the effect of tidal locking. We should be aware of the issues but not exclude stars on that base, as the theories are not conclusive yet.

❑ Distance limit of stars

The important point concerning the limiting distance of the host star is the flux level we receive from the planet. We have limited the distance at 25pc for the host stars and have a significant number of stars within that distance. An IR free flyer interferometer can adjust the configuration to the target star observed, thus we can adjust to the extent of the stellar disk and HZ for individual stars.

❑ Metallicity: unconventional thought

One should keep in mind that we should not exclude stars with lower metallicity, as we do not know yet if stars with a terrestrial planetary system form like the systems hosting EGP. The possibility exists that the presence of EGPs destabilizes the orbit of terrestrial planets.

❑ Binaries

Due to the technique we will use to detect planets (whether it is nulling interferometry or coronography using a mask) a second star at a certain distance will swamp the planetary signal we try to detect. For different stellar types that distance varies and thus binaries should be investigated carefully. On this issue we have to take the detection technique into account, because whether planets exist around the binary will not be an issue as we will simply not be able to detect them.

❑ Stability

Most of the stars closest to our sun are variable star. It will be an issue to determine what variability we can accept and still keep the conditions habitable for Earth-like planets. That issue will have to be investigated with atmospheric models of planets. We should be ready for surprises on what reactions exist to stabilize habitable condition on planets including reactions we might not be aware of now as it is outside of the temperature pressure range we are able to reproduce. It is a very important issue we need to investigate.

❑ Age

Young stars determined trough x-ray emission and e.g. Li abundance should be tagged in further studies in the target catalogue. The point that the zodiacal cloud is bright if the system is young is correct but also the planets are still brighter especially for missions in the IR, thus it should not provide a problem for detection. The formation of Earth-like planets takes time thus young stars should be secondary targets. As we also will investigate planets in the different formation/evolution stages to understand planet formation young stars are an essential part of our target sample.

## 2.3 Habitable distance

The Habitable Zone HZ around a star is defined as the zone around a star within which starlight is sufficiently intense to maintain liquid water at the surface of the planet, without initiating runaway greenhouse conditions that dissociate water and sustain the loss of hydrogen to space, see J. Kasting et al. (Kasting, 1997) see section 6.4. The planet's effective temperature $T_p$ depends on the temperature $T_*$ and thus brightness of the star, the planet's albedo $A$ and its distance to the star $a_p$.

$$T_p = \frac{(1-A)^{1/4} T_*}{\sqrt{2}} \left( \frac{R_*}{a_p} \right)^{1/2} \tag{2.3.1}$$

We determine the Habitable Distance (Hd), as the distance where a planet like Earth-would encounter the same effective temperature, equivalent to a distance of 1AU around the sun. The calculation for the radius of the Hd is based on the luminosity of the star. It is a simplified model based on work by Kasting et al. (Kasting, 1993).

$$a_{HD} = \left( \frac{T_*}{T_\Theta} \right)^2 \frac{R_*}{R_\Theta} = \left( \frac{L_*}{L_\Theta} \right)^{1/2} \tag{2.3.2}$$

For the Hd of the target star presented we use the distance corresponding to Earth's distance around the Sun, 1AU, and Earth's albedo for calculations. The HZ is defined as a zone that extends further outwards as well as inwards from that distance. While a debate about the extent of the HZ around different stars is ongoing, Kasting's work is used in this document to determine the standard HZ. Other models are being established. Once they become conclusive the calculations can be updated.

## 2.4 Detailed information on the stars

The main characteristics included in the star catalogue are listed below

| | |
|---|---|
| HIP | HIP Identifier (HIP number) |
| HD | Henry Drapper catalogue identifier (HD number) |
| d[pc] | distance in parsec |
| SpType | Spectral Type |
| B-V | B-V color to ensure main sequence type |
| RAJ2000 | Right ascension Equinox=J2000.0 Epoch=J2000, |
| DEJ2000 | Declination Equinox=J2000.0 Epoch=J2000 |
| el lat | ecliptic latitude of star |
| el lon | ecliptic longitude of star |
| Glon | Galactic longitude Epoch=J2000, |
| Glat | Galactic latitude Epoch=J2000, |
| Rahms | Right ascension in h m s, ICRS (J1991.25) |
| Dedms | Declination in deg ' ", ICRS (J1991.25) |
| Vmag | Magnitude in Johnson V (H5) |
| RA(ICRS) | degrees (ICRS, Epoch=J1991.25) |
| DE(ICRS) | degrees (ICRS, Epoch=J1991.25) |
| Plx | Trigonometric parallax |
| MultFlag | Multiple star flag to identify multiple systems |

Linear interpolation between stellar luminosity classes were used to calculate the effective temperature, stellar radius, stellar luminosity magnitude in K, L, L', M and Hd (in rad, AU), the distance where the sensitivity of the transmission map should be set to constructive interference.

| | |
|---|---|
| L/Lo | luminosity in units of sun luminosities |
| R/Ro | stellar radius in units of sun luminosities |
| $T_{eff}$ | Effective Temperature of the Star |
| K, L, L', M | K, L, L', M magnitude of the star |
| Rstar [rad] | stellar radius in radians |
| Hd [AU][rad] | Distance equivalent to 1AU around the Sun in AU and radians |

The multiple flag in the Hipparcos catalogue indicates that further details are given in the Double and Multiple Systems Annex: C : solutions for the components, G : acceleration or higher order terms, O : orbital solutions, X : stochastic solution (probably astrometry binaries with short period).

# 3 Details on the DARWIN target stars and overall statistics

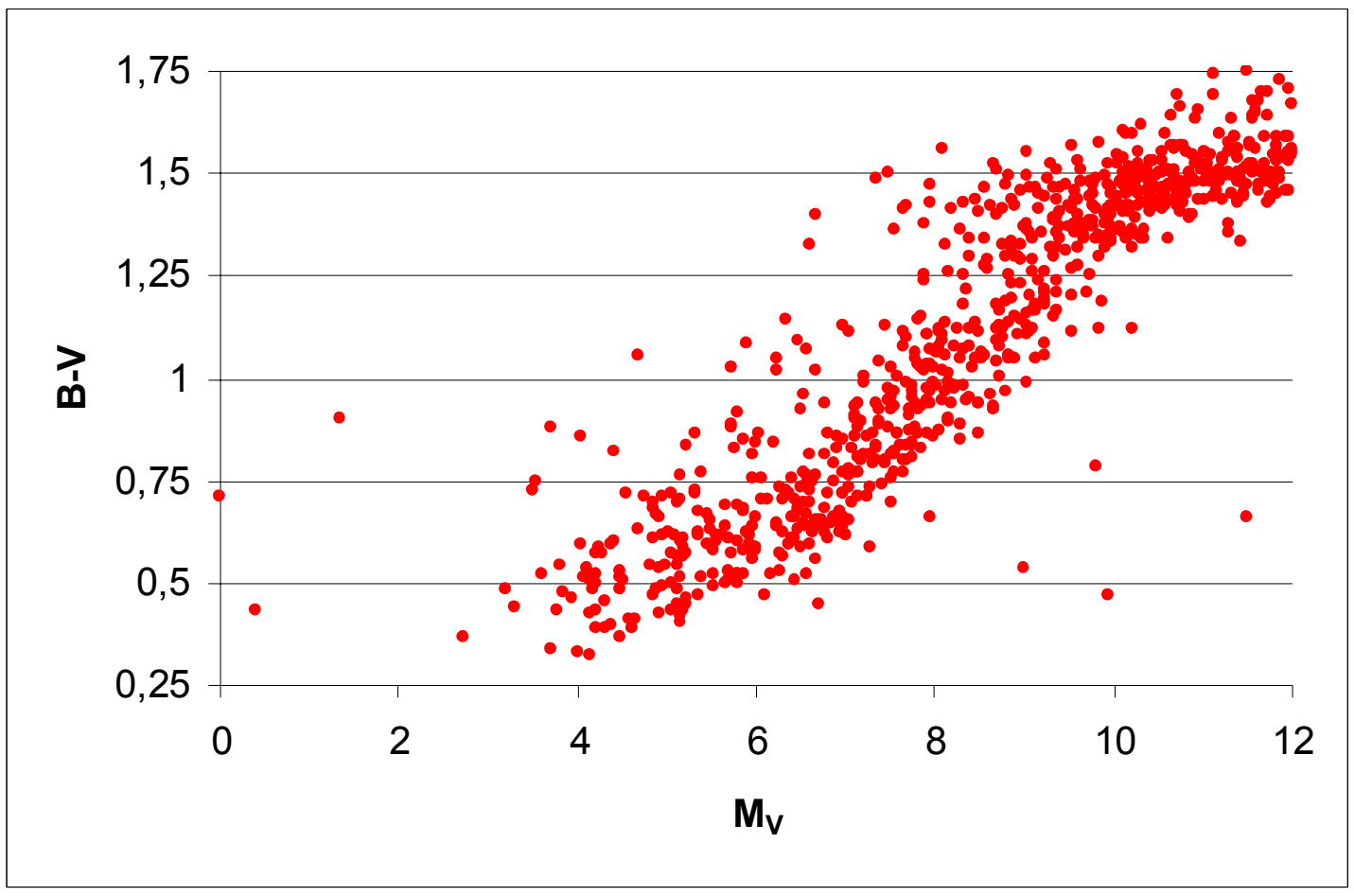

**Fig. 27: Color-magnitude diagram for the DARWIN target stars.**

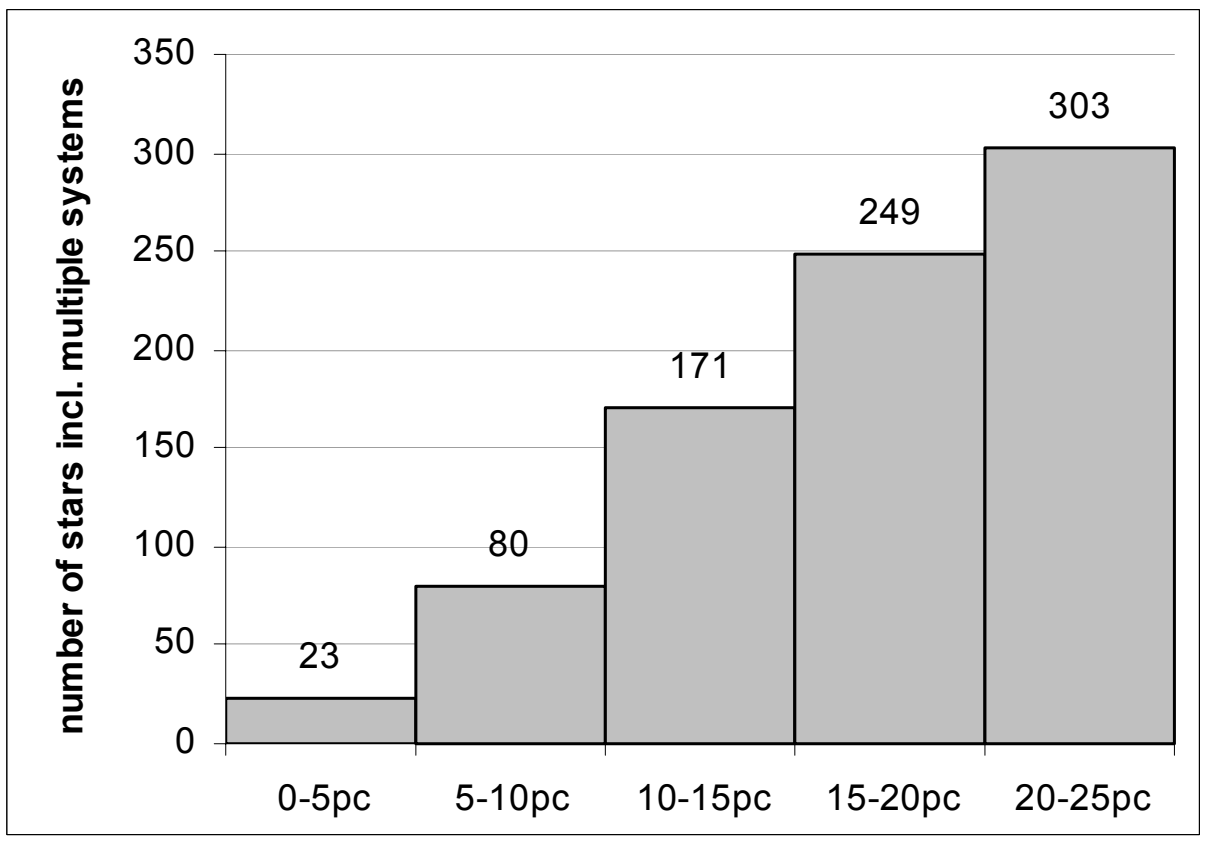

**Fig. 28: Number of target stars as a function of distance**

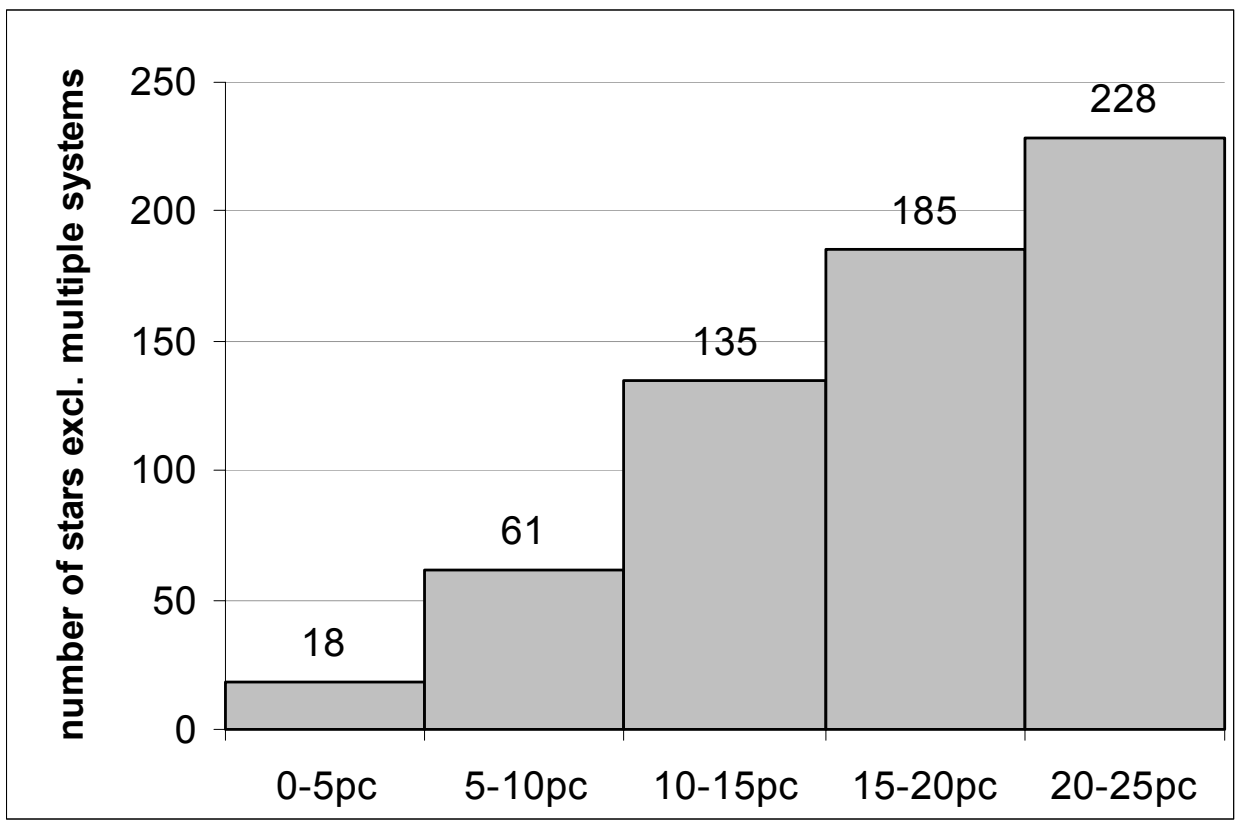

**Fig. 29: Number of target stars as a function of distance excluding multiple stars**

Overall the target lists consists of 826 stars including multiple systems and 625 stars excluding multiple systems tagged by HIPPARCOS. Further investigation of the sample using the CCDM catalogue was also conducted in the scope of this PhD see section 3.1.

74 F stars within 25pc, B-V color for main sequence stars, the closest star 3.48 pc,
143 G stars within 25pc, B-V color for main sequence stars, the closest star 1.35 pc
309 K stars within 25pc, B-V color for main sequence stars, the closest star 3.47 pc
300 M stars within 25pc, B-V color for main sequence stars, the closest star 1.29 pc.

Most of the M and K stars do not have a subclass classification, thus a main part should turn out to be non-main sequence stars as brighter stars (giant stars) are picked up more easily by observations. Note that the closest 40 target stars are not equally distributed in the different stellar classes: 1x F star, 2x G stars, 9x K stars, 28 x M stars.

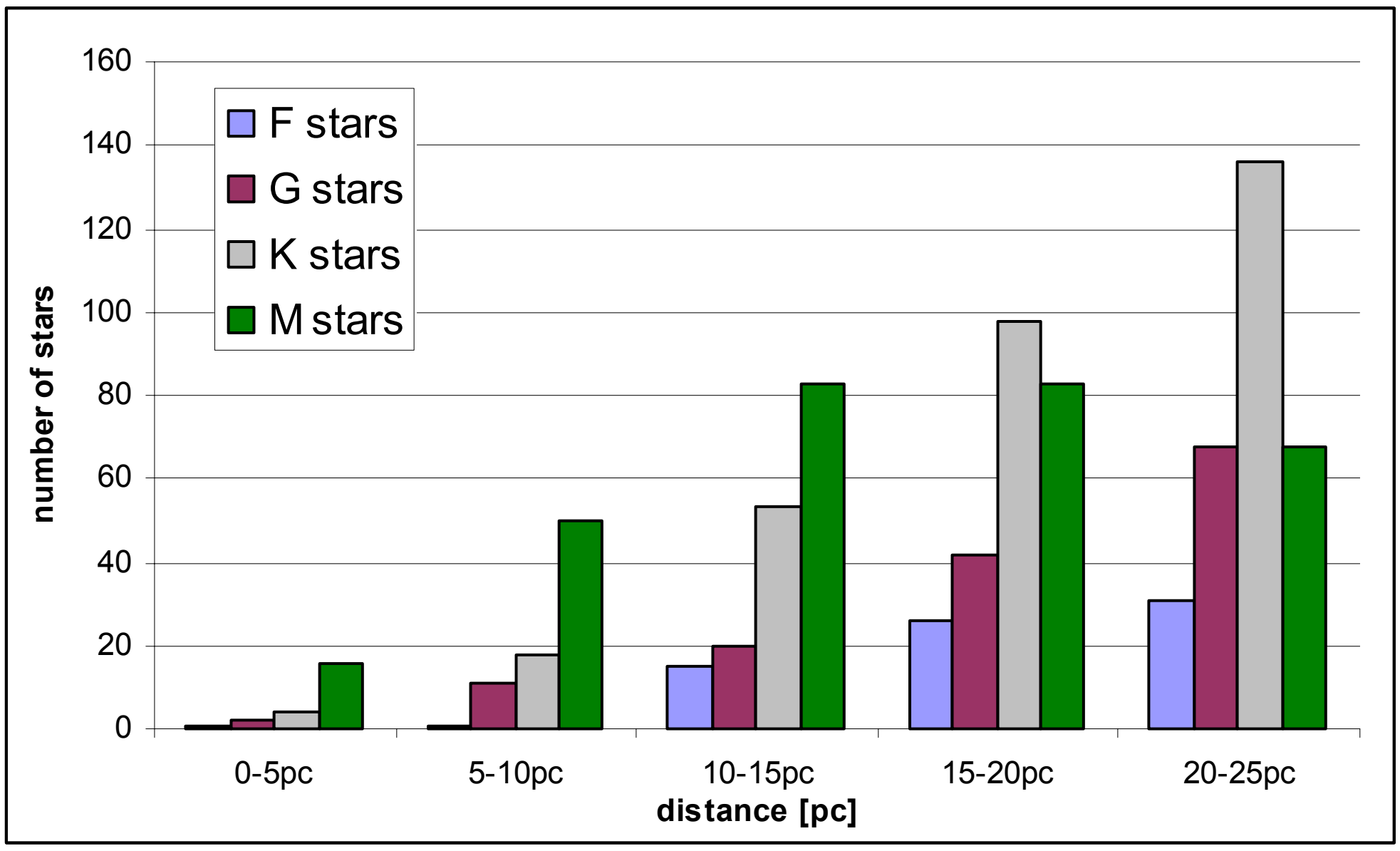

**Fig. 30: Number of prime target stars per spectral type as a function of distance**

## *3.1 Additional selection criteria*

This section shows the influence of further selection criteria on the target star sample. Changing the observation cone from +/-45 degrees anti-sun direction to +/- 25 degrees around the ecliptic, what would allow using fixed sunshields, but would reduce the star sample from 625 prime target stars to 451 stars. Prime target stars from the DARWIN target list exclude multiple stars tagged by Hipparcos. The major concern is that it would remove 5 of the 7 sun-like target stars from the list of targets. Additionally it eliminates 59 of the 112 G-type stars from the list of observable targets, thus one would loose about half of the G-type target stars.

By decreasing the angle of the observation cone one selects target stars further away to get a large enough sample of target stars, currently the baseline of the scientific mission has 165 stars. As an effect of the increased distance the required integration time increases for a given SNR. Thus fewer stars will be observable in a given time. Note that the dominant noise source, the Local zodiacal (LZ) cloud, remains constant over distance of the targets star sample. The LZ noise contribution decreases with an off angle from the ecliptic, thus additionally targets at high angles from the ecliptic are observed with a lower level of LZ noise, but the decrease is marginal.

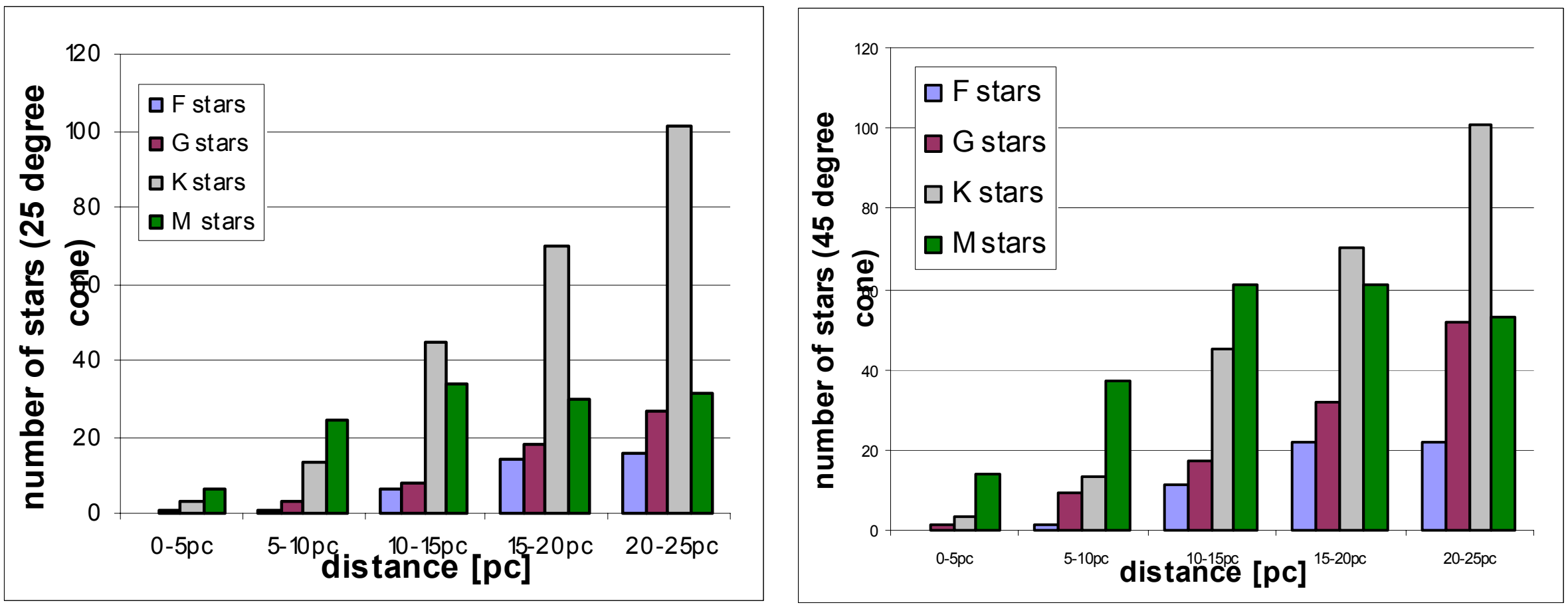

**Fig.31: DARWIN prime target star distribution versus distance for a 45 degree (left) and 25 degree cone (right)**

Further investigation of the target star sample using the CCDM catalogue to exclude all multiple components from the sample has reduces the sample to 488 prime target stars in a 45 degree cone and 280 stars in a 25 degree cone.

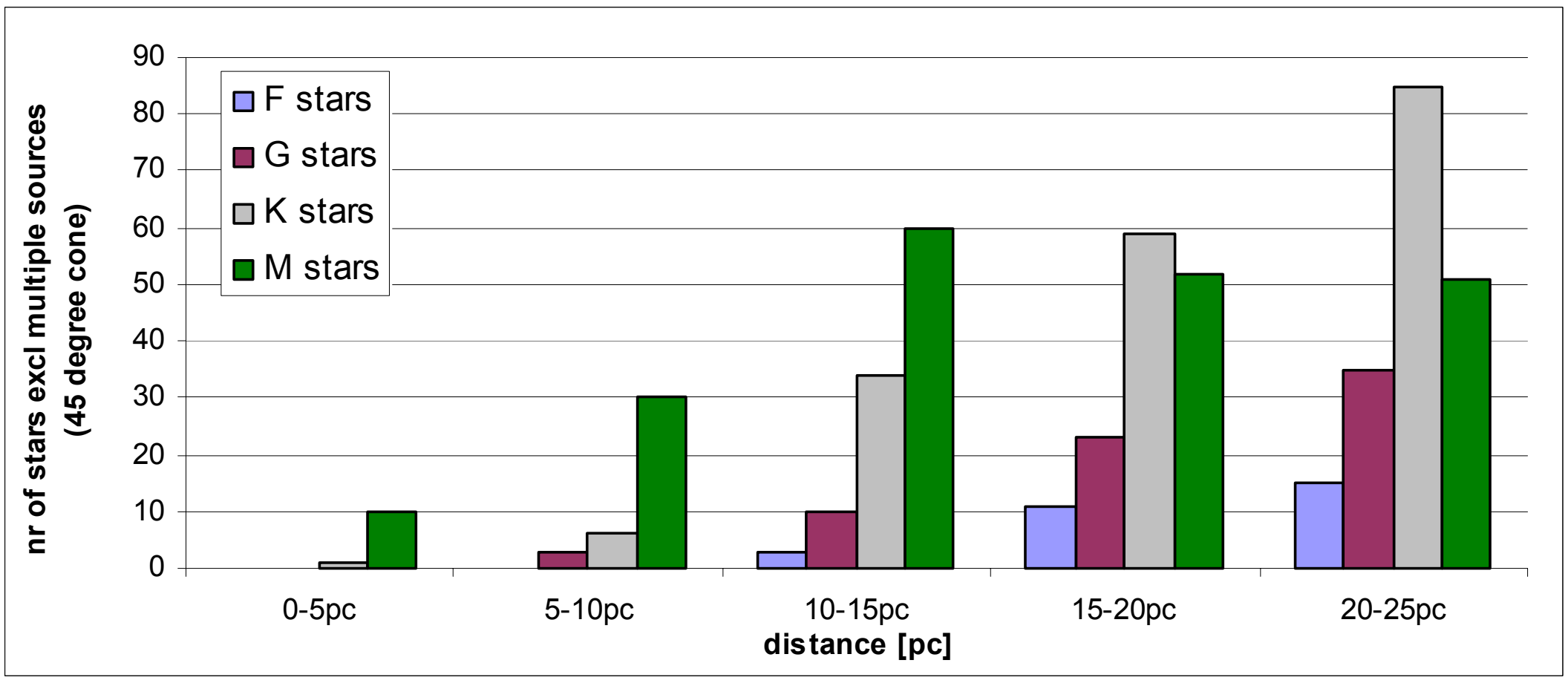

**Fig. 32: DARWIN target star distribution versus distance excluding multiple sources for a 45-degree cone**

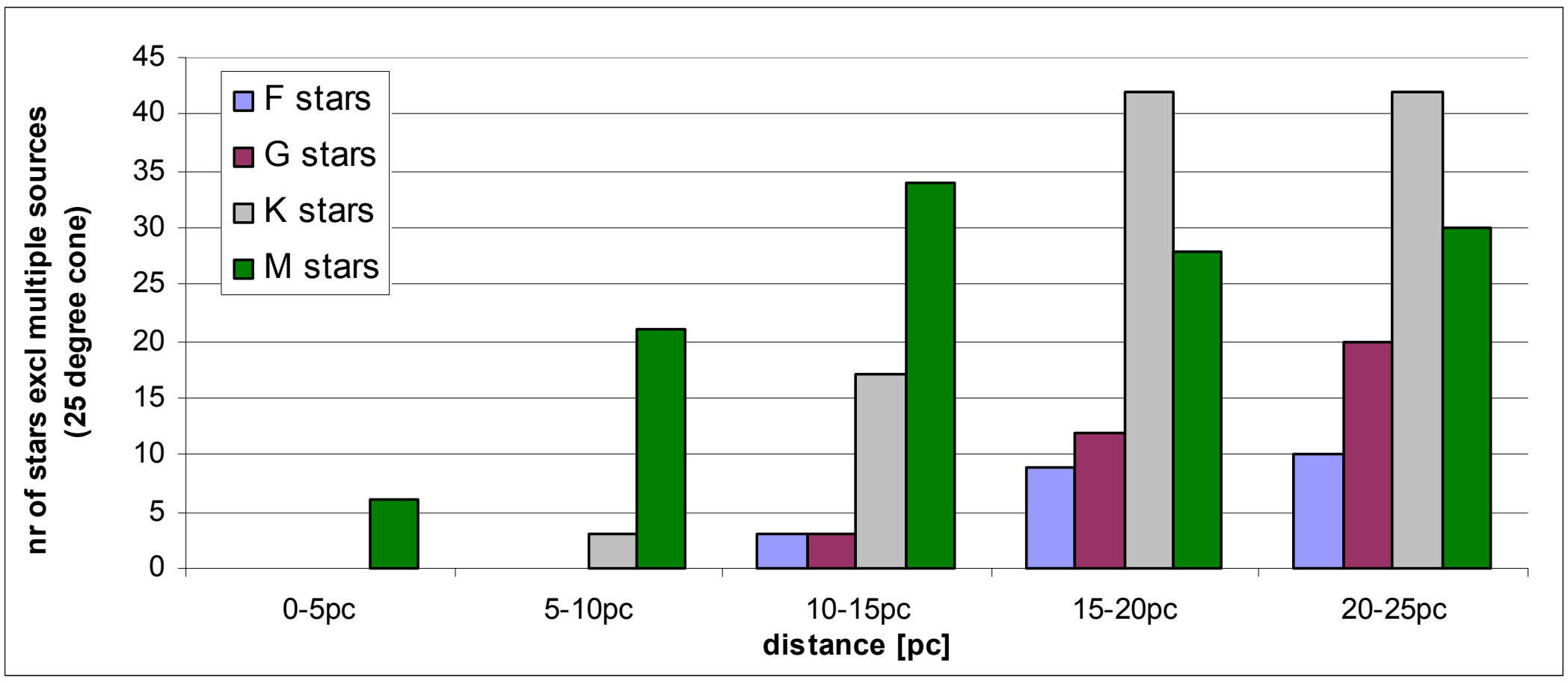

**Fig. 33: DARWIN target star distribution versus distance excluding multiple sources for a 25-degree cone**

## *3.2 Overall statistics*

The DARWIN target star sample has been compiled from the Hipparcos catalogue thus a realistic sample can now be used for simulations. Below the angular extend of the stars of the different luminosity classes, their Habitable distance (Hd), the equivalent of 1AU around other stars and also the extent of a 3AU disk around those stars is calculated. The zodiacal cloud in our own solar system has an extend of 3.2AU (Kasting, 1998). Disks around other stars have been found to extend further outwards, a few tens of AU (Chiang, 1997) (D'Allessio, 2001). We adopted 3AU as a standard value for the summary and 30AU for the following calculations. We do not model the disks, thus one has to keep in mind that disks around other stellar types will have different extents.

**Table 1: Maximum, minimal and mean values for luminosity and angular extend of the whole DARWIN target stars**

| F stars | L' [mag] | V [mag] | Hd [rad] | SR [rad] | 3AU [rad] | Hd [mas] | SR [mas] | 3AU [mas] |
|---|---|---|---|---|---|---|---|---|
| max | -0.74 | 0.4 | 2.47E-06 | 1.67E-08 | 4.15E-06 | 5.09E-01 | 3.44E-03 | 8.56E-01 |
| Min | 8.56 | 9.95 | 2.70E-07 | 2.11E-09 | 5.84E-07 | 5.57E-02 | 4.35E-04 | 1.20E-01 |
| Mean | 3.65 | 4.87 | 5.07E-07 | 3.41E-09 | 8.65E-07 | 1.05E-01 | 7.03E-04 | 1.78E-01 |
| **G stars** | **L' [mag]** | **V [mag]** | **Hd [rad]** | **SR [rad]** | **3AU [rad]** | **Hd [mas]** | **SR [mas]** | **3AU [mas]** |
| max | -1.52 | -0.001 | 3.96E-06 | 3.43E-08 | 1.08E-05 | 8.17E-01 | 7.07E-03 | 2.23E+00 |
| Min | 9.84 | 11.48 | 1.48E-07 | 1.61E-09 | 5.82E-07 | 3.05E-02 | 3.32E-04 | 1.20E-01 |
| Mean | 4.55 | 6.15 | 3.18E-07 | 2.90E-09 | 9.47E-07 | 6.56E-02 | 5.98E-04 | 1.95E-01 |
| **K stars** | **L' [mag]** | **V [mag]** | **Hd [rad]** | **SR [rad]** | **3AU [rad]** | **Hd [mas]** | **SR [mas]** | **3AU [mas]** |
| max | -0.81 | 1.35 | 1.60E-06 | 2.49E-08 | 1.08E-05 | 3.30E-01 | 5.14E-03 | 2.23E+00 |
| Min | 9.82 | 11.84 | 6.30E-08 | 1.17E-09 | 5.82E-07 | 1.30E-02 | 2.41E-04 | 1.20E-01 |
| Mean | 5.78 | 8.35 | 1.52E-07 | 2.20E-09 | 9.29E-07 | 3.14E-02 | 4.54E-04 | 1.92E-01 |
| **M stars** | **L' [mag]** | **V [mag]** | **Hd [rad]** | **SR [rad]** | **3AU [rad]** | **Hd [mas]** | **SR [mas]** | **3AU [mas]** |
| max | 2.61 | 6.69 | 4.20E-07 | 9.40E-09 | 1.12E-05 | 8.66E-02 | 1.94E-03 | 2.31E+00 |
| Min | 8.7 | 12 | 2.10E-08 | 5.00E-10 | 5.80E-07 | 4.33E-03 | 1.03E-04 | 1.20E-01 |
| Mean | 6.05 | 10.53 | 9.16E-08 | 1.83E-09 | 1.27E-07 | 1.89E-02 | 3.77E-04 | 2.62E-01 |

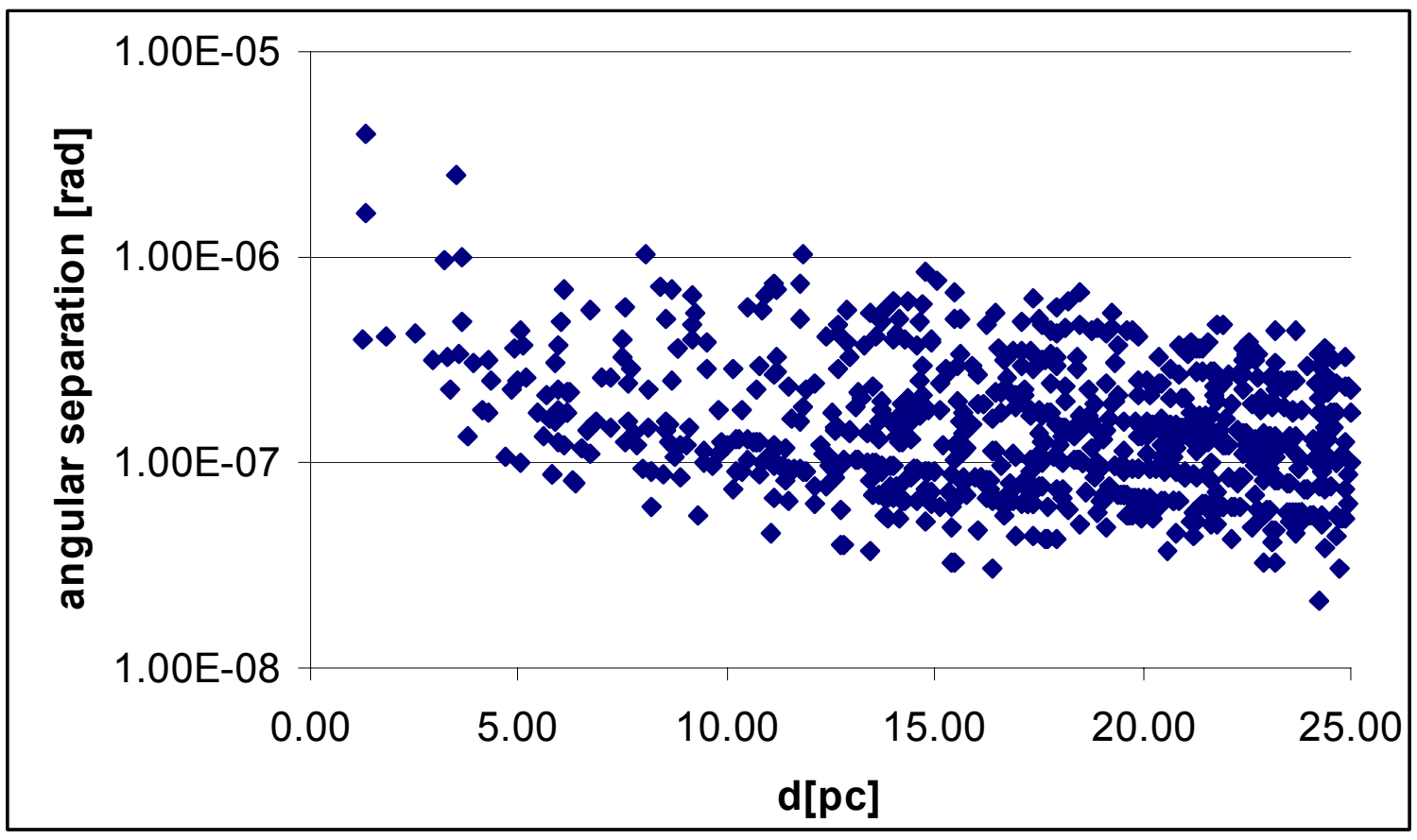

**Fig. 34: Angular separation of the orbit of Earth-like planets from the target stars for the complete DARWIN target star sample**

Fig. 34 shows that DARWIN needs to detect a planet at an orbit of down to 2 $10^{-8}$ rad if it should detect planets at 1AU equivalent orbits around all it's target stars, not including any inclination effects into the calculations. Inclination effects will reduce that value.

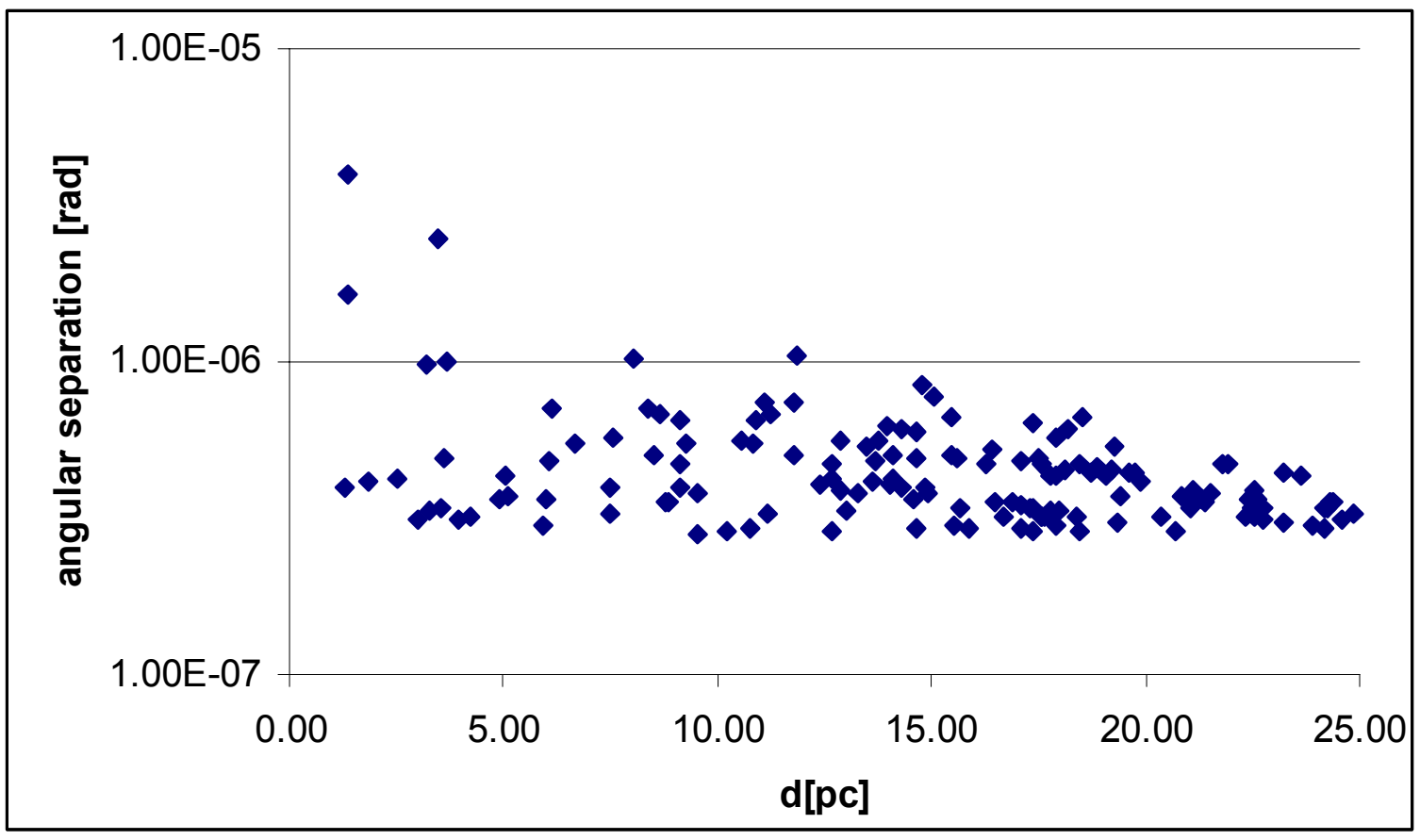

**Fig. 35: Angular separation of the 150 widest orbits of Earth-like planets from the target stars list**

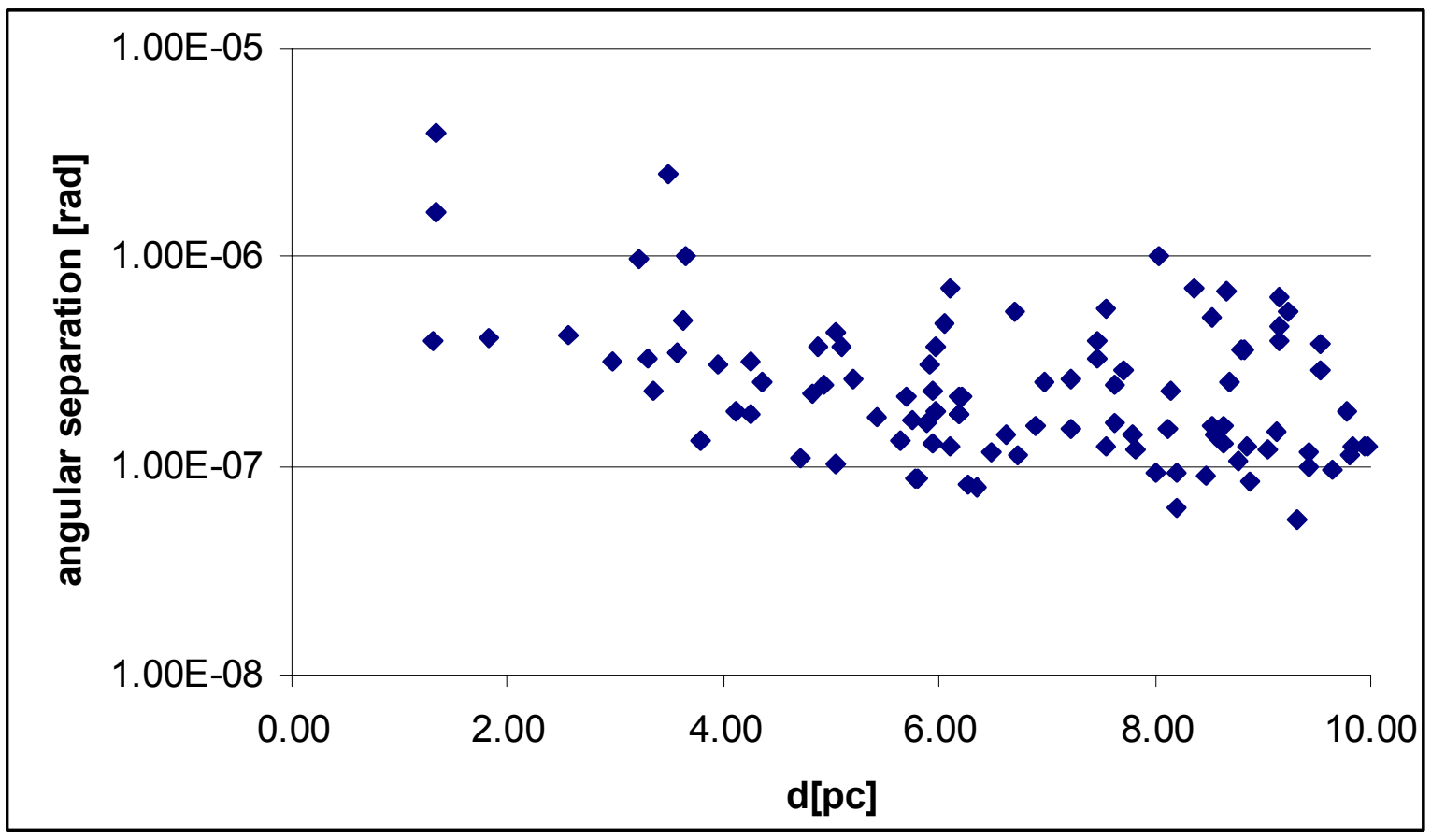

**Fig. 36: Angular separation of the orbit of Earth-like planets for the closest 150 from the target stars**

Fig. 35 shows that DARWIN needs to detect a planet at an orbit of down to 2 $10^{-7}$ rad if it should detect planets at 1AU equivalent orbits around the closest 150 target stars. Fig. 36 shows that DARWIN needs to detect a planet at an orbit of down to 1 $10^{-7}$ rad if it aims to detect planets at 1AU equivalent orbits around all it's target stars, again not including any inclination effects.

As the G stars in the target list are primary targets we have investigated their characteristics further (Eiroa, 2003). The stellar metallicity is used by different as a selection criteria for host stars for extrasolar giant planets by different groups and seems to influence the probability to find planets.

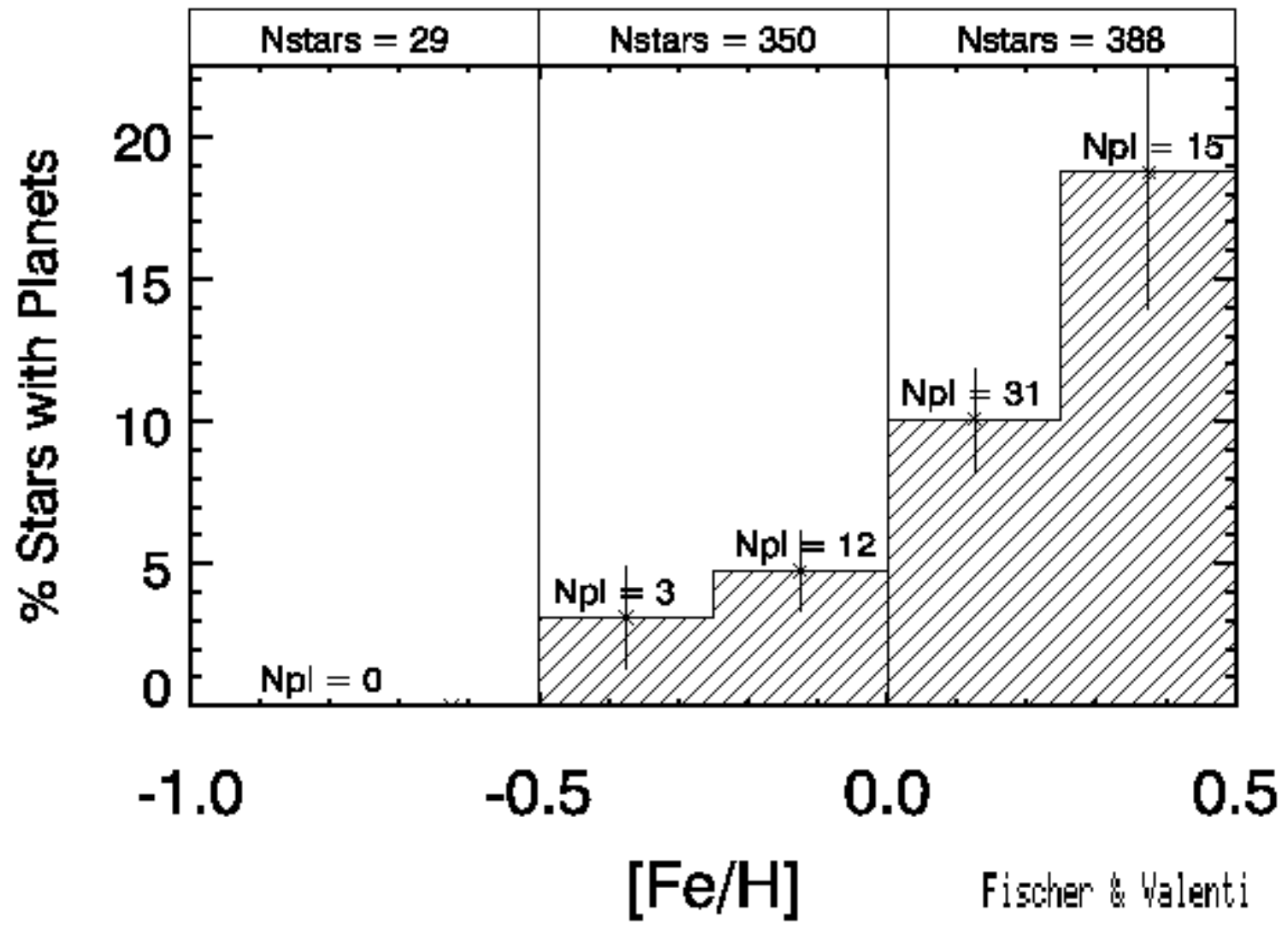

**Fig. 37: Metallicity contribution of the host stars to extrasolar planets found by RV search (Marcy, 2004)**

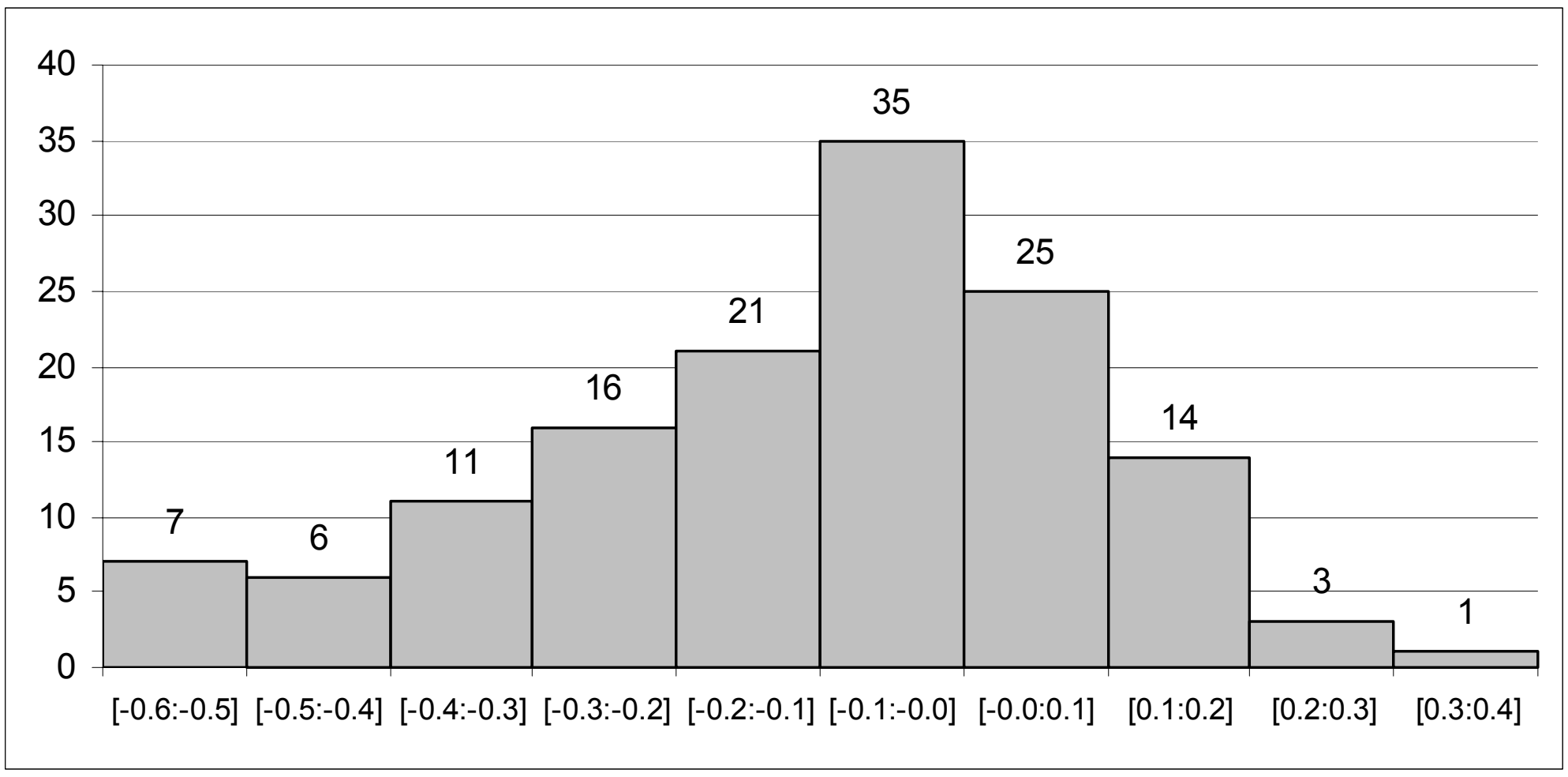

**Fig. 38: Metallicity contribution of the G star target sample (data courtesy C. Eiroa) (Eiroa, 2003)**

The metallicity in the DARWIN G type target sample Fig. 38 is comparable to the metallicity of the host stars for extrasolar planets found by the radial velocity search. The lower limit of the metallicity of the host stars for extrasolar planets found so far is often used as selection criteria for target stars. 9 of the selected 143 G-type target stars would have to be excluded from our list based on that criterion. The difference in metallicity is not high enough for us to exclude them from the sample. One should also keep in mind that the metallicity relation has been established empirically from a sample of host stars to EGPs and might not necessarily reflect the characteristics of host stars of Earth-like planets.

## 3.3 Binaries and high proper motion stars:

The majority of the target stars are high proper motion stars or part of a binary system. The binary systems have to be investigated further to see whether nulling can be applied. If the second star in the binary is too close, its photons can swamp the planet's signal. Different argumentation has been brought forward on one hand arguing that there are stable orbits in a multiple star system what makes it likely that they can harbor planets around each or one of the stars while on the other hand theories exist that find it difficult to create a stable orbit in a multiple star system. These issues need to be investigated, but our major concern is not the planet formation and its orbital stability, but whether or not we can detect the light of the planet when we only null the light of one component of the multiple star system.

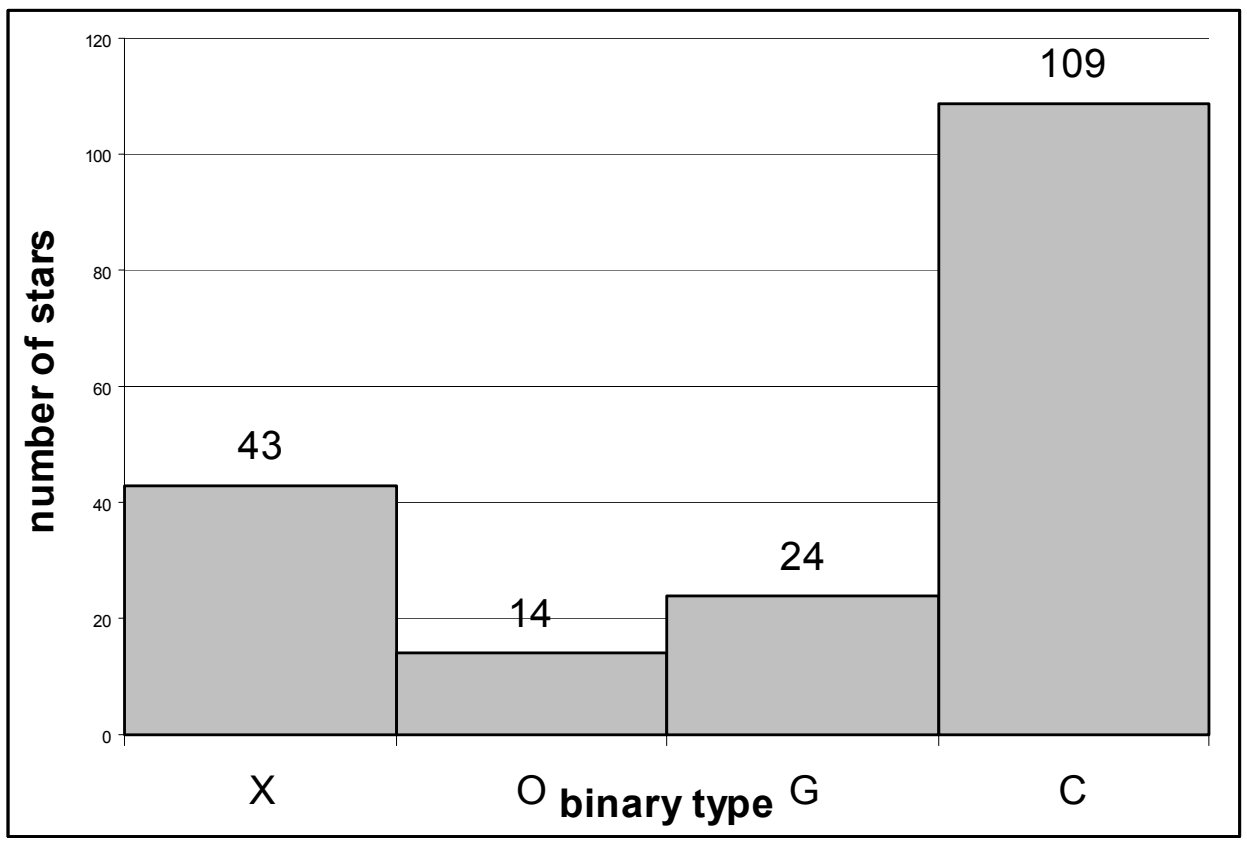

**Fig. 39: Type of multiple star system in the DARWIN target star sample, C : solutions for the components, G : acceleration or higher order terms, O : orbital solutions, X : stochastic solution (probably astrometric binaries with short period)**

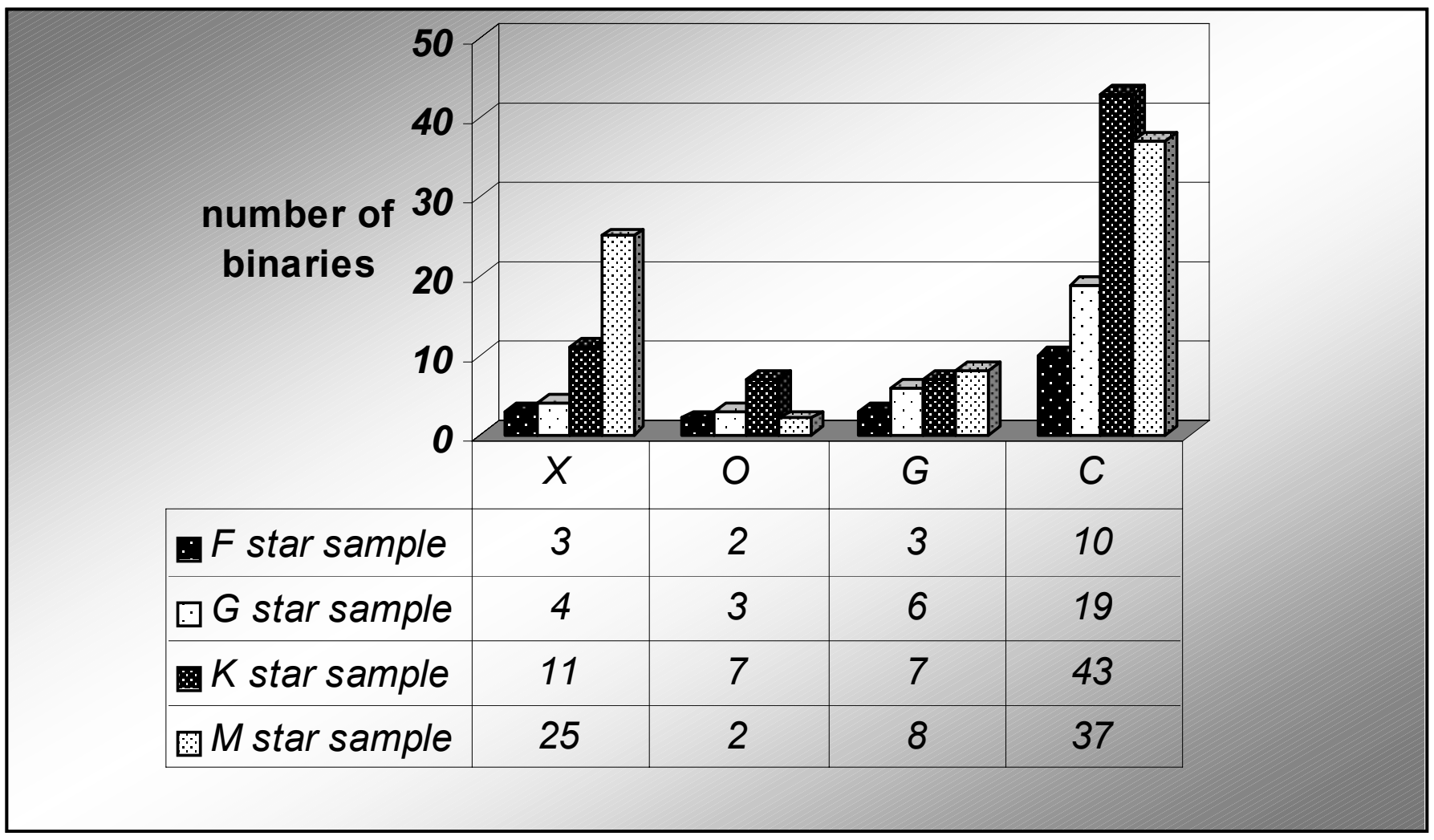

| | X | O | G | C |
|---|---|---|---|---|
| F star sample | 3 | 2 | 3 | 10 |
| G star sample | 4 | 3 | 6 | 19 |
| K star sample | 11 | 7 | 7 | 43 |
| M star sample | 25 | 2 | 8 | 37 |

**Fig. 40: Number of multiple systems in the target star sample**

## 3.4 Angular extend of the closest 10 stars in each stellar class and the corresponding Habitable distance

Below the angular extend of the stars of the different luminosity classes, their Habitable distance (Hd), the equivalent of Earth's 1AU orbit around other stars, as well as the extent of a 3AU disk around those stars similar to our zodiacal system is calculated. We want to focus first on the closest stars, as they will be prim targets for the search for planets. We split the selection of the closest target stars in 1) the 10 closest stars of each stellar type we investigate, 2) the 40 closest stars and finally 3) we concentrate on the closest 40 G stars (same stellar type as the Sun). All stars in the DARWIN target list are chosen by the selection criteria shown in section 2.1. The closest 40 target stars are not equally distributed in the different stellar classes.

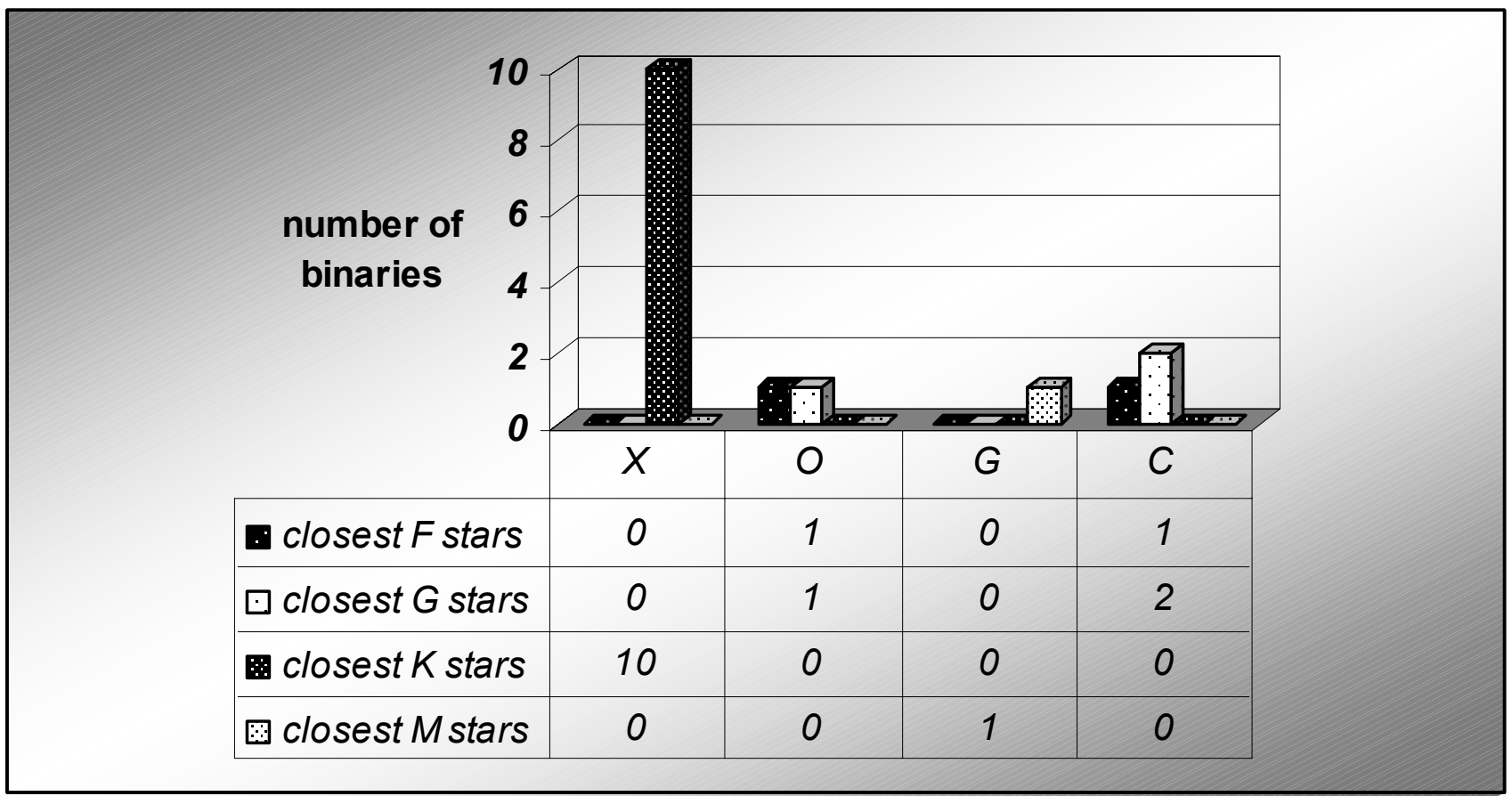

**Fig. 41: Number of multiple systems in the closest 10 star of each stellar type target sample**

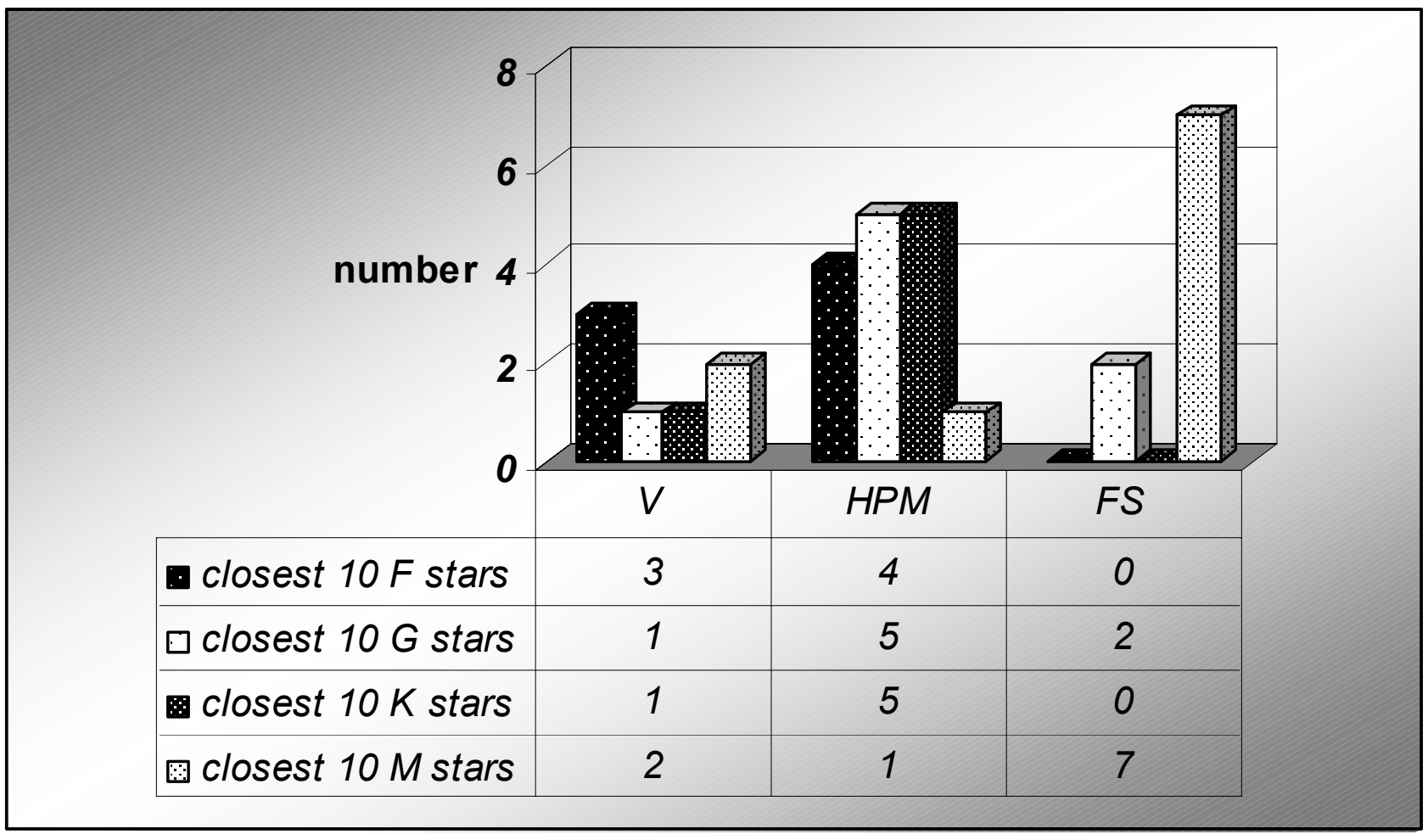

**Fig. 42: Characteristics of 10 closest target stars in each star subclass (tags from the SIMBAD interface: V = variable star, HPM = high proper motion star, F = Flare star)**

**Table 2: Maximum, minimal and mean values for luminosity and angular extend of the closest 10 DARWIN F type target stars**

| F stars | L' [mag] | V [mag] | Hd [rad] | SR [rad] | 3AU [rad] |
|---|---|---|---|---|---|
| max | -0.74 | 0.4 | 2.47E-06 | 1.67E-08 | 4.15E-06 |
| Min | 2.86 | 4.29 | 4.79E-07 | 3.74E-09 | 1.06E-06 |

**Table 3: Maximum, minimal and mean values for luminosity and angular extend of the 10 closest DARWIN G type target stars**

| G stars | L' [mag] | V [mag] | Hd [rad] | SR [rad] | 3AU [rad] |
|---|---|---|---|---|---|
| max | -1.52 | -0.001 | 3.96E-06 | 3.43E-08 | 1.08E-05 |
| Min | 9.84 | 6.42 | 3.99E-07 | 4.32E-09 | 1.59E-06 |

**Table 4: Maximum, minimal and mean values for luminosity and angular extend of the 10 closest DARWIN K type target stars**

| K stars | L' [mag] | V [mag] | Hd [rad] | SR [rad] | 3AU [rad] |
|---|---|---|---|---|---|
| max | -0.81 | 1.35 | 2.18E-06 | 2.76E-08 | 1.08E-05 |
| Min | 5.41 | 7.70 | 4.3E-07 | 5.74E-09 | 2.32E-06 |

**Table 5: Maximum, minimal and mean values for luminosity and angular extend of the 10 closest DARWIN M type target stars**

| M stars | L' [mag] | V [mag] | Hd [rad] | SR [rad] | 3AU [rad] |
|---|---|---|---|---|---|
| max | 4.42 | 11.01 | 3.92E-07 | 9.40E-09 | 1.12E-05 |
| Min | 5.47 | 11.12 | 1.83E-07 | 3.67E-09 | 3.53E-05 |

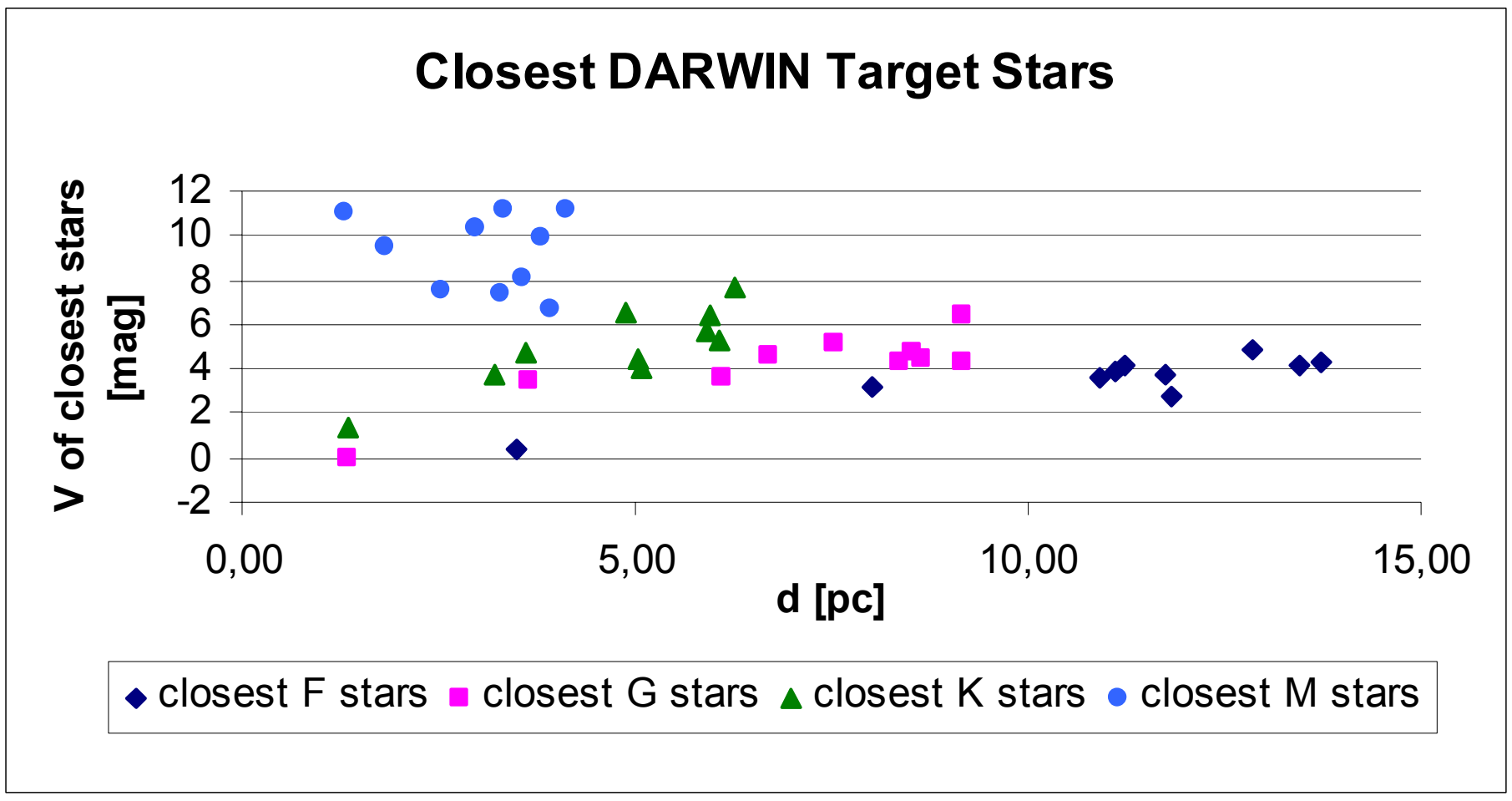

**Fig. 43: V magnitude of the closest DARWIN target stars by stellar class.**

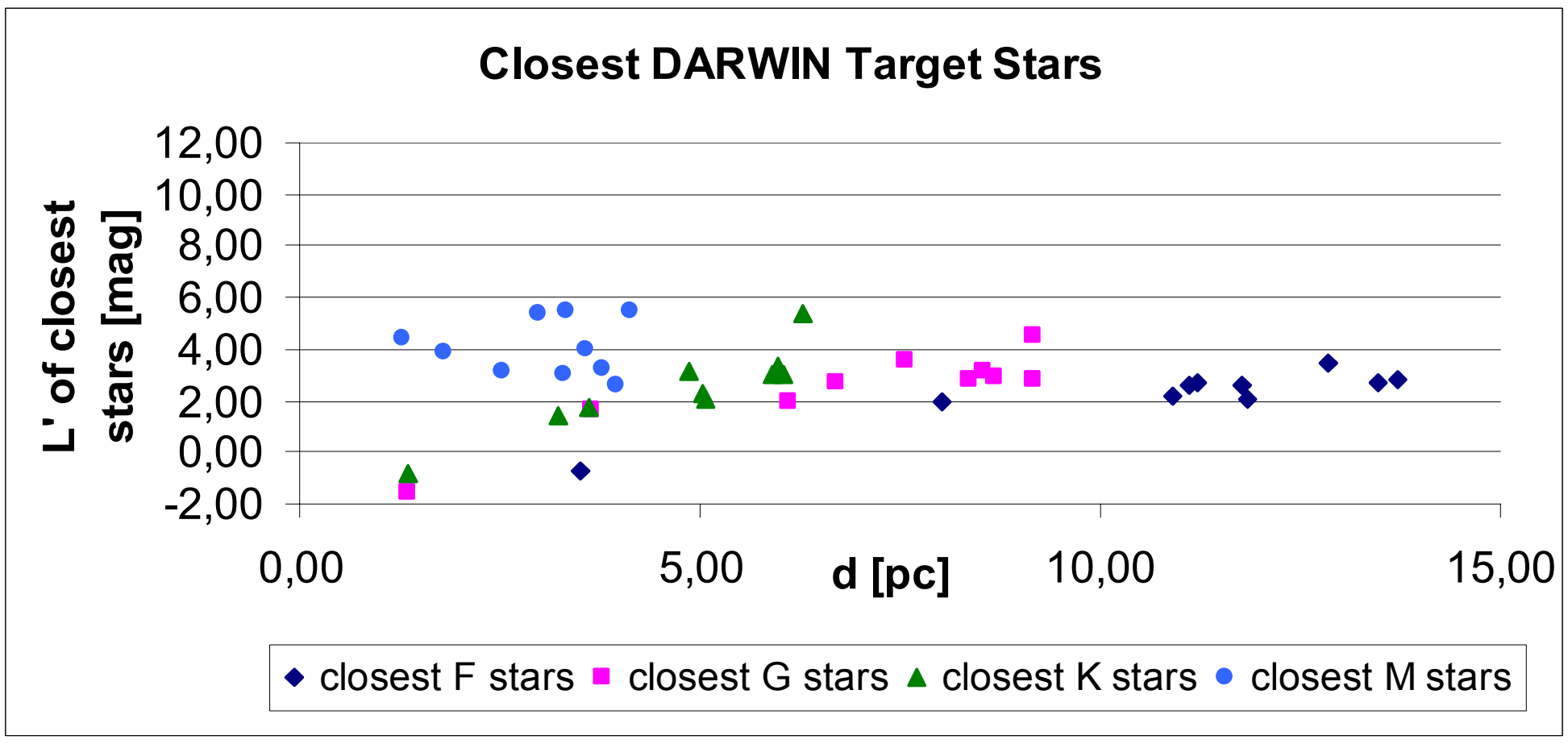

**Fig. 44: L' magnitude of the closest DARWIN target stars by stellar class.**

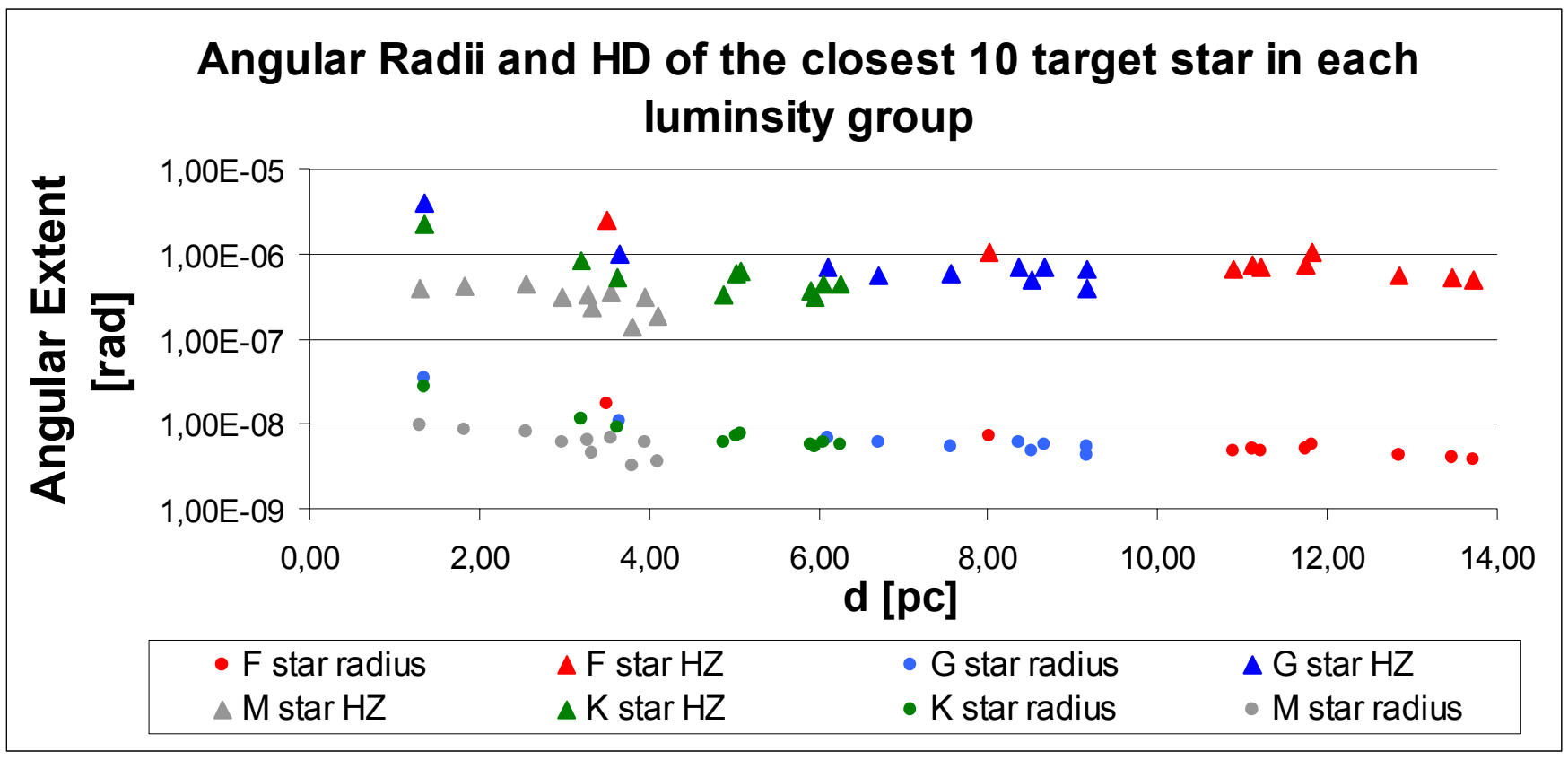

**Fig. 45: Angular extent of the radius of the star and their corresponding Hd for the closest 10 target stars for each stellar type.**

We calculated the extent of a 30AU disk around the target stars. The zodiacal cloud in our own solar system has an extend of 3.2AU (Kasting, 1997) but disks around other stars have been found to extend further outwards, a few tens of AU thus we adopted 30AU as a standard value for the calculations. Dividing the value by 10 leaves you with the angular extend of our zodiacal cloud around the target stars. We do not model the disks, thus one has to keep in mind that disks around other stellar types will have different extents.

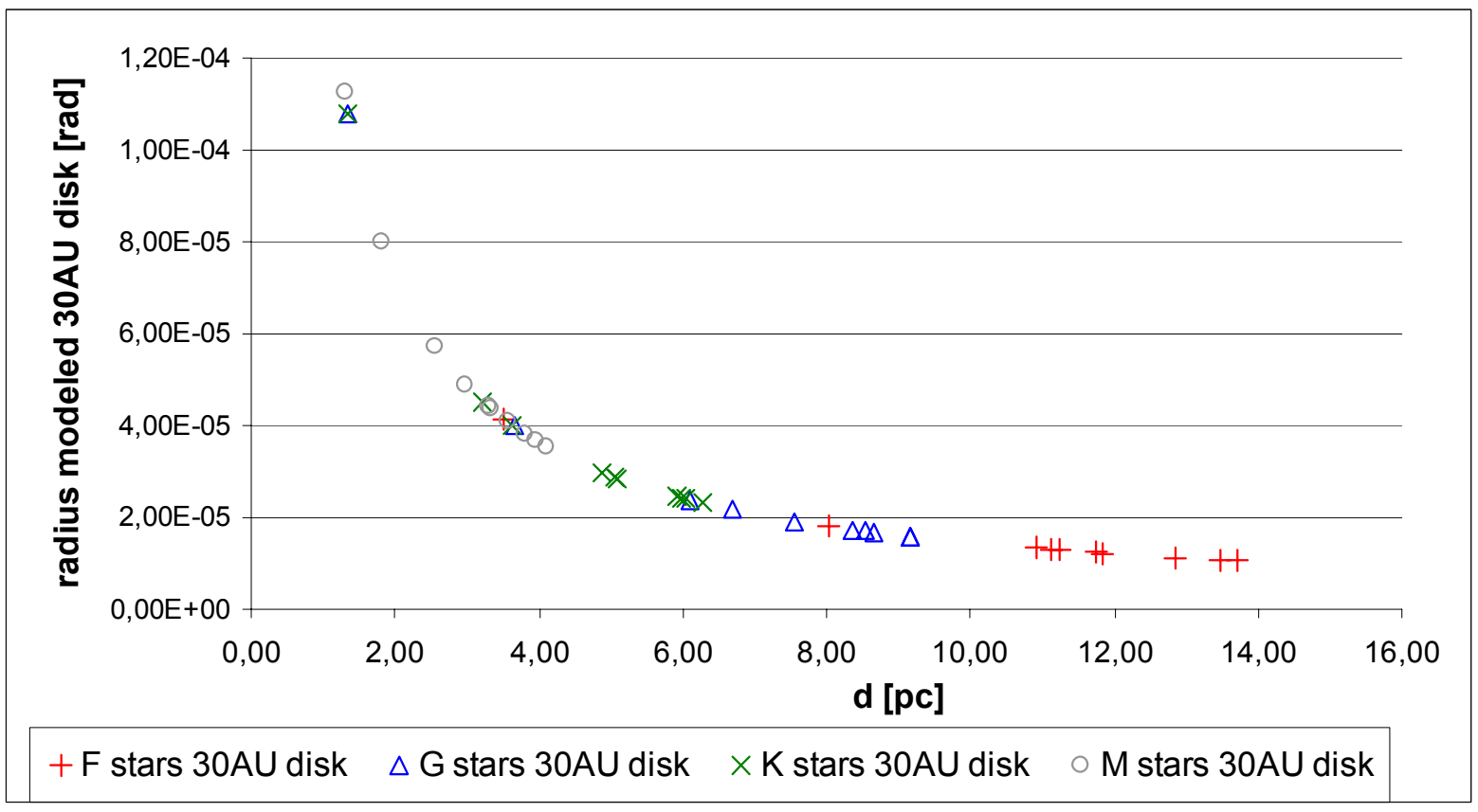

**Fig. 46: Angular extent of the radius of a 30AU disk around the closest 10 target stars per stellar type.**

**Table 6: Data on the closest 10 F target stars**

| 10 closest F stars DARWIN | | | | | | | | | |
|---|---|---|---|---|---|---|---|---|---|
| D[pc] | Vmag | SpType | L/Lo | R/Ro | Teff | L'(cal) | Rstar (rad) | Hd [AU] | Hd [rad] |
| 3.50 | 0.4 | F5IV-V | 3.2 | 1.3 | 6440 | -0.74 | 1.67E-08 | 1.79 | 2.48E-06 |
| 8.03 | 3.19 | F6V | 2.9 | 1.26 | 6358 | 1.94 | 7.07E-09 | 1.69 | 1.02E-06 |
| 10.90 | 3.59 | F8V | 2.2 | 1.18 | 6194 | 2.20 | 4.88E-09 | 1.48 | 6.57E-07 |
| 11.12 | 3.85 | F6V | 2.9 | 1.26 | 6358 | 2.60 | 5.10E-09 | 1.69 | 7.37E-07 |
| 11.23 | 4.1 | F7V | 2.5 | 1.22 | 6276 | 2.74 | 4.89E-09 | 1.59 | 6.85E-07 |
| 11.76 | 3.77 | F5V | 3.2 | 1.3 | 6440 | 2.63 | 4.98E-09 | 1.79 | 7.38E-07 |
| 11.83 | 2.74 | F0V+... | 6.5 | 1.5 | 7200 | 2.01 | 5.71E-09 | 2.55 | 1.04E-06 |
| 12.85 | 4.82 | F8V | 2.2 | 1.18 | 6194 | 3.43 | 4.13E-09 | 1.48 | 5.57E-07 |
| 13.47 | 4.1 | F8V | 2.2 | 1.18 | 6194 | 2.71 | 3.94E-09 | 1.48 | 5.31E-07 |
| 13.72 | 4.29 | F9V | 1.8 | 1.14 | 6112 | 2.86 | 3.74E-09 | 1.36 | 4.79E-07 |

**Table 7: Data on the closest 10 G target stars**

| 10 closest G stars DARWIN | | | | | | | | | |
|---|---|---|---|---|---|---|---|---|---|
| D[pc] | Vmag | SpType | L/Lo | R/Ro | Teff | L'(cal) | Rstar (rad) | Hd [AU] | Hd [rad] |
| 1.35 | -0.01 | G2V | 1.216 | 1.03 | 5926 | -1.52 | 3.43E-08 | 1.10 | 3.97E-06 |
| 3.65 | 3.49 | G8V | 0.57 | 0.88 | 5458 | 1.63 | 1.08E-08 | 0.75 | 1E-06 |
| 6.11 | 3.55 | G5IV-Vvar | 0.79 | 0.92 | 5770 | 1.91 | 6.79E-09 | 0.89 | 7.06E-07 |
| 6.70 | 4.54 | G8V | 0.57 | 0.88 | 5458 | 2.68 | 5.90E-09 | 0.75 | 5.45E-07 |
| 7.55 | 5.17 | G5Vp | 0.79 | 0.92 | 5770 | 3.53 | 5.49E-09 | 0.89 | 5.71E-07 |
| 8.37 | 4.24 | G0V | 1.5 | 1.10 | 6030 | 2.78 | 5.92E-09 | 1.22 | 7.09E-07 |
| 8.53 | 4.74 | G5V | 0.79 | 0.92 | 5770 | 3.1 | 4.86E-09 | 0.89 | 5.05E-07 |
| 8.66 | 4.39 | G0V | 1.5 | 1.10 | 6030 | 2.93 | 5.72E-09 | 1.22 | 6.85E-07 |
| 9.15 | 4.23 | G0V | 1.5 | 1.10 | 6030 | 2.77 | 5.41E-09 | 1.22 | 6.49E-07 |
| 9.16 | 6.42 | G8Vp | 0.57 | 0.88 | 5458 | 4.56 | 4.32E-09 | 0.75 | 3.99E-07 |

**Table 8: Data on the closest 10K target stars**

| 10 closest K stars | | | | | | | | | |
|---|---|---|---|---|---|---|---|---|---|
| D[pc] | Vmag | SpType | L/Lo | R/Ro | Teff | L' (cal) | Rstar (rad) | Hd [AU] | Hd [rad] |
| 1,35 | 1,35 | K1V | 0,37 | 0,8 | 5070 | -0,81 | 2,76E-08 | 0,60 | 2,18E-06 |
| 3,22 | 3,72 | K2V | 0,31 | 0,8 | 4890 | 1,43 | 1,12E-08 | 0,55 | 8,42E-07 |
| 3,63 | 4,69 | K5V | 0,15 | 0,7 | 4350 | 1,73 | 8,95E-09 | 0,38 | 5,18E-07 |
| 4,87 | 6,6 | K8V | 0,11 | 0,6 | 4050 | 3,14 | 5,99E-09 | 0,32 | 3,24E-07 |
| 5,04 | 4,43 | K1V | 0,37 | 0,8 | 5070 | 2,27 | 7,36E-09 | 0,60 | 5,81E-07 |
| 5,09 | 4,03 | K0V SB | 0,42 | 0,9 | 5250 | 2,01 | 7,53E-09 | 0,64 | 6,18E-07 |
| 5,91 | 5,72 | K4V | 0,2 | 0,7 | 4530 | 2,99 | 5,69E-09 | 0,45 | 3,71E-07 |
| 5,97 | 6,33 | K5V | 0,15 | 0,7 | 4350 | 3,37 | 5,44E-09 | 0,38 | 3,15E-07 |
| 6,05 | 5,32 | K2V | 0,31 | 0,8 | 4890 | 3,03 | 5,94E-09 | 0,55 | 4,47E-07 |
| 6,27 | 7,7 | K2 | 0,31 | 0,8 | 4890 | 5,41 | 5,74E-09 | 0,55 | 4,32E-07 |

**Table 9: Data on the closest 10 M target stars**

| 10 closest M stars | | | | | | | | | |
|---|---|---|---|---|---|---|---|---|---|
| D[pc] | Vmag | SpType | L/Lo | R/Ro | Teff | L' (cal) | Rstar (rad) | Hd [AU] | Hd [rad] |
| 1.29 | 11.01 | M5Ve | 0.01 | 0.27 | 3240 | 4.42 | 9.40E-09 | 0.10 | 3.93E-07 |
| 1.82 | 9.54 | sdM4 | 0.02 | 0.336 | 3362 | 3.89 | 8.31E-09 | 0.16 | 4.14E-07 |
| 2.55 | 7.49 | M2V | 0.05 | 0.468 | 3606 | 3.15 | 8.27E-09 | 0.22 | 4.27E-07 |
| 2.97 | 10.37 | M3.5Ve | 0.03 | 0.402 | 3484 | 5.4 | 6.09E-09 | 0.19 | 3.15E-07 |
| 3.29 | 7.35 | M2/M3V | 0.05 | 0.468 | 3606 | 3.01 | 6.41E-09 | 0.22 | 3.3E-07 |
| 3.34 | 11.12 | M4.5V | 0.02 | 0.336 | 3362 | 5.47 | 4.53E-09 | 0.16 | 2.26E-07 |
| 3.57 | 8.09 | M1V | 0.06 | 0.534 | 3728 | 4.01 | 6.74E-09 | 0.25 | 3.43E-07 |
| 3.80 | 9.84 | M5 | 0.01 | 0.27 | 3240 | 3.25 | 3.20E-09 | 0.10 | 1.34E-07 |
| 3.95 | 6.69 | M1/M2V | 0.06 | 0.534 | 3728 | 2.61 | 6.09E-09 | 0.25 | 3.1E-07 |
| 4.12 | 11.12 | M4.5Ve | 0.02 | 0.336 | 3362 | 5.47 | 3.67E-09 | 0.16 | 1.83E-07 |

## *3.5 Angular extent of the closest 40 stars and the corresponding Habitable distance*

The closest 40 stars are not equally distributed in the different stellar classes. The closest 40 stars are: 1x F star, 2x G stars, 9x K stars, 28 M stars. Their characteristics are shown in Table 11.

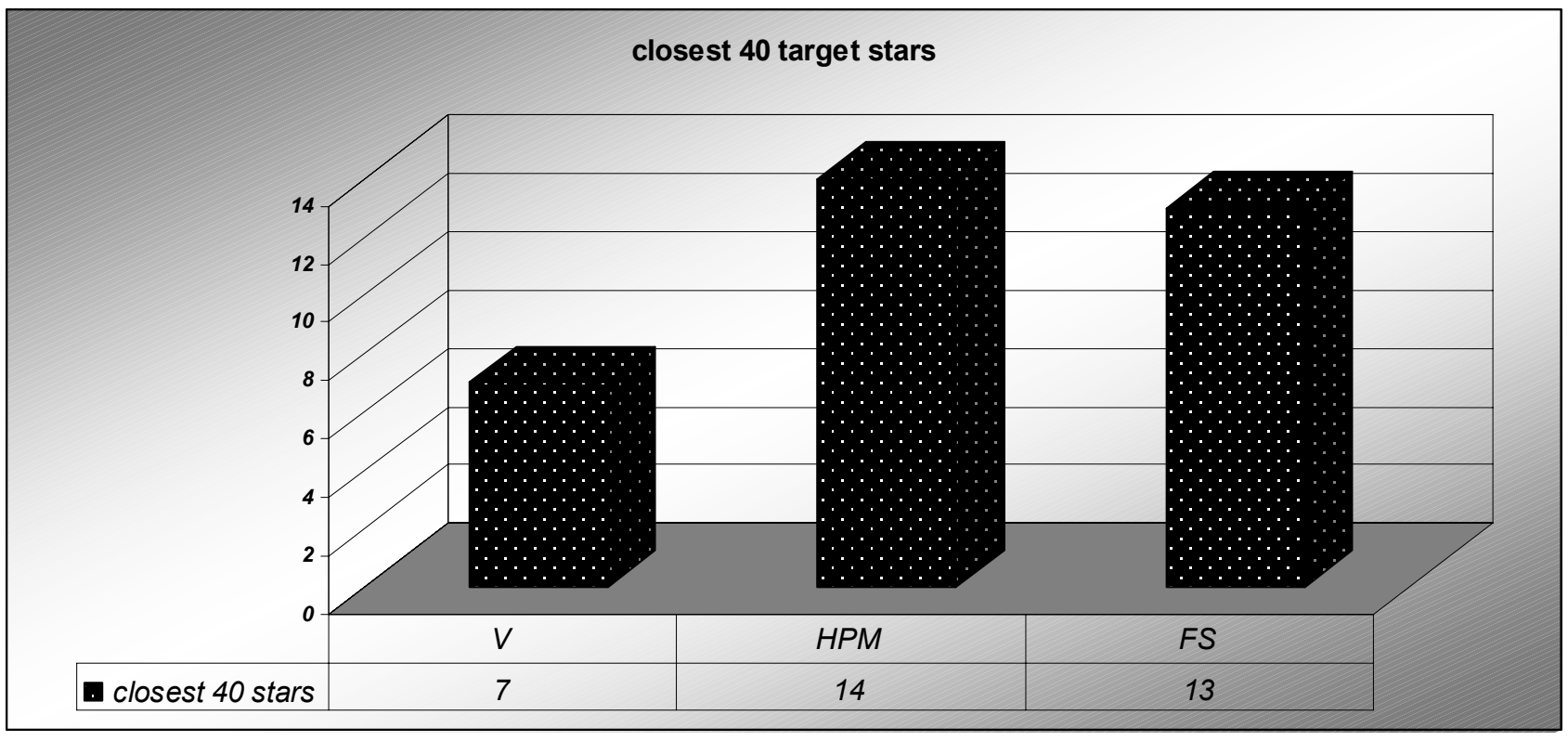

**Fig. 47: Characteristics of the 40 closest target stars (tags from the SIMBAD interface: V = variable star, HPM = high proper motion star, F = Flare star).**

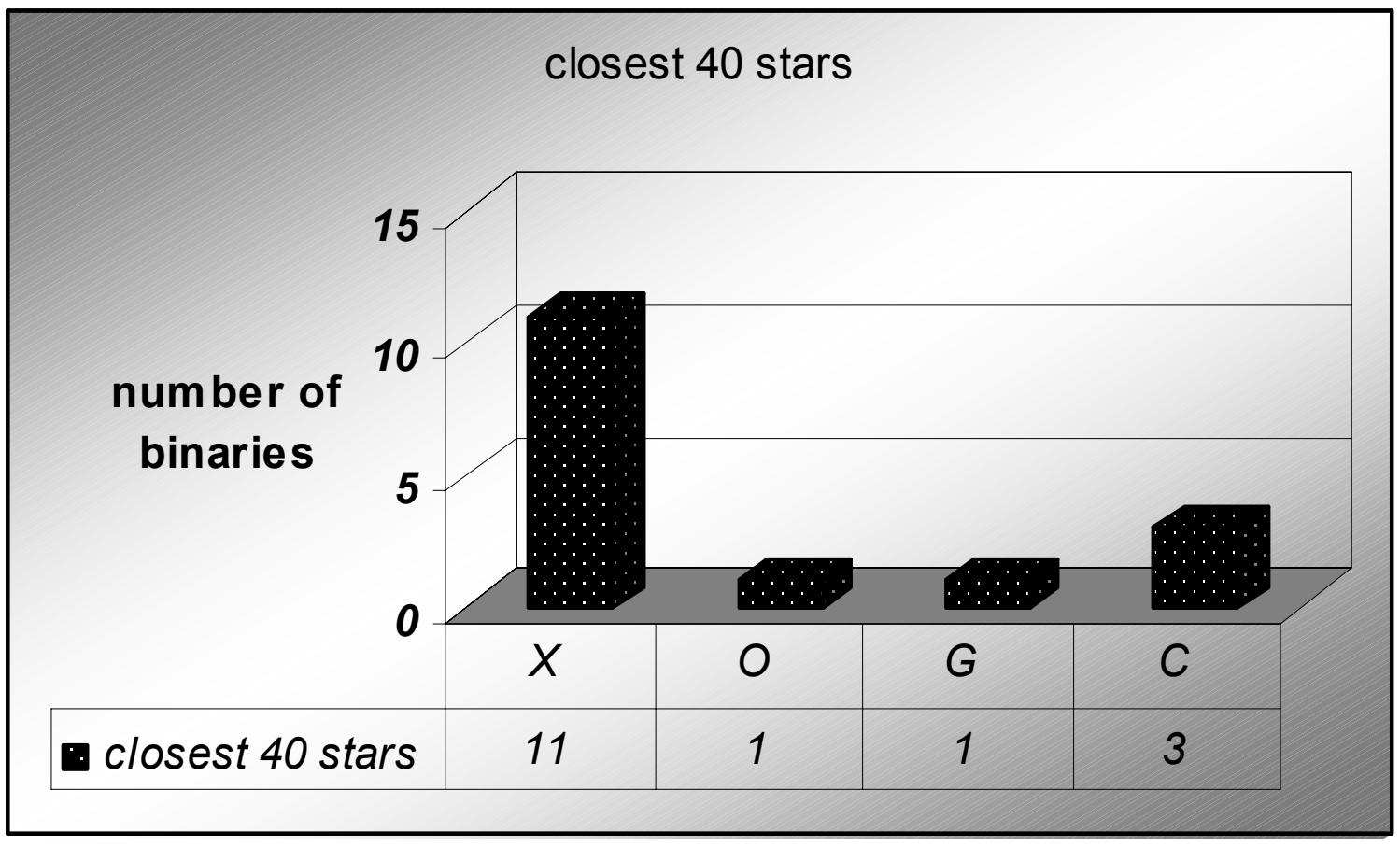

**Fig. 48: Number of multiple systems in the closest 40 star target sample.**

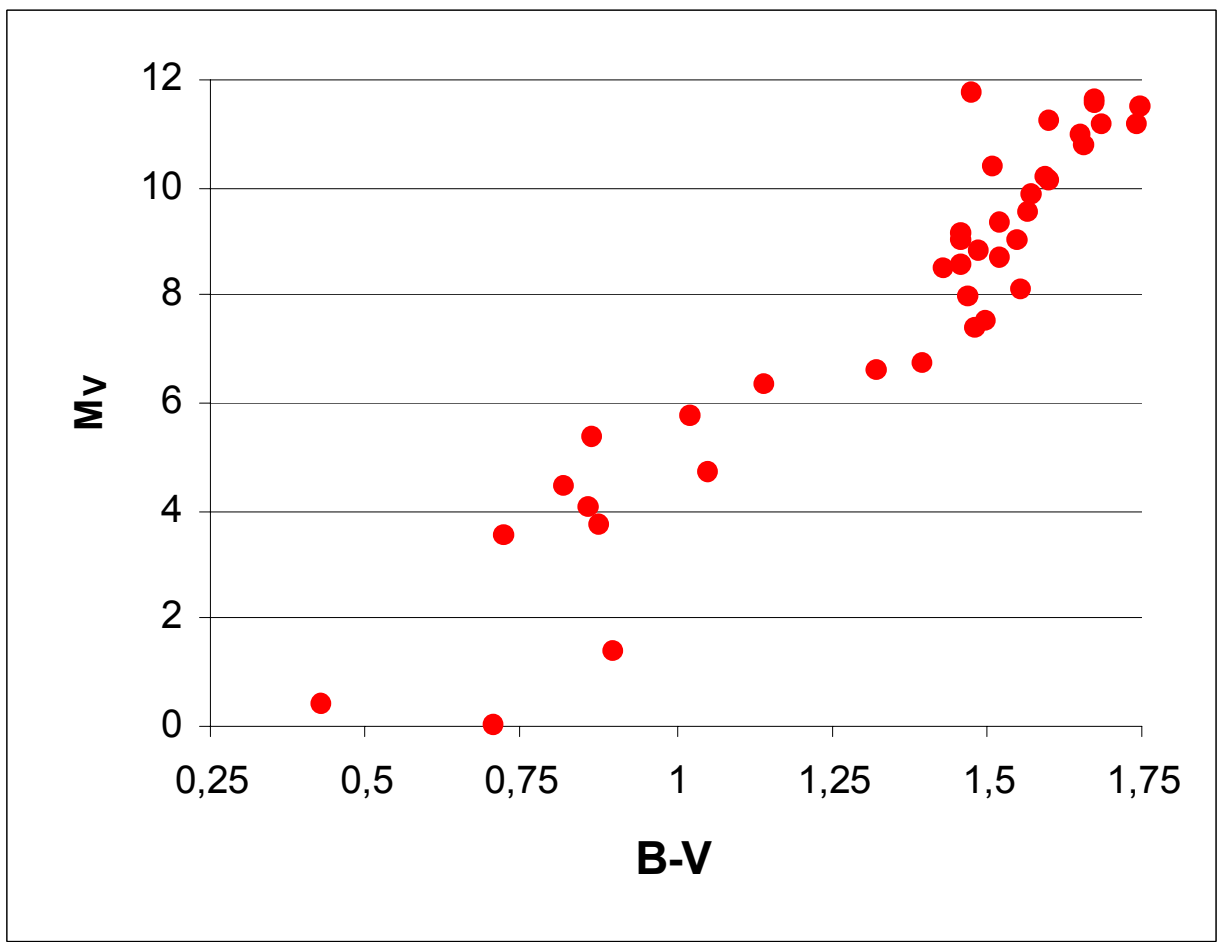

**Fig. 49: Color-magnitude diagram for the closest 40 target stars.**

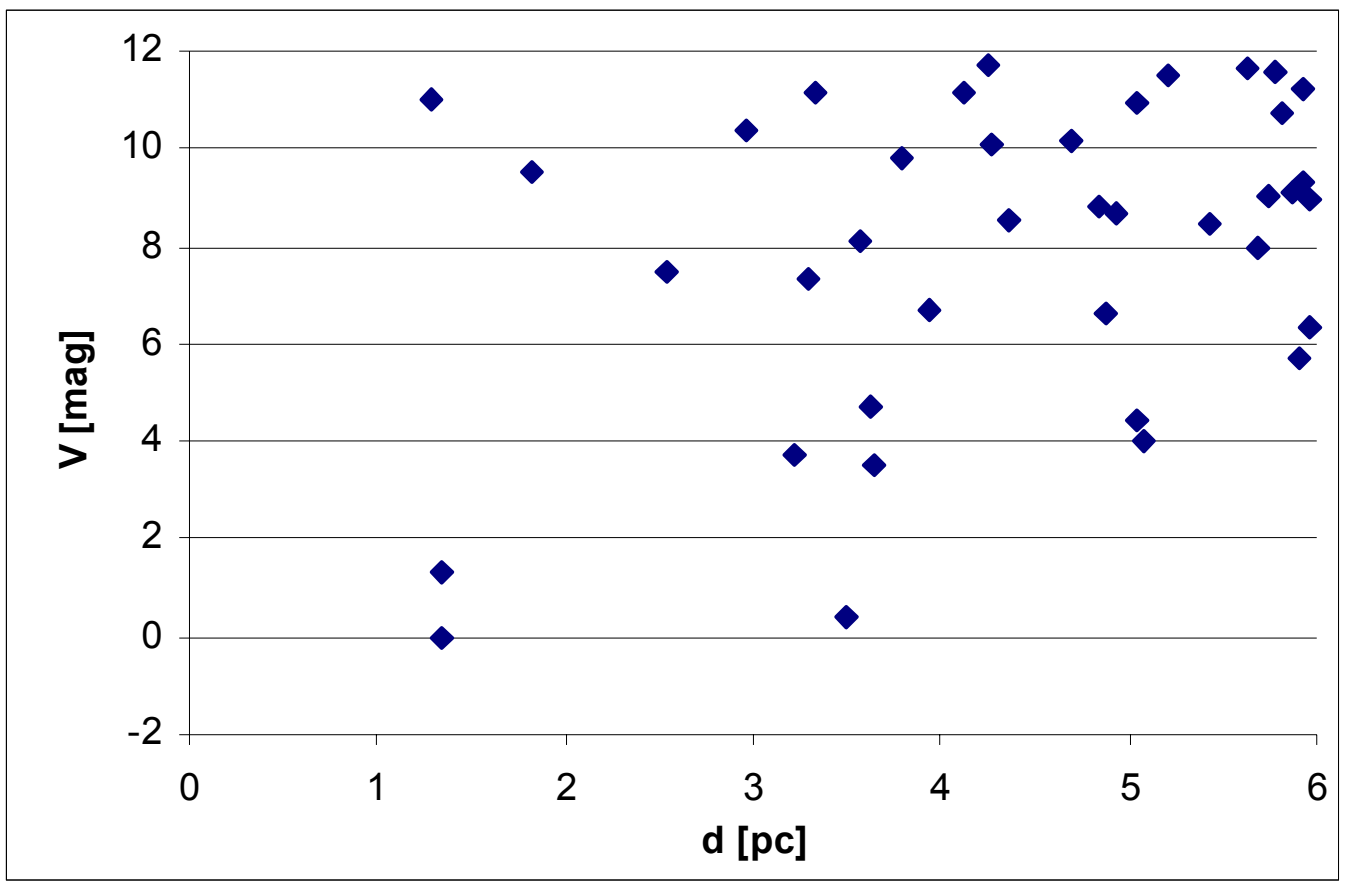

**Fig. 50: V magnitude of the closest 40 DARWIN target stars.**

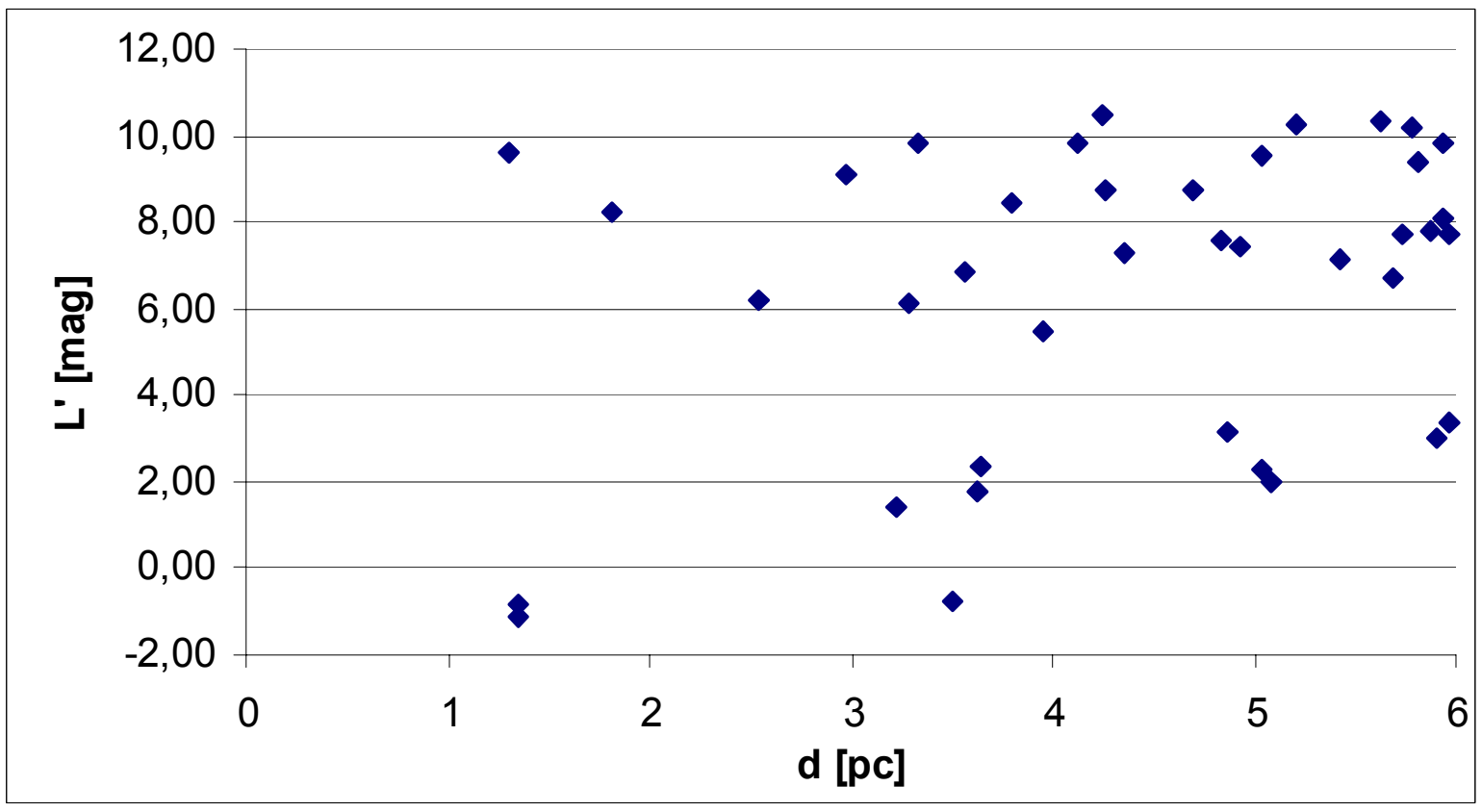

**Fig. 51: L' magnitude of the closest 40 DARWIN target stars.**

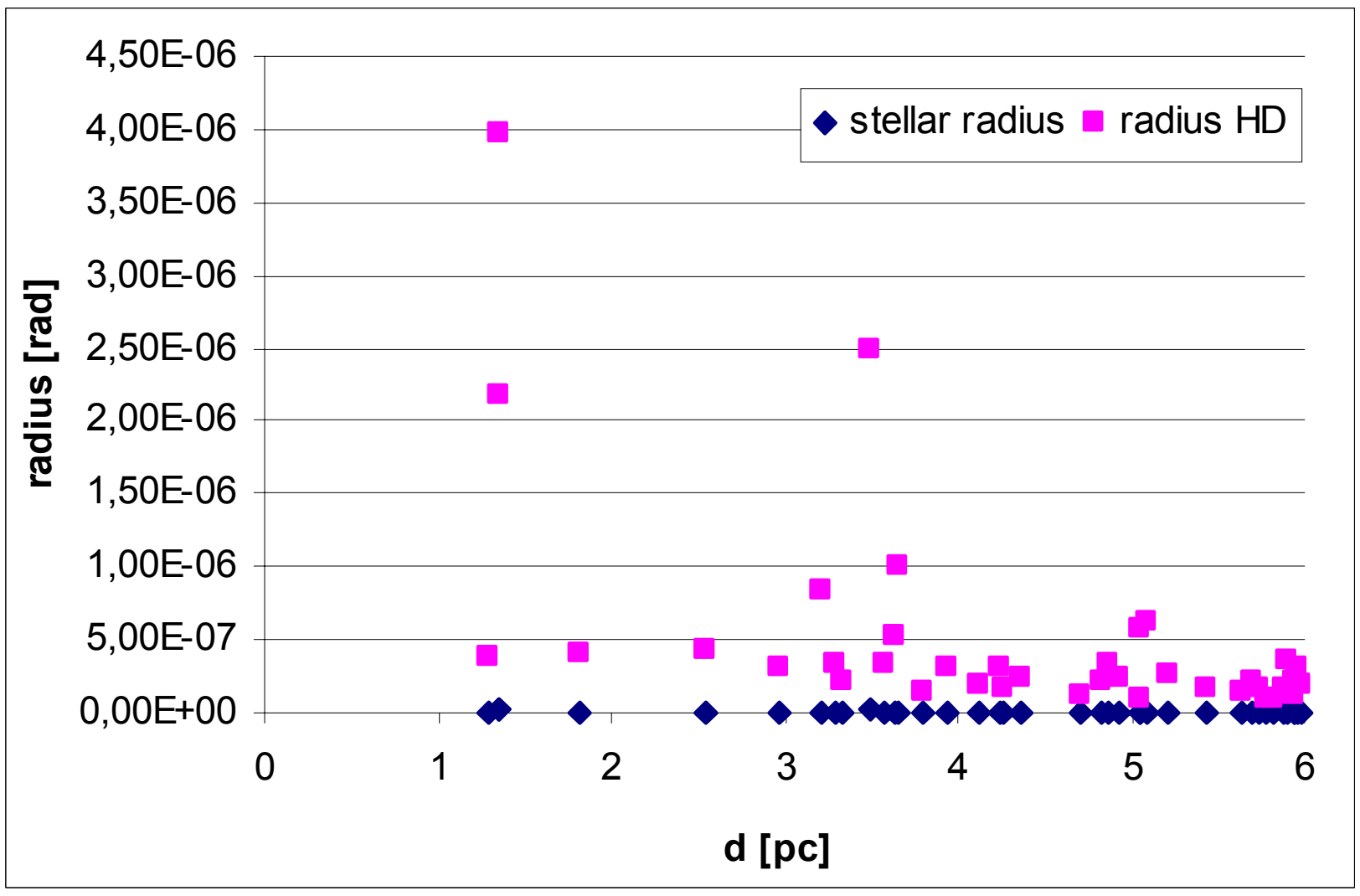

**Fig. 52: Angular extent of the radius of the star and corresponding Hd for the closest 40 target stars.**

**Table 10: Data on the 40 closest DARWIN target stars**

| Closest DARWIN target stars | | | | | | | | | | |
|---|---|---|---|---|---|---|---|---|---|---|
| **d[pc]** | **Vmag** | **SpType** | **L/Lo** | **R/Ro** | **Teff** | **L'(cal)** | **Rstar (rad)** | **Hd (AU)** | **Hd (RAD)** | **30AU (RAD)** |
| 1,29 | 11,01 | M5Ve | 0,01 | 0,27 | 3240 | 4,42 | 9,40E-09 | 0,10 | 3,93E-07 | 1,12E-04 |
| 1,35 | -0,01 | G2V | 1,21 | 1,03 | 5926 | -1,52 | 3,43E-08 | 1,10 | 3,97E-06 | 1,08E-04 |
| 1,35 | 1,35 | K1V | 0,37 | 0,80 | 5070 | -0,81 | 2,76E-08 | 0,60 | 2,18E-06 | 1,08E-04 |
| 1,82 | 9,54 | sdM4 | 0,02 | 0,33 | 3362 | 3,89 | 8,31E-09 | 0,15 | 4,14E-07 | 7,99E-05 |
| 2,55 | 7,49 | M2V | 0,05 | 0,46 | 3606 | 3,15 | 8,27E-09 | 0,22 | 4,27E-07 | 5,71E-05 |
| 2,97 | 10,37 | M3.5Ve | 0,03 | 0,40 | 3484 | 5,4 | 6,09E-09 | 0,19 | 3,15E-07 | 4,89E-05 |
| 3,22 | 3,72 | K2V | 0,31 | 0,80 | 4890 | 1,43 | 1,12E-08 | 0,55 | 8,42E-07 | 4,52E-05 |
| 3,29 | 7,35 | M2/M3V | 0,05 | 0,46 | 3606 | 3,01 | 6,41E-09 | 0,22 | 3,30E-07 | 4,42E-05 |
| 3,34 | 11,12 | M4.5V | 0,02 | 0,33 | 3362 | 5,47 | 4,53E-09 | 0,15 | 2,26E-07 | 4,36E-05 |
| 3,50 | 0,4 | F5IV-V | 3,20 | 1,30 | 6440 | -0,74 | 1,67E-08 | 1,78 | 2,48E-06 | 4,16E-05 |
| 3,57 | 8,09 | M1V | 0,06 | 0,53 | 3728 | 4,01 | 6,74E-09 | 0,25 | 3,43E-07 | 4,08E-05 |
| 3,63 | 4,69 | K5V | 0,15 | 0,70 | 4350 | 1,73 | 8,95E-09 | 0,38 | 5,18E-07 | 4,01E-05 |
| 3,65 | 3,49 | G8V | 0,57 | 0,88 | 5458 | 1,63 | 1,08E-08 | 0,75 | 1.00E-06 | 3,99E-05 |
| 3,80 | 9,84 | M5 | 0,01 | 0,27 | 3240 | 3,25 | 3,20E-09 | 0,10 | 1,34E-07 | 3,83E-05 |
| 3,95 | 6,69 | M1/M2V | 0,06 | 0,53 | 3728 | 2,61 | 6,09E-09 | 0,25 | 3,10E-07 | 3,69E-05 |
| 4,12 | 11,12 | M4.5Ve | 0,02 | 0,33 | 3362 | 5,47 | 3,67E-09 | 0,15 | 1,83E-07 | 3,53E-05 |
| 4,25 | 11,72 | M | 0,07 | 0,60 | 3850 | 7,9 | 6,36E-09 | 0,27 | 3,16E-07 | 3,42E-05 |
| 4,26 | 10,1 | M4 | 0,02 | 0,33 | 3362 | 4,45 | 3,55E-09 | 0,15 | 1,77E-07 | 3,41E-05 |
| 4,36 | 8,56 | M2V | 0,05 | 0,46 | 3606 | 4,22 | 4,83E-09 | 0,22 | 2,49E-07 | 3,34E-05 |
| 4,70 | 10,16 | M5 | 0,01 | 0,27 | 3240 | 3,57 | 2,58E-09 | 0,10 | 1,08E-07 | 3,09E-05 |
| 4,83 | 8,82 | M2Vvar | 0,05 | 0,46 | 3606 | 4,48 | 4,36E-09 | 0,22 | 2,25E-07 | 3,01E-05 |
| 4,87 | 6,6 | K8V | 0,11 | 0,60 | 4050 | 3,14 | 5,99E-09 | 0,32 | 3,24E-07 | 2,98E-05 |
| 4,94 | 8,66 | M1V | 0,06 | 0,53 | 3728 | 4,58 | 4,87E-09 | 0,25 | 2,48E-07 | 2,95E-05 |
| 5,04 | 10,94 | M5 | 0,01 | 0,27 | 3240 | 4,35 | 2,41E-09 | 0,10 | 1,01E-07 | 2,88E-05 |
| 5,04 | 4,43 | K1V | 0,37 | 0,80 | 5070 | 2,27 | 7,36E-09 | 0,60 | 5,81E-07 | 2,88E-05 |
| 5,09 | 4,03 | K0VSB | 0,42 | 0,90 | 5250 | 2,01 | 7,53E-09 | 0,64 | 6,18E-07 | 2,86E-05 |
| 5,21 | 11,49 | M: | 0,07 | 0,60 | 3850 | 7,67 | 5,18E-09 | 0,27 | 2,58E-07 | 2,79E-05 |
| 5,43 | 8,46 | M3V | 0,03 | 0,40 | 3484 | 3,49 | 3,33E-09 | 0,19 | 1,73E-07 | 2,68E-05 |
| 5,64 | 11,64 | M4: | 0,02 | 0,33 | 3362 | 5,99 | 2,68E-09 | 0,15 | 1,34E-07 | 2,58E-05 |
| 5,69 | 7,97 | M1V | 0,06 | 0,53 | 3728 | 3,89 | 4,23E-09 | 0,25 | 2,15E-07 | 2,56E-05 |
| 5,74 | 9,02 | M3Ve | 0,03 | 0,40 | 3484 | 4,05 | 3,15E-09 | 0,19 | 1,63E-07 | 2,53E-05 |
| 5,79 | 11,56 | M5 | 0,01 | 0,27 | 3240 | 4,97 | 2,10E-09 | 0,10 | 8,79E-08 | 2,51E-05 |
| 5,81 | 10,75 | M5 | 0,01 | 0,27 | 3240 | 4,16 | 2,09E-09 | 0,10 | 8,75E-08 | 2,50E-05 |
| 5,87 | 9,12 | M3.5V | 0,03 | 0,40 | 3484 | 4,15 | 3,08E-09 | 0,19 | 1,60E-07 | 2,48E-05 |
| 5,91 | 5,72 | K4V | 0,20 | 0,70 | 4530 | 2,99 | 5,69E-09 | 0,45 | 3,71E-07 | 2,46E-05 |
| 5,93 | 11,19 | M4.5Ve | 0,02 | 0,33 | 3362 | 5,54 | 2,55E-09 | 0,15 | 1,27E-07 | 2,45E-05 |
| 5,93 | 9,31 | M0 | 0,07 | 0,60 | 3850 | 5,49 | 4,55E-09 | 0,27 | 2,27E-07 | 2,45E-05 |
| 5,97 | 6,33 | K5V | 0,15 | 0,70 | 4350 | 3,37 | 5,44E-09 | 0,38 | 3,15E-07 | 2,44E-05 |
| 5,97 | 8,98 | M2V | 0,05 | 0,46 | 3606 | 4,64 | 3,53E-09 | 0,22 | 1,82E-07 | 2,44E-05 |
| 6,05 | 5,32 | K2V | 0,31 | 0,80 | 4890 | 3,03 | 5,94E-09 | 0,55 | 4,47E-07 | 2,40E-05 |

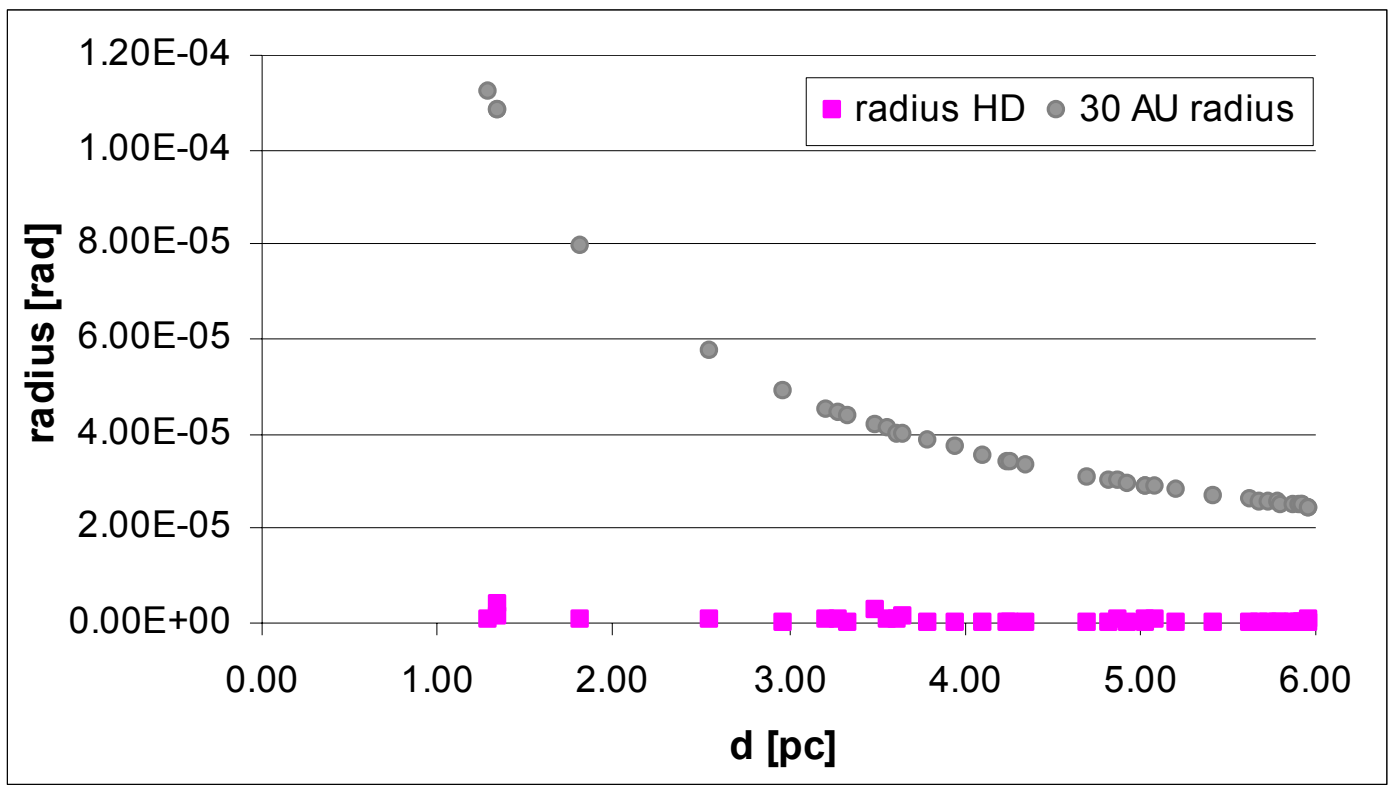

**Fig. 53: Angular extent of the radius of a 30AU disk around the closest 40 target stars**

To model realistic observation scenarios for nulling interferometry, the luminosity, radius of the star and the corresponding Hd has to be taken into account. Table 10 gives the data for the closest 40 stars according to the Hipparcos catalogue.

## *3.6 Angular extend of the closest 40 G stars and the corresponding Habitable distance*

Fig. 54 to Fig. 58 show the data on the closest 40 G stars. Our sun is a G-type star thus G-stars will be prime targets for the search for extra-solar planets.

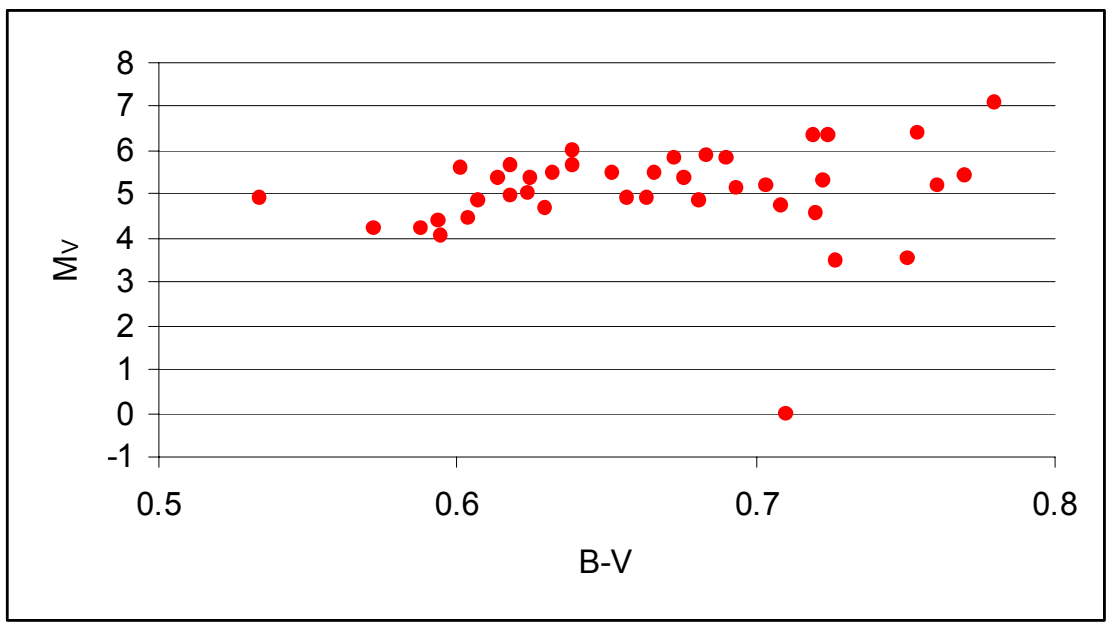

**Fig. 54: Color-magnitude diagram for the 40 closest G target stars.**

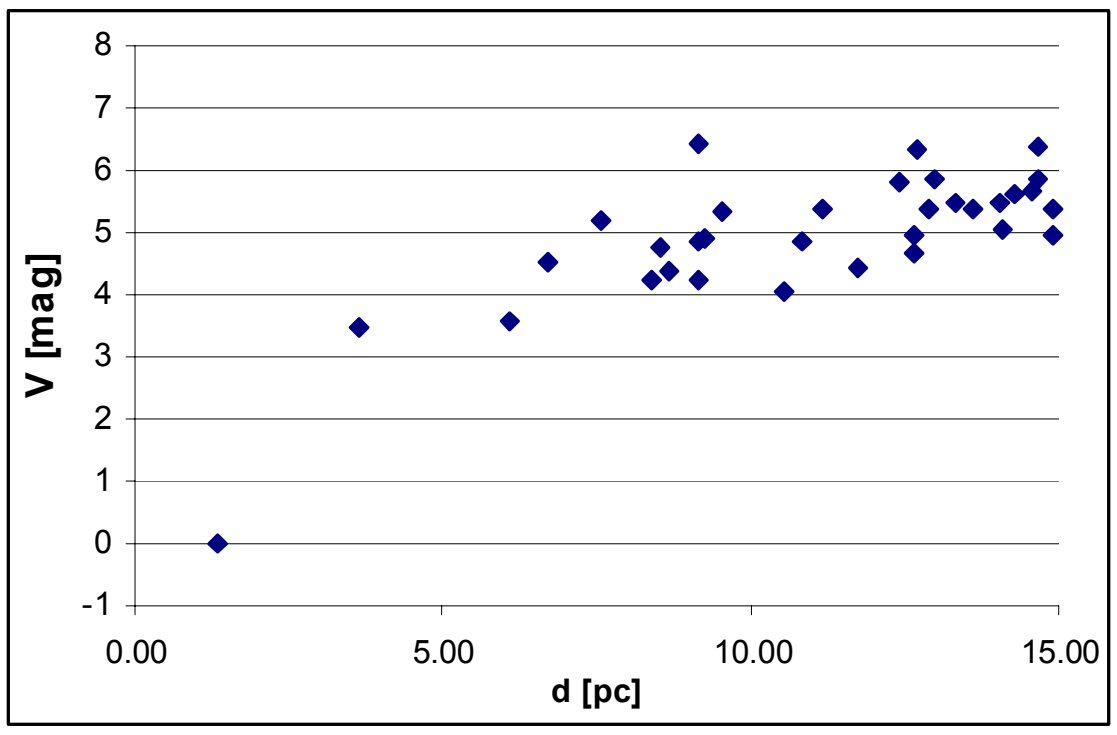

**Fig. 55: V magnitude of the closest 40 DARWIN G target stars.**

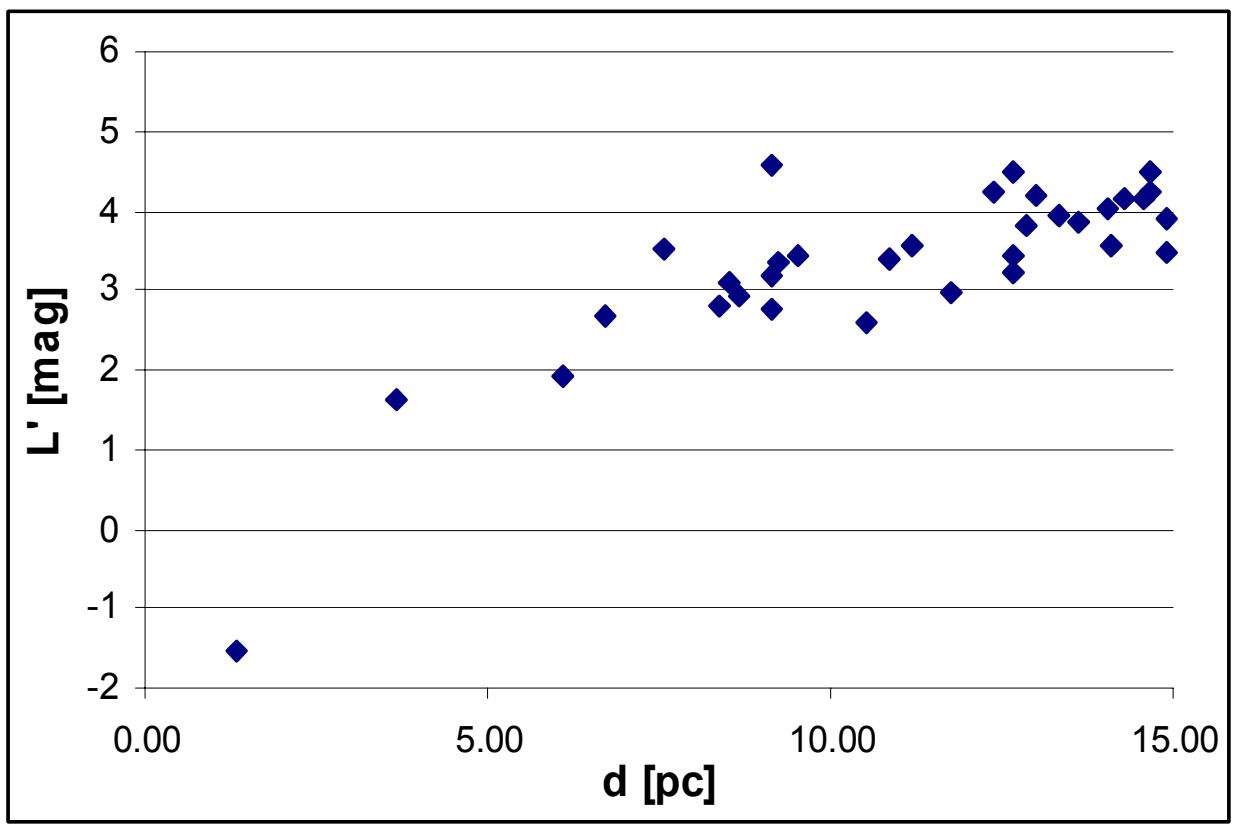

**Fig. 56: L' magnitude of the closest 40 DARWIN G target stars.**

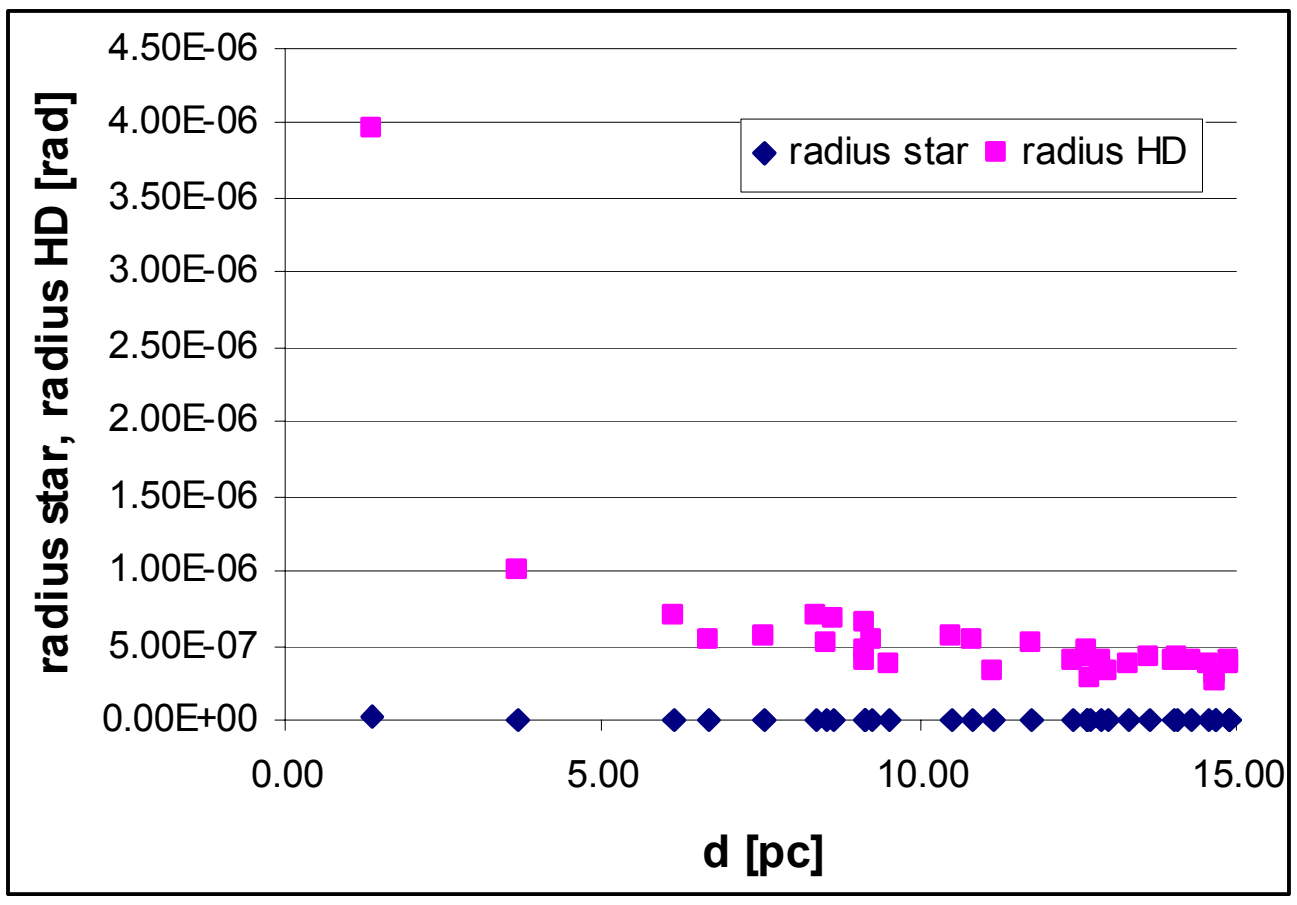

**Fig. 57: Angular extent of the radius of the star and their corresponding Hd for the closest 40 DARWIN G target stars.**

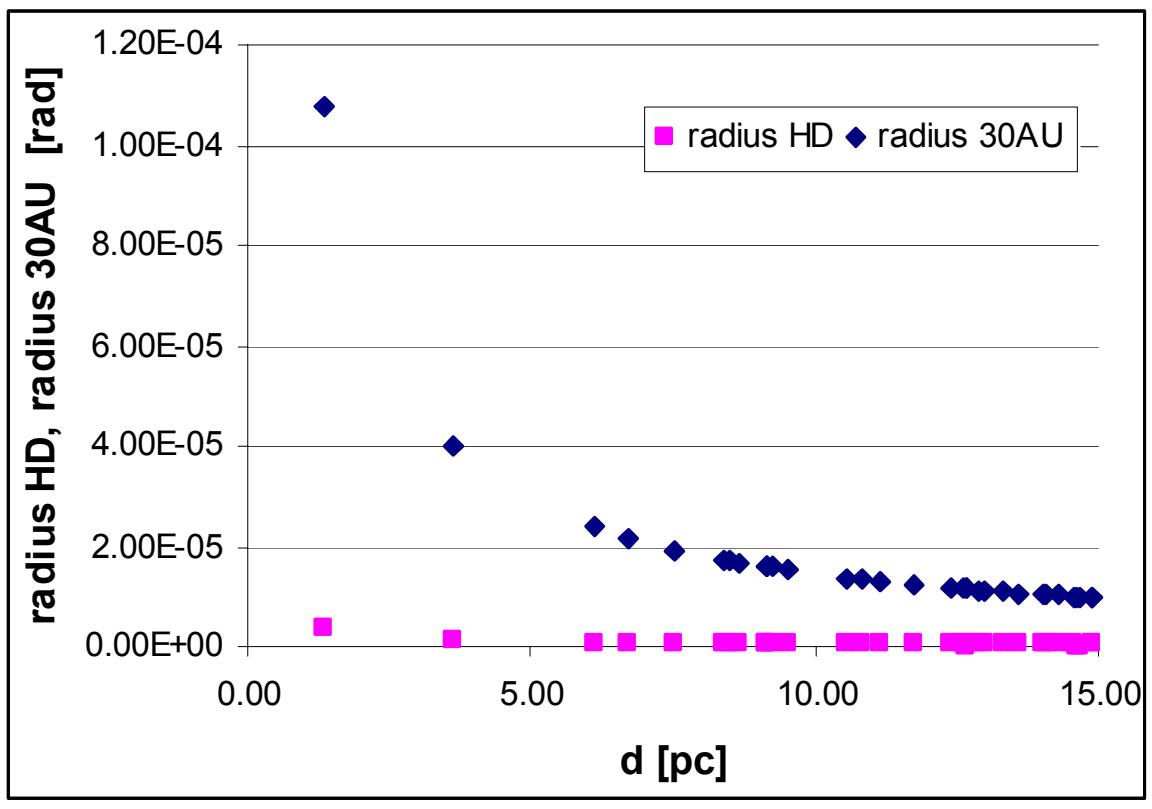

**Fig. 58: Angular extent of the radius of a 30AU disk around the closest 40 DARWIN G target stars**

## 3.7 *Angular extend of the complete sample of target stars and the corresponding Habitable distance*

For realistic simulations the maximum, minimum and mean values of the angular extend of the star as well as the Hd for the DARWIN target stars is essential. From the star target catalogue, these values are extracted (see Table 11).

**Table 11: Maximum, minimal and mean values for luminosity and angular extend of the DARWIN target stars**

| F stars | L' [mag] | V [mag] | Hd [rad] | SR [rad] | 30AU [rad] | Hd [mas] | SR [mas] | 30AU [mas] |
|---|---|---|---|---|---|---|---|---|
| max | -0.74 | 0.4 | 2.47E-06 | 1.67E-08 | 4.15E-05 | 5.09E-01 | 3.44E-03 | 8.56E+00 |
| Min | 8.56 | 9.95 | 2.70E-07 | 2.11E-09 | 5.84E-06 | 5.57E-02 | 4.35E-04 | 1.20E+00 |
| Mean | 3.65 | 4.87 | 5.07E-07 | 3.41E-09 | 8.65E-06 | 1.05E-01 | 7.03E-04 | 1.78E+00 |
| **G stars** | **L' [mag]** | **V [mag]** | **Hd [rad]** | **SR [rad]** | **30AU [rad]** | **Hd [mas]** | **SR [mas]** | **30AU [mas]** |
| max | -1.52 | -0.001 | 3.96E-06 | 3.43E-08 | 1.08E-04 | 8.17E-01 | 7.07E-03 | 2.23E+01 |
| Min | 9.84 | 11.48 | 1.48E-07 | 1.61E-09 | 5.82E-06 | 3.05E-02 | 3.32E-04 | 1.20E+00 |
| Mean | 4.55 | 6.15 | 3.18E-07 | 2.90E-09 | 9.47E-06 | 6.56E-02 | 5.98E-04 | 1.95E+00 |
| **K stars** | **L' [mag]** | **V [mag]** | **Hd [rad]** | **SR [rad]** | **30AU [rad]** | **Hd [mas]** | **SR [mas]** | **30AU [mas]** |
| max | -0.81 | 1.35 | 1.60E-06 | 2.76E-08 | 1.08E-04 | 3.30E-01 | 5.14E-03 | 2.23E+01 |
| Min | 9.72 | 11.74 | 6.34E-08 | 1.17E-09 | 5.82E-06 | 1.31E-02 | 2.41E-04 | 1.20E+00 |
| Mean | 5.62 | 8.20 | 1.55E-07 | 2.22E-09 | 9.33E-06 | 3.20E-02 | 4.58E-04 | 1.92E+00 |
| **M stars** | **L' [mag]** | **V [mag]** | **Hd [rad]** | **SR [rad]** | **30AU [rad]** | **Hd [mas]** | **SR [mas]** | **30AU [mas]** |
| max | 2.61 | 6.69 | 4.20E-07 | 9.40E-09 | 1.12E-04 | 8.66E-02 | 1.94E-03 | 2.31E+01 |
| Min | 8.7 | 12.00 | 2.10E-08 | 5.00E-10 | 5.80E-06 | 4.33E-03 | 1.03E-04 | 1.20E+00 |
| Mean | 6.05 | 10.53 | 9.16E-08 | 1.83E-09 | 1.27E-05 | 1.89E-02 | 3.77E-04 | 2.62E+00 |

# 4 Details of target stars by stellar subgroup

## 4.1 *F target stars*

74 F stars within 25pc, B-V color for main sequence stars, the closest star 3.48 pc.

**Table 12: Maximum, minimal and mean values for luminosity and angular extend of the DARWIN F type target stars**

| F stars | L' [mag] | V [mag] | Hd [rad] | SR [rad] | 30AU [rad] | Hd [mas] | SR [mas] | 30AU [mas] |
|---|---|---|---|---|---|---|---|---|
| max | -0.74 | 0.40 | 2.47E-06 | 1.67E-08 | 4.15E-05 | 5.09E-01 | 3.44E-03 | 8.56E+00 |
| Min | 8.56 | 9.95 | 2.70E-07 | 2.11E-09 | 5.84E-06 | 5.57E-02 | 4.35E-04 | 1.20E+00 |
| Mean | 3.65 | 4.87 | 5.07E-07 | 3.41E-09 | 8.65E-06 | 1.05E-01 | 7.03E-04 | 1.78E+00 |

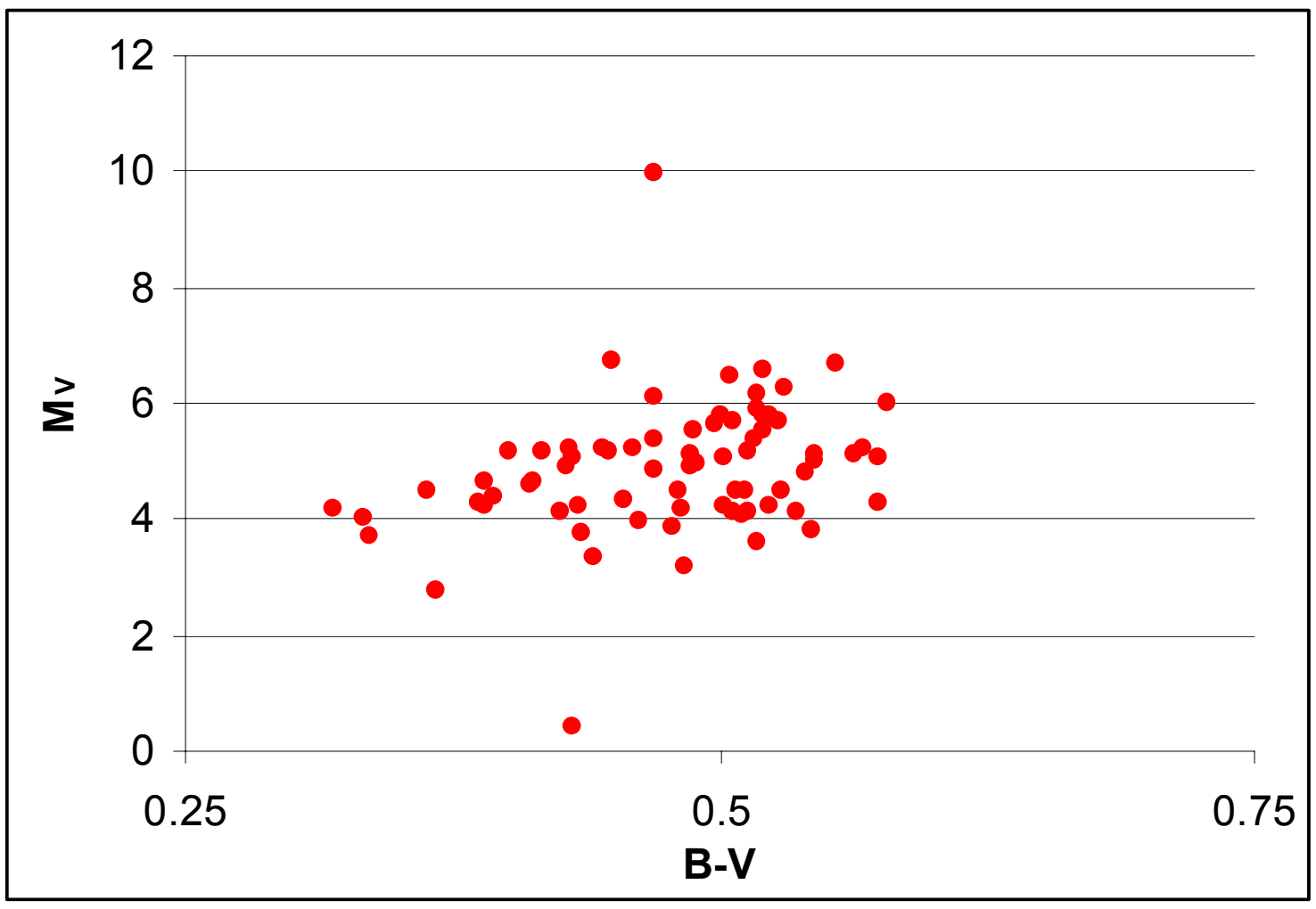

**Fig. 59: Color-magnitude diagram for the F target stars.**

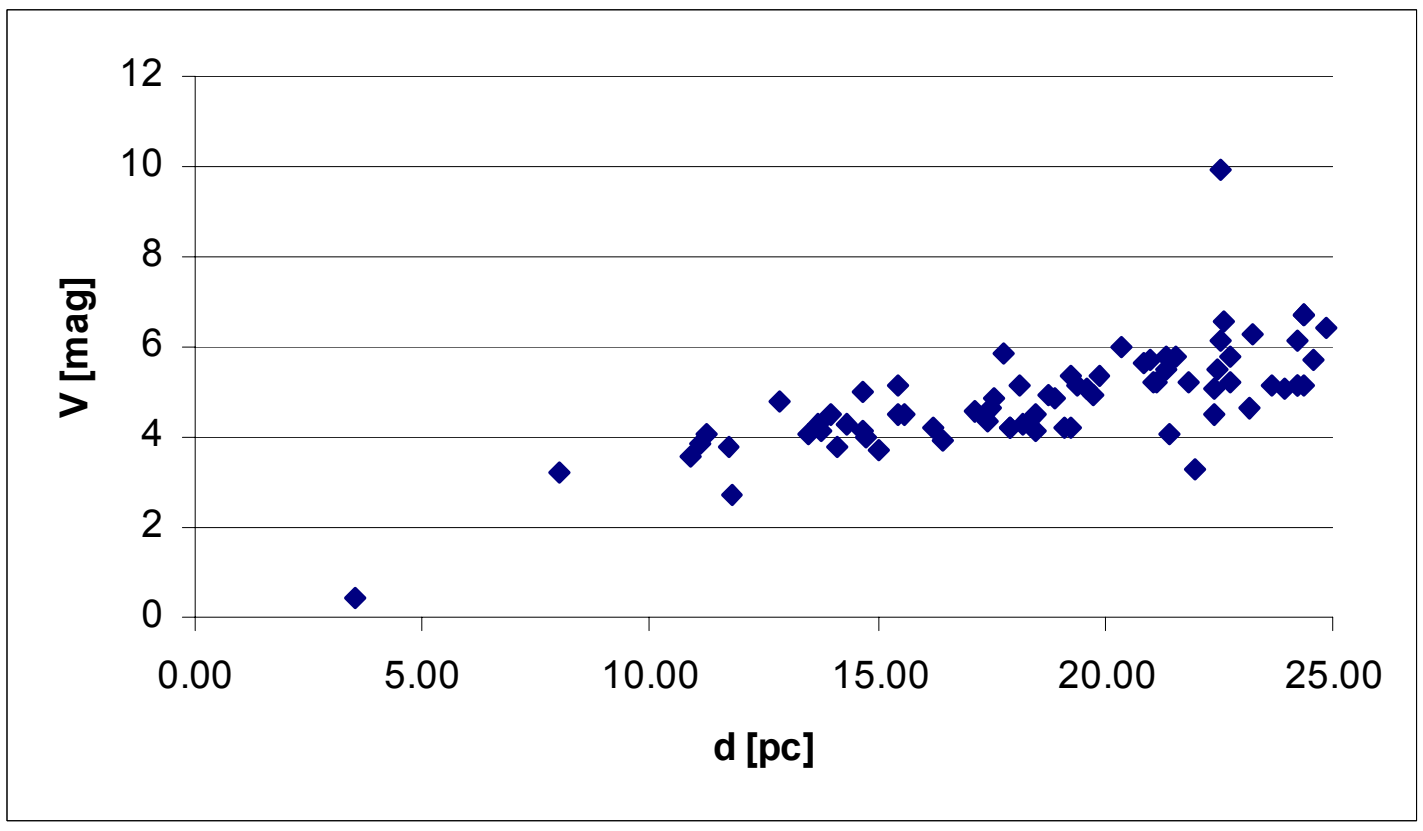

**Fig. 60: V magnitude of the F target stars.**

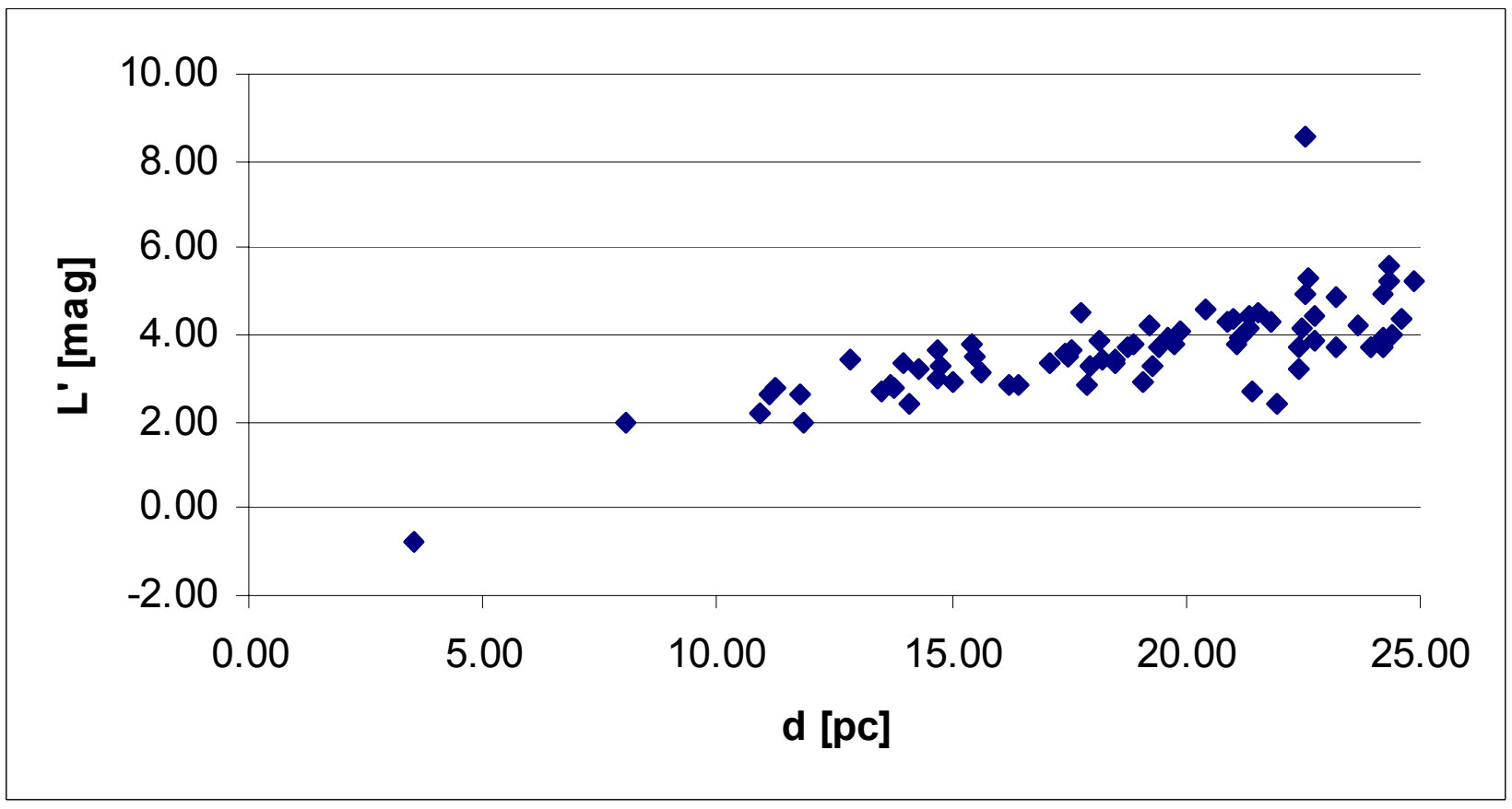

**Fig. 61: L' magnitude of the F target stars.**

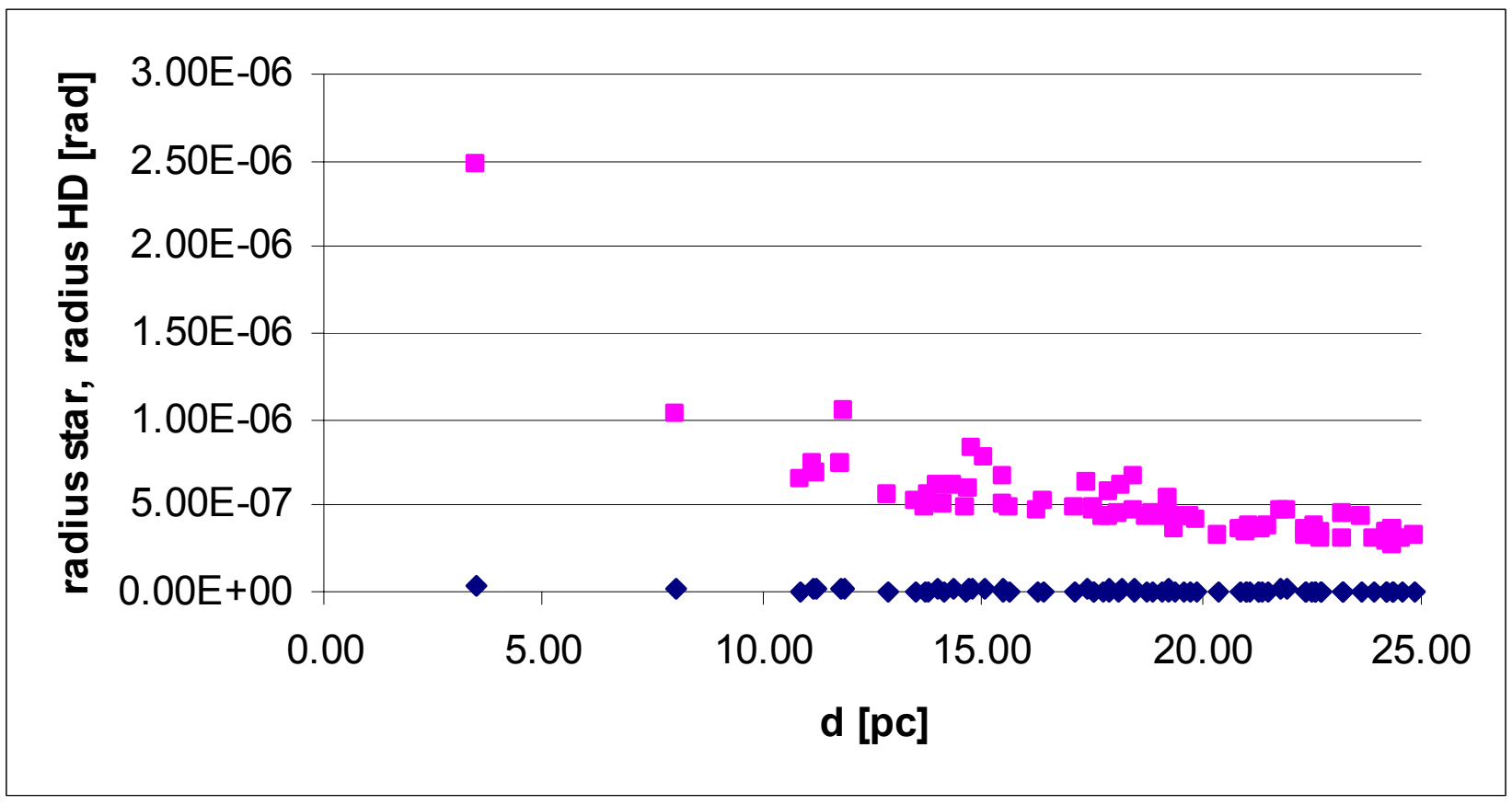

**Fig. 62: Angular extent of the radius of the star and their corresponding Hd of the F target stars.**

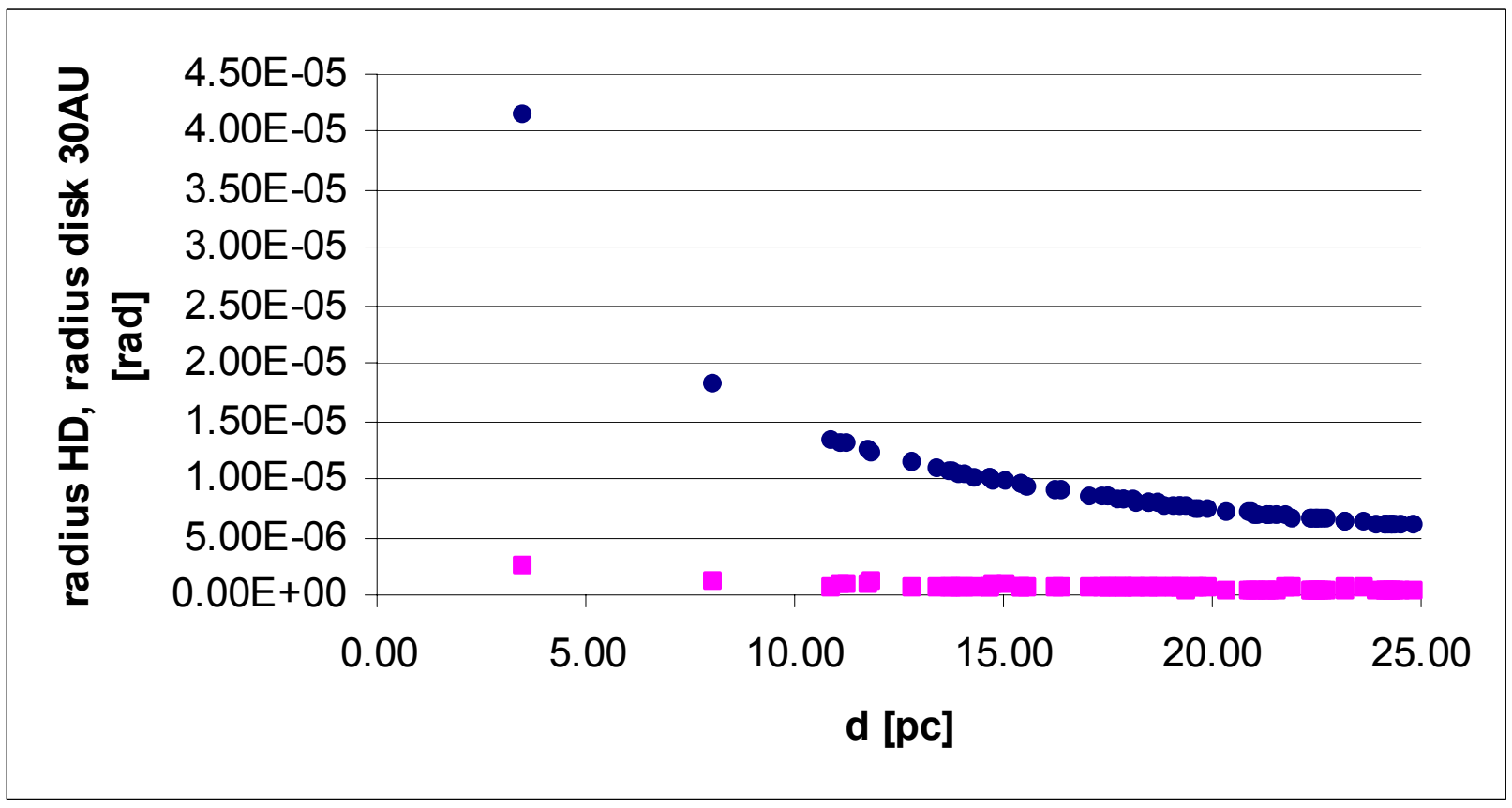

**Fig. 63: Angular extent of the radius of Hd and 30AU of the F target stars.**

## *4.2 G target stars:*

143 G stars within 25pc, B-V color for main sequence stars, the closest star 1.35 pc.

**Table 13: Maximum, minimal and mean values for luminosity and angular extend of the DARWIN F type target stars**

| G stars | L' [mag] | V [mag] | Hd [rad] | SR [rad] | 30AU [rad] | Hd [mas] | SR [mas] | 30AU [mas] |
|---|---|---|---|---|---|---|---|---|
| max | -1.52 | -0.001 | 3.96E-06 | 3.43E-08 | 1.08E-04 | 8.17E-01 | 7.07E-03 | 2.23E+01 |
| min | 9.84 | 11.48 | 1.48E-07 | 1.61E-09 | 5.82E-06 | 3.05E-02 | 3.32E-04 | 1.20E+00 |
| mean | 4.55 | 6.15 | 3.18E-07 | 2.90E-09 | 9.47E-06 | 6.56E-02 | 5.98E-04 | 1.95E+00 |

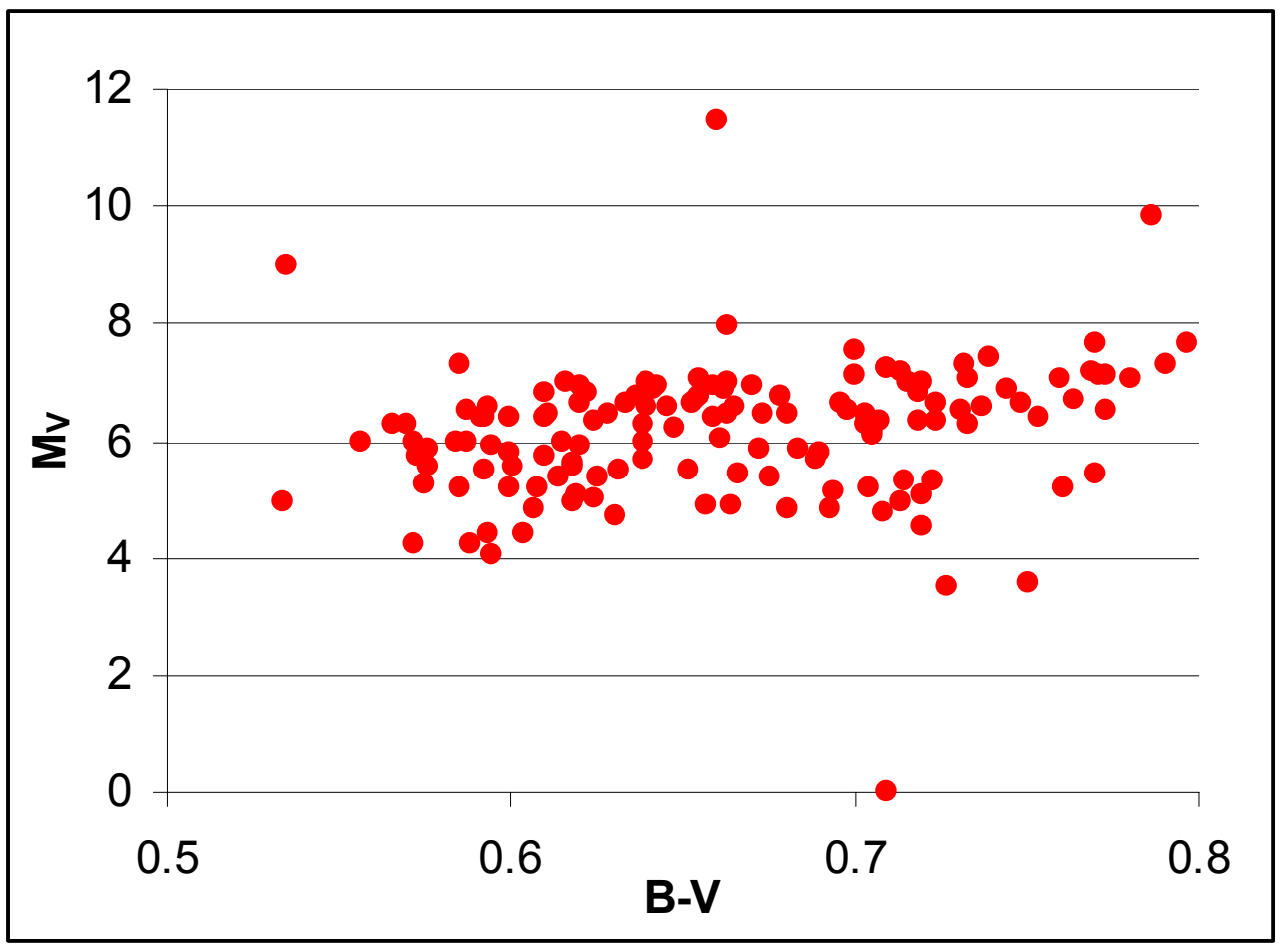

**Fig. 64: Color-magnitude diagram for the G target stars.**

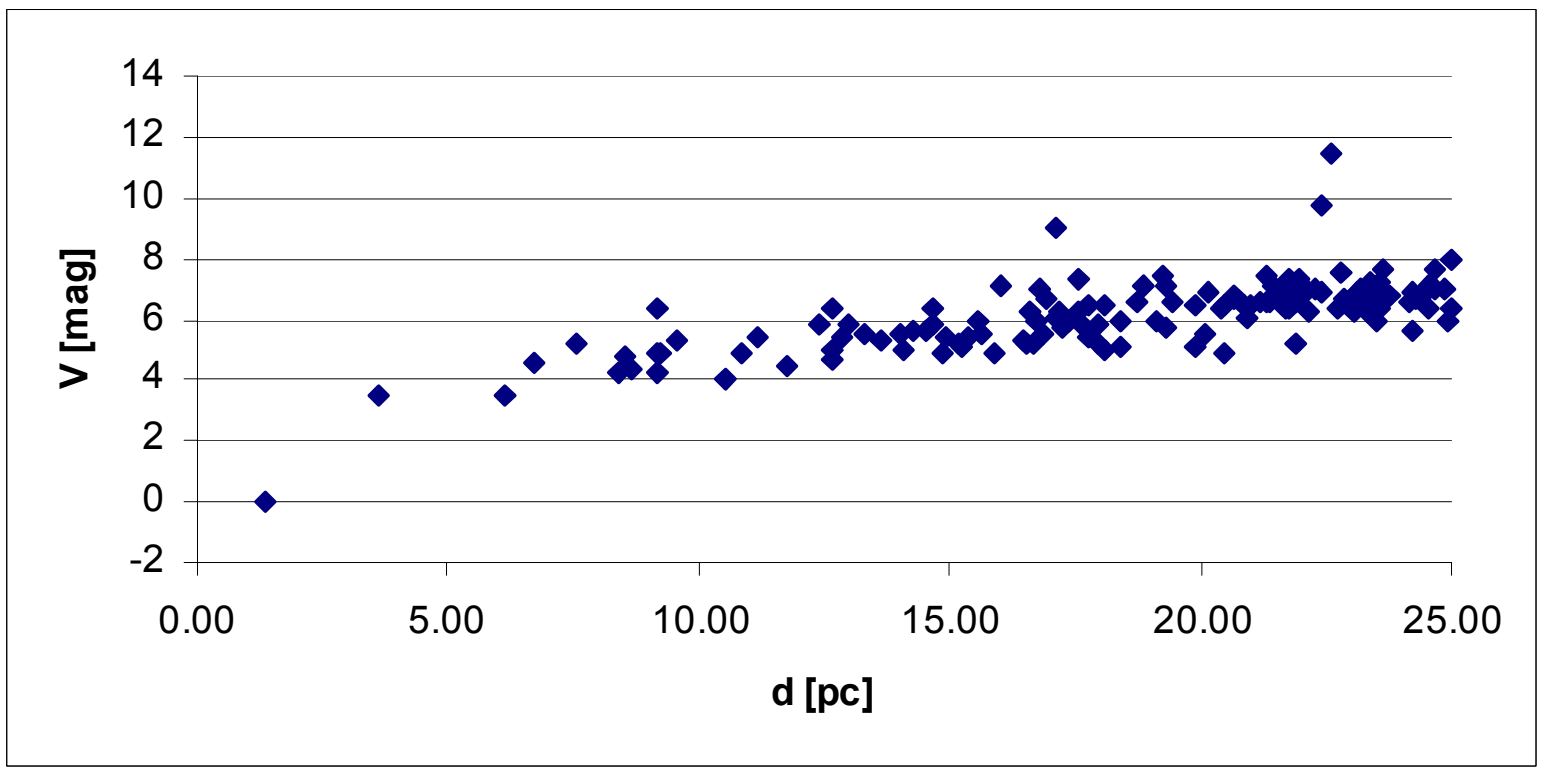

**Fig. 65: V magnitude of the G target stars.**

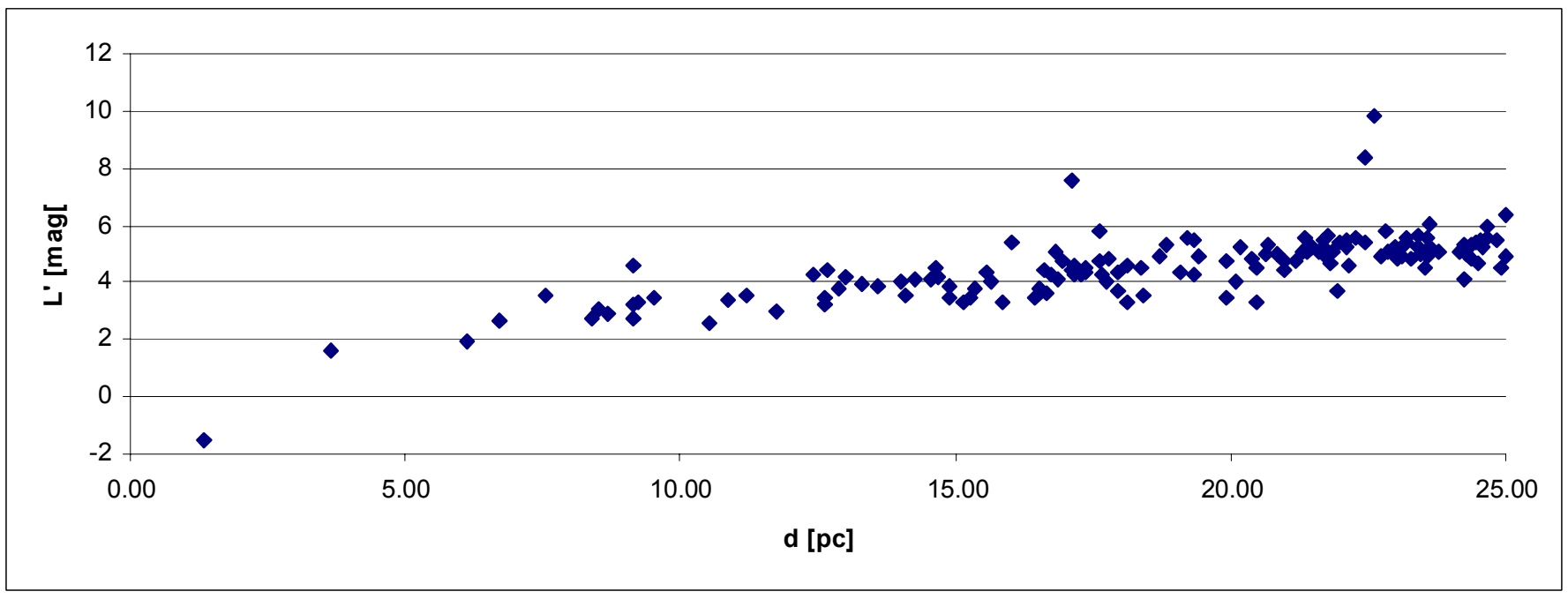

**Fig. 66: L' magnitude of the G target stars.**

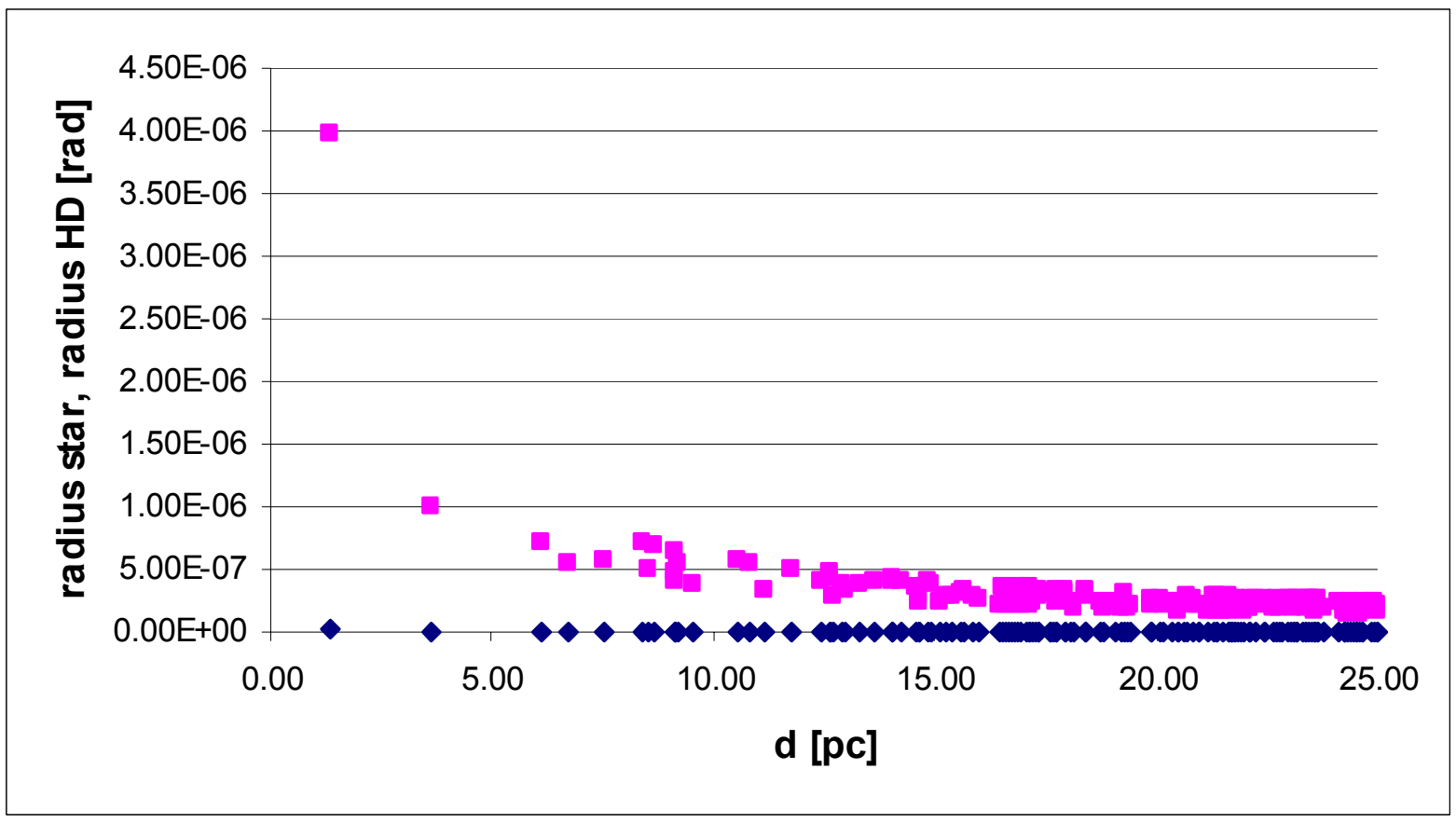

**Fig. 67: Angular extent of the radius of the G target stars and their Hd**

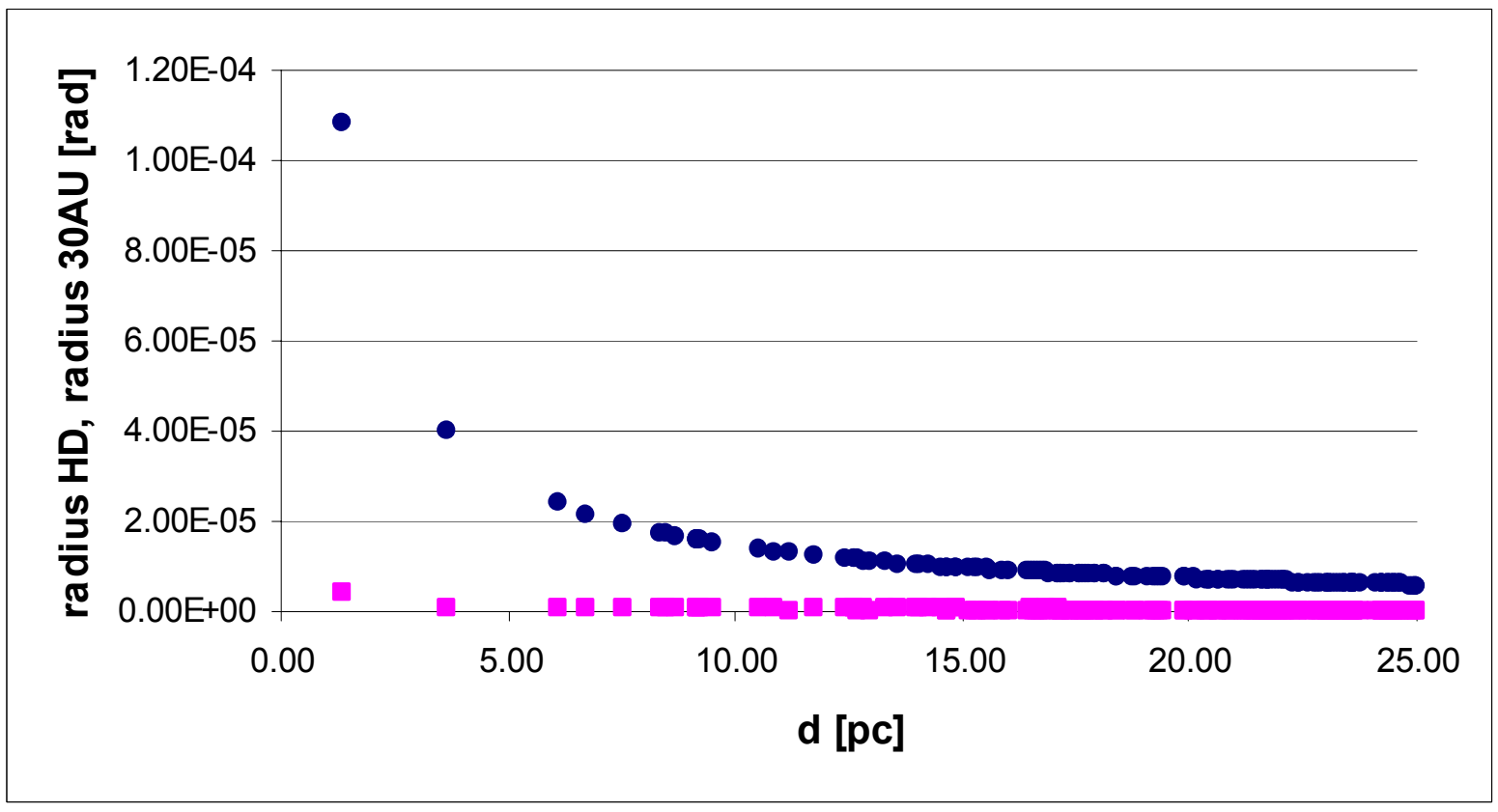

**Fig. 68: Angular extent of the radius of Hd and 30AU of the G target stars.**

## 4.3 K target stars

309 K stars within 25pc, B-V color for main sequence stars, the closest star 3.48 pc (of these stars most do not have a subclass classification, thus a main part should turn out to be non-main sequence stars as brighter stars (giant stars) are picked up more easily by observations).

**Table 14: Maximum, minimal and mean values for luminosity and angular extend of the DARWIN K type target stars**

| K stars | L' [mag] | V [mag] | Hd [rad] | SR [rad] | 30AU [rad] | Hd [mas] | SR [mas] | 30AU [mas] |
|---|---|---|---|---|---|---|---|---|
| max | -0.81 | 1.35 | 1.60E-06 | 2.76E-08 | 1.08E-04 | 3.30E-01 | 5.14E-03 | 2.23E+01 |
| Min | 9.72 | 11.74 | 6.34E-08 | 1.17E-09 | 5.82E-06 | 1.31E-02 | 2.41E-04 | 1.20E+00 |
| Mean | 5.62 | 8.2 | 1.55E-07 | 2.22E-09 | 9.33E-06 | 3.20E-02 | 4.58E-04 | 1.92E+00 |

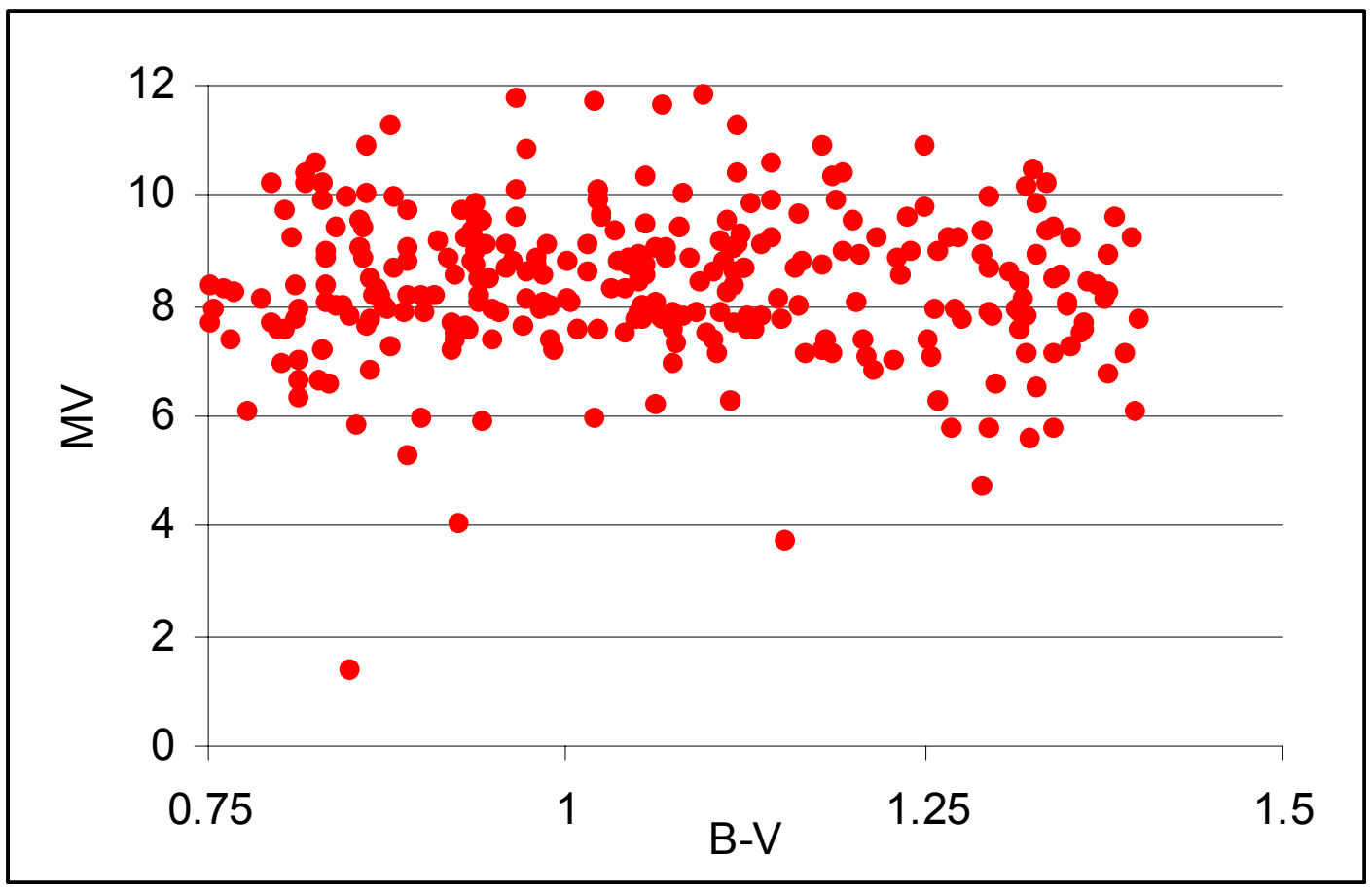

**Fig. 69: Color-magnitude diagram for the K star target sample.**

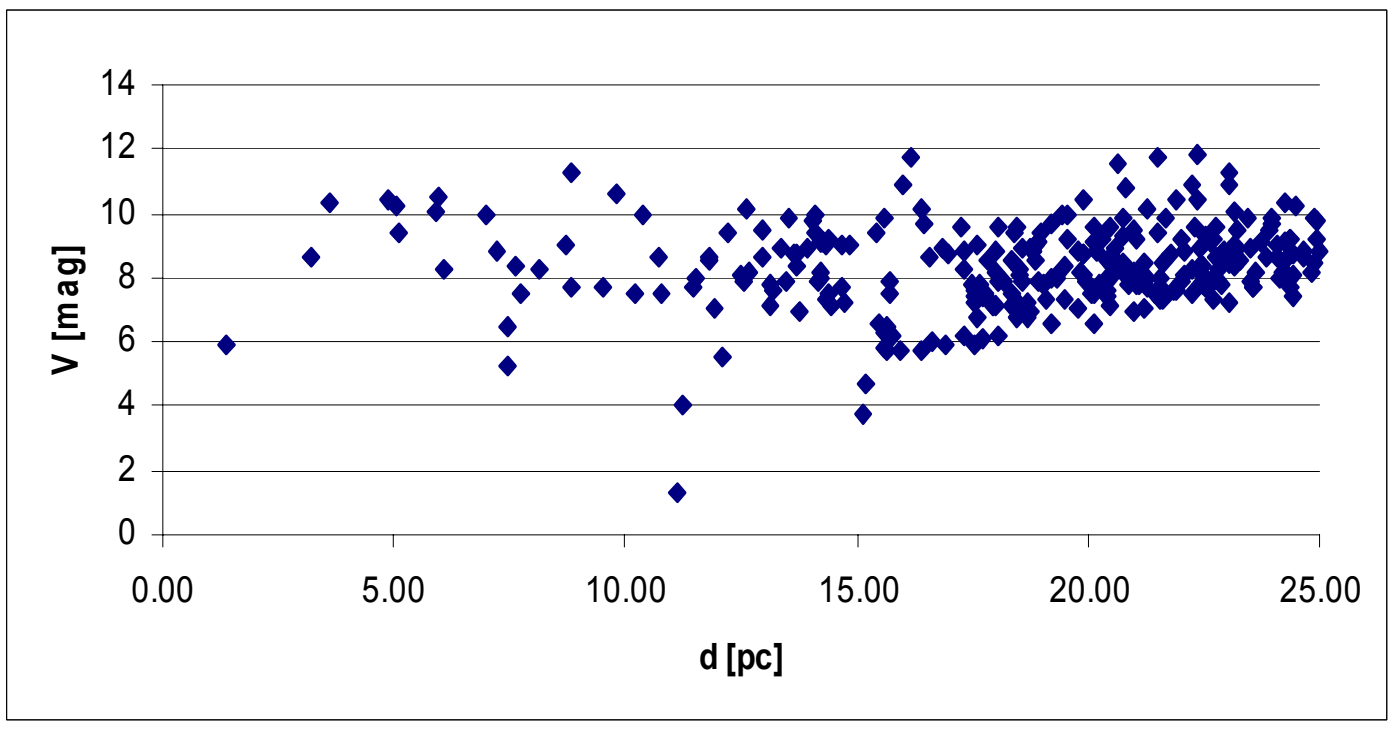

**Fig. 70: V magnitude of the K target stars.**

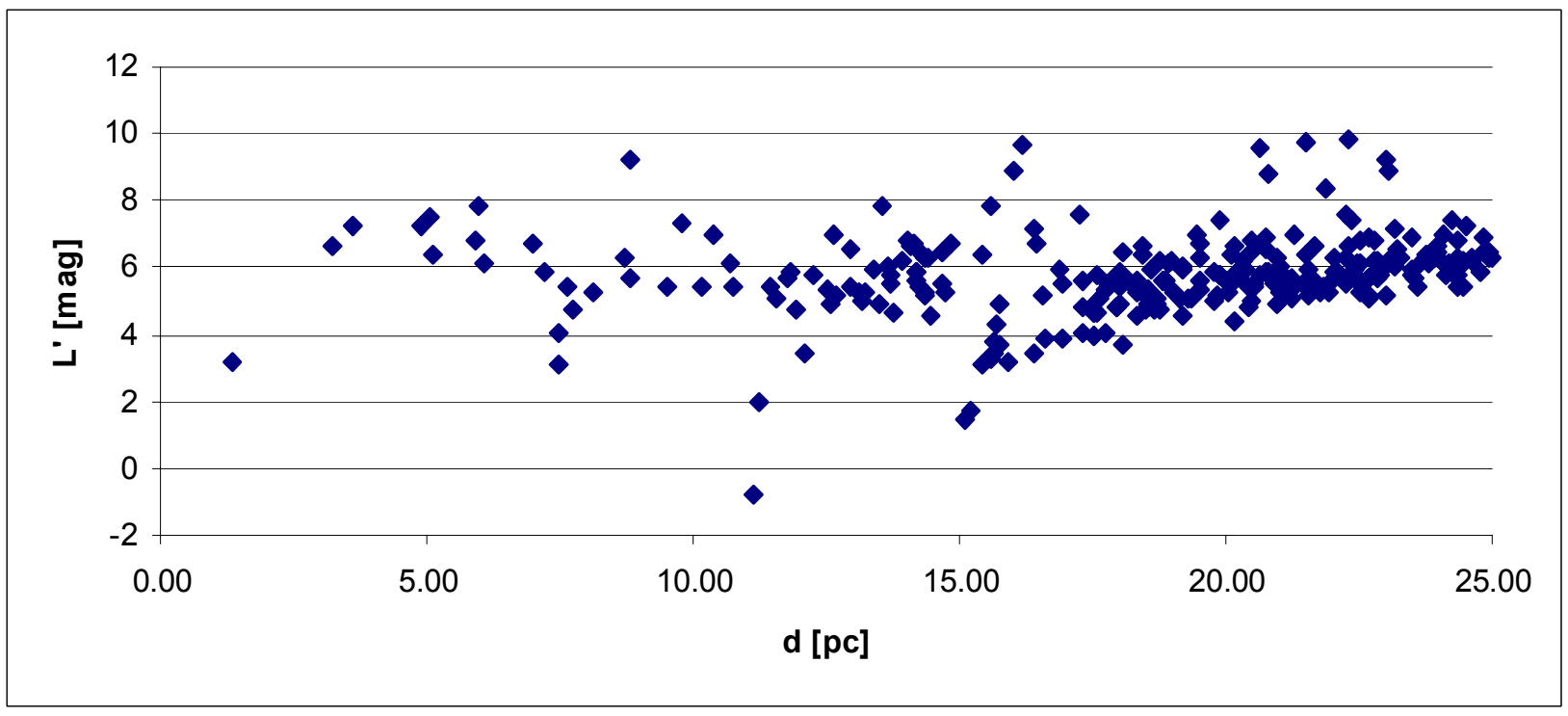

**Fig. 71: L' magnitude of the K target stars.**

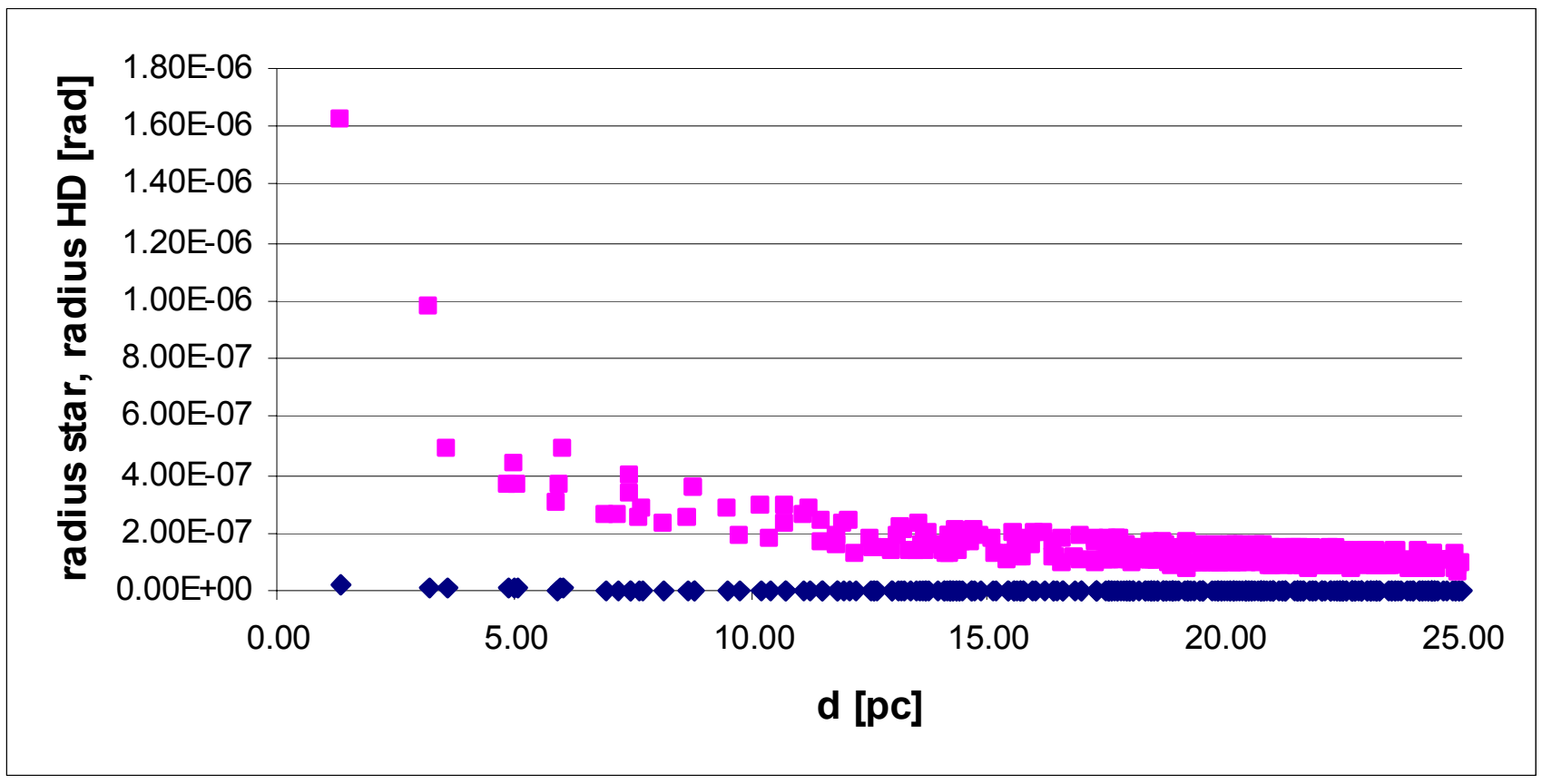

**Fig. 72: Angular extent of the radius of the K target stars and their Hd**

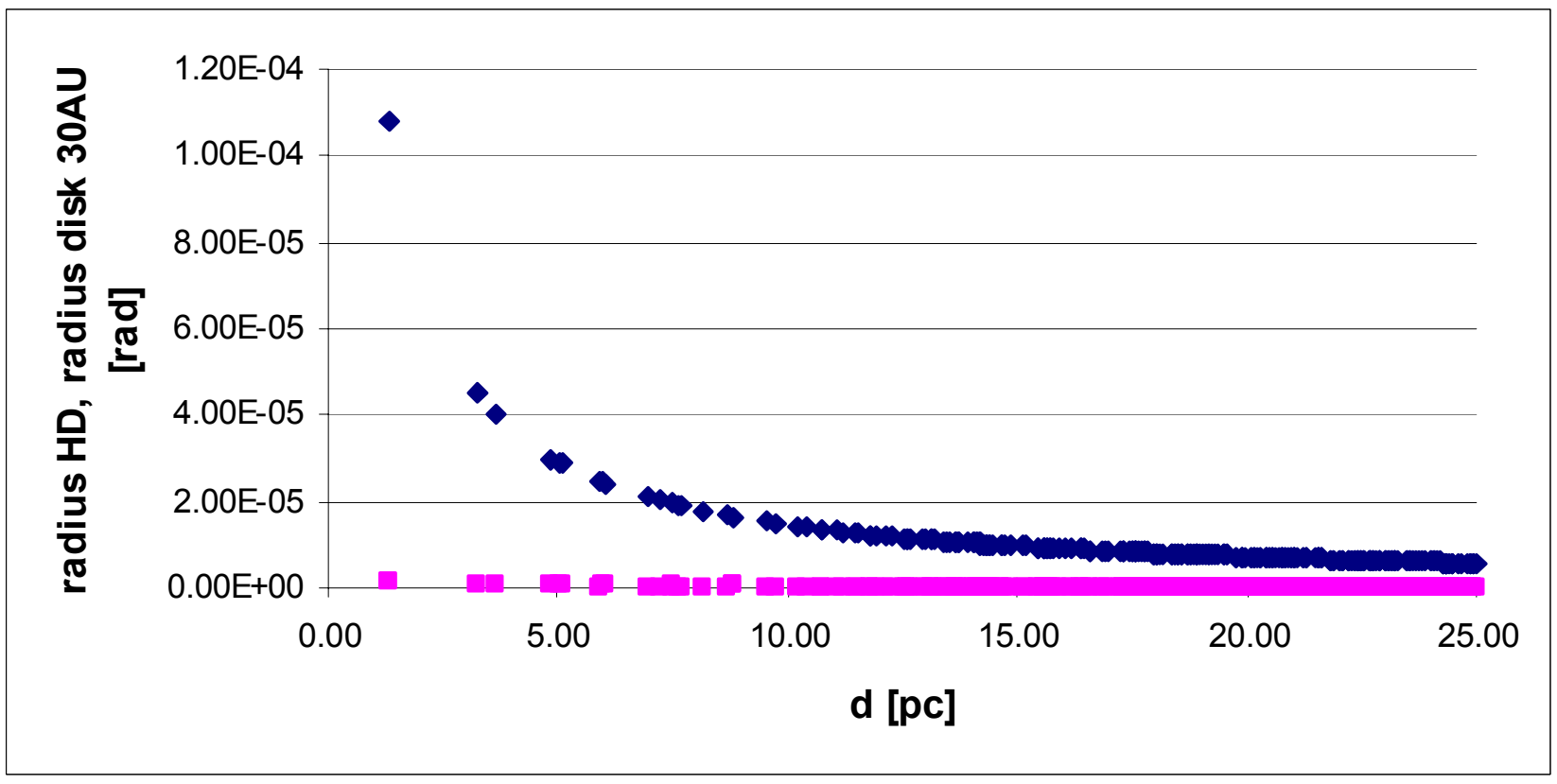

**Fig. 73: Angular extent of the radius of Hd and 30AU of the K target stars.**

## *4.4 M target stars*

300 M stars within 25pc, B-V color for main sequence stars, the closest star 1.29 pc (of these stars most do not have a subclass classification, thus a main part should turn out to be non-main sequence stars as brighter stars (giant stars) are picked up more easily by observations).

**Table 15: Maximum, minimal and mean values for luminosity and angular extend of the DARWIN M type target stars**

| M stars | L' [mag] | V [mag] | Hd [rad] | SR [rad] | 30AU [rad] | Hd [mas] | SR [mas] | 30AU [mas] |
|---|---|---|---|---|---|---|---|---|
| max | 2.61 | 6.69 | 4.20E-07 | 9.40E-09 | 1.12E-04 | 8.66E-02 | 1.94E-03 | 2.31E+01 |
| Min | 8.70 | 12.00 | 2.10E-08 | 5.00E-10 | 5.80E-06 | 4.33E-03 | 1.03E-04 | 1.20E+00 |
| mean | 6.05 | 10.53 | 9.16E-08 | 1.83E-09 | 1.27E-05 | 1.89E-02 | 3.77E-04 | 2.62E+00 |

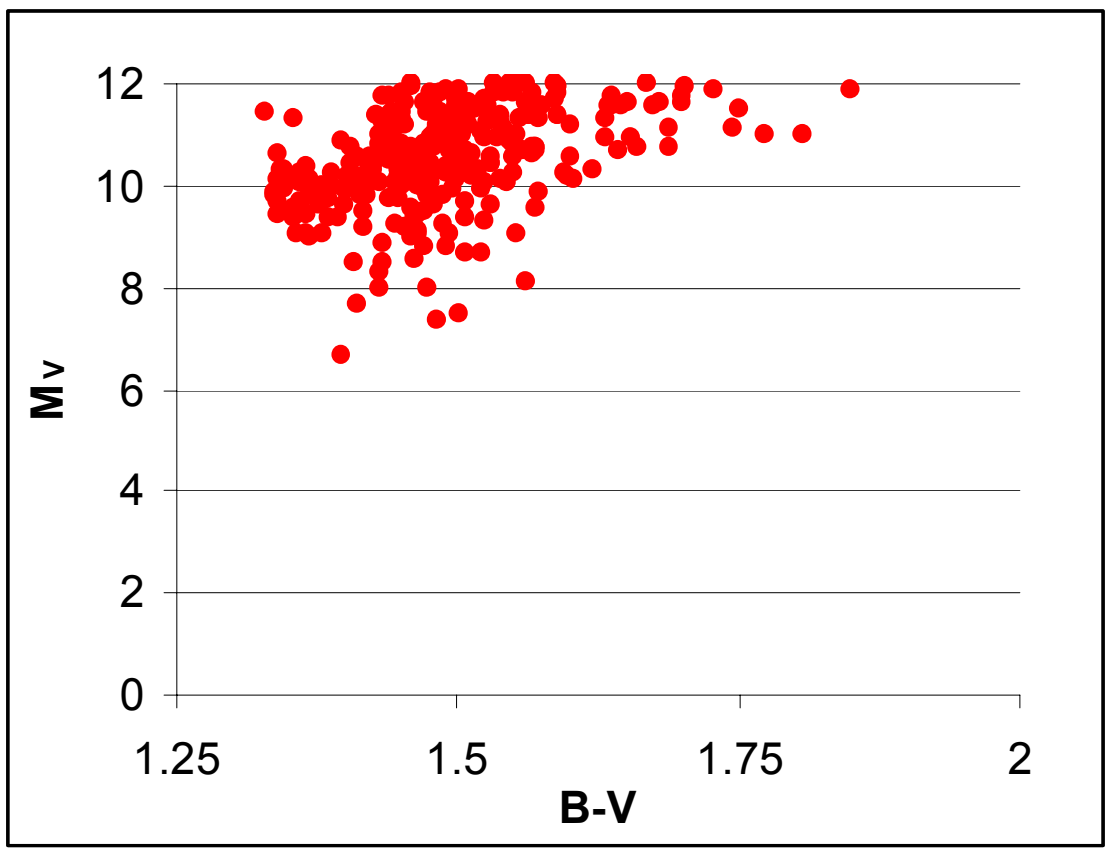

**Fig. 74: Color-magnitude diagram for the M target stars.**

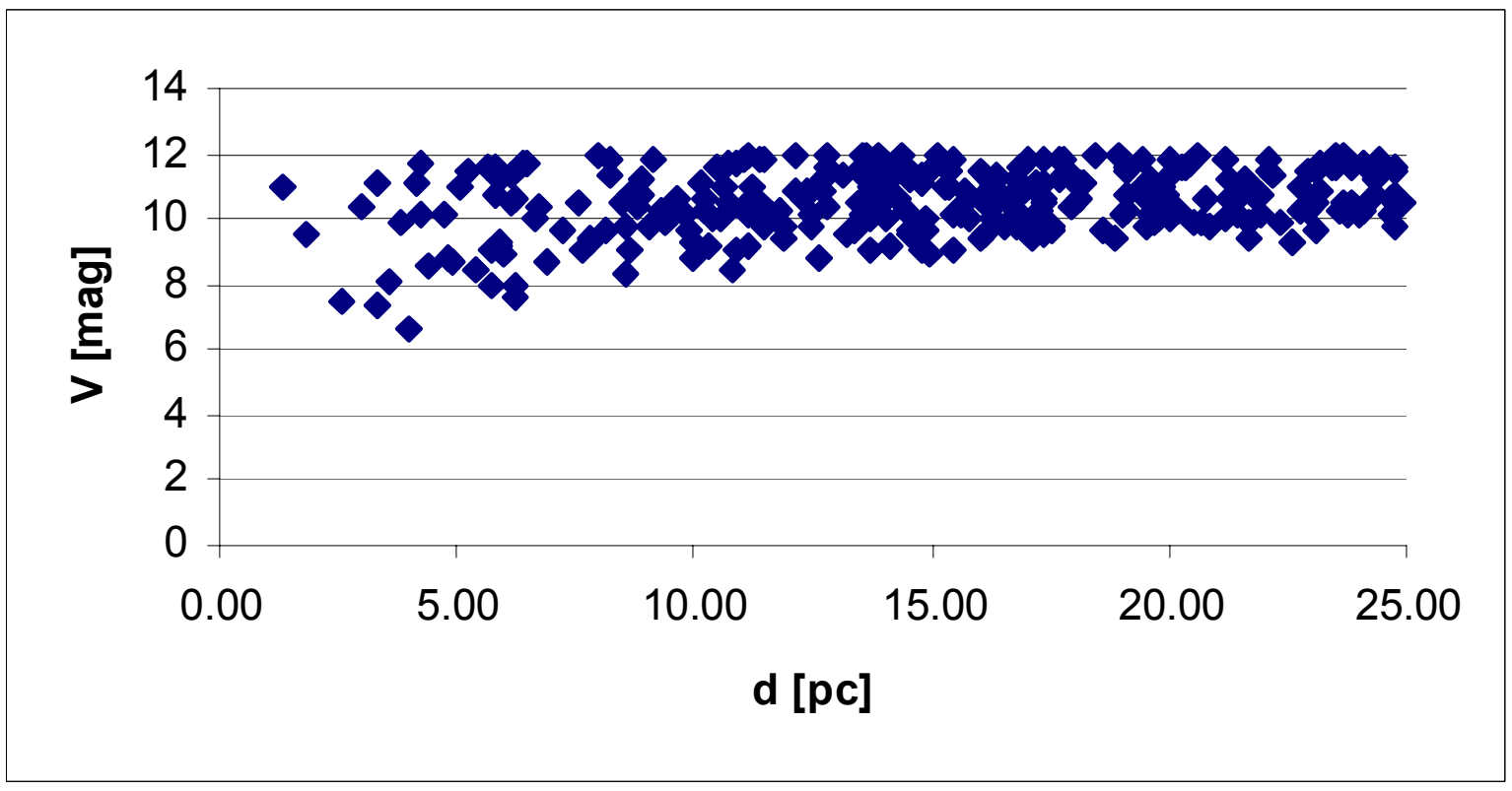

**Fig. 75: V magnitude of the M target stars.**

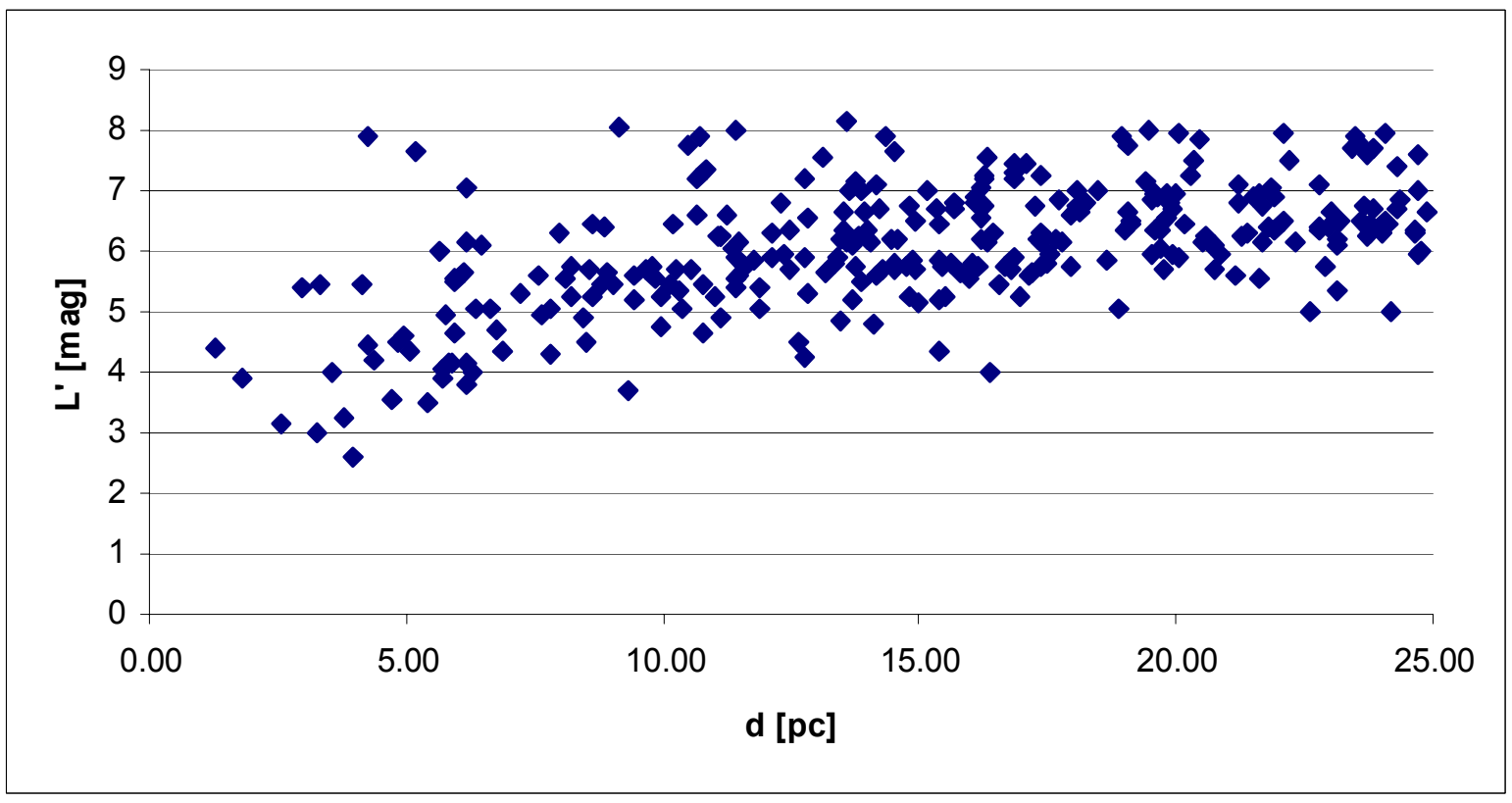

**Fig. 76: L' magnitude of the M target stars.**

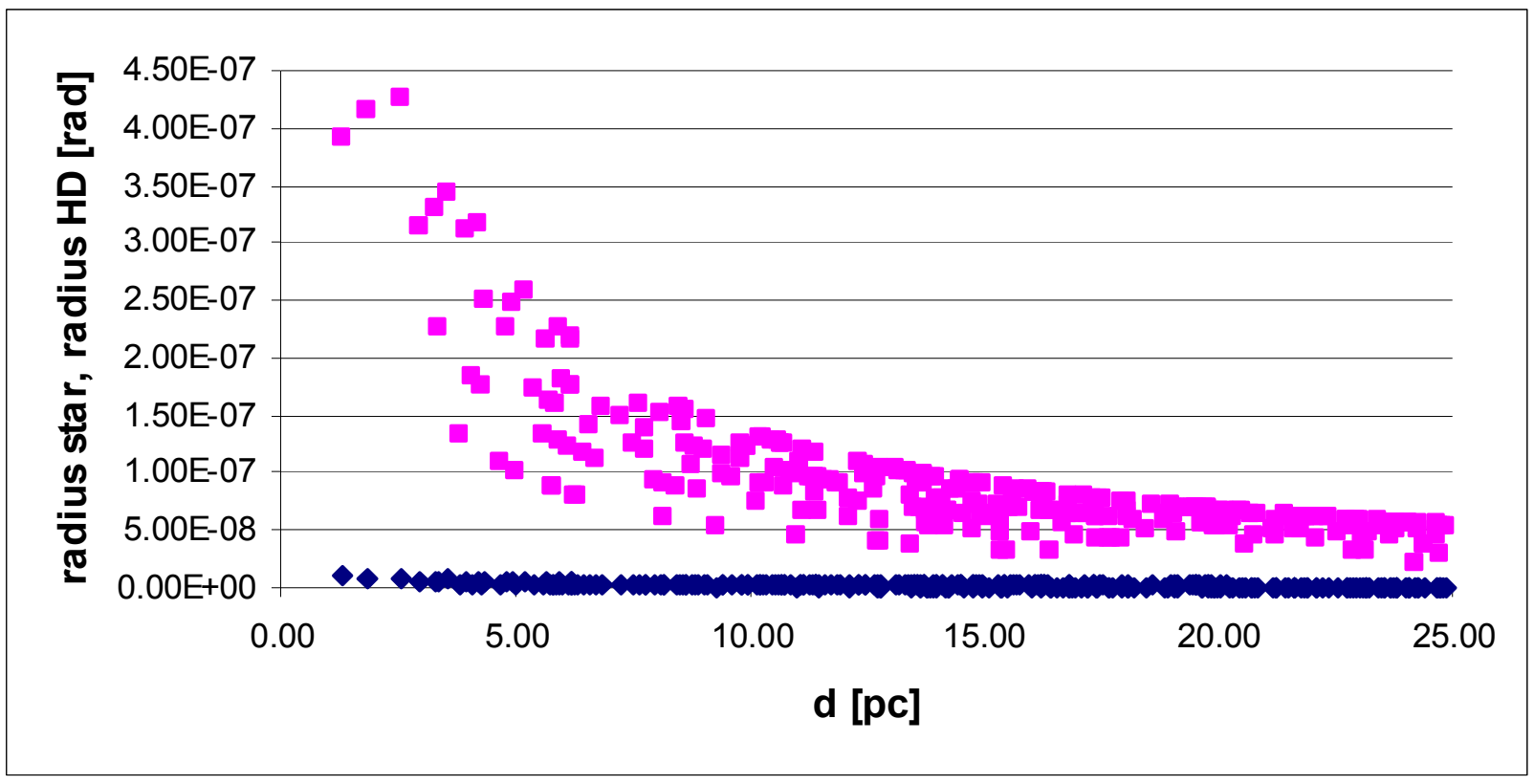

**Fig. 77: Angular extent of the radius of the M target stars and their Hd**

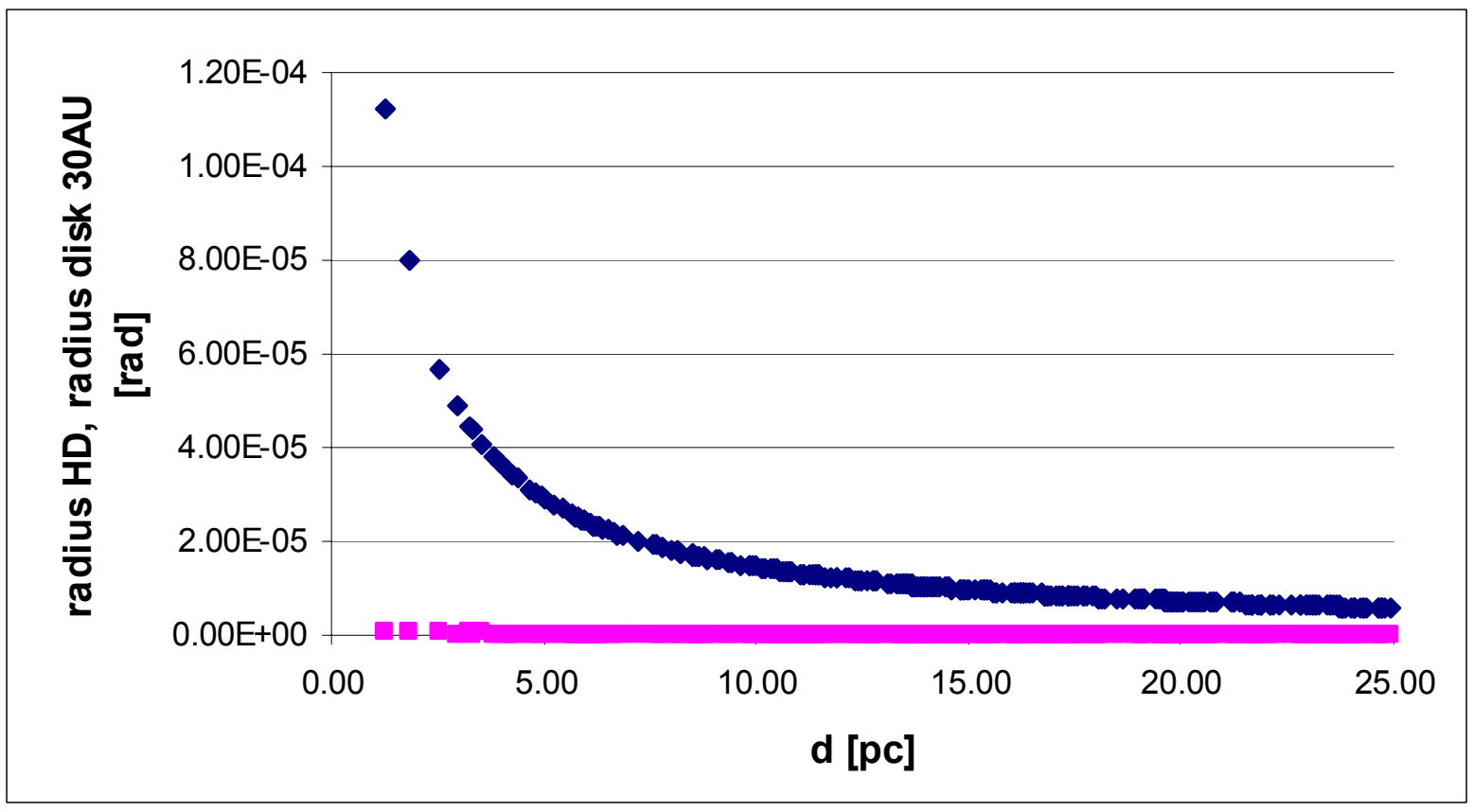

**Fig. 78: Angular extent of the radius of Hd and 30AU of the M target stars.**

## 4.5 Calculation of basic stellar data: Interpolation model

To determine the temperature, luminosity, mass and radius of the different stellar types we use linear interpolation between the luminosity classes of the stars (Cox, 2000). The curves are not straight lines because the luminosity classes of stars are determined by the stellar spectra, so the classes do not correspond to an equal temperature range

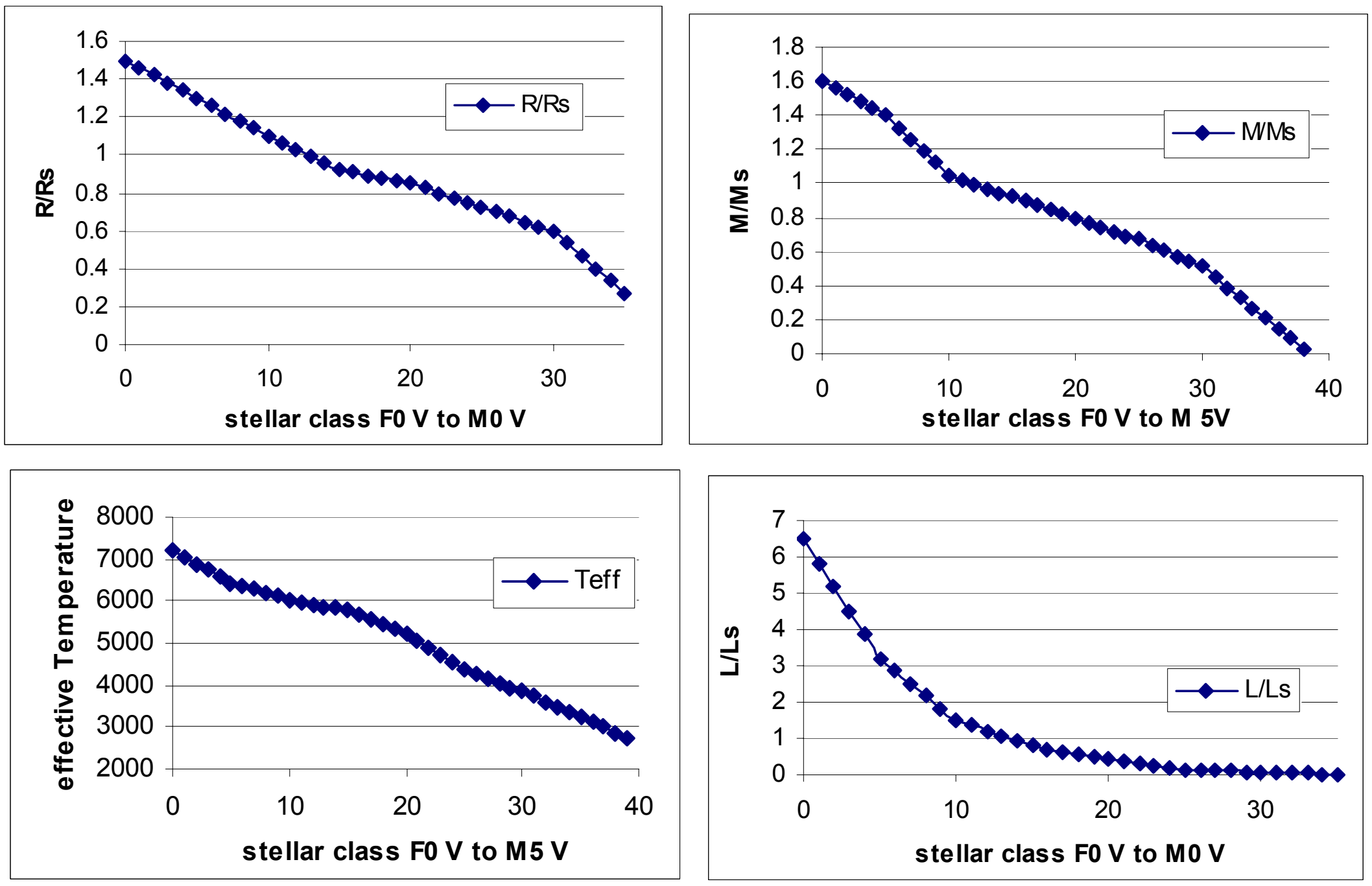

**Fig. 79: Linear Interpolation between luminosity classes.**

**Table 16: Data for the linear interpolation model between stellar luminosity classes used to determine luminosity, temperature and radius (Cox, 2000)**

| luminosity class | L/Ls | R/Rs | Teff | rho/rhos | M/Ms | g/gs |
|---|---|---|---|---|---|---|
| F0 V | 6.5 | 1.5 | 7200 | 0.5 | 1.6 | 0.8 |
| F5 V | 3.2 | 1.3 | 6440 | 0.63 | 1.4 | 0.8 |
| G0 V | 1.5 | 1.1 | 6030 | 0.8 | 1.05 | 0.9 |
| G5 V | 0.79 | 0.92 | 5770 | 0.8 | 0.92 | 1.12 |
| K0 V | 0.42 | 0.85 | 5250 | 1.26 | 0.79 | 1.12 |
| K5 V | 0.15 | 0.72 | 4350 | 1.78 | 0.67 | 1.26 |
| M0 V | 0.077 | 0.6 | 3850 | 2.24 | 0.51 | 1.41 |
| M5 V | 0.011 | 0.27 | 3240 | 10 | 0.21 | 3.16 |

# 5 DARWIN mission targets

The Infra Red Space Interferometer Darwin is an integral part of ESA's Cosmic Vision 2020 plan, intended for a launch towards the middle of next decade. It has been the subject of a feasibility study and is now undergoing technological development. Darwin aims to detect other Earth-like worlds, analyze their characteristics, determine the composition of their atmospheres and – investigate their capability to sustain life, as we know it. As a secondary objective, it will also provide interferometric imaging of astrophysical objects in the wavelength range of the thermal infrared with unprecedented resolution (Fridlund, 2004B).

The most challenging of these objectives consist of the recording of infrared spectra of Terrestrial extrasolar planets that could detect signs of biological activity at distances up to 25pc. In order to do this, the Darwin project is constructed around the new technique of 'nulling interferometry', which exploits the wave nature of light in order to extinguish the light from a bright object, the central star in this case. At the same time the light from a nearby source, the planet, is enhanced. The contrast between planets and stars is the smallest in the infrared wavelength region, thus that region has been chosen for this mission.

The DARWIN mission targets are (Fridlund, 2004): in nulling mode:

- Direct detection of an Earth like planet at a black body temperature close to 275K, circling a list of target stars at a distance of at least 10 pc (desirable 25 pc) with a SNR $\geq$ 5-10 in a reasonable integration time $\leq$ 30 hours
- To characterize detected planets physically through a determination of its orbital elements through observations at several epochs.
- Obtaining the planets thermal spectrum with a spectral resolution large enough ($\geq$ 20) to determine its atmospheric composition

High resolution imaging is a secondary goal of the DARWIN mission and consists in

- Spatial resolution of at least 0.1 arc sec at 20 $\mu$m, at operating wavelength regime 5 $\mu$m to at least 20 $\mu$m. The dynamical range should be at least 100.
- Spectroscopic capability with a $\delta\lambda/\lambda$ allowing a resolution of at least 100 km $s^{-1}$ - 1000 km $s^{-1}$.

Several major trade offs were carried out in the beginning of the study including the rejection of a connected structure option.

The challenges when trying to achieve these goals include:

1. Trying to distinguish between extrasolar planets and extrasolar zodiacal light through rotation of the complete interferometer and through internal or inherent modulation of the signal.
2. Stellar light rejection to better than $10^{-5}$. This put stringent requirements on the optical path length compensation; the achromatic equality of the combined wave fronts (Optical Path Delays (OPD), tilt, Wave Front Error (WFE); The achromatic equality of wave front intensity (flux balance, polarization); Severe stray light requirements).
3. Thermal IR observations imply telescopes and structure at 40K; Detectors cooled to below 6K.
4. Extreme complexity of system.

The current baseline of the model mission -- which is fully presented in the final report of the European Space Agency/Alcatel (ESA SCI2000(12)) – consist of 6 free-flying 1.5 meter telescopes, flying in a hexagonal configuration. In the center of the array a beam-combining satellite equipped with optical benches for both the `nulling' and the `imaging' part of the mission, is located. Additionally to that concept other architectures are under study see section 10.

The six telescopes, the hub beam combiner satellite and a separate power and communications satellite are all foreseen to be launched by a single Ariane 5 launch vehicle into a direct transfer orbit to the L2 Lagrangian point in the Sun-Earth system. The area of the sunshields of the telescopes limit the observation area to a ±45° cone around the anti-Sun axis see section 2. The gravity gradient is relatively flat in L2 and consequently, the greatest disturbing force on the array during one observation will be differential solar photon pressure. Several layers of metrology will be used to control the formation flying array incl. radio frequencies, high precision lasers and the tracking of interferometric fringes from the guide star. The guide star will be the central star observed through a separate channel on the 'short' side of the wavelength band used for science data. μN-thrusters will fire continuously in order to keep the array in phase. This technology is extremely complex, and new in space related applications. It will therefore be a prime candidate for flight validation on a precursor mission, like a possible SMART3 mission.

# 6 Habitability

The DARWIN concept is based on the assumption that one can screen extrasolar planets for habitability spectroscopicaly. Observations over interstellar distances can only distinguish those planets whose habitability and biological activity is apparent from observations of the reflected or emitted radiation. Planets with habitable surfaces that are hidden by totally opaque atmospheres as well as biospheres with liquid water only in the subsurface probably cannot be recognized as habitable. Even if that means we may miss some habitable planets, we are more concerned about concluding that a planet is habitable, when it is not. Basing the search for life on the carbon chemistry assumption allows establishing criteria for habitable planets in terms of their size and distance from their stars. DARWIN also has the unique capacity to investigate the physical properties and composition of a broader diversity of planets to understand the formation of planets and interpret potential biosignature compounds. Atmospheric features can provide clues of possible life forms for at least 2 billion years (Owen, 1980), more than $10^7$ times longer than radio signals reveal the presence of an advanced civilization on our planet. NASA also develops a similar concept for an IR free flying interferometer, called TPF. The design study has not been finalized yet and an alternative concept based on a connected structure and a third option, a mission in the visible using a coronograph are still evaluated.

## 6.1 Development of life on Earth

The Earth formed about 4.5 billion years ago. The release of gases from the interior formed the primitive atmosphere, most likely dominated by carbon dioxide, with nitrogen being the second most abundant gas and trace amounts of methane, ammonia, sulphur dioxide, hydrochloride acid and oxygen (Kasting, 1984). Carbon dioxide and/or methane played a crucial role for the development of an early greenhouse effect that counteracted the lower solar output. Two major processes have changed the primitive atmosphere, the reduction of $CO_2$ and the increase of $O_2$ (Kasting, 1988). A huge amount of $CO_2$ must have been removed from the atmosphere, most likely by the burial of carbon into carbonate rocks, though the process is still debated. Primitive cyanobacteria are believed to have produced oxygen. After most of the reduced minerals were oxidized, about two billion years ago, atmospheric oxygen could accumulate.

## 6.2 Search for life

The search for life requires a general definition of life. Leger (Leger, 1996) defines a living being according to biochemistry and biophysics as a system that:

- contains information
- is able to replicate itself
- undergoes random changes in its information package that allow Darwian evolution to proceed

The DARWIN team currently uses this definition. It has to be refined further see e.g. Ehrenfreund et al. (Ehrenfreund, 2003). A final definition of life would give an answer, how to search for different life forms – but the structure of life forms unlike our own is unknown so far, as well as the signals that show their present in a planets atmosphere.

*"Our imagination in this domain is both too narrow and too wide. It is too narrow to make us sure that we do not miss interesting types of livings. And it is too wide, so that we do not know in which direction we should start our search for other forms of life."(Schneider, 1999)*

To define search criteria, we presume that life has a carbon-based structure not too dissimilar from our own. Basing our assumptions on carbon organic compounds based life seems reasonable. Out of Earth's $10^7$ known molecules, excluding DNA, there are only about 10% that do not contain carbon (Ollivier, 1998). Carbon is a highly abundant element and exhibits a remarkable ability to form a host of complex molecules. The advantage of carbon is that it can move easily between a fully oxidized and a fully reduced state, whereas other elements like silicon tend to form large stable polymers that make it very difficult to recover the element. Thus rocks constitute a very significant sink of silicon. Carbon compounds and structures appear to be unrivalled in their potential to obtain high information content.

Arguments in favor of water as a medium for life are that it is highly abundant, an excellent solvent, and remains liquid over a broad temperature range. The temperature range is high enough for chemical reactions to occur rapidly but also to allow large molecules to form. It has an extremely high dielectric constant, $\varepsilon = 80$, that allows salt ionization and the capability of building H-bonds with dissolved molecules. Specific conformation of macromolecules can form when in solution in water due to the solvent's attraction of hydrophilic groups (OH, CO, COOH,...) and repulsion of their hydrophobic ones (CH, $CH_3$,...). These conformations allow specific chemical reactions to build reproducible complex structures (Des Marais, 2001). Other solvents cannot match the strong polar-nonpolar dichotomy that water maintains with certain organic components. Dichotomy is essential to maintain stable biomolecular and cellular structure (Des Marais, 2002). T. Owen (Owen, 1980) argues that carbon chemistry with water, as a solvent is the most likely base of life. Ammonia is often suggested as a substitute for water and is nearly as good in many categories but it lacks the capability of protecting itself from destruction by ultraviolet light.

Water and ammonia behave quite differently from a photochemical standpoint. When $H_2O$ is photodissociated to form H and OH, these byproducts usually react to reform $H_2O$. This is not necessarily true for stratospheric water, which is why photodissociation at high altitudes can lead to hydrogen escape and water loss. $NH_3$, by contrast, photolysis to form $NH_2$ and H. The $NH_2$ radicals react to form hydrazine, $N_2H_4$, and ultimately $N_2$. $N_2$ is stable against photolysis because of its strong triple bond. Thus, over time, an ammonia-rich atmosphere will be converted to $N_2$ (Des Marais, 2001).

## 6.3 Origin of life

"*Life as we know it is assertive, demanding, and unstoppable*" (Ehrenfreund, 2003)

Life as we know it is a chemical phenomenon. What kind of chemistry can produce self-organizing systems is the central question. There is a consensus that life started during the first billion years after the Earth formed (Schopf, 1993). Light carbon isotope data from putative microfossils from 3.5 billion year-old Australian deposits present the first preserved signs of life on early Earth although there is now some debate as to the biological origin of this evidence (Van Zuilen, 2002) (Brasier, 2002). The assumed 3.8 billion year-old Greenland rock inclusions (Mojzsis, 1996) that have been assumed to be the first signs of life on Earth have been backdated to be about 1 billion year old. Current understanding of the emergence of life on Earth has been recently summarized (Bada, 2002) (Ehrenfreund, 2003) (see Fig. 80).

There are several competing theories how life may have formed: in a prebiotic “soup”, in submarine hydrothermal vents, or by extraterrestrial origin of organic compounds. In spite of their diversity they have the common assumption that abiotic organic compounds were necessary for the emergence of life. Evidence suggests that a combination of exogenous and endogenous sources provided the first building blocks of life on early Earth. The simple molecules reacted and assembled on catalytic surfaces like minerals, forming more complex structures that later developed into primitive cells (Ehrenfreund, 2003). Most theories for the origin of life suggest that life began with an organic chemical system that developed the ability to propagate itself, genetically or metabolically. There are three principal theories for the source of the organic compounds (Deamer, 1994). 1) Atmospheric synthesis via Miller-Urey type synthesis from electric discharges, UV light, or other forms of high-energy radiation on reduced or neutral atmospheric gases, 2) extraterrestrial delivery via dust, comets, or carbonaceous chondrites, 3) mineral catalyzed synthesis in and around hydrothermal vents.

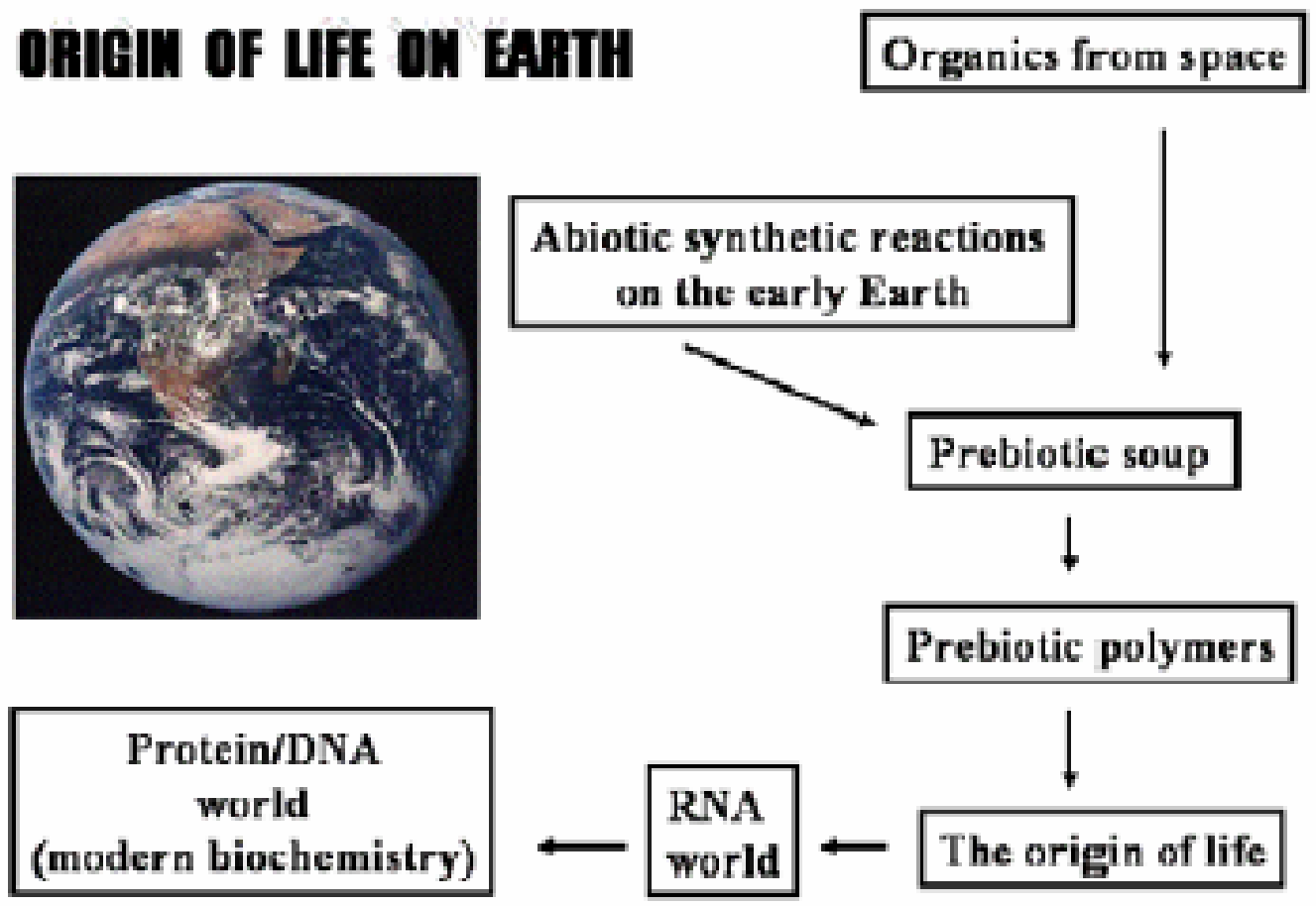

**Fig. 80: Various steps thought to be involved in the origin of life on Earth (Bada, 2002)**

There are three competing theories for the chemical origin of life (Deamer, 1994): 1) metabolism-first theories, that speculate that life began with mutually catalytic chemical reactions, 2) membrane-first theories, that speculate that the appearance of encapsulating lipid bilayers made the origin of energy harvesting membranes possible, 3) gene-first theories that suggest that the origin of life is intrinsically tied to the appearance of the ability to pass on acquired catalytic abilities. There are two different approaches to investigate the origin of life: The bottom-up approach investigates which molecules are likely to have been available on the primitive Earth, and how these can be assembled into a replicating or catalytic structure under plausible geochemical conditions. This approach is difficult due to many uncertainties. Another method is the top-down approach, which takes modern biochemistry and attempts to deduce the most fundamental property or structure of a living cell without which life could not exist.

Recognition of the central role of RNA in protein biosynthesis and genetic information transfer, as well as the discovery of catalytic RNA, gave rise to the concept of the RNA World, a postulated period in which RNA functioned as both catalyst and information carrier. Later, coded protein synthesis was developed and the catalytic duties relegated to proteins, while the more stable DNA molecule developed and was given the task as storage medium (Dworkin, 2003). Two fundamental hurdles remain in understanding the chemical origin of life: What is the nature of the first catalyst/replicas and by which method did these become organized into polymers of sufficient length to adopt three-dimensional structures capable of forming binding and catalytic sites (Ehrenfreund, 2003)? The conditions at which the origin of life occurred remain controversial. Extremophiles, organisms dubbed for their ability to live in extreme environments, have been found in practically every terrestrial niche that allows for the existence of water: environments of extreme pH (0.6 - 12.5), temperature (-2 - 115° C), pressure (to 110 MPa) and salinity (to 37.5% NaCl) (Cavicchioli, 2002). This has led some to conclude that the range of environments where life may arise is extremely wide. However extremophiles are highly evolved organisms providing evidence for the incredible adaptability of life, but may ultimately tell us little about the actual conditions required for the origin of life (Arrhenius, 1999). Laboratory experiments show that iron sulphide at elevated temperature and pressure can facilitate the natural generation of organomettalic compounds by an autocatalytic process. The products of reaction catalyze the next cycle of reaction, making the process likely to be self-sustaining (Cody, 2000) (Wachtershauser, 2000). Iron and sulfur atoms are at the center of many enzymes and the process provides a simple explanation of their presence in living cells (Traub, 2002). The experiments suggest that the origin of life on Earth may have been deep underground, away from the possibly changing conditions of the early Earth.

Kasting et al. (Kasting, 2002) present arguments why microorganisms have probably determined the basic composition of Earth's atmosphere since the origin of life. That might have generated a titan-like atmosphere during the first half of Earth's history and a breathable $O_2$ atmosphere during the second half. Oxygen is produced during photosynthesis as $CO_2 + H_2O \rightarrow CH_2O + O_2$, where $CH_2O$ denotes complex forms of organic matter. Higher plants carry out most photosynthesis on land. But that is nearly balanced by the reverse process of respiration and decay. The authors argue that by contrast marine photosynthesis is a net source of $O_2$ because a small fraction, about 0.1%, of the organic matter synthesized in the oceans is buried in sediments. That leak in the marine organic carbon cycle is responsible for most of our atmospheric $O_2$, another convincing argument that water is an essential factor in the development of life.

## 6.4 Circumstellar habitable zones

The definition of the habitable zone (HZ) is mostly based on work by Kasting et al. (Kasting, 1993) (Kasting, 1997) and relies on the greenhouse effect due to $CO_2$. Forget et al. (Forget, 1997) and Selsis et al. (Selsis, 2002) have also investigated the issue in detail. Providing that habitability requires liquid water on the planets' surfaces, for planets of the size and mass of the Earth, containing large $H_2O$ and $CO_2$ reservoirs with $CO_2$-$H_2O$-$N_2$ atmospheres, criteria of the HZ are defined by Kasting et al.. The HZ is defined as the zone around a star within which starlight is sufficiently intense to maintain liquid water at the surface of the planet, without initiating runaway greenhouse conditions that dissociate water and sustain the loss of hydrogen to space (Kasting, 1997).

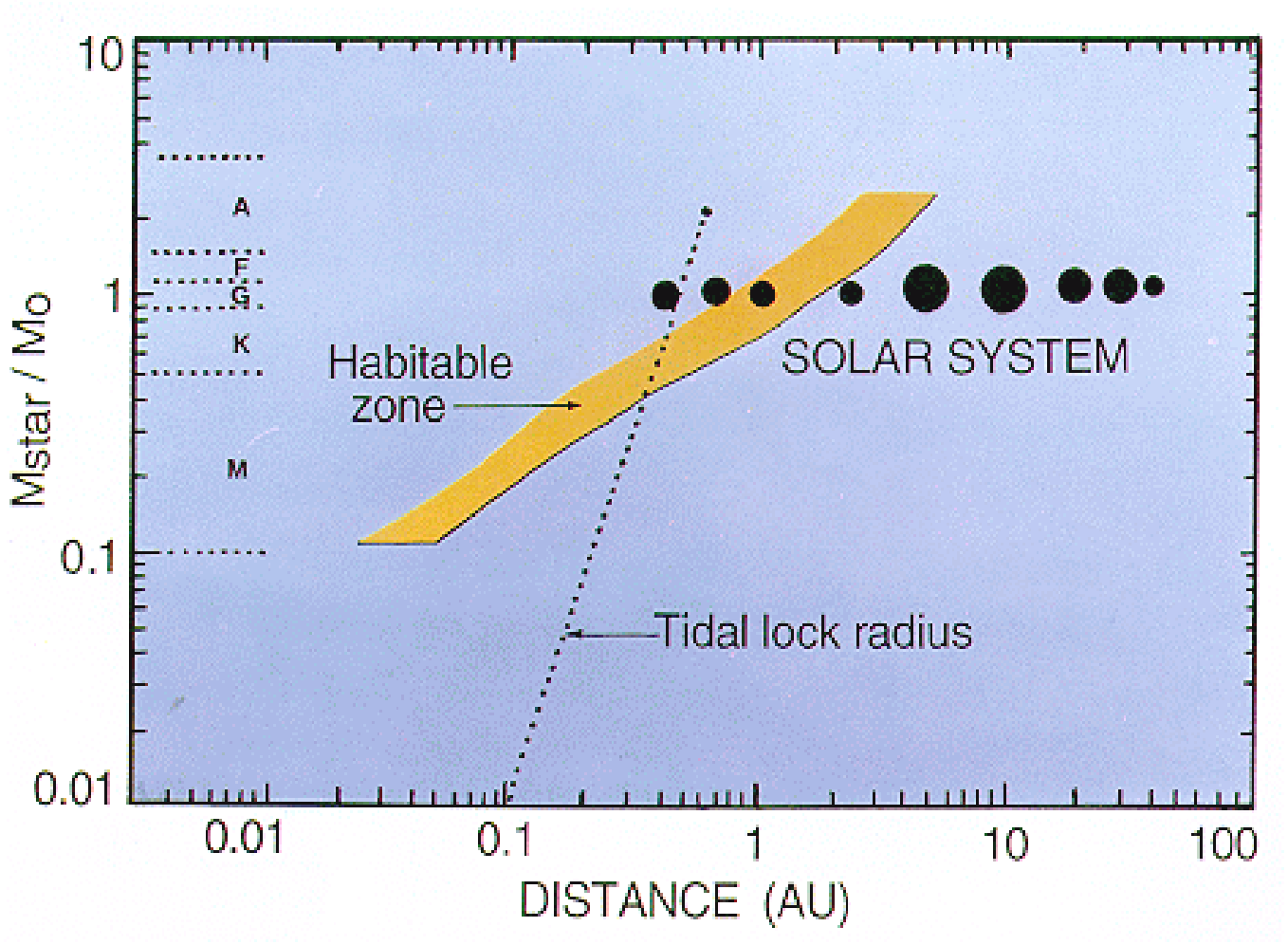

**Fig. 81: Habitable Zone of various star types as function of mass and orbital distance (Kasting, 1997).**

Liquid water would be stable on Earth's surface for temperatures up to 647 K, the critical point for water, because the atmosphere acts like a pressure cooker. If there is a level atmospheric $CO_2$ able to sustain stable liquid $H_2O$ at the surface, the distance $D$ belongs to the HZ.

Table 17 translates that distance into radial extent from the star for different distances to the host system.

**Table 17: Angular extent of the radii of stars and 1AU at different distances**

| | **5pc** | **10pc** | **15pc** | **20pc** |
|---|---|---|---|---|
| Radius sun [rad] | $4.51\ 10^{-9}$ | $2.26\ 10^{-9}$ | $1.50\ 10^{-9}$ | $1.13\ 10^{-9}$ |
| 1AU [rad] | $9.71\ 10^{-7}$ | $4.85\ 10^{-7}$ | $3.23\ 10^{-7}$ | $2.42\ 10^{-7}$ |

The inner edge of the HZ is determined by water loss via hydrogen escape and photolysis. If $D$ is below a critical distance $D_{in}$, the planet would experience a "runaway greenhouse effect": the water evaporates, changes the adiabatic lapse rate of the atmosphere and allows vapour to reach the upper atmosphere without condensing. In the upper atmosphere $H_2O$ is photolysed and the light hydrogen atoms are lost to space. Within a period of the order of 10 to 100 Myr, the planet looses the hydrogen to space and becomes dry. The runaway process was studied (Kasting, 1988) in the case of Venus with a 1-D radiative convective model. A critical distance of 0,84AU (for the present solar luminosity) was estimated by the author for a clear-sky atmosphere. With clouds that limit could be shifted inwards to about 0.5AU (Selsis, 2004). The surface temperature of the planet is assumed to be stabilized above 0ºC through the carbon-silicate cycle (Walker, 1981). The outer edge of the HZ is determined by the formation of $CO_2$ clouds, which increase a planet's albedo, leading to runaway glaciation. The external limit, again according to Kasting, is reached when it is no longer possible to increase the surface temperature by adding $CO_2$ to the atmosphere. The increase in density of the atmosphere implies also an increase of the atmospheric albedo due to Rayleigh back-scattering. Kasting estimated an external limit of 1.67AU for the present solar HZ.

Recent work (Forget, 1997) (Mischna, 2000) indicates that $CO_2$ cloud cover could conceivably warm a planet surface by Mie scattering in the infrared provided the clouds are cold, cover the entire surface and are optical thick. Both groups found that under optimal conditions surface temperatures on an Earthlike planet could be maintained above 273 K out to 2.4AU, but Mischna et al. also found less warming under a less-optimized cloud cover. The results are still debated. Recent photochemical calculations show that a dense $CO_2$ atmosphere induces the formation of a warm ozone layer, which may prevent the formation of $CO_2$ clouds and therefore extent the HZ considerably. This effect has to be addressed, as the HZ could be wider than found in the Kasting model. The radiative properties of a high-pressured $CO_2$ gas are not well established. A further influence to consider is the existence of a giant planet that renders planet formation outside of a certain radius inefficient. For a sun-like star the distance between 0.6AU and 1.7AU should cover most of the habitable zone (Selsis, 2004).

In these limits climate stability is provided by a feedback mechanism in which atmospheric $CO_2$ concentrations vary inversely with planetary surface temperature. Concerning $CO_2$, the estimation of the external frontier of the HZ requires taking the collision induced absorption (self broadening) into account that produces a continuum of absorption on the whole infrared spectrum at high pressure and therefore induces a very efficient greenhouse warming. This collision induced absorption continuum is only known for wavelength above about 40µm were it can be calculated theoretically (Gruszka, 1997). This effect is usually included in models as an extrapolation of the few available measurements. It is therefore extremely important to improve our knowledge of this effect, experimentally and/or theoretically. The continuum may cover the features of the other compounds like $H_2O$, $O_3$, $CH_4$ or even $CO_2$. Also collision with other molecules has an influence on the width of the spectral lines.

With this definition, the HZ only depends on the spectral energy distribution of the star. Obtaining accurate stellar data is important because it is luminosity along with star-planet separation that most strongly influences whether a terrestrial world might harbor water-dependent life. The HZ moves outwards with time around all main sequence stars as they increases in luminosity with age. The region that provides habitability around stars during their lifetime on the main sequence is called the Continuously Habitable Zone (CHZ). It depends on the time that the planet is required to remain habitable and is narrower than the HZ. For our own Sun the corresponding boundaries on the 4.6-Gyr-CHZ are at least 0.95AU < d < 1.15AU (Kasting, 1997). Stars more massive than about F0 have narrow, short-lived CHZ. That excludes planets orbiting massive stars with a main sequence lifetimes less than 2Gyr from the target list in the search for life. The difference for a HZ for planets with different size and mass still has to be investigated using atmosphere models. The definition of the habitable zone and the estimation of its limits give an extremely useful guideline but many issues still need to be investigated: A complete definition of the HZ should tell us that outside of its limits, habitable planets cannot be found, but it highly debatable if we have sufficient information to achieve this even for Earth-like planets. The inner limit could be smaller due to a dense atmosphere, low greenhouse effect and strong back scattering to space. $CO_2$ clouds may play an important role concerning the outer limit (Forget, 1997) as the main process contributing to the warming at high pressure is the poorly characterized $CO_2$-$CO_2$ collision induced absorption (Lammer, 2003B). Also other greenhouse gases like $CH_4$, $N_2O$, $NO_X$, $NH_3$, $SO_2$ may expand the HZ by warming the surface. Most of the stars belong to a binary or multiple system, which could affect the long-term stability of the orbits within the HZ.

We define a habitable planet as a planet with an atmosphere and liquid water on its surface. One has to keep in mind that there are open questions like how we could determine whether signs of habitability are present over geological time or limited time periods, except by assumption? The range of characteristics of planets is likely to exceed our experience with the planets and its satellites in our own solar system. Earth like planets orbiting stars of different spectral type might evolve differently (Lammer, 2003C). Some planets might also not be coplanar and thus their orbit might slowly migrate in and out of the habitable zone. We have absolutely no experience with such bodies and atmosphere models are our only guides to unfamiliar and fascinating cases.

### 6.4.1 HZ around different stellar types

The HZ around different stellar types will vary as it only depends on the spectral energy distribution of the star. The HZ around F stars is larger and occurs farther out, the HZ for K and M stars is smaller and situated farther in than around our Sun. Its width is slightly greater for planets that are larger than Earth or have higher $N_2$ partial pressure.

High ultraviolet fluxes could be a problem for life around F stars. For stars later than about K5 spectral class the HZ falls within the 4.6-Gyr tidal locking radius of the star. The planetary companions may rotate synchronous with their orbits as the Moon does around the Earth. On the cold dark side of habitable planets the atmosphere may therefore condense (Kasting, 1997). The resulting temperature difference could generate winds that efficiently reduce the temperature gradient on the planet if the atmosphere were thick enough. Thick CO2 atmospheres will very likely cover bio-signatures like $O_3$, what makes M stars a secondary target. The stability of the atmospheres has to be tested by observations.

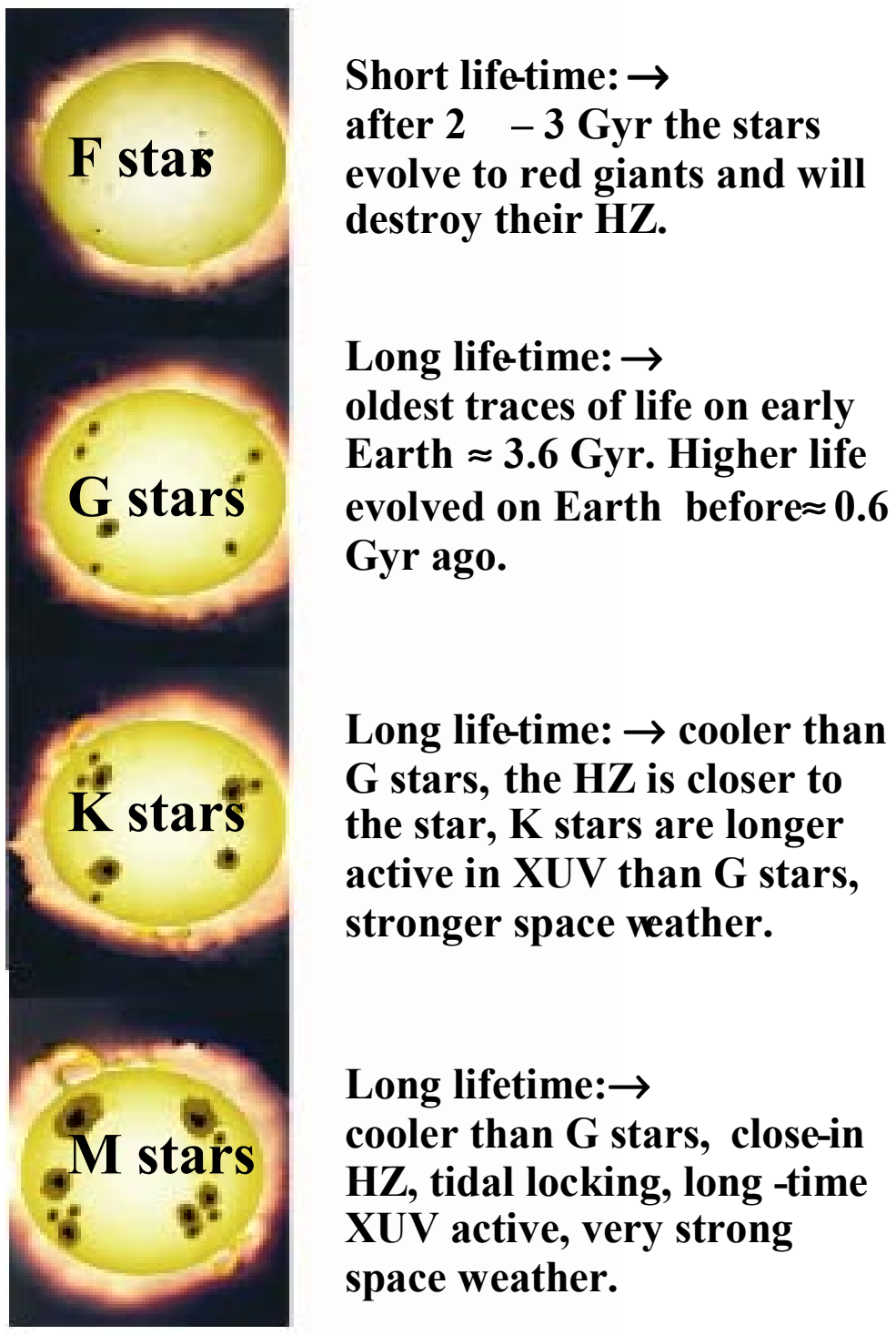

**Fig. 82: HZ around different stellar types (Lammer, 2003B)**

Model calculations show that $CO_2$-rich atmospheres of big terrestrial planets may survive strong XUV warming of the upper atmosphere. What kind of stable atmosphere of a terrestrial extrasolar planet is possible? Models already exist, although they are usually developed for a specific case like a given planet. Applying such a model to another case than the one considered, has proved to be misleading (Lammer, 03D). One can use the well-known model MODTRAN to compute the spectrum of an Earth with 300 mbar of $CO_2$ instead of $O_3$ and the model will give a result. However, in this specific case, the computed spectrum will not be consistent because the parameterization for $CO_2$-$CO_2$ absorption in MODTRAN cannot be used for high $CO_2$ pressures. Many examples like this can be found. Therefore, developing or adapting a model applicable to broad range of cases is a difficult but important step.

Observations of a planet will determine whether it is inside a predefined habitable zone around its parent star, but that only provides a rough estimate of the temperature. Two temperatures are of interest to determine the habitability of a planet, the effective temperature based on a blackbody having the same surface area and the same total radiated thermal power, and surface temperature at the interface between any atmosphere and a possible solid surface. In case a greenhouse effect is present the surface temperature will be warmer than the effective temperature.

### 6.4.2 Greenhouse gasses

Greenhouse gasses let stellar Visible-NIR radiation penetrate to the surface of the planet, but infrared light cannot radiate from the planet. $CO_2$ (red), $H_2O$ (blue) and $SO_2$ (yellow) absorb the light in special wavelengths. Without these gasses, Venus surface would have a constant temperature of about -20° Celsius. The equilibrium temperature of the Earth with its present albedo, solar flux and no green house gases would be about 246K, below the freezing point of water. The greenhouse effect is driven at temperatures below the freezing point of water by $CO_2$ and aided at higher temperatures by water vapor. The combination of present levels of $CO_2$ and $H_2O$ is sufficient to warm the surface to 290K, about the current average surface temperature (Traub, 2003). The greenhouse effect can be defined as the warming due to the energy radiated by the atmosphere to the surface.

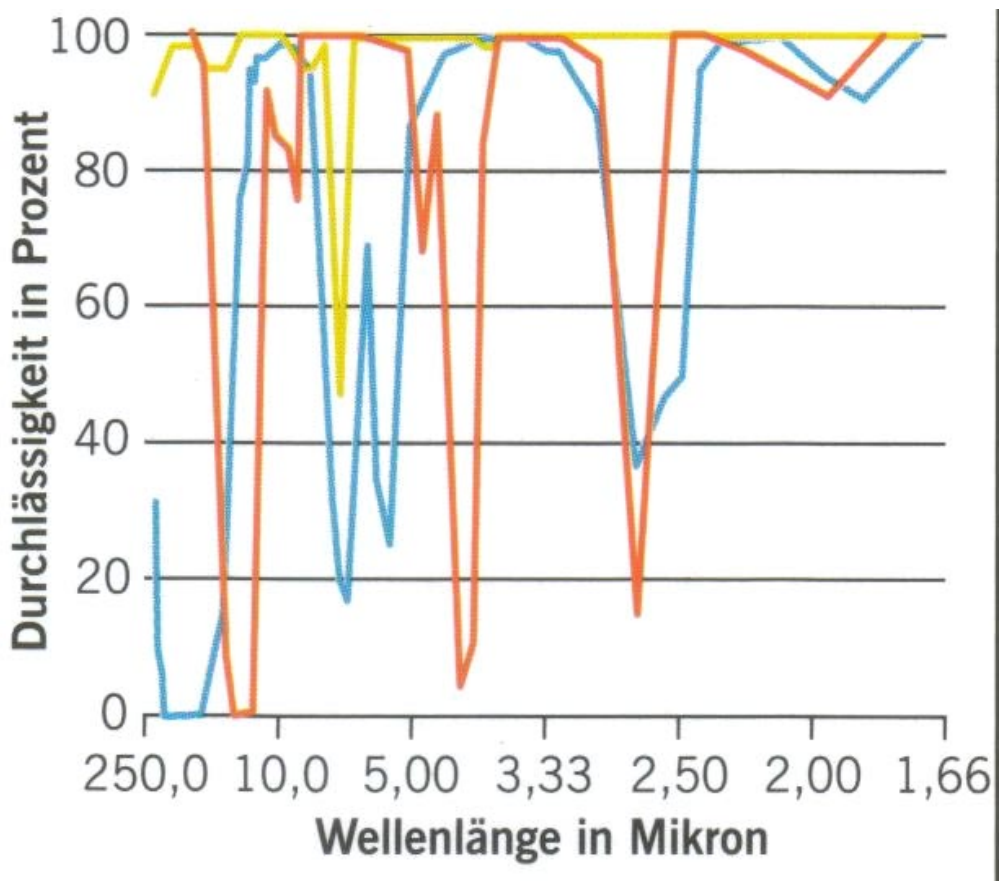

**Fig. 83: Greenhouse gas absorption in the atmosphere over wavelength in µm (Lammer, 2003)**

### 6.4.3 Neoproterozoic Earth

A weaker sun, about 0.71 times the present luminosity, leading to a lower surface temperature, illuminated the early Earth. Almost certainly the $CO_2$ abundance also dropped significantly for major glaciations to occur. Traub et al. (Traub, 2002) simulated the infrared spectra of a neoprotetozoic hothouse and icehouse Earth. The icehouse state uses the present arctic temperature profile, 1% PAL $O_2$ (present atmospheric level), modified abundances of $O_3$, $H_2O$, $N_2O$ and $CH_4$ for 1%$O_2$ (Kasting, 1980), 100ppm $CO_2$ and 1bar $N_2$. To simulate the hothouse state, present tropical temperature with 10% $O_2$ and corresponding abundance of $O_3$, $H_2O$, $N_2O$ and $CH_4$ is used along with 120.000ppm $CO_2$. The 15µm $CO_2$ feature can be seen in emission in the hothouse spectrum while an absorption feature in the icehouse spectrum (Fig. 84). The emission is due to a warm inversion layer in the stratosphere due to $O_3$ heating combined with $CO_2$ mixing ratio.

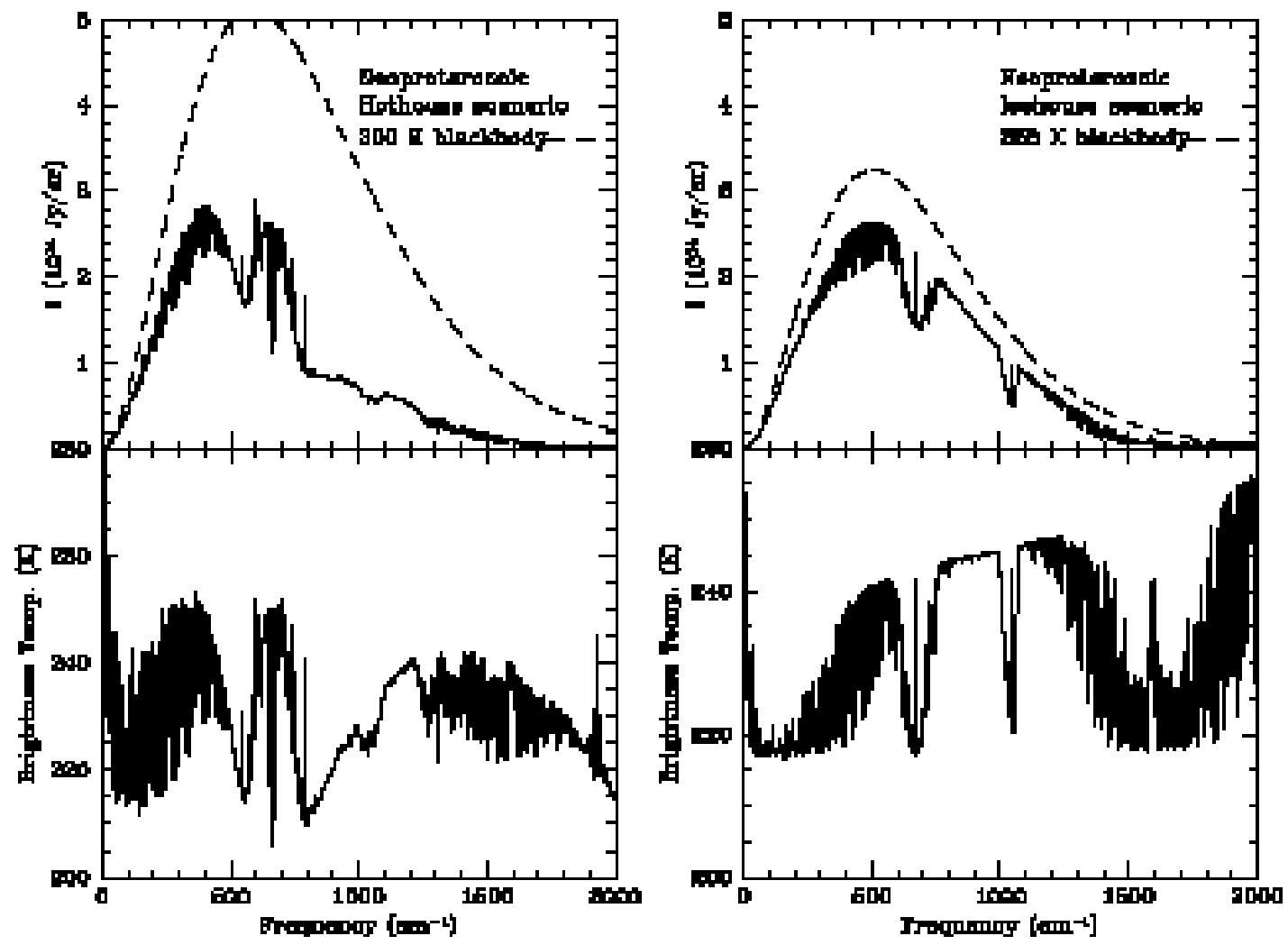

**Fig. 84: Calculated thermal IR spectra for a neoproterozoic hothouse condition (left) and icehouse (right) (Traub, 2002)**

# 7 Features of a habitable planet

## *7.1 Characterization of a planet*

Lovelock (Lovelock, 1975) concluded that the simultaneous presence of large amounts of reducing and oxidizing gases in an atmosphere out of thermodynamic equilibrium is a criterion for the presence of biological activity. There is no definite physical basis for this general criterion. All planetary atmospheres are out of thermodynamic equilibrium because their photochemistry is driven by UV photons from their parent star. However on our planet the only processes that actually produce large quantities of gas out of equilibrium in the atmosphere are biological ones.

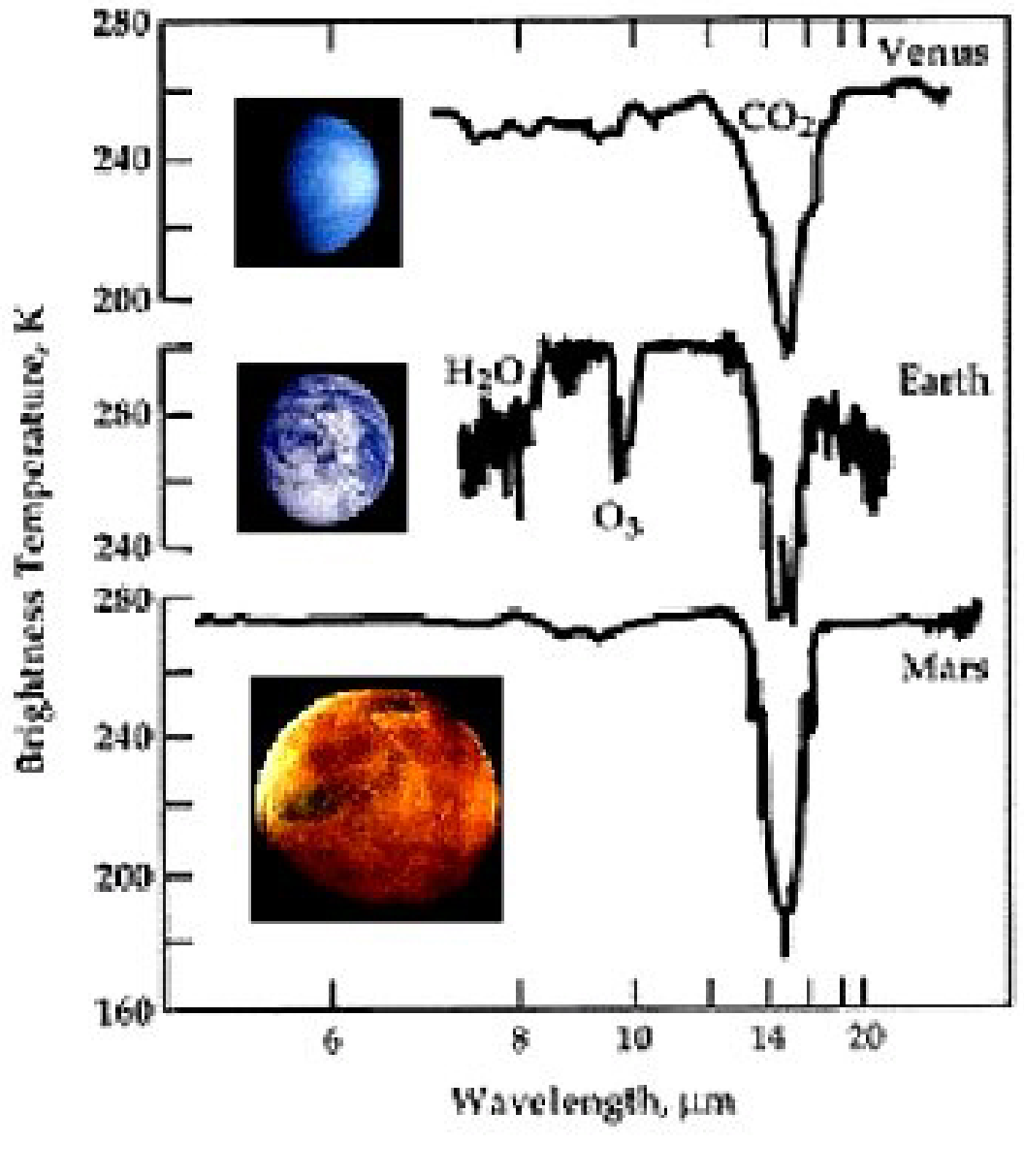

**Fig. 85: Thermal infrared spectra of Venus, Earth and Mars. The 15µm $CO_2$ band is seen in all three planets. Earth also shows evidence of $O_3$ and $H_2O$ (Fridlund, 2000)**

Indirect methods of detection like radial velocity, astrometry, and observation of planetary transits allow getting morphological information: period of revolution, semi major axis, eccentricity of the orbit, planet diameter and mass. The planet is only deduced from its effect on the parent star under survey. DARWIN will collect photons from the planet allowing a spectral analysis opening the unique capacity to investigate the properties of the planet itself.

## 7.2 Spectra

The spectral characteristics of a planet are very important to identify the wavelength region to search for a planet as well as characterize the spectra once it has been detected. The radiation of a planet consists of two parts, reflected starlight that retains the spectral shape of the star's radiation and thermal emission because absorbed starlight heats the planet's surface. Using black body radiation laws, the lowest ratio of the light received from a star to that detected from a planet is found in the IR, where the thermal emission of the planet is at its maximum. The direct detection of a planet close to its parent star is challenging because the signal detected from the parent star is between $10^9$ and $10^6$ times brighter than the signal of a planet in the visual and IR respectively. Future space based missions like DARWIN and TPF concentrate on the region between 6μm to 18μm, a region that contain the $CO_2$, $H_2O$, $O_3$ spectral features of the atmosphere. The presence or absence of these spectral features would indicate similarities or differences with the atmosphere of telluric planets. In order to interpret spectroscopic data in the context of possible biological activity we need to know the planet's size and temperature. Estimates of planet size and albedo, which will be crucial to determine the characteristics of a planet, can be determined directly from mid-IR observations, but involve atmospheric modeling in the visible or near IR range because of the possible albedo range (Des Marais, 2001).

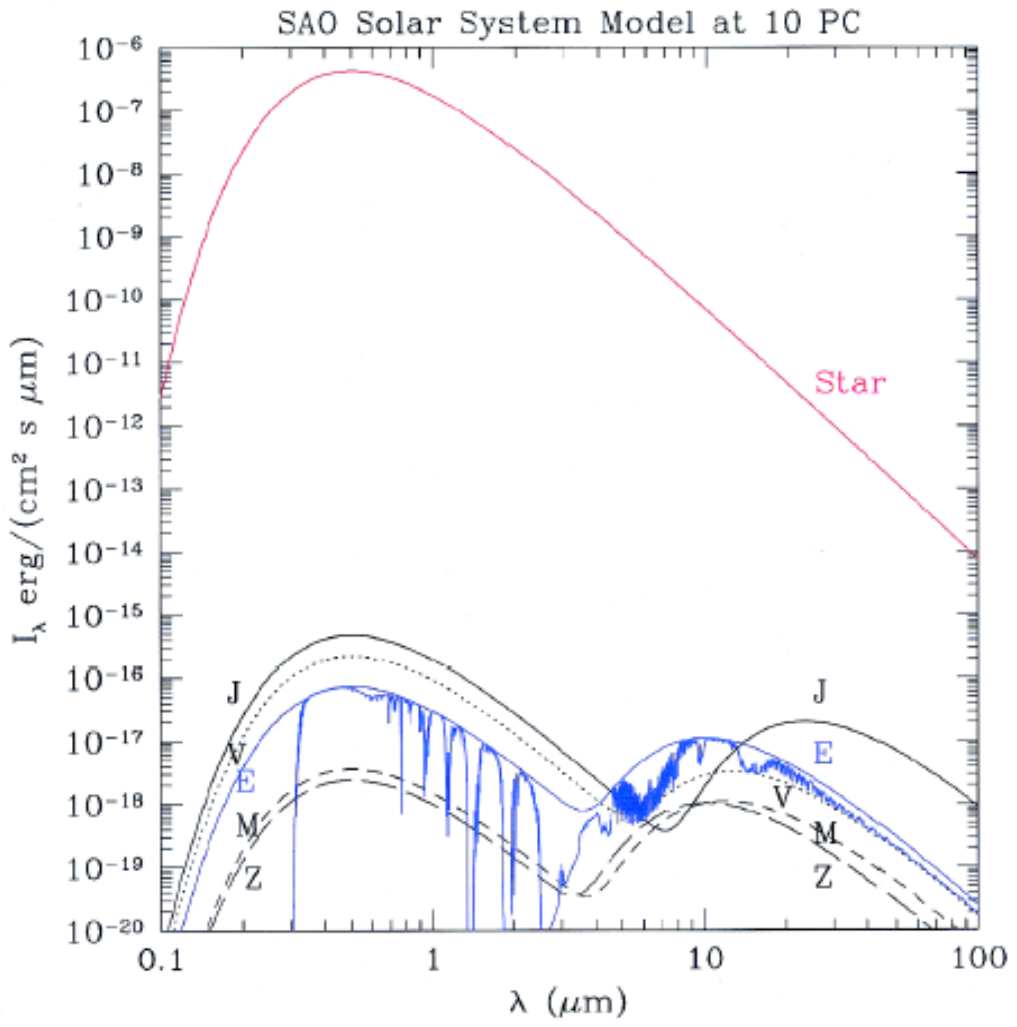

**Fig. 86: Model spectrum of the sun and planets as seen from a distance comparable to that of a nearby star (10 pc), shown in physical units. Simple Planck emission and wavelength-independent albedo reflectance components are shown. For Earth, a pure molecular absorption spectrum is superimposed for reference (Traub, 2002).**

Simple Planck emission and wavelength independent albedo reflection components are shown (see Fig. 86), a pure molecular absorption spectra is superimposed for reference. Fig. 87 shows model spectra in the Visible and IR respectively. The advantage of operating in the thermal IR instead of the visible is the $10^3$ times larger ratio of the planet/star emission. Planets emit IR signatures in the form of heat at locations where the heat from the parent star is falling off, while in the visible light the planets glow with dimly reflected starlight in contrast to the bright visible light from the star itself.

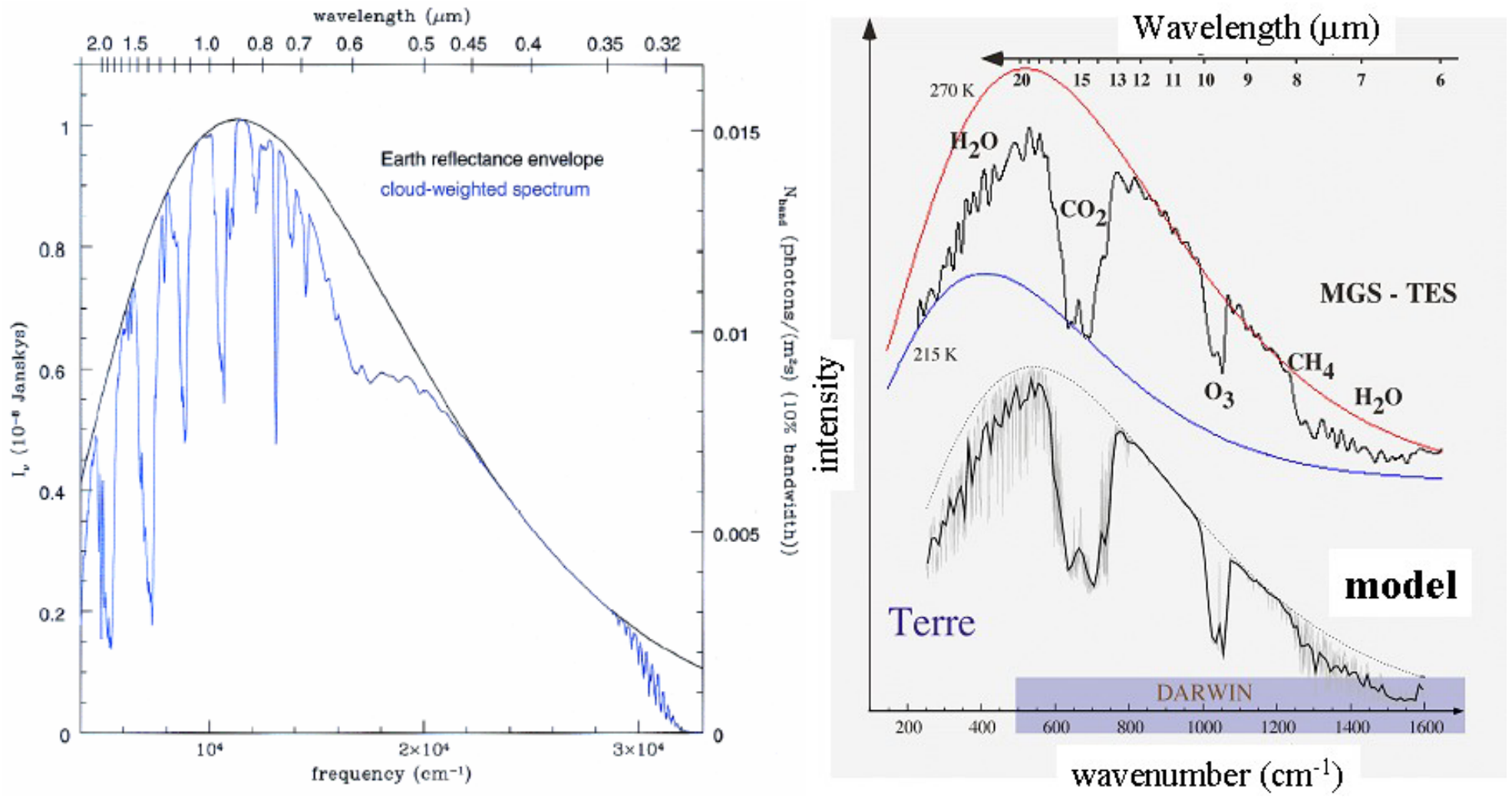

**Fig. 87: (left) Visible and near-IR portion of the Earth model spectrum, showing a reflection spectrum of the Earth's surface and clouds in the absence of an atmosphere (upper curve) and the net spectrum of a realistic model atmosphere mixed in with a model distribution of opaque clouds distributed over several altitudes (lower curve). (right) Earth spectra in the IR (Selsis, 2002).**

### 7.2.1 Influence of cloud coverage

Clouds are a very important component of extrasolar spectra as their high reflection is relatively flat with wavelength (Traub, 2003). Clouds hide the atmospheric molecular species below them, essentially weakening the spectral lines in both the IR and the visible.

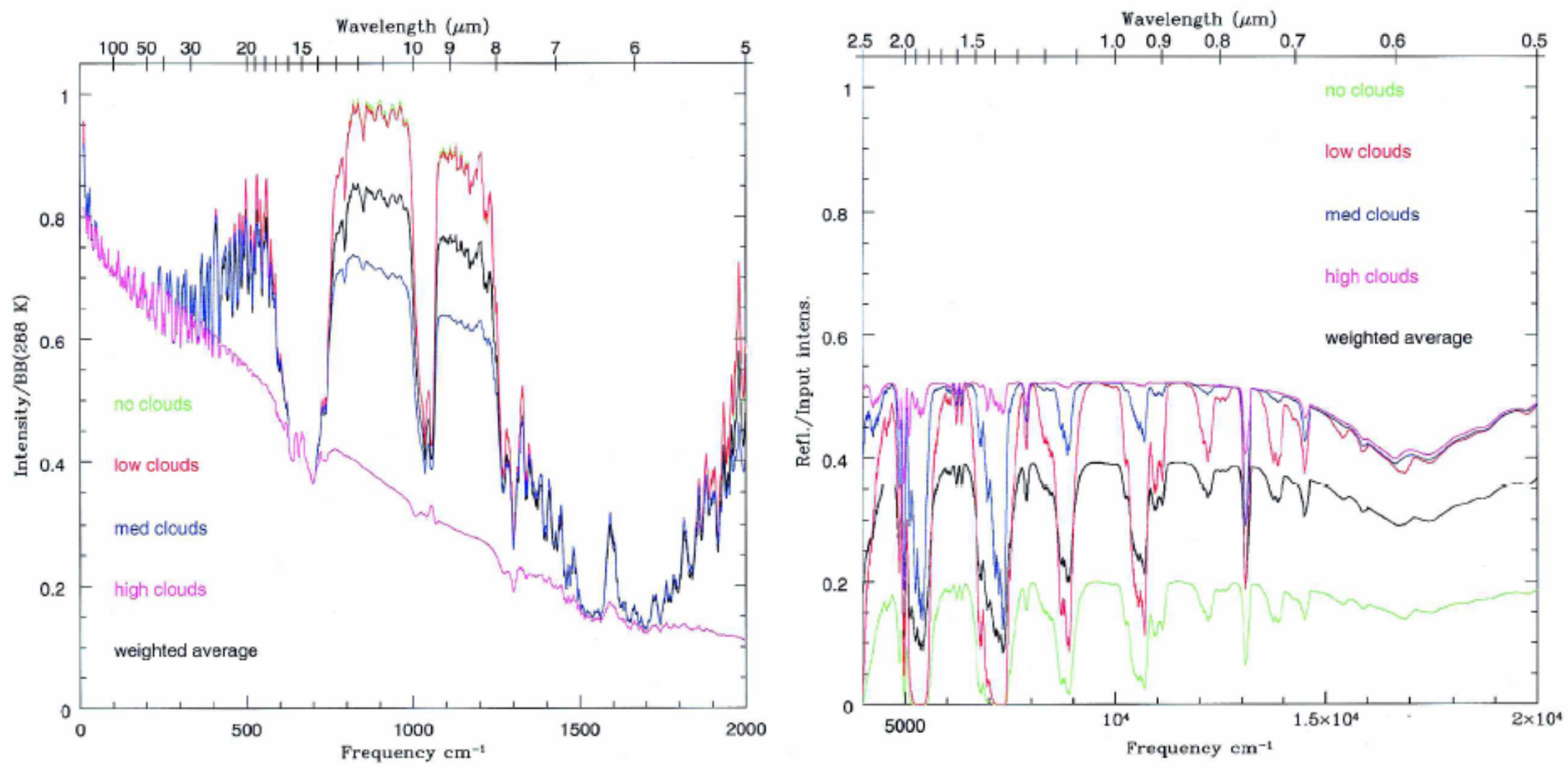

**Fig. 88: Normalized IR thermal emission and visible reflection spectral models of the Earth for five cloud conditions. Each spectrum is calculated for the present Earth's altitude-dependent, midlatitude temperature structure and gas species mixing ratio profiles to simulate a whole Earth integration. Each spectrum is normalized to a blackbody (BB) spectrum at 288K (Des Marais, 2002).**

Fig. 88 shows the normalized Visible Reflection Spectral Models of the Earth and normalized Infrared Thermal Spectral Models of the Earth shown for 5 cloud conditions (Des Marais, 2002). The cloud altitudes and covering fraction in the weighted average spectra in Fig.83 are: low cloud, 1 km, 20%; medium cloud, 6 km, 30%; and high cloud, 12 km, 10%. In the IR model the top curve represents a cloud-free atmosphere, the other curves represent complete overcast conditions with clouds at low, medium, and high altitudes, plus a curve for a nominally realistic mixture of these four cases. In the IR spectrum, the main effect of adding clouds is to decrease the emitted continuum flux and reduce the relative depths of spectral features. In the reflected visible spectra the clear atmosphere case represented by the lower curve has the smallest continuum level because the surface reflectivity is low compared with clouds, and thus the overcast cloud cases have high continuum fluxes. The weighted average is a linear combination of the extreme cases and was calculated for an overall albedo of 39%. In the reflected spectrum, the main effect of adding clouds is to increase the reflected continuum flux and reduce the relative depths of absorption features.

## 7.3 Visible & IR

The observation wavelength range for the search for extrasolar planet search is based on a trade off between the visible and the infrared spectral range. The decision has not been made for the American Terrestrial Planet Finder (TPF) mission studied by NASA that has the same goals as DARWIN and thus might lead to an US/European collaboration. Here we review some of the most interesting capabilities of the thermal infrared and visual spectral range particularly in terms of sensitivity of the detection, gaseous spectral features and environmental data. Both visible and IR measurements can give the projected orbit.

The TPF Science Working Group (TPF-SWG) (Des Marais, 2001) identified the waveband between 8.5µm to 20µm and preferably 7µm to 25µm, respectively, for the search for biomarkers in the mid-IR region and 0.7µm to 1.0µm and preferably 0.5µm to 1.1µm for the visible to near IR region. The DARWIN project concentrates on the waveband between 6µm to 18µm. In the thermal part of the spectrum, the shape gives a measure of the temperature of the object examined. Observations from 8µm to 12µm of the $H_2O$ continuum allow estimations of the surface temperature of Earth-like planets. However the atmosphere of planets warmer than about 310K will be opaque in this region because of continuum absorption by water vapor. Visible to near-IR spectra offer higher spatial resolution and are minimally affected by temperature. They could be used to determine the abundance of atmospheric species if an independent determination of the pressure and high resolution to access the shape of the line is available depending on the atmospheric species (Selsis, 2003B). In the visible using low resolution, the orbital distance is the only information related to the temperature of a planet. Modeling the albedo of the detected planet in the visible will be an important step to determine the temperature.

**Table 18: Main spectral features of atmospheric compounds in the thermal infrared and corresponding resolution. Features positions are rounded (Ollivier, 2003) (Selsis, 2003B).**

| Compound | Features (µm) | Resolution |
|---|---|---|
| $NH_3$ | 11, 10, 8.6, 5.8 | 10, 20, 10, 25 |
| $SO_2$ | 19, 8.6, 7.7 | 3,20,25 |
| $CH_4$ | 8, 7.6 | 6, 13 |
| $N_2O$ | 16.5, 8.6, 8 | 3, 20, 20 |
| NO | 5.4 | 50 |
| $NO_2$ | 6.3 | 30 |
| CO | 4.5 | 100 |
| $0_3$ | 9.6 | 17 |
| $H_20$ | 19, 6 | 3, 3 |
| $CO_2$ | 15 | 4 |
| $CO_2$ (50 mbar) | 15, 10.4, 9.3 | 4, 16, 19 |
| $CO_2$ (1 bar) | 15, 10.4, 9.3, 8, 7.3 | 4, 16, 19, 20, 20 |

For a S/N ratio of 20, ozone, $CO_2$ and water can be clearly identified. Other compounds are below the detection threshold. The disadvantage of operating in the thermal IR is that cooling of the telescope is required on arrays of smaller apertures of 1.5 – 3.5m configured as an interferometer. The discussed strategy is a free flying array of 3 to 7 spacecrafts. The baseline design for DARWIN is an array of 6 telescopes with beam combination on a central beam-combiner and an out-of the plane reference spacecraft. In the visible a single 8-16m diameter coronograph needs to be used for planet characterization depending on how many stars should be observed (30 – 150 stars), providing an equal technical challenge (Fridlund, 2000).

The much longer wavelength of the IR compared to the visible allows for greater tolerance. The broad absorption band would allow for a lower spectral resolution. Carbon dioxide can only be found at low spectral resolution in the infrared spectrum (Owen, 1980). Due to instrumental limitations like low spectral resolution for space missions - the gases that could be identified in the IR are $H_2O$, $CO_2$ and $O_3$. IR spectral band profiles can be strongly influenced by details of the thermal structure of the atmosphere. The spectral band can be used to show the presence of atmospheric constituents and constrain the details of the thermal structure of the atmosphere. Table 18 (Ollivier, 2003) shows that a modest spectral resolution of 20-25 in the IR allows identifying many atmospheric compounds. This is not always the case if these compounds were looked for in the visible spectral range.

### 7.3.1 S/N ratio of a direct detection of an Earthlike planet

The angular distance between the planet and the star is typically 0.1 arcsec for an Earthlike planet around a solar type star at 10 parsecs. The contrast between the planet and its parent star is about $7\ 10^6$ at 10µm and $5\ 10^9$ in the visible spectral range. In order to minimize the integration time or increase the S/N ratio the detection is assumed to be broadband. Even with this assumption, the amount of photons is still not much, about 10 photons per second per square meter of collecting area in the 6-20µm spectral range. As a comparison, trying to detect a planet in the visible and NIR is still more difficult. The amount of photons is about 0.3 photon per second and square meter of collecting area in the 0.5–2µm spectral range (30 times less than in the thermal IR) (Ollivier, 2003).

### 7.3.2 The effective and surface temperatures

Two temperatures are of interest to determine the habitability of a planet, the effective temperature based on a blackbody having the same surface area and the same total radiated thermal power, and the surface temperature at the interface between any atmosphere and any solid surface. In case a greenhouse effect is present the surface temperature will be warmer than the effective temperature.

On Venus, Earth and Mars the two temperatures are very different. While the effective temperature of Venus and Earth are nearly identical, 220K and 255K respectively, the planets surface temperatures are 730K and about 288K respectively, due to the different greenhouse gas abundance (Goode, 2001) (Kieffer, 1977) (Rosenqvist, 1995) (Tomasko, 1980). The effective and surface temperature is only the same on Mercury, due to the absence of an atmosphere. The close orbit around the Sun and its weak intrinsic magnetic field are responsible that Mercury could not hold an atmosphere. There are indications from observations of solar proxies that the early solar wind was more than 1000 times denser disrupting the development of stable atmospheres so close to the Sun. Mercury's average dayside surface temperature is about 550K and cools down on the nightside to about 90K. The quantity of energy received per unit area is based on the luminosity of its star and the semi-major axis of the planetary orbit. Assuming the thermal equilibrium between incoming and outgoing flux, the effective temperature $T_{eff}$ can be deduced from incoming and outgoing radiation:

$$[R_*^2\ \sigma\ T_*^4\ (1-A)\ \pi\ R_p^2]\ /\ a_p^2 = m\ \pi\ R_p^2\ \sigma\ T_{peff}^4, \qquad (7.3.1)$$

$$T_{peff} = T_*\left(\frac{1-A}{m}\right)^{1/4}\left(\frac{R_*}{a_p}\right)^{1/2} \qquad (7.3.2)$$

where, A the albedo, σ the Stefan Boltzmann constant, $a_p$ the distance of the planet from the host star, $R_p$ the planetary radius, *m* a factor related to planetary rotation (m=4 for a fast rotating planet and 2 for a slow rotating planet). The planet's effective temperature $T_{peff}$ depends on the temperature $T_*$ and thus brightness of the star, the planet's albedo A and its distance to the star. The unknown value of the planet albedo prevents determining $T_{peff}$. In the IR the effective temperature is apparent from the spectra while modeling is needed in the visible to determine the temperature and radius of the detected planet. *A* could in principle be determined using a spectra of the planet in a large spectral range including UV and visible spectral range.

In the case of Earth, the total planetary spectrum is a mixture of surface emission and cloud emission at lower temperature. The temperature given by the envelope of the spectrum is thus slightly lower than the average surface temperature. This temperature also depends on the observation geometry, as well as seasons. Another influence on the effective temperature is the day-night, D-N, temperature difference. In our planetary system Venus has an albedo of 0.80 ± 0.02 at 0.72AU (Tomasko, 1980), Earth an albedo of 0.297 ± 0.005 (Goode, 2001) at 1AU and Mars an albedo of 0.214 ± 0.063 (Kieffer, 1977) at 1.52AU. On Venus and Earth D-N is small while on Mars D-N is large. That translates into uncertainty of the temperature of the detected planet and thus the habitable conditions on the surface. The NASA biomarker reports concludes, using results of atmospheric models, that if all $R_p$, A and D-N are unknown the predicted effective temperature spans a range of 202K (Des Marais, 2001). If either one $R_p$, A or D-N is known, the range drops to 123K, 159K and 169K respectively, if $R_p$ and A are known the range drops to 70K, while known values of $R_p$ and D-N give a range of 75K. The actual uncertainties will always be larger when the error bars in the measurements are taken into account. Thus the modeling of albedo required for visual wavelength observations will strongly influence detection results. Observations with DARWIN will give $R_p$ from physical observations. *A* can be determined by measuring IR flux and, D-N by measuring the flux at two or more points in the orbit (day and night side). Results by Selsis show that the variations of the IR flux over an orbital period is very different for planets with or without an atmosphere and could be used to discriminate these two scenarios.

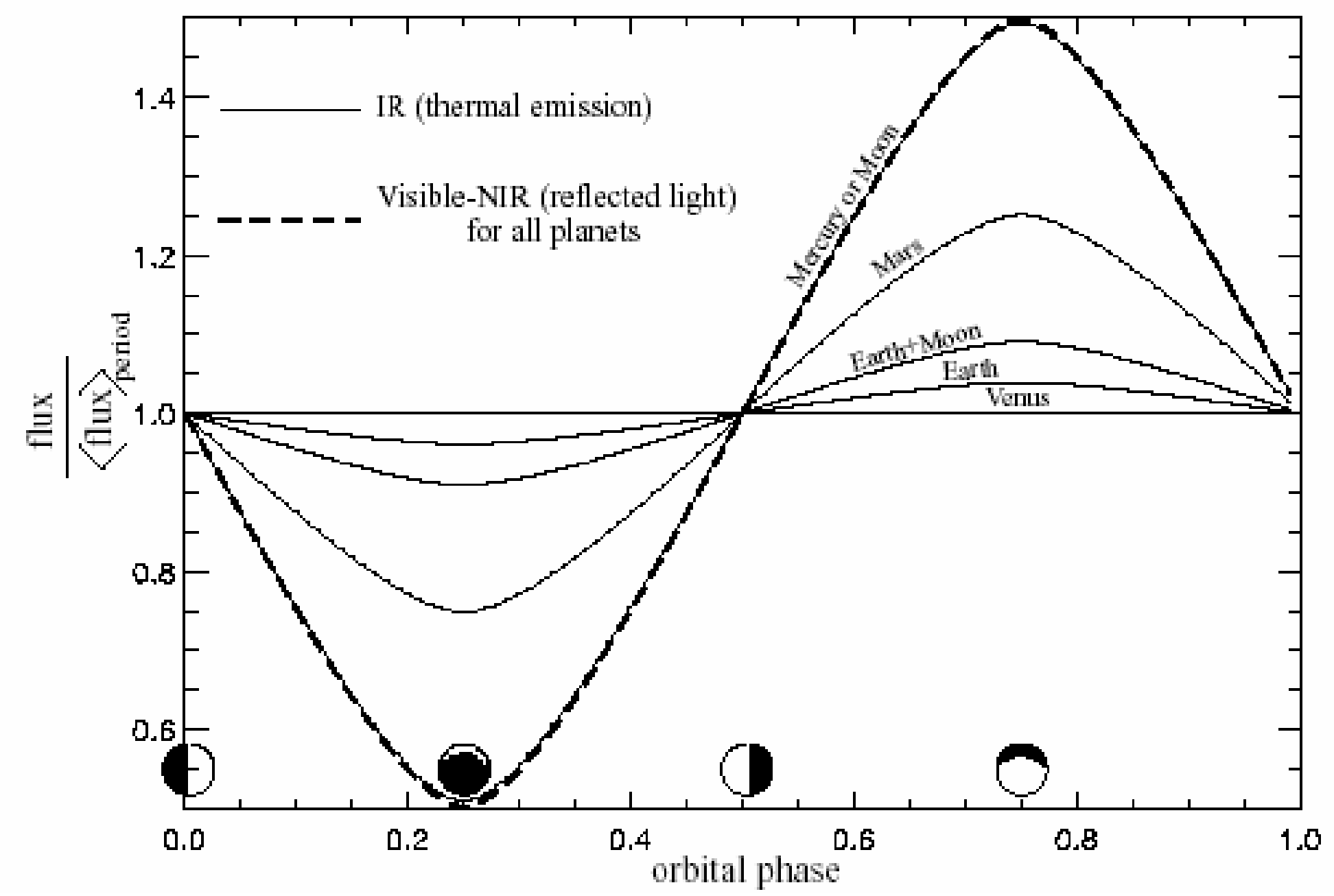

**Fig. 89: Variation of IR and Visible-NIR flux over an orbital period of different planets (Selsis 2004)**

One thing to stress is that in the past, Earth has been trough cold phases in which it had high albedo due to partial ice coverage, and low surface temperature. If the albedo is high we will know that we either determine cloud-top characteristics or we are seeing an ice-bound planet and the decision which one it is will be very difficult. Observations should be able to discriminate between the different cases, but will be unable to penetrate clouds, therefore surface conditions may be difficult to derive. Planets with small fractions of habitable surfaces or habitable surfaces that are hidden by deep, totally opaque material cannot be detected. For small and intermediate cloud coverage, we can determine the surface characteristics fairly well, while for extensive cloud cover we can only determine the characteristics of the cloud layer and above. Attributes of the planet can also be discerned from its spectra.

## 7.4 Determining planet characteristics

We can measure the quantity of energy coming from the planet in the spectral range of the instrument and derive the planetary characteristics. The following sections show the implications for the IR and visual waveband.

### 7.4.1 Planet characteristics in the IR

It is straightforward to estimate planetary radius from the mid-IR flux. IR observations of the continuum measure a color temperature from spectral shape. By equating color temperature for selected regions with physical temperatures and using the observed flux, we can use Planck's law to obtain the surface area and hence radius of the planet. The presence of an opaque atmosphere does not prevent that. The diameter is deduced from the black body fit. Using size and temperature, we can calculate the total surface emission and derive the unknown albedo of the emitting layer (Des Marais, 2001). The black body temperature can be deduced from the wavelength at the maximum of emission and knowing the surface emission from the planet the radius of the planet and the unknown albedo can be inferred. However a large atmospheric band can absorb at this maximum and therefore make the spectrum uncertain (see Fig. 90). For present Earth one can determine the surface temperature by observing the 8-12μm region. There is some absorption by the wings of water vapor bands in the window region.

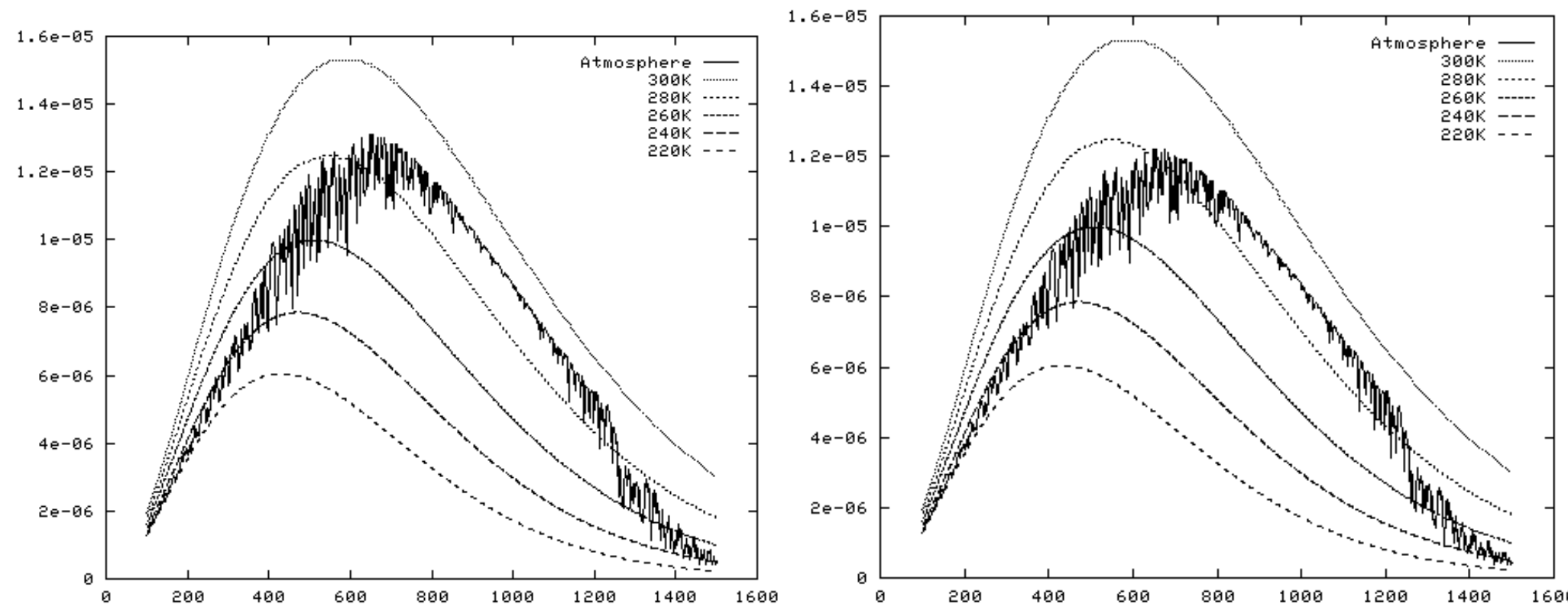

**Fig. 90: These two spectra, showing the absorption of water vapor, are obtained with the thermal and pressure profile of the Earth (mid-latitude summer, $T_{surf}$=290 K), but the signature of $CO_2$ and $O_3$ have been removed. The spectrum on the right has been computed with a double amount of water in the atmosphere. One can see that the signature would not be easily identified at low resolution. Moreover, the absorption of $H_2O$ produces a "blackbody-like" curve with a maximum of emission displaced to higher frequencies. If the water vapor were not identified, this would produce an overestimation of the temperature and an underestimation of the planetary radius (Selsis, 2003).**

The spectrum of Venus reveals strong $CO_2$ absorption features that induces a big different between the temperature obtained by a black body fit and the real surface temperature but the presence of strong atmospheric effect can be detected from observations (Ollivier, 2003). This problem can be solved if at least two atmospheric windows are identified and used where no absorption seems to occur. The same black body emission law gives the intensity of these bands depending on the same unknown parameters: the temperature and the planet radius. The integrated intensity in each window depends on the radius of the planet and the temperature of the emitting level. The emitting level is assumed to be the same for all atmospheric windows (Selsis, 2002). Other planets might exhibit different atmospheres, thus almost all the window can, in theory, be affected by absorption. It is thus difficult to predict "universal atmospheric windows". To interfere the likely mass we can use the observed Solar System relations between mass, radius and thermal environment. Although the presence of rings and satellites might distort the calculations due to additional flux, a planet's radius can be determined to an accuracy of about 10% (Des Marais, 2002).

### 7.4.2 Planets characteristics in the visible

The total reflected light by the planet of our solar system is negligible, however EGPs are found much closer to their host star and thus could give a stronger signal. The detection or non-detection of such a signal can be used to validate EGP evolution models as an upper limit of the reflected flux can be derived from observations of EGP (Charbonneau, 1999). The fraction of the luminosity of the planet to the stellar luminosity is given by (Charbonneau, 1998):

$$f_\lambda(\phi,i) = (reflectivity)x(sky\ fraction\ occupied\ by\ the\ planet)x(phase)$$

$$f_\lambda(\phi,i) = p\left(\frac{R_p}{a_p}\right)^2\left[\frac{1-(\sin i \sin 2\pi\phi)}{2}\right]$$

where i [0, p/2] is the orbital inclination, $\phi$[0, 1] the orbital phase, $R_p$ the planetary radius and $a_p$ the distance to the host star. The value of the geometric albedo p depends on the amplitude and angular dependence of the various sources of scattering in the planetary atmosphere, integrated over the surface of the sphere. For a Lambert law sphere, p=2/3, whereas for a semi-infinite purely Rayleigh scattering atmosphere, p=3/4. The geometric albedos at 480 nm of Jupiter, Saturn, Uranus, and Neptune are 0.46, 0.39, 0.60, and 0.58, respectively (Karkoschka, 1994). The phase functions of the gas giants of our solar system are well approximated as Lambert spheres (e.g. Pollack, 1986).

At visible and near-IR wavelengths a planet's radius and mass can be estimated from its apparent brightness or colors using an atmospheric model as input (Des Marais, 2002). The Solar System could be used as a framework to establish such relations. Assuming extrasolar planets show similar characteristics, confusion between giant planets and Earth-like planets is unlikely. Using a rough estimate of radius from the apparent brightness and assuming some density, the photometric mass can be constrained. Alternatively planet mass might be estimated from low-resolution photometry, leading to a color mass. The key idea is that a planet's color can indicate whether it is a giant or terrestrial type planet, based on the experience of planets in our Solar System. Traub (Traub, 2003) proposed to use planet color to establish a planet sub-type using low-resolution spectroscopy and thus limit the possible planet albedo through modeling. By measuring the flux of an object in three equal-widths wavelength bands and remove the color contribution of the host star, Traub establishes a color-color diagram that shows the differences of giant and terrestrial planets.

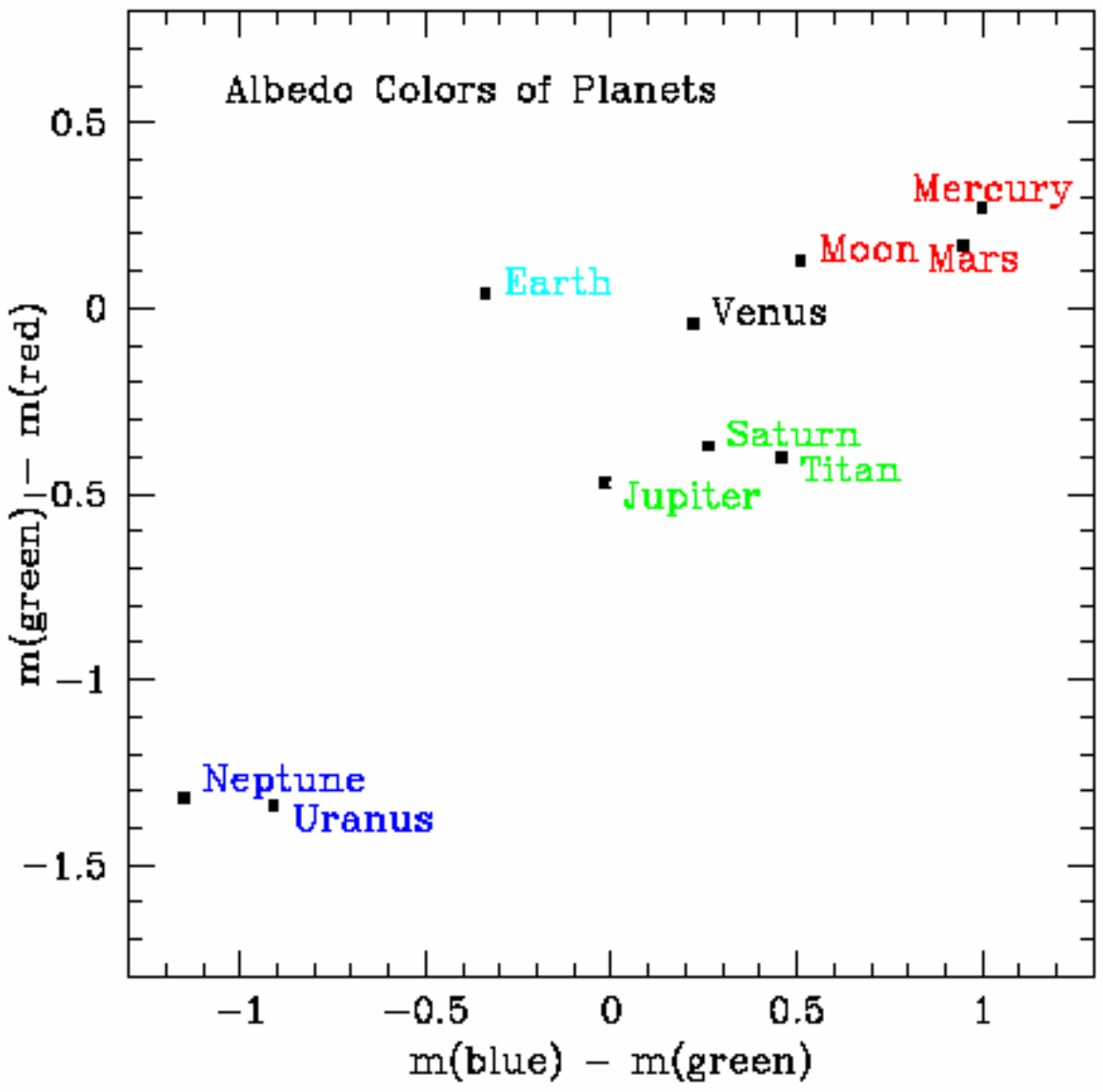

**Fig. 91: A color-color plot of solar system objects suggesting that one can constrain the albedo of extrasolar planets in the visible (Traub, 2003)**

Traub (Traub, 2003) concluded that rocky planets are found in the red-red end of the plot, as rocky surfaces and dust tend to reflect long wavelength better than short wavelengths. Deep molecular atmospheres of Uranus and Neptune can be found in the blue-blue part of the plot. Cloudy outer planets like Jupiter, Saturn and Titan also cluster in the plot. One can estimate the albedo from the planet's position on the color-color plot, assigning the numerical albedo of the corresponding solar system object. From the absolute albedo and the planet's apparent brightness, one can thus assign a planet radius. From the assigned planet type and its probable density, one can derive mass from its area. Knowing the orbital radius, one can estimate the absorbed radiation and hence its effective temperature. From mass and radius one can derive the surface pressure and by taking a guess of the lapse rate one can estimate the surface temperature.

The intensity of visible/near-IR absorption features is not affected by thermal structure: thus a direct measurement of the planet's temperature is difficult, while the planet's spectrum is suitable for abundance determination if pressure can be independently estimated. For problems encountered limiting the abundance calculations see Des Marais et al. (Des Marais, 2002). While our Solar System provides a good template for interpreting brightness and color mass, it is likely that extrasolar planets show a much greater diversity and thus challenge this way of estimating planetary radii and masses. Dealing with other planetary atmospheres is challenging: The set of reaction rates used may not be complete and the data measured for 200-300K may have to be extrapolated far from this range. Even when the reaction rates are available in the required temperature range, it always holds some uncertainty. Photochemical models typically use several hundreds of reaction rates, numerous data for radiative transfer and photolysis rates (absorption and scattering cross sections, quantum yields). The way that eventually affects the result of the model is all but negligible. One way to estimate the error on the result is to use a Monte-Carlo method. The final result is a series of profiles defining a domain that contains the solution (Selsis, 2002).

## 7.5 Biomarkers

Biomarkers are features whose presence or abundance requires a biological origin (Des Marais, 2002). They are created either during the acquisition of energy and/or the chemical ingredients necessary for biosynthesis (e.g. atmospheric oxygen and methane) or are products of the biosynthesis (e.g. complex organic molecules and cells).

The detection of $O_2$ or its product $O_3$ simultaneously with liquid water and $CO_2$ is our most reliable biomarker. While the existence of $H_2O$ in liquid state on the surface of a planet is considered essential for the development of life, it is not by itself a bio-indicator. As signs of life in themselves $H_2O$ and $CO_2$ are secondary in importance because although they are raw materials for life, they are not unambiguous indicators of its presence. Taken together with the molecular oxygen, abundant $CH_4$ can indicate biological processes. Although depending on the degree of oxidation of a planet's crust and upper mantel also non-biological origins can produce large amounts of $CH_4$ under certain circumstances. Oxygen is a chemically reactive gas. Reduced gases and oxygen have to be produced concurrently to be detectable in the atmosphere, as they react rapidly with each other. $N_2O$ is interesting because it is produced in abundance by life but only in trace amounts by natural processes. But it can only be detected in regions strongly overlapped by $CH_4$ and $H_20$, so it is unlikely to become a prime target. Ozone is detectable in the infrared with a resolution of 10 while oxygen's spectral features at 720 and 760 nm in the visible spectral range requires a spectral resolution of about 70 to be detected. In the visible spectral range, the photon flux is also 30 times lower and the contrast with the parent star is 1000 times higher.

### 7.5.1 Biomarkers in the IR

In the mid-IR the classical signatures of biological activity are the 9.6µm $O_3$ band, the 15µm $CO_2$ band and the 6.3µm $H_2O$ band or its rotational band that extends from 12µm out into the microwave region. In the same spectral region, the 7.7µm band of $CH_4$ is a potential biomarker for early-Earth type planets. The 9.6µm $O_3$ band is highly saturated and is thus a poor quantitative indicator, but an excellent qualitative indicator for the existence of even traces of $O_2$. Ozone is a very nonlinear indicator of $O_2$, the ozone column depth remains nearly constant as $O_2$ increases from 0.01 present atmosphere level (PAL) to 1 PAL.

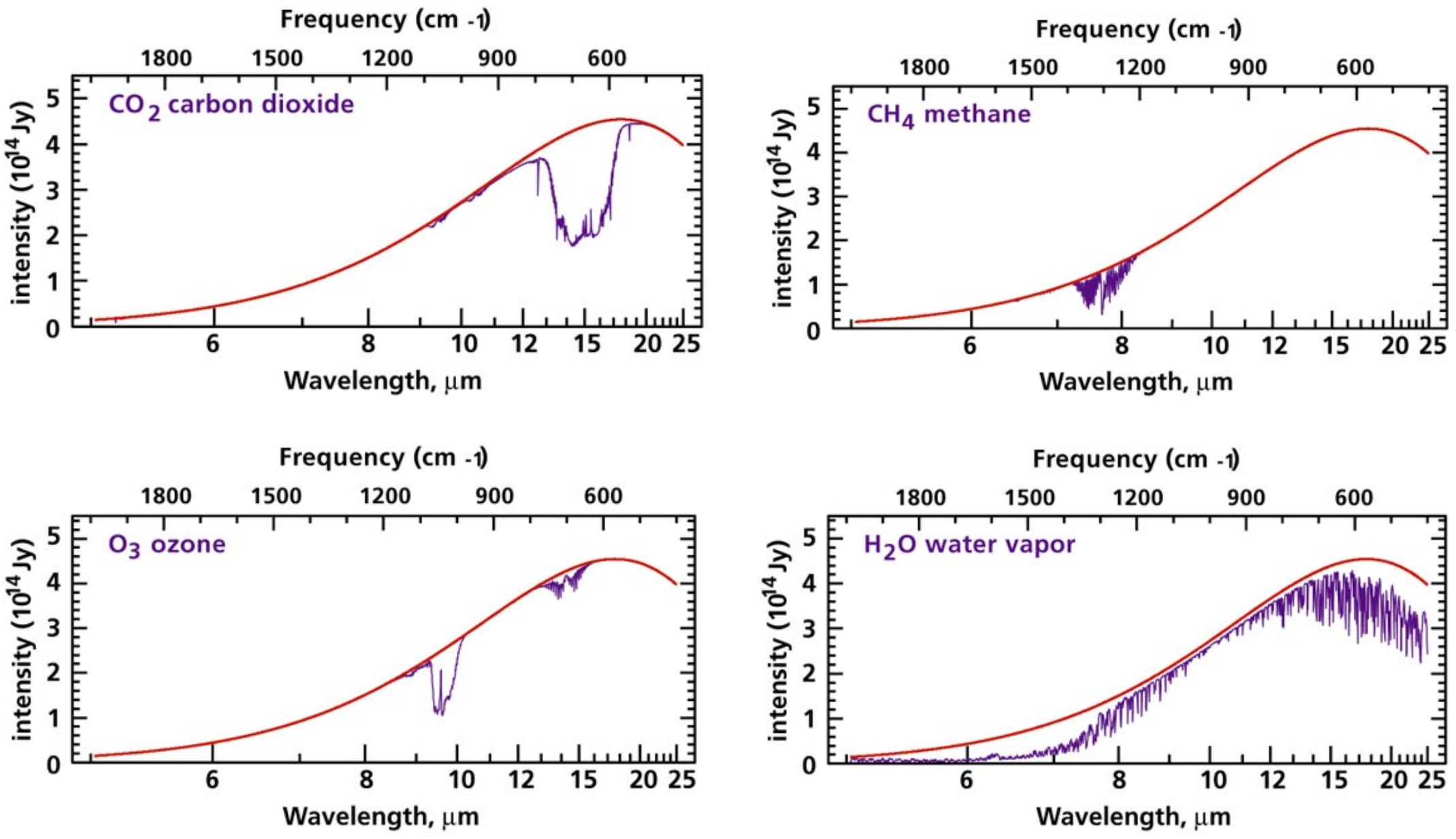

**Fig. 92: Calculated thermal emission spectrum of the individual species of present Earth (Traub, 2003)**

In the mid-IR $N_2O$ has a band near 8µm, comparable in strength to the adjacent $CH_4$ band but weak compared with the overlapping $H_2O$ band. The absorption bands of those three species are different, so in principle their features may be separated. Spectral features of $N_2O$ become apparent in atmospheres with less $H_2O$ vapor (Des Marais, 2002). Freon 11 and Freon 12 have spectral features between 9 and 12µm but require a resolution of about 200 to be evidenced. Observations from 8µm to 12µm of the $H_2O$ continuum allow estimations of the surface temperature of Earth-like planets. However the atmosphere of planets that are warmer than about 310K will become opaque in this region because of continuum absorption by water vapor. Fig. 92 shows the Earth's thermal emission spectra with its major molecular species ($H_2O$, $O_3$, $CH_4$, $CO_2$, $N_2O$) as well as minor contributors ($H_2S$, $SO_2$, $NH_3$, $SF_6$, CFC-11, CFC-12) as discussed in Traub et al. (Traub, 2002) with a resolution of 100 like the visible spectra in Fig. 95. $CO_2$ shows a strong 15µm feature even at 5% of its present abundance. Methane and nitrous oxide are shown at twice natural abundance and have significant features nearly overlapping in the 7µm region, lying in the red wing of the 6µm water band, therefore not readily separable, but in principle measurable.

#### 7.5.1.1 SILICATE IN THE IR

Silicate minerals and $O_3$ both have strong features in the 10µm region, but there is little chance that the two could be confused because their spectral shape is quite distinctive (Des Marais, 2002). Silicates produce a signature around 10µm that may mask or mimic the detection of $O_3$ at low resolution due to an absorption overlapping the ozone band. The thermal emission of a planet can present 3 types of silicate signature: one due to the dust-disk of the planetary system, the extrasolar zodiacal light, one due to dust in suspension and one due to surface silicates. Observing extrasolar systems without a planet and subtracting that dust spectrum from the planet's spectrum may remove the first one. The absorption by atmospheric dust is clearly visible in Martian spectra taken during dust storms. Its signature appears to be wider than ozone's and it is therefore the resolution of the instrument that may allow one to distinguish the origin of the feature (Selsis, 2002).

7.5.1.2 SIMULATED IR SPECTRA AROUND DIFFERENT HOST STARS

The DARWIN target catalogue consists of different stellar types. The characteristics of the star influence the atmosphere of a planet orbiting it. Selsis et al. have shown that influence (Selsis, 2002B).

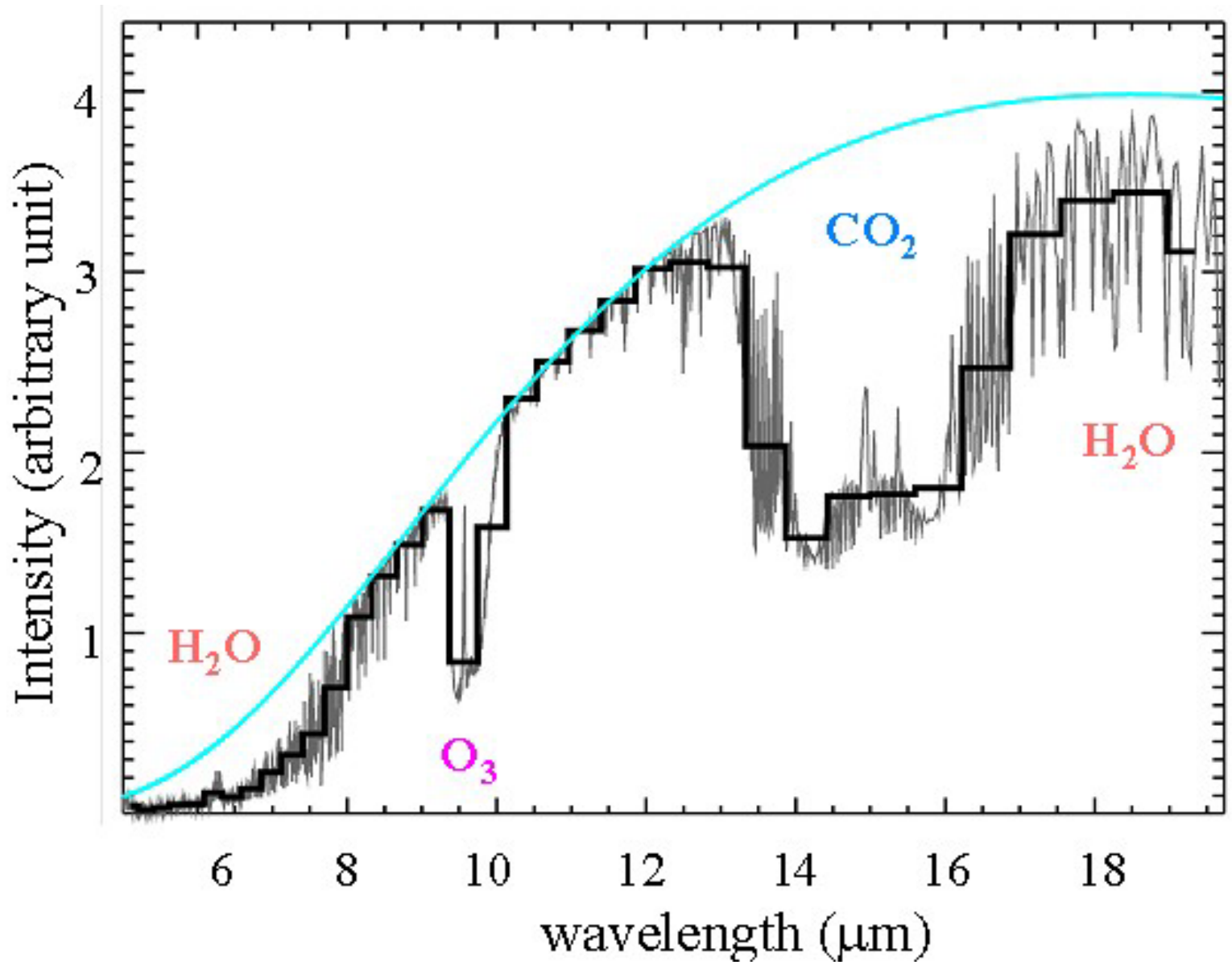

**Fig. 93: Calculated IR spectrum detectable with DARWIN for a resolution of 25 around a G star (Selsis, 2002B)**

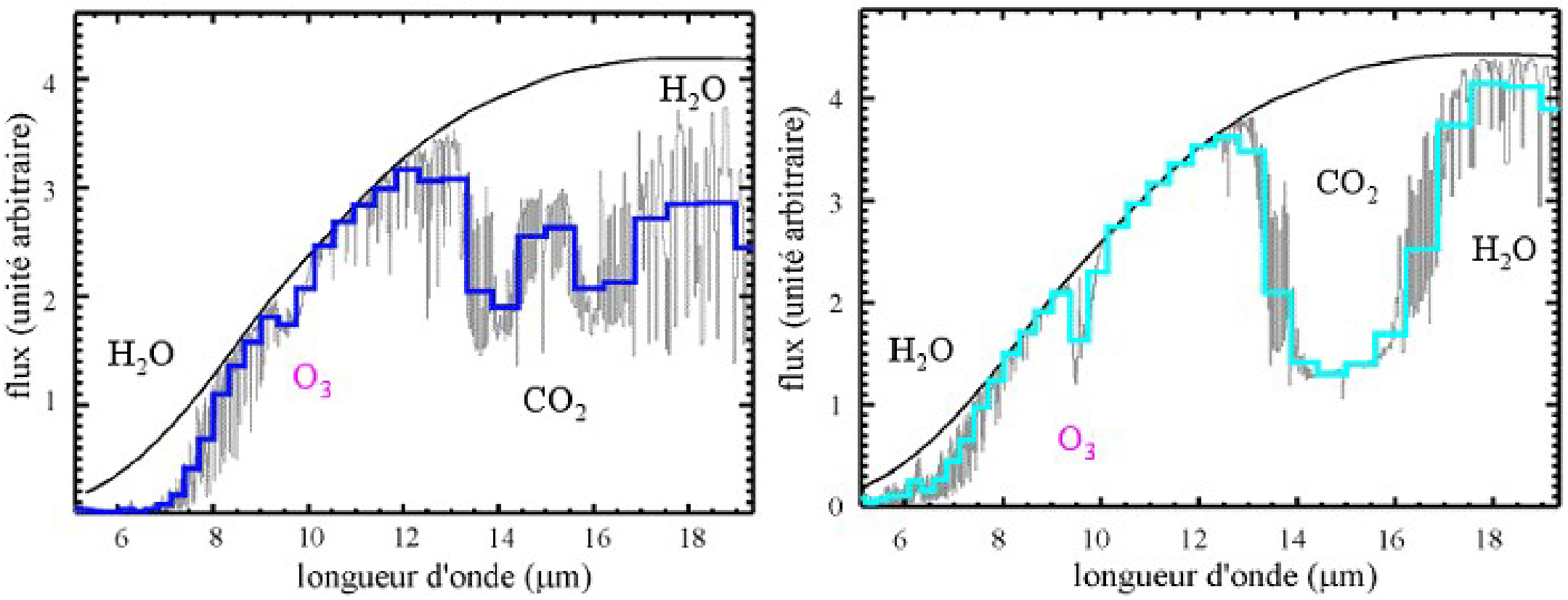

**Fig. 94: Calculated IR spectrum detectable by the DARWIN mission resolution of 25 around an F and K star (Selsis, 2002B)**

## 7.5.2 Biomarkers in the visible and near-IR

In the near-IR one can see a strong $O_2$ absorption feature at 0.76μm, a broadband $O_3$ absorption at 0.45μm to 0.75μm and a strong $H_2O$ band at 0.94μm. Using the $O_2$ and $O_3$ feature, the quantitative abundance of $O_2$ can be determined. If the planet is $CO_2$ rich, the weak 1.06μm band can show its abundance, but requires high spectral resolution. The best $CH_4$ band, short wards of 1.0μm, is at 0.88μm, but only shows absorption when the concentration is considerable greater than on Earth.

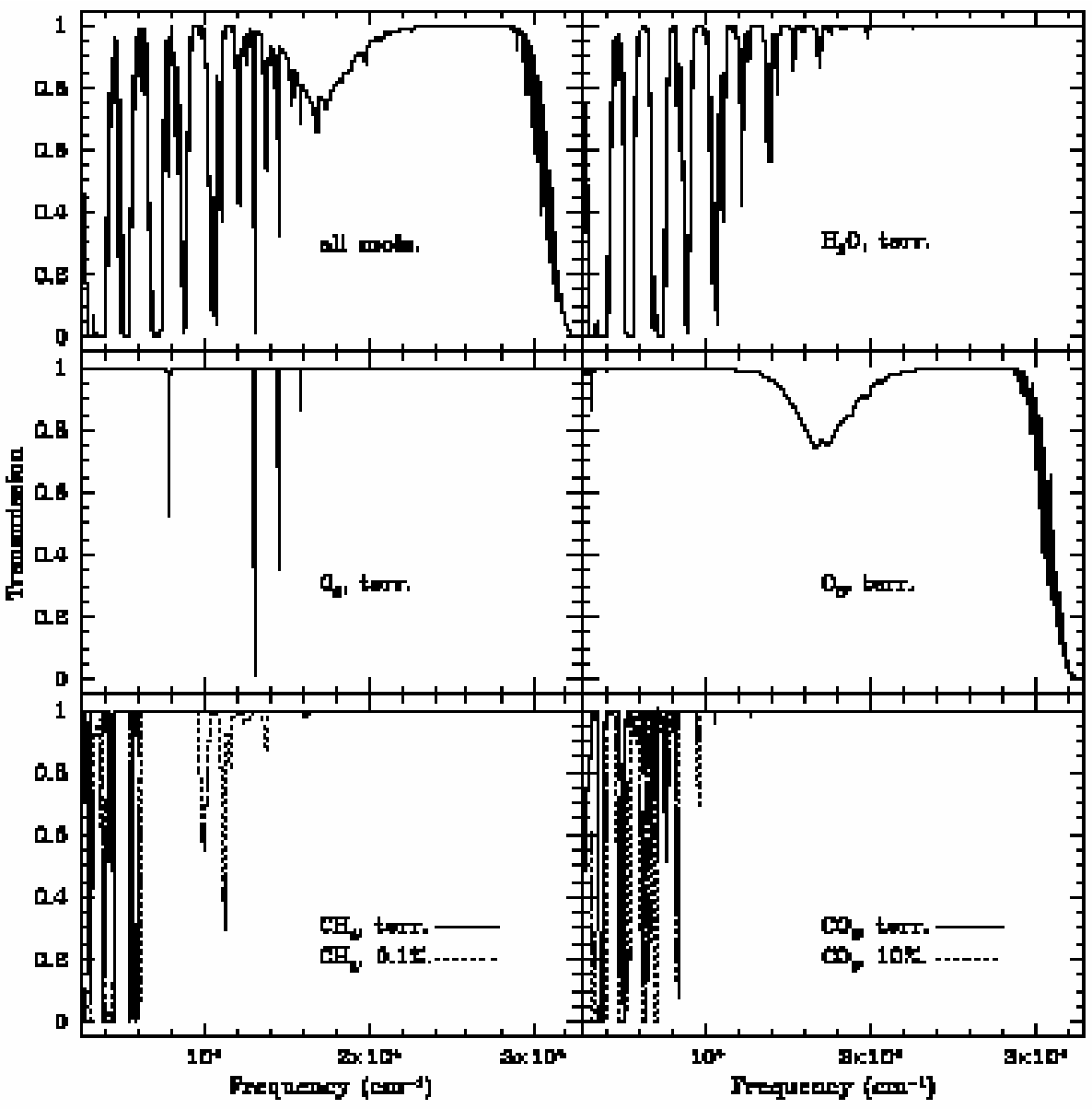

**Fig. 95:Reflectivity of the Earth with present atmospheric abundance (upper left) clouds, aerosols and Rayleigh scattering are ignored. For $CH_4$ and $CO_2$ greatly enhances abundances are also shown at a resolution of 100 (Traub, 2002)**

Fig. 95 shows the reflection spectra of the Earth with present atmospheric abundance (upper left) clouds, aerosols and Rayleigh scattering are ignored. For $CH_4$ and $CO_2$ greatly enhanced abundances are also shown at a resolution of 100 (Traub, 2002). The $H_2O$ absorption bands increase in strength toward the near infrared. They are essential independent of temperature, but increase in proportion to the abundance of water and the square root of air pressure. However since the lines are relatively saturated, the average band depth will only increase as the square root of band strength, making them less useful as quantitative indicators of water mixing ratio unless we have some independent knowledge of temperature and pressure.

The strongest $O_2$ feature is the saturated Frauenhofer A-band at 0.76µm that will still be relatively strong for significant smaller mixing ratios than present Earth's. $O_3$ has two broad features, the extremely strong Huggins band in the UV at 0.33µm and the Chappius band which shows as a broad triangular dip in the middle of the visible spectrum from about 0.45µm to 0.74µm. Methane at present terrestrial abundance (1.65ppm) has no significant visible absorption feature, but at high abundance 0.1% it has strong visible bands at 0.9µm and 1.0µm. $CO_2$ has negligible visible features at present abundance, but in a high $CO_2$-atmosphere of 10% it would have a significant band at 1.2µm and even stronger ones at longer wavelengths (Traub, 2002).

### 7.5.3 Surface features as biomarkers in the visible

Atmospheric biomarkers such as $O_2$, $O_3$, $H_2O$, $CO_2$, and $CH_4$ are the focus of our attention but to complete the section we will also discuss another species of biomarkers: surface biomarkers. Their spectra may be detectable in an extrasolar planet's spectrum. To detect and study surface properties we can only use wavelengths that penetrate to the planetary surface. On a cloud-free Earth, the diurnal flux variation caused by different surface features rotating in and out of view could be high (Seager, 2002). The planet's flux in the visible is an addition of the planet's own flux and scattered starlight. When the planet is only partially illuminated more concentrated signal from surface features could be detected as they rotate in and out of view but most surface features like ice or sand show very small or very smooth continuous opacity changes with wavelength.

#### 7.5.3.1 THE RED EDGE

An interesting example for surface biomarkers on Earth is the red edge signature from photosynthetic plants at about 750 nm. Photosynthetic plants have developed strong infrared reflection as cooling mechanism to prevent overheating and chlorophyll degradation (Des Marais, 2002). The primary molecules that absorb the energy and convert it to drive photosynthesis ($H_2O$ and $CO_2$ into sugars and $O_2$) are chlorophyll A (0.450 µm) and B (0.680 µm). The reflectivity changes by almost an order of magnitude, a much larger change than caused by chlorophyll absorption that is negligible in a hemispherically averaged spectrum (Seager, 2002). Phytoplankton can also cause temporal change in large areas of the ocean, but is not detectable due to its weak signal in the presence of strong water absorption at these wavelengths

Several groups (Woolf, 2002) (Arnold, 2002) have measured the integrated Earth spectrum via the technique of Earthshine, sunlight reflected from the un-illuminated side of the moon. Earthshine measurements have shown that detection of Earth's vegetation-red edge is difficult due to smearing out by other atmospheric and surface features. Trying to identify such small features at unknown wavelengths in an extrasolar planet spectrum may be very difficult (Seager, 2002). A variable signal around 700 nm of 4 to 10 ±3% has been measured in the Earth albedo (Woolf, 2002) (Arnold, 2002). It is interpreted as being due to the vegetation red edge, expected to be between 2 to 10% of the Earth albedo at 700 nm, depending on models. The detection of a VRE index between 0 and 10% requires a photometric precision better than 3%.

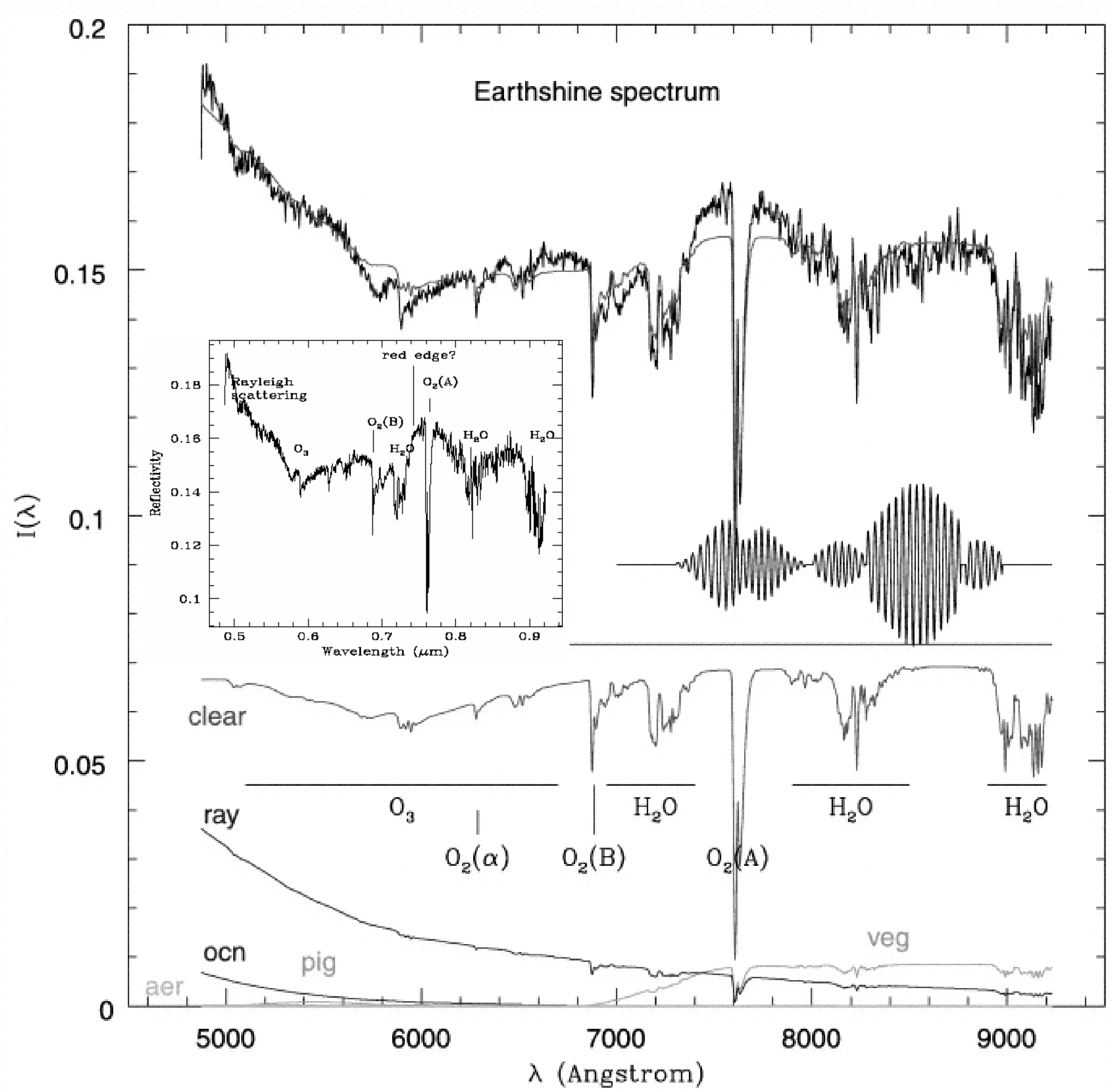

**Fig. 96: Observed reflectivity spectrum of the integrated Earth at a resolution of 600, as determined from Earthshine with a superposed model spectrum. The reflectivity scale is arbitrary (Woolf, 2002) (Seager, 2002).**

The spectrum shown in Fig. 96 was derived from observation of lunar Earthshine in the range 480nm–920nm at a spectral resolution of about 600 by Woolf et al. (Woolf, 2002). It shows increased reflectivity in the 720–790nm region, suggestive of a vegetation signal. However, the data did not show continued high reflectivity at longer wavelengths (790–920nm), as would be expected for a vegetation reflection profile, making the interpretation ambiguous. Seven component spectra were fitted and summed to produce the model spectrum seen in Fig. 96 (Woolf, 2002): "High": reflectivity from a high cloud. "Clear": the clear-atmosphere transmission. "Ray": Rayleigh-scattered light. "Veg": the vegetation reflection spectrum from land chlorophyll plants. "Ocn": the blue spectrum from subsurface ocean water. "Aer": aerosol scattered light, here negligible. "Pig": the green-pigmented phytoplankton reflection of ocean waters, also negligible here

.

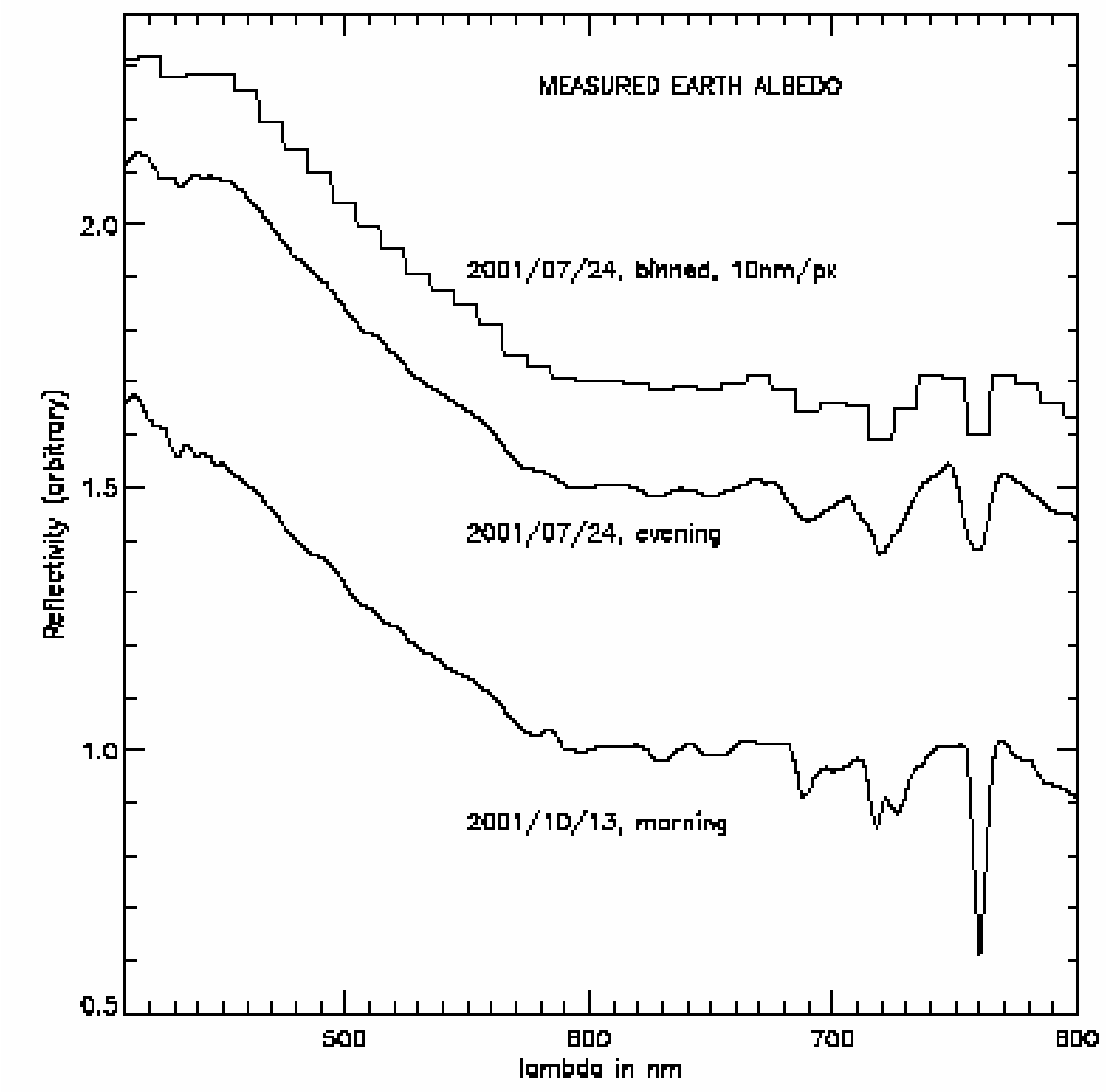

**Fig. 97: Examples of measured Earth albedo spectra. Both spectra are normalized to 1 at 600 nm, but the July spectrum is shifted upwards by 0.5 for clarity see Arnold et al. (Arnold, 2003).**

In Fig. 97 the spectral resolution for the July measurements is 50, for the October measurements it is 240. The July spectrum has been binned to 10 nm=px to mimic the low resolution that might be used for the first extrasolar planet spectrum (Arnold, 2003). Around 690nm and 720nm $H_2O$ bands as well as a narrower $O_2$ band at 760nm are clearly visible with a resolution of 50. The slope variation at about 600nm is partially the signature of the deepest zone of the broad ozone absorption band from 440nm to 760nm (Arnold, 2003).

Our knowledge of different surface reflectivities on Earth, like deserts, ocean and ice, help assigning the VRE of the Earthshine spectrum to terrestrial vegetation. For extrasolar planets, a VRE might be as difficult to measure as for Earth due to variable cloud cover of the planet. A model of the extrasolar planets atmosphere is necessary to be able to remove the absorption bands that may partially hide the vegetation. Even if a clear VRE-like spectral signal were shown, its use as a biosignature still raises some questions (Arnold, 2003): For several organisms (Blankenship, 1995) the “red edge" is not at 700nm, but at 1100nm also some rocks like schists, may have a similar spectral feature. Spectra of Mars show a similar spectral feature at 3.5µm, which were erroneously interpreted as vegetation due to their similarity with lichen spectra (Sinton, 1957). The exact wavelength and strength of the spectroscopic “red edge” depends on the plant species and environment. The feature is very strong for an individual plant leaf but averaged over a spatially unresolved hemisphere of Earth, the spectral feature is reduced from this high reflectivity down to a few percent (Seager, 2001). The main effects are forest canopy architecture, soil characteristics, non-continuous coverage of vegetation across Earth's surface, and presence of clouds, which prevent view of the surface. In addition the reflectance of vegetation is anisotropic and so the illumination conditions and viewing angle are important. Keeping in mind that the chances are very small that another planet has developed the exact same vegetation as Earth, finding the same signatures would be thrilling, but might be unlikely.

Observations of extrasolar Earth-like planets at wavelengths that penetrate the planet's atmosphere could be very useful, especially for planets with much lower cloud cover than Earth's 50%. A time series of spectra or broadband photometry could reveal surface features. Earth's hemispherically integrated vegetation red-edge signature is weak, but planets with different rotation rates, obliquities, land-ocean fraction, and continental arrangement may have lower cloud-cover. The reflectivity of the Earth has not been static throughout the past 4.5Gyr (Woolf, 2003). Oxygen and ozone became abundant about roughly 2.3Gyr ago, affecting the atmospheric absorption component of the reflection spectrum. About 2.0Gyr ago, a green phytoplankton signal developed in the oceans and about 0.44Gyr ago, an extensive land plant cover developed, generating the red chlorophyll edge in the reflection spectrum (Lunine, 1999). The oxygen and ozone absorption features could have been used to derive the presence of life on Earth anytime during the past 50% of the age of the solar system, while the chlorophyll red-edge reflection feature evolved during the most recent 10% of the age of the solar system.

### 7.5.4 Oxygen and ozone as spectral signature of life

It is thought that carbon in the primitive Earth atmosphere was mostly fully oxidized. Building of organic material required the reduction of $CO_2$ (Ollivier, 1998).

$$C\,O_2 + 2nH_2O + \text{energy} \rightarrow (CnH_2O) + O_2$$

This reaction replenished free oxygen, which is a highly reactive gas. If not continuously supplied, it would disappear. The presence of free $O_2$ and $H_2O$ and $CO_2$ in a planet's atmosphere seems to indicate the presence of carbon-based life. On Earth, the rate of biological $O_2$ production is much larger than the abiotic one. The two main photochemical ways to produce $O_2$ are initiated by the photolysis of $H_2O$ and $CO_2$ by UV radiation. The first is responsible for the detection of molecular oxygen on Europa, the second for the molecular oxygen in the Martian atmosphere.

$$H_2O + h\upsilon \rightarrow OH + H$$

$$OH + OH \rightarrow H_2O + O$$

$$O + O \rightarrow O_2$$

$$CO_2 + h\upsilon \rightarrow CO + O^*$$

$$O^* + M \rightarrow O + M$$

$$O + O \rightarrow O_2$$

A star hotter than the Sun emits more UV radiation and could therefore induce high photo dissociation rates. On Earth-like planets with surface temperatures around 300K the first process is strongly reduced by the existence of cold traps at the tropopause. They block most of the ascending water vapor. It returns to the planet's surface as rain or ice. Then $H_2O$ is only a minority gas in the UV rich upper atmosphere. According to J. Rosenqvist et al. (Rosenqvist, 1995) the photo dissociation of $CO_2$ can only produce small amounts of $O_2$.

$O_3$ forms photochemically from Photolysis of $O_2$:

$$O_2 + h\upsilon \rightarrow O + O,$$

$$O + O_2 + M = O_3 + M$$

M is a third molecule necessary to carry off the excess energy of the collision. $O_3$ is a sensitive indicator of $O_2$ as it has a logarithmic dependence upon the $O_2$ abundance. Ozone has a strong absorption feature at about 9.6µm with a weaker band at 14µm. However, if $CO_2$ is present in even small amounts the 14µm feature will be mostly blocked. The 9.6µm $O_3$ band is highly saturated and stays essentially unchanged while the $O_3$ abundance varies from 1 to 6ppm (Des Marais, 2001). It is thus a poor quantitative, but an excellent qualitative indicator for the existence of even traces of $O_2$. Photochemical models show that a biologicaly effective UV protection can be established at an $O_2$ concentration of about 0.01 – 0.1 times the present atmospheric level. After 10% of the present oxygen was produced, a surface protecting $O_3$ layer (ozone absorption between $\lambda$ = 200 nm to 300 nm) could be produced in the atmosphere. The short wavelength protecting ozone layer could be compared to a protecting layer of about 10 m $H_2O$ (Lammer, 2003).

#### 7.5.4.1 FALSE POSITIVE DETECTIONS

A planet that has a much smaller source of reduced volcanic gases would not remain habitable over a long period because of the missing mechanism for recycling $CO_2$. Could it have an $O_2$-rich atmosphere? A planet, bigger than Mars, but not big enough to maintain active volcanism might be able to retain its $O_2$ (Kasting, 1997). If this abiotic planet had liquid water at its surface, it should have a vanishing small atmospheric $O_2$ concentration. If not, $O_2$ might be detectable.

A planet that has lost a lot of water could also accumulate a large amount of $O_2$. If $H_2O$ were transported to the top of the atmosphere, where it is dissociated, hydrogen would be lost to space in a rapid rate while oxygen could be retained. That effect might allow $O_2$ to surpass the rate of outgassing of reduced gases and possibly create a transient $O_2$-rich atmosphere (Kasting, 1988). This signature would eventually disappear on a volcanically active planet, but might generate a substantial ozone layer during the time it exists. If oxygen sinks on a planet are inactive, abiotic $O_2$ would accumulation due to UV photolysis and hydrogen escape. The lack of liquid water slows down the rates of both physical and chemical weathering. A planet between the size of Earth and Mars could retain a thick atmosphere but not its volcanism (Kasting, 1997) (Kasting, 1998). The sinks due to water weathering would also be missing in the case of a low surface temperature. Consequently a planet outside of the HZ could have an atmosphere rich in abiotic $O_2$. Such a planet would be small and its surface temperature would be below the freezing point of water; thus, it should be identifiable because by its location outside of the HZ and the fact that its water vapor bands would be weak.

Selsis et al. (Selsis, 2002) have investigated the risk of 'false positive' detection of life with new photochemical and radiative convective models of terrestrial planet atmospheres. The authors concluded that the simultaneous detection of $O_3$, $CO_2$ and $H_2O$ is a robust way to discriminate photochemical $O_2$ production from biological photosynthesis. If $O_2$ is produced by $H_2O$ photolysis at high altitude, several hydrogen components are produced. They attack $O_3$ and prevent its accumulation. The only way to have a significant amount of $O_3$ in the atmosphere spectrum is producing $O_2$ at low altitude and no $H_2O$ at high altitude where UV is present. Consequently, as on terrestrial planets (Selsis, 2002), the simultaneous presence of $O_3$, $H_2O$ and $CO_2$ in the atmosphere appears to be a reliable biosignature. This points out the superiority of $O_3$ as a biosignature with respect to $O_2$.

## 7.6 Effects of Dust Rings and Moons

The albedo of a planet can be derived from its radius and flux:

$$F_r = A \pi R_{pl}^2 \phi(t) \quad (7.6.1)$$

where $\phi(t)$ is an orbital phase factor assuming the planet is spherical. When circum-stellar rings or companions like large Moons are present, the assumption does not hold. The planetary albedo, estimated from the visual reflected flux or the radius estimated from a guess albedo will be incorrect in these cases. High-resolution observations in the visual could distinguish between the effects of rings versus effects of moons, as they will express a distinctive shape in their phase effect. In the mid-IR range, the effects of large moons would result in spectral features whose strength vary throughout the year and thus the temperature of the detected planet would appear to vary in this cycle. A planet in an eccentric orbit would show the temperature variation, but not the washing out of the spectral features. The spectral resolution is a key issue in those special cases and it is unlikely that DARWIN achieves such high spectral resolution.

## 7.7 Eccentric orbits and habitable moons

Williams et al. (Williams, 2002) investigated the climates of either bound or isolated Earths on extremely elliptical orbits near the HZ using a three-dimensional general climate model and a one-dimensional energy-balance model. The planet's features like spin axis, continental topography and atmospheric composition resemble our own planet. The models yield an eccentricity limit for preventing water from reaching the stratosphere at 0.42 and the limit associated with a runaway greenhouse at 0.70 for planets at 1.0AU around a star with a luminosity of 1.0 $L_{solar}$. Generally, results of climate models that investigate extreme conditions have to be viewed critically as model parameters might exceed their valid range of definition. Nevertheless critically reviewed results can give indications of the critical processes and changes involved.

Long-term climate stability depends primarily on the average stellar flux received over an entire orbit, not the length of the time spent within the HZ. In an eccentric orbit, the ratio of periastron to apoastron flux $[(1+e)/(1-e)]^2$ becomes comparable to the ratio of summer to winter solstice flux at mid-latitude (35°) and present obliquity (23±5°) once eccentricity reaches 0.2. Seasonal cycles of moons around EGPs (many with $e>0.4$) will be more strongly affected by the shapes of their orbits than their spin-axis tilts, provided their axial tilts are small. Most of the planets within the narrow HZ limits between $0.95<d'<1.4$AU do not remain there over a full orbit. $d'$ is the equivalent solar distance. The planet HD222582b ($e$=0.71), for example, spends most of its time well outside the HZ margins, heated stronger than Mercury near periastron ($d'$=0.347AU) and far less than Mars near apoastron ($d'$=2.05AU). Still, moons belonging to planets such as HD222582b might be suitable for life if they possess enough volatiles in their oceans or atmospheres to moderate the climatic extremes caused by high eccentricity (Williams, 2002).

## 7.8 Rocky moons in the HZ

Rocky moons orbiting extrasolar planets or BDs could also be habitable if the planet moon system orbits within the HZ. Such a moon would have to be large enough, bigger than 0.12 Earth-masses, to retain a long-lived atmosphere (Williams, 1997). It also would need to posses a strong magnetic field in order to prevent its atmosphere from being sputtered away by the bombardment of energetic ions from the planet's magnetosphere. The high eccentric orbit of their planets would also have a negative effect on their atmospheres. The detection of habitable moons is still far in the future as their signals are extremely faint even in comparison with planetary signals. The influence of a Moon-like satellite, 1/4 Earth diameter and ½ Earth diameter in an on-orbit system, on the infrared spectrum has been estimated (Des Marais, 2001): When we see the un-illuminated side of the Moon with a temperature of 200K, the spectrum of the system does not differ essentially from the spectrum of the planet itself leading to a slight overestimation of the Earth's temperature to 279K. When we see the bright side of the Moon with a temperature of 375K, the Earth temperature would be overestimated to 315K, instead of the assumed 285K surface temperature using the 9/12µm ratio. The strength of the $O_3$ band is reduced to 84% and 57% respectively, a washing out of the $O_3$ band at 9.6µm. The two effects are very small and are most likely undetectable.

Mars has no large moon to stabilize its obliquity resulting in large climate effects, which may affect the evolution of life. Studies show that the obliquity of the present Earth, with a rotation period of 24 hours, would have a chaotic behaviour but not with a faster rotation of about 10 hours (Lammer, 2003). The rotation of the Earth slows down as the Moon moves away so its initial rotation was faster, what might have been caused by the Moon-forming impact. Barnes et al. (Barnes, 2002) investigated the long-term dynamic stability of hypothetic moons orbiting EGPs, putting constraints on the mass of possible extrasolar moons. Worlds around other stars are likely to be very different from Earth and further modelling has to be done.

## 7.9 Earth

*"If distance from the star were the only thing to consider, Earth's moon would have liquid water (Marcy, 1998)."*

### 7.9.1 Carbonate-silicate cycle

Planetary habitability is believed to be critically dependent on atmospheric $CO_2$ and its control by the carbonate-silicate cycle that stabilizes the surface temperature by negative feedback mechanisms. The argument that carbonates could also form through direct carbonatization when the volcanic activity on a planet is very high (Sleep, 2001) has started an interesting discussion on whether or not such a cycle has to exist on all planets to provide habitability conditions. A destabilizing effect is the greenhouse effect that increases with surface temperature because of increased atmospheric water vapor. Cold surface temperatures on the other hand enhance the planetary albedo by increased snow and ice cover, cooling the surface further.

The size of the planet can determine its capacity to sustain habitable conditions. Large planets sustain higher levels of tectonic activity that also persist for a longer time. Tectonic activity sustains volcanism, heats crustal rocks, recycles $CO_2$ and other gases back into the atmosphere. These outgassing processes are required to sustain climate stability over geologic timescales on Earth. The $CO_2$ concentration in Earth's atmosphere is controlled mostly by the carbonate-silicate cycle, a slow interaction with the crustal rock reservoir. An increase in surface temperature produces more water vapor in the atmosphere, enhances dissolution of $CO_2$ in rain and thus the fixation of $CO_2$ into carbonates and the subsequent precipitation and burial of carbonate sediments. The Greenhouse effect is weakened due to the lower concentration of $CO_2$ in the atmosphere and the surface of the planet cools. If the surface temperature of a planet reaches values high enough to dissociate carbonates into $CO_2$ and oxides, the mechanism fails. A massive input of $H_2O$ into the stratosphere can occur and the UV photolysis of water becomes a powerful source of $O_2$. An abiotic $O_2$-rich atmosphere can build. Kasting (Kasting, 1993) (Kasting, 1988) gives an estimate of the distance to the star where that moist greenhouse effect occurs. It is also the inner limit of the Continuous HZ (CHZ):

$$a_{\min} = 0.95\left(\frac{L}{L_{sun}}\right)^{1/2} AU \qquad (6.11.1)$$

The recycling process on Earth is termed carbonate metamorphism. It occurs when seafloor sediments are subducted along certain oceanic plate margins. The higher temperature and pressures cause carbonate rocks to decompose and release their trapped $CO_2$ (Kasting, 1997). These arguments suggest that another requirement for the habitability of a planet is that it is large enough to maintain active plate tectonics or volcanism throughout its lifetime. A non-equilibrium should have been seen during the first billion years of life on Earth - that both carbon dioxide and methane coexist in an atmosphere requires some unusual chemistry that might be based on life. Between 2.2Gyr and 3.5Gyr ago Earth may have had a $CH_4$-rich atmosphere produced by methanogenic bacteria. During the first billion years of life on Earth, our planet may have passed through a phase when it contained large amounts of biogenic methane, no biogenic oxygen, a large amount of primarily outgassing carbon dioxide and some fraction of nitrogen (Owen, 1980). Similar conditions could also be produced by submarine volcanism on a planet with highly reduced crust, and thus has to be viewed with caution (Kasting, 1997).

### 7.9.2 Earth changed place with Venus

If Earth were much smaller, it would not be able to produce and sustain an atmosphere sufficiently massive to permit the continued existence of liquid water. If it were much larger it would not have lost hydrogen so readily, thus preventing the atmosphere from moving from reducing to oxidizing conditions. The ranges for the size variations are difficult to specify. If Earth changed place with Venus, our planet would experience a runaway greenhouse effect: the oceans would heat up, providing more water vapor in the atmosphere, leading to an increased greenhouse effect (Kasting, 1984). Finally all water would be in the atmosphere, which would then be so warm that there would be no cold trap to confine the vapor to lower levels where it could be protected from photolysis. Incident ultraviolet sunlight would dissociate the water molecules, allowing the hydrogen to escape and the oxygen to make carbon dioxide and combine with other elements in the planet's crust. In the absence of water and life the carbon dioxide would remain in the atmosphere, providing an efficient greenhouse effect such that the mean temperature would approach about 750K (Owen, 1980).

### 7.9.3 Migrating Neptune planets

The cores of Uranus and Neptune are similar to terrestrial rocky planets, which are covered with thick ice-layers, where $CH_4$ and $NH_3$ is trapped in clathrate (Podolak, 1991) (Hubbard, 1995) (Marley, 1995) (Guillot, 1999). A Neptune-type planet that migrates from its origin in the outer solar system to the HZ of its parent star could loose its dense hydrogen atmosphere due to XUV-driven hydrodynamic escape and melt the water-ice shell of these bodies. The outgassing of reduced gasses like $CH_4$, $NH_3$ etc. could build up a secondary reduced atmosphere (Lammer, 2003). A similar process is suggested for the origin of the atmosphere of Saturn's large satellite Titan (Yung, 1984). Titan is also thought to have a rocky core, which is hidden below an $H_2O$ ice-layer (Griffith, 2003) with trapped $CH_4/NH_3$ and other volatiles. During the formation of Saturn and Titan the $H_2O$ ice-clathrate melted so that the trapped $CH_4$ and $NH_3$ could be released in large amounts and form the present reduced atmosphere. The out-gassed $CH_4$ reacted with $NH_3$ and produced Titan's dense $N_2/CH_4$ atmosphere. After the body cooled, the orbital distance of about 10AU is responsible for a surface temperature of about 100K so that $H_2O$ exists on Titan's surface layers only in the form of ice.

In recent studies of close-in EGP the radiative effective temperature, was used to estimate atmospheric evaporation rates (Konacki, 2003) (Sasselov, 2003). New studies by Lammer et al. (Lammer, 2003C) investigated the atmospheric loss processes and concluded that these studies lead to significant underestimations of thermal atmospheric escape rates effecting the conclusions of long-term atmospheric stability, because the exosphere temperature controls the thermal escape in an upper atmosphere. The exosphere temperature is usually much higher than the effective temperature as the upper planetary atmospheres are mainly controlled by absorption of X-rays and extreme ultraviolet (XUV) radiation (Bauer, 1971). Table 19 (Lammer, 2003C) shows the planetary effective temperature $T_{eff}$ of various planets in our solar system compared to their dayside surface temperature $T_S$, mesopause temperature $T_0$ at the base of the thermosphere and exosphere temperature $T_{:}$, based on exospheric XUV heating contribution. Observations by various spacecraft indicate that $T_{:}$ depends solely on the XUV flux, $I_{XUV}$, into planetary thermospheres on terrestrial planets but additional thermospheric heating sources caused by accelerated particles and atmospheric gravity waves become important for giant planets in the outer Solar System.

Jupiter's $T_{:}$ is about 1000K, close to $T_{eff}$ of 51 Peg b of about 1300K. Even Earth's $T_{:}$ can reach $T_{eff}$ of 51 Peg b during high solar activity. In their study, Lammer et al. (Lammer, 2003C) used a scaling relation from solar system planets to estimate the exospheric temperature for extrasolar planets. If planets 1 and 2 have thermospheres with comparable gas compositions, the following scaling relation can be used (Bauer, 1989).

**Table 19: $T_{eff}$, $T_S$, $T_0$ and $T_{:}$. on various planets (Lammer, 2003C).**

| Planet | $T_{eff}$ [K] | $T_S$ [K] | $T_0$ [K] | $T_{:}$ [K] |
|---|---|---|---|---|
| **Venus** | 232 | 750 | 160 | 300 |
| **Earth** | 255 | 288 | 180 | 1000-1500 |
| **Mars** | 217 | 225 | 120 | 220 |
| **Jupiter** | 124 | - | 125 | 700-1000 |
| **Saturn** | 95 | - | 95 | 800-900 |
| **Uranus** | 59 | - | 52 | 750 |
| **Neptune** | 59 | - | 52 | 750 |

$$\frac{\left(T^s_{XUV\infty} - T^s_0\right)_I}{\left(T^s_{XUV\infty} - T^s_0\right)_{II}} \approx \frac{I_{EUV\,I}\, g_{II}}{I_{EUV\,II}\, g_I} \qquad (7.9.3)$$

Eq. (6.11.2) shows that $T_{XUV.}$ depends on the gravity g, which is related to the mass of the planet, as well as on $I_{XUV}$, which decreases with distance from the star. It is based on the assumption of equilibrium between the XUV heat input and downward heat transport by conduction.

### 7.9.4 Evolution of the XUV flux with time

The time-dependence of the $I_{XUV}$ flux is very critical for the evolution of thermal escape during the history of a planetary system. Estimates of the solar high-energy flux evolution are indirectly possible by comparison with solar proxies. The *Sun in Time* program has collected multiwavelength data for a sample of solar proxies, containing stars that represent most of the Sun's main sequence lifetime from 130 Myr to 8.5Gyr (Guinan, 2002). The resulting relative XUV fluxes yield a power-law relationship for Lyman-α (λ= 1215.6 ± 10 Å) $I_{L\alpha}(t)/I_{L\alpha}$ and for a waveband from 1 Å to 1000 Å $I_E$ (t)/$I_E$. (Lammer, 2003C).

$$I_{L\alpha}(t)/I_{L\alpha} = 3.17 \times [t\,(Gyr)]^{-0.75}$$
$$I_E(t)/I_E = 6.13 \times [t\,(Gyr)]^{-1.19}$$
$$I_{XUV}(t)=I_0\,(t)+I_{L\alpha}(t)$$

These expressions are valid for ages between 0.1 and 8Gyr. $I_0$ and $I_{L\alpha}$ are the present integrated fluxes at 1AU. $I_0(t)$ and $I_{L\alpha}(t)$ are the integrated fluxes as function of time. These relations lead to estimated fluxes of $6I_0$ and $3I_{L\alpha}$ about 3.5Gyr ago, and $100I_0$ and $20I_{L\alpha}$ about 100 Myr after the Sun's arrival on the main sequence. Lammer et al. (Lammer, 2003C) concluded that the Jean's approach is not valid for close-in hydrogen rich extrasolar planets to estimate atmospheric mass loss rates and hydrodynamic escape must be considered. Their calculation yield an energy-limited mass loss rate for HD 209458b in the order of $10^{10}gs^{-1}$- $10^{11}gs^{-1}$, which is in agreement with the inferred observational based estimation of about $10^{10}$ $gs^{-1}$ (Vidal-Madjar, 2003). In contrast, the value resulting from Jeans escape at $T_{eff}$ is < $1gs^{-1}$.

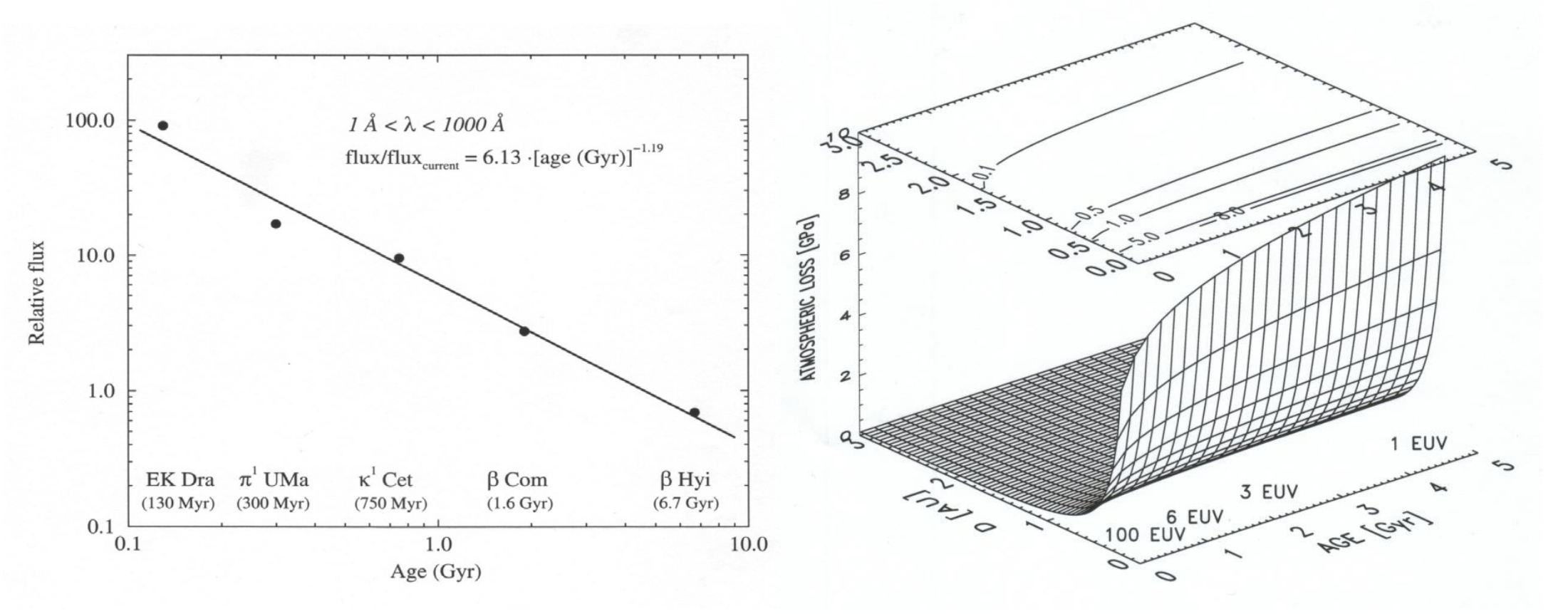

**Fig. 98: Evolution of the XUV flux obtained from solar proxies inside the "Sun in Time" program for G-type Sun-like stars (Lammer, 2003C) (left) Calculated atmospheric loss of a planet over the age of its host M star (right).**

They found that Uranus-size planets might loose their entire hydrogen atmospheres by thermal and non-thermal atmospheric escape processes due to higher stellar $I_{EUV}$ irradiance and/or closer distance to the host star and can evolve into a new type of terrestrial planet. Secondary atmospheres can evolve by out-gassing their remaining ice-rocky cores. Fig. 98 shows that Uranus-size planets lose ≈ 10 GPa corresponding to their present hydrogen atmosphere at orbital distances of about 0.06AU (Lammer, 2003C). Leger et al. (Leger, 2003) investigated the possible characteristics of ocean planets between 1 and 8 times the mass of the Earth made out of equal amounts of metals, rocks and water ice. A possible formation in an ice-rich environment and migration into the HZ of their host star could allow them to possess a surface water ocean.

## 7.10 Summary: Indicators of habitability and biological activity

The detection of $O_2$ or its product $O_3$ is our most reliable biomarker so far but only for planets well within the HZ. Mars-like planets have small $O_2$-sinks and Venus-like planets large abiotic $O_2$ sources. The existence of $H_2O$ in liquid state on the surface of a planet is considered essential for the development of life, even so, it is not a bioindicator. $CO_2$ indicates an atmosphere, and abundant $CH_4$ can indicate biological sources, although depending on the degree of oxidation of a planet's crust as well as upper mantle non-biological sources could also produce large amounts of $CH_4$. An independent temperature measurement will be needed to make sure that the planet is not in the middle of a runaway greenhouse stage, with water rapidly dissociating to produce the oxygen we observe. To constrain the alternative scenarios, the determination of the albedo and size of the planet and its distance from the parent star is very important.

A planet with $O_3$ and $H_2O$ and $CO_2$ absorption bands in its spectrum that lies within the HZ is the main target of a mission that searches for habitable planets. We point out the possibility that life exists on planets that do not show $H_2O$ and $O_3$, if the production of $O_2$ by photosynthesis is not able to overcome the oxygen sinks. On Earth that describes the situation up to 2Gyr ago (Leger, 1996). Estimates of planet size and albedo that will be crucial to determine the characteristics of a planet, can be determined from mid-IR observations. The visible/near-IR continuum does not give direct indication of the planet size because of the possible albedo range. However visible to near-IR spectra offer higher spatial resolution for the same collecting area, are minimally affected by temperature and may therefore be able to determine the abundance of atmospheric species. We now have the fascinating opportunity to simulate atmospheric features on diverse planets around their host stars, define biomarkers for different evolution steps of planetary atmospheres and investigate the resulting features in a real DARWIN mission scenario, using the real target catalogue and evaluating the different proposed concepts.

The HZ around different stellar types will vary as it only depends on the spectral energy distribution of the star. The HZ around F stars is larger and occurs farther out, the HZ for K and M stars is smaller and situated farther in than around our Sun. High ultraviolet fluxes could be a problem for life around F stars. For stars later than about K5 spectral class the HZ falls within the 4.6-Gyr tidal locking radius of the star. The planetary companions may rotate synchronous with their orbit. The resulting temperature difference could generate winds that efficiently reduce the temperature gradient on the planet if the atmosphere were thick enough. A thick $CO_2$ atmosphere would very likely cover bio-signatures like $O_3$, what makes M stars a secondary target.

Darwin is a mission aimed at the search for, and the study of terrestrial exo-planets orbiting within the Habitable Zone (HZ) around nearby (< 25pc) stars.

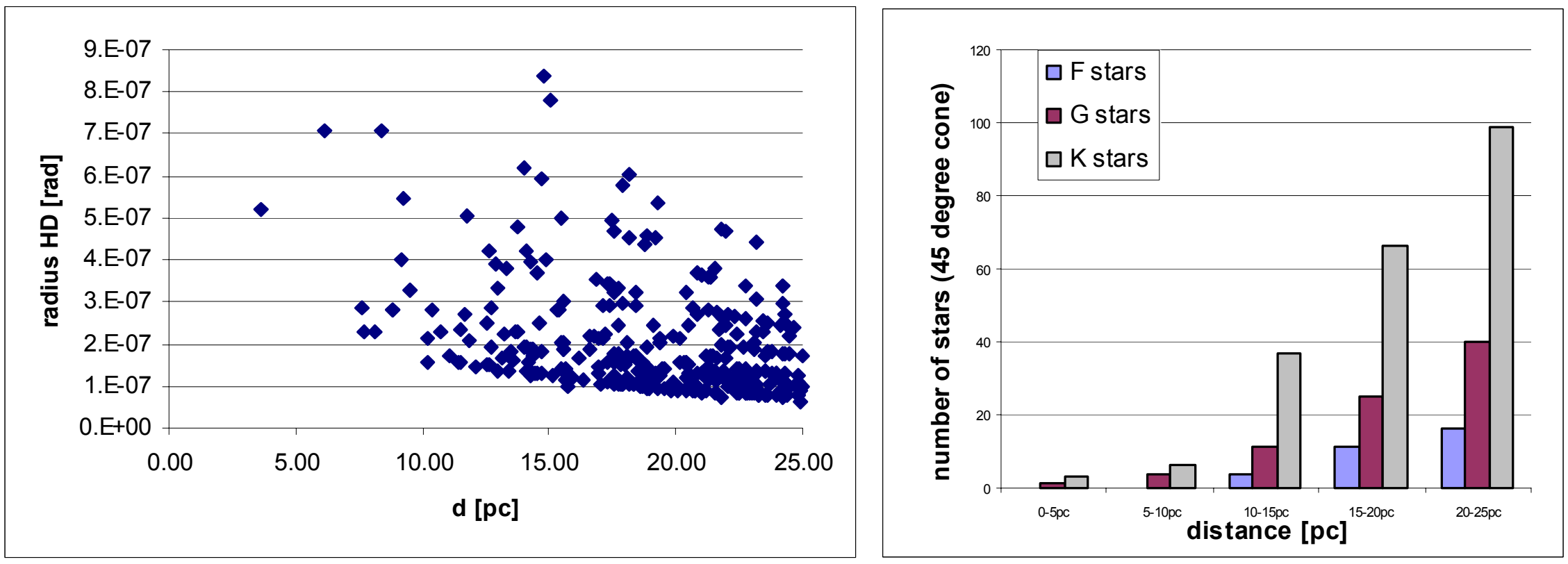

**Fig. 99: Habitable distance around the prime target stars of the DARWIN mission in rad (left), distribution of the stars versus distance (right)**

# 8 Detection methods

*"Small and cool, planets are poor emitters of radiation, combined with an annoying tendency to be expected very close to a luminous star (Provencal, 1997)"*

A planet orbiting a star with the brightness $L_*$, has a brightness $L_p$ due to the reflected light.

$$L_p = \frac{AL_*}{8}\left(\frac{R_p D}{a_p}\right)^2 \phi(t) \tag{8.1}$$

*D* gives the distance from our solar system, $P_p$ is the orbital period, $\phi(t)$ is an orbital phase factor given by

$$\phi(t) = 1 - \sin(i)\sin\left(\frac{2\pi \cdot t}{P_p}\right) \tag{8.2}$$

where *i* is the inclination of the orbit with respect to the sky plane. An orbit viewed edge on corresponds to *i* = 90 degrees. When a companion is too faint or too close to its bright parent star we rely on indirect methods to detect it, using the reflex motion of the star due to gravitational interaction – so far. Space based interferometry missions like DARWIN will provide information on the planets by destructively interfering the light of the host star and collecting the photons from the planet itself. So far we rely on Radial velocity search to interfere the existence of a planet from the movement of its host star. When a planet makes a revolution around its parent star, both objects are in orbit around their common center of mass. As a consequence the star makes a small circular orbit with radius $a_p$.

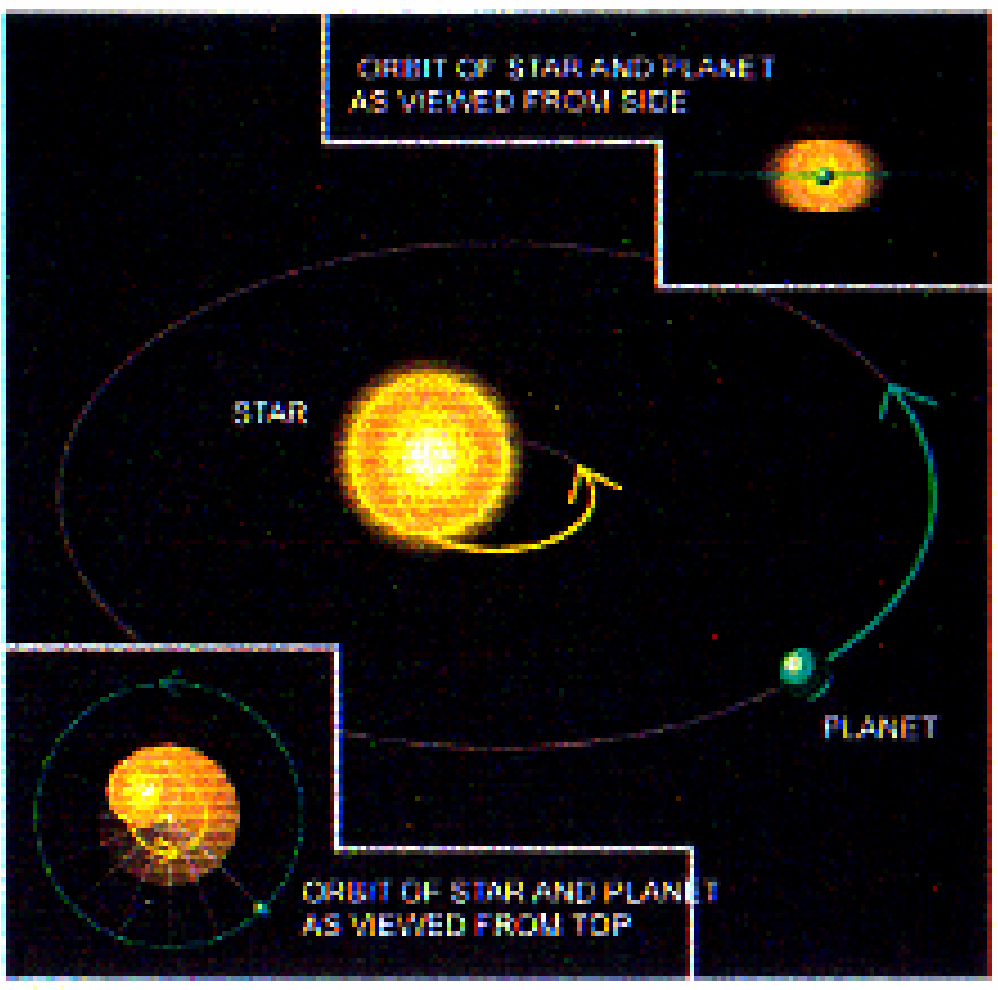

**Fig. 100: Basis of Radial velocity search, detecting the movement of the planet's host star, as both objects are in orbit around their common center of mass (Marcy 1996).**

A planet orbiting around its parent star influences its motion see Fig. 100.

$$a_* = a_p\left(\frac{M_p}{M_*}\right) \tag{8.3}$$

Its period is equal to the planet period. Resulting, the perturbation of three observables can be measured (Schneider, 1996), the radial velocity $\delta v_{r*}$.

$$\delta v_{r*} = \frac{2\pi a_*}{P_p} \tag{8.4}$$

the angular position $\alpha_*$

$$\alpha_* = \frac{a}{D} \tag{8.5}$$

and the time of arrival (TOA) of the signals $T_*$:

$$\delta T_* = \frac{a_*}{c} \tag{8.6}$$

Searching techniques look for effects of a planet on

- the motion (radial velocity)
- the position (astrometry)
- the apparent brightness (occultation) of its parent star or
- the apparent brightness of random background sources (gravitational microlensing).
- the TOA of the signals (pulsar timing)
- direct imaging (Point Spread Function Subtraction, interferometry, coronography)
- indirect evidence such as gaps in circumstellar disks

For an overview of the different detection methods see e.g. Schneider (Schneider, 1996), Kaltenegger (Kaltenegger, 1999).

## 8.1 Direct imaging

A star seen with a telescope makes a diffraction peak with an angular radius of

$$\theta = \frac{1.22 \cdot \lambda}{d} \tag{8.1.1}$$

where $d$ denotes the diameter of the telescope. The faint signal of planets is immersed in the photon noise of the wings of the diffraction spot (Hinz, 1998). To eliminate the effect of the diffraction peak the surface of the telescope secondary mirror can be modified or a nulling interferometer array can be used. The starlight coming from different parts of the mirror or different telescopes of the interferometer interferes destructive - but not the planet's signal. Piezoelectric actuators can obtain several $d$ at different points of the mirror. Interferometers use introduced phaseshifts in different arms to provide the nulling condition.

## 8.2 DIRECT IMAGING

The small difference of intensity between stars and their faint environment prevents the detection of small companions or low gas levels by direct imaging technique from the ground. A faint companion is difficult to detect against a high stellar background, as the image of a planet is difficult to distinguish from fluctuations in the stellar background that are comparable in angular size to the planet. The competing PSF of the star makes detection difficult. In the visible, Jupiter's luminosity is about $10^{-9}$ that of its host star. As a benchmark: a solar type star located 10pc away would have a magnitude of V=5 while a Jupiter at 5AU would have a V=27, separated by 0.5arcsec (Marcy, 1999). In the thermal IR from 20 -100 μm, the contrast is about $10^{-6}$. This improvement to the visible is due to the reduced stellar flux and the thermal radiation from the gravitational contraction of a giant planet.

PSF wings are caused by seeing, micro-roughness of the mirror, and diffraction. The detection problem does not lie in the peculiarities of the diffraction pattern caused by spider supports or other optics, since that will be nearly constant in time or mirror micro-roughness that will also be traceable. The atmospheric seeing for observations from the ground and the extrasolar zodiacal and local zodiacal light for ground as well as space based observations are the biggest challenge in detecting extra-solar planets (Marcy, 1999). Ground based telescopes with very high-resolution adaptive optics (AO) have the potential to detect and confirm EGP. Direct detection imaging through reflected light near 1 micron should allow detections within an orbital radius of 0.3-2 arcsec. Although BDs and hot EGP could be detected by an AO system from the ground, terrestrial planetary companions that are a million times fainter in the K band need a suppression of the stellar halo by a factor of 1 million relative to the peak in the Airy pattern and are therefore not detectable by AO systems (Langlois, 1998).

### 8.2.1 "PSF subtraction"

At young ages giant planets and BDs are relatively hot and luminous and radiate most of their gravitational energy at infrared wavelengths. A new method to detect substellar objects from the ground through direct imaging was developed at the IAC (Rebolo, 1998) (Kaltenegger, 1999) (Rebolo, 2003). Ground based direct imaging search focuses on young EGP and BD. The higher luminosity of young objects provides a higher probability to detect them. Companions discovered by imaging techniques can be further investigated by spectroscopy that allows obtaining information on its atmospheric conditions and evolutionary status. Young, cool dwarf stars are ideal search targets for sub-stellar mass companions using direct imaging technique. The basic idea is to suppress the starlight by rotating the recorded image of the star around the centre of the star and subtract the original from the rotated image.

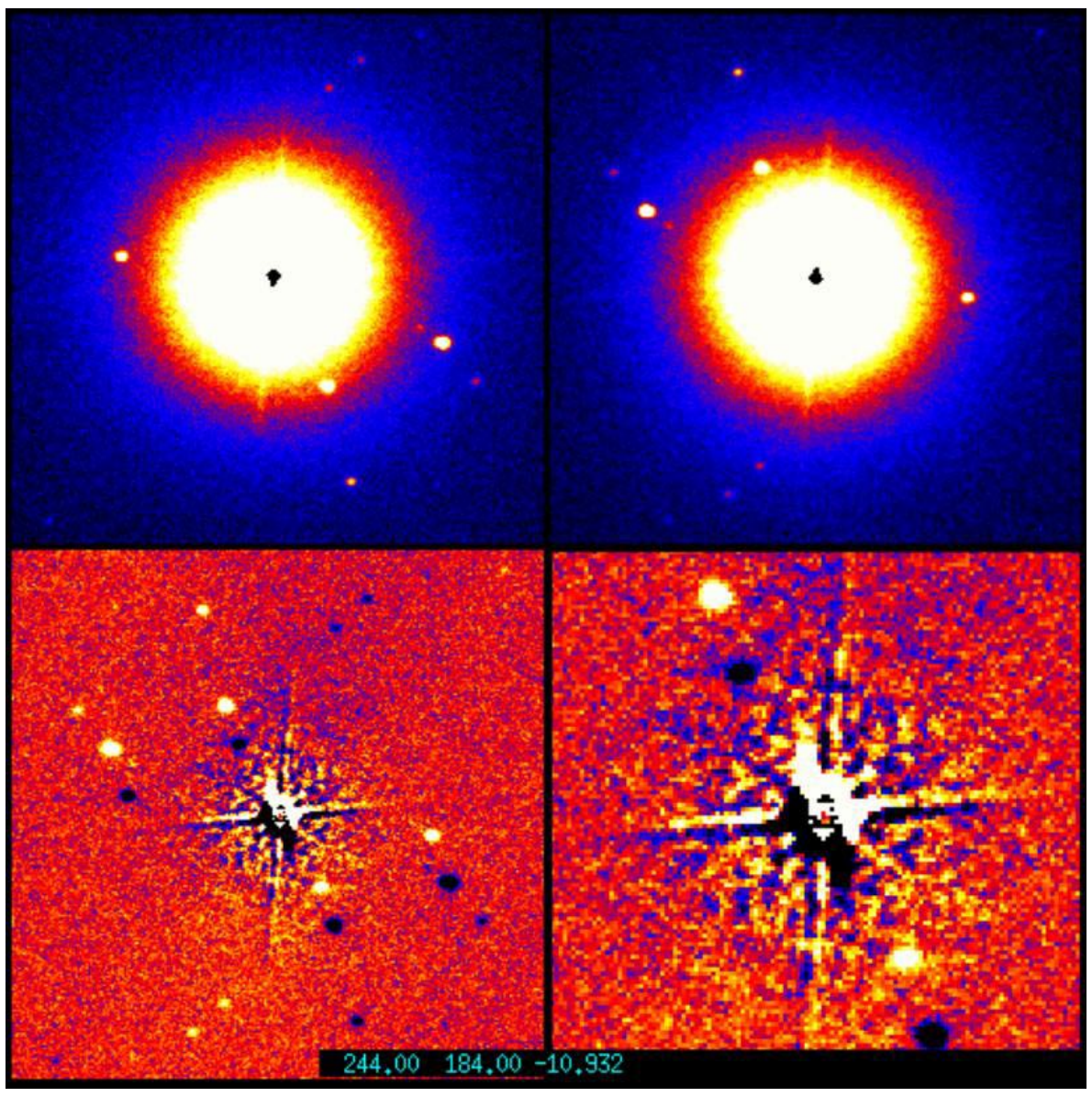

**Fig. 101: PSF subtracted image of a star showing close by background sources (Kaltenegger, 1999)**

Theoretically the starlight should cancel out completely. Only the residual flux of the planetary companion or surrounding stars should be visible. Due to the asymmetric form of the PSF the starlight does not cancel out completely. By modelling the PSF of the telescope it should be possible to detect objects even fainter and closer to the host star. Recent results by Rebolo et al. (Rebolo, 2004) using this method include the detection of a dim isolated object. Models predict it to be Jupiter sized. Adaptive optics and the use of bigger telescope should allow even higher sensitivity to detect substellar companions.

## 8.3 Interferometer

The direct detection of a planet close to its parent star is challenging because the signal detected from the parent star is between $10^9$ and $10^6$ times brighter than the signal of a planet in the visual and IR respectively. Future space based missions like DARWIN and the IR concept of TPF concentrate on the region between 6μm to 20μm, a region that contain the $CO_2$, $H_2O$, $O_3$ spectral features of the atmosphere. The presence or absence of these spectral features would indicate similarities or differences with the atmosphere of telluric planets.

### 8.3.1 Nulling Interferometry

A nulling interferometer interferes light from an on-axis source destructively. The basic concept is to sample the incoming wavefront from the star and its planet with several telescopes that individually do not resolve the system. Nulling interferometry differs from 'normal' imaging interferometry (Michelson interferometry) in that one attempts to obtain a dark fringe (or fringe pattern) in the centre of the field, by introducing phase shifts in the light paths of one or several of the interferometer arms. By keeping a star in the centre of the image plane, coronography is realized without the presence of a physical mask. In its simplest form, a two-telescope nulling interferometer introduces a 180-degree phase-shift between the two apertures, resulting in destructive interference along the line of sight. At the same time, light at small off-axis angles from the line of sight will experience constructive interference, thus allowing a faint object close to a bright star to be discernible. The observable area around the central null depends on the separation of the telescopes – the baseline. The star can be centered on the deep central null. Specific off-axis locations like the position of an Earth-like planet orbiting the host star can be constructively interfered by adjusting the distances between telescopes. The interferometer can be rotated around the observed star-interferometer axis or the signal can be chopped to modulate the intensity signal of a companion by a known periodic function. Such a technique can differentiate a companion from a symmetric zodiacal cloud because of different modulation of their respective signals. Nulling Interferometry was originally proposed by Bracewell and McPhie in 1979 (Bracewell, 1979).

A nulling interferometer can best be described by considering two telescopes (called a Bracewell interferometer). The beams coming from each of the telescopes are parallel and restricted to the diameter of the point-spread function of each individual telescope. By introducing a phase shift of $\varphi = \pi$ in one of the ray paths, we will achieve destructive interference on the optical axis of the system in the combined beam. At the same time, we can arrange the baseline of the two telescopes so that there is constructive interference a small angle $\theta$ off the optical axis. This angle $\theta$ depends on the distance between the two input telescopes.

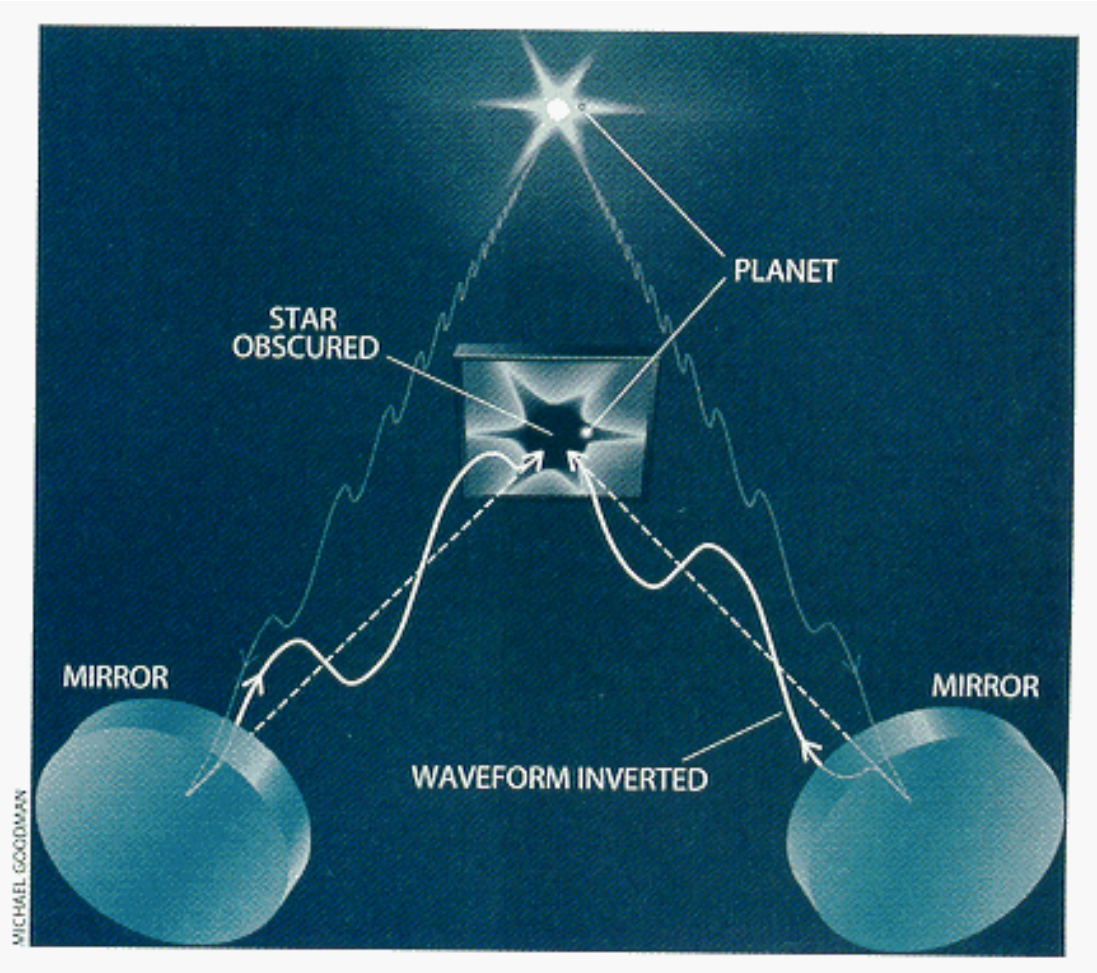

**Fig. 102: Concept of a nulling interferometer (Angel, 1996)**

The output of the system can be described by an angular transmission map (TM) featuring interference fringes, with a sharp null (destructive interfered area) in the center of the map. The stellar signal is nulled out only on the optical axis. A leakage of photons out of the central null exists because the star has a finite photospheric disk. That leakage is a very important additional noise source. No spatial information is extracted in a single exposure. Information about the distribution of planets in the target system can be recovered by modulation of that signal like rotating the interferometer around its axis. Rotation modulates the interferometer output intensity as a planet passes in and out of the dark fringes. From the intensity and actual pattern of this modulation one can derive the planet's parameters.

The use of more telescopes achieves a symmetric pattern around the star, with a deep central null placed on the star's position. The actual shape and transmission properties of the pattern are a function of the number of telescopes, configuration, and the distance between the telescopes, see section 10. In the Darwin baseline concept, 6 apertures of 1.5m diameter are used in different combinations (sub-interferometers – typically in groups of 2-3 or 3-4 sub-apertures) producing a pattern on the sky with transmission peaks and valleys in an essentially ring-formed pattern surrounding a deep 'null' in the center (Absil, 2000). In this case, the pattern is then modulated by switching between the different sub apertures, by rotating the array, and by expanding the array. This allows searching for planets in pre-determined areas surrounding a star, the calculated habitable zone. The output beam comprises all information from a specific set of sources on the sky (consisting of star, planet(s) and zodiacal dust surrounding the target star), as well as the background, that is expected to be dominated by local zodiacal dust emission. The current baseline consists of free-flying telescope units in an L2 orbit. The baseline mission duration is 5 years, extendable to 10 years.

In an ideal scenario with only a single planet around its host star and no other disturbing sources, such as extrasolar zodiacal dust in the target system, the detection of a positive flux would imply that a planet is present, if the star is well and truly `nulled out'. In real observations, several factors affect the signal to noise in a detrimental way. This has been analyzed e.g. by Beichman & Velusamy (Beichmann, 1999) and Kaltenegger and Karlsson (Kaltenegger, 2004B). There is a significant amount of background radiation coming from dust in our solar system and the unknown target system. The zodiacal dust temperature within the HZ will also be close to 300K and thus the peak of the emission is radiated around 10µm. The zodiacal dust levels in our own Solar system would outshine a terrestrial planet by about a factor of 400 at these wavelengths (Legér, 1996). An extrasolar planet thus will most likely be embedded in a very bright background. In order to separate out the signal of the planet from this background one needs to modulate the signal from the planet and that from the zodiacal dust at different frequencies. The planetary signal is separated out from that of any extrasolar zodiacal dust because of its temporal and spectroscopic behavior (Bracewell, 1979), (Leger, 1996) (Menesson, 1997) (Absil, 2000) due to modulation.

Transmission of the telescope and optics, detector efficiency, sky background, solar system zodiacal light background, imperfect nulling on the on-axis central source, thermal emission of telescope and optics, straylight in the optical system and detector noise are relevant factors for the performance. Using current optical technology, the required rejection rates of $10^5$ – $10^6$, cannot be reached unless the interfering wave fronts are filtered. Two methods of achieving this are pinhole filtering (Ollivier, 1997) and the use of single-mode wave guides (Mennesson, 2000) limiting the FoV of the Interferometer. Optical quality is critical in nulling interferometers designed for extrasolar planet detection. Effects like amplitude mismatching and phase defects will limit the rejection rate of the stellar photons.

The first practical demonstration of nulling was undertaken in February 1998 (Hinz, 1998). Using the Multiple Mirror telescope on Mount Hopkins, Arizona, Angel and his team were able to cancel out the image of a star: α-Orionis. The test was made using two of the 1.8m Cassegrain telescopes of the Multiple Mirror Telescope, separated by 5m. As long as the pathlength difference is larger than the coherence length (26µm for 10% bandwidth) the flux of the star is quite steady, but it blinks on and off once the pathlength are close to equal. This effect is caused by atmospheric turbulence, which includes pathlength changes of about 5µm, enough to shift the interference randomly between constructive and destructive states. The ability of the interferometer to suppress the entire Airy pattern was shown in that performance. The nulled image had a peak in intensity of 4.0% and the total integrated flux of 6.0% of the constructive image. The small residue is due to the high order aberrations in the wavefront that do not cancel out. The residual flux was mainly caused by atmospheric aberrations (Hinz, 1998). Fig. 103 demonstrates that dust around a star can be detected using nulling interferometry as the stellar image cancels out but the residual flux from the dust is measured as seen for HD100546. To improve the quality of the null from the ground atmospheric aberration it must be corrected by adaptive optics.

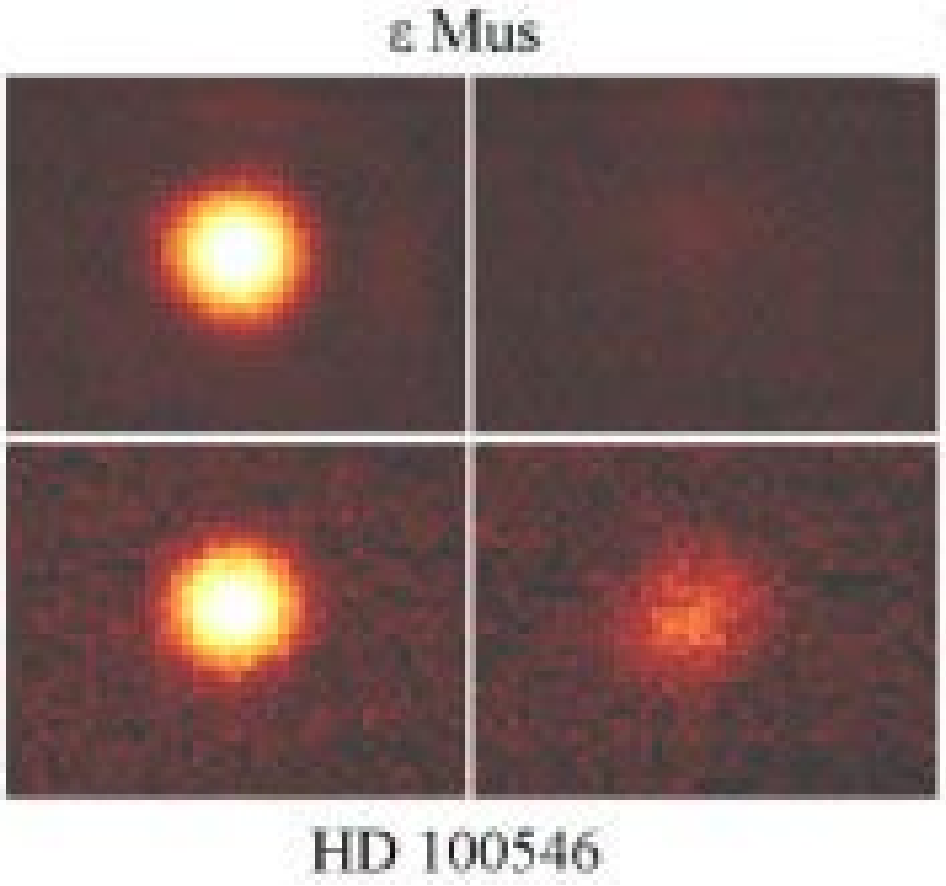

**Fig. 103: Nulling demonstration from the ground for a star with and without a strong dust disk (Hinz, 1998)**

## 8.4 Ground based versus space based

In the MOFFIT study (ESA SCI(96)7) two 6-element interferometers were compared (Fridlund, 2004). One located on the ground consisted of 8m units equipped with adaptive optics – thus more powerful than any system currently under construction. The space based array utilized 1m aperture units. At wavelengths shorter than 2µm the ground-based interferometer is superior because of larger collecting area and a better filling of the uv-plane. At all longer wavelengths, the space-based system will be significantly more sensitive, the difference being about 4 magnitudes. This can be translated into observing time; the space interferometer is about 40 times faster for bright targets (such as those expected in the visible), while for objects in the near and thermal infrared, the space-based system is 1600 times faster. Using the DARWIN telescope, as an example, the integration time to detect the Earth at interstellar distances (5pc – 10pc) would be about 10 – 30 hours, depending on the actual distance. Using the ground based system assumed to be larger than the VLTI; it would take 1600 times longer which is equivalent to several years per object. This is also assuming that the ground-based array observes 24 hours a day and is kept stable during the whole observing time.

# 9 Interferometry

One has to distinguish two different techniques of nulling interferometry, e.g. (Traub, 1999):

- Pupil-plane, or Michelson interferometry, which uses half-transparent mirrors for the combination of on-axis beams.
- Image-plane, or Fizeau interferometry, can also be applied to combine more than two beams simultaneously in the focal plane.

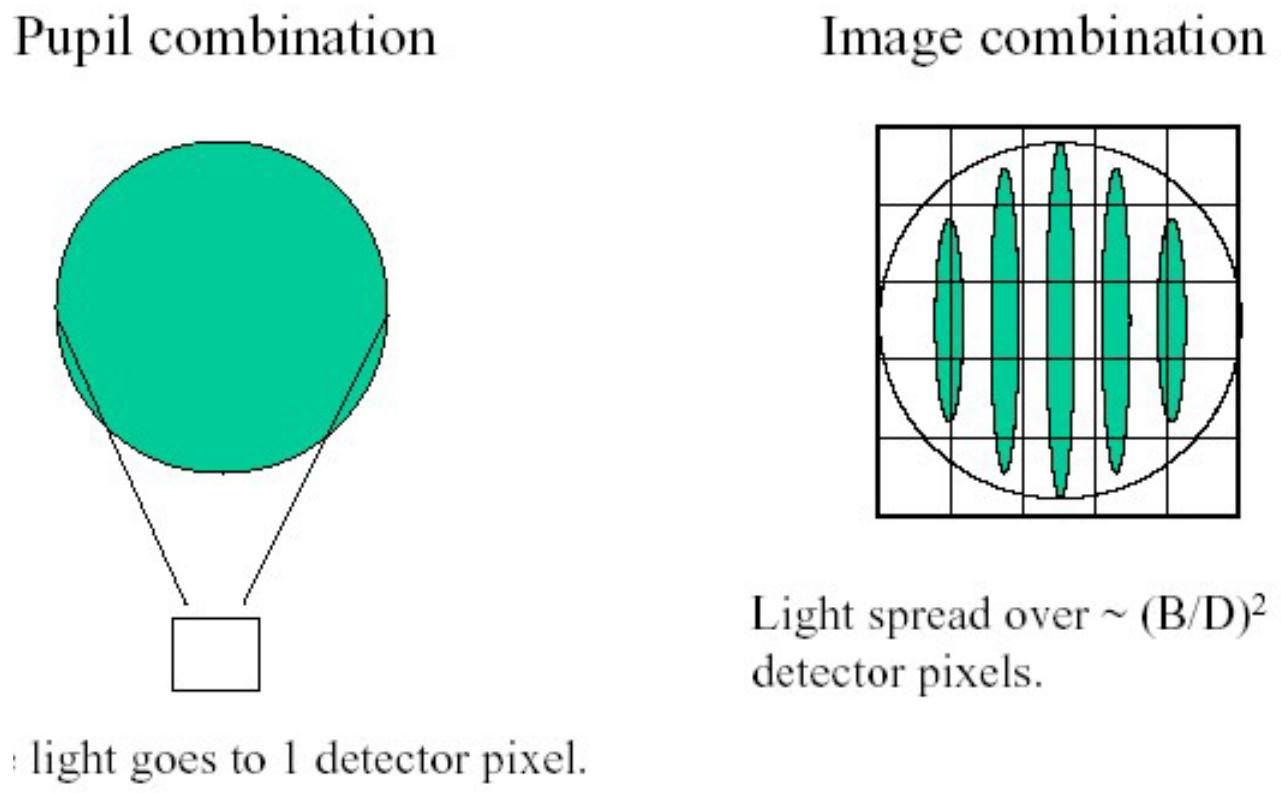

**Fig. 104: Output signal of a pupil-plane (left) and image-plane (right) combination (Swain, 2003)**

## 9.1 Pupil plane interferometry

In absence of optical aberrations, the overall complex amplitude $A(\vec{\theta},\vec{r},\lambda)$ received at the circular output pupil from a point source is given by equation (9.7). It is proportional to the interferometric response, R ($\vec{\theta}$,λ). The response of the interferometer is scaled to the maximal achievable value, that is the maximum response value in the generate transmission map (TM), $TM_{max}$ (given by coherent addition of all waves) see equation (9.9). Note that if x TM maps are used for modulation of the detected signal (see section on modulation) the maximum achievable intensity in the modulation map is given by x $TM_{max}$. Normally two TM are used to modulate a signal.

## 9.2 Basic equations for an interferometer array

The interferometer configurations consist of $N$ telescopes with relative amplitude $A_k$ that are located in a common plane, perpendicular to the line of sight. Their position is determined by the polar coordinates $L_k$ and $\delta_k$ in a reference frame of arbitrary origin *0*, used as a common phase reference (Absil, 2000) (Mennesson, 1999). $\vec{r}$ is the position vector on the circular output pupil with outer radius $R$, $\Pi(r/R)$ is a rectangular box function that is 1 if $r \in [0,1]$ for $r \langle 0$. A detectable point like source in the sky, like a planet, is located trough its angular offset $\theta$ with respect to the line of sight and its azimuth $\phi$. ($\theta$, $\phi$) are the apparent source coordinates after projection on the sky plane denoted as $\vec{\theta}$ .

The beams coming from the $N$ telescopes are recombined in a common output pupil. An additional internal phase shift $\phi_k(\lambda)$ is applied to each of them prior to recombination by phase shifters. In the most general case, they depend on wavelength, although there is a strong effort to develop achromatic phase shifters. In absence of optical aberrations, the overall complex amplitude $A(\vec{\theta},\vec{r},\lambda)$ received at the circular output pupil of diameter $R$ from a point source is given by equation (9.1).

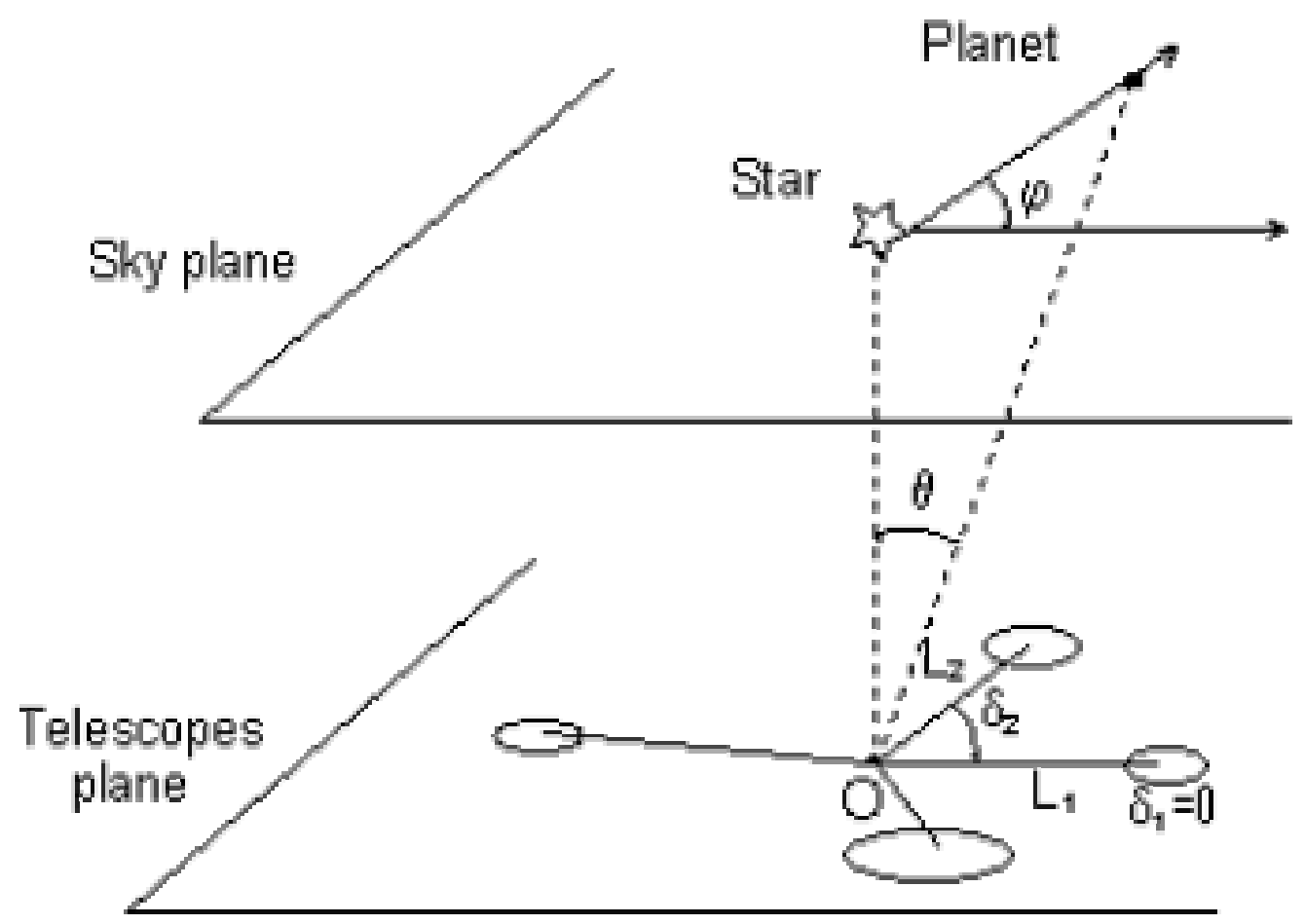

**Fig. 105: Overview of the coordinates used in the calculations**

$$A(\vec{\theta},\vec{r},\lambda) = \prod(r/R)\left[\sum_{k=1}^{N} A_k e^{i\phi_k(\lambda)}\right] \quad (9.1)$$

$$\begin{aligned} &\prod(r/R) = 1 \quad if\; r \in [0,R] \\ &\prod(r/R) = 0 \quad if\; r \notin [0,R] \\ &r \geq 0 \end{aligned} \quad (9.2)$$

where $\vec{r}$ is the position vector on the circular output pupil with outer radius R, the complex amplitudes $A_k$ are proportional to the telescope diameter $D_k$, provided that the same transmission affects all $k$ beams. Taking the external delays $\vec{L}_k\vec{\theta}$ in respect to the origin $O$ into account, equation (9.3) gives the overall complex amplitude received.

$$A(\vec{\theta},\vec{r},\lambda) = \prod(r/R)e^{i2\pi\vec{r}\vec{\theta}/\lambda}\left[\sum_{k=1}^{N} A_k e^{i2\pi(L_k\theta/\lambda)\cos(\delta_k-\phi)} e^{i\phi_k(\lambda)}\right] \quad (9.3)$$

In the focal plane of angular coordinate $\vec{\theta}_f$, the complex amplitude distribution is obtained by Fourier transform. Assuming a pupil magnification of 1 for simplification that leads to equation (9.4).

$$\tilde{A}(\vec{\theta},\vec{\theta}_f,\lambda) \propto \frac{2J_1(\pi D\theta/\lambda)}{\pi D\theta/\lambda} \otimes \delta(\vec{\theta}-\vec{\theta}_f)\cdot\left[\sum_{k=1}^{N} A_k e^{i2\pi(L_k\theta/\lambda)\cos(\delta_k-\phi)} e^{i\phi_k(\lambda)}\right] \quad (9.4)$$

$$I(\vec{\theta},\vec{\theta}_f,\lambda) = \left|\tilde{A}(\vec{\theta},\vec{\theta}_f,\lambda)\right|^2 \quad (9.5)$$

The intensity distribution (9.6) in the focal plane is a product of a classical diffraction term and a constant interferometer term that only depends on the aperture configuration.

$$I(\vec{\theta},\vec{\theta}_f,\lambda) \propto \left[\frac{2J_1(\pi D\theta/\lambda)}{\pi D\theta/\lambda}\cdot\delta(\vec{\theta}-\vec{\theta}_f)\right]^2 \cdot \left[\left|\sum_{k=1}^{N} A_k e^{i2\pi(L_k\theta/\lambda)\cos(\delta_k-\phi)} e^{i\phi_k(\lambda)}\right|^2\right] \quad (9.6)$$

The monochromatic intensity distribution in the focal plane is the product of two terms.

- The first term is the classical diffraction pattern from each telescope, constant over the whole focal plane and only depending on the point source coordinate $\vec{\theta}$, the entrance pupil geometry and phase shift $\phi_k(\lambda)$.
- The second term is the interferometric response $R(\vec{\theta},\lambda)$, often refereed to as the transmission map of the interferometer. It acts as transmission grit on the sky plane.

Using a monopixel detector and neglecting diffraction terms, the energy signal obtained is proportional to the interferometer response (9.9).

$$\overrightarrow{A_g}(\vec{\theta},\vec{r},\lambda) = \prod(r/R)e^{i2\pi\vec{r}\vec{\theta}/\lambda}\left[\sum_{k=1}^{N} A_k e^{i2\pi(L_k\theta/\lambda)\cos(\delta_k-\phi)} e^{i\phi_k(\lambda)}\right] \quad (9.7)$$

$$\vec{A}(\vec{\theta},\vec{r},\lambda) = \sum_{k=1}^{N} A_k e^{i2\pi(L_k\theta/\lambda)\cos(\delta_k-\phi)} e^{i\phi_k(\lambda)} \quad (9.8)$$

$$R(\vec{\theta},\lambda) = \left|\sum_{k=1}^{N} A_k e^{i2\pi(L_k\theta/\lambda)\cos(\delta_k-\phi)} e^{i\phi_k(\lambda)}\right|^2 \quad (9.9)$$

$$TM_{\max} = \left(\sum_{k=1}^{N} A_k\right)^2 \quad (9.10)$$

$A_k$ are the fraction of the amplitudes collected by the k-th telescope that will be used for beam combination. The factors $\alpha_k$ are determined by the configuration itself.

$$A_k = \alpha_k A_{in} \quad (9.11)$$

Due to energy-conservation one has to normalize the intensity by 1/N introducing a for-factor of 1/N in the intensity calculations.

$$I(\vec{\theta},\lambda) \propto \frac{1}{N} R(\vec{\theta},\lambda) = \frac{1}{N}\left|\sum_{k=1}^{N} A_k e^{i2\pi(L_k\theta/\lambda)\cos(\delta_k-\phi)} e^{i\phi_k(\lambda)}\right|^2 \quad (9.12)$$

## 9.3 Evaluation of a transmission map

The map of transmission of an interferometer array is a function of the coordinates of the source in the sky (θ, φ) in reference to the telescope plane and shows the response of an interferometer at a sky position. We optimize the baseline of the array so that the maximum of that response coincides with a planet's position. As the response is wavelength depended that is only possible for certain wavebands. The transmission map can be used as a criterion to select the geometry of the array configuration. The quality parameters of a transmission map are:

1) a deep and wide central null to reject the star light
2) a high average and peak modulation of a planetary signal, that influences the achievable SNR
3) center symmetry properties, that influence the background rejection using modulation techniques
4) a small rotation angle needed to evenly cover the interesting area of the FOV that influences the planet search time

5) the accuracy to determine a detected planet's position

The null close to the line of sight, in the central area of the transmission map, has to be deep and wide enough to prevent light from the star's disk from swamping the planet's signal. A high ratio between the planet's signal and the star's signal is needed to allow detection of a faint companion. Modulation between different sub-configurations is currently an integral step of the mission allowing detection of the planetary signal against the huge background from extended sources e.g. local (LZ) and extrasolar zodiacal clouds (EZ). Modulation between different sub-arrays allows subtracting the physical stellar leakage, which is only related to the angular size of the star and the size of the array. Note that instrumental leakage, associated to pointing errors, OPD errors, etc, is not the same in both outputs and thus won't be cancelled by modulation.

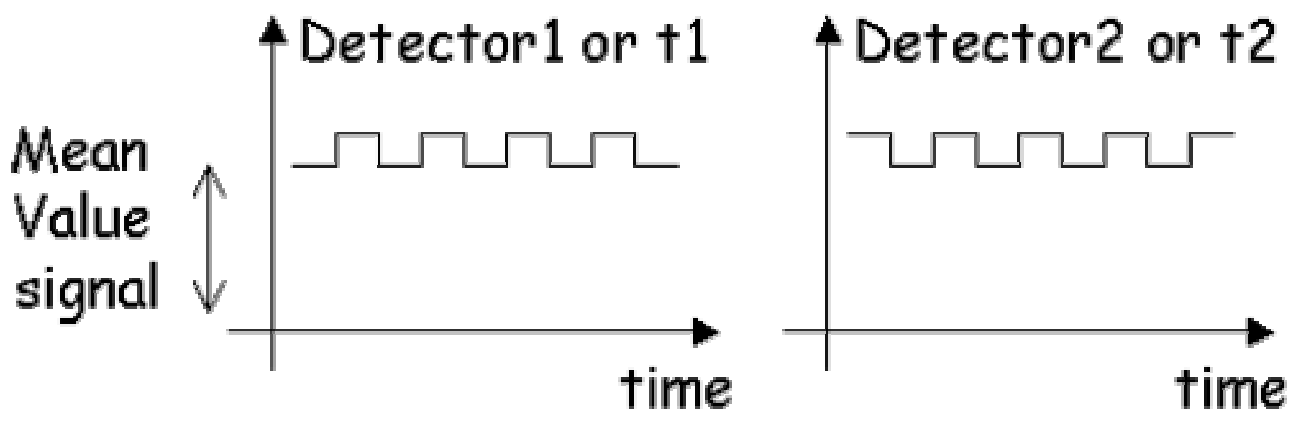

**Fig. 106: Modulated signal detected at t1 and t2 or detector 1 and detector2**

Asymmetry of a transmission map is important because it allows us to distinguish the planetary signal from the signal of a symmetric dust disk around the star by rotation or modulation of the transmission map. Extended sources with central symmetry have the same contribution during the whole rotation of the array or on the output of each sub-interferometer while a point source has a modulated signal because it alternatively crosses a bright or dark fringe in the transmission map during rotation or falls on areas of high and low transmission at the different transmission maps generated by each sub-interferometer.

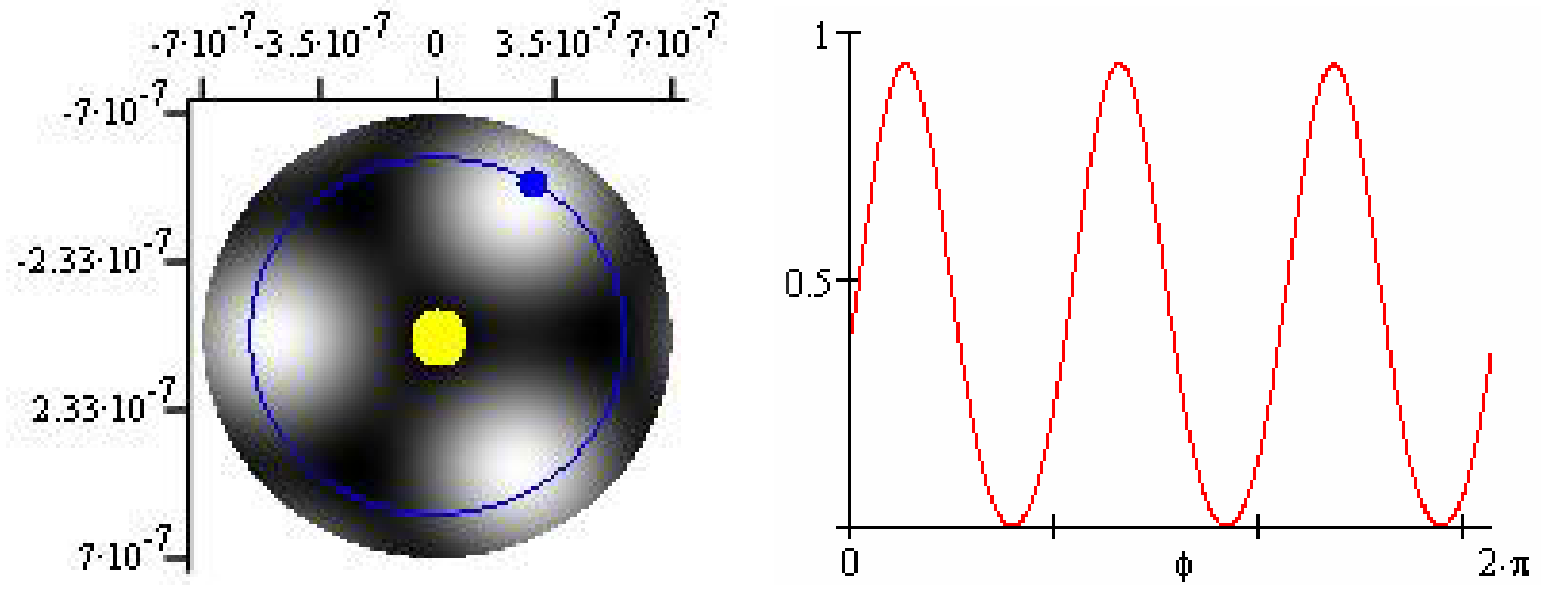

**Fig. 107: (left) Concept of modulation through rotation of the array: a point-like source's signal is modulated as it crosses low and high transmission areas, (right) detected signal of a calculated point-source moving along the circle shown on the left.**

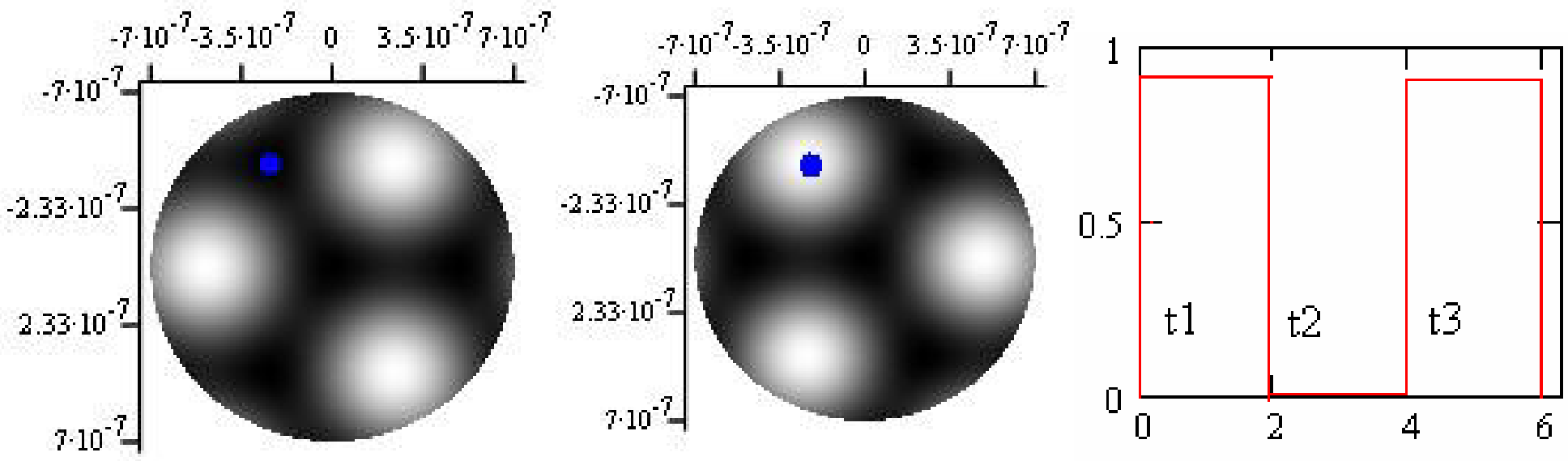

**Fig. 108: (left) Concept of internal or inherent modulation: the signal is modulated by generating different transmission maps so that a point-like source's signal once falls onto a low and once on a high transmission area (Absil, 2000), (right) detected signal of a calculated point-source shown on the left by switching from one modulation map to the other.**

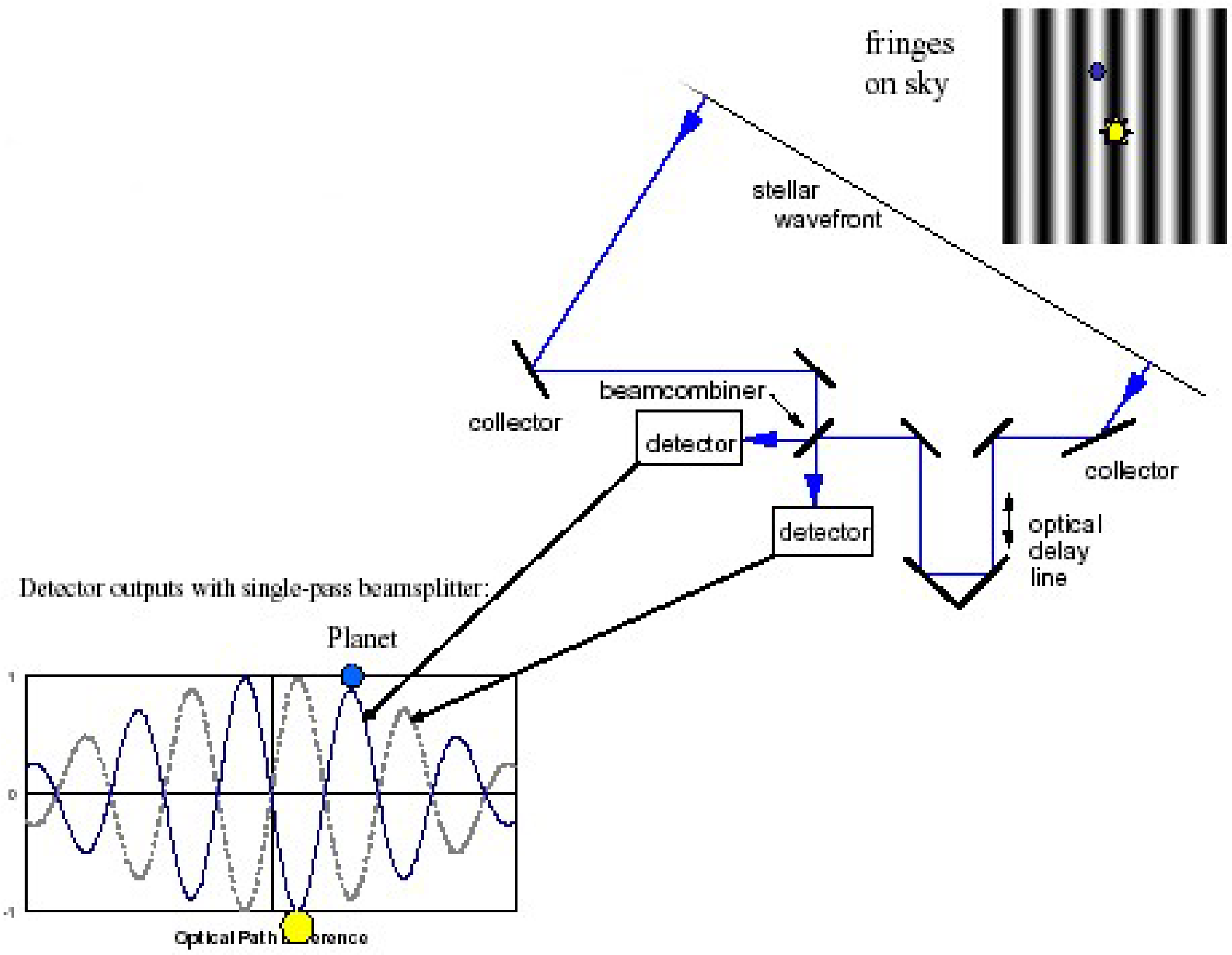

**Fig. 109: Overview of an interferometer nulling array and its response (polychromatic light) (Serabyn, 2003)**

The use of sub-interferometers, instead of using all telescopes as one single array, lowers the strength of the collected signal as well as the starlight rejection of the array as fewer telescopes are used simultaneously in each sub-interferometer generating a shallower central null shape. On the other hand it allows distinguishing the planetary signal from the signal of an extended symmetric source and additionally provides a better coverage of the orbit of the planet under the pattern where high modulation is possible in the modulation map. The maximum modulation value translates into the number of photons of a planet detected. If only 50% maximum modulation is possible only half the number of photons will arrive at the detector than if 100% planmod can be achieved. Currently the rotation of the array as well as inherent and internal modulation technique (Mariotti, 1997) (Absil, 2003) is discussed to modulate a planetary signal.

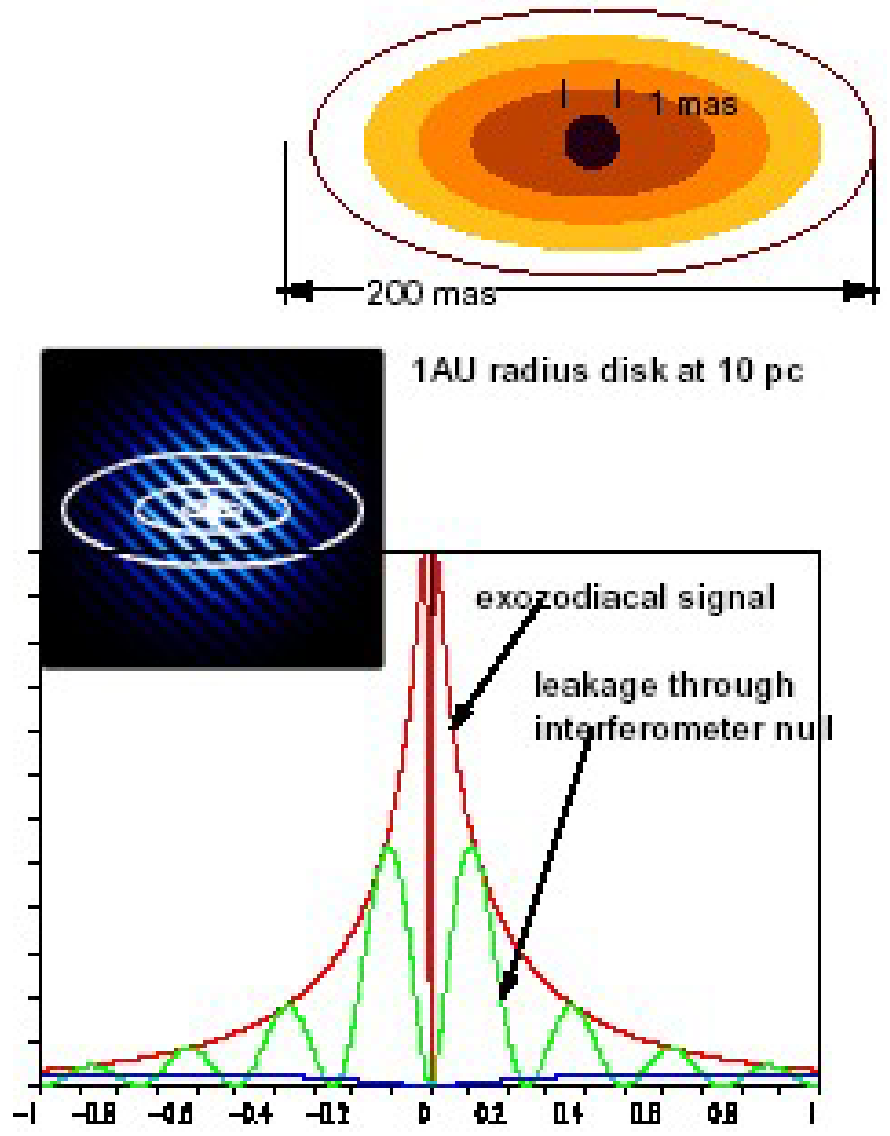

**Fig. 110: Leakage of the Extra-solar cloud signal through the null (Serabyn, 2003)**

## 9.4 Modulate a planetary signal.

### 9.4.1 Signal modulation

To distinguish a point like source from a spatially extended, circular symmetric source we have to look at the detected overall signal. If the generalized pupil function of each subnulling interferometer is real the signal is proportional to equation (9.4.1) (Absil, 2003).

$$S_j(t) \propto a + b\cos(\phi_{im}(t)) \tag{9.4.1}$$

When combining the nulled outputs of the subinterferometers on a beamsplitter one obtains two complementary outputs $S_j(\vec{\theta})$ and $S_k(\vec{\theta})$.

$$\begin{aligned} S_j(\vec{\theta}) &= \left|\frac{1}{2}\sqrt{R_k(\vec{\theta})}\right|^2 + \left|\frac{1}{2}\sqrt{R_j(\vec{\theta})}\right|^2 + \left|\sqrt{R_k(\vec{\theta})R_j(\vec{\theta})}\right| \cdot \\ &\left[\cos(Arg\sqrt{R_k(\vec{\theta})} - Arg\sqrt{R_j(\vec{\theta})})\cos(\phi_{im}) + \sin(Arg\sqrt{R_k(\vec{\theta})} - Arg\sqrt{R_j(\vec{\theta})})\sin(\phi_{im})\right] \end{aligned} \tag{9.4.2}$$

Only part of the signals is specific to the planet's modulation. If $\phi_{im}(t)$ can be switched from $\phi_0$ to $-\phi_0$, symmetric signals from extended sources are not modulated with time. The part of the modulated signal, specific to a planetary point-like source, is given by Equation (9.4.3).

$$\left|\sqrt{R_k(\vec{\theta})R_j(\vec{\theta})}\right| \cdot \left[\sin(Arg\sqrt{R_k(\vec{\theta})} - Arg\sqrt{R_j(\vec{\theta})})\right] \tag{9.4.3}$$

It shows that the maximum modulation efficiency for a point source signal is given when $\phi_{im}(t)$ is set alternatively to $\pm\pi/2$, while the signal of an extended source is not modulated. The modulation signal only depends on the planetary location and entrance pupil geometry.

### 9.4.2 Internal modulation

Mariotti and Mennesson (Mariotti, 1997) proposed internal modulation, a technique using fast signal multiplexing to isolate the planetary signal from the noise sources. The outputs of two sub-interferometers are recombined on a lossless beam combiner, which induce a $\pi/2$ phase shift between the input beams at the first output and a $-\pi/2$ phase shift at the second output. That adds an additional level of signal modulation to the expected planetary signal in the face of varying backgrounds and detector drifts. Furthermore the transmission maps associated with the two outputs $S_j$, $S_k$ have to be asymmetric, although symmetric to each other with respect to their centre. Internal modulation can be seen as a traditional chopping performed on ground based astronomical observations, with the additional requirement to keep the star nulled and on-axis.

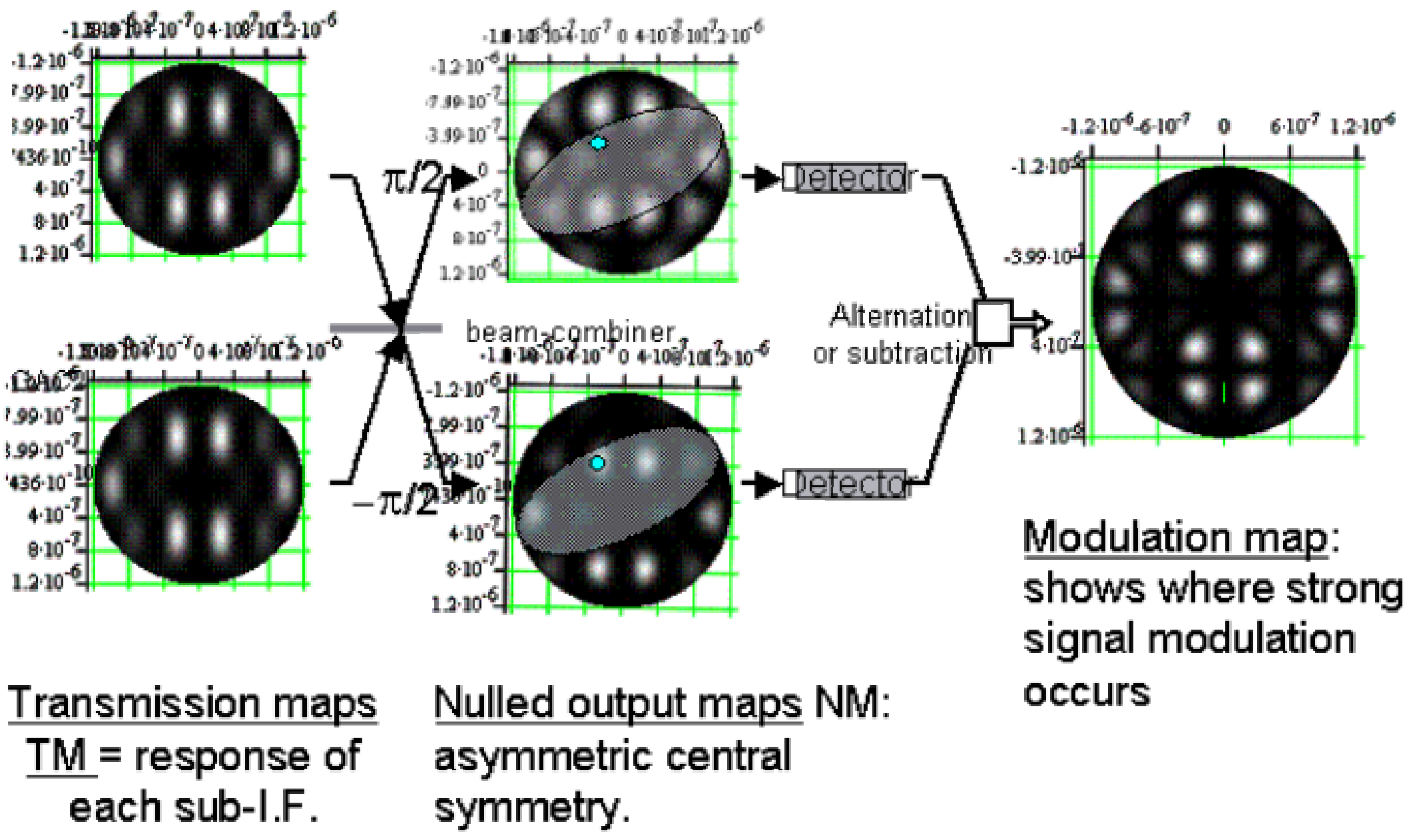

**Fig. 111: Internal modulation generates a centre-symmetric transmission map from each sub-interferometer using π/2 phasehifters, that result in two nulled outputs.**

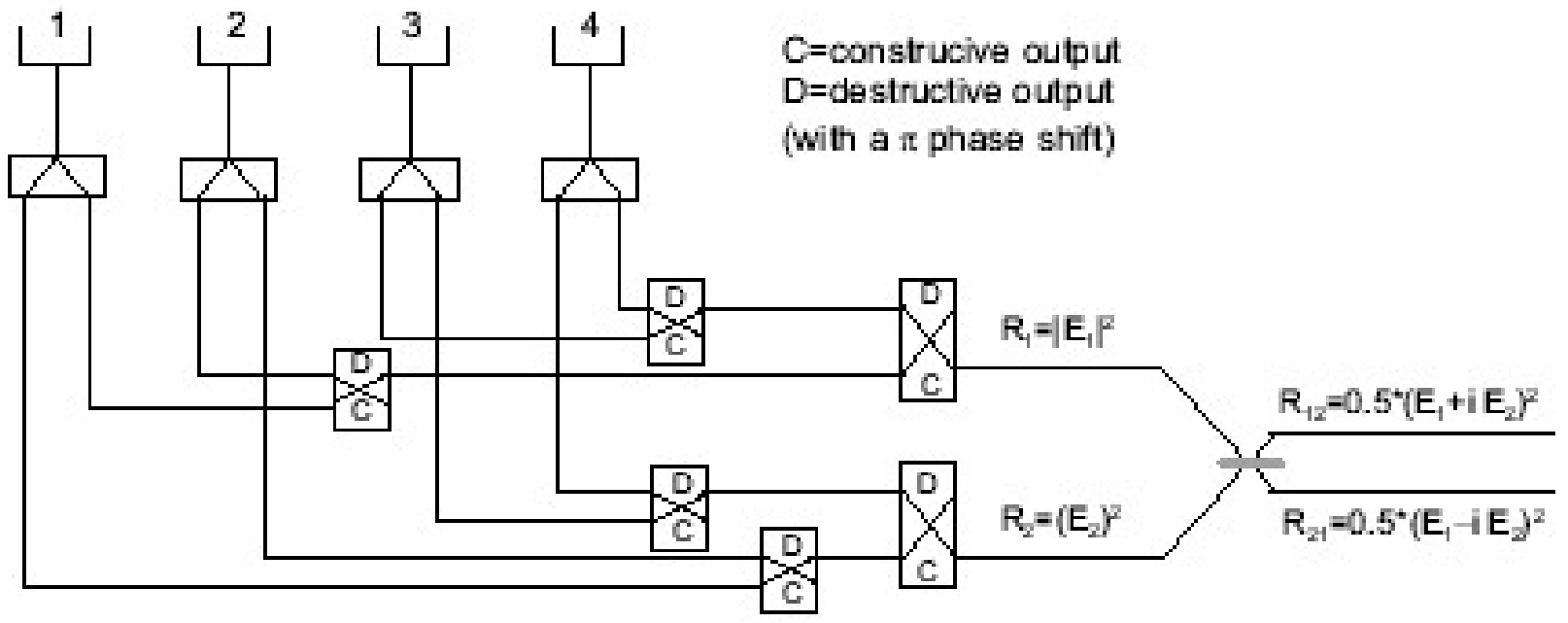

**Fig. 112: Beam combination scheme for 4 telescopes using internal modulation (Absil, 2003)**

### 9.4.3 Inherent modulation

Absil et al. (Absil, 2003) proposed the concept of inherent modulation. The concept combines a number of input beams with different applied phase shifts that are normally fractions of π, such that the transmission map is obtained directly. Additionally by combining the input beams with different phaseshifts, transposed transmission maps can be obtained, generating a modulation map (MM). One can also use the whole array to generate the nulled output NM1 at time t1, and generate NM2 at time t2 with a different set of phaseshifts, allowing the signal to be time mutiplexed between the two sub-interferometers. Thus one can use the whole array at time t1 as one sub-interferometer to create NM1 and at time t2 using a different set of phaseshifts to create NM2. Inherent modulation creates NM1 and NM2 by changing the sign of the achromatic phase shifts from $\phi_k$ to $-\phi_k$ what mimics a π rotation of the array (Absil, 2003).

In order to derive the sets of phase shifts and amplitudes for inherent modulation, starting from a known configuration, one calculates the contribution, in terms of amplitude and phase, from each input telescope to the output signal on the detector. The achromatic phase shifts, which are applied, are generally not π/2. Configurations using inherent modulation do not necessarily have a counterpart based on modulation between sub-interferometers, e.g. TTN (Karlsson, 2004).

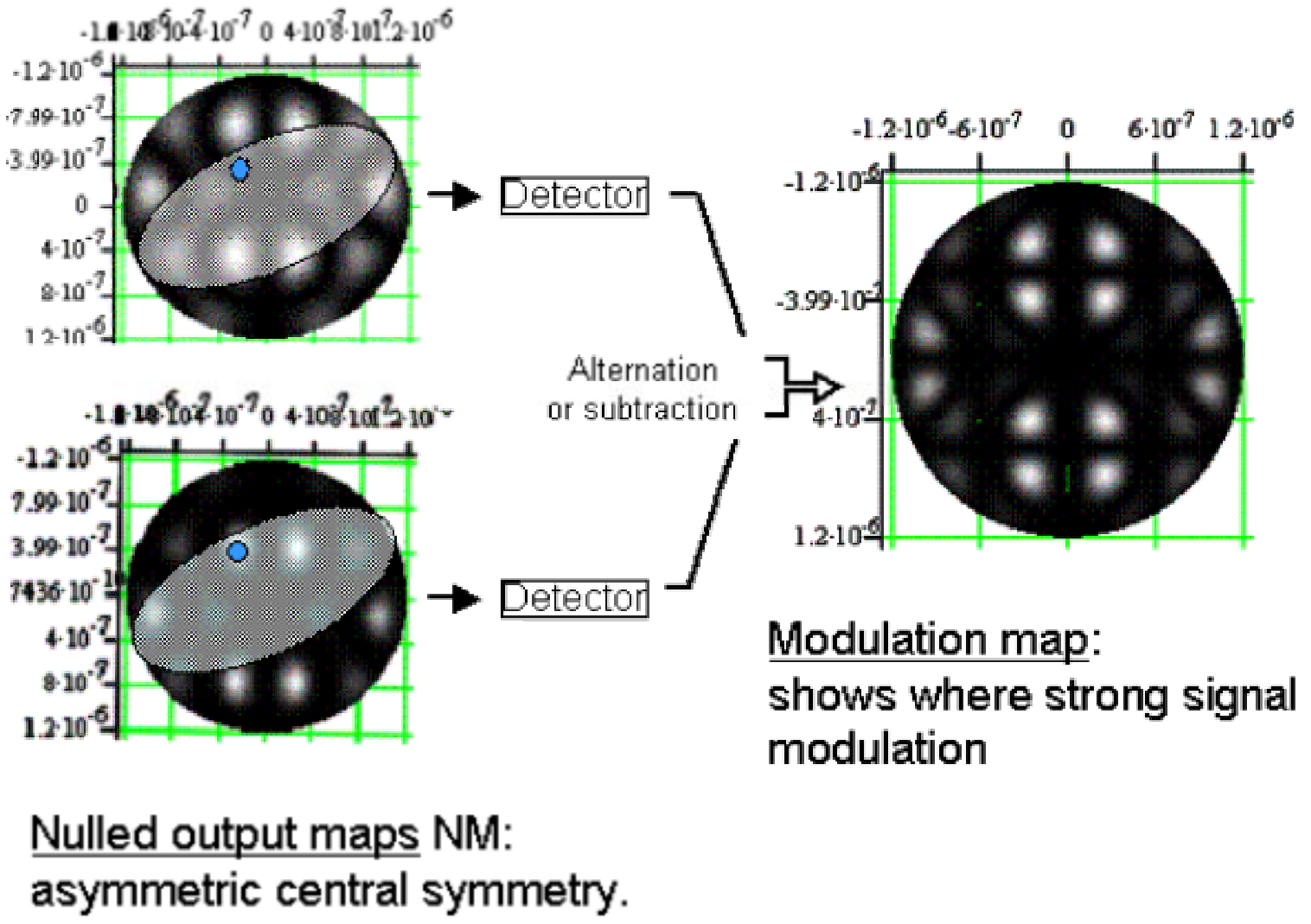

**Fig. 113: Inherent modulation uses two sets of phase shifts to the interferometer's input beams to generate two different centre-symmetric nulled outputs. Concept by (Mariotti, 1997) (Absil, 2003).**

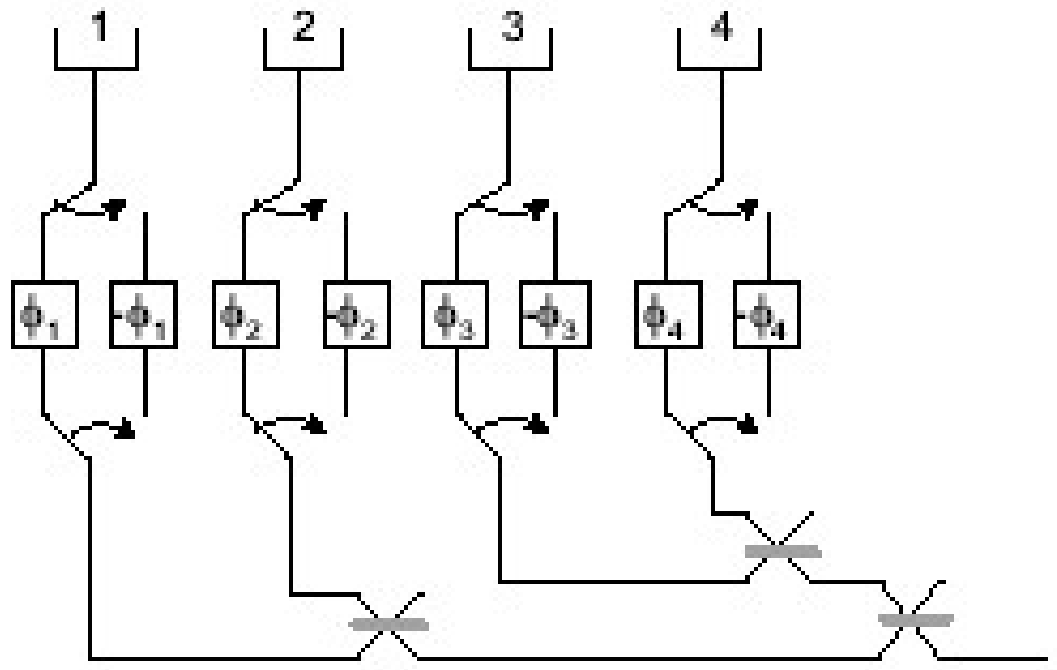

**Fig. 114: Beam combination scheme for 4 telescopes using inherent modulation (Absil, 2003)**

Fig. 114 shows a scenario where the whole array is used as sub-array 1 at time t1 and as sub-array 2 at time t2. The beams can also be split and transmitted to the two beam paths at the same time, generating the two nulled output maps at the same time.

## 9.5 Modulation of a planetary signal $\xi$ (planmod)

The collected signal is modulated to detect the planetary signal against the huge background from extended sources. Modulation between different sub-arrays allows subtracting the physical stellar leakage, which is only related to the angular size of the star and the size of the array. Note that instrumental leakage, associated to pointing errors, OPD errors, etc, is not the same in both outputs and thus won't be cancelled by modulation.

Modulation compares the value at the nulled outputs (NM) of the signal at position $(\vec{\theta})$ between two NMs. The NMs are generated at combination of the response of two sub-interferometers at the same (internal modulation, inherent modulation) or consecutive time (rotation, inherent modulation). Equation (10.1) defines the so generated modulation map.

$$M(\vec{\theta},\lambda) = |NM1(\vec{\theta},\lambda) - NM2(\vec{\theta},\lambda)|, \qquad (9.5.1)$$
$$0 \le NM(\vec{\theta},\lambda) \le 1$$
$$0 \le M(\vec{\theta},\lambda) \le 1$$

where *NM1* and *NM2* are two nulled outputs generated at combination of two different transmission maps (internal modulation) on a beam combiner or due to two different set of phase-shifter using inherent modulation. The two NMs are the response of two sub-interferometers depending on the number, size and arrangement of the telescopes. The maximum modulation value shows the ability of the array to modulate a point-source signal by switching between the NM. Depending on the symmetry of the NM the max modulation value can reach 100% at certain positions $(\vec{\theta})$. For more sub-interferometers equation (9.5.1.) can be generalized, even so for modulation only two maps are generally used. If there are more sub-interferometers used, there is generally no place in the sky for which all beams are in phase.

$$M(\vec{\theta},\lambda) = \frac{2}{N_{NM}(N_{NM}-1)} \sum_{i=2}^{N_{NM}} \sum_{k=1}^{i-1} \left| NM_i(\vec{\theta},\lambda) - NM_k(\vec{\theta},\lambda) \right| \qquad (9.5.2)$$

Modulation between different sub-configurations is currently an integral step of the mission allowing detection of the planetary signal against the huge background from extended sources e.g. EZ and LZ clouds.

### 9.5.1 Modulation using sub-interferometers

Switching between two achromatic phase shifts $\phi_k$ can generate two different centre-symmetric nulled output maps, such that the transmission peak coincides on the planet's location in NM1 while NM2 has a transmission low at the same position. To $\phi_k$ calculate the modulation efficiency for internal (using beamsplitters) or inherent modulation (no TM maps are generated), one combines the nulled outputs of two sub-interferometers on a beamsplitter and obtains two complementary outputs $S_j(\vec{\theta},\lambda)$ and $S_k(\vec{\theta},\lambda)$ of a signal that is still nulled. Modulation between 2 nulling sub-interferometers leads to a detected intensity proportional to $S_j(\vec{\theta},\lambda)$ and $S_k(\vec{\theta},\lambda)$ on each side of the beam combiner.

$$S_j(\vec{\theta},\lambda) = \frac{1}{2}\left( \left| \vec{A}_j(\vec{\theta},\lambda) + e^{i\phi_{im}(t)} \vec{A}_k(\vec{\theta},\lambda) \right| \right)^2 \qquad (9.5.1.1)$$

$$S_k(\vec{\theta},\lambda) = \frac{1}{2}\left( \left| \vec{A}_j(\vec{\theta},\lambda) + e^{i\phi_{im}(t)+\pi} \vec{A}_k(\vec{\theta},\lambda) \right| \right)^2 \qquad (9.5.1.2)$$

where $A(\vec{\theta},\lambda)$ is given by

$$\vec{A}(\vec{\theta},\lambda) = \sum_{k=1}^{N} A_k e^{i2\pi(L_k\theta/\lambda)\cos(\delta_k-\phi)} e^{i\phi_k(\lambda)} \qquad (9.5.1.3)$$

Due to the beamsplitter that is used to combine the signal, the phase difference between the two outputs is π. The factor ½ in equation (9.5.1.1) and (9.5.1.2) is introduced because the fractions of the beams transmitted to each of the two outputs are added.

Inherent modulation also allows using the whole interferometer as sub-array. The principle is different from the scheme shown above, as no sub-interferometers are used *at the same time*. The concept combines a number of input beams with time varying phase shifts, such that each nulled map used for signal modulation is obtained directly at different time. The switching implies that only one output can be sampled at a time resulting in the fore-factor of 1/(number of nulled output maps generated). The planmod efficiency is normally defined for two nulled output maps, one of which NM1 has the planet on a dark, the other one NM2 on the corresponding bright fringe, thus the fore-factor becomes ½.

$$S_j(\vec{\theta},\lambda) = \frac{1}{2}\left|\vec{A}_j(\vec{\theta},\lambda)\right|^2 \qquad (9.5.1.4)$$

$$S_k(\vec{\theta},\lambda) = \frac{1}{2}\left|\vec{A}_k(\vec{\theta},\lambda)\right|^2 \qquad (9.5.1.5)$$

The modulation of the collected signal, planmod, of a configuration using two sub-interferometers can be calculated using equation (9.5.1.6). The maximum value of the planmod matrix $\xi(\vec{\theta})$ gives the maximum achievable value of planmod.

$$\xi\left(\vec{\theta},\lambda\right) = \frac{\left|S_2\left(\vec{\theta},\lambda\right) - S_1\left(\vec{\theta},\lambda\right)\right|}{NM_{\max}} \qquad (9.5.1.6)$$

$$\xi_{\max} = \max \xi\left(\vec{\theta},\lambda\right) \qquad (9.5.1.7)$$

The mean and rms of the signal modulation is the relevant figure for the detection phase, when the planetary location is unknown. It corresponds to the mean value of the modulation of a signal from a point source generated by switching between transmission maps of the sub-interferometers. The peak modulation efficiency is relevant, once the planet has been located. Spectroscopy at a given wavelength is then possible with this efficiency if the planet can be put on the peak in the MM.

The maximum detected intensity is calculated by Energy-conservation. That introduces a normalization factor of 1/N to calculate the output intensity. Tracing the intensity by stepping through the beam combination system leads to the same result if one considers that at each beamsplitter half of the intensity of the incoming beam is lead into one of the two paths after the beamsplitter introducing a fore-factor ½. at each beam-combiner.

## 9.6 Detected signal and transmission value

The intensity detected at the output of the interferometer depends on the Transmission factor of the array. $T_{TP}$ denotes losses introduced in the system due to components in the light path, not taking the interference of the beams into account. $T_{BC}$ accounts for the technical implementation of the beam combination schemes. Each component in the optical path reduces the intensity through e.g. reflection, imperfections and intrinsic efficiency. Losses due to errors in e.g. OPD control, pointing etc. also have to be included in this value. To include those losses and thus get the absolute intensity detected at the output, one has to multiply the modulation efficiency with the transmission value of the system, $T = T_{BC}\, T_{TP}$. The factor $T_{BC}$ has to be included to take the possible inability to combine 100% of all incoming light due to BC implementation schemes (see Table 20). The two factors are connected, as e.g. the number of optical elements in the beam path will depend on the BC scheme chosen. $T_{BC}$ is determined by the beam combination scheme implemented for a certain configuration. As all configurations can be implemented using different beam combination schemes, that transmission factor is a separate characteristic of the array. It depends on the beam combination scheme chosen as well as on the number of telescopes used.

$$signal_{general}\left(\vec{\theta},\lambda\right) = \xi\left(\vec{\theta},\lambda\right)\cdot TM_{\max}\cdot T_{TP}(\lambda)\cdot T_{BC}(\lambda)\cdot BD_{system}\left(\vec{\theta},\lambda\right)\cdot\frac{1}{N} \qquad (9.6.1)$$

$$signal_{planet}(\lambda) = \xi_{mean}(\lambda)\cdot TM_{\max}\cdot T_{TP}(\lambda)\cdot T_{BC}(\lambda)\cdot BD_{planet}(\lambda)\cdot\frac{1}{N} \qquad (9.6.2)$$

$BD_{system}$ stands for the brightness distribution of the target star-planetary system, where 1/N is the normalization factor derived through energy-conservation. For image plane interferometry (see next section) the Fourier transform (FT) of the brightness distribution of the target system has to be convoluted with the FT of the transmission map to give the complex amplitude at the image plane.

## 9.7 Beam-combining scheme efficiency (BCS efficiency)

For equal-size telescopes using internal modulation, 100% efficiency for $T_{BC}$ can only be reached if the number of telescopes is equal to a power of 2. In the current proposed schemes using internal modulation, the beams are combined pairwise using a Modified Mach Zehnder (MMZ) to provide complete symmetry of the beams.

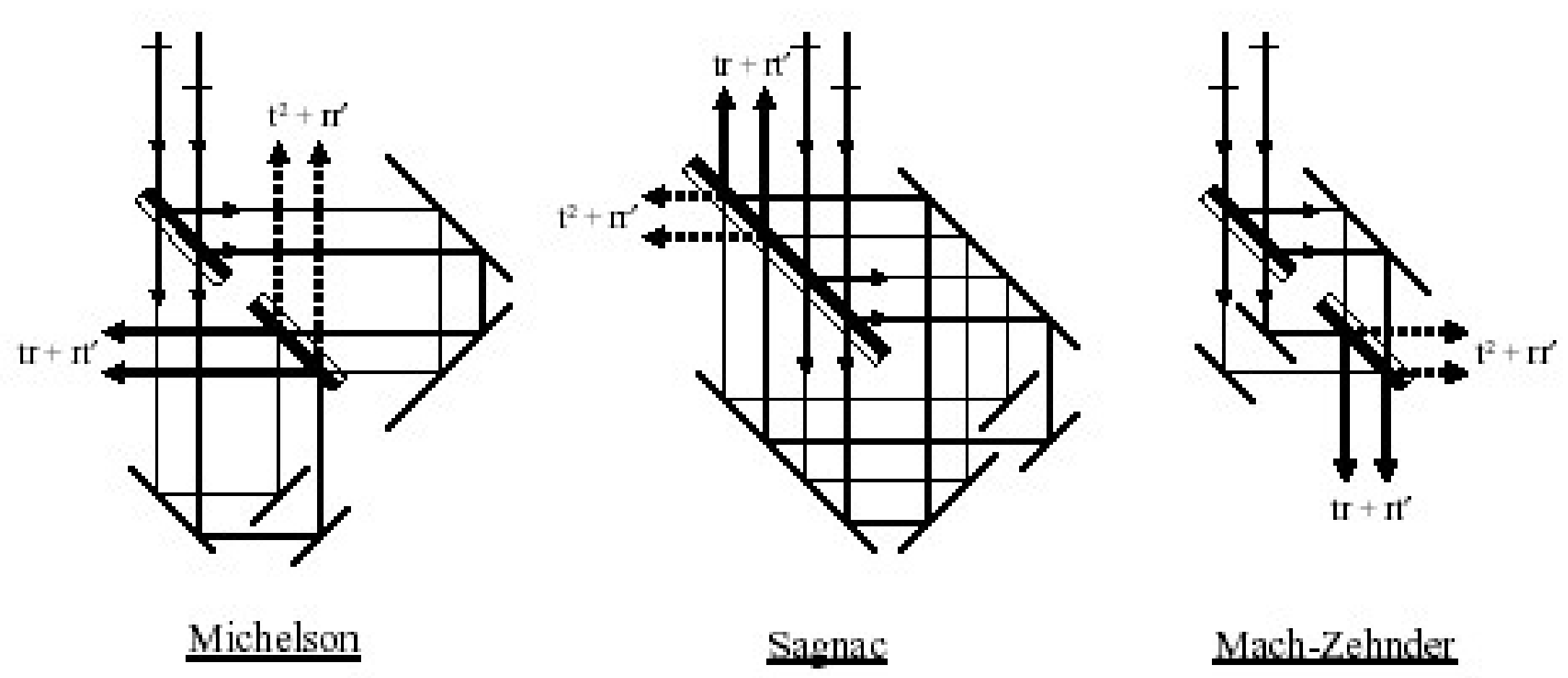

**Fig. 115: Different beam combiners providing complete symmetry for beam combination (Serabyn, 2003)**

For equal sized telescopes the ideal factors for multi-axial beam injection given in Table 20. If the telescopes are not identical, the scheme has to be recalculated.

**Table 20: BCS efficiency depending on the number of equal telescopes**

| BCS | 2 tel | 3 tel | 4 tel | 5 tel | 6 tel |
|---|---|---|---|---|---|
| Classical | 100%<br>80%* | 75%<br>60%* | 100%<br>80%* | 62.5%<br>50%* | 75%<br>60%* |
| MBI | 56%* | 67%* | 62%* | 55%* | 49%* |

*including coupling losses into a single mode fibre assumed 20%

For multi-beam injection into a fiber, we assume perfect circular beams with constant intensity, most compact multi-axial beam configuration before coupling optics that is optimised for the number of injected beams (Wallner, 2004). Note that for multi-axial beam injection polarization effects might be an issue. For inherent modulation we imagine an image-plane combination scheme. We propose to feed the light into one single mode fibre. When beams are coherently combined in the focal plane through a lens, in the case of an ideal overlap between the input field and the fundamental mode of the SMW, one can at best achieve a coupling efficiency proportional to the relative surface area of the beam.

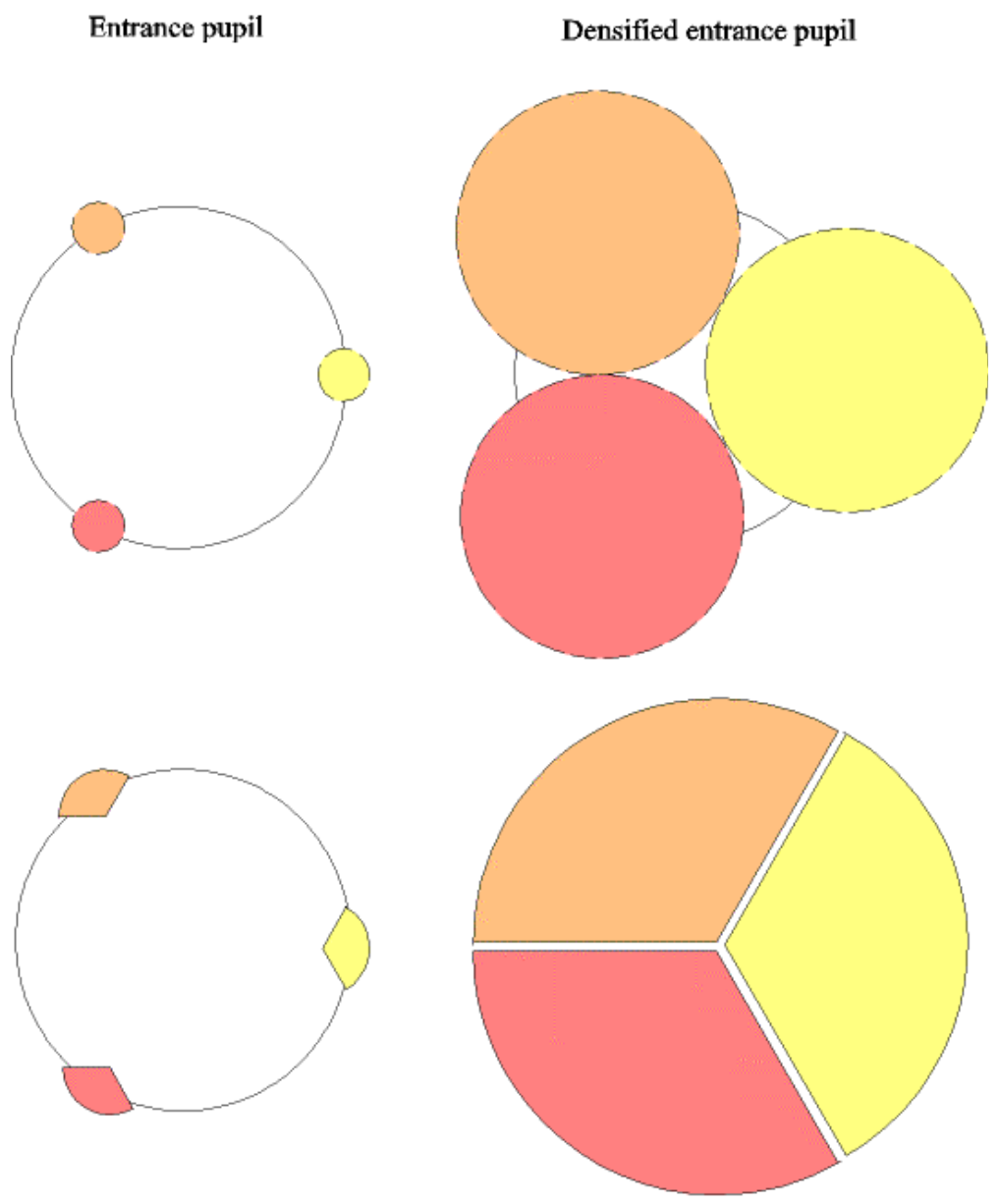

**Fig. 116: Concept of multi-beam injection (Absil, 2003B)**

If alternative methods like integrated optics (I.O.) become available, characteristics of the I.O. device, like throughput and coupling efficiency, determine the strength of the detected signal. I.O. devices have the potential to improve the intensity substantially for high numbers of telescopes combined.

## *9.8 Optical filtering of the beams*

One key point of nulling interferometry is the high rejection of star light needed. High rejection implies great similarities of all incoming wavefronts that have to be combined, concerning their amplitude, phase and polarization. However the different optical beampaths have different effects on the wavefronts and thus decreases the nulling performance. In order to get a $10^6$ rejection rate with a two-beam interferometer, the difference from one wavefront to another cannot exceed 0.1% in amplitude and $10^{-3}$ rad in phase on each point of the beam recombiner. The phase requirements are the most stringent and translate into a wavefront quality better than $\lambda/6000$ at the working wavelength of 10μm (Perrin, 2000). J.M. Mariotti (Mariotti, 1997) proposed the use of small pieces of single mode fibers to perform a perfect modal filtering. The solution is based on the fact that individually each telescope does not resolve the star/planet system because of diffraction. The whole of the useful information is thus included in the Airy disk.

In the focal plane of the telescope, the Fourier plane, one can access the diffraction pattern formed by the primary mirror. The diffraction pattern is the Fourier transform of the entrance pupil. In a perfect system, the diffraction pattern is a perfect Airy figure. In case of a defect wavefront, the Fourier transform is the algebraic sum of the ideal pupil and the contribution of the defects. Because the size of the defect is much smaller than the diffraction pattern, the Fourier transform of the defect's contribution is spread over a larger area in the focal plane than the ideal pupil. Using a pinhole one can keep the central part of the diffraction pattern only, which keeps about 85% of the energy in the Airy disk while most of the energy of the defect is eliminated. The smaller the defect, the more efficient the filtering is, as long as the defect is smaller than the size of the pupil. Spatial filtering is an inefficient method to correct low order defects such as differential piston, tilt or defocus. Ollivier et al. (Ollivier, 1997) showed through numerical calculations that optical constrains on defects from polishing errors could be reduced from $\lambda/300$ (rms) to $\lambda/20$(rms) when a pinhole is used. The size of a pinhole is a compromise between filtering efficiency and optical throughput. The diffraction pattern at the focus of a lens is about $f\lambda/D$. The filtering efficiency of a pinhole with a fixed size thus depends on the wavelength used. For optical filtering we need an achromatic device.

For single mode fibers the size of the propagating mode varies as λ. In a first approximation the device is achromatic. The radius of the fiber's fundamental mode scales almost proportionally with wavelength. Since the size of the Airy pattern at the focus of the telescope also scales proportionally to the wavelength, the coupling efficiency is roughly achromatic between the cut-off wavelength of the fundamental mode and about twice this wavelength (Menneson, 1999). For symmetry reasons (Ruilier, 1999) all defects of non-zero azimuthal order, such as tilt or astigmatism introduce no phase shifts at the injection into the guide. Also, low spatial frequency defects can be eliminated, using single-mode fibers. Whatever the shape of the wavefront at the input, the output beam is the fundamental mode of the fiber and the phase is the mean phase at the injection added to an injection phase shift. When a wavefront is injected into an optical fiber, its structure is decomposed over the different modes of the fiber. Each mode propagates with its own characteristics (Neumann, 1988). In the particular case of a single mode fiber, only the fundamental mode propagates with a mean phase. The other modes are diffused in the fiber cladding. The intensity of the signal at the output of a single mode fiber is proportional to the overlap integral of the incoming wavefront and the mode of the fiber. The irregularities of the incoming wavefronts are converted into amplitude fluctuations, which can be easily monitored. At the output of the fiber the amplitude and phase defects are theoretically completely transformed into a global amplitude and phase variation of the Gaussian mode of the fiber. The fact that the recombination and the optical filtering are two linear operations allows filtering before or after recombination. Using a waveguide after recombination of the beams corrects phase defects after the foci and omits differential effects due to propagation in various waveguides. A single mode fiber could also be directly inserted in the cold part of the detector block and thus reduce the amount of thermal background seen by the detector. Mennesson et al. (Mennesson, 1999) give a detailed study of optical filtering and comparison of the two methods.

Main losses due to coupling into a fiber come from the match between propagation modes in vacuum and the single mode of the fiber. Up to 78% of the energy of a perfect telescope can be coupled into a single-mode fiber (Shaklan, 1988). Drawbacks of optical fibers are the chromatic dispersion that arises if the optical paths through the different fibers used are not matched, the mechanical and thermal sensitivity and the birefringence of some materials. However Reynaud and Lagorceix have shown that these difficulties can be overcome by controlling the fiber length actively, polishing the fiber ends carefully and by using polarization-maintaining fibers (Malbet, 1999). A source off axis by more than an Airy disk will not be seen by the fiber. Its field of view is limited to λ/D, where D is the diameter of the telescope (Mennesson, 1999) also limiting the FOV of the interferometer array. All spatial information contained in the PSF of individual telescopes is lost when single mode waveguides are used. Sources from different directions in the sky are not angularly separated, but are seen with a characteristic external differential phase due to the interferometer baseline and the corresponding interferometric transmission efficiency. The signal of the interferometer is modulated to recover the position of an off-axis companion.

### 9.8.1 Single mode coupling theory

The coupled efficiency is defined as the overlap integral between the electrical field distribution in the focal plane of the telescope and in the guided mode, or the transmission of the spatial filter (Coude Du Foresto, 1996).

$$\rho = \frac{\left|\int_{A\infty} E_{focus} E^{*}_{fiber}\, dA\right|^2}{\int_{A\infty} \left|E_{focus}\right|^2 dA \cdot \int_{A\infty} \left|E_{fiber}\right|^2 dA} \qquad (9.8.1)$$

where the integration domain extends at infinity in a transverse plane. The symbol * denotes a complex conjugate. The squared modulus of the amplitude A is often used to quantify the amount of injected energy *E*:

$$E \propto \left|A^2\right| \qquad (9.8.2)$$

Light in an optical fiber is decomposed into waveguide modes. The different modes propagating through the fiber are described by Maxwell's equation with applicable bounding conditions. In a single mode fiber only the fundamental mode propagates through. Its profile is a Bessel function in the core and a Hankel function in the cladding. Under some conditions the profile can be approximated by a Gaussian (Ruilier, 1999):

$$F_{01}(r) \approx e^{-\left(\frac{r}{\omega_0}\right)^2} \tag{9.8.3}$$

where $\omega_0$ is the fundamental mode radius. It is chromatic and linked to the physical radius of the fiber. Assuming a circular pupil with diameter D and central circular obscuration $\alpha$, the effect of a single-mode fiber in the focal plane is the filtering of the Point Spread Function. The coupling efficiency is given by (Ruilier, 1999):

$$\rho_\alpha(\beta) = 2\left(\frac{e^{-\beta^2}\left(1-e^{\beta^2(1-\alpha^2)}\right)}{\beta\sqrt{1-\alpha^2}}\right)^2 \tag{9.8.4}$$

with $\beta = \frac{\pi}{2}\frac{D}{f}\frac{\omega_0}{\lambda}$

$f$ is the focal length of the telescope. For an unobstructed pupil a maximal coupling efficiency of about 81% can be achieved, with $\beta = 1.12$. Taking the Fresnel reflections on the head of the fiber into account, the maximum is 78%. That leads to an optimal fundamental mode radius $\omega_{0,opt.}$

$$\omega_{0opt} = 0.71\frac{f\lambda}{D} \tag{9.8.5}$$

Converted to FWHM, this means a Gaussian that is 18% wider than the Airy disk. The fundamental mode receives less energy when there is a central obscuration. An obscuration the diameter of 20% of the pupil, results in 70% of coupling efficiency while 40% of the diameter results in a coupling efficiency of 50% only.

## 9.9 Integrated optics beam combiner

Since the 80's, I.O. components on planar substrates have become available for telecommunication applications. The principle of integrated optics is similar to that of fiber optics since the light propagates in optical waveguides, except that the former propagates inside a planar substrate. Integrated optics can provide an interferometric combination unit on a single optical chip. The compactness of the optical layout of an integrated optics device provides stability, low sensitivity to external constrains like temperature, pressure or mechanical stresses, no necessity for optical alignment except for coupling, simplicity and intrinsic polarization control (Malbet, 1999). It has been demonstrated that spatial filtering before the interferometric combination significantly improves the quality of visibility measurements (Kern, 1996) as single mode propagation performs spatial filtering. The major advantages of integrated optics components are their simplicity, small size and reliability for space-based missions, as no internal alignments are needed. Beam combiners for N telescopes need only a single component. The size of waveguides is similar to the size of pixels in infrared arrays. Thus matching of the planar component with an infrared detector would lead to a completely integrated instrument with no relay optics between the beam combiner and the detector (Kern, 1996). The compactness also opens the possibility of a fully cooled instrument. Transmission properties of the chosen material and minimization of losses intrinsic to integrated optics structures should be the major drivers of research in this area for future missions.

## 9.10 Starlight rejection $\rho_{star}$

The starlight rejection $\rho_{star}$ is a unitless value, defined as the flux of the stellar disk with radius $s$ divided by the flux received from that same region through the transmission map of the nulling interferometer. It is a measure for the quality of the null achieved.

$$\rho := \frac{\pi \cdot s^2}{\int_0^{2\pi}\int_0^{s} \left[ \mathrm{Re}\left[ \left( \left| \sum_k A_k \cdot e^{1i\cdot 2\cdot \pi \cdot L \cdot \frac{\theta}{\lambda}\cdot \cos(\delta_k - \phi)} e^{1i\phi_k} \right| \right)^2 \right] \cdot \theta \right] d\theta \, d\phi} \tag{9.10.1}$$

In the spectroscopy phase where the array is arranged so that the planet lies on a modulation maximum, the maximum planmod efficiency scales the number of photons detected from the planet in respect to its host star. The extent of the planet is small in comparison to the transmission features, thus as a first approximation the max of the modulation map can be used for the whole planet surface once the planet is detected. In the detection phase the mean modulation can be used to estimate the expected planetary flux. Star and planet are assumed to radiate like black bodies (BB(T,λ)). Once the planet is detected and its position determined, the array can be arranged so that the $max\xi$ is located at that sky position. For a quantitative photon count of the planet the intensity at the output depending on the intensity at the input $I_{in}$ at each telescope can be calculated see equation (9.10.2). Losses due to the detector are not included here.

$$BB(T_{planet},\lambda)\ \max(\xi)\ r_{pl}^2\ \pi\ T_{BC}\ T_{TP}(\lambda)\ N\ I_{in}(\lambda) \tag{9.10.2}$$

## 9.11 Image-plane interferometry

In the case of image-plane interferometry, beams are combined using an imaging telescope. In this case, one must consider that the pupil position $\vec{\ell}_j$ at recombination differs from the original positions $\vec{L}_j$ of the apertures. This induces an additional $\exp(-ik\vec{\ell}_j \cdot \vec{\theta})$ field-dependent phase delay to the beam. Furthermore, the size of the beams at recombination is R/M and the field angle is $M\left|\vec{\theta}\right|$, where $M$ is the beam compression ratio of the system (Traub, 1999). Beam compression increases the physical size of the PSF at the detector. The complex amplitude of the superposed electrical field, measured at the entrance pupil of the beam combiner telescope for a point source is:

$$A_{XP}(\vec{\theta},\vec{r},\lambda) = \sum_{k=1}^{N} \Pi\left( \frac{\left|\vec{r} - \vec{\ell}_k\right|}{R/M} \right) e^{i(2\pi/\lambda)M(\vec{r}-\vec{\ell}_k)\vec{\theta}} A_k e^{i(2\pi/\lambda)\vec{L}_k\vec{\theta} + i\varphi_k(\lambda)} \tag{9.11.1}$$

The first phasor represents the overall tilt within each wavefront segment. This is M times larger than the slope of the wavefront hitting the apertures. The phasor $e^{i2\pi\vec{L}_i \cdot \vec{\theta}/\lambda}$ describes the (field-dependent) phase difference between the beams collected at the various telescopes. As a result of the change in the pupil position, the wavefront segments are no longer coplanar. In other words, the system is anisoplanatic: the fringe center at the final image plane does not coincide with the image center.

A nulling interferometer must be anisoplanatic, since ideally its transmission must change from zero to unity within a few tenths of arcsec in the sky. In general, it is advantageous to position the recombining beams adjacent to each other (densified pupil), or $\vec{\ell}_k \approx R/M$. With this choice, the interference pattern consists of one tall and broad peak (where most of the photons are concentrated) with small adjacent sidelobes. It is also apparent that focal-plane interferometry requires careful coordination between telescope configuration $\vec{L}_k$ and the beam injection configuration $M\vec{\ell}_k$ in order to ensure the desired interferometric properties. The condition $\vec{L}_k = M\vec{\ell}_k$ is known as homothetic mapping. When this is satisfied, the various beams will be imaged in phase at the detector plane, for all sources in a wide field of view. Homothetic mapping is particularly interesting since it may enable wide-field interferometric imaging.

The electrical field in the focal plane is obtained by taking the Fourier transform of eq. (9.11.1), going from the pupil coordinates $\vec{r} - \vec{\ell}_k$ to a single focal plane coordinate $\vec{\varphi}$. In this coordinate system, a source at sky position $\vec{\theta}$ results in a spot of diameter $1.22Mf\,\lambda/R$, centered at $\vec{\varphi} = Mf\vec{\theta}$. With the application of the shift theorem this yields:

$$\tilde{A}(\vec{\theta},\vec{\varphi},\lambda) = Airy^{\frac{1}{2}}\left(\frac{R\,2\pi}{\lambda f M}\left|\vec{\varphi} - fM\vec{\theta}\right|\right)\sum_{k=1}^{N} A_k e^{i2\pi(\vec{L}_j - M\vec{\ell}_j)\cdot\vec{\theta}/\lambda + i\phi_k(\lambda)} \qquad (9.11.2)$$

with

$$Airy(x) = \left[\frac{2J_1(x)}{x}\right]^2 \qquad (9.11.3)$$

being the squared Fourier transform of a circular pupil and $f$ the focal length of the telescope.

The measured intensity distribution as a function of position in the sky, and wavelength, is obtained by integration of the squared complex amplitude over the focal plane, collecting all the power under the envelope of the PSF:

$$A_{XP}(\vec{\theta},\vec{r},\lambda) = \sum_{k=1}^{N} \Pi\left(\frac{\left|\vec{r} - \vec{\ell}_k\right|}{R/M}\right) e^{i(2\pi/\lambda)M(\vec{r}-\vec{\ell}_k)\vec{\theta}} A_k e^{i(2\pi/\lambda)\vec{L}_k\vec{\theta} + i\phi_k(\lambda)} \qquad (9.11.4)$$

$$I(\vec{\theta},\vec{\varphi},\lambda) = \mathrm{Re}\left\{\int_{focal\,plane} \tilde{A}(\vec{\varphi},\vec{\theta},\lambda)\tilde{A}^*(\vec{\varphi},\vec{\theta},\lambda)d\vec{\varphi}\right\} = Airy\left(\frac{Rk}{fM}\left|\vec{\varphi} - fM\vec{\theta}\right|\right)$$
$$\mathrm{Re}\left\{\sum_{m=1}^{N} F_m e^{i2\pi(\vec{L}_m - M\vec{\ell}_m)\cdot\vec{\theta}/\lambda + i\phi_m(\lambda)} \sum_{n=1}^{N} F_n e^{-i2\pi(\vec{L}_n - M\vec{\ell}_n)\cdot\vec{\theta}/\lambda - i\phi_n(\lambda)}\right\} \qquad (9.11.5)$$

This formula can be rewritten as:

$$I(\vec{\theta},\vec{\varphi},\lambda) = Airy\left(\frac{Rk}{fM}\left|\vec{\varphi} - fM\vec{\theta}\right|\right)\left\{\sum_{m=1}^{N} F_m^2 + 2\sum_{m=1}^{N}\sum_{n>m}^{N} F_m F_n \cos\left[\begin{array}{l}\frac{k}{fM}(\vec{\ell}_m - \vec{\ell}_n)\vec{\varphi} + \\ \frac{k}{f}(\vec{L}_m - \vec{L}_n - M(\vec{\ell}_m - \vec{\ell}_n))\cdot\vec{\theta} + (\phi_m - \phi_n)\end{array}\right]\right\} \qquad (9.11.6)$$

This formula shows the appearance of fringes in the focal plane, whose spacing is governed by the coordinates $\vec{\ell}_k$ of the reconfigured pupil. A field-dependent phase offset is also present, which is responsible for the anisoplanatic behavior of the interferometer. If $M\vec{\ell}_k << \vec{L}_k$ (compact reconfigured pupil) this result is almost identical to eq. (9.11.1), and hence the derivation of the transmission map is completely analogous.

## 9.12 Modulation using rotation of the array

The time needed for rotation of the array to cover a circular planetary orbit under the region of highest modulation is an important feature of a configuration as the mission lifetime is limited and numerous target stars have to be investigated. The thrusters used determine the time needed for the rotation. It is not clear yet whether observations can be continued during rotation. Table 23 shows the rotation angle needed to cover a circular planetary orbit under the region of highest modulation assuming continuous rotation. If that is not possible Table 23 shows the number of different modulation maps generated through rotation of a set angle and consecutive observation at that position, that are needed to provide a min of 50% of the maximum achievable planmod of that configuration for the whole search region. To cover the whole interesting region (taking into account the inclination of an orbit) the baseline of the array will most likely have to be adjusted. The actual distance between a planet and its host star is unknown, since the orientation of the system is unknown. Only repeated observations can determine the orbit of the planet and thus the real separation.

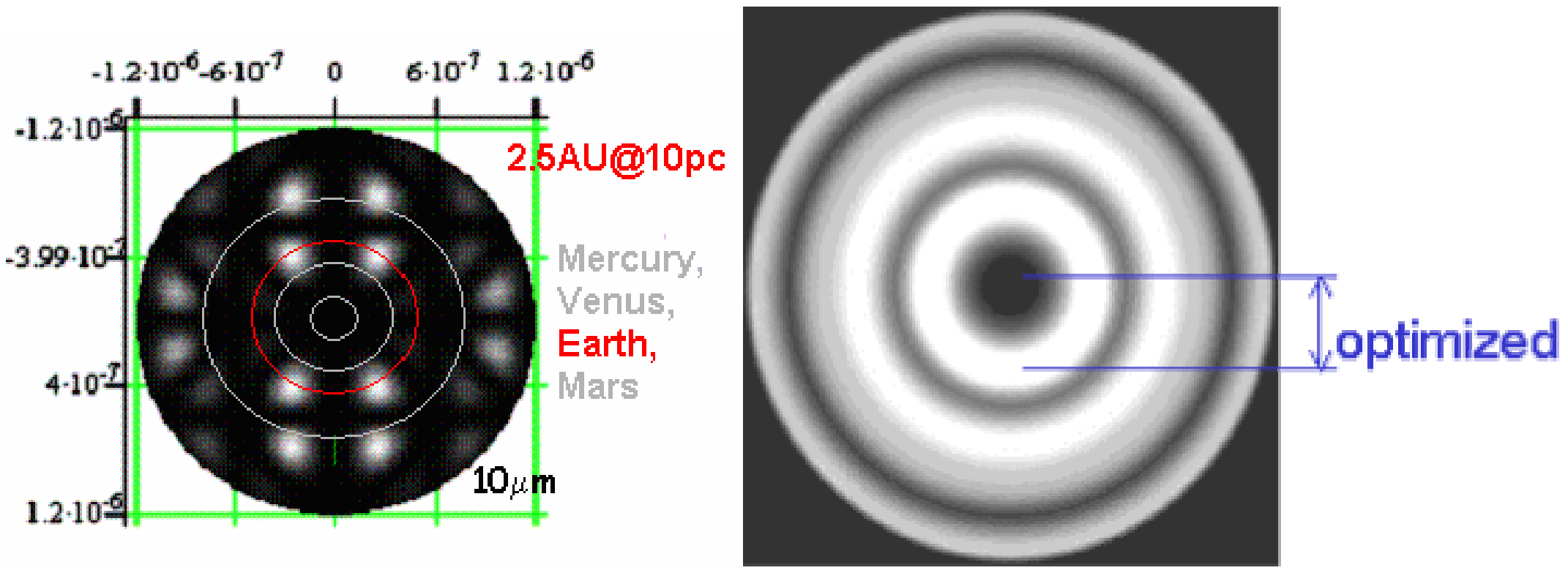

**Fig. 117: Modulation map (left) and rotated modulation map (right) of the array to cover the full torus of a 1AU equivalent orbit**

We intent to observe during rotation or at least keep fringes while rotating. Thus rotation is slow because only small thrust levels are investigated. Note that for symmetrical transmission maps like the Bracewell configuration the position of the planet can not be determined through rotation, and the signal of two point sources on either side of the star will generate the same signal as one point source with double brightness.

**Table 21: Time needed for 360-degree rotation of the array (data courtesy of M. Kilter)**

| time (hours) for 360 degree rotation | | | | |
|---|---|---|---|---|
| | Circle radius of the array | | | |
| Thrust level | 25m(min) | 50m | 100m | 250m(max) |
| 150 µN | 15.93 | 22.53 | 31.87 | 50.38 |
| 1 mN | 6.17 | 8.73 | 12.34 | 19.51 |
| 5 mN | 2.76 | 3.9 | 5.52 | 8.73 |
| 50 mN | 0.87 | 1.23 | 1.75 | 2.76 |
| (likely thrust-level marked in orange) | | | | |

In concurrent designs some rotation of the array has to be performed to cover the whole orbit of a planet in the detection phase. The dependence of rotation angle versus time is not linear (Kilter, 2004) thus the values for a 30-degree rotation are given in Table 22.

**Table 22: Time needed for 30-degree rotation of the array (data courtesy of M. Kilter)**

| time (hours) for 30 degree rotation | | | | |
|---|---|---|---|---|
| | Circle radius (baseline/2) | | | |
| Thrust level | 25m(min) | 50m | 100m | 250m(max) |
| 150 µN | 1.33 | 1.88 | 2.65 | 4.20 |
| 1 mN | 0.52 | 0.73 | 1.03 | 1.63 |
| 5 mN | 0.23 | 0.33 | 0.47 | 0.73 |
| 50 mN | 0.07 | 0.10 | 0.15 | 0.23 |
| (likely thrust-level marked in orange) | | | | |

## 9.13 Notes on modulation of a planetary signal

If only one detector is used, the maximal modulation efficiency is limited to 100% divided by the number of different nulled output maps $N_{TM}$ (generated by each of the subinterferometers), as each signal is only observed $100/N_{TM}$ % of the time. If more than one detector is used that limitation does not apply to modulation schemes where both nulled outputs are generated at the same time e.g. internal modulation, thus 100% planmod can be achieved. For schemes where the two nulled outputs are not generated simultaneously, a possibility using inherent modulation, modulation between two sub-interferometer maps can only produce a maximum of 50% planmod.

Sub-interferometers use only a fraction of the incoming light/amplitude per telescope. They can be implemented by dividing the amplitudes of the light beams collected and/or using the collected light from subsets of the array. The proposed DARWIN configurations are optimized for utilizing equal sized telescopes. An example is the Bowtie configuration, where six telescopes are used as two sub-interferometers of 4 telescopes each. Each sub-interferometer is generated using amplitude fractions equal to $1/\sqrt{2}$, 1, 1, $1/\sqrt{2}$ of the collected amplitude per telescope, respectively (Absil, 2000).

## 9.14 Normalization

Configurations using inherent modulation do not necessarily have a counterpart based on internal modulation between sub-interferometers, but each configuration using internal modulation can be implemented using inherent modulation. The normalization for internal modulation can be calculated by tracing the intensity of the beam through the beam combination process. One has to consider that at each beamsplitter half of the intensity of the incoming beam is lead into one of the two paths after the beamsplitter introducing a fore-factor ½ at each beam-combiner. This can be a highly complex process. As each configuration can be implemented using inherent modulation, one can use the intensity calculation on base of inherent modulation to normalize the intensity of any configuration. The maximum detected intensity is also calculated through Energy-conservation. That introduces a normalization factor of 1/N.

Combination of $N$ waves with amplitudes $A_k$ to combined wave $E$ with amplitude $A_0$. Each wave can be written as:

$$E_k = A_k e^{i(\omega_k t + \phi_k(\lambda))} = A_k e^{i\phi_k(\lambda)} e^{i\omega_k t} \quad (9.14.1)$$

$$E = \sum_{k=1}^{N} E_k \quad (9.14.2)$$

$$E = \sum_{k=1}^{N} A_k e^{i\phi_k(\lambda)} e^{i\omega_k t} \tag{9.14.3}$$

$$E = A_0 e^{i(\omega_0 t + \phi_0(\lambda))} \tag{9.14.4}$$

the square of the complex amplitude $A_0$ of the combined wave is given by

$$A_0^{\ 2} = \sum_{k=1}^{N} A_k^{\ 2} + 2\sum_{j>k}^{N}\sum_{k=1}^{N} A_k A_j \cos(\phi_k - \phi_j) \tag{9.14.5}$$

The max value of the cos-term is 1, thus the $A_0^2$ is given by

$$A_0^{\ 2} \le \sum_{k=1}^{N} A_k^{\ 2} + 2\sum_{j>k}^{N}\sum_{k=1}^{N} A_k A_j = \left(\sum_{k=1}^{N} A_k\right)^2 \tag{9.14.6}$$

The energy-conservation laws introduce a normalization factor 1/N into the calculations to determine the detected intensity. If all amplitudes are equal to $A_{01}$ for $N$ telescopes that leads to

$$I_{\max} = \frac{1}{N}(NA_{01})^2 = \frac{N^2 A_{01}^{\ 2}}{N} \tag{9.14.7}$$

e.g. for two telescopes $N=2$ that leads to a maximum intensity of

$$\begin{aligned} I = |E_0|^2 = \left|\left(A_0 e^{i\phi_0(\lambda)}\right)\left(A_0 e^{i\phi_0(\lambda)}\right)^*\right| = \\ \left|\left(A_1 e^{i\phi_1(\lambda)} + A_2 e^{i\phi_2(\lambda)}\right)\left(A_1 e^{-i\phi_1(\lambda)} + A_2 e^{-i\phi_2(\lambda)}\right)\right| \end{aligned} \tag{9.14.8}$$

$$\begin{aligned} I = |E_0|^2 &= \left|A_1^{\ 2} + A_2^{\ 2} + A_1A_2\left[e^{i(\phi_1(\lambda)-\phi_2(\lambda))} + e^{-i(\phi_1(\lambda)-\phi_2(\lambda))}\right]\right| \\ &= \left|A_1^{\ 2} + A_2^{\ 2} + A_1A_2 e^{i(\phi_1(\lambda)-\phi_2(\lambda))} + A_1A_2 e^{i(\phi_2(\lambda)-\phi_1(\lambda))}\right| \\ &= \left|A_1^{\ 2} + A_2^{\ 2} + A_1A_2\left(e^{i(\phi_1(\lambda)-\phi_2(\lambda))} + e^{-i(\phi_1(\lambda)-\phi_2(\lambda))}\right)\right| \xrightarrow[e^{-i\phi}=\cos\phi - i\sin\phi]{e^{i\phi}=\cos\phi + i\sin\phi} \\ &= \left|A_1^{\ 2} + A_2^{\ 2} + 2A_1A_2\cos(\phi_1(\lambda) - \phi_2(\lambda))\right| \end{aligned} \tag{9.14.9}$$

$$I_{\max} = \frac{(A_1 + A_2)^2}{2} \tag{9.14.10}$$

If $A_1=A_2=A_{01}$

$$I_{\max} = \frac{(2A_{01})^2}{2} = 2A_{01}^{\ 2} = I_{in} \tag{9.14.11}$$

# 10 Characterization of different proposed configurations

DARWIN could be implemented in a wide variety of configurations, constrained by the number of telescopes and the necessary background and starlight suppression. For a fixed number of telescopes there is a performance trade-off: The ability to provide a deep and wide null to suppress starlight combining all telescopes into one single array versus the ability to suppress large-scale diffuse emissions form a zodiacal cloud through modulation using subinterferometer arrays. Sub-arrays generate a narrower null shape because not all the output from each telescope can be used for each subinterferometer. Simulations show that modulation between different sub-configurations is currently an integral step of the mission allowing detection of the planetary signal against the huge background from extended sources. This section shows the characteristics for five proposed schemes to implement the DARWIN mission: Dual chopped Bracewell (DCB, 4 telescopes) configurations, 3 telescope configurations (TTN equal triangular, TTN linear and TCB) and the BOWTIE configuration (baseline design of DARWIN using 6 telescopes in the Alcatel study).

A number of alternative mission architectures have been evaluated here on the basis of interferometer response as a function of wavelength, achievable modulation efficiency, number of telescopes and starlight rejection capabilities. The study has shown that a reduced science mission goal should be achievable with a lower level of complexity as compared to the baseline configuration of 6 telescopes. If the size of the telescope were larger than 1.5 m (baseline design for the mission), not only a core sample of stars (reduced scientific return) could be observed. The size of the telescopes limits the detected flux, thus for a mission with less telescopes and complexity, the overall surface area should be kept constant to fulfil the mission requirements. A telescope diameter of 3.5m is suggested.

## *10.1 Background Flux*

The detected signal is governed by the contribution of background noise to the interferometer output signal. The local zodiacal cloud (LZ) provides the foreground through which DARWIN will observe. Because zodiacal background is diffuse, it cannot be avoided by interferometry. Thermal emission of the zodiacal dust is the strongest source of background photons at short wavelengths if we assume that the Exo-zodiacal cloud (EZ) is similar to our own LZ cloud. The LZ diminishes with distance from the Sun due to decreased temperature and decreased dust density. Large coherent structures such as wakes and clumps behind planets can masquerade as planets. Those structures could also serve as markers for their presence if their location in respect to a planet were understood. As exo-zodical clouds are not uniform, a planet must be detected against a non-flat field of corrugations. In our own cloud, they are assumed to have roughly less than 0.1% of the amplitude of the total cloud brightness and are thus no source of confusion (Beckmann, 1998).

The stellar leakage is relevant for the small wavelength end of the observation band, assuming a rejection of at least $10^5$. At the large wavelength end the thermal background level due to the emission of the optics can approach that of the zodiacal dust for temperatures above ~40K. Recent work suggests that the shape of the null can be shallower like $\theta^2$ due to noise introduced by instrument errors like OPD (Kaltenegger 2004) (Dubovitsky 2004) thus configurations with fewer telescopes can be investigated as candidates for the DARWIN mission reducing complexity and cost of the mission.

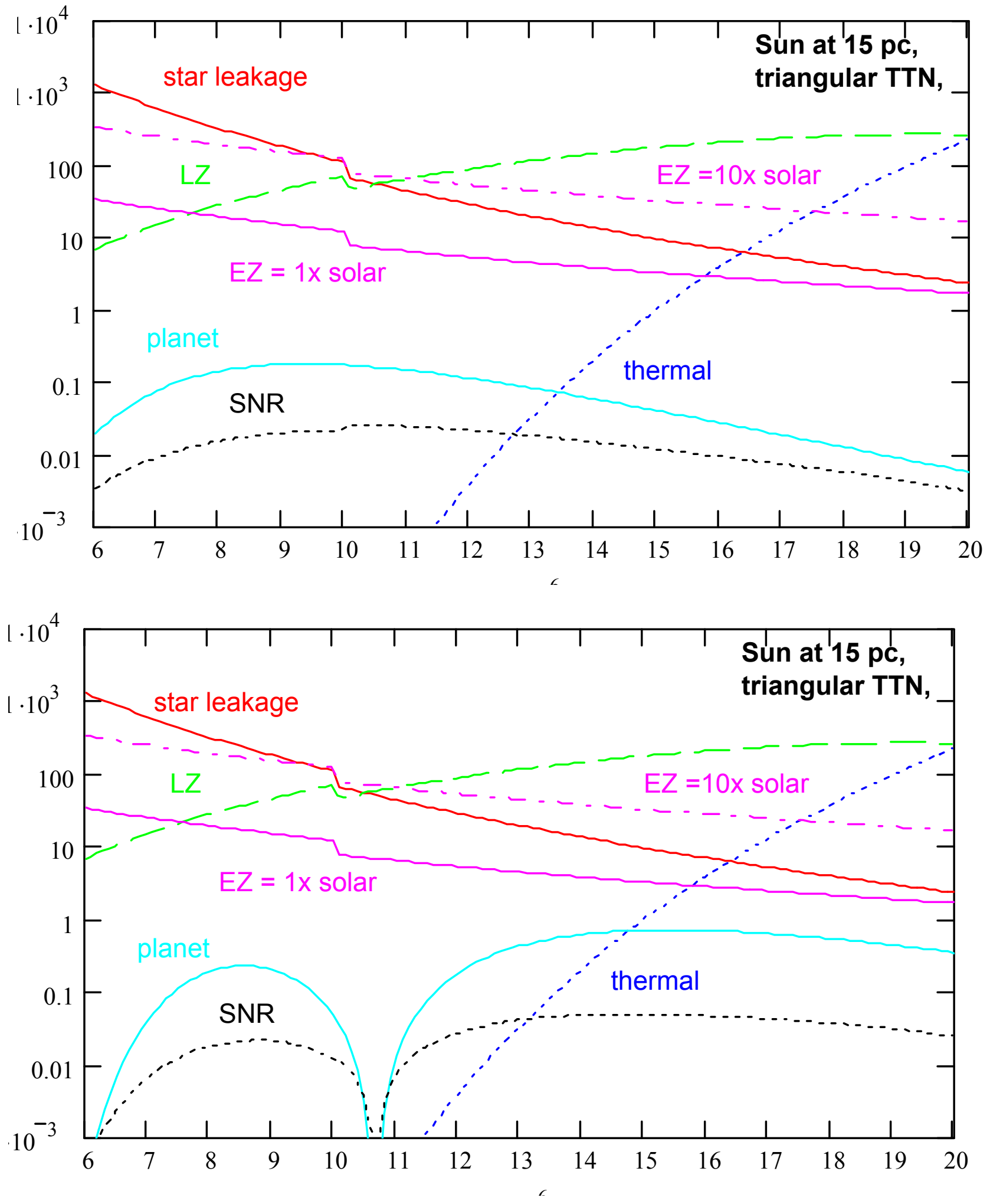

**Fig. 118: Background and earth-sized planet for a G star system at 15pc as a function of wavelength. The graphs show the modulated signal as obtained for an equal size triangular TTN configuration, including detector characteristics and fibre coupling effects. The planet is located on a modulation maximum. The diagrams show the output signal of the array when optimized for (upper panel) total signal, the 6.5 micron feature (lower panel). An extrasolar dust disk of 10 times the solar intensity is the worst case scenario we consider.**

## *10.2 Transmission and modulation properties*

In this section we will go over a general case, showing what the different visualisations of the array characteristics mean. Fig. 119 shows the response of each sub-array as a contour map for an array of 3 telescopes (section 10.7.4). The transmission map (TM) of an interferometer array is its response as a function of the sky coordinates of the source relative to the telescope plane. To maximize the response the baseline and the rotation of the array are chosen such that planet's position coincides with a peak in the TM, at a certain wavelength.

The optimized baseline for an array for the distance of a given target star puts the planet at the position of the first modulation maximum ata  certain wavelength. It can be determined by choosing the baseline so that the tilted wavefronts coming from a planet encounter no phase difference at the different telescopes and are constructively interfered (see section 10). In an optimum scenario the planet lies on the first interference maximum for a given wavelength or wavelength-band. Due to collision avoidance the inter-satellite distance (ISD) will constrain the minimum possible baseline, thus for the closest target stars the planet is put on a secondary maximum or one of the following maxima. As the transmission and thus also the modulation pattern repeats, the planet can be put on a following maximum, not the first one. Note that that will increase the stellar leakage, as the extent of the central null naturally also scales with the baseline. For the optimized distance for different proposed configurations see Table 24. The optimized baseline scales linearly and can be calculated taking into account that the star furthest away in the target sample is at 25pc.

In the detection phase the overall signal is maximized, while in the spectroscopy phase the baseline is chosen so that the transmission at the spectral band of interest is at maximum. The optimal baseline has the characteristic that the phase difference resulting from the fact that the planet is off the interferometer's optical axis, is cancelled by the applied phase shifts. The planet is put on one of the MM's interference maxima for a given waveband.

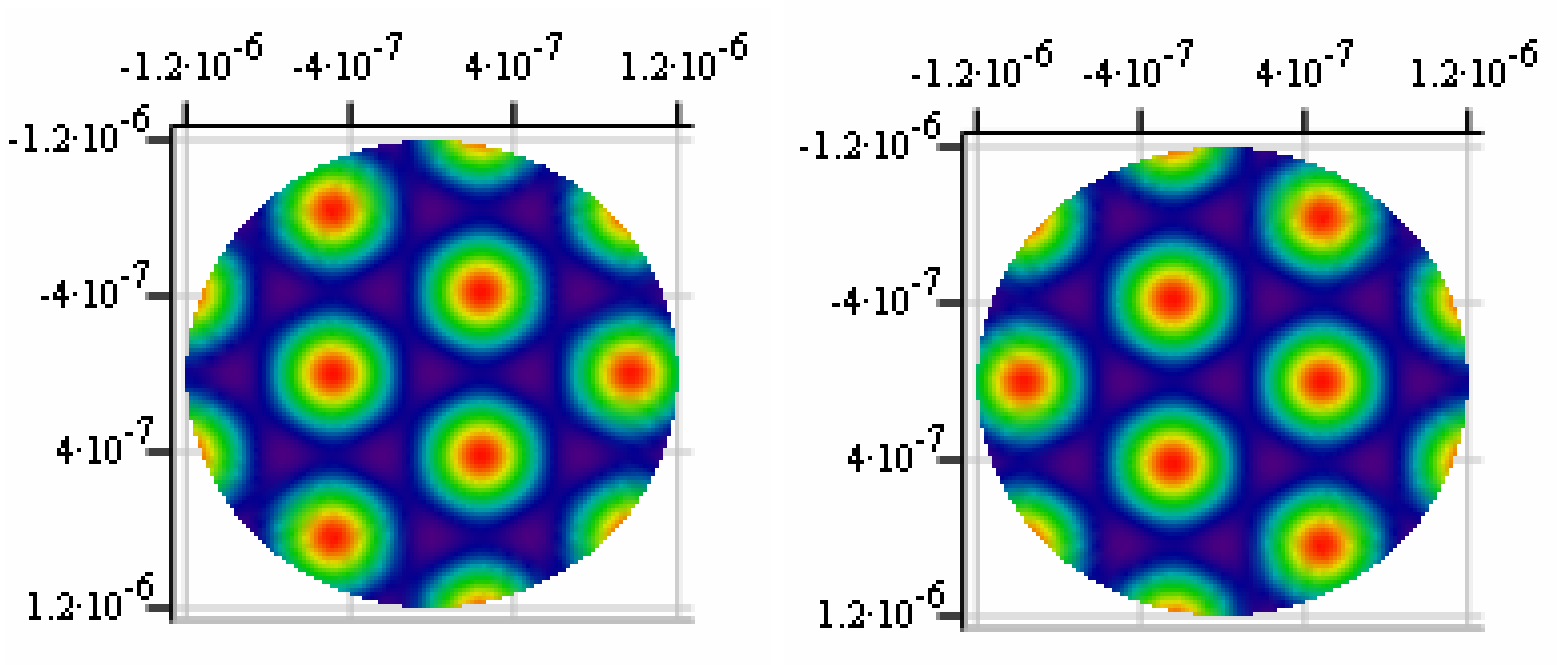

**Fig. 119: Response of the interferometer in an equilateral triangular aperture configuration, showing the two TMs. Red indicates peak transmission areas.**

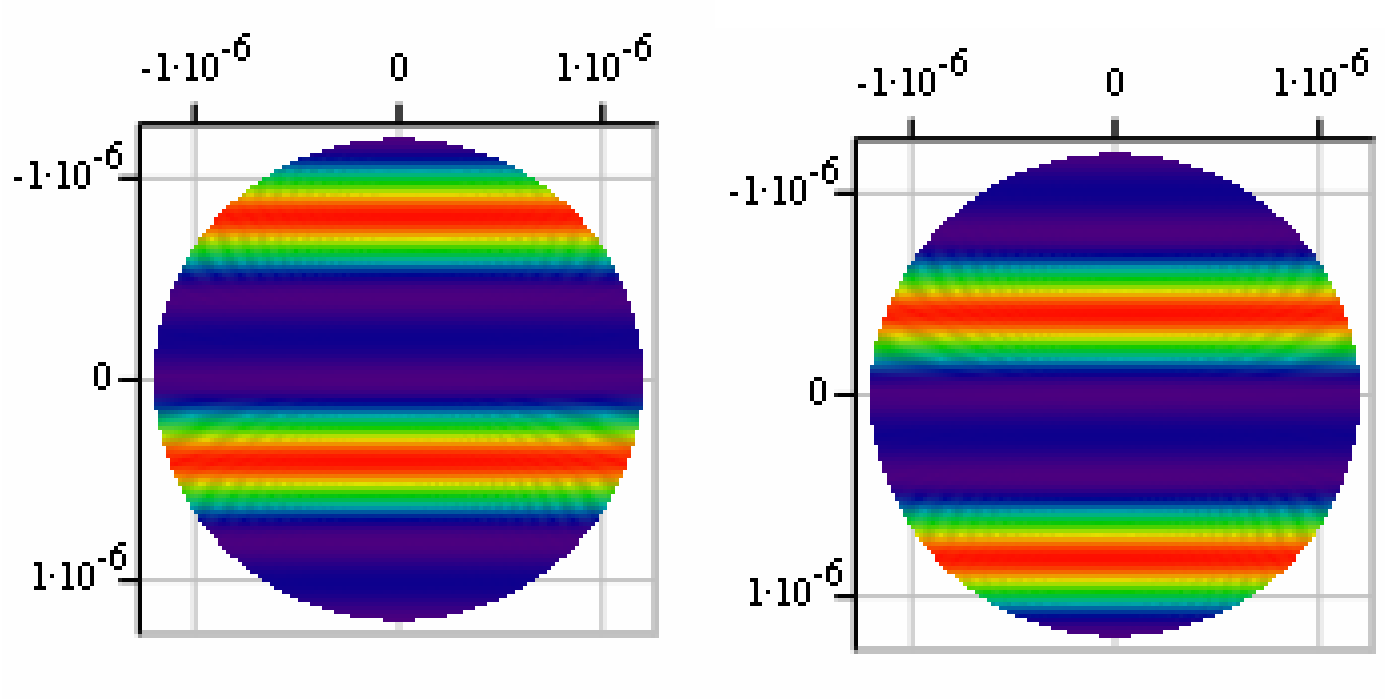

**Fig. 120: Response of the interferometer in a linear aperture configuration, showing the two TMs. Red indicates peak transmission areas.**

Modulation between different sub-configurations is essential for the operation of the instrument, allowing detection of the planetary signal against the background sources e.g. local and exo-zodiacal clouds. The modulation map is created post-detection by subtraction of two TMs. Fig. 121 shows the modulation maps of the linear and triangular configurations. The position of an assumed planet as the array is rotated is indicated with a black circle. At the peaks the planetary signal is modulated with the maximum planetary modulation efficiency of 93.3%.

The detected planetary signal is proportional to the planet signal convolved with the modulation at that location times the efficiency of the optical system, see Fig. 128. Modulation between TMs removes the geometrical stellar leakage from the output signal. The geometrical stellar leakage is a function of the angular size of the star and the size of the array, while the instrumental leakage, associated with pointing errors, OPD errors, etc, is not necessary the same in both outputs and thus cannot be cancelled by modulation. A 360-degree rotation of the interferometer array around its axis will modulate the signal according to Fig. 123 as the planet goes through transmission maxima and minima. Taking the difference between the two generated output signals on the two detectors shows how strong the planetary signal is modulated.

Fig. 121 shows different presentations of the modulation maps used: the left figure shows the orbit of a planet represented by a circle (in polar coordinates), while on the right side the orbit is represented by a straight vertical line at the according radial distance (in a Cartesian coordinate frame). Both graphs show the same scenario, a Sun-like star seen from 10pc distance and a planetary orbit at 1 AU. Fig 130 shows that depending on the field of view (FoV) also other planets will be detected simultaneous and the signals from multiple planets need to be disentangled through multiple observations of the same system at different times in their orbits. Alternatively one can resize the array to be able to distinguish the signal from planets at different orbital distances and signal strengths, see section 10.4. The FoV will be determined by the single mode fibre used to filter the wavefront.

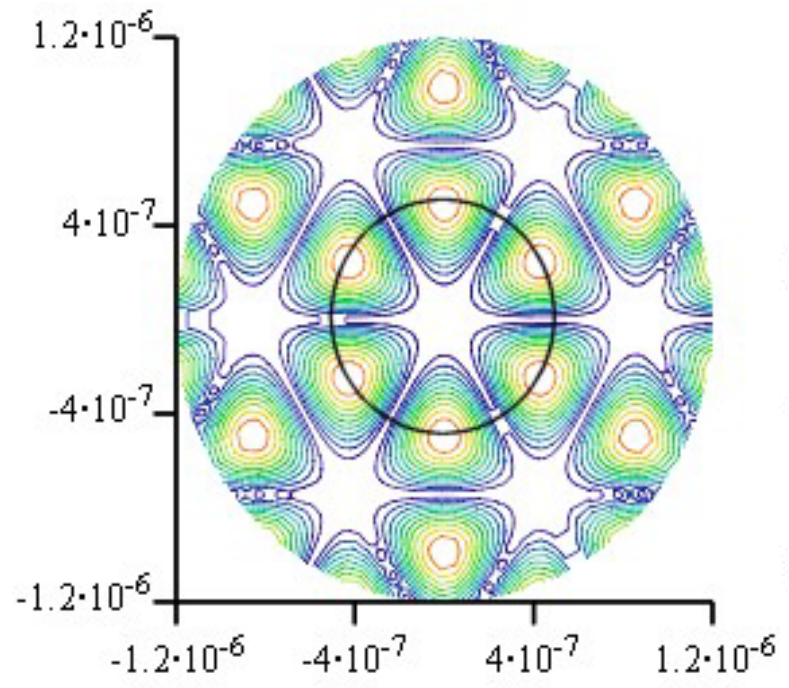

**Fig. 121: Modulation Map of the equilateral triangular TTN interferometer optimized for a 1AU orbit at 10pc at 10µm with a FOV of 2.5AU. Seen at a distance of 10pc 1AU is about 5 $10^{-7}$ rad. The black circle and line respectively, show the orbit of a planet at 1AU**

## 10.3 Planet location

To determine a planets location one needs to modulate its signal e.g. by rotation of the array, what results in a characteristic modulation of the planetary signal. For the linear TTN configuration can be determine the location of the planet by rotating the array by 180º. The symmetry of the interferometer response of the equilateral triangular configuration in comparison leads to three possible planet locations creating exactly the same output signal, but only needs a 60º rotation to cover the torus of the planetary orbit see Fig. 122. Note that the possible inclination of the system is not taken into account.

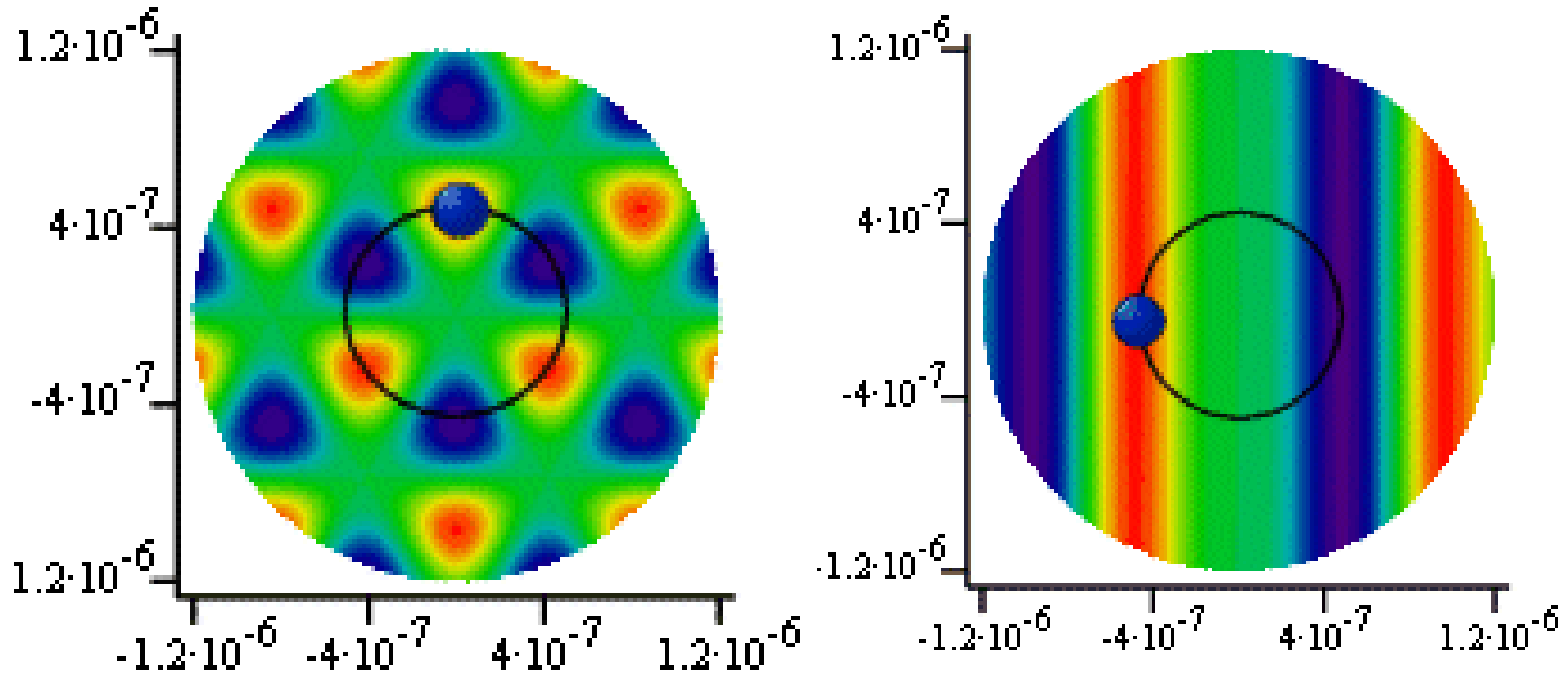

**Fig. 122: Modulation map generated by switching between the two nulled outputs generated on each of the two detectors transmission maps for the triangular and linear configuration of the TTN configuration (FoV 2.5AU at10µm). Note that green shows the zero modulation level**

This ambiguity in the determination of the planet's position due to the symmetry of the interferometer response for the equilateral triangular configuration is illustrated in Fig. 123. When using the linear configuration this ambiguity is removed. However, the MM of the linear configuration has only low spatial frequencies and does not allow accurate determination of planet position. The modulation map of the equilateral triangular configuration, on the other hand, has higher spatial frequencies.

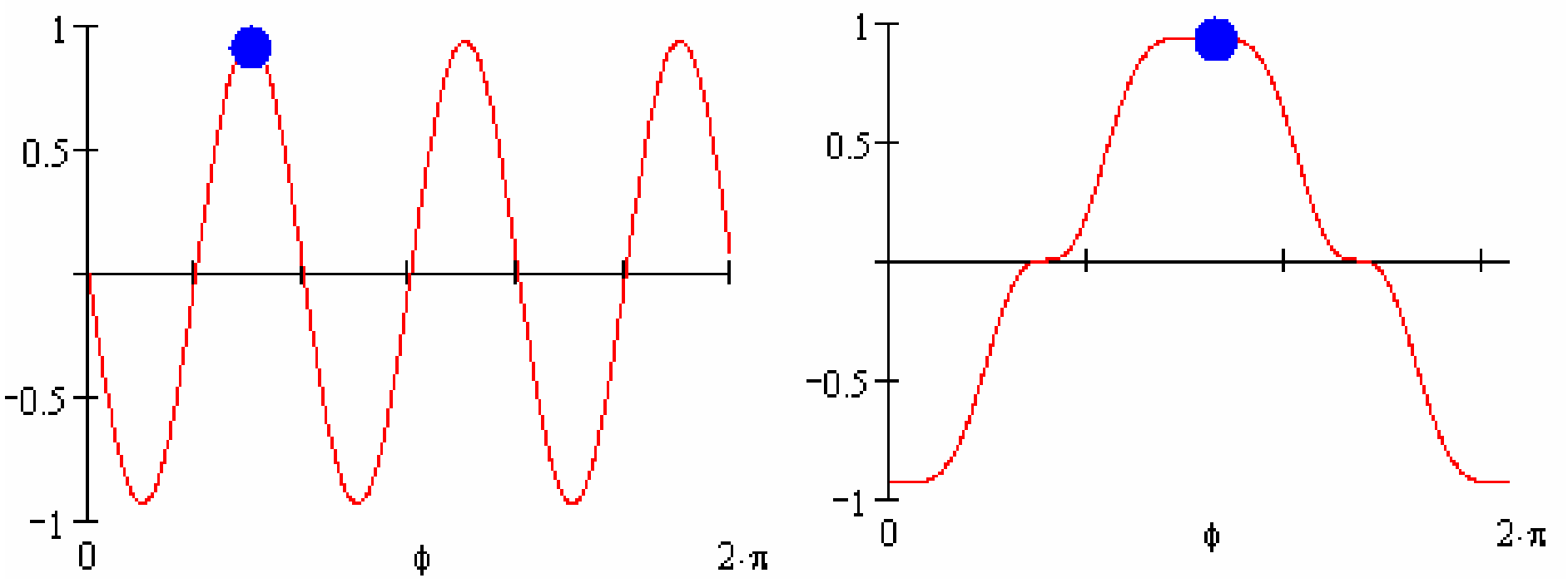

**Fig. 123: The modulated response of the triangular array when rotated 360° around its optical axis, considering a point source as shown in Fig.119 (left). The modulated response of the linear array when rotated 360° around its optical axis, considering a point source as shown in Fig.120 (right). The x-axis shows the array rotation while the y-axis shows intensity in arbitrary units.**

Note that the spacecrafts can be distributed arbitrarily around the beam combiner spacecraft, i.e. the telescope spacecrafts do not necessarily have to form an equilateral triangle.

Note that one has to break the central symmetry of the MM to be able to determine the planet's real location, thus e.g. an equilateral triangular TTN shows ambiguity in respect to the planet's position. To determine the exact planetary location one can move one telescope in respect to the rest of the array to disentangle the signal from planets at different orbital distance. The response of the interferometer when doing that is well known and creates a new modulation pattern that with the old one, can determine the planets position unambiguously.

## 10.4 Signals of multiple planets

If multiple planets are present, what we assume, all the planets will be detected simultaneous, contributing to the overall signal with different intensity. It is reasonable to assume that a target system will, in case it has planets, host multiple planets, i.e. a target system with a single planet forms an exception. All the planets will contribute to the overall signal of the interferometer, with different intensities. Note that even though Jupiter type planets are very large they are outside the FoV of the interferometer. On the other hand close in hot Jupiter type planets would contribute strongly to the signal.

The signal from multiple planets needs to be disentangled. One could use multiple observations of the same target system at different times assuming that the movement of the planets will differ due to their orbits. The observations should take place months apart, on base of our knowledge of our solar system. A potential problem could be the unknown variability of the planetary flux due to the planet's position in respect to the star at each consecutive observation. For a planet with an atmosphere that variation should be a minor factor, but for a planet with no atmosphere the effect will be notable, see Fig. 89. Fig. 124 shows the results of a simulation considering our own solar system with Venus, Mars and Earth, as seen from 10 pc. A strong planetary signal at the orbit of Venus can easily be confused with an Earth signal if only one spectral channel is used. Note that all figures show a monochromatic signal for simplicity.

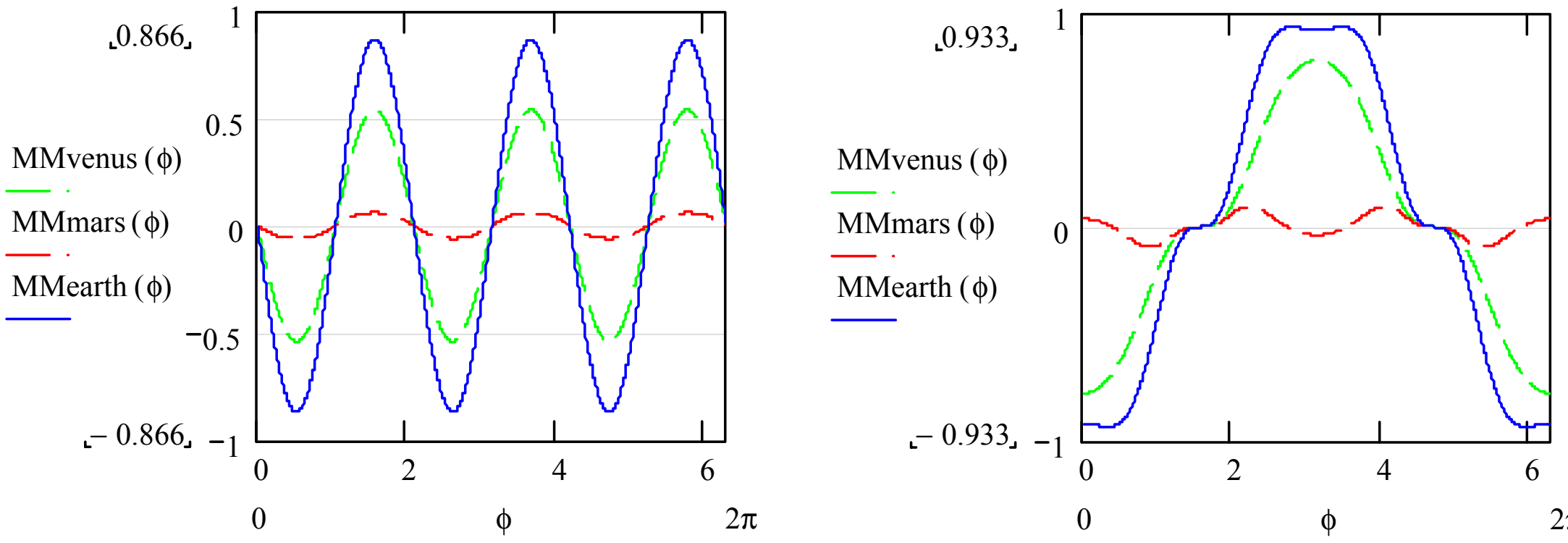

**Fig. 124: Modulated signal response of a triangular (left) and linear TTN (right) architecture of the signal of Venus, Earth and Mars for a Sun-equivalent star at 10pc, seen at 10µm. The symmetry of the detected signal shows that, depending on the relative positions of the planets their signals may be difficult to separate in one spectral channel. Note that the planets were assumed to be aligned, i.e. the worst case for any configuration.**

One unknown component in the calculations is the flux received from each of the multiple planets. Even so that parameter will be unknown, it will be constant as the planet will in first approximation not move while we e.g. resize the baseline of the array if the information from different spectral channels is not sufficient to disentangle the signal. If we revisit the system after a few month that variation will have an additional effect of variability on the detected signal.

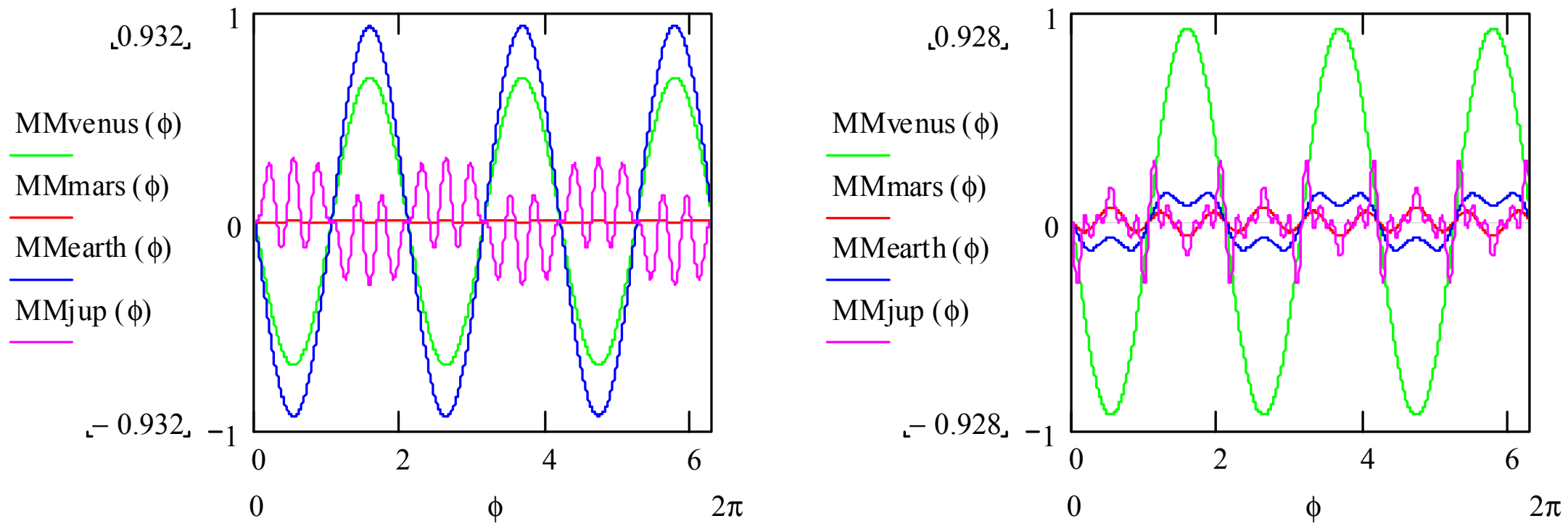

**Fig. 125: Response of an equilateral triangular TTN array in two different spectral channels to multiple planets (signal of Venus, Earth and Mars for a Sun-equivalent star at 10pc). The spectral information allows us to disentangle the signals from the planets in our solar system.**

Disentangling the signals from multiple planets has yet not been studied in detail. However, it is expected that a number of techniques could be used to separate the signals, e.g.

- repeated observations separated by several months,
- using the spectral information recognizing that the interferometer's modulation map scales with wavelength,
- using different configurations, and thus modulation maps,
- resizing the array, i.e. scaling the modulation map

A potential problem could be the unknown variability of the planetary flux due to the planet's position in respect to the star at each consecutive observation. For a planet with an atmosphere that variation should be a minor factor, but for a planet with no atmosphere the effect is expected to be notable.

## 10.5 Wavelength dependence

One has to keep in mind that the characteristics of an interferometer like the transmission and modulation map depend on observation wavelength as well as the baseline $\propto\lambda/B$. Fig. 126 and Fig. 127 show that dependence.

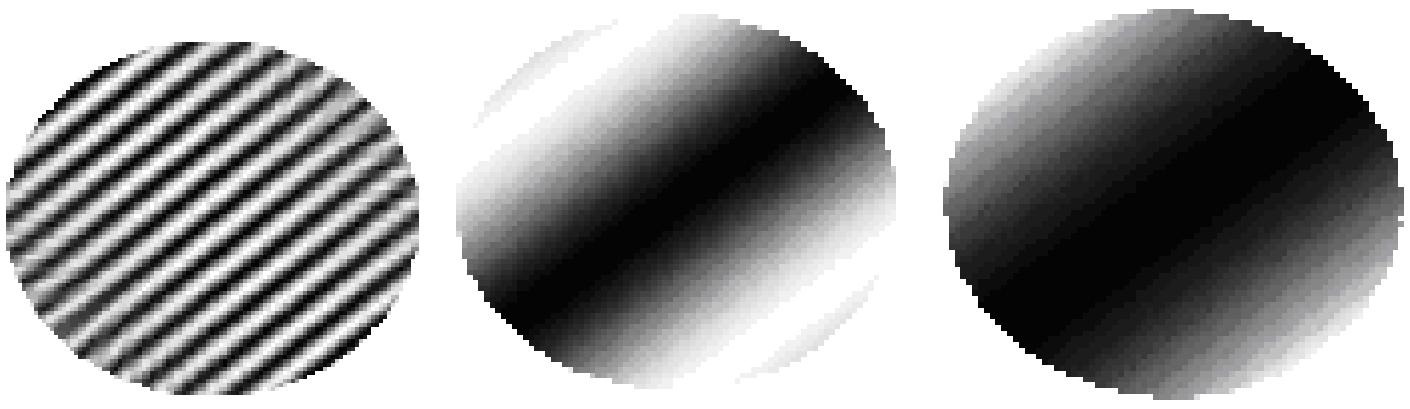

**Fig. 126: Transmission map of a Bracewell array at different wavelengths (0.1µm, 1µm, 10µm) for a fixed baseline.**

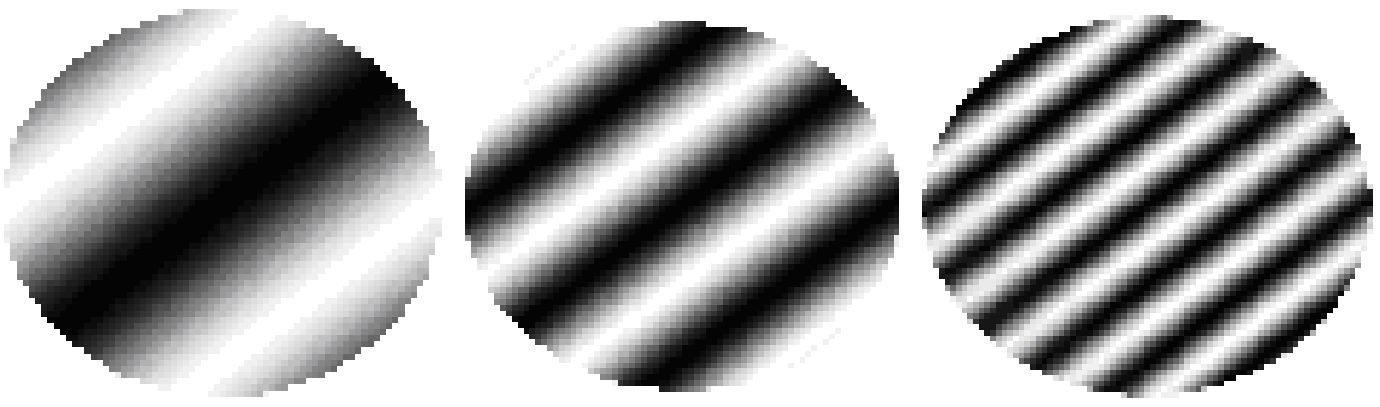

**Fig. 127: Transmission map of a Bracewell array at 10µm for different baselines (10m, 20m, 40m)**

Due to the influence of the observation wavelength and baseline of the interferometer, it is important to optimize the configuration for the individual target systems. This is one of the advantages of a free flying array in respect to a structural connected array, the configuration can be optimized for each observed target system. Using the DARWIN target catalogue we calculate the optimized baselines for a wavelength of 10 µm for the proposed configurations.

Fig. 128 shows response of a planet at a 265K BB at a 1AU orbit over wavelength, it also shows that Jupiter and Earth-like planets will show very different characteristics, even if only limited colour information from 2 or 3 spectral channels is initially available (e.g. in the detection phase, two spectral channels or observations with two different baselines would yield that information).

The interferometer will be rotated around the line of sight (LoS) during observations, in the detection phase possibly also in the spectroscopy phase of the mission. For all aperture configurations, a rotation of 180° will be sufficient to produce the characteristic modulation, see Fig. 122. The equilateral triangular configuration would only require a 60° rotation.

Depending on the planned observation the choice of baseline is different, e.g. during the planet detection phase of the mission a baseline that allows detection of planets in the whole HZ is desired, using all wavelengths, while in the spectroscopy phase of the mission the planet location is known and an optimal baseline can be sought for a specific waveband, e.g. the $H_2O$ absorption line.

During the detection phase the baseline is adjusted such that a source in the HZ produces a signal over the whole waveband, i.e. the strongest possible total signal, see Fig. 129. For nearby targets, the optimal baseline leads to a distance shorter than the min. ISD. Therefore the second ring of maxima in the MM has to be used, corresponding to larger ISDs. Unfortunately a large baseline means for a certain waveband the transmission will be zero. To cover the complete wavelength band, for the nearby targets, at least two baseline settings are required for a central symmetric configuration like e.g. the equilateral triangular TTN, see Fig. 128. Note that Fig. 128 represents only one 2D cut through the 3D modulation map. If the TM and thus also the MM is not central symmetric, rotation of the array can improve the modulation at certain wavelengths. For instrument performance calculations that will e.g. show an advantage of the linear TTN versus the equilateral triangular TTN configuration. The MM of the linear TTN is not central symmetric while the MM of the equilateral triangular TTN is. These two cases are only the two extreme cases of many possible 3 telescope arrangements of telescopes, the telescope spacecrafts can be arranged anywhere around the beam combiner. Configuring them on a circle will avoid the need for reflections or a long delay line on board of one or two of the telescope spacecrafts.

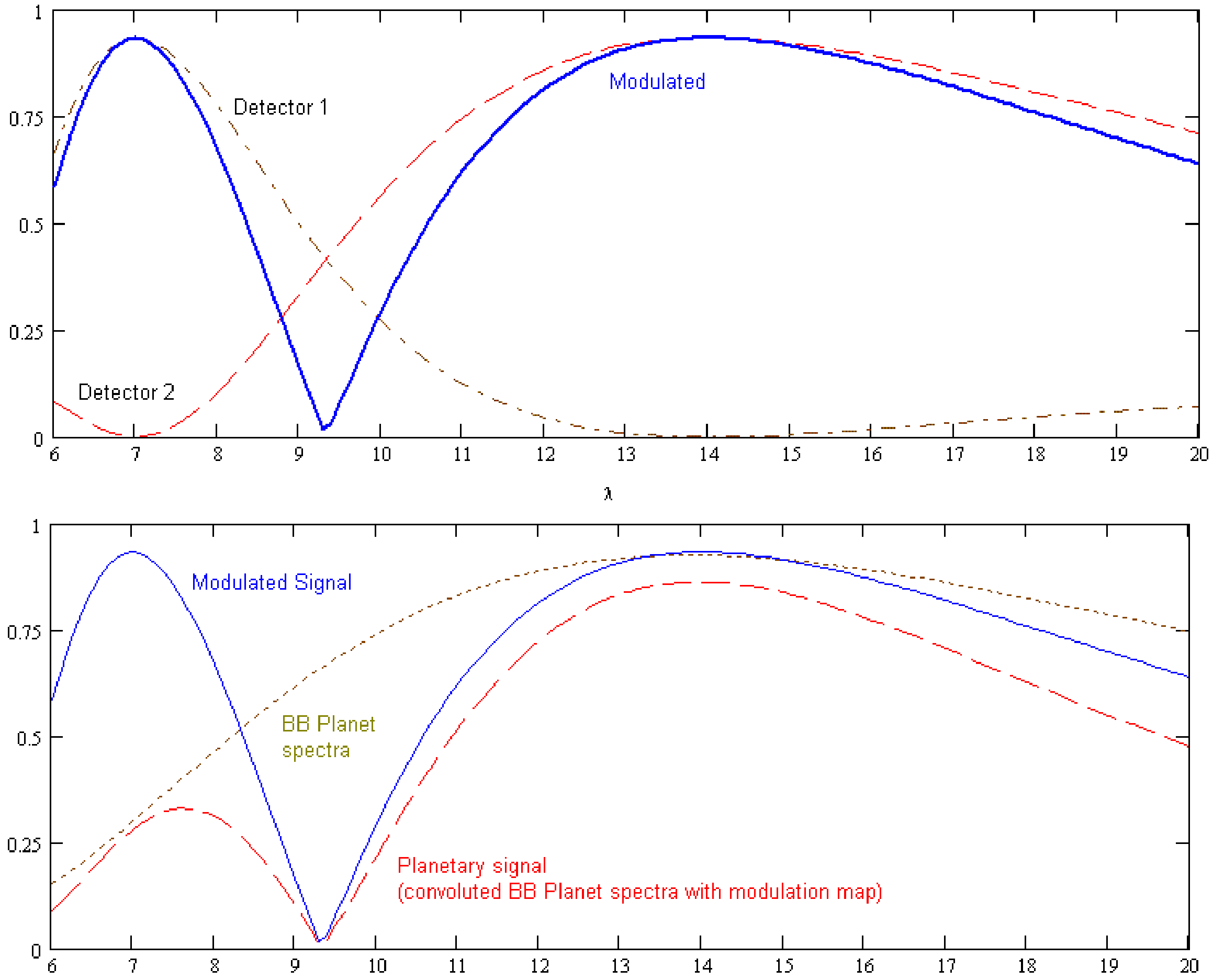

**Fig. 128: Response on each detector and resulting modulation signal of the array over wavelength (upper panel). Modulation signal and resulting planetary signal assumes a BB spectrum of the planet. The short wavelengths give less signal due to the BB curve of the planet. Note that the array is optimized for high transmission at small wavelength (where very low flux levels are detected due to the BB spectra of the planet), as a result the array does not provide full transmission over the whole waveband. Transmission over the full waveband can be provided if the array is not optimized for flux at the lower wavelength.**

Coverage of the whole wavelength band with high transmission might not be necessary as part of the waveband is only needed to establish the Black Body temperature of the planet. In that part a small region with low transmission could be tolerated.

The MM of the nulling interferometer is scaled by adjustment of the baselines, i.e. increasing the baselines shrinks the features of the MM. This could be used as a means of modulation, possibly in combination with array rotation. However, this modulation has not been addressed in the present work. Adjustment of the baseline is possible without loosing fringe lock, avoiding repetition of the procedure of fringe lock acquisition.

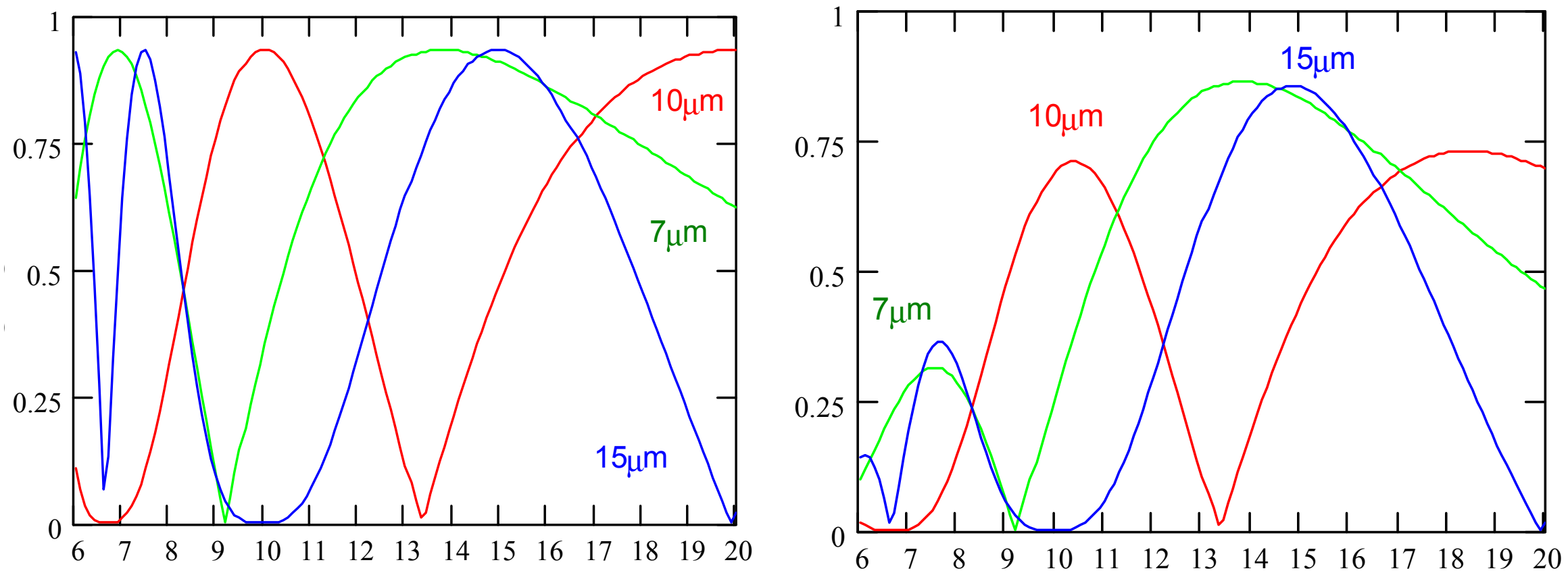

**Fig. 129: Modulation response of the Interferometer array for a point source in the HZ, for a nearby target system. Due to ISD considerations the baseline is non-ideal, producing a waveband with low response. Therefore a set of three baselines is chosen (blue, red, green lines). The modulated response of the interferometer (left) and the planetary flux detected (right) is shown. The same effect is seen for any stellar distance if the array is optimized for the flux at small wavelengths.**

Fig. 129 shows that the optimized baseline of the array is for one wavelength and its multiple only, thus the maximum transmission of the planetary photons will only be achieved over a certain wavelength band. A planet in Fig. 129 lies on a modulation maximum at e.g. 7.5μm and at 15μm (blue line). For most of the waveband the modulated strength of the planetary signal is lower, what would lead to lower signal level at those wavebands, an effect that could be countered with longer integration time. Fig. 129 shows Zero modulation for a wavelength of 10μm (blue line). Thus to get any signal at that wavelength the baseline has to be changed if the MM is central symmetric, if not, the array can be rotated to get a signal at that wavelength.

Due to this effect the flux detected by the array using a certain baseline will correspond to the flux from certain wavebands. That could be used to retain very crude colour information on the detected planets. The colour information could be used to make first estimates of the planet's composition. Time will be needed to reconfigure the array, thus an architecture of the array that needs the least number of different baselines to detect a signal over the complete observed wavelength range will be favourable compared to one that needs more than that. Fig. 129 demonstrates that for the detected photons from a planet also its behaviour over wavelength has to be considered. It also shows that in the shorter wavelength range, more time will be needed to achieve a certain SNR.

## 10.6 Angular extent of the Habitable zone for the prime DARWIN targets

The array architecture has to be optimized to detect planets in the Habitable zone around the selected target stars. For calculations of the HZ around the prime DARWIN target stars (excl multiple systems) we use: $0.8\ a_{Hd} < d < 1.7\ a_{Hd}$.

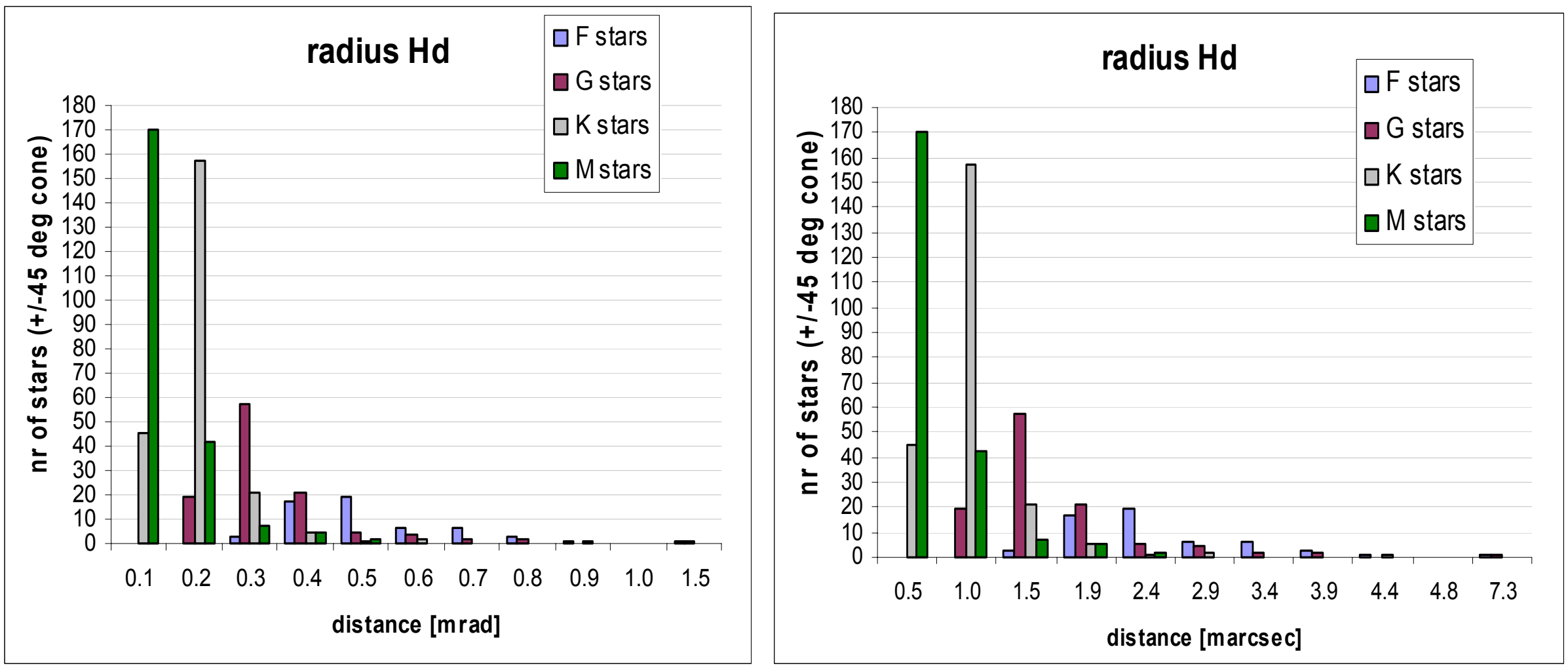

**Fig. 130: Radius of the Habitable Distance of the DARWIN prime target stars (Kaltenegger 2004)**

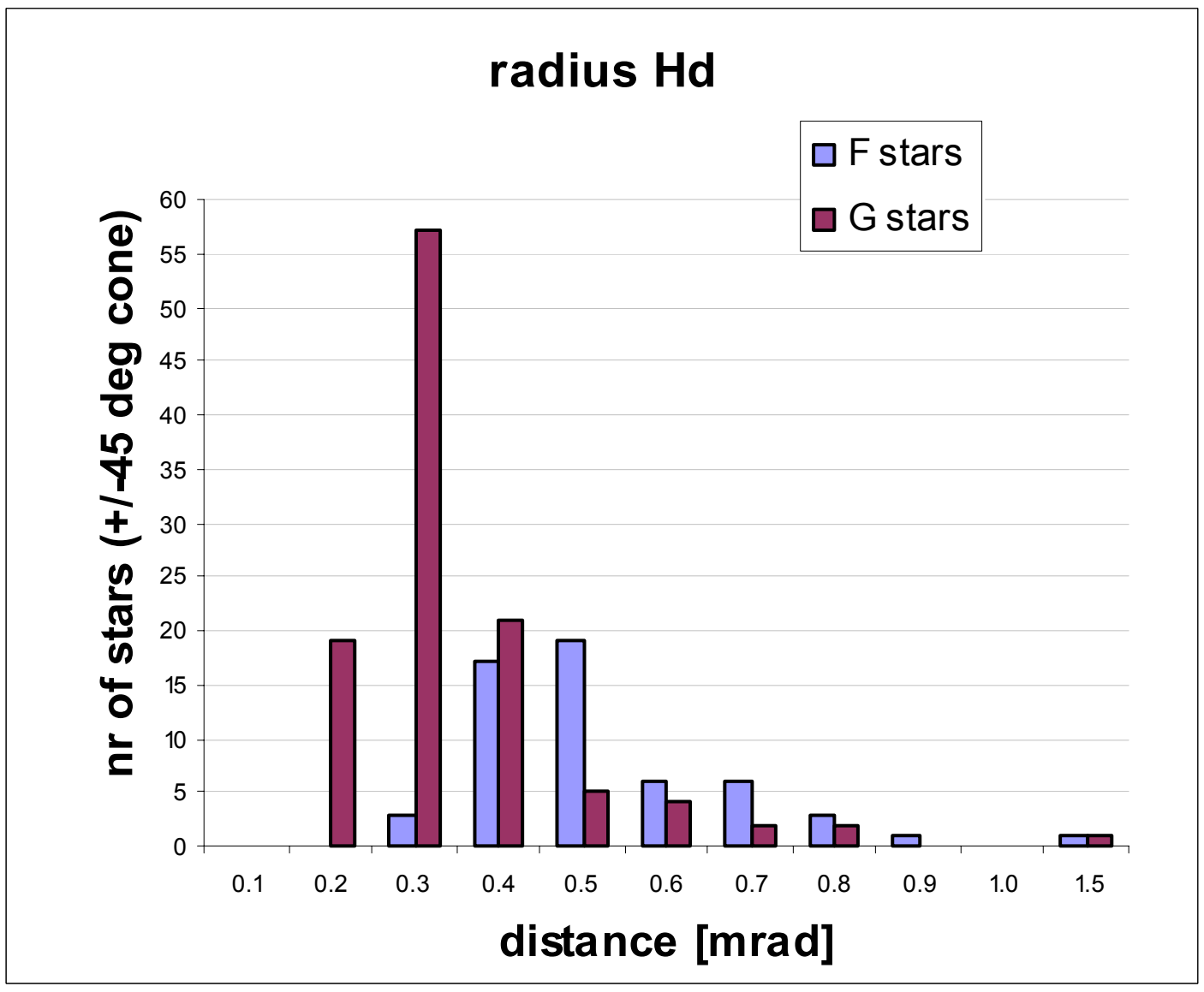

**Fig. 131: Radius of the Habitable Distance of the DARWIN prime F and G target stars (Kaltenegger 2004) (1arcsec = 4.848 mrad)**

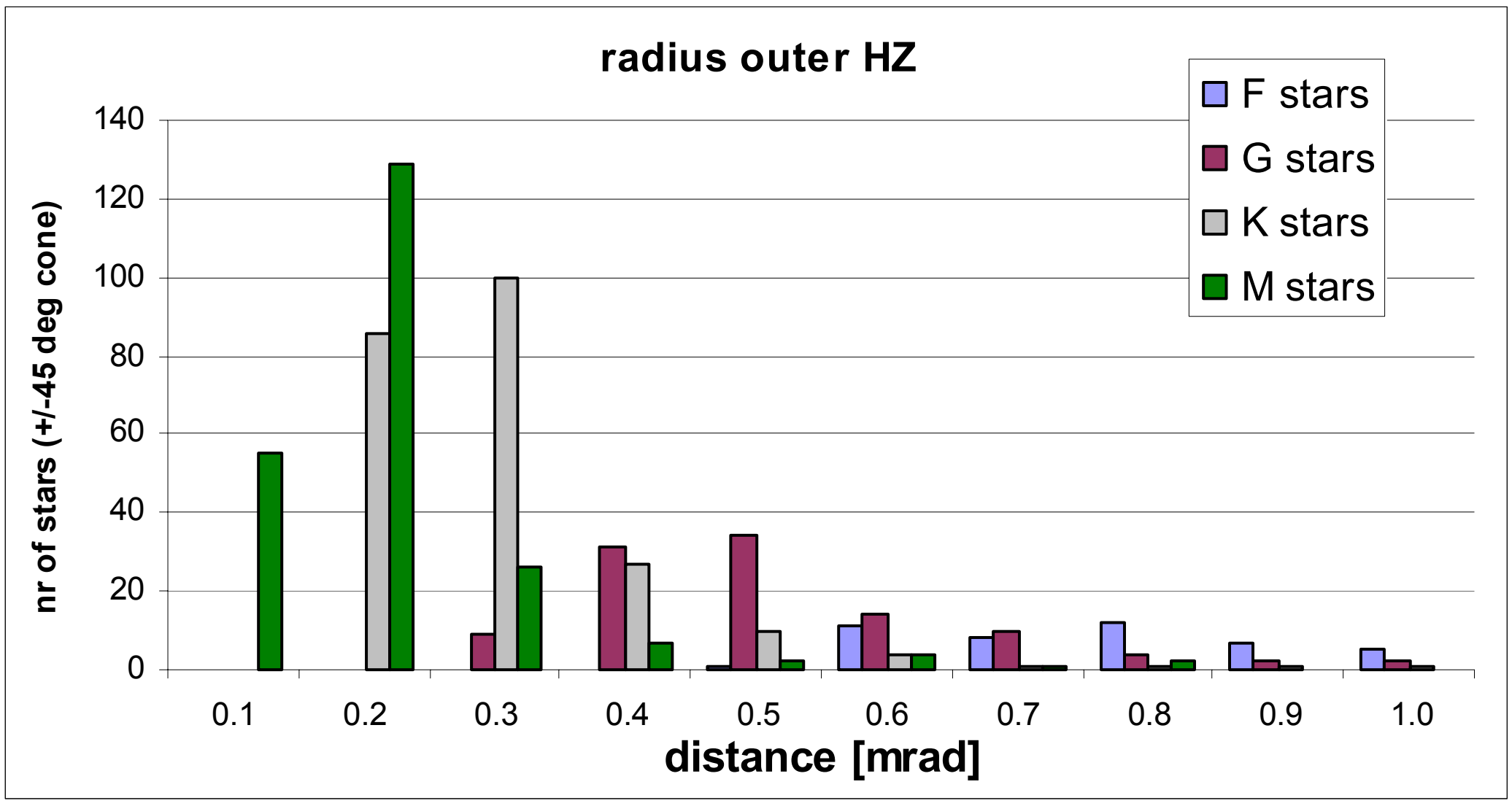

**Fig. 132: Outer radius of the Habitable Zone of the DARWIN prime target stars (Kaltenegger 2004) (1arcsec = 4.848 mrad)**

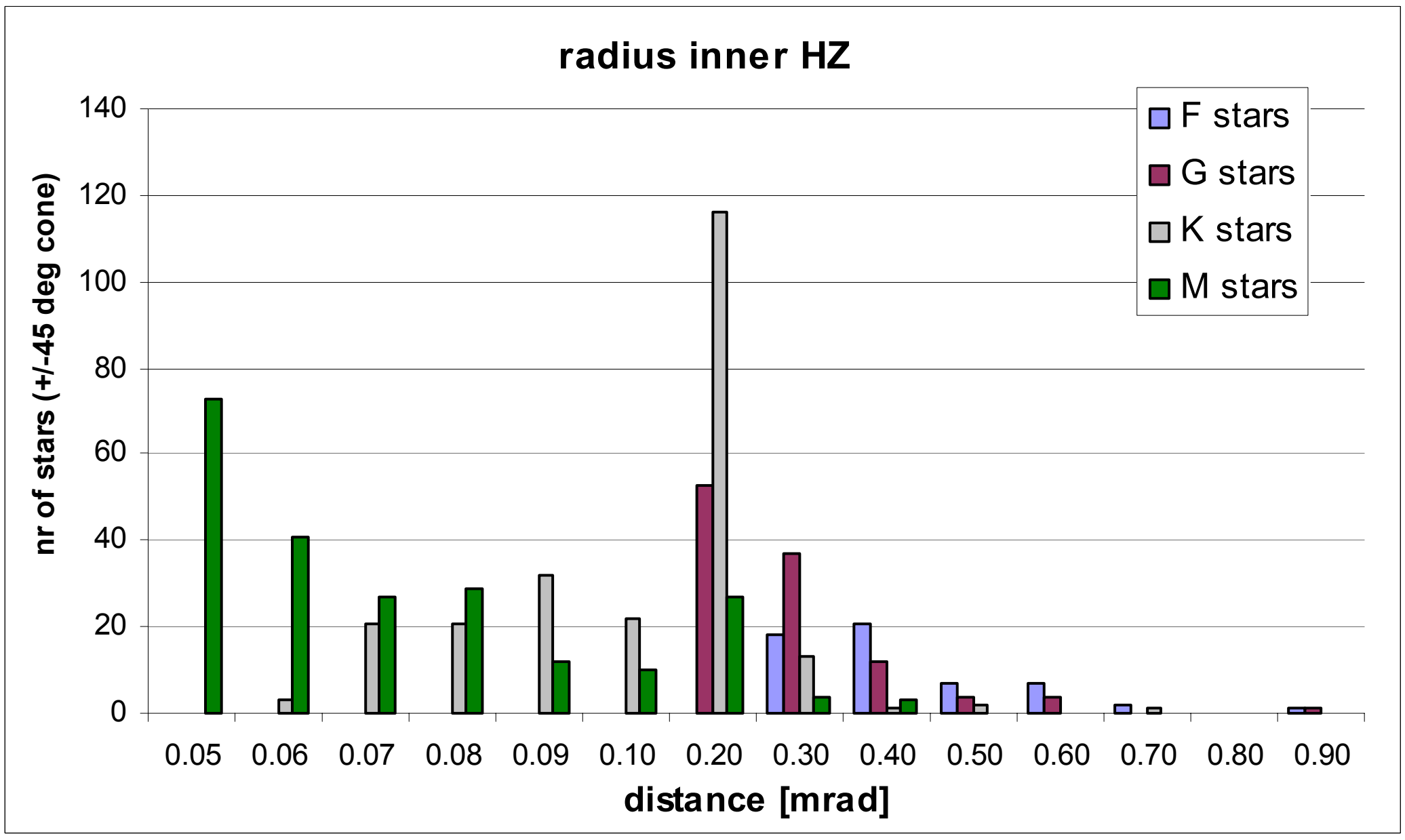

**Fig. 133: Inner radius of the Habitable Zone of the DARWIN prime target stars (Kaltenegger 2004) (1arcsec = 4.848 mrad)**

## 10.7 Overview Configurations

This section shows the characteristics for 5 different proposed DARWIN configurations giving comparable values for the efficiency for planetary signal modulation, starlight rejection, necessary rotation angle to cover the HZ evenly etc. The configurations using 3 telescopes are: Three Telescope Nuller (TTN) and Triple chopped Bracewell (TCB), using 4 telescopes: Linear Double Chopped Bracewell (LinDCB) and Modified Double Chopped Bracewell (MDCB). Note that for simplicity all TM and MM in this chapter are drawn as their absolute value. Thus a planet on the height of transmission on detector one will show the same positive value as the planet when it is on a height of transmission on detector two. In the previous chapter we had shown detector one as the positive and detector two as the negative value (as we subtract the detected signal from detector two from the signal from detector one to generate the modulation pattern). In the following graphs 0 is the minimum value.

An overview of different proposed concepts is shown in Table 23 and Table 24. Tel denotes the number of telescopes used for each configuration. An array with 6 telescopes like the Bowtie achieves better performance in starlight rejection than an array with 3 or 4 telescopes, an effect that can be seen in Table 23 but adds complexity and cost to the mission. If continuous observation during rotation of the array is not possible we determined the number of different modulation maps generated through rotation of a set predetermined angle, that are needed to provide a min transmission of 50% of the maximum achievable modulation for a planetary signal of that configuration.

Baselines for the DARWIN investigated architectures are:

1) telescopes on a circle/ passing beams via third spacecraft to equalize optical path length (OPL) to avoid long delay lines
2) equal size telescopes to reduce manufacturing costs
3) starlight suppression proportional to $\theta^x$ ($\theta^4$ better than $\theta^2$) to minimize stellar leakage and thus minimize integration time per target (effect important for close target stars)
4) high mean modulation efficiency for planetary photon transmission in the detection phase, important when the planet position is not known, to screen effectively
5) high peak modulation efficiency for planetary photon transmission in the spectroscopy phase, when location of the planet is known and can be put on high transmission peak (torus if the array is rotated)
6) narrow modulation peaks to locate a detected planet
7) simple BC scheme, higher order leads to higher complexity and light loss

Bowtie: Industrial study (2000) baseline configuration, based on:

- telescopes on a circle to avoid long delay lines
- equal size telescopes
- starlight suppression proportional to $\theta^4$ for each subinterferometer
- additional spacecraft for beam combination
- 70% peak modulation efficiency
- 16% mean modulation efficiency
- 6 telescopes and one beam combiner
- 2 subinterferometers
- internal modulation: 3rd order pupil-plane recombination scheme
- high beam combination complexity

MDCB: Modification of the standard DCB scheme, based on:

- parallelogram configuration (60º angles),
- equal size telescopes
- hub implemented in one telescope free flyer,
- optical path lengths equalized by passing beams via third spacecraft
- starlight suppression proportional to $\theta^2$ for each subinterferometer
- 100% peak modulation efficiency

- 26% mean modulation efficiency
- 4 telescopes and one beam combiner
- 2 subinterferometers
- internal modulation: 2nd order pupil-plane recombination scheme
- increased beam relay complexity, magnification of attitude errors

LinDCB: Modification of the standard DCB scheme, based on:
- linear arrangement of telescopes
- equal size telescopes
- starlight suppression proportional to $\theta^2$ for each subinterferometer
- 100% peak modulation efficiency
- 35% mean modulation efficiency
- 4 telescopes and one beam combiner
- 2 subinterferometers
- internal modulation: 2nd order pupil-plane recombination scheme

TTN, no beam combiner spacecraft: based on:
- equilateral triangle configuration,
- telescopes on a circle
- equal size telescopes
- hub implemented in one telescope free flyer
- optical path lengths equalized by passing beams via third spacecraft
- inherent modulation using multibeam injection into two single-mode fiber
- starlight suppression proportional to $\theta^2$ for each subinterferometer
- 93% peak modulation efficiency
- 32% mean modulation efficiency
- no hub, efficient beam combination
- Increased beam relay complexity, magnifies attitude errors
- 3 telescopes
- 2 subinterferometers

TTN with a beam combiner spacecraft: based on:
- equilateral triangle configuration,
- telescopes on a circle to avoid long delay lines
- equal size telescopes
- inherent modulation using multibeam injection into two single-mode fiber
- starlight suppression proportional to $\theta^2$ for each subinterferometer
- 93% peak modulation efficiency
- 32% mean modulation efficiency
- 3 telescopes and one beam combiner
- 2 subinterferometers

TCB: based on:
- equilateral triangle configuration,
- hub implemented in one Telescope Free Flyer,
- optical path lengths equalized by passing beams via third spacecraft
- internal modulation: 2nd order pupil-plane recombination scheme
- no hub, efficient beam combination
- starlight suppression proportional to $\theta^2$ for each subinterferometer
- Increased beam relay complexity, magnifies attitude errors
- 65% peak modulation efficiency
- 20% mean modulation efficiency
- 3 telescopes
- 3 subinterferometers

TTN configurations: outputs can also be generated consecutively, in that case modulation efficiency is ½ of the values quoted above, as each output can only be observed 50% of the time. The telescopes in the TTN configuration can also e.g. be aligned in a linear configuration that has a none central symmetric MM see Fig. 124.

**Table 23: Overview and comparison of different proposed DARWIN configurations**

| Configurations | Tel | Planet signal mod. | | | Starlight rejection |
|---|---|---|---|---|---|
| | | max | mean | Sdv | |
| BOWTIE | 6 | 0.67 | 0.16 | 0.26 | 1.49E+09 |
| MDCB | 4 | 1.00 | 0.26 | 0.28 | 1.87E+04 |
| TTN inherent(t1,t2) | 3 | 0.50 | 0.16 | 0.14 | 2.10E+04 |
| TTN inherent (t1) | 3 | 0.93 | 0.32 | 0.27 | 2.10E+04 |
| TCB | 3 | 0.65 | 0.20 | 0.19 | 1.49E+05 |
| LinDCB | 4 | 1.00 | 0.35 | 0.36 | 1.87E+04 |

| Configurations | Tel. | Coverage of full planet orbit mod > 50% maxmod | | optimized baseline (10μm, system at 10pc) | |
|---|---|---|---|---|---|
| | | rotation angle | nr of maps | No BC | BC |
| BOWTIE | 6 | 5π/6 | 4 | 23.8 | 11.9 |
| MDCB | 4 | π | 3 | 13.6 | 6.8 |
| TTN inherent(t1,t2) | 3 | π/3 | 2 | 13.7 | 7.9 |
| TTN inherent (t1) | 3 | π/3 | 2 | 13.7 | 7.9 |
| TCB | 3 | π/3 | 2 | 8.4 | 5.9 |
| LinDCB | 4 | π | 2 | 5.1 | 5.1 |

Table 23 shows that the optimized baseline and thus the minimum intersatellite distance (ISD) of the array is wavelength dependent, thus the maximum transmission of the planetary photons will only be achieved over a certain wavelength band. Note that the BC is smaller than a telescope thus the ISD could be smaller than between the telescope free flyers.

Table 23 shows the evaluation of the alternative mission architectures. Even so the starlight rejection properties of the BOWTIE configuration is outstanding in the comparison, the higher mean modulation efficiency of e.g. the TTN configuration counteracts that superiority. An additional factor in mission design and complexity is the number of telescopes used, here the BOWTIE has a big disadvantage, as it needs 6 telescopes. Its beam combination scheme is also highly complex because it needs 3 steps. Especially architectures with less telescope spacecrafts allow for bigger telescopes to be launched in a similar launch scenario. Table 23 shows that a reduced science mission goal should be achievable with a lower level of complexity compared to the BOWTIE configuration.

In more detailed simulations we have compared these architectures for integration time and number of stars observable in a 5 years mission lifetime. The detailed simulations have confirmed that a reduced science mission goal is achievable with a lower level of complexity as compared to the current baseline configuration of 6 telescopes of 1.5m diameter. If the diameter of the telescopes is increased the full science goal can be achieved. We propose a diameter of 3.5m for e.g. a TTN configuration.

**Table 24: Basic Parameters of the DARWIN configurations**

| **Configuration** | | **Tel.1** | **Tel.2** | **Tel.3** | **Tel.4** | **Tel.5** | **Tel.6** |
|---|---|---|---|---|---|---|---|
| BOWTIE (6Tel) | | | | | | | |
| | **L** | 1 | 1 | 1 | 1 | 1 | 1 |
| | $\phi$ | 7π/4 | 0π/4 | 1π/4 | 3π/4 | 4π/4 | 5π/4 |
| *SUB-IF 1* | ***A1*** | *1* | *1/√2* | *0* | *0* | *1/√2* | *1* |
| *SUB-IF 2* | ***A2*** | *0* | *1/√2* | *1* | *1* | *1/√2* | *0* |
| *Applied phaseshift* | $\sigma_k$ | *0* | *π* | *0* | *π* | *0* | *π* |

| **Configuration** | | **Tel.1** | **Tel.2** | **Tel.3** | **Tel.4** |
|---|---|---|---|---|---|
| ModifiedDCB (4Tel) | | | | | |
| | **L** | 1 | √3/2 | 1 | √3/2 |
| | $\phi$ | 0 | π/2 | π | 3π/2 |
| *SUB-IF 1* | ***A1*** | *1* | *0* | *0* | *1* |
| *SUB-IF 2* | ***A2*** | *0* | *1* | *1* | *0* |
| *Applied phaseshift* | $\sigma_k$ | *0* | *0* | *π* | *π* |

| **Configuration** | | **Tel.1** | **Tel.2** | **Tel.3** | **Tel.4** |
|---|---|---|---|---|---|
| LinDCB | | | | | |
| | **L** | 1.5 | 0.5 | 0.5 | 1.5 |
| | $\phi$ | 0 | 0 | π | π |
| *SUB-IF 1* | ***A1*** | *1* | *0* | *1* | *0* |
| *SUB-IF 2* | ***A2*** | *0* | *1* | *0* | *1* |
| *Applied phaseshift* | $\sigma_k$ | *π* | *π* | *0* | *0* |

| **Configuration** | | **Tel.1** | **Tel.2** | **Tel.3** |
|---|---|---|---|---|
| TTN inherent | | | | |
| | **L** | 1 | 1 | 1 |
| | $\phi$ | 0 | 2π/3 | 4π/3 |
| *SUB-IF 1* | ***A1*** | *1* | *1* | *1* |
| *Applied phaseshift t1* | $\sigma_{k1}$ | *0* | *2π/3* | *4π/3* |
| *SUB-IF 2* | ***A2*** | *1* | *1* | *1* |
| *Applied phaseshift t2* | $\sigma_{k2}$ | *4π/3* | *2π/3* | *0* |

| **Configuration** | | **Tel.1** | **Tel.2** | **Tel.3** |
|---|---|---|---|---|
| TCB | | | | |
| | **L** | 1 | 1 | 1 |
| | $\phi$ | 0 | 2π/3 | 4π/3 |
| SUB-IF 1 | ***A1*** | 1/√2 | 1/√2 | 0 |
| Applied phaseshift | $\sigma_{k1}$ | 0 | π | 0 |
| SUB-IF 2 | ***A2*** | 0 | 1/√2 | 1/√2 |
| Applied phaseshift | $\sigma_{k2}$ | 0 | π | 0 |
| SUB-IF 3 | ***A3*** | 1/√2 | 0 | 1/√2 |
| Applied phaseshift | $\sigma_{k3}$ | 0 | π | π |

*Distance from center…L, Angular position of Tel...$\phi$, Amplitude used of each telescope in the subIFx…Ax, Applied phaseshift per telescope in subIFx…$\sigma_k$*

### 10.7.1 Characteristics BOWTIE

The Bowtie configuration consists of six equal sized telescopes that form two sub-interferometers of four telescopes each, which amplitude fractions equal to $1/\sqrt{2}$ , 1, 1, $1/\sqrt{2}$ of the collected amplitude per telescope, respectively (Absil, 2000). The BOWTIE concept consists of two sub-interferometers and yields a starlight rejection of $1.49 \times 10^9$ at 10µm for a sun-like star at 10pc.

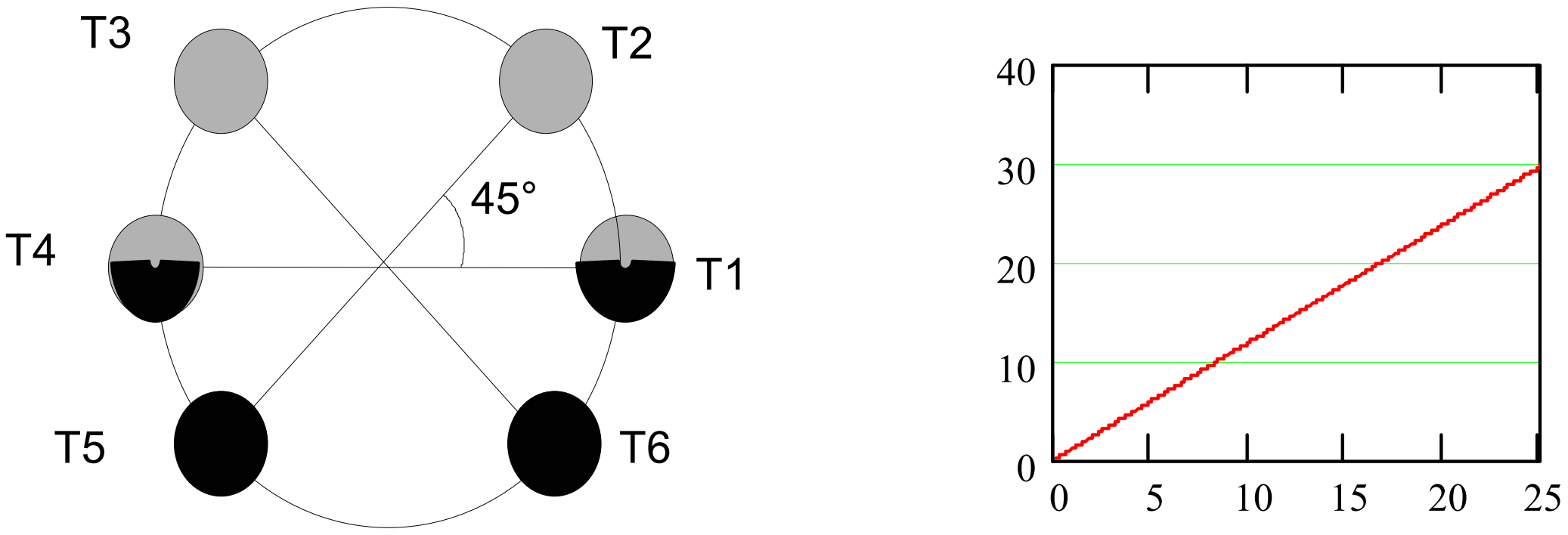

**Fig. 134: BOWTIE configuration used for calculation of the optimized distance of the telescopes from the central beam combiner that puts the planet on the first maxima of the Transmission map (right) (see text)**

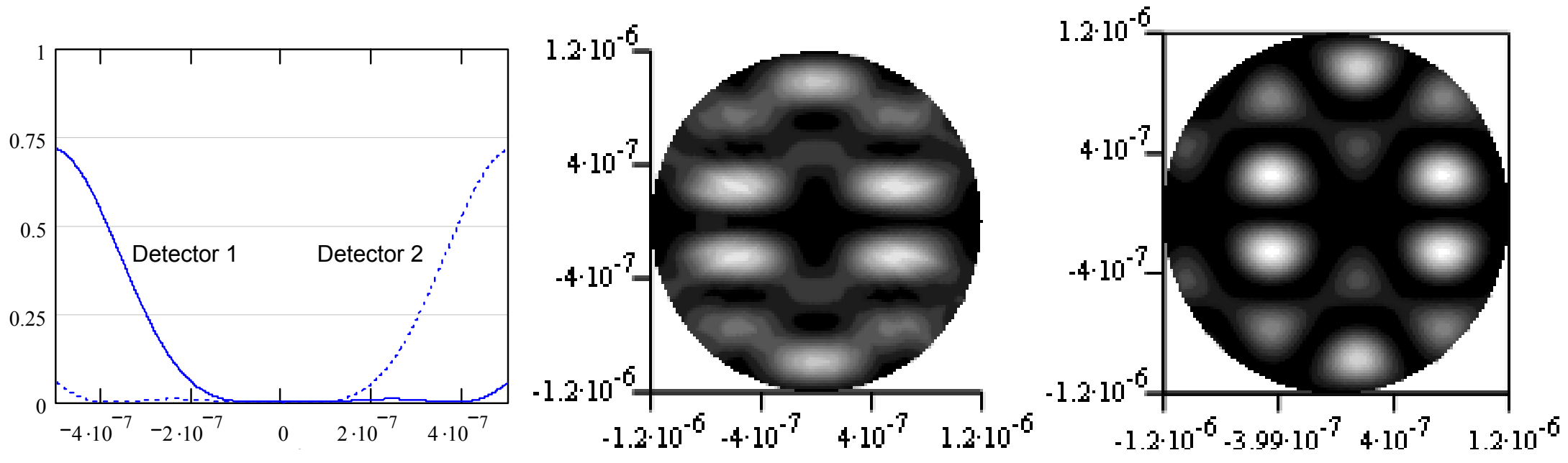

**Fig. 135: Shape of the Null along thedirection of the maximum over angular extend from the on-axis position to 1AU for each sub interferometer (left), transmission map used to generate the modulation map out to 2.5AU (middle), nulled output on one detector (right)**

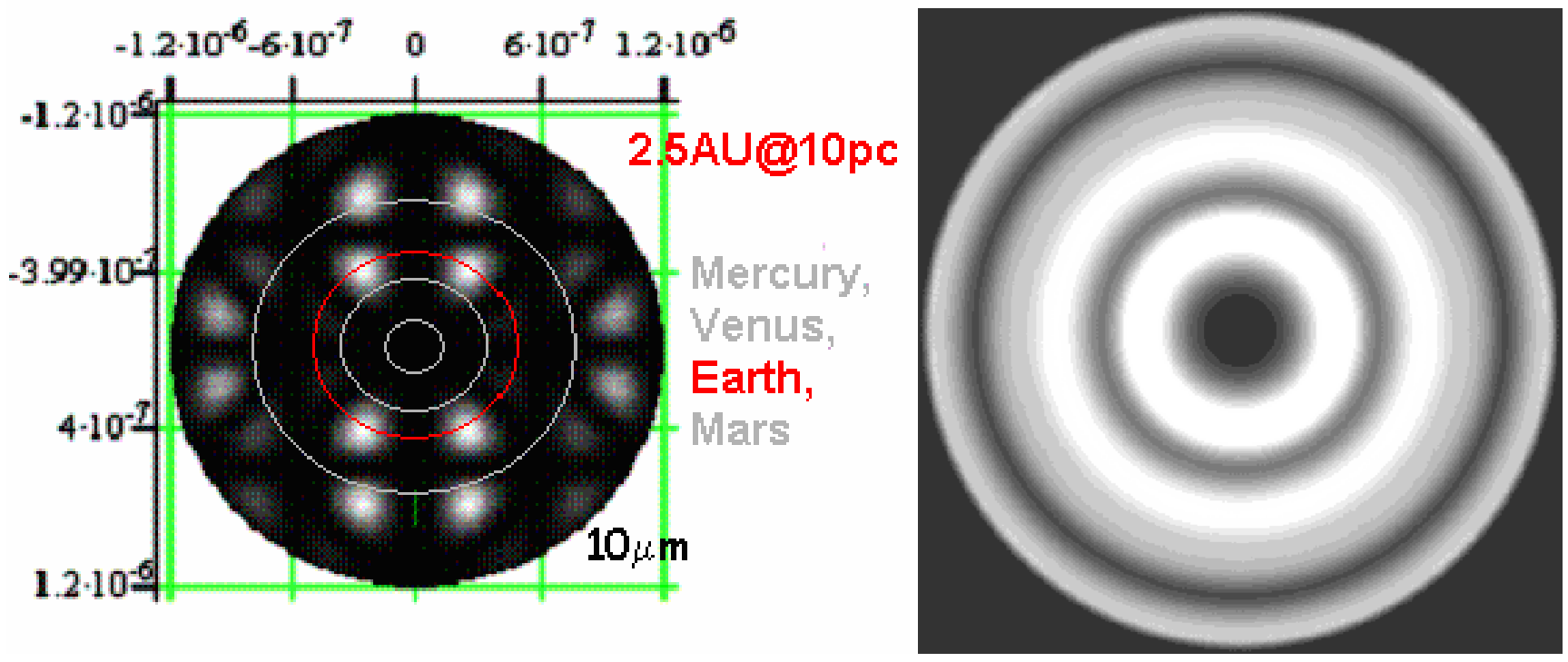

**Fig. 136: Coverage of the planetary orbit by rotation of the array: Observation without rotating the array (left, circles showing the orbit of Mercury, Venus, Earth and Mars), coverage assuming continuous observation during rotation of 2π/3 (right)**

Section 10 points out different issues involved in the evaluation of array architecture. The circles in Fig. 136 show the radial distance of planets from the star for a certain distance of the planetary system (here we show Mercury, Venus, Earth and Mars around our Sun seen from 10pc distance). The baseline of the array is chosen so that the peak of the modulation map is positioned at the location of an Earth-equivalent orbit for a wavelength of 10µm. For other wavelengths the peak of the transmission will not coincide with the Earth-equivalent orbit for the same baseline (see Fig.120). Depending on the field of view also other planets will be detected simultaneous and the signal from multiple planets needs to be disentangled through multiple observations of the same system at different times in their orbits. The modulation maps can be visualized in different coordinate systems. Fig. 137 and Fig. 138 shows different presentations of the modulation maps. In Fig. 137 the orbit of a planet is represented by a circle, while in Fig. 138 the orbit is represented by a straight vertical line at the according radial distance.

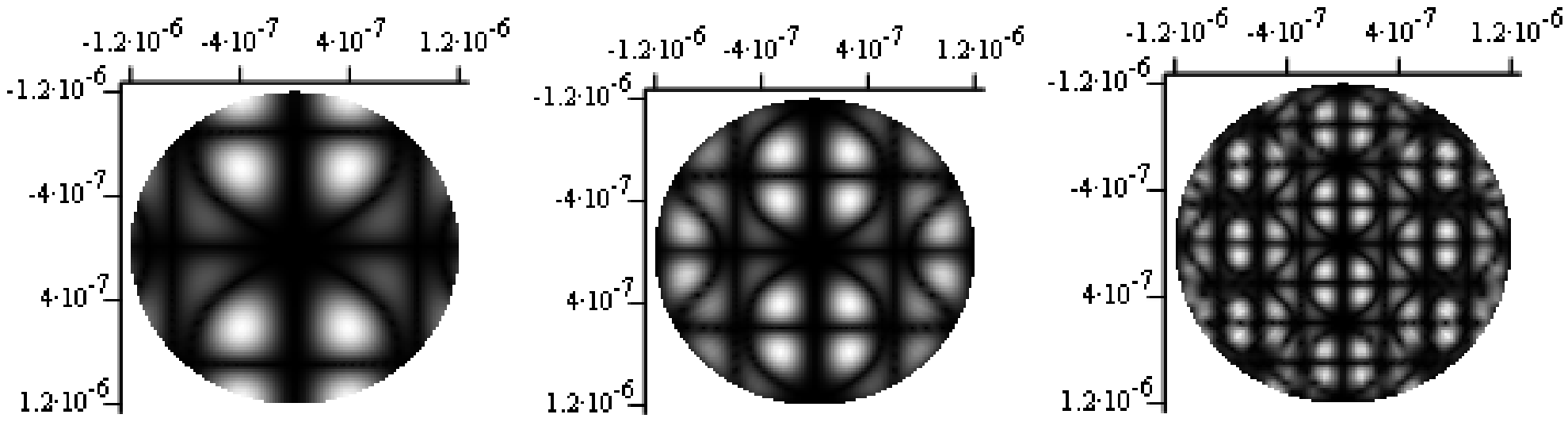

**Fig. 137: Dependences of the polar coordinate modulation map on wavelength in polar coordinates (left 15µm, middle 10µm, right 5µm) in this plot, the star is at the centre position. Seen at a distance of 10pc, 1AU translates into about 5 $10^{-7}$ rad. The modulation maps shown are optimized for 10µm.**

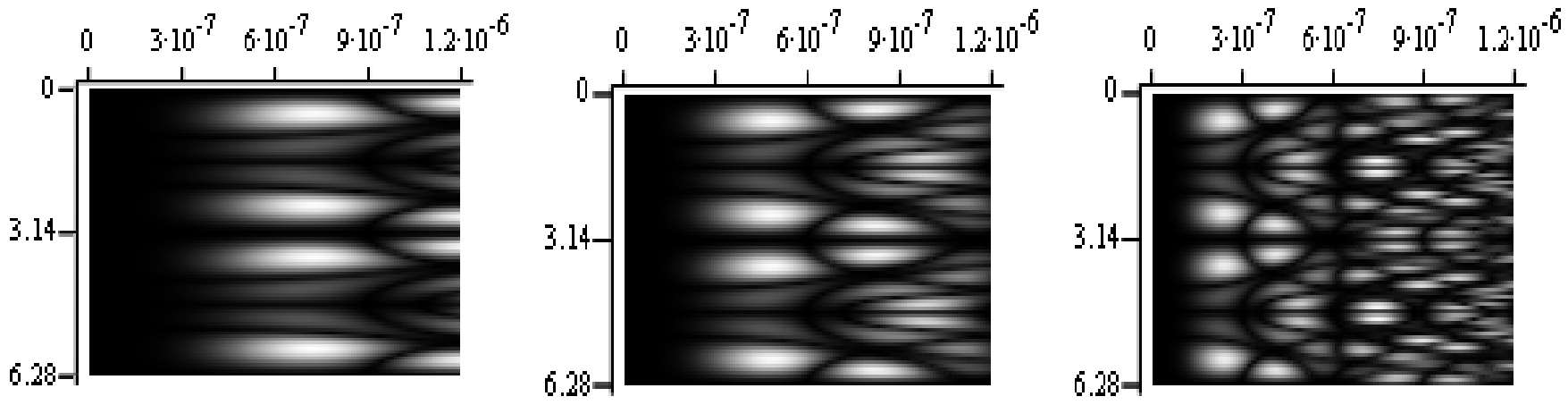

**Fig. 138: Dependences of the modulation map on wavelength (left 15µm, middle 10µm, right 5µm). The star is at position 0(left line of the graph), to go from the top to bottom of the plot in a straight line equals a 2π rotation. Seen at a distance of 10pc, 1AU is about 5 $10^{-7}$ rad.**

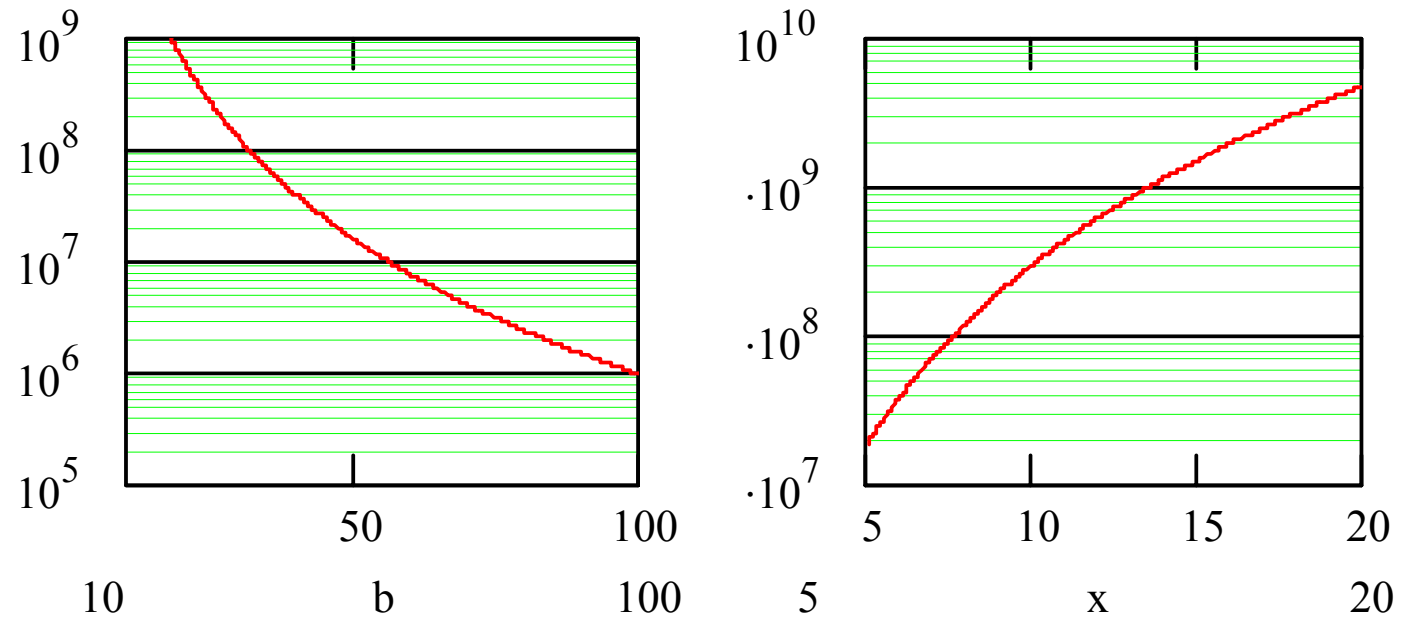

**Fig. 139: Rejection of starlight as a function of baseline at 10 µm for a sun-like star at 10pc (left), at the optimized baseline at 10 µm as a function of wavelength (right).**

Fig. 140 shows that the first maximum in a modulation map does not necessarily need to be the highest maxima for all configurations. In the case of the BOWTIE configuration, the second maximum is 3% higher than the first maximum, for the other configurations discussed there is no higher maximum than the first.

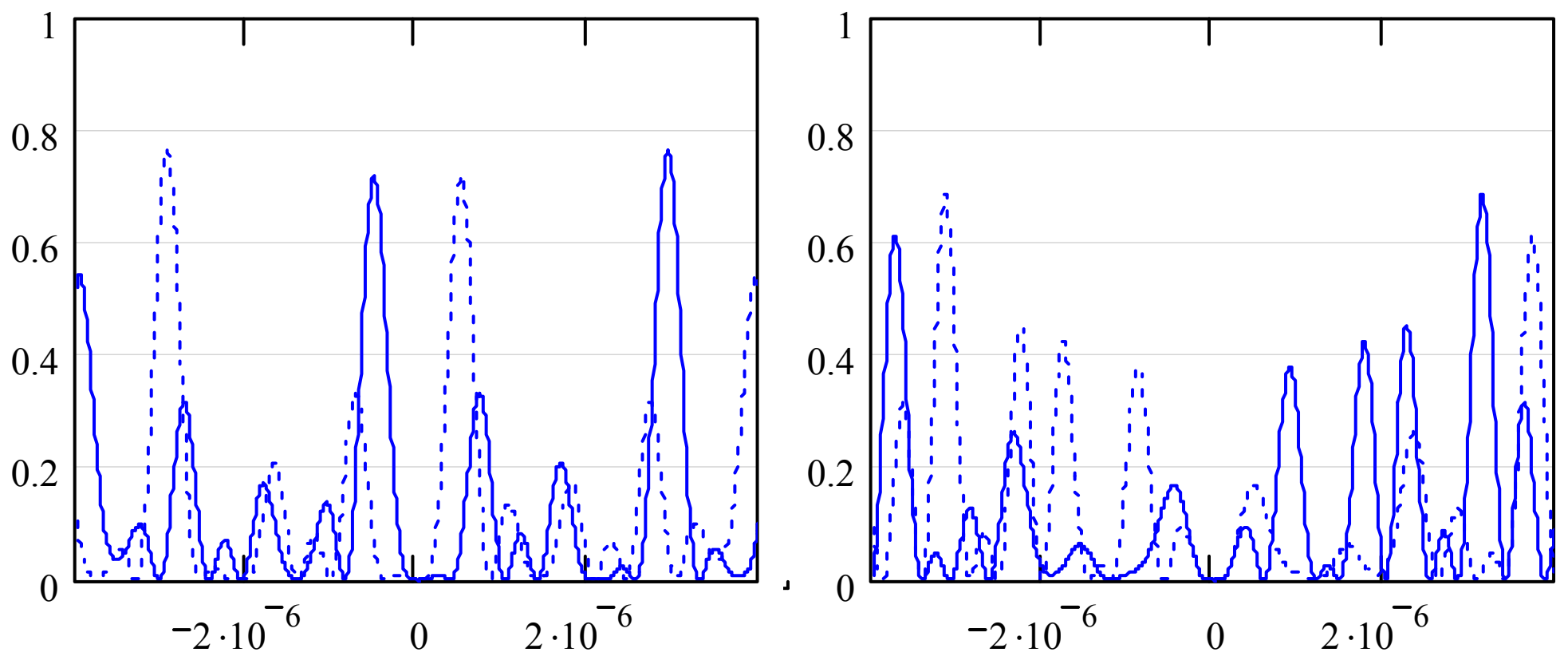

**Fig. 140: 2D cut along the direction of the maximum through the modulation map. Note that the second maximum is 3% higher than 1st maxima (left) and a 2D cut perpendicular to that direction (right).**

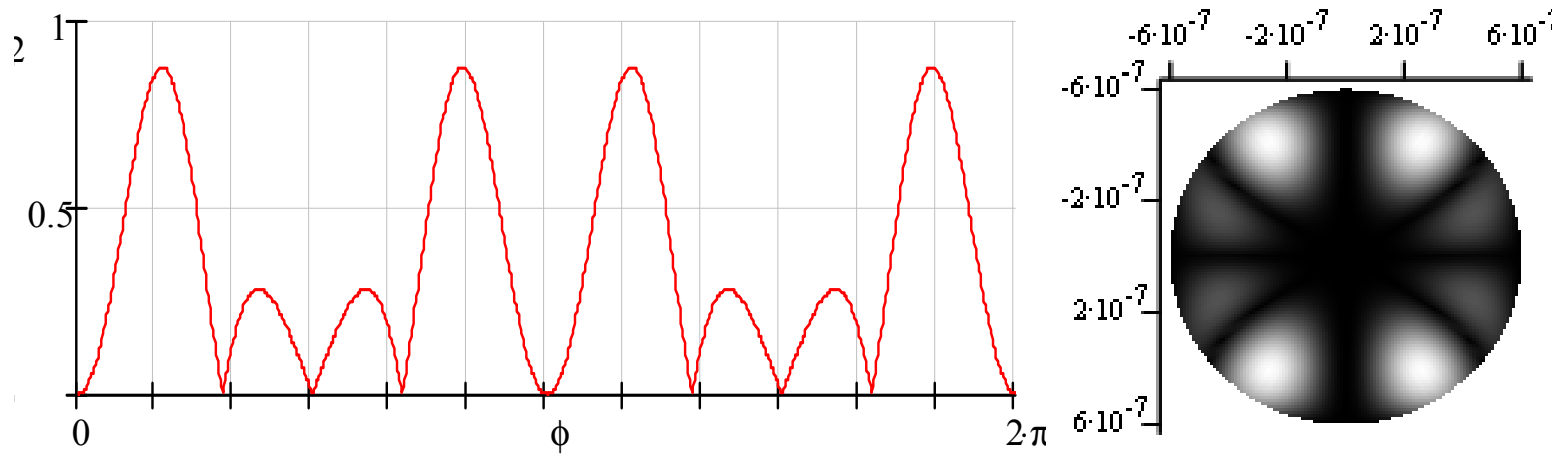

**Fig. 141: Modulation of a planetary signal over a full circular orbit of the planet/ a full rotation of the array (left) and resulting modulation map (right)**

If one can not – for technical reasons -observe during rotation of the array one needs to observe, rotate the array, stop, observe and repeat those step as often as needed to cover the whole area of a possible planetary orbit with a high transmission value. Fig. 142 shows this approach. To achieve a transmission of 50% of the modulation maximum over the whole area of the possible planetary orbit, one has to use four different positions of the array. The individual curves in Fig. 142 show the response of the array over a full rotation starting at each of the four different physical positions. One can see that the response is only shifted. If one connects the upper crests of all the curves, one sees that a min transmission value of 50% of the modulation maximum is given over the whole circle of the possible planetary orbit. Note that Fig. 142 only shows the monochromatic behavior. In different spectral channels this response will look slightly different, but Fig. 142 is a representation for that. We only investigated slow rotation for the arrays (M Kilter and L. dArcio, 2004) to be able to keep the fringes during rotation. The ideal scenario is to keep observing continuously while rotating the array.

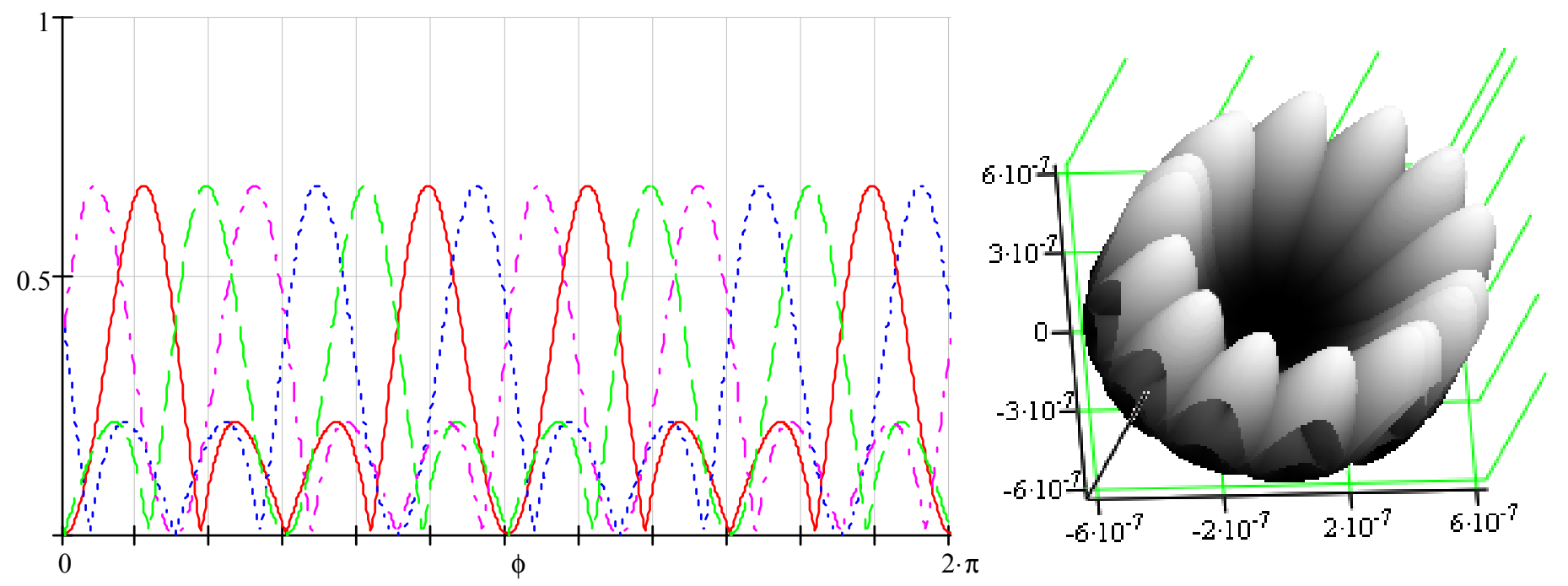

**Fig. 142: Modulation of a planetary signal using 4 modulation maps to provide sensitivity at the full circular orbit of the planet if continuous observation should not be possible (see text for explanation) (left) and resulting covered modulation map (right)**

| BOW TIE | | | | | | | |
|---|---|---|---|---|---|---|---|
| **Nulling architecture** | Symbols | Aperture index | | | | | |
| | | $k$=1 | 2 | 3 | 4 | 5 | 6 |
| **Aperture location (norm. radius)** | $L_k$ | | 1 | 1 | 1 | 1 | 1 |
| **Aperture angles** | $\delta_k$ | | $\pi/4$ | $3\pi/4$ | $\pi$ | $5\pi/4$ | $7\pi/4$ |
| **Sub-interferometers, nr., /type** | $N = 2$, GAC ($\theta^4$ null) | | | | | | |
| Sub-Interferometer 1: amplitudes | $a_{1,k}$ | $1/\sqrt{2}$ | 1 | 1 | $1/\sqrt{2}$ | 0 | 0 |
| phases | $\psi_{1,k}$ | 0 | $\pi$ | 0 | $\pi$ | 0 | 0 |
| Sub-Interferometer 2: amplitudes | $a_{2,k}$ | $1/\sqrt{2}$ | 0 | 0 | $1/\sqrt{2}$ | 1 | 1 |
| phases | $\psi_{2,k}$ | 0 | 0 | 0 | $\pi$ | 0 | $\pi$ |
| **Implementation** | | | | | | | |
| Aperture diameter [m] | $D_k$ | 1.5 | 1.5 | 1.5 | 1.5 | 1.5 | 1.5 |
| Array diameter | $L$ | Variable | | | | | |
| Modulation method | Internal modulation | | | | | | |
| Beam combination | Pupil-plane | | | | | | |
| Number of outputs/ports | 2 | | | | | | |
| **Platform** | Free Flyer Interferometer | | | | | | |
| **Science performance** | Full mission | | | | | | |

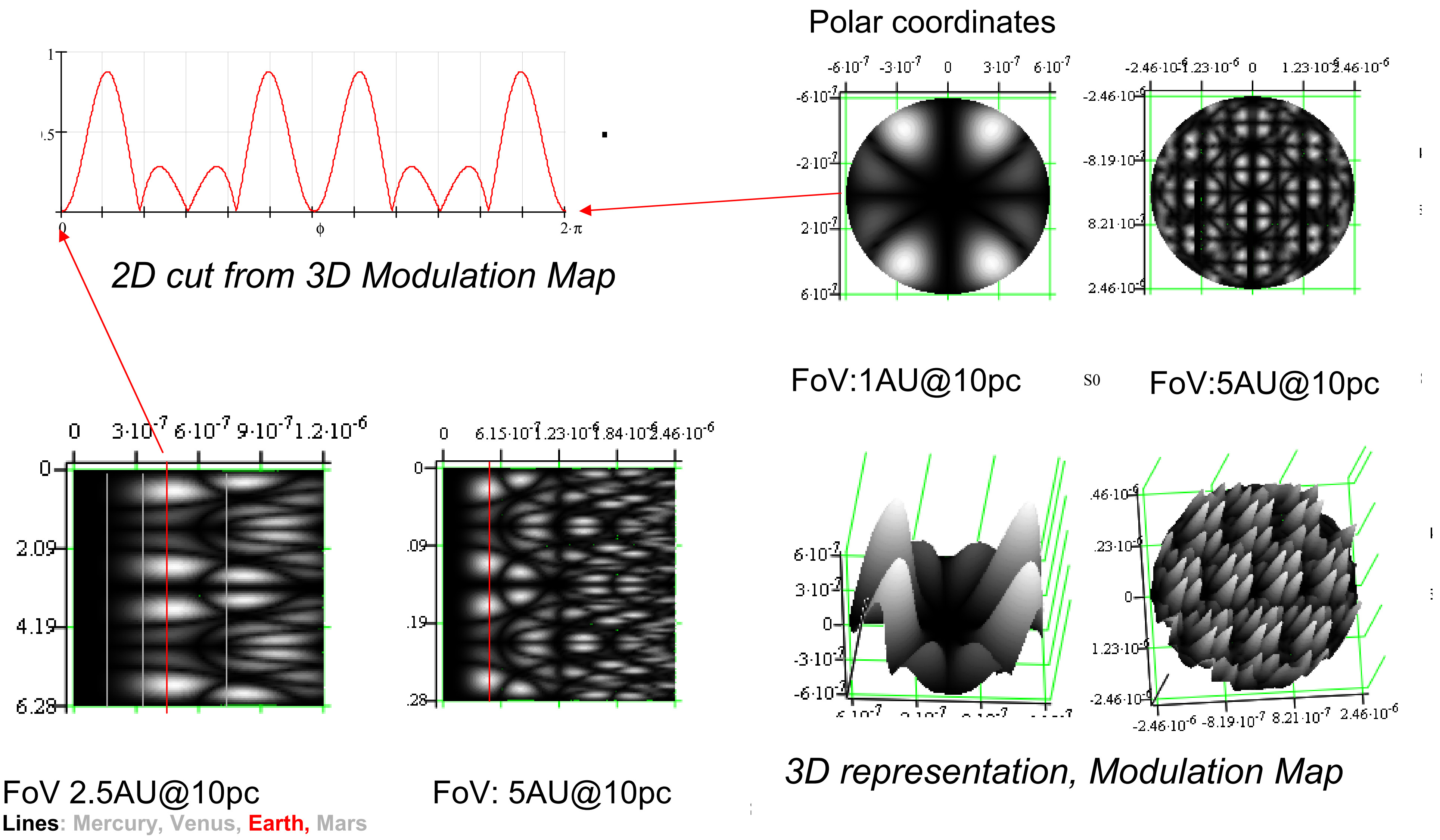

**Fig. 5143: Representation of the characteristics of the modulation map of the Bowtiel**

### 10.7.2 Modified Dual chopped Bracewell (Detailed example calculations)

The modified Dual Chopped Bracewell (MDCB) concept consists of two sub-interferometers and yields a starlight rejection of $1.87 \times 10^4$ for a sun-like star at 10pc. The values are calculated using the optimized baseline for the array, assuming a system around a sun-like star at 10pc. The modulation of the planetary signal $\xi(\vec{\theta})$ leads to the mean and max value of the achievable modulation. The mean modulation efficiency is 26.6%; the maximum value is 100%. The optimized baseline puts the maximum transmission region of the MM on the orbit of a planet at 1AU, a distance that translates into about 5 $10^{-7}$rad as $\theta p$ for sun-like star at 10pc. The equation to optimize the baseline of the array to detect such a planet can be derived from that plot by choosing a baseline that will have the planetary light interfere constructively $\sin\theta_p = \dfrac{\lambda}{2*S}$ leading to an expression for $L_{opt}$.

Fig. 144 shows that for a min ISD of 12m, 1AU equivalent regions for target stars with a distance smaller than 15pc can not be put on the first maxima, increasing the stellar leakage for those systems and thus their integration time for a certain SNR.

$$\text{Lopt} := \frac{\lambda}{2 \cdot \left[\sqrt{1 + \left(\frac{\sqrt{3}}{2}\right)^2}\right] \cdot \sin(\theta p)}$$

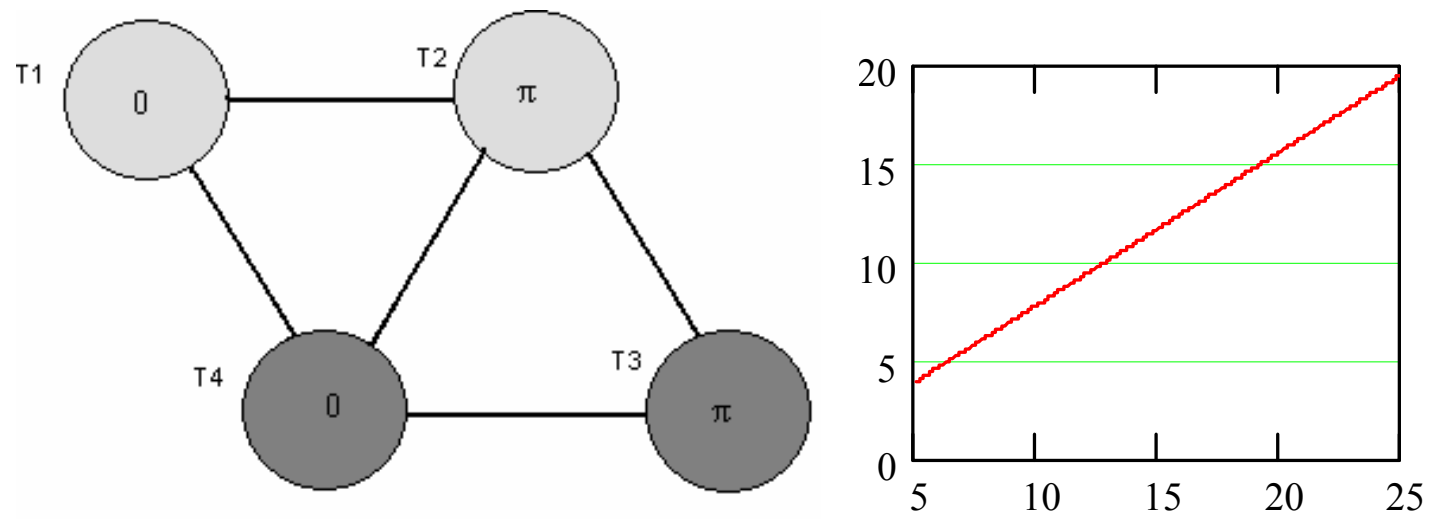

**Fig. 144: MDBW configuration used for calculation of the optimized baseline that puts the planet on the first maxima of the Transmission map (right) (see text)**

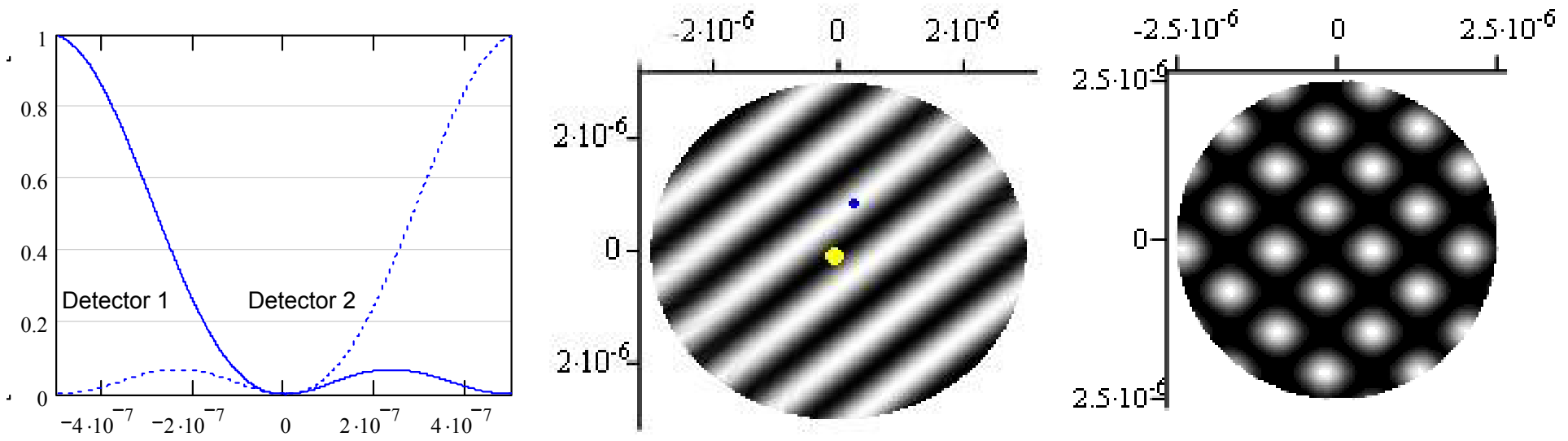

**Fig. 145: Shape of the null over angular extend along thedirection of the maximum from the on-axis position to 1AU for each sub interferometer (left), transmission map used to generate the modulation map 5AU(middle), nulled output on one detector 5AU (right)**

The transmission and modulation map are shown for the optimized baseline. For F, K and M stars the values will vary, but not significantly.

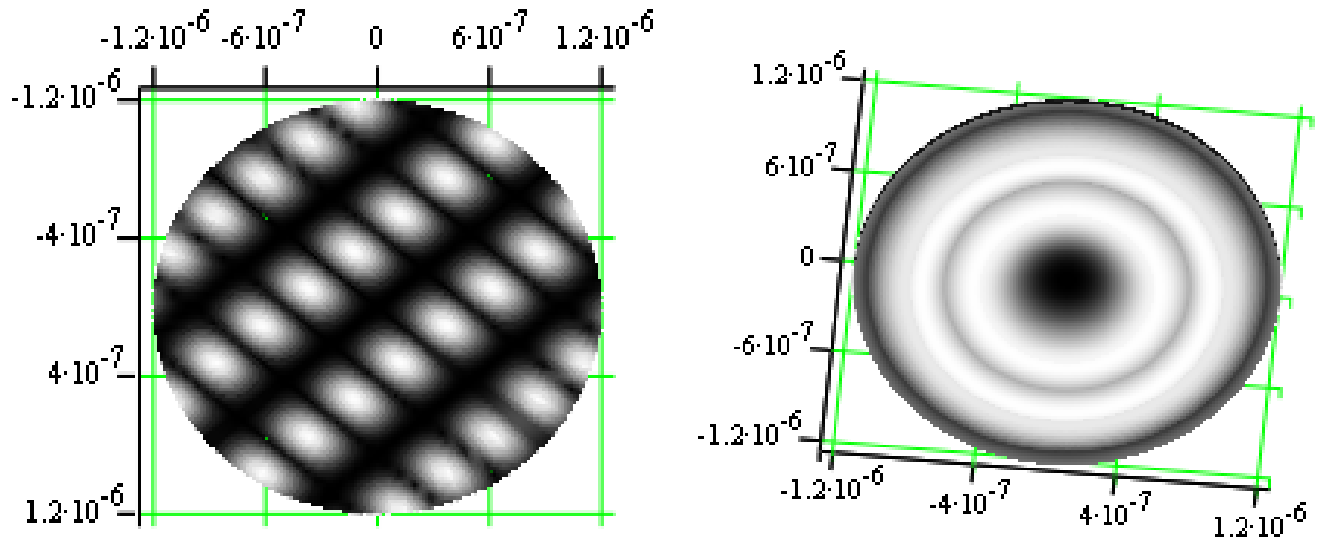

**Fig. 146: Coverage of the planetary orbit by rotation of the array: Observation without rotating the array (left), coverage assuming continuous observation during rotation of π (right)**

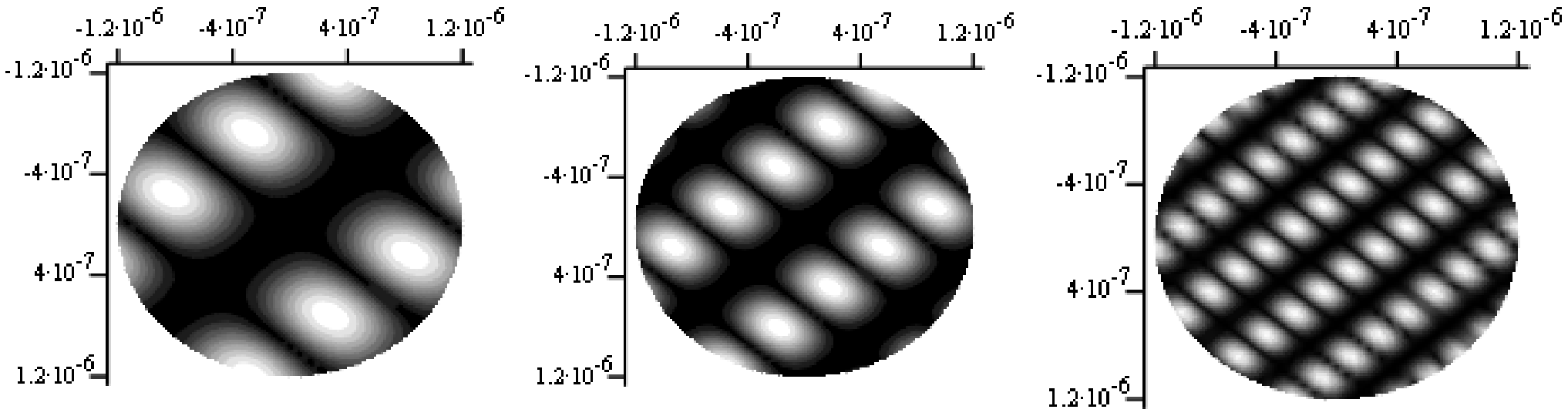

**Fig. 147: Dependences of the modulation map on wavelength in polar coordinates (left 15µm, middle 10µm, right 5µm) in this plot, the star is at the centre position. Seen at a distance of 10pc 1AU is about 5 $10^{-7}$ rad. The modulation maps shown are optimized for 10µm.**

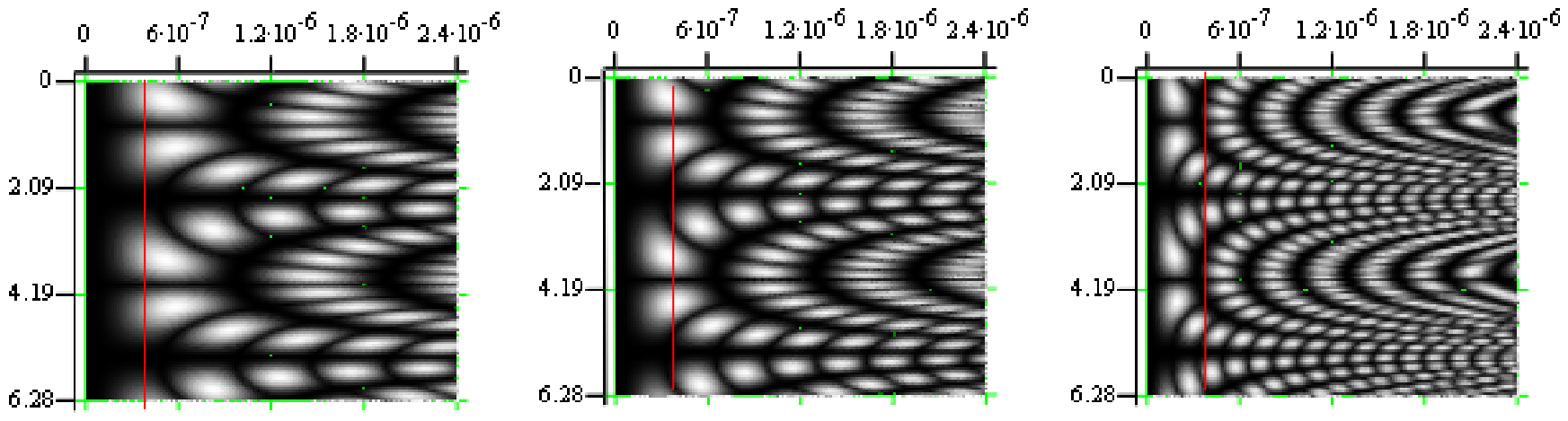

**Fig. 148: Dependences of the modulation map on wavelength (left 15µm, middle 10µm, right 5µm) in this plot, the star is at position 0 (left line of the graph), to go from the top to bottom of the plot in a straight line equals a 2π rotation. Seen at a distance of 10pc 1AU is about 5 $10^{-7}$ rad, the red line shows that orbit for a configuration optimized for 10µm.**

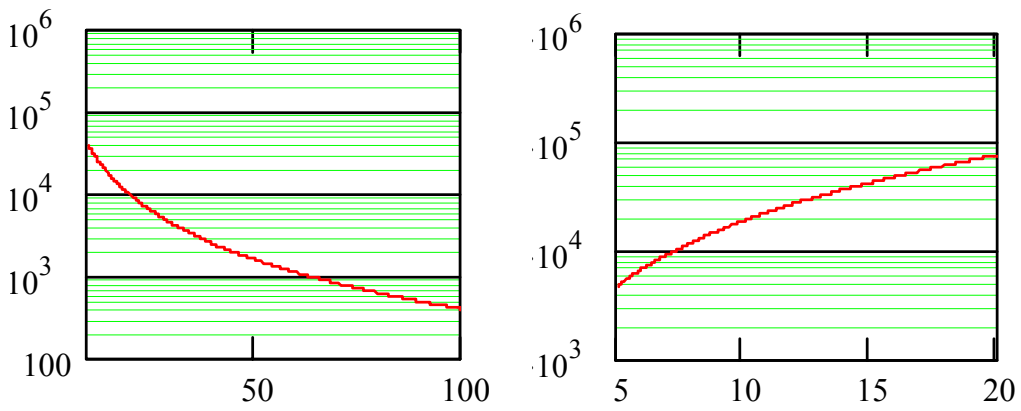

**Fig. 149: Rejection of starlight as a function of baseline at 10 µm for a sun-like star at 10pc (left), at the optimized baseline at 10 µm as a function of wavelength (right).**

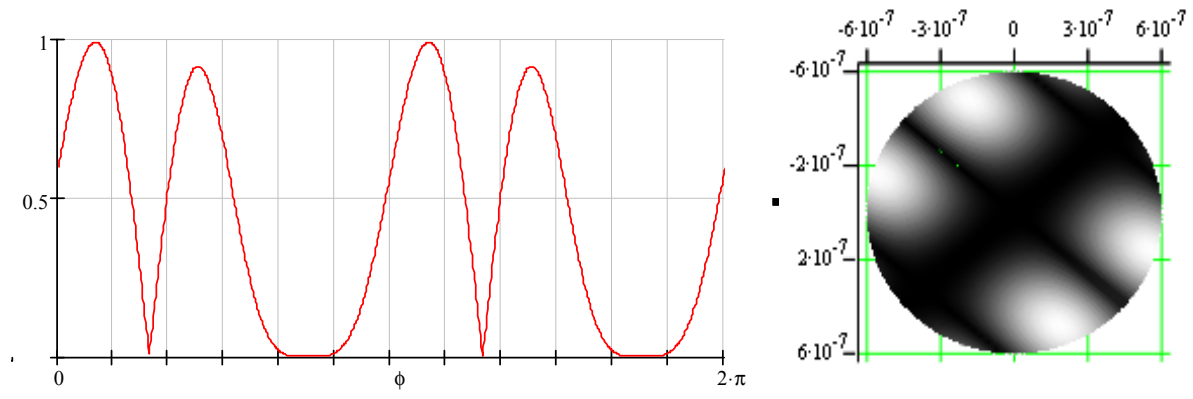

**Fig. 150: Modulation of a planetary signal over a full circular orbit of the planet/ a full rotation of the array (left) what corresponds to the according modulation map (right)**

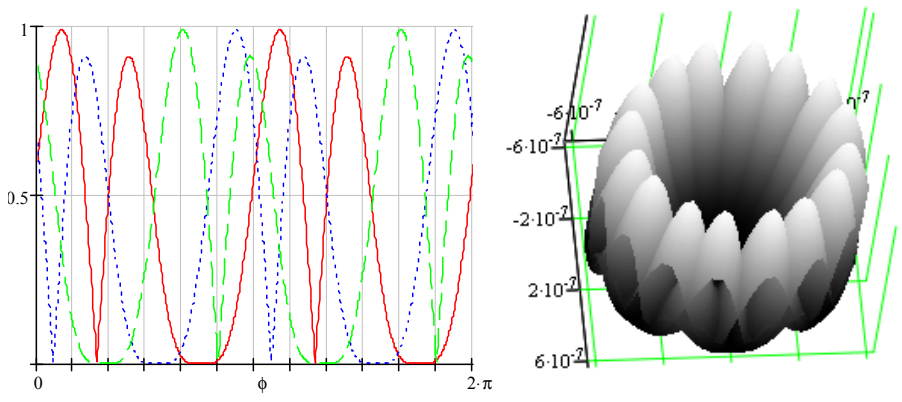

**Fig. 151: Modulation of a planetary signal using 4 modulation maps to provide 70% transmission over the full circular orbit of the planet if continuous observation should not be possible (see text for explanation) (left) and resulting covered modulation map (right)**

| Modified DCB | | | | | | | |
|---|---|---|---|---|---|---|---|
| **Nulling architecture** | Symbols | Aperture index | | | | | |
| | | $k$=1 | 2 | 3 | 4 | | |
| **Aperture location (norm. radius)** | $L_k$ | √3/2 | 1 | √3/2 | 1 | | |
| **Aperture angles** | $\delta_k$ | 0 | π/2 | π | 3π/2 | | |
| **Sub-interferometers, nr., /type** | $N = 2$, Bracewell ($\theta^2$ null) | | | | | | |
| Sub-Interferometer 1: amplitudes | $a_{1,k}$ | 1 | 1 | 0 | 0 | | |
| phases | $\psi_{1,k}$ | π | 0 | 0 | 0 | | |
| Sub-Interferometer 2: amplitudes | $a_{2,k}$ | 0 | 0 | 1 | 1 | | |
| phases | $\psi_{2,k}$ | 0 | 0 | 0 | π | | |
| **Implementation** | | | | | | | |
| Relative aperture diameter | $D_k$ | 1 | 1 | 1 | 1 | | |
| Array diameter | $L$ | Variable | | | | | |
| Modulation method | Internal modulation | | | | | | |
| Beam combination | Pupil-plane | | | | | | |
| Number of outputs/ports | 2 | | | | | | |
| **Platform** | Free Flyer Interferometer | | | | | | |
| **Science performance** | Full mission | | | | | | |

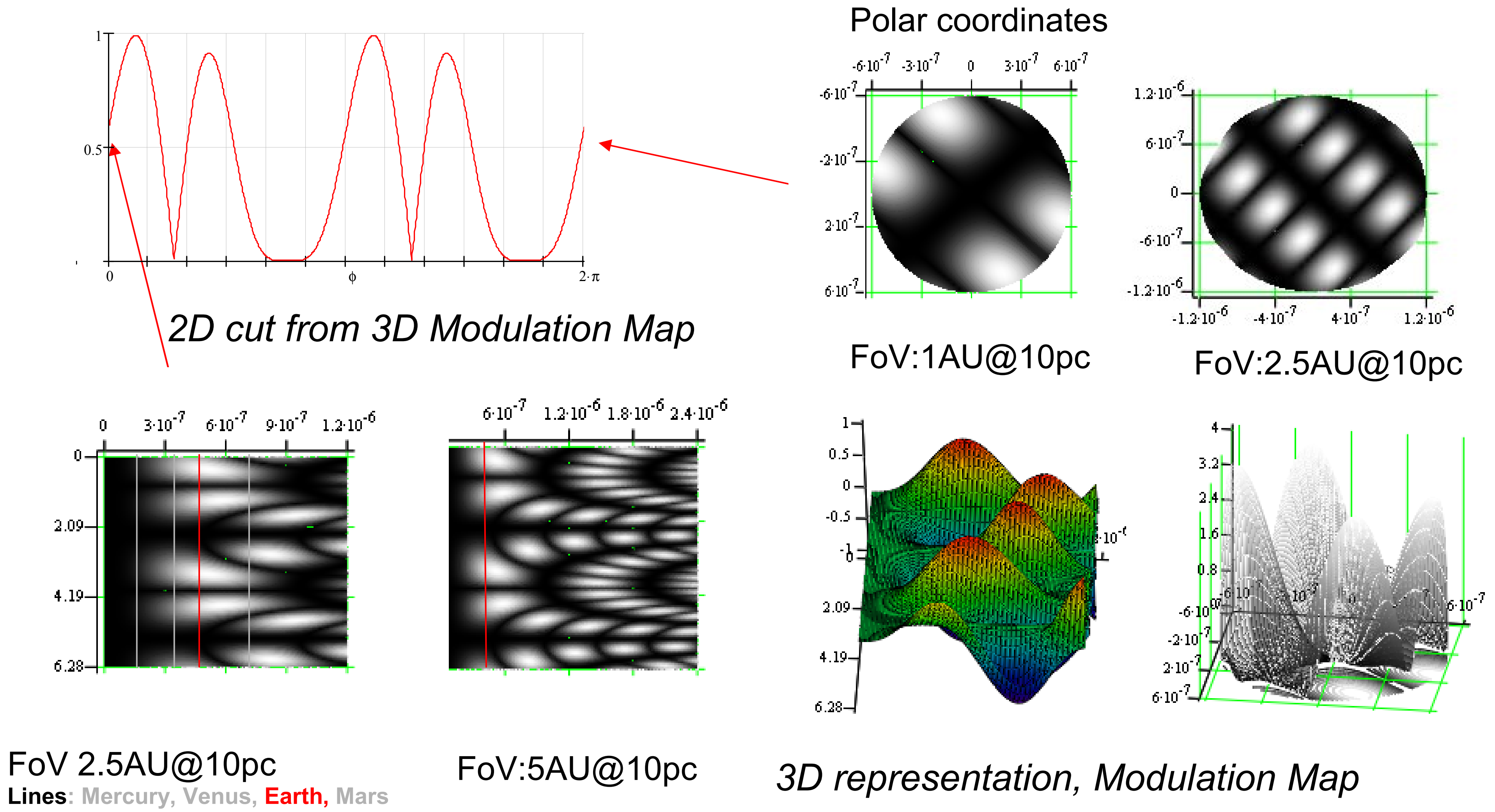

**Fig. 152: Representation of the characteristics of the modulation map of the modified dual chopped Bracewell**

### 10.7.3 Linear Double Chopped Bracewell

The linear Double Chopped Bracewell (LDCB) consists of a linear array of 4 telescopes. It uses two pair of telescopes as subinterferometers. The LDCB concept consists of two sub-interferometers and yields a starlight rejection of $3.32 \times 10^4$ at 10 µm for a sun-like star at 10pc.

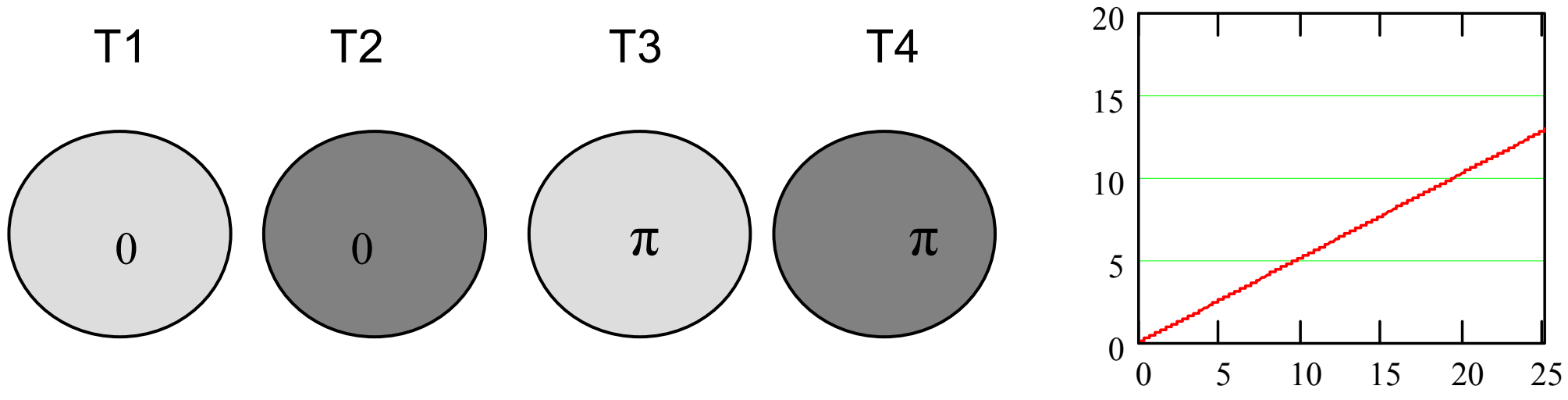

**Fig. 153: Linear DCB configuration used for calculation of the optimized baseline that puts the planet on the first maxima of the Transmission map (right) (see text)**

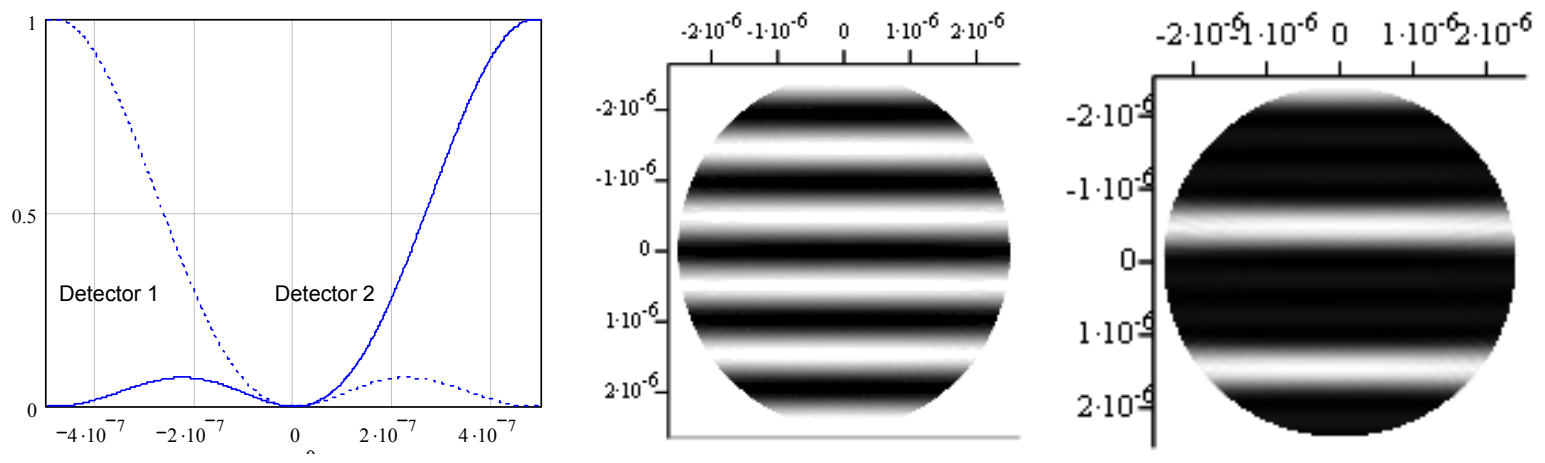

**Fig. 154: Shape of the null along thedirection of the maximum over angular extend from the on-axis position to 1AU at 10pc for each sub-interferometer (left), transmission map used to generate the modulation map 5AU (middle), nulled output on one detector 5AU (right)**

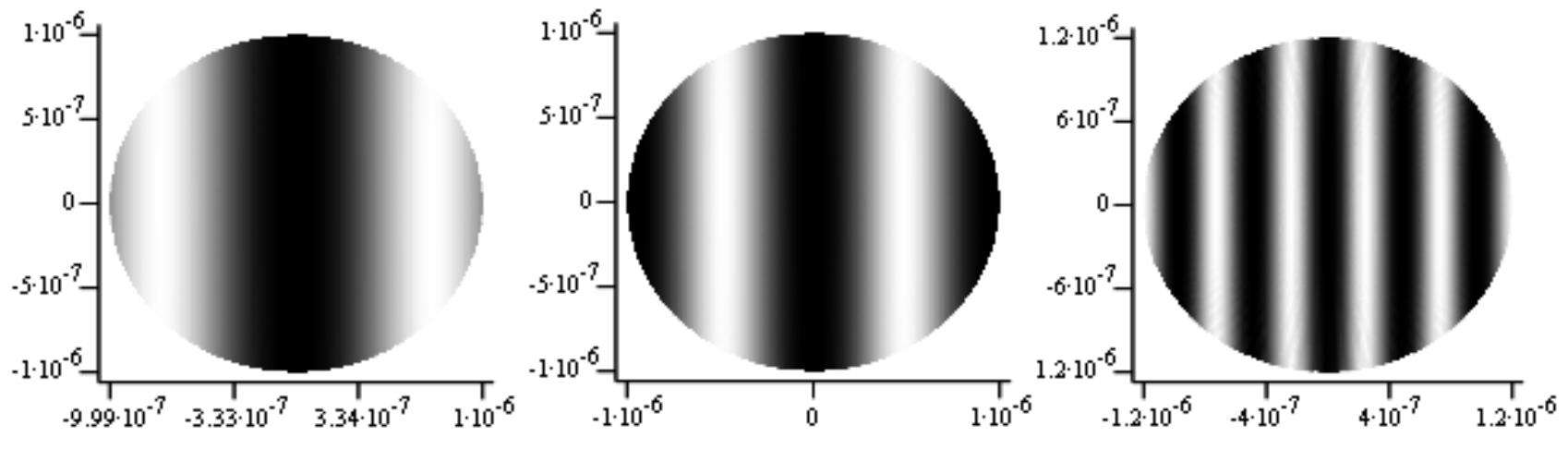

**Fig. 155: Dependences of the modulation map on wavelength in polar coordinates (left 15µm, middle 10µm, right 5µm) in this plot, the star is at the centre position. Seen from 10pc 1AU is about 5 $10^{-7}$ rad. The modulation maps shown are optimized for 10µm.**

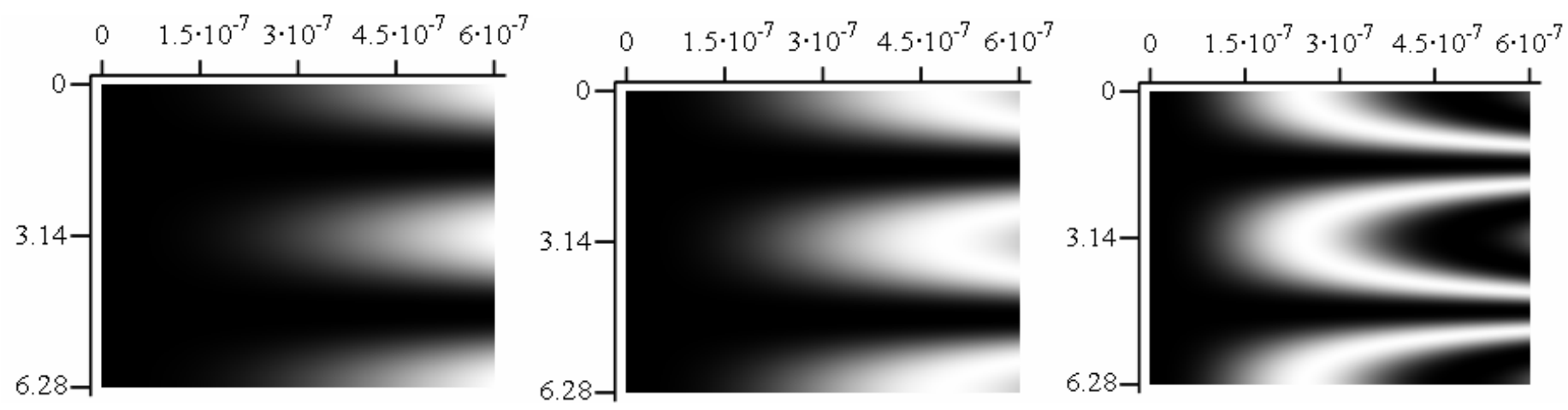

**Fig. 156: Dependences of the modulation map on wavelength (left 15µm, middle 10µm, right 5µm) in this plot, the star is at position 0 (left line of the graph), top to bottom of the plot shows the response of the transmission map over a 2π rotation. Seen at a distance of 10pc 1AU is about 5 $10^{-7}$ rad, the red line shows that orbit for a configuration optimized for 10µm.**

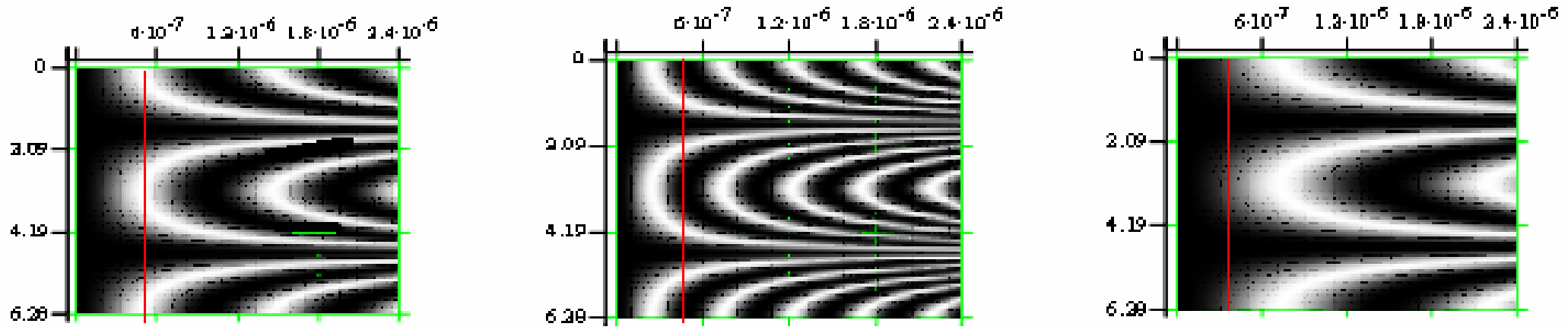

**Fig. 157: Dependences of the modulation map on wavelength (left 15µm, middle 10µm, right 5µm) in this plot, the star is at position 0(left line of the graph), to go from the top to bottom of the plot in a straight line equals a 2π rotation. Seen at a distance of 10pc 1AU is about 5 $10^{-7}$ rad, the red line shows that orbit for a configuration optimized for 10µm.**

Fig. 154 shows that planets outside of the habitable zone will also show signatures as the maxima in the MM repeat. For each configuration that has to be investigated carefully as planets are thought to form as multiple bodies.

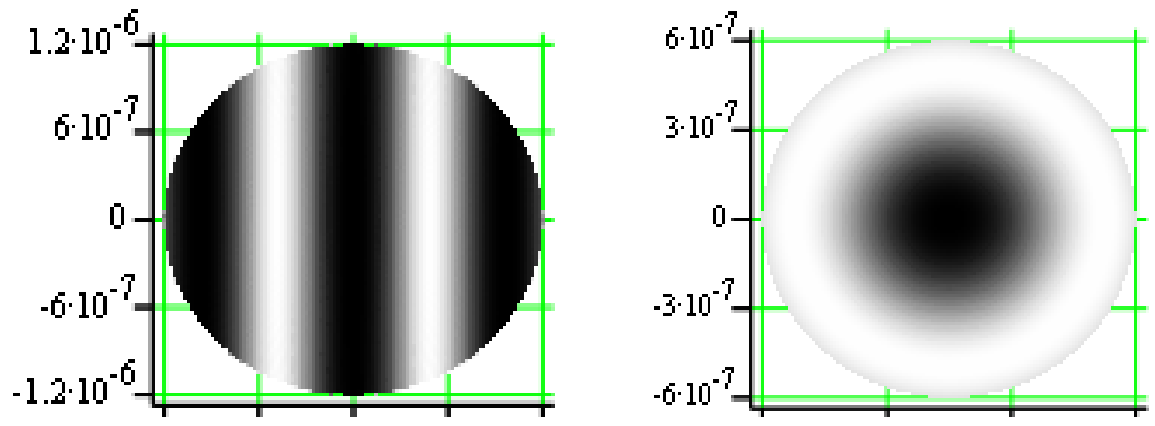

**Fig. 158: Coverage of the planetary orbit by rotation of the array: Observation without rotating the array (left), coverage assuming continuous observation during rotation of π (right).**

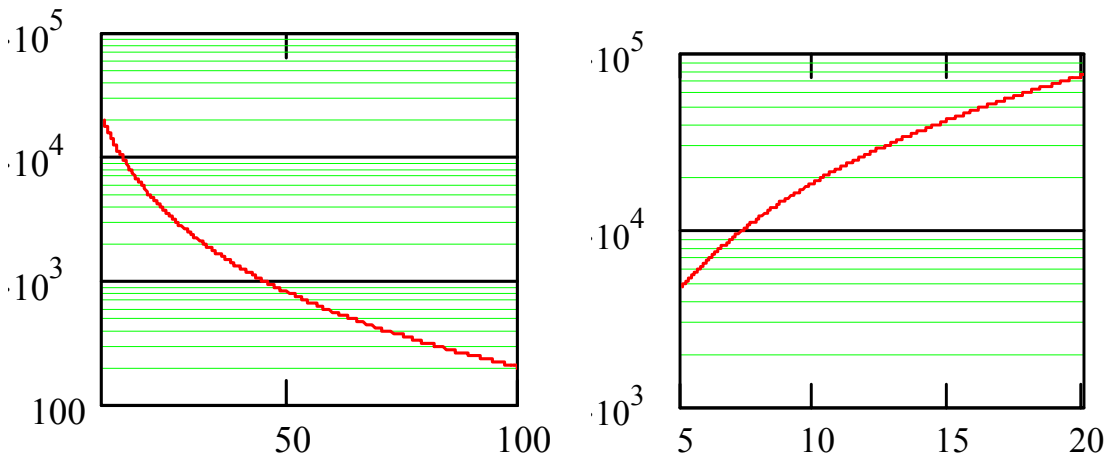

**Fig. 159: Rejection of starlight as a function of baseline at 10 µm for a sun-like star at 10pc (left), at the optimized baseline at 10 µm as a function of wavelength (right).**

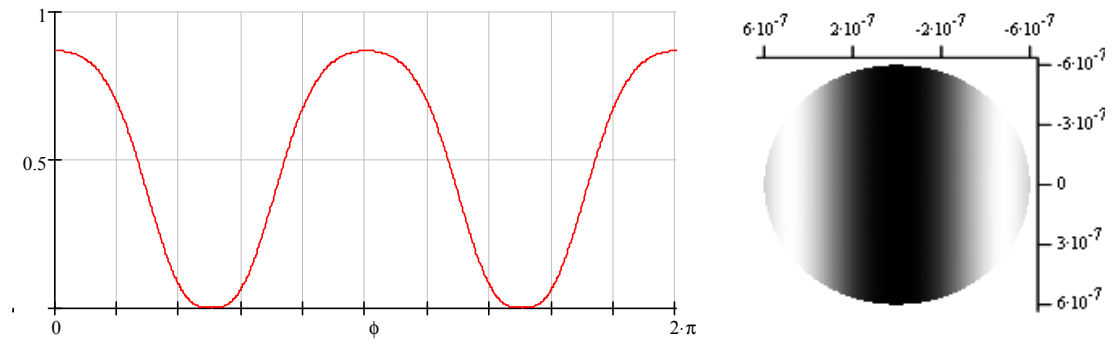

**Fig. 160: Modulation of a planetary signal over a full circular orbit of the planet/ a full rotation of the array (left) and resulting modulation map (right)**

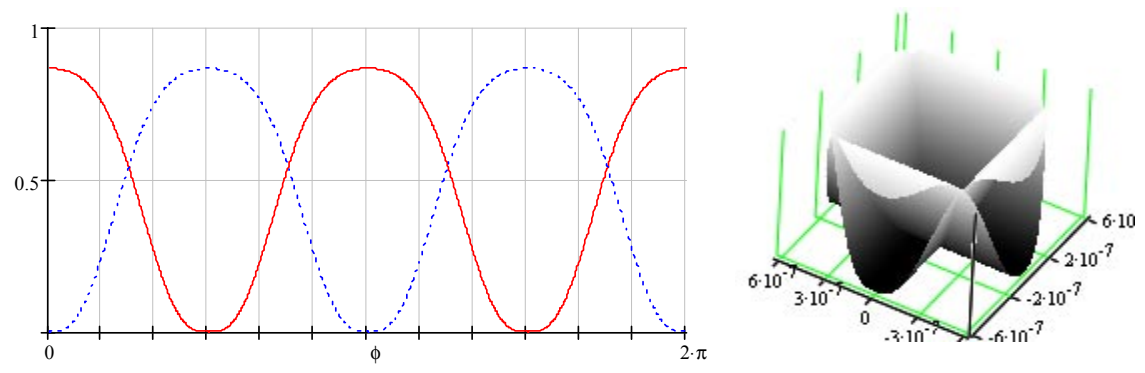

**Fig. 161: Modulation of a planetary signal using 2 modulation maps to provide sensitivity at the full circular orbit of the planet (left) and resulting covered modulation map**

| Linear DCB | | | | | | | | |
|---|---|---|---|---|---|---|---|---|
| **Nulling architecture** | Symbols | Aperture index | | | | | | |
| | | $k$=1 | 2 | 3 | 4 | | | |
| **Aperture location (norm. radius)** | $L_k$ | 1.5 | 0.5 | 0.5 | 1.5 | | | |
| **Aperture angles** | $\delta_k$ | 0 | 0 | π | π | | | |
| **Sub-interferometers, nr., /type** | *N = 2, Bracewell ($\theta^2$ null)* | | | | | | | |
| Sub-Interferometer 1: amplitudes | $a_{1,k}$ | 1 | 1 | 0 | 0 | | | |
| phases | $\psi_{1,k}$ | π | 0 | 0 | 0 | | | |
| Sub-Interferometer 2: amplitudes | $a_{2,k}$ | 0 | 0 | 1 | 1 | | | |
| phases | $\psi_{2,k}$ | 0 | 0 | 0 | π | | | |
| **Implementation** | | | | | | | | |
| Aperture diameter [m] | $D_k$ | 1 | 1 | 1 | 1 | | | |
| Array diameter | $L$ | Variable | | | | | | |
| Modulation method | Internal modulation | | | | | | | |
| Beam combination | Pupil-plane | | | | | | | |
| Number of outputs/ports | *2* | | | | | | | |
| **Platform** | Free Flyer Interferometer | | | | | | | |
| **Science performance** | Full mission | | | | | | | |

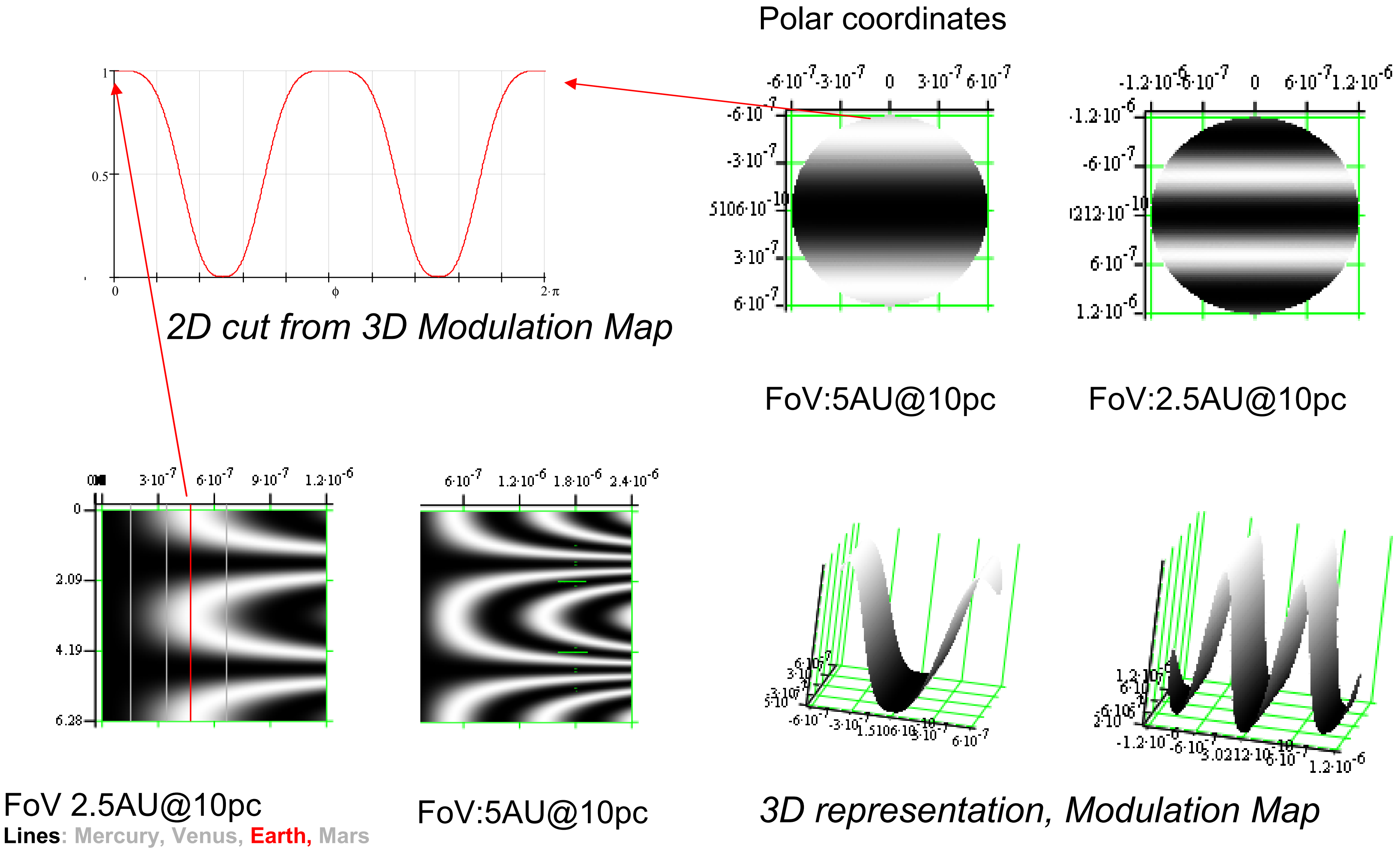

**Fig. 162: Representation of the characteristics of the modulation map of the linear TCB**

### 10.7.4 TTN

The Three Telescope Nulling Configuration proposed by A. Karlsson (Karlsson, 2003) consists of 3 Telescopes and of two sub-interferometers 1) using half of the output of each telescopes to generate two MM with two separate beam combination schemes as the same time t1 or 2) using the whole array to generate thw two MM at two different times t1 and t2. It yields a starlight rejection of 2.10x$10^4$at 10µm for a sun-like star at 10pc. Inherent modulation is used to generate two TM maps from the same array at time t1 and t2. The two maps are used to generate the modulation map shown. If the whole array is used to generae the MMs at time t1 and t2, the max of planmod cannot surpass 50%. If the signal of the input telescopes is split and the two modulation maps are generated at the same time on two detectors, the maximum of planmod is 93.3% and a mean modulation efficiency of 32%. Fig. 164 shows that for an ISD of 12m, all targets stars at distances smaller than 9pc have an optimum baseline smaller than the ISD, thus the planetary orbit must be put on a following maximum in the MM.

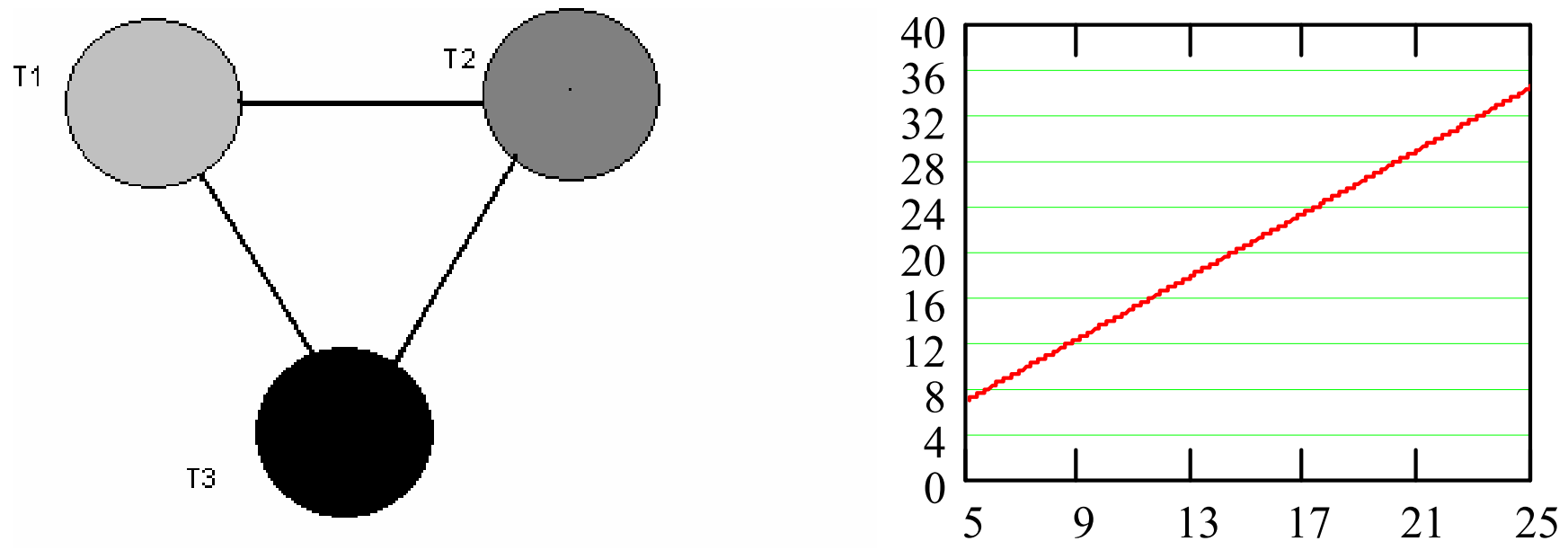

**Fig. 163: TTN configuration (left) used for calculation of the optimized baseline (right) (see text)**

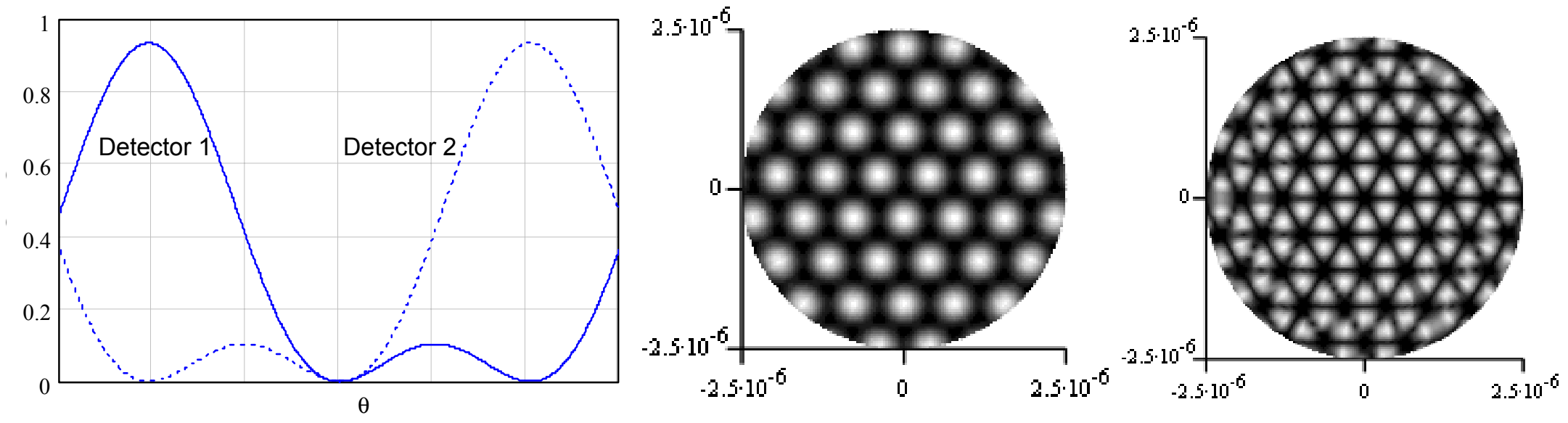

**Fig. 164: Baseline optimized to constructively interfere at planet's position. Shape of the null along the direction of the maximum over angular extend from 0 to 1.5AU at 10pc (left), TM to 5AU (middle), nulled output on one detector out to 5AU (right) (note that the null is not symmetric).**

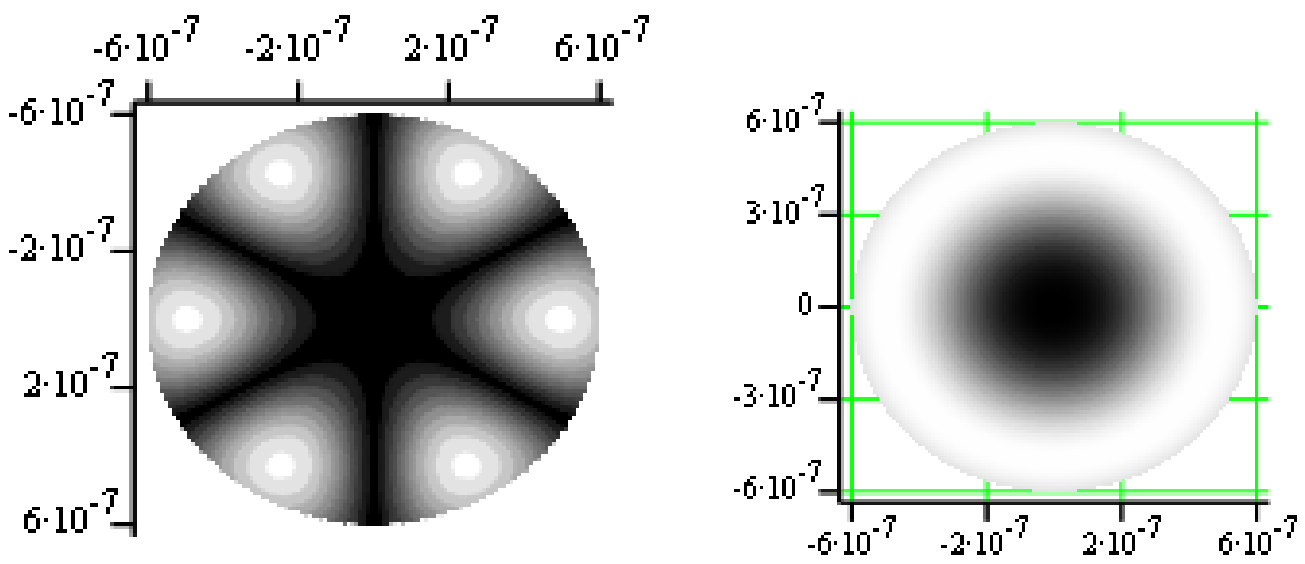

**Fig. 165: Coverage of the planetary orbit by rotation of the array: Observation without rotating the array (left), coverage assuming continuous observation during rotation of π (right)**

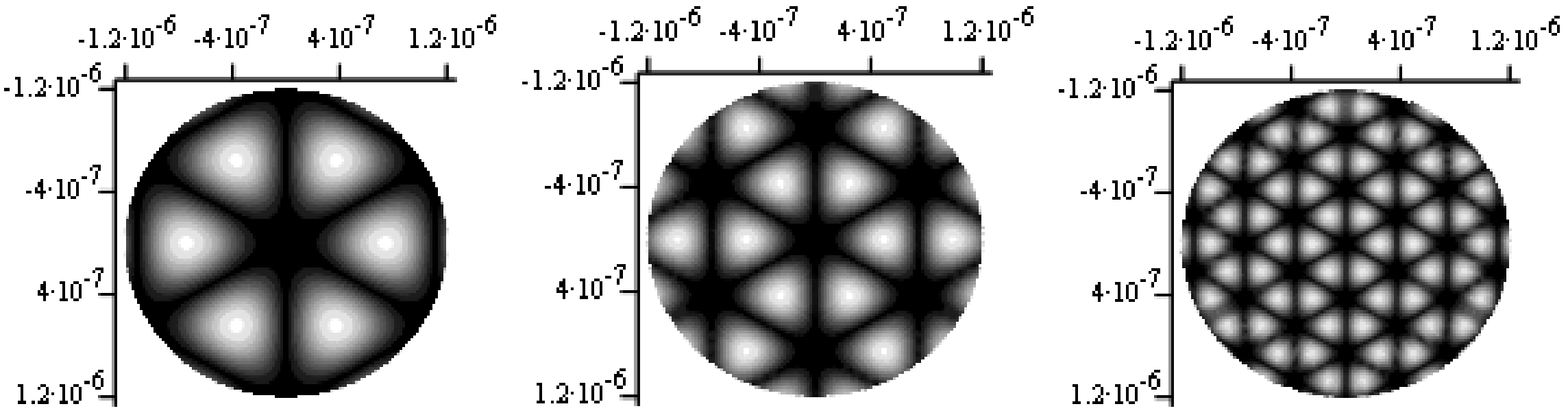

**Fig. 166: Dependences of the modulation map on wavelength in polar coordinates (left 15µm, middle 10µm, right 5µm) in this plot, the star is at the centre position. Seen from 10pc 1AU is about 5 $10^{-7}$ rad.**

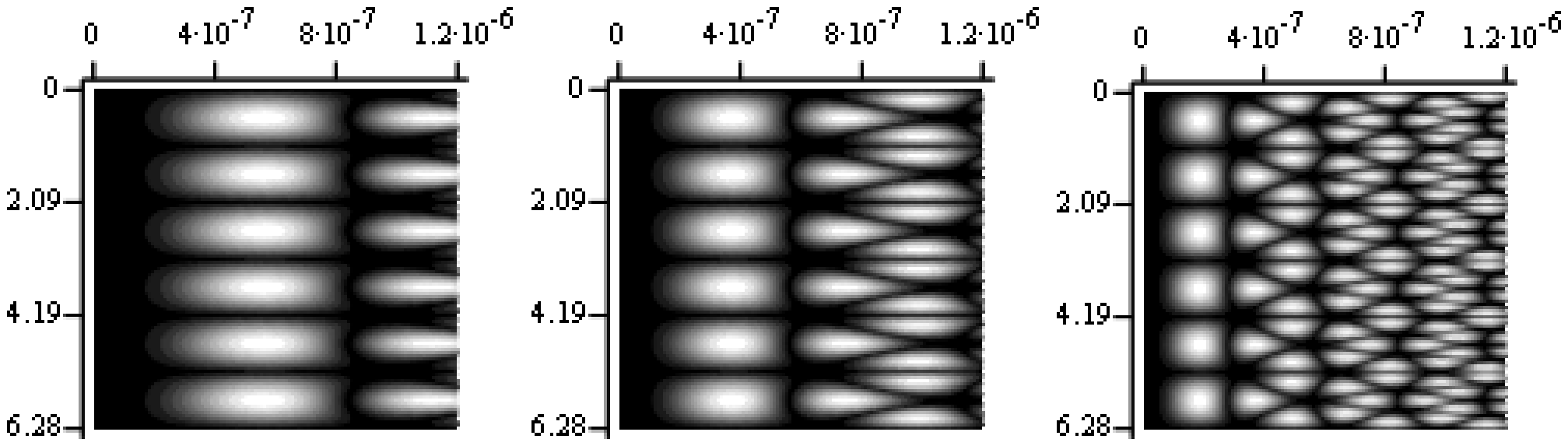

**Fig. 167: Dependences of the modulation map on wavelength (left 15µm, middle 10µm, right 5µm) in this plot, the star is at position 0 (left line of the graph), to go from the top to bottom of the plot in a straight line equals a 2π rotation. Seen at a distance of 10pc 1AU is about 5 $10^{-7}$ rad.**

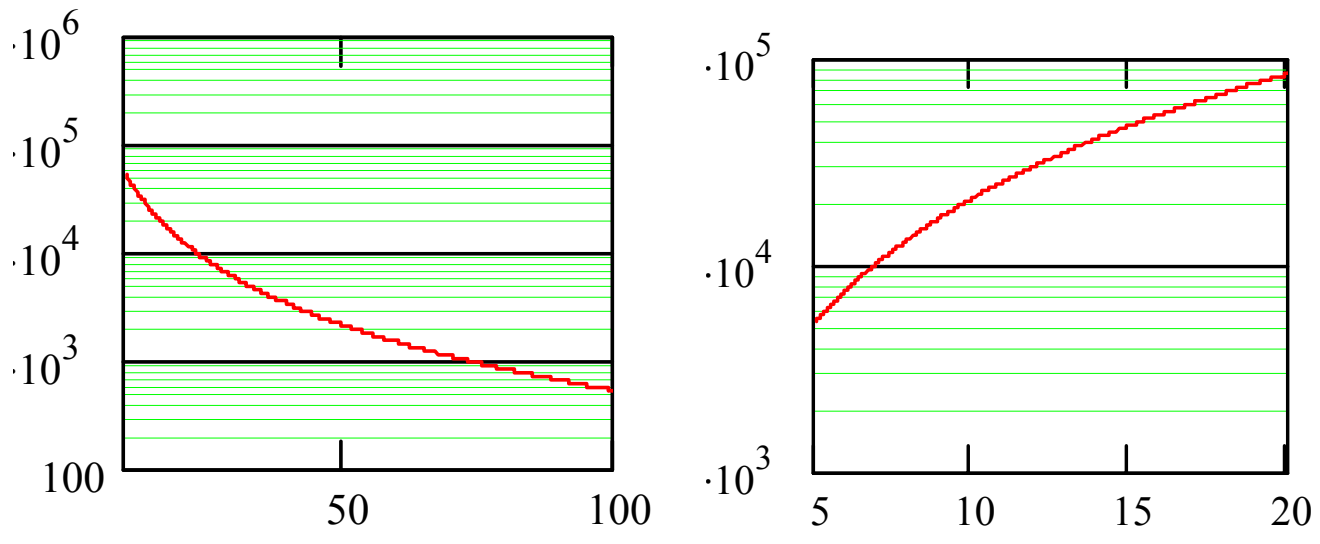

**Fig. 168: Rejection of starlight as a function of baseline at 10 µm for a sun-like star at 10pc (left), at the optimized baseline at 10 µm as a function of wavelength (right).**

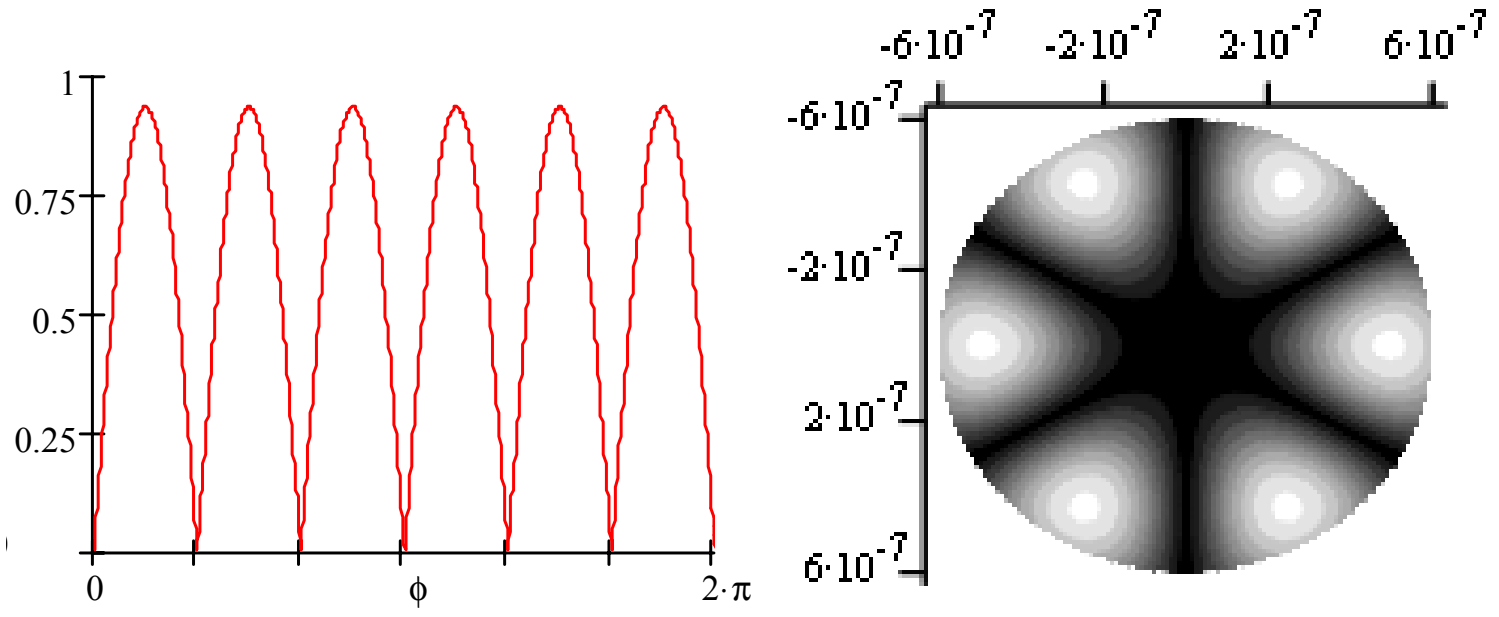

**Fig. 169: Modulation of a planetary signal over a full circular orbit of the planet/ a full rotation of the array (left) and resulting modulation map (right)**

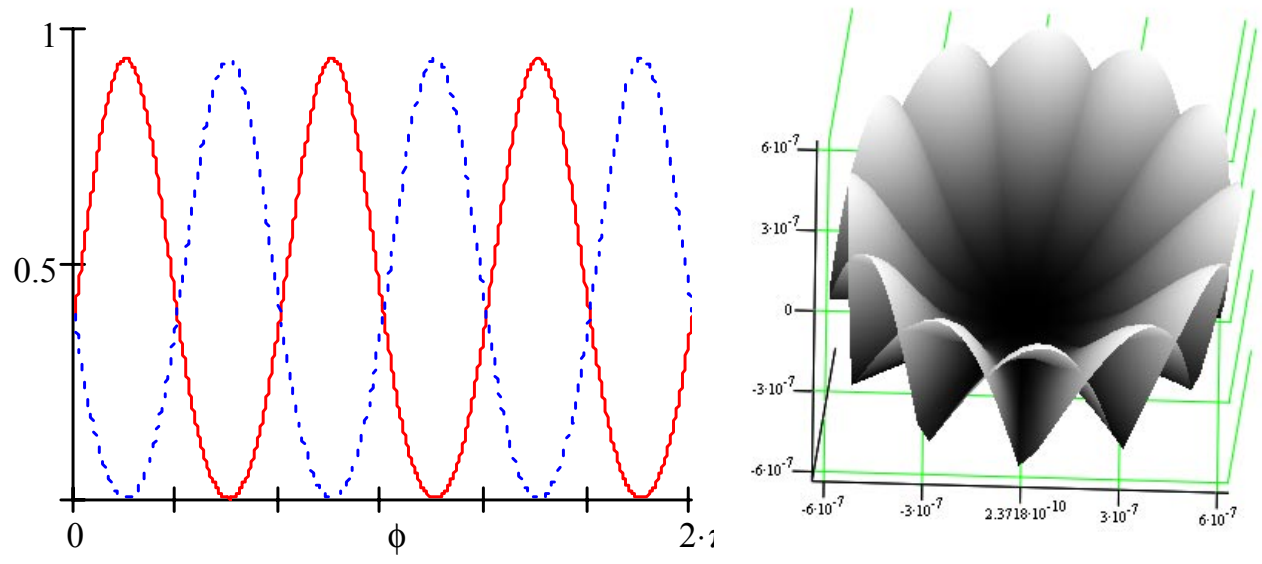

**Fig. 170: Modulation of a planetary signal using 2 modulation maps to provide sensitivity at the full circular orbit of the planet (left) and resulting covered modulation map (right)**

Note that the spacecrafts in the TTN configuration can be distributed arbitrarily around the beam combiner spacecraft, i.e. the telescope spacecrafts do not necessarily have to form an equilateral triangle. The linear TTN configuration consists of a linear arrangement for the three telescopes applying the same phase shifts to each beam.

The nullshape of the linear TTN along the direction of the maximum is similar to the null shape shown for the equilateral TTN configuration but the MM pattern is different. The linear TTN array achievesa higher mean modulation efficiency because the maximum is highly localized. The mean modulation efficiency is 40%; the maximum value is 93.3%. Note that a higher mean modulation efficiency decreases the observation time in the detection phase. Fig. 172 shows that the MM of the linear TTN configuration is not central symmetric thus by rotation of the array the modulation efficiency will not be zero at any point of the planetary orbit (see section 10.5)

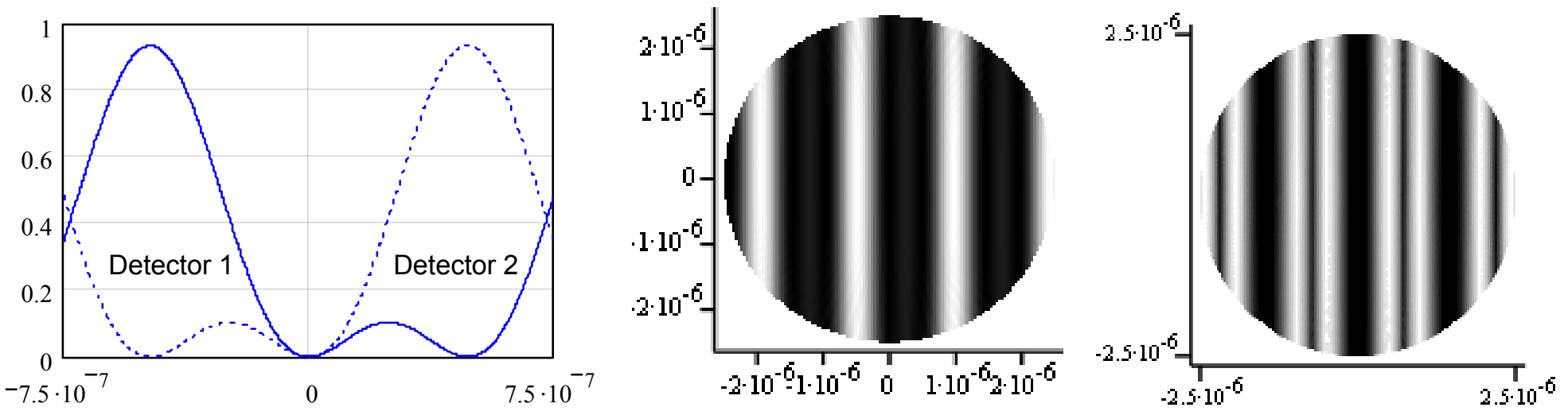

**Fig. 171: Baseline optimized to constructively interfere at planet's position. Shape of the null along the direction of the maximum over angular extend from 0 to 1.5AU at 10pc (left), TM to 5AU (middle), nulled output on one detector out to 5AU (right) (note that the null is not symmetric).**

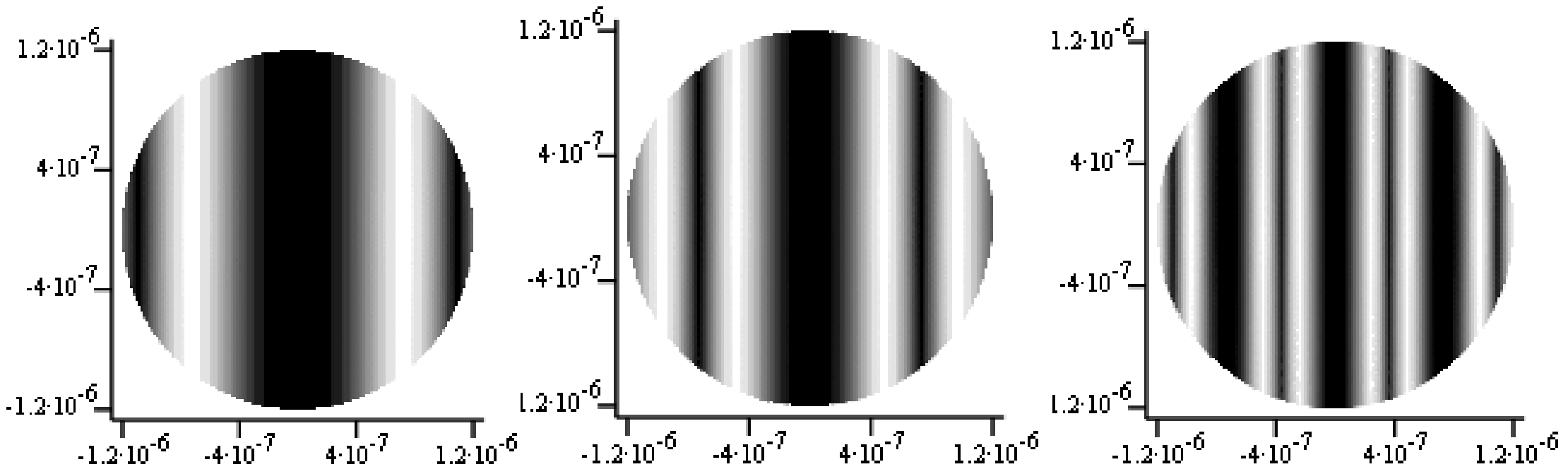

**Fig. 172: Dependences of the modulation map on wavelength in polar coordinates (left 15µm, middle 10µm, right 5µm) in this plot, the star is at the centre position. Seen from 10pc 1AU is about 5 $10^{-7}$ rad.**

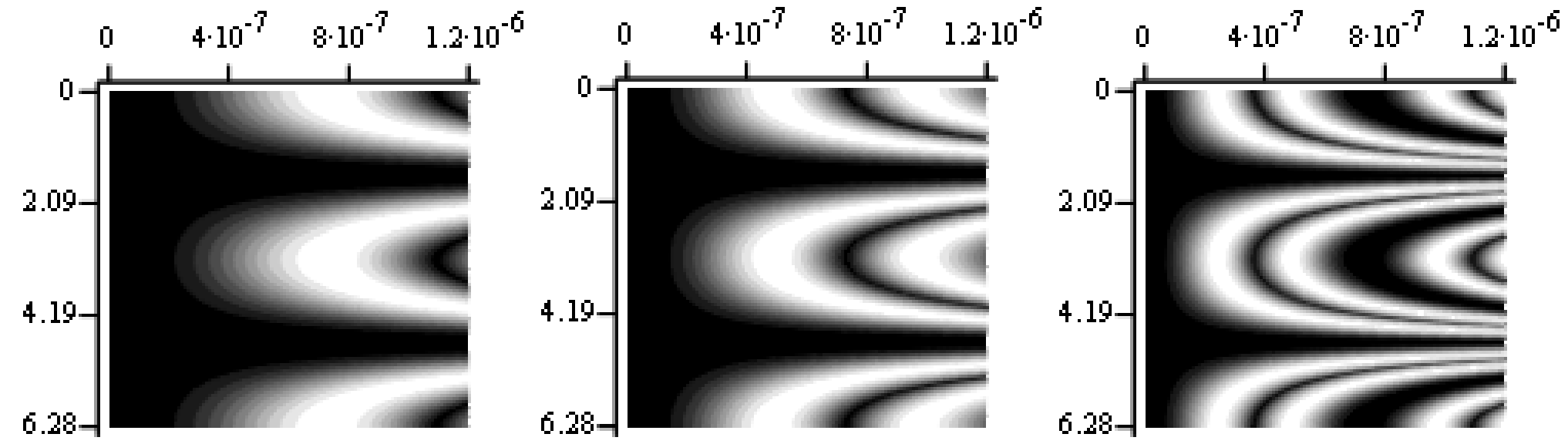

**Fig. 173: Dependences of the modulation map on wavelength (left 15µm, middle 10µm, right 5µm) in this plot, the star is at position 0 (left line of the graph), to go from the top to bottom of the plot in a straight line equals a 2π rotation. Seen at a distance of 10pc 1AU is about 5 $10^{-7}$ rad.**

The MM of the linear configuration has only low spatial frequencies and does not allow accurate determination of planet position see section 10.4

### 10.7.5 TCB

The Triple Chopped Bracewell (TCB) Configuration proposed by A. Karlsson consists of 3 Telescopes. They are used to generate 3 subinterferometers, each yields a starlight rejection of $1.49x10^5$ at 10µm for a sun-like star at 10pc. Internal modulation is used to generate 3 TM maps. Two TM maps are used to generate the modulation map. The characteristics of the TM and MM are similar to the TTN array and thus are not reproduced here.

| TTN | | | | | | | |
|---|---|---|---|---|---|---|---|
| **Nulling architecture** | Symbols | Aperture index | | | | | |
| | | $k$=1 | 2 | 3 | | | |
| **Aperture location (norm. radius)** | $L_k$ | 1 | 1 | 1 | | | |
| **Aperture angles** | $\delta_k$ | 0 | $2\pi/3$ | $4\pi/3$ | | | |
| **Sub-interferometers, nr., type** | $N = 2\ (\theta^2\ null)$ | | | | | | |
| Sub-Interferometer 1: amplitudes | $a_{1,k}$ | | 1 | 1 | | | |
| phases | | 0 | $2\pi/3$ | $4\pi/3$ | | | |
| Sub-Interferometer 2: amplitudes | | 1 | 1 | 1 | | | |
| phases | | 0 | $4\pi/3$ | $2\pi/3$ | | | |
| **Implementation** | | | | | | | |
| Aperture diameter [m] | $D_k$ | 1 | 1 | 1 | | | |
| Array diameter | $L$ | Variable | | | | | |
| Modulation method | Inherent modulation | | | | | | |
| Beam combination | Image-plane | | | | | | |
| Number of outputs/ports | *1* | | | | | | |
| **Platform** | | | | | | | |
| **Science performance** | reduced mission | | | | | | |

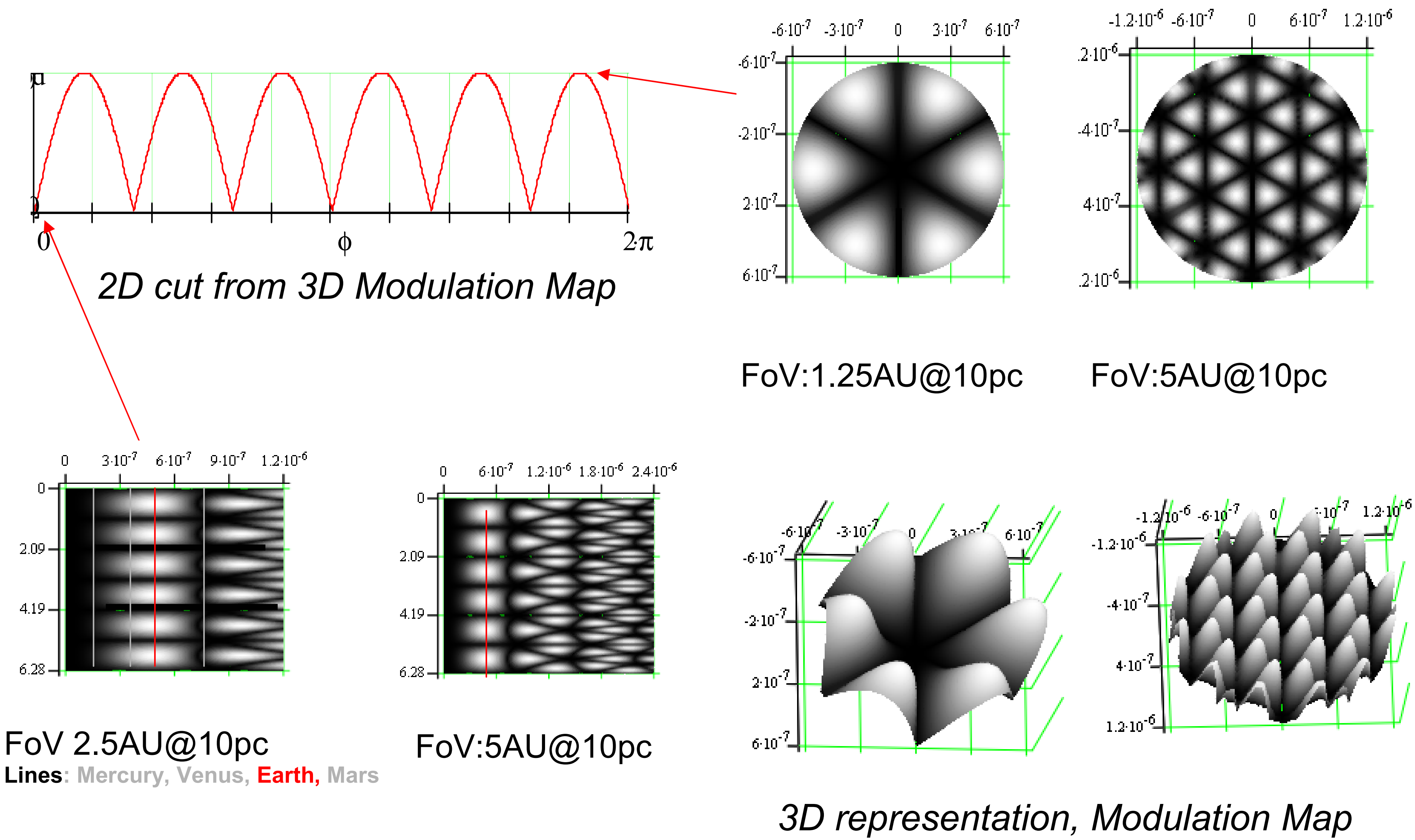

**Fig. 174: Representation of the characteristics of the modulation map of the linear TCB**

| TCB | | | | | | | |
|---|---|---|---|---|---|---|---|
| **Nulling architecture** | Symbols | Aperture index | | | | | |
| | | $k$=1 | 2 | 3 | | | |
| **Aperture location (norm. radius)** | $L_k$ | 1 | 1 | 1 | | | |
| **Aperture angles** | $\delta_k$ | 0 | 2π/3 | 4π/3 | | | |
| **Sub-interferometers, nr., type** | *$N$ = 3 Bracewell ($\theta^2$ null)* | | | | | | |
| Sub-Interferometer 1: amplitudes | $a_{1,k}$ | 1/√2 | 1/√2 | 0 | | | |
| phases | $\psi_{1,k}$ | 0 | 1 | 0 | | | |
| Sub-Interferometer 2: amplitudes | $a_{2,k}$ | 0 | 1/√2 | 1/√2 | | | |
| phases | $\psi_{1,k}$ | 0 | 0 | 1 | | | |
| Sub-Interferometer 3: amplitudes | $a_{2,k}$ | 1/√2 | 0 | 1/√2 | | | |
| phases | $\psi_{1,k}$ | 1 | 0 | 0 | | | |
| **Implementation** | | | | | | | |
| Aperture diameter [m] | $D_k$ | 1 | 1 | 1 | | | |
| Array diameter | $L$ | Variable | | | | | |
| Modulation method | Inherent modulation | | | | | | |
| Beam combination | Image-plane | | | | | | |
| Number of outputs/ports | *3* | | | | | | |
| **Platform** | Free Flyer Interferometer | | | | | | |
| **Science performance** | reduced mission | | | | | | |

# 11 Conclusions

This PhD work focused on two crucial aspects of the DARWIN mission. Firstly, a DARWIN target star list has been established that includes characteristics of the target star sample that will be critical for final mission design, such as, luminosity, distance, spectral classification, stellar variability, multiplicity, location and radius of the star. Constrains were applied as set by planet evolution theory and mission architecture. The catalogue contains nearby stars that might harbor planets that are potentially habitable to complex life. On the basis of theoretical studies, the angular separations of potential habitable planets from their parent stars for the target systems have been established. The resulting target list allows to model realistic observation scenarios for nulling interferometry and translates into architectural constraints for the mission shown for different mission architectures.

Secondly, a number of alternative mission architectures have been evaluated on the basis of interferometer response as a function of wavelength, achievable modulation efficiency, number of telescopes and starlight rejection capabilities. Table 23 shows the evaluation of the alternative mission architectures. Even so the starlight rejection properties of the BOWTIE configuration is outstanding in the comparison, the higher mean modulation efficiency of e.g. the TTN configuration counteracts that superiority. An additional factor in mission design and complexity is the number of telescopes used, here the BOWTIE has a big disadvantage, as it needs 6 telescopes. Its beam combination scheme is also highly complex because it needs 3 steps. Especially architectures with less telescope spacecrafts would allow for bigger telescopes to be launched in a similar launch scenario. Table 23 shows that a reduced science mission goal should be achievable with a lower level of complexity compared to the BOWTIE configuration. If the telescope size is increased from 1.5m diameter to e.g. 3.5m for the TTN mission architecture, the whole science mission goal can be achieved. Further work has to be done e.g. in respect to the beam combiner technologies to verify technological assumptions.

Extrasolar planet search is a very young, rapidly evolving field. A lot of questions need to be answered and a lot of questions still need to be asked, making it for me one of the most fascinating subjects in our century.

*"Good planets are hard to find" (Marcy, 1998)*

but we are working on it…

"With 10 to the 11th stars in our galaxy and 10 to the 9th other galaxies, there are at least 10 to the 20th stars in the universe. Most of them may be accompanied by solar systems. If there are 10 to the 20th solar systems in the universe and the universe is 10 to the 10th years old [..] on the average, a million solar systems are formed in the universe each hour (Bell, 1998)."

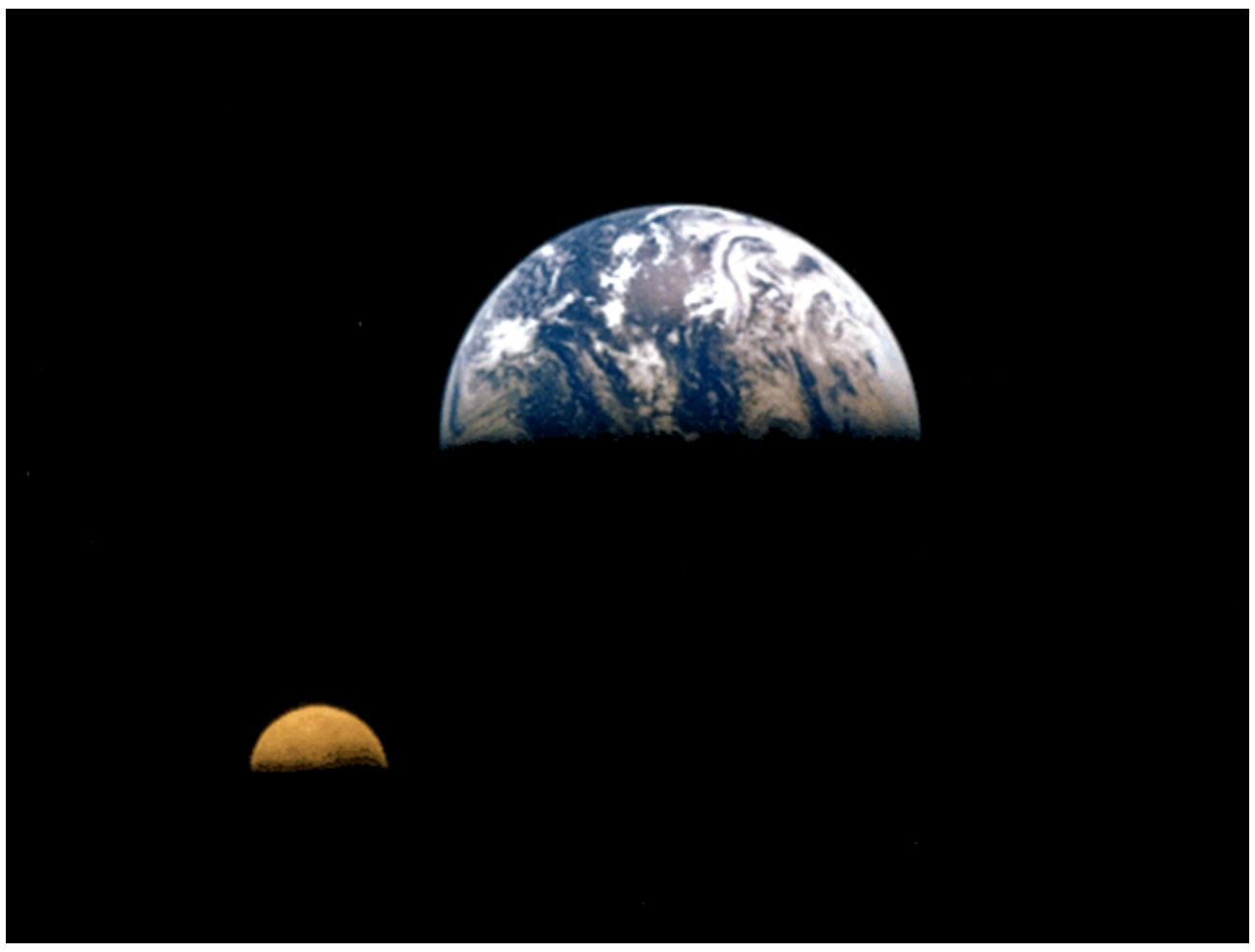

**Useful web-pages:**

http://www.obspm.fr/encycl/encycl.html (news and a profound archive by J. Schneider)
http://cfa-www.harvard.edu/planets (US mirror site)
http://sci.esa.int/science-e/www/area/index.cfm?fareaid=28 (The DARWIN mission)
http://planetquest.jpl.nasa.gov (JPL site, incl. TPF mission)
http://obswww.unige.ch/~udry/planet/planet.html (Geneva Extrasolar Planet Search)
http://exoplanets.org/science.html (California & Carnegie Planet Search)
http://exoplanets.org/science.html (Astrobiology magazine)